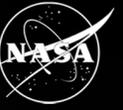

# L U V O I R

## F I N A L   R E P O R T

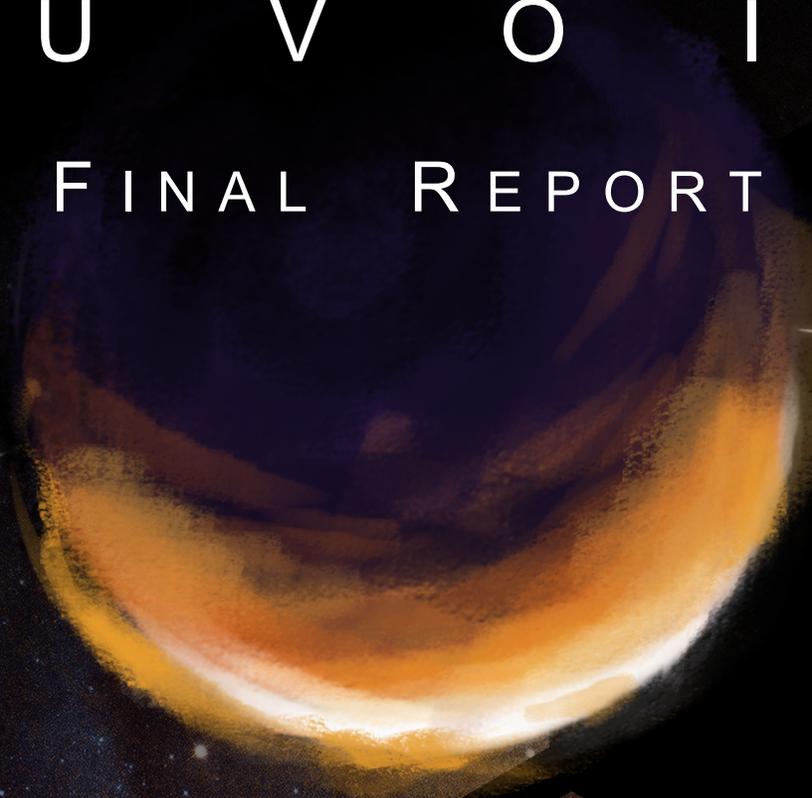

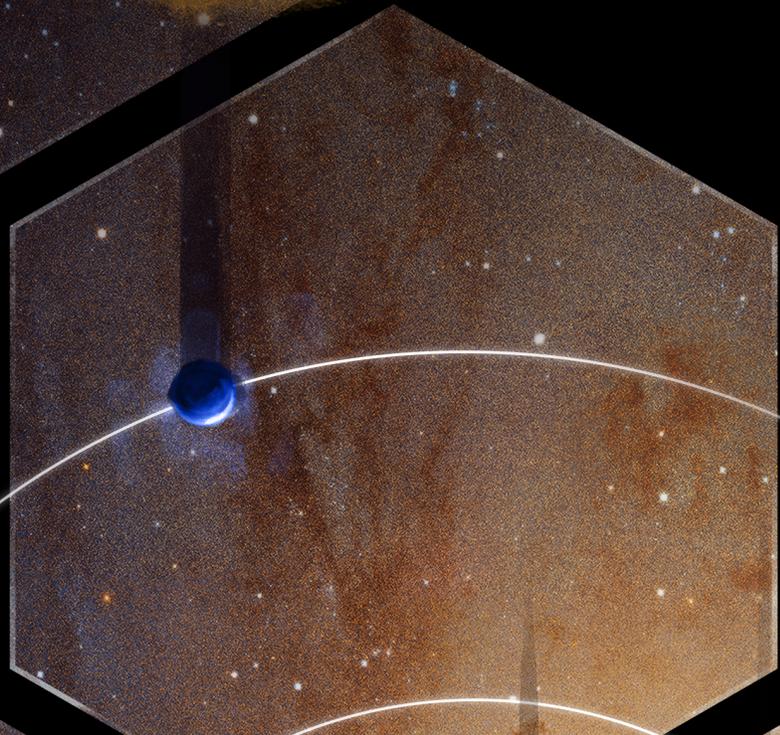

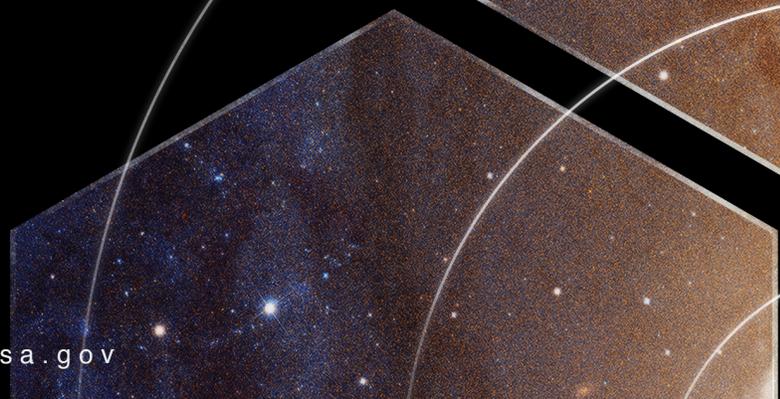

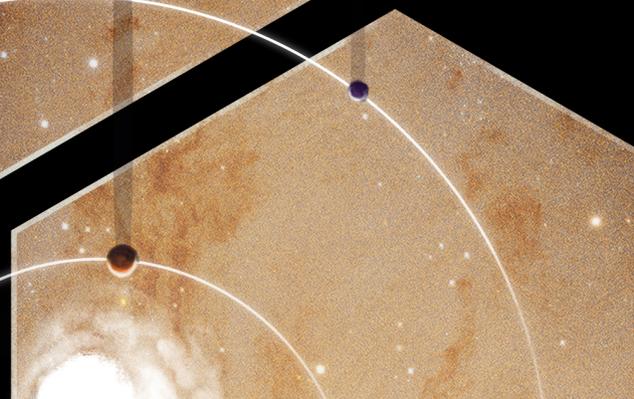

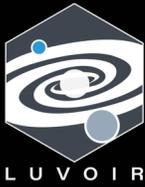

# LARGE UV / OPTICAL / INFRARED SURVEYOR

## TELLING THE STORY OF LIFE IN THE UNIVERSE

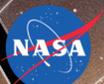

https://asd.gsfc.nasa.gov/luvoir/

## FIND EARTH-LIKE WORLDS AND SEARCH FOR SIGNS OF LIFE

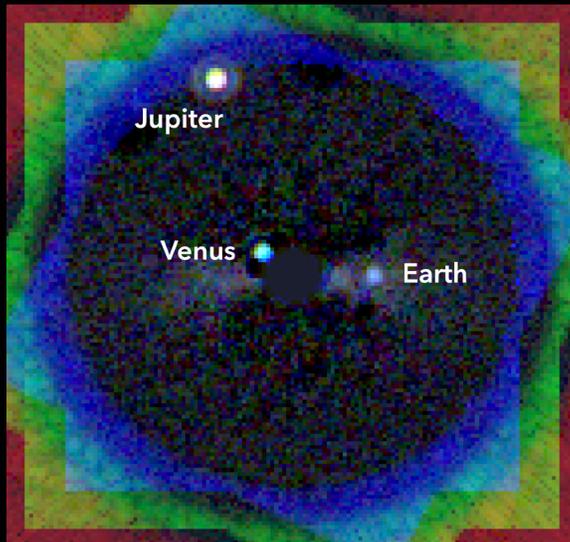

## CHARACTERIZE LARGE NUMBERS OF DIVERSE EXOPLANETS

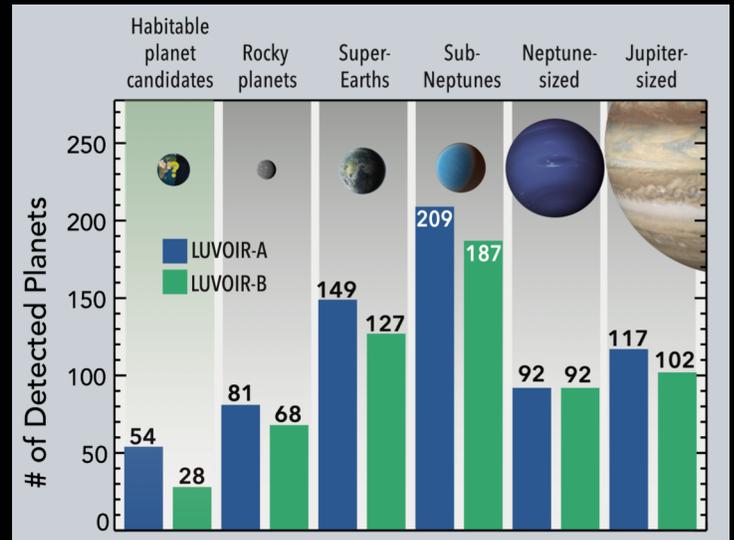

## REVEAL THE FORMATION OF COSMIC STRUCTURE AND THE EVOLUTION OF GALAXIES

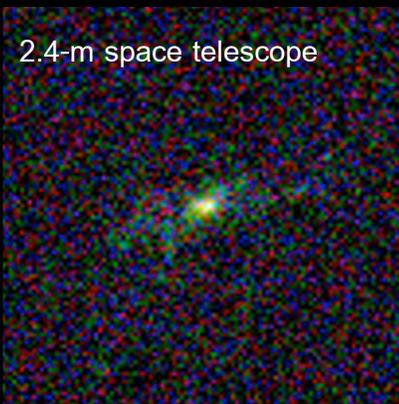

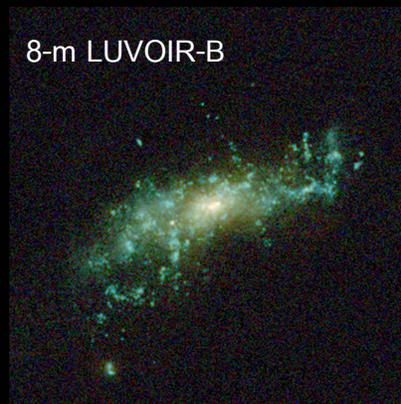

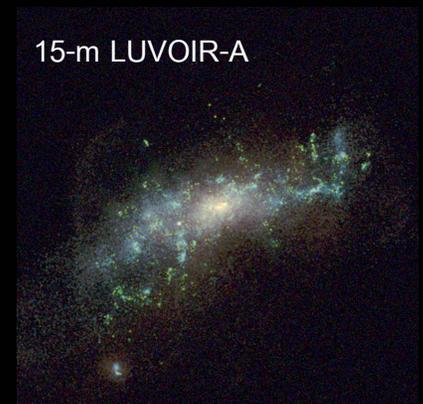

## MONITOR THE SOLAR SYSTEM IN HIGH DEFINITION

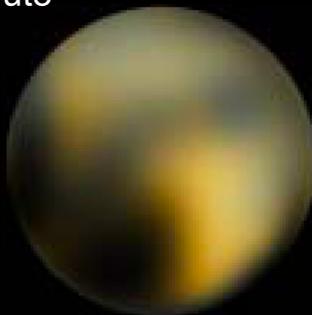

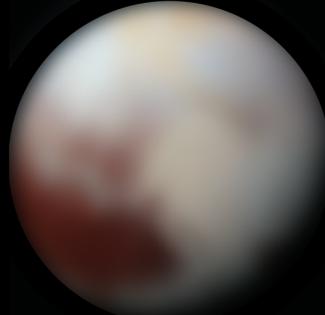

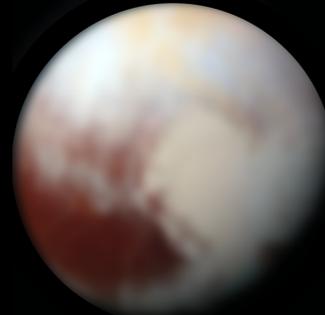

**HST (2.4 m)**
Buie et al. 2010

**LUVOIR-B (8 m)**

**LUVOIR-A (15 m)**

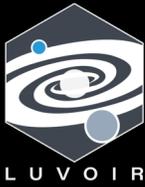
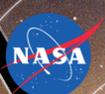

# Large UV / Optical / Infrared Surveyor

## Telling the Story of Life in the Universe

https://asd.gsfc.nasa.gov/luvoir/

## Two powerful and scalable space observatories, responsive to different future landscapes, to answer the questions of the 2030s and beyond

**LUVOIR-A**
15-m

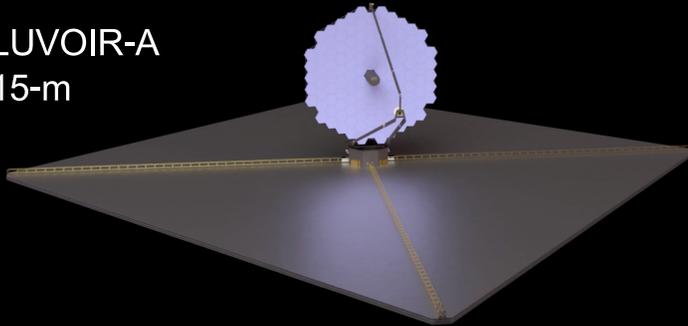

**LUVOIR-B**
8-m

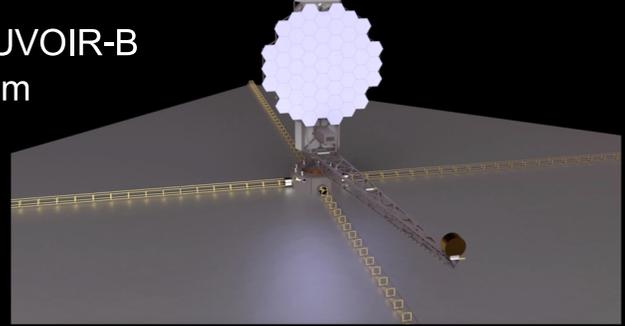

### Observatory Characteristics

Community-driven observing program

Serviceable and upgradable modular design

Sun-Earth L2 orbit

Late 2030s launch date

5-year prime mission; 10 yrs. consumables; 25-year lifetime goal for non-serviceable components

Diffraction limited at 500 nm; 270 K telescope operating temp.

Field-of-regard: Sun-Telescope-Target angles > 45 degrees ($3\pi$ steradians)

Tracking speed: 60 mas/sec (2x JWST)

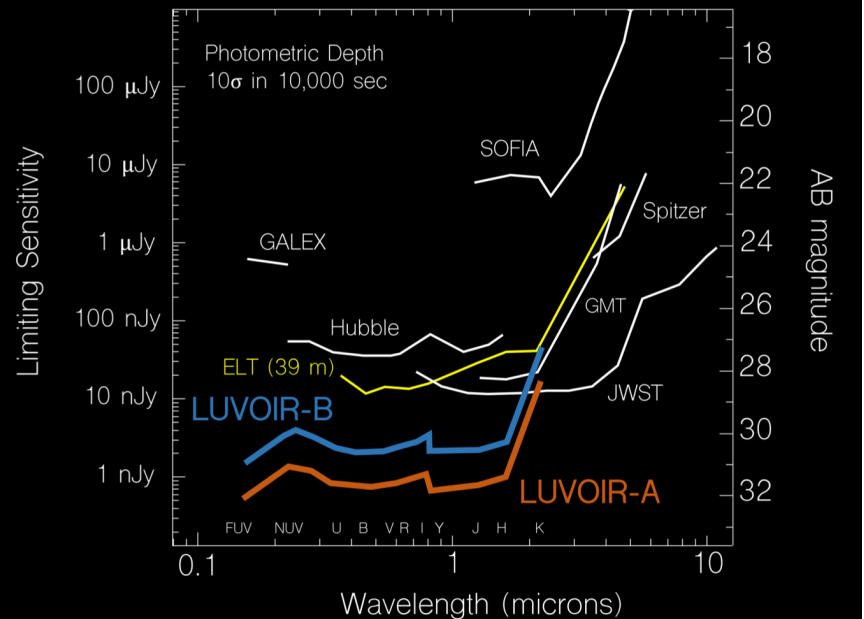

## Candidate Instruments Studied

### ECLIPS

Coronagraph with imaging and imaging spectroscopy

| Bandpass | 200–2000 nm |
|---|---|
| Contrast | $1 \times 10^{-10}$ |
| IWA | 3.5 λ/D |
| OWA | 64 λ/D |
| $R\,(\lambda/\Delta\lambda)$ | Vis: 140 NIR: 70, 200 |

### HDI

Wide field imager with simultaneous UV/Vis and NIR coverage

| Bandpass | 200–2500 nm |
|---|---|
| FoV | 3′× 2′ |
| 67 science filters + grism | |
| Nyquist sampled | |
| High-precision astrometry | |

### LUMOS

UV/Vis multi-object spectrograph and FUV imager

| Bandpass | 100–1000 nm |
|---|---|
| MOS FoV | 2′× 2′ |
| Apertures | 840 × 420 |
| $R\,(\lambda/\Delta\lambda)$ | 500–50,000 |

### POLLUX

Point-source UV spectropolarimeter (European study for LUVOIR-A only)

| Bandpass | 100–400 nm |
|---|---|
| $R\,(\lambda/\Delta\lambda)$ | 120,000 |
| Circular + linear polarization | |

# CONTENTS

**Fact sheet**

**List of figures and tables**

## LUVOIR MISSION CONCEPT STUDY TEAM
*Credit for cover art: NASA GSFC / Adriana Manrique Gutierrez (CI Labs)*

| Science and Technology Definition Team – Voting Members |
| --- |
| Debra Fischer (Yale University), *Community Chair* |
| Bradley Peterson (The Ohio State University), *Community Chair* |
| Jacob Bean (University of Chicago) |
| Daniela Calzetti (University of Massachusetts – Amherst) |
| Rebekah Dawson (Pennsylvania State University) |
| Courtney Dressing (University of California – Berkeley) |
| Lee Feinberg (NASA GSFC) |
| Kevin France (University of Colorado – Boulder), *LUMOS Instrument Lead* |
| Olivier Guyon (University of Arizona) |
| Walter Harris (University of Arizona / LPL), *Solar System Working Group Lead* |
| Mark Marley (NASA Ames), *Exoplanets Working Group Lead* |
| Victoria Meadows (University of Washington) |
| Leonidas Moustakas (JPL) |
| John O'Meara (Keck Observatory), *Cosmic Origins Working Group Lead* |
| Ilaria Pascucci (University of Arizona / LPL) |
| Marc Postman (STScI), *HDI Instrument Lead* |
| Laurent Pueyo (STScI), *ECLIPS Instrument Lead* |
| David Redding (JPL), *Technology Working Group Lead* |
| Jane Rigby (NASA GSFC) |
| Aki Roberge (NASA GSFC), *Study Scientist* |
| David Schiminovich (Columbia University) |
| Britney Schmidt (Georgia Institute of Technology) |
| Karl Stapelfeldt (JPL) |
| Christopher Stark (STScI) |
| Jason Tumlinson (STScI), *Simulations Working Group Lead* |
| Science and Technology Definition Team – International Non-Voting Members |
| Martin Barstow (University of Leicester, UK) |
| Lars Buchhave (DTU Space, National Space Institute, Denmark) |
| Nicolas Cowan (McGill University, Canada) |
| José Dias do Nascimento Jr. (Brazilian Federal University, Brazil) |
| Marc Ferrari (Laboratoire d'Astrophysique de Marseille, France) |
| Ana Gomez de Castro (Universidad Complutense de Madrid, Spain) |
| Kevin Heng (University of Bern, Switzerland) |
| Thomas Henning (Max Planck Institute for Astronomy, Germany) |
| Michiel Min (Netherlands Institute for Space Research, Netherlands) |
| Antonella Nota (STScI, European Space Agency) |
| Takahiro Sumi (Osaka University, Japan) |

## Science and Technology Definition Team – Ex-Officio Non-Voting Members

Shawn Domagal-Goldman (NASA GSFC), *Deputy Study Scientist*

Mario Perez (NASA HQ), *Program Scientist*

Michael Garcia (NASA HQ), *Deputy Program Scientist*

Susan Neff (NASA GSFC), *COR Program Scientist*

Erin Smith (NASA GSFC), *COR Deputy Program Scientist*

## Study Office & Engineering Team

Julie Crooke (NASA GSFC), *Study Manager*

Matthew Bolcar (NASA GSFC), *Lead Engineer*

Jason Hylan (NASA GSFC), *Deputy Lead Engineer*

Giada Arney (NASA GSFC), *Science Support Analysis Team Lead*

Lisa May (Murphian Consulting), *Final Report Manager*

| | |
|---|---|
| Steve Aloezos (NASA GSFC) | Adrienne Beamer (NASA GSFC) |
| Carl Blaurock (Night Sky Systems) | Vince Bly (NASA GSFC) |
| Ginger Bronke (NASA GSFC) | Christine Collins (NASA GSFC) |
| Knicole Colón (NASA GSFC) | James Corsetti (NASA GSFC) |
| Don Dichmann (NASA GSFC) | Jean-Etienne Dongmo (NASA GSFC) |
| Lou Fantano (NASA GSFC) | Thomas Fauchez (NASA GSFC / USRA) |
| Joseph Generie (NASA GSFC) | Gene Gochar (NASA GSFC) |
| Qian Gong (NASA GSFC) | Tyler Groff (NASA GSFC) |
| Kong Ha (NASA GSFC) | William Hayden (NASA GSFC) |
| Andrew Jones (NASA GSFC) | Roser Juanola Parramon (NASA GSFC/UMBC) |
| Ravi Kopparappu (NASA GSFC) | Irving Linares (NASA GSFC) |
| Alice Liu (NASA GSFC) | Eric Lopez (NASA GSFC) |
| Avi Mandell (NASA GSFC) | Bryan Matonak (NASA GSFC) |
| Sang Park (SAO) | Shannon Rodriguez (NASA GSFC) |
| Lia Sacks (NASA GSFC) | Lisa Smith (NASA MSFC) |
| Hari Subedi (NASA GSFC/Princeton) | Steve Tompkins (NASA GSFC) |
| Geronimo Villanueva (NASA GSFC) | Garrett West (NASA GSFC) |
| Dewey Willis (NASA GSFC) | Kan Yang (NASA GSFC) |
| Neil Zimmerman (NASA GSFC) | |

## POLLUX Team

Coralie Neiner (Observatoire de Meudon, France), *Co-Lead*

Jean-Claude Bouret (Laboratoire d'Astrophysique de Marseille, France), *Co-Lead*

Stéphane Charlot (IAP, France), *Extragalactic Working Group Lead*

Jean-Yves Chauffray (LATMOS, France), *Solar System Working Group Lead*

Chris Evans (University of Edinburgh, UK), *Stellar Physics Working Group Co-Lead*

Luca Fossati (Space Research Institute, Austria), *Exoplanet Working Group Lead*

Ana Gomez de Castro (Comp. U of Madrid, Spain), *Stellar Physics Working Group Co-Lead*

| Cecile Gry (LAM, France), *ISM/IGM Working Group Lead* | |
| Pasquier Noterdaeme (IAP, France), *Cosmology Working Group Lead* | |
| Eduard Muslimov (LAM, France), *Lead Engineer, Optical Designer* | |
| Arturo Lopez Ariste (IRAP, France), *Polarimeter Designer* | |
| Martin Barstow (University of Leicester, UK), *Detector Working Group Manager* | |
| David Montgomery (UK Astronomy Technology Center, UK), *Mechanical Designer* | |
| Pierre Royer (Catholic University of Leuven, Belgium), *Calibration Unit Manager* | |
| Udo Schuehle (MPS, Germany), *Coatings Working Group Manager* | |
| Louise Lopes (CNES, France), *CNES Manager* | |
| Marc Ferrari (LAM, France), *CNES Representative on LUVOIR STDT* | |
| Dietrich Baade (ESO, Germany) | Richard Desselle (CSL/U of Liège, Belgium) |
| Boris Gäensicke (U of Warwick, UK) | Miriam Garcia (CAB/CSIC, Madrid, Spain) |
| Serge Habracken (U of Liège, Belgium) | Jon Lapington (University of Leicester, UK) |
| Vianney Lebouteiller (AIM/CEA, France) | Maelle Le Gal (LESIA, France) |
| Frédéric Marin (Strasbourg Obs., France) | Fabrice Martins (U of Montpelier, France) |
| Yael Nazé (University of Liège, Belgium) | Hadi Rahmani (GEPI, France) |
| Hugues Sana (Catholic U of Leuven, Belgium) | Steve Shore (University of Pisa, Italy) |
| Daphne Stam (TU Delft, Netherlands) | Luca Teriaca (MPS, Germany) |
| Jorick Vink (Armagh Observatory, UK) | Huirong Yan (DESY, Germany) |
| **Community Working Group Members / Additional Contributors** | |
| **Cosmic Origins** | |
| Nate Bastian (Liverpool John Moores U) | Brendan Bowler (Univ. of Texas – Austin) |
| Charlie Conroy (Harvard University) | Paul Crowther (University of Sheffield) |
| Selma de Mink (University of Amsterdam) | Ruobing Dong (University of Arizona) |
| Bruce Elmegreen (IBM) | Steven Finkelstein (U of Texas – Austin) |
| Brian Fleming (U of Colorado – Boulder) | Andrew J. Fox (STScI) |
| Michele Fumagalli (Durham University) | Jay Gallagher (U of Wisconsin – Madison) |
| Melissa Graham (U of California – Berkeley) | Eva Grebel (University of Heidelberg) |
| Gregory Herczeg (KIAA/University of Peking) | Benne Holwerda (University of Louisville) |
| Christopher Howk (Univ. of Notre Dame) | Shotara Kikuchihara (Univ. of Tokyo) |
| Mark Krumholz (Australian National U) | Søren S. Larsen (Radboud University) |
| Tod Lauer (NOAO) | Janice Lee (IPAC) |
| F. Jay Lockman (Green Bank) | Yoshiki Matsuoka (Ehime University) |
| Stephan McCandliss (Johns Hopkins U) | Stella Offner (University of Texas – Austin) |
| Masami Ouchi (Univ. of Tokyo) | Ian Roderer (University of Michigan) |
| Elena Sabbi (STScI) | P. Christian Schneider (U of Hamburg) |
| Vicky Scowcroft (University of Bath) | Paul Scowen (Arizona State University) |
| Warren Skidmore (Thirty Meter Telescope) | Linda Smith (ESA/STScI) |
| Russell Smith (Durham University) | Rachel Somerville (Rutgers University) |
| Harry Teplitz (IPAC) | Christy Tremonti (U of Wisconsin – Madison) |
| Kate Whitaker (U of Mass – Amherst) | Gerard Williger (University of Louisville) |
| Rosemary Wyse (Johns Hopkins University) | Allison Youngblood (NPP/NASA GSFC) |

### Exoplanets

| | |
|---|---|
| Dorian S. Abbot (Univ of. Chicago) | Eric Agol (University of Washington) |
| Daniel Apai (University of Arizona) | Natalie Batalha (NASA Ames) |
| Thomas G. Beatty (U. Arizona) | Ruslan Belikov (NASA Ames) |
| Svetland Berdyugina (Leibniz Institute) | David Catling (University of Washington) |
| Jade Checlair (Univ. of Chicago) | Emilio Enriquez (UC Berkely) |
| Keigo Enya (Japan Aerospace Explor. Agency) | Y. Katherina Feng (UC Santa Cruz) |
| Luca Fossati (Austrian Academy of Sciences) | Yuka Fujii (Tokyo Institute of Technology) |
| Hideki Fujiwara (Subaru) | Ryan Garland (Oxford) |
| Claire Marie Guimond (McGill University) | Jacob Haqq-Misra (BMSIS) |
| Eric Hébrard (University of Exeter) | Masahiro Ikoma (Univ of Tokyo) |
| Patrick Irwin (Oxford) | Tiffany Jansen (Columbia University) |
| Shingo Kameda (Rykkyo Univ.) | Eliza Kempton (U of Maryland – College Park) |
| Tadayuki Kodama (Subaru) | Joshua Krissansen-Totton (UC Santa Cruz) |
| Brianna Lacy (Princeton University) | Andrew Lincowski (University of Washington) |
| Jacob Lustig-Yeager (Univ. of Washington) | Timothy Lyons (U of California – Riverside) |
| Johan Mazoyer (JPL) | Gijs Mulders (LPL) |
| Gen Murakami (Inst. Of Space & Astro Sci) | Norio Narita (National Astro Obs of Japan) |
| Stephanie Olson (Univ. of Chicago) | N. Osada (Rikkyo Univ.) |
| Daria Pidhorodetska (UMBC) | Ramses Ramirez (Tokyo Institute of Tech) |
| Christopher Reinhard (Georgia Tech) | Tyler Robinson (Northern Arizona University) |
| Edward Schwieterman (U of CA – Riverside) | Stoney Simons (National Institutes of Health) |
| Dillon Teal (UMD) | N. Terada (Subaru) |
| Guadalupe Tovar (University of Washington) | Sara Walker (Arizona State University) |
| Robert J. Webber (New York Univ.) | Diana Windemuth (Univ. of Washington) |

### Solar System

| | |
|---|---|
| Alvaro Alvarez-Candal (Obs. Nacional Brazil) | Gerbs Bauer (JPL) |
| Dennis Bodewits (U of MD – College Park) | Richard Cartwright (SETI Institute) |
| Valeria Cottini (NASA GSFC) | Lori Glaze (NASA HQ) |
| Jeff Morgenthaler (PSI) | Marc Neveu (NASA GSFC) |
| Lucas Paganini (NASA GSFC) | Alex Parker (SWRI) |
| Noah Petro (NASA GSFC) | Noemi Pinilla-Alonso (U of Central Florida) |
| Silvia Protopapa (U of MD – College Park) | Andrew Rivkin (JHU APL) |
| Sara Walker (Arizona State University) | |

### Technology

| | |
|---|---|
| Lynn Allen (Harris Corporation) | David Allred (Brigham Young University) |
| Jon Arenberg (Northrop Grumman) | Kunjithapatham Balasubramanian (JPL) |
| Allison Barto (Ball Aerospace) | Scott Basinger (JPL) |
| Ruslan Belikov (NASA Ames) | Ray Bell (Lockheed Martin) |
| Paul Bierdan (Boston Micromachines) | Ron Broccolo (Harris Corporation) |
| Eric Cady (JPL) | Kerri Cahoy (MIT) |
| Jeff Cavaco (Northrop Grumman) | Alberto Conti (Ball Aerospace) |
| Jim Contreras (Ball Aerospace) | Laura Coyle (Ball Aerospace) |
| Brendan Crill (JPL) | Javier Del Hoyo (NASA GSFC) |
| Larry Dewell (Lockheed Martin) | Jeanette Domber (Ball Aerospace) |
| Matthew East (Harris Corporation) | Mike Eisenhower (SAO) |
| Michael Feinberg (Boston Micromachines) | Greg Fellers (Lockheed Martin) |
| James R. Fienup (University of Rochester) | Don Figer (Rochester Institute of Technology) |
| Brian Fleming (U of Colorado – Boulder) | Kevin Fogarty (STScI) |
| Thomas Greene (NASA Ames) | Pascal Hallibert (ESA) |
| Erika Hamden (Caltech) | Alex Harwit (Ball Aerospace) |
| Michael Helmbrecht (Iris AO) | John Hennessy (JPL) |
| Peter Hill (Zygo / Ametek) | Joe Ho (Ball Aerospace) |
| Sona Hosseini (JPL) | Tony Hull (University of New Mexico) |
| Jeffrey Jewell (JPL) | Jeremy Kasdin (Princeton University) |
| Scott Knight (Ball Aerospace) | Mary Li (NASA GSFC) |
| Paul Lightsey (Ball Aerospace) | Chris Lindensmith (JPL) |
| Sarah Lipscy (Ball Aerospace) | John Lou (JPL) |
| Makenzie Lystrup (Ball Aerospace) | Taro Matsuo (Osaka University) |
| Gary Matthews (ATA Aerospace) | Stephan McCandliss (Johns Hopkins U) |
| Ted Mooney (Harris Corporation) | Dustin Moore (JPL) |
| Didier Morancais (Airbus) | Mamadou N'Diaye (Université Côte d'Azur) |
| Shouleh Nikzad (JPL) | Joel Nissen (JPL) |
| Alison Nordt (Lockheed Martin) | Jim Oschmann (Ball Aerospace) |
| Sang Park (SAO) | Enrico Pinna (Istituto Nazionale di Astrofisica) |
| Bill Purcell (Ball Aerospace) | Manuel Quijada (NASA GSFC) |
| Bernie Rauscher (NASA GSFC) | Norman Rioux (NASA GSFC) |
| Michael Rodgers (Synopsys/JPL) | Garreth Ruane (Caltech) |
| Derek Sabatke (Ball Aerospace) | John Sadleir (NASA GSFC) |
| Babak Saif (NASA GSFC) | Eric Schindhelm (Ball Aerospace) |
| Gene Serabyn (JPL) | Stuart Shaklan (JPL) |
| Michael Shao (JPL) | Chris Shelton (JPL) |
| Fang Shi (JPL) | Evgenya Shkolnik (Arizona State University) |
| Nick Siegler (JPL) | Remi Soummer (STScI) |
| Joe Sullivan (Ball Aerospace) | Phil Stahl (NASA MSFC) |
| Kathryn St Laurent (STScI) | Kiarash Tajdaran (Lockheed Martin) |
| John Trauger (JPL) | Steve Turley (Brigham Young University) |

| John Vayda (Northrop Grumman) | Michael Werner (JPL) |
| Scott Will (University of Rochester) | Marco Xompero (Nat Inst of Astro, Italy) |
| Toru Yamada (JAXA) | Tomoyasu Yamamuro (OptCraft) |
| David Yanatsis (Harris Corporation) | John Ziemer (JPL) |

**Acknowledgements**

The STDT would like to extend special thanks to the following people and groups who helped make this report possible.

- For their major contributions to the science cases: Steven Finkelstein (University of Texas – Austin), Luca Fossati (Austrian Academy of Sciences), Jacob Lustig-Yeager (University of Washington), Tyler Robinson (Northern Arizona University), and Diana Windemuth (University of Washington)

- For their extensive participation in instrument design work: Brian Fleming (University of Colorado – Boulder), John MacKenty (STScI), and Stephan McCandliss (Johns Hopkins University)

- For their efforts in developing early instrument and observatory concepts: The Goddard Integrated Design Center teams, led by Jennifer Bracken, Tammy Brown, Rachel Rivera, James Sturm, and Frank Kirchmann

- For their diligent review and valuable feedback: The Interim and Final Report Red Teams, chaired by Jacqueline Townsend (NASA GSFC)

- For their hard work laying out this document, creating graphics, and writing data simulation tools: Pat Tyler, Jay Friedlander, T. Britt Griswold (NASA GSFC), Chad Smith, and Graham Kanarek (STScI)

- We are very grateful to our cooperative agreement partners: Ball Aerospace, Harris Corporation, Lockheed Martin, and Northrop Grumman

- Finally, we thank B. Scott Gaudi (Ohio State University) for his unflagging efforts to promote cooperation and collaboration between the LUVOIR and HabEx study teams

# LIST OF FIGURES AND TABLES

## FIGURES

### Chapter 1.  LUVOIR final report summary

### Chapter 2.  Roadmap to the report

### Chapter 3.  Is there life elsewhere? Habitable exoplanets and solar system ocean worlds

## Chapter 4. How do we fit in? Comparative planetary science

## Chapter 5. What are the building blocks of cosmic structure?

## Chapter 6.  How do galaxies evolve?

## Chapter 7.  Science mission traceability

## Chapter 8.  Observatory segment

## TABLES

### Chapter 1.  LUVOIR final report summary

### Chapter 2.  Roadmap to the report

### Chapter 3.  Is there life elsewhere? Habitable exoplanets and solar system ocean worlds

### Chapter 4.  How do we fit in? Comparative planetary science

### Chapter 5.  What are the building blocks of cosmic structure?

## Chapter 6.   How do galaxies evolve?

## Chapter 7.   Science mission traceability

## Chapter 8.   Observatory segment

## Chapter 9.   Mission operations and ground segment

## Chapter 10. Launch vehicle ..........................................................10-1

## Chapter 11. Technology development ...............................................11-1

## Chapter 12. Management and systems engineering ....................................12-1

## CHAPTER 1. LUVOIR FINAL REPORT SUMMARY

Humanity is driven by the quest to know about the world around us, an endeavor of immeasurable value to our species. Ages-old questions and investigations earned us the revelations that the "wanderers" in the night sky are other worlds and the stars are Suns swirling in a vast galaxy, itself one of a myriad of islands in an expanding cosmos. Now we have crossed another threshold of discovery: there are planets around other stars (e.g., Mayor & Queloz 1995). At this key point in human history, tracing a path from the dawn of the universe to life-bearing worlds is within our grasp.

This monumental objective demands powerful and flexible new tools, and different ways of combining scientific skills. The abundance and diversity of worlds is far greater than imagined (e.g., Winn & Fabrycky 2015), but the vast majority of known exoplanets are "small black shadows" indirectly detected through their effects on their host stars. Our knowledge of exoplanet properties is largely limited to orbits, masses, and sizes. Astronomers have just begun to measure the atmospheres of giant exoplanets; such studies will greatly expand in the coming years.

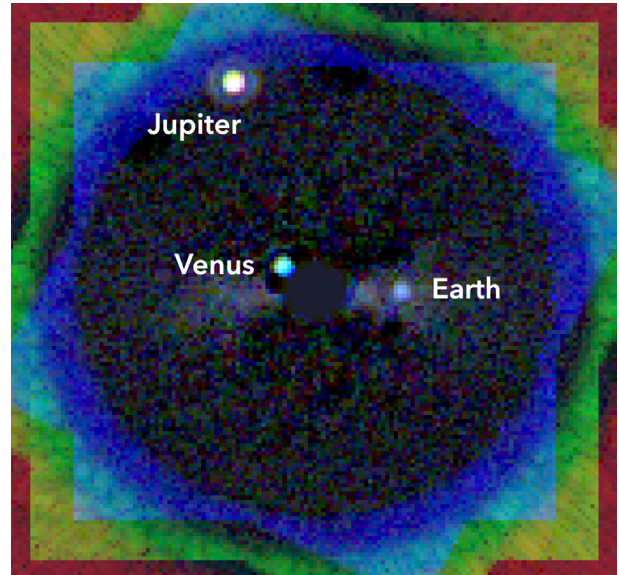

**Figure 1-1.** *Imaging another Earth. Simulation of the inner solar system in visible light viewed from a distance of 12.5 parsec with the 15-m LUVOIR-A space telescope concept. The enormous glare from the central star has been suppressed with a coronagraph so the faint planets can be seen. Each planet's atmosphere can be probed with direct spectra to reveal its composition. The simulation assumes realistic noise sources, wavefront errors, and post-processing. Credit: R. Juanola Parramon, N. Zimmerman, A. Roberge (NASA GSFC)*

The next frontier is to extend characterization capabilities to rocky planets and find the "pale blue dots" in the solar neighborhood (**Figure 1-1**). With a large enough sample size, scientists can determine whether habitable, Earth-like conditions are rare or common on nearby worlds and then probe them for signs of life. Focusing on the planetary systems most like the solar system, those with Earth-size exoplanets orbiting Sun-like stars, increases the chances of finding and recognizing biosignatures. Concurrently, we will nurture a new discipline—comparative exoplanetology—by studying a huge range of exoplanets, thereby gaining invaluable information for placing our own system in a broader context. A vital part of establishing that context is deeper understanding of the bodies within the solar system.

Our drive to know goes beyond asking the question "what exists?" to "why does it exist?" and pushes us to understand the origins of all we see around us. The boundary of what we can see now stretches all the way to the dawn of the universe, but like our first steps in the study of new worlds, our current view lacks completeness, precision, and context. We seek to understand the processes and environments that gave rise to a life-supporting universe: from the formation of the earliest structures, to the assembly and evolution of galaxies, to the detailed mechanisms of star and planet formation. The boundaries of physics will be tested while exploring the birth and evolution of the cosmos.

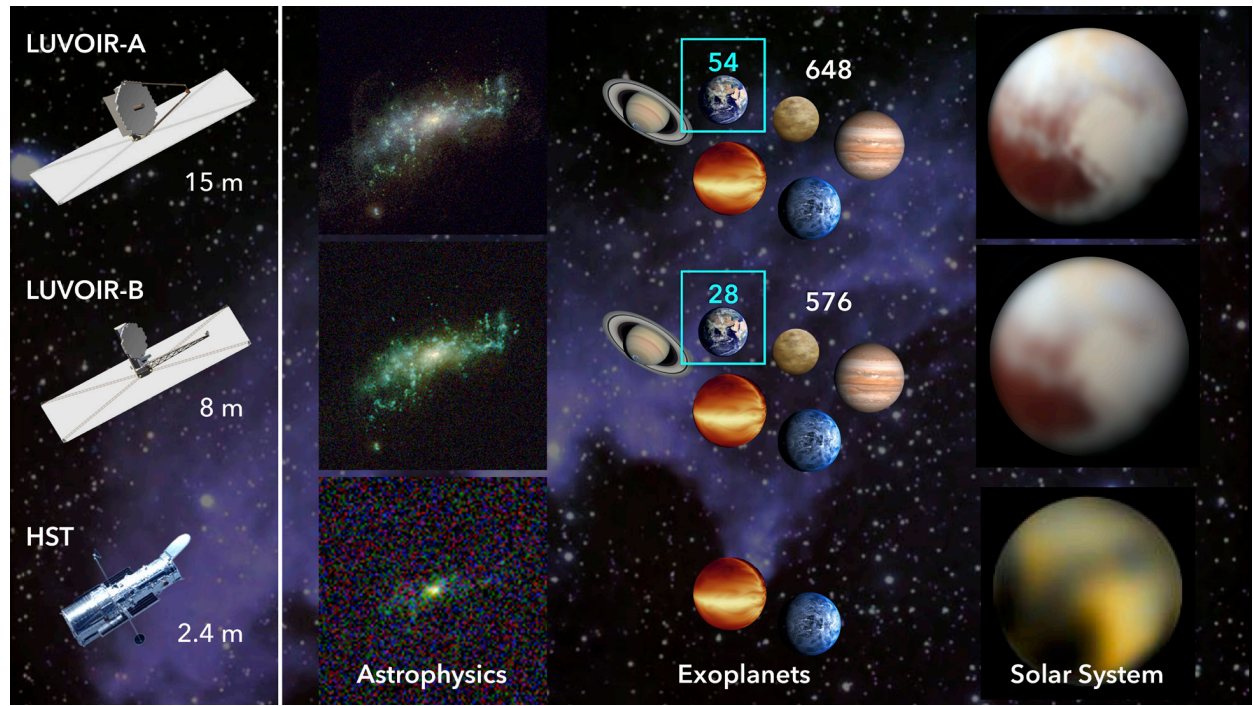

**Figure 1-2.** *LUVOIR will revolutionize huge areas of space science. Its sensitivity and spatial resolution open the door to the ultra-faint and ultra-distant regime, enabling detailed observations of the full variety of galaxies. LUVOIR will dramatically increase the sample size and diversity of exoplanets that can be studied, providing dozens of Earth-like exoplanet candidates that can be probed for signs of life (54 with LUVOIR-A and 28 with LUVOIR-B) and hundreds of non-habitable exoplanets (648 with LUVOIR-A and 576 with LUVOIR-B). Finally, LUVOIR will provide near-flyby quality observations of solar system bodies. Hubble Space Telescope (HST) Pluto image from Buie et al. 2010. Credits: NASA / New Horizons / M. Postman (STScI) / A. Roberge (NASA GSFC)*

This report delves into far greater detail on all these scientific investigations, which demand an observatory beyond any in existence or in development. They require a large aperture to capture the very faintest objects and study their structure at the ~100 pc scale across cosmic time. They require the ability to block the blinding light from hundreds of stars and observe dozens of small, faint planets orbiting them. They require access to a range of wavelengths broad enough to read the fingerprints of matter across all temperatures and densities. These requirements define the necessary tool, the Large Ultraviolet / Optical / Infrared Surveyor (LUVOIR).

LUVOIR is a vision for a community facility with the power and flexibility to answer scientists' questions across the full portfolio of Exoplanet, Cosmic Origins, Physics of the Cosmos, and Solar System Exploration science (**Figure 1-2**). LUVOIR's main features are:

- Two distinct concepts spanning a wide range of design space (**Figure 1-3**)

- Segmented, deployable telescopes each designed for a single launch

- Scalable architecture for a serviceable observatory at Sun-Earth L2

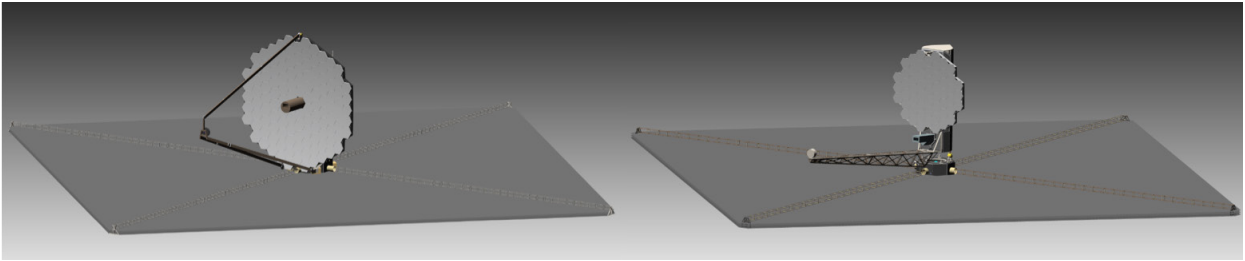

**Figure 1-3.** *The LUVOIR observatory concepts. LUVOIR-A (left) has a 15-m diameter on-axis primary mirror and four instrument bays. LUVOIR-B (right) has an 8-m diameter off-axis primary mirror and three instrument bays. Animations of observatory deployment and pointing may be viewed at* https://asd.gsfc.nasa.gov/luvoir/design/. *Credit: A. Jones (NASA GSFC)*

- UV-capable telescopes that are compatible with high-constrast exoplanet observations (total wavelength range of 100–2500 nm; diffraction limited at 500 nm)

- Extreme Coronagraph for LIving Planetary Systems (ECLIPS): A high-contrast coronagraph with imaging cameras and integral field spectrographs spanning 200–2000 nm, capable of directly observing a wide range of exoplanets

- High Definition Imager (HDI): A high spatial resolution camera covering 200–2500 nm, incorporating high precision astrometry capability

- LUVOIR Ultraviolet Multi-Object Spectrograph (LUMOS): A far-UV imager and multi-resolution, multi-object spectrograph covering 100–1000 nm, capable of simultaneous observations of up to hundreds of sources

- POLLUX: A high-resolution, point-source UV spectropolarimeter covering 100–400 nm, designed for LUVOIR-A. This instrument study was contributed by a consortium of European institutions, with leadership and support from the French Space Agency

A transformative facility like LUVOIR with its large aperture and need for high-contrast observations of Earth-like exoplanets requires new approaches to designing and realizing an observatory. Far more than previous space telescopes, LUVOIR must be considered as an integrated system and was designed for adaptability in multiple ways:

- Scalable architecture to leverage a rapidly changing launch vehicle landscape

- Modular design to facilitate on-orbit servicing and integration & testing on the ground

- Several layers of active wavefront control on the telescope and within instruments to relax constraints on hardware precision and on-the-ground verification

- Minimization and isolation of vibrations and thermal disturbances throughout the observatory

Development of some key technologies, including coronagraph design, wavefront sensing and control, and ultra-stable structures, will also be needed. With input from NASA, the aerospace industry, and other partners, this study has identified the technological gaps and created a plan to close them before approval for mission formulation (start of Phase A). The technology development plan takes advantage of:

- Segmented mirror technologies from the ground and the James Webb Space Telescope (JWST)

- Coronagraph development from the ground and the Wide-Field Infrared Survey Telescope (WFIRST)

- UV instrument and mirror coating technologies developed via sub-orbital payloads and smallSats

LUVOIR is an unquestionably ambitious project and, like past Great Observatories, it will come with unique management challenges. The LUVOIR Study Team has developed a plan to meet those challenges—and reduce cost and schedule risk—using a combination of strategies, including:

- A well-funded, coordinated plan for activities prior to Phase A start, which will develop LUVOIR's technologies and architecture in concert

- A Science Steering Committee that will ensure timely creation of well-defined science requirements and guard against late changes

- Suggestions for changes to project management and funding policies to increase effectiveness and efficiency

LUVOIR has its genesis in the 2013 NASA Astrophysics Roadmap "Enduring Quests, Daring Visions." The current study builds upon earlier telescope concepts (e.g., ATLAST, High Definition Space Telescope) but goes far beyond them in both scientific and technical detail. LUVOIR will provide the powerful observing capabilities required to execute the revolutionary investigations discussed here. Furthermore, just as the Hubble Space Telescope (Hubble or HST) is today doing science not envisioned at the time of its design or launch, LUVOIR's power, flexibility, and longevity will allow it to execute the as-of-yet unknown science of the 2040s and beyond.

## 1.1  How to use this report

The LUVOIR Study Team has organized this report in tiered layers of detail as an aid to the reader. The first level starts with an extended summary (**Chapter 1**), providing an overview of the entire Final Report. **Chapter 2** contains a roadmap to the report that identifies the locations of key material. In the next level, **Chapters 3**–**6** more fully describe the science cases; while **Chapters 7**–**12** present engineering details, the technology development plan, and the management strategy. **Chapter 13** contains the science goals, design, and technology for the European-designed POLLUX instrument. The third level provides supporting information and calculations (**Appendices A**–**J**); these documents are available in a separate PDF document. ***Orange buttons within Chapter 1 link the reader to the relevant explanatory sections.***

The scope of science enabled by LUVOIR is truly vast and multi-disciplinary, encompassing all the topics addressed by Hubble and more. The LUVOIR Science and Technology Definition Team (STDT) therefore decided to focus on a set of "Signature Science Cases" (**Table 1-1**). In **Chapters 3–6**, we explain the motivations for the Signature Science Cases, identify key measurements, and set out needed observations. The telescope and instrument characteristics required for the observations are also identified. We have developed concrete observing programs for each Signature Science Case to ensure that the LUVOIR designs can execute this compelling science within the 5-year prime mission lifetime (**Appendix B**).

The Signature Science Cases represent some of the most compelling types of observing programs that scientists could do with LUVOIR at the limits of its performance. As compelling as they are, they should not be taken as a complete specification of LUVOIR's future scientific potential. Additional science cases contributed by the LUVOIR STDT and the broader community appear in **Appendix A**. We fully expect that our creative community, empowered by the revolutionary capabilities of LUVOIR, will ask questions, acquire data, and solve problems far beyond those discussed here.

**Table 1-1.** *LUVOIR's Signature Science Cases*

| Report Section | Signature Science Case |
|---|---|
| 3.2 | # 1 – Finding habitable planet candidates |
| 3.3 | # 2 – Searching for biosignatures and confirming habitability |
| 3.5 | # 3 – The search for habitable worlds in the solar system |
| 4.1 | # 4 – Comparative atmospheres |
| 4.2 | # 5 – The formation of planetary systems |
| 4.3 | # 6 – Small bodies in the solar system |
| 5.1 | # 7 – Connecting the smallest scales across cosmic time |
| 5.2 | # 8 – Constraining dark matter using high precision astrometry |
| 5.3 | # 9 – Tracing ionizing light over cosmic time |
| 6.1 | # 10 – The cycles of galactic matter |
| 6.2 | # 11 – The multiscale assembly of galaxies |
| 6.3 | # 12 – Stars as the engines of galactic feeback |

## 1.2 LUVOIR frequently asked questions

The following boxes contain short answers to a number of frequently asked questions (FAQs) about LUVOIR. For most FAQs, the location in the report where the reader can find more information is identified.

## 1.3 The hunt for other Earths and life

Chapter 3

We have long speculated about the existence of planets outside the solar system. The names of imagined worlds are familiar to us: Tatooine, Vulcan, Dune. Our own planetary system is largely bimodal in nature, with small rocky planets close to the Sun and massive gas giants in the cold outer reaches. Scientists naturally expected that other systems would be similar—if they existed at all.

In the mid-1990s, decades of persistence and technological innovation were finally rewarded with the discovery of the first exoplanets. We have been astonished and delighted to find that exoplanets are common and diverse—reality far surpassed our scientific predictions. We can now look into the sky and realize that most of the stars we see harbor worlds. Soon, we will more fully reveal the character of giant planets, which will further stretch the bounds of our theories. The next steps in this quest are even more amazing.

## LUVOIR FAQs (Part 1 of 3)

**How big is LUVOIR?**
We have developed two concepts, LUVOIR-A and LUVOIR-B. LUVOIR-A has an on-axis primary mirror with a diameter of 15 meters. This size corresponds to the largest observatory that can be launched in a SLS Block 2 vehicle. LUVOIR-B has an off-axis primary mirror with a diameter of 8 meters, which is the largest observatory that can be launched in a 5-meter fairing similar to those in use today. [**Chapters 7** & **8**]

**What if the SLS Block 2 launch vehicle is not available in the 2030s?**
LUVOIR's segmented telescope architecture facilitates scalability, as demonstrated with LUVOIR-B. LUVOIR can take advantage of future opportunities in heavy lift launch vehicles with large fairings (e.g., NASA's SLS Block 1B Cargo, Blue Origin's New Glenn rocket, and SpaceX's Starship). [**Chapter 10**]

**When will LUVOIR launch?**
In the mission development schedules created for this study, the total time from approval for formulation (start of Phase A) to launch & commissioning (end of Phase D) is 15.6 years for LUVOIR-A and 15.3 years for LUVOIR-B. With an assumed Phase A start in Jan 2025, LUVOIR-A would launch in late 2039 and LUVOIR-B in mid 2039. [**Chapter 12**]

**How long will LUVOIR last?**
The prime mission is 5 years long, the standard prime mission duration for a large space mission, with 10 years of on-board consumables. LUVOIR is designed to be serviceable and upgradable, with a lifetime goal of 25 years for non-serviceable components. [**Chapter 7**]

**How much of LUVOIR's time will be for community observers?**
LUVOIR is envisioned as a facility in the tradition of NASA's Great Observatories (Hubble, Compton, Chandra, Spitzer), with guest observer programs of all sizes.

**What are LUVOIR's key science goals?**
This report describes a number of "Signature Science Cases" that demand LUVOIR. We also collected community input on additional LUVOIR science and have developed a number of public on-line tools to help create science programs (https://asd.gsfc.nasa.gov/luvoir/tools/). Finally, LUVOIR is designed to be flexible and powerful enough to enable the as-of-yet unknown science of the 2040s and beyond. [**Chapters 3–6**, **Appendix A**]

**What is LUVOIR's primary technological challenge?**
Achieving the wavefront stability needed for high contrast direct observations of Earth-like exoplanets using a large telescope. LUVOIR has been designed with this in mind and a detailed plan to mature technologies has been developed. [**Chapters 7**, **8**, & **11**]

### LUVOIR FAQs (Part 2 of 3)

**What unique capabilities will LUVOIR offer compared to future ground-based observatories?**
LUVOIR will a) provide continuous wavelength coverage from the far-UV to the near-IR, b) achieve the $10^{-10}$ contrast levels required for direct spectroscopy of Earth-like planets around Sun-like stars, c) provide ultra-sensitive observations of objects that are 300 times fainter than the sky background, d) provide diffraction limited imaging resolution over wide fields of view and a wide wavelength range. [**Chapters 3–6**]

**Why is direct spectroscopy of habitable exoplanet candidates a key goal for LUVOIR?**
Both scientists and the public are eager to find habitable planets and see if any show signs of life as we know it. Achieving this goal for the exoplanet systems most like the solar system—ones with Earth-like planets around Sun-like stars—requires direct spectroscopy of light coming from the planets using a coronagraph or starshade paired with a space-based telescope. [**Chapter 3**]

**Why is LUVOIR focused on exoplanet spectroscopy at near-UV/optical/near-IR wavelengths?**
This wavelength range contains key markers of habitable surface conditions (e.g., water vapor, $CO_2$), atmospheric biosignatures (e.g., $O_2$, ozone), and other major atmospheric constituents (e.g., methane). The variety of molecules available enables study of a wide range of exoplanets, including habitable planets that are not identical to the modern Earth (e.g., Proterozoic Earth, Archean Earth). [**Chapters 3 & 4**]

**What is the benefit of knowing the locations of specific habitable exoplanet candidates in advance of LUVOIR?**
Finding such candidates around Sun-like stars is currently beyond the capabilities of indirect detection techniques. Direct imaging, as planned for LUVOIR, may be the most effective means of discovery. However, if radial velocity techniques advance to the needed precision, LUVOIR can use knowledge of the locations of potentially habitable exoplanets to decrease the time needed to execute LUVOIR's exoplanet programs. [**Chapter 3**]

**Can LUVOIR do exoplanet transit spectroscopy?**
Yes. LUVOIR can provide high quality UV/optical/near-IR transit spectroscopy of a wide range of exoplanets, including some small habitable zone planets around nearby M dwarf stars. [**Chapters 3 & 4**]

**Can LUVOIR observe solar system objects?**
Yes. LUVOIR is designed for moving target capability with tracking speed of 60 milliarcsec/sec, about two times faster than JWST. Both LUVOIR-A and B can point 45° towards the Sun, resulting in a field-of-regard covering most of the sky. [**Chapter 8**]

## LUVOIR FAQs (Part 3 of 3)

**Why make LUVOIR capable of observing so far into the ultraviolet (< 115 nm)?**
To understand the history of the universe's matter, it must be observed over a wide range of physical conditions. Many key atomic and molecular transitions, spanning the full range of temperatures and densities, lie at wavelengths < 115 nm. [**Chapters 4–6**]

**Can you do high-performance coronagraphy with a UV-capable telescope?**
Yes. Lab studies have shown that UV-compatible coated aluminum mirrors provide coronagraph performance over broad wavelength ranges comparable to that of UV-incompatible silver mirrors. [**Chapter 8**]

**Can you do high-performance coronagraphy with obscured, segmented telescopes?**
Yes. While coronagraphs generally perform better with un-obscured telescopes, recent advances have shown that high-performance coronagraphs can be designed for ob-scured apertures. Other studies have shown that telescope segmentation has very minor impact on coronagraph performance. [**Chapters 8 & 11**]

**Why is the telescope warm?**
Wavefront stability is easier to achieve by actively stabilizing the telescope temperature. Maintaining a temperature near or above the water freezing point avoids risk of optics contamination that would harm the telescope's UV sensitivity. [**Chapter 8**]

**How big are the LUVOIR sunshades?**
The LUVOIR-A sunshade is 56-m × 56-m, while the LUVOIR-B sunshade is 48-m x 48-m. Both are simpler than the JWST sunshield: LUVOIR's have three layers instead of JWST's five and relaxed requirements on layer positioning after deployment. [**Chapter 8**]

**What is the difference between LUVOIR and HabEx?**
These large mission concepts share similar science goals, but differ in ambition and quantitative scientific returns. LUVOIR enables tight constraints on the occurrence rates of habitable conditions and signs of life, as well as a large portfolio of general astrophysics and solar system observations. The two study teams have collaborated since their initiation, and are presenting a buffet of options for UV/optical/near-IR observatories that enable exoplanet studies and general astrophysics.

**How much will LUVOIR cost?**
Accurate cost estimates for LUVOIR are critical desired results of this study. We aimed to achieve greater cost fidelity through increased design and engineering detail, as well as a technology development plan that includes cost and schedule. This report presents several cost estimates performed by NASA's Goddard Space Flight Center. [**Section 1.13 & Appendix J**]

### 1.3.1 The census of Earth-like exoplanets

Section 3.2

Indirect planet discovery techniques have shown that small, rocky planets are not rare. Results from NASA's Kepler mission indicate that $\eta_{Earth}$, the fraction of Sun-like stars with roughly Earth-size planets in their habitable zones, is about 24% (Kopparapu et al. 2018). These zones span the range of distances from the stars where models indicate rocky planets can have liquid water—an essential material for all life on Earth—on their surfaces. This revelation drives scientists to proceed to the next step: build the capabilities to reveal the character of rocky exoplanets and measure the fraction that are truly Earth-like (a.k.a. exoEarths). We need to probe rocky planet atmospheres and determine their thermal/chemical states by measuring molecular abundances, including water vapor and other greenhouse gases.

Furthermore, measuring the frequency of habitable conditions requires observations of

> **Signature Science Case #1:**
> **Finding habitable planet candidates**
>
> **Science Objective**
> Determine the occurrence rate of Earth-like conditions on rocky worlds around Sun-like stars.
>
> **Description**
> Find and study $\gtrsim 28$ habitable planet candidates to discover at least 1 Earth-like planet orbiting an FGK star (at 95% confidence), for occurrence rates of habitable conditions $\gtrsim 10\%$.
>
> **Key Functional Requirements**
>
> | Inscribed telescope diameter | $\gtrsim 6.7$ m |
> |---|---|
> | Total time | 2 years |
> | Inner working angle | $\lesssim 4\,\lambda/D$ |
> | Raw contrast | $1 \times 10^{-10}$ |
> | Wavelength range | $\lesssim 500$ nm to $\gtrsim 950$ nm |
> | Spectral resolution & SNR (near 940 nm for $H_2O$ detection) | R $\gtrsim 70$, SNR $\gtrsim 5$ |

a large number of candidate exoplanets (dozens; **Figure 1-4**). To find and study that many candidates, an even larger number of stars must be observed (hundreds). A systematic survey of exoplanets most similar to the Earth—rocky planets in the habitable zones of Sun-like stars—guarantees that whatever we find will be meaningful. In the absence of water vapor detections, we would know that global-scale surface oceans are rare on rocky worlds in the habitable zone. This null result would still transform our understanding of habitability as a planetary process, and further confirm the special nature of our home. On the other hand, if we find that Earth-like conditions are common on rocky exoplanets, then a stunning vista of hospitable new worlds will be unveiled.

Scientists have calculated that to study the habitability of rocky planets around Sun-like stars, we must collect and analyze light from the planets themselves, i.e. obtain direct images and spectra (e.g., Kaltenegger & Traub 2009; Snellen et al. 2015; Exoplanet Science Strategy Report). But doing this for exoEarth candidates around Sun-like stars is made extremely challenging by the fact that the Earth is 10 billion times fainter than the Sun at visible wavelengths and orbits close to its host star. Viewed from an interstellar distance of 10 pc, the apparent maximum separation of the Earth from the Sun (0.1″) is only about the width of a human hair seen from a distance of two football fields.

The first direct images of giant exoplanets have been obtained with ground-based observatories, by suppressing the bright light from the central stars using coronagraph instruments (e.g., Marois et al. 2010). High-performance coronagraphs on future 30-m-class ground-based telescopes (Extremely Large Telescopes, or ELTs), supported by adaptive optics systems (AO), may be able to study the atmospheres of rocky planets orbiting in the habitable

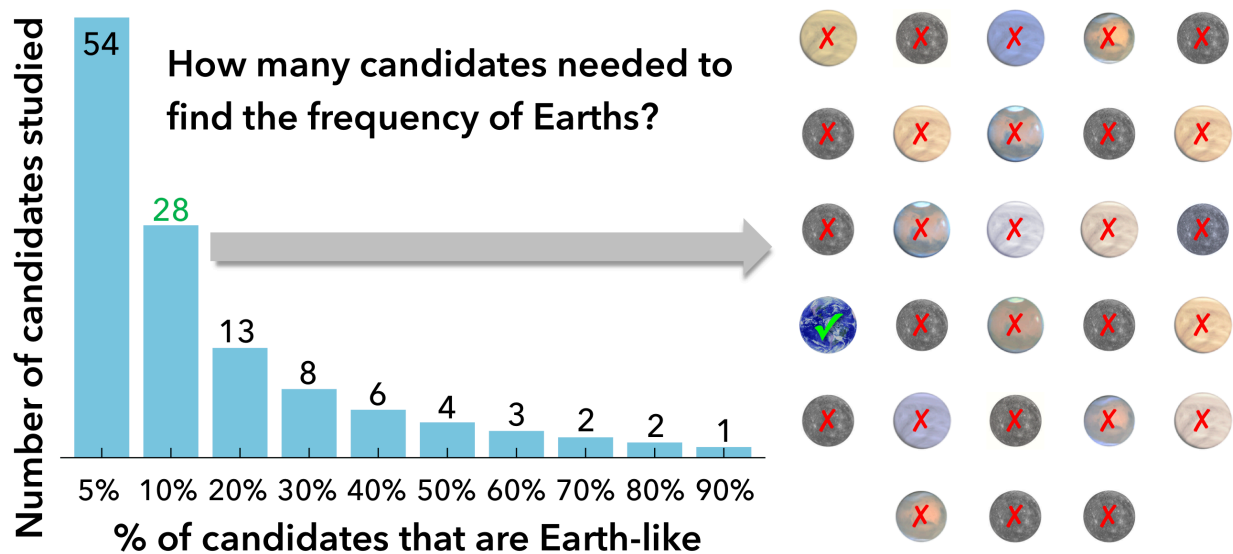

**Figure 1-4.** *How many rocky habitable zone exoplanets need to be studied? The bar chart shows the number of candidates needed to discover one Earth-like planet (at the 95% confidence level), as a function of the percentage of temperate rocky planets that are actually Earth-like. If the percentage is 10%, then 28 candidates need to be observed. To put it another way, if 28 candidates are observed and no water vapor is detected, then we learn that <10% of rocky planets in stellar habitable zones are Earth-like. Credit: C. Stark (STScI) / A. Roberge (NASA GSFC)*

zones of the nearest cool red dwarf stars (e.g., Wang et al. 2017). However, direct observations of rocky exoplanets around Sun-like stars demand the very high contrast possible only with a space telescope coupled to a high-performance starlight suppression system (Exoplanet Science Strategy Report).

The challenges in executing a direct imaging and spectroscopic survey for habitable worlds around Sun-like stars go beyond the technological ones of starlight suppression. While the target stars have been known for many decades, careful prioritization is needed to execute an efficient survey. The levels of interplanetary dust coming from comets and asteroids in the system (exozodiacal dust) impact required exposure times and affect target prioritization. Fortunately, constraints on typical dust levels around nearby Sun-like stars have been provided by the Large Binocular Telescope Interferometer (LBTI; Ertel et al. 2018). The possibility of confusion between habitable planet candidates and background sources (galaxies and stars), as well as non-habitable planets, must also be considered.

**Figure 1-5** summarizes the survey strategy developed by the LUVOIR Study Team that obtains the needed measurements while dealing with all of these complexities in an efficient manner, as explained in **Chapter 3**. The steps in the initial 2-year habitable planet survey are colored blue in **Figure 1-5**. At each step, stars and exoplanets that do not show the characteristics we are looking for are set aside; they can be followed up in comparative planetology studies outside of the habitable planet survey.

The survey begins with high-contrast imaging of hundreds of nearby stars to discover point sources. Images at multiple wavelengths are obtained to measure colors and begin confusion discrimination. Repeat imaging visits are performed to confirm which point sources are background objects via proper motion. Once it appears that a point source is likely to be a planet (plausible color and co-moving with the target star), additional images

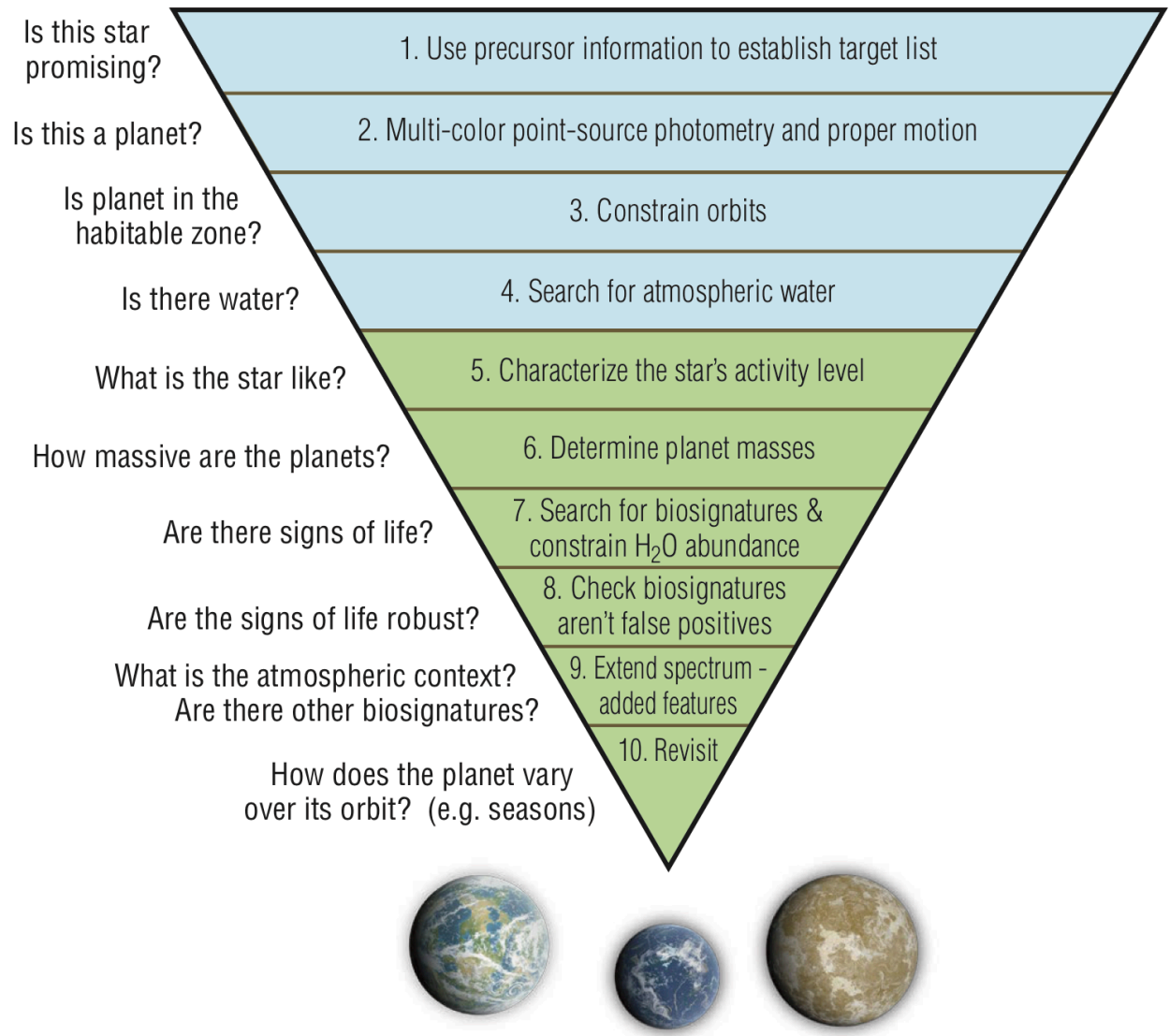

Is this star promising?
1. Use precursor information to establish target list

Is this a planet?
2. Multi-color point-source photometry and proper motion

Is planet in the habitable zone?
3. Constrain orbits

Is there water?
4. Search for atmospheric water

What is the star like?
5. Characterize the star's activity level

How massive are the planets?
6. Determine planet masses

Are there signs of life?
7. Search for biosignatures & constrain $H_2O$ abundance

Are the signs of life robust?
8. Check biosignatures aren't false positives

What is the atmospheric context? Are there other biosignatures?
9. Extend spectrum - added features

How does the planet vary over its orbit? (e.g. seasons)
10. Revisit

**Figure 1-5.** *The LUVOIR strategy for the search for life. Blue steps at the top of the figure represent an initial survey optimized to discover habitable planets. Green steps at the bottom of the figure refer to characterization of those planets, confirming habitability and searching for biosignatures. Credit: T. B. Griswold (NASA GSFC)*

(6 per star in total) separated by carefully chosen time intervals are obtained to measure the planet's orbit. Planets with orbits within the star's habitable zone are then followed up with a partial spectrum centered near 940 nm, covering a water vapor absorption band. A detection of water vapor marks the planet as a habitable planet candidate. The wavelength range of the partial spectrum also contains an absorption band of methane that is seen in giant planet spectra, adding scientific returns to the reconnaissance spectroscopy.

An advanced code to determine the optimal direct imaging survey plan and calculate its scientific returns has been developed over several years (Stark et al. 2014, 2015, 2016). The starting target star list was created from the Hipparcos and Gaia catalogs, limited to stars within 50 pc. The code accounts for the complex interplay between astrophysical constraints (e.g., exoplanet occurrence rates, habitable zone boundaries, planet albedos, exozodiacal dust levels), hardware capabilities (e.g., coronagraph inner working angle, contrast, spatial resolution, throughput), and observing strategy. **Section 3.4.1** contains an explanation of how this code was used in the LUVOIR study and the code itself is fully described in **Appendix B.2**.

The expected observations and exoplanet yields obtained in initial 2-year surveys with LUVOIR-A and -B returned by our analysis appear in **Table 1-2** and **Figure 1-6**. The LUVOIR-B observatory meets the requirement set by the LUVOIR STDT for a statistically significant exoEarth candidate survey (28 habitable planet candidates). With 54 habitable planet candidates, LUVOIR-A surpasses

**Table 1-2.** *Properties of stars and planets observed in the initial exoEarth survey*

|  | LUVOIR-A | LUVOIR-B |
|---|---|---|
| **Stars** | | |
| Total # of stars observed | 287 | 158 |
| A types | 3 | 1 |
| F types | 99 | 48 |
| G types | 98 | 55 |
| K types | 57 | 39 |
| M types | 29 | 15 |
| Max distance | 27.7 pc | 23.0 pc |
| Range of $m_v$ | 2–11 | 0–11 |
| **Exoplanets** | | |
| Habitable candidates | $54^{+61}_{-34}$ | $28^{+30}_{-17}$ |
| Total non-habitable | $648^{+251}_{-312}$ | $576^{+166}_{-260}$ |

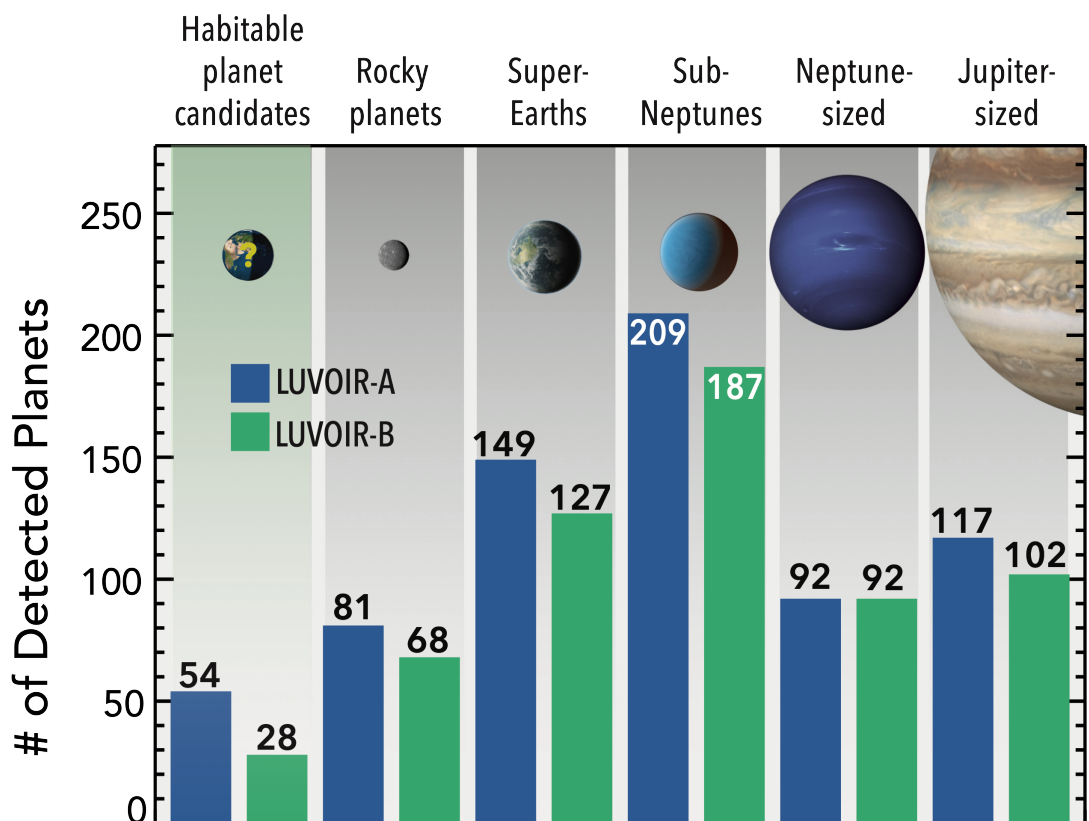

**Figure 1-6.** *LUVOIR will discover dozens of habitable planet candidates and hundreds of other kinds of exoplanets. The chart shows exoplanet detection yields from an initial 2-year survey optimized for habitable planet candidates with LUVOIR-A (blue bars) and -B (green bars). The first column shows the expected yields of habitable planet candidates. Non-habitable planets are detected concurrently during the 2-year survey. Color photometry is obtained for all planets. Orbits and partial spectra capable of detecting water vapor and/or methane are obtained for all habitable planet candidates. Credit: C. Stark (STScI) / J. Friedlander (NASA GSFC)*

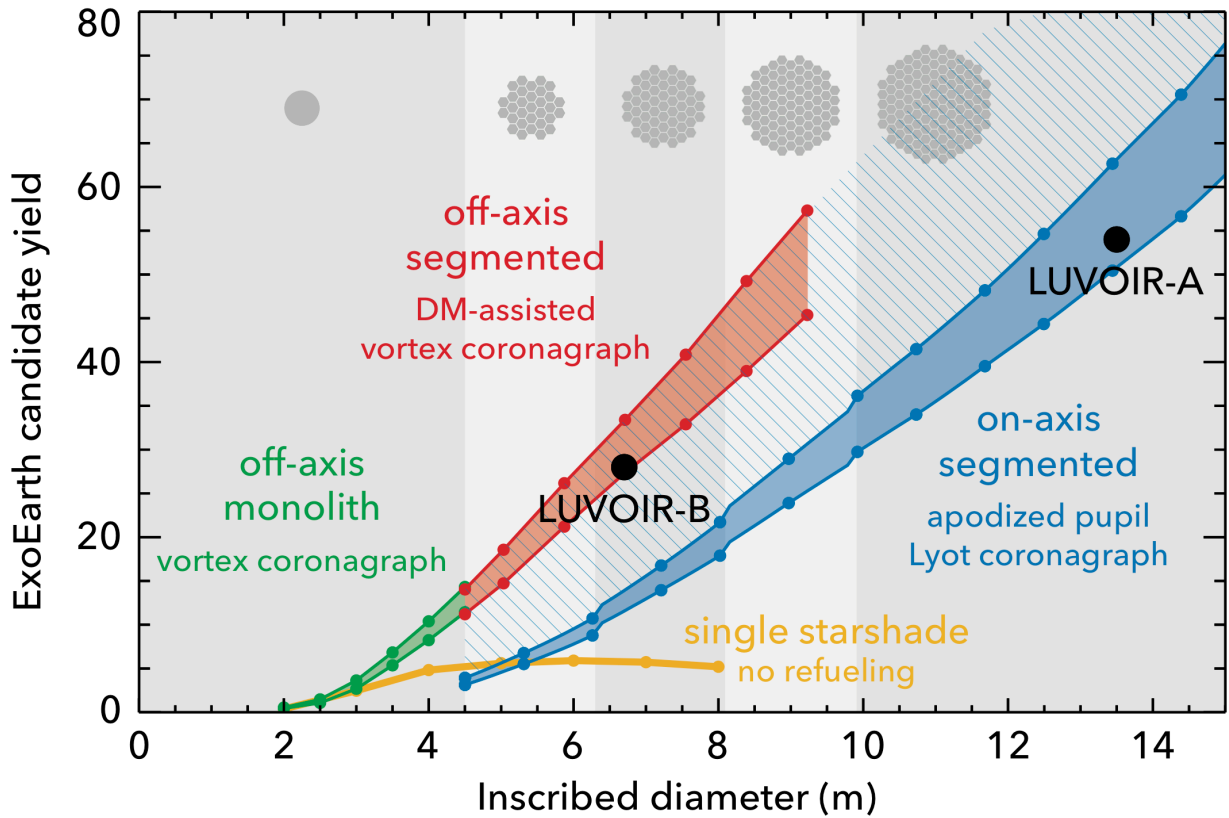

**Figure 1-7.** *Telescope size, aperture geometry, and coronagraph type all affect the expected detection yields of exoEarth candidates. On the x-axis, the inscribed diameter is the diameter of the largest circle completely contained within the telescope aperture. The green, red, and blue curves show yields for different combinations of telescope aperture geometry and coronagraph type, more fully explained in the main text. The yellow curve shows the yields for a single starshade paired with a 4-m telescope. Credit: Stark et al. (2019)*

the requirement and provides ample margin against astrophysical and technological uncertainties. The same code with the same astrophysical input assumptions was used to calculate exoplanet yields for the Habitable Exoplanet Observatory (HabEx) mission concept.

A summary of the set of stars observed also appears in **Table 1-2** (details in **Section 3.4.1**). The most favorable spectral types for detection of a habitable planet candidate are F, G, and K; significant numbers of M stars and a few A types are also observed. For the yield calculations, we adopted a fraction of Sun-like stars with habitable exoplanet candidates $\eta_{Earth} = 0.24^{+0.46}_{-0.16}$ (Kopparapu et al. 2018). The expected yield of such exoplanets from a direct imaging survey is proportional to $\eta_{Earth}$ (Stark et al. 2015). Therefore, it is easy to scale expected yields using updated $\eta_{Earth}$ values. A more complete discussion of how yields depend on different astrophysical and hardware parameters appears in **Appendix B.2**.

**Figure 1-7** shows that the expected yield of habitable planet candidates is a strong function of telescope diameter and changes with aperture geometry; these realities drove the design of the LUVOIR concepts. The inscribed diameter is the diameter of the largest circle completely contained within the telescope primary aperture; this is the parameter that has the single greatest impact on yields. The yields also depend on whether the aperture is unobscured (off-axis) or obscured (on-axis), which affects the choice of coronagraph type.

In **Figure 1-7**, off-axis monolithic and segmented telescopes are shown with the green and red curves, respectively. Both sets of curves end at the approximate largest size that appears feasible for that type of telescope. The range between the upper and lower curves spans different approaches to observing strategy. The off-axis telescopes are paired with vector vortex charge 6 (VC6) coronagraphs. The smooth joining of the green and red curves show that new coronagraph designs have eliminated any penalty for segmentation.

On-axis segmented telescopes are shown with blue curves. They are paired with apodized pupil Lyot coronagraphs (APLC), which perform better with obscured apertures than VC6. The yield gap between red and blue curves shows that there remains a penalty for obscuration. The two black points show the expected yields for LUVOIR-B (6.7-m inscribed diameter, off-axis segmented paired with VC6) and LUVOIR-A (13.5-m inscribed diameter, on-axis segmented paired with an APLC). The yellow curve shows the expected yield from a single starshade paired with a 4-m telescope and without refueling of the starshade's propulsion system. This curve demonstrates that, unlike coronagraph-only systems, the yield from starshade-only systems does not continue increasing with telescope diameter.

### 1.3.2 The search for life on exoplanets

**Section 3.3**  LUVOIR will not only achieve the exoEarth census but will probe the atmospheres of those planets for biosignature gases, possibly revealing the first signs of life outside our home planet. The spectrum of the Earth contains features arising from gases of biological origin, like $O_2$ and methane, in addition to features from water vapor and other greenhouse gases (**Figure 1-8**). The Earth is the only planet we know of teeming with surface life that strongly affects the planet's atmosphere. Thus, there are good reasons to focus serious searches for life outside the solar system on exoplanets that are as much like the Earth as possible. By targeting exoplanets with sizes and orbits similar to

---

**Signature Science Case #2:**
**Searching for biosignatures and confirming habitability**

**Science Objective**

Search for signs of global biospheres on rocky worlds around Sun-like stars by assessing the chemical state of planet atmospheres. Confirm liquid water.

**Description**

Measure abundances of key molecules (e.g., $H_2O$, $O_2$, $O_3$, $CH_4$, $CO_2$, CO) in the atmospheres of rocky exoplanets via direct spectroscopy. Perform other characterization observations to place the planets in context.

**Key Functional Requirements**

| Inscribed telescope diameter | $\gtrsim 6.7$ m |
|---|---|
| **Total time** | |
|     LUVOIR-A | 6.5 months |
|     LUVOIR-B | 6.2 months |
| Inner working angle | $\lesssim 4\,\lambda/D$ |
| Raw contrast | $\lesssim 1\times 10^{-10}$ |
| Planet spectroscopy wavelength range | 200 nm to $\gtrsim 1800$ nm |
| **Spectral resolution & SNR** | |
|     $O_2$ (760 nm) | $R \approx 140$, SNR $\gtrsim 8.5$ |
|     $H_2O$ (1120 nm) | $R \approx 70$, SNR $\gtrsim 8.5$ |
|     $CO_2$ (1600 nm) | $R \approx 200$, SNR $\gtrsim 8.5$ |
| Astrometric precision for planet masses | $<1\ \mu$as |
| Stellar spectroscopy wavelength range | 100 nm to 1000 nm |

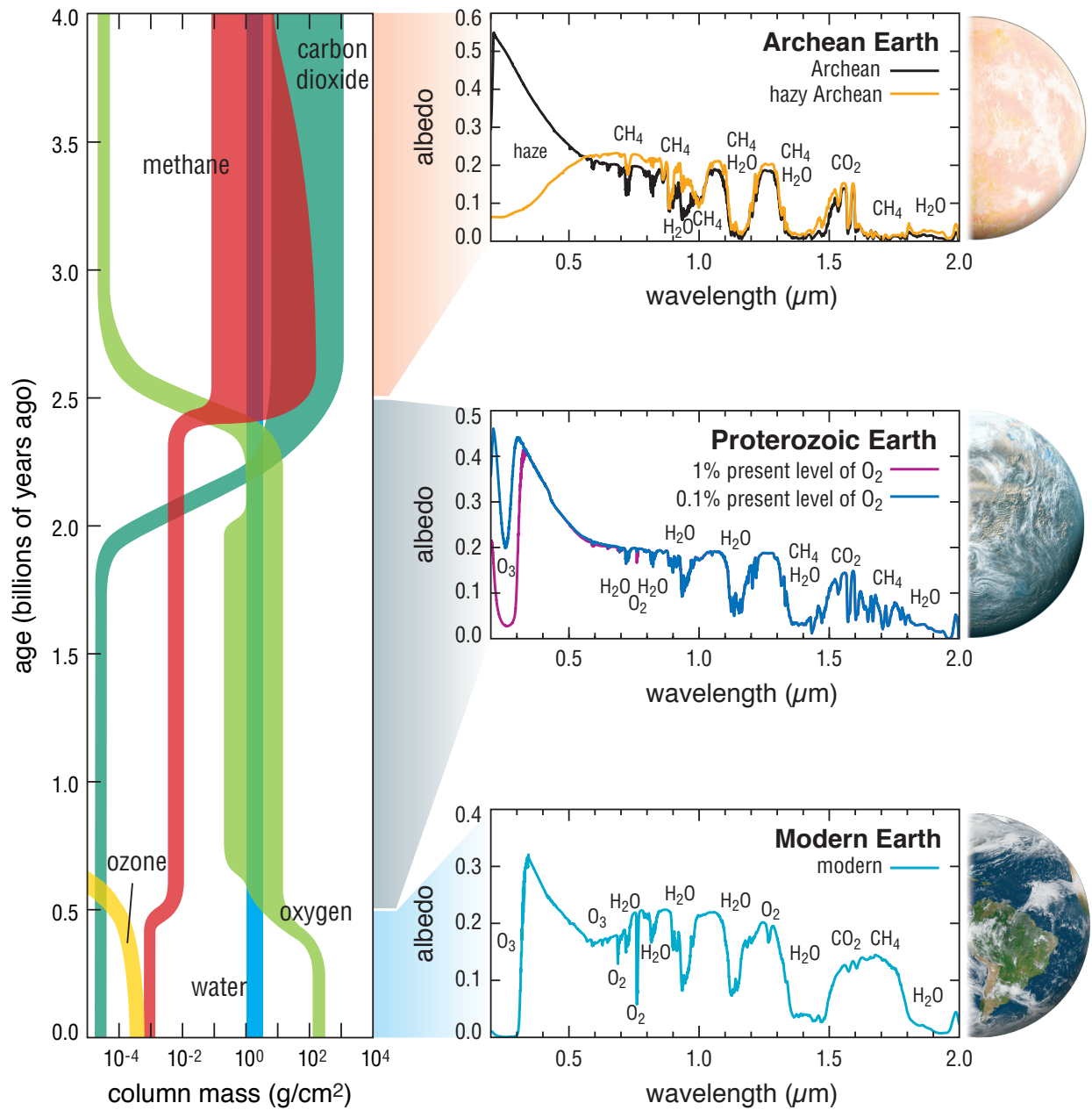

**Figure 1-8.** *Earth's atmospheric composition has changed significantly over its inhabited history (left), as has its spectral features (right). For instance, the canonical biosignatures of modern Earth (e.g., $O_2$, $O_3$) cannot be observed in the Archean eon. A wide wavelength range enables higher fidelity spectral characterization by providing observations of many atmospheric species, and often multiple absorption bands of a given species. This breaks degeneracies between overlapping features (e.g., $CH_4$ and $H_2O$ absorption bands in the near-infrared) and enables the search for non-traditional biosignatures. Credit: G. Arney, S. Domagal-Goldman, T. B. Griswold (NASA GSFC)*

Earth's and host stars similar to the Sun, we hope to increase our chances of finding—and recognizing—extraterrestrial life.

However, it is important to understand that no one molecule (or pair of molecules) is automatically a biosignature in every context. Many studies over the last decade have identified pathways to atmospheres containing abundant oxygen without life, especially for planets

## Why LUVOIR for habitable exoplanets?

The most powerful tool to characterize habitable exoplanets most like the Earth is a large space-based telescope, which can conduct direct imaging and spectroscopy of candidates around a wide range of stars over a broad wavelength range.

*Why direct observations?* Understanding the surface and near-surface environment of a potentially habitable planet is the key to discovering whether that planet can support liquid water. Direct observations provide a short line of sight down through the planet's atmosphere to the surface. This enables crucial measurements of near-surface physical conditions, water vapor abundance, and biosignature gases, including those that are not abundant at higher altitudes. Transmission spectroscopy (a.k.a. transit spectroscopy) cannot sample the near-surface atmospheres of Earth-like planets around Sun-like stars, due to the longer line of sight through the atmosphere, the effects of aerosols, and refraction.

*Why space? Why NUV-Optical-NIR wavelengths?* Beyond achieving the high contrast needed for direct observations of Earth-like exoplanets around Sun-like stars, LUVOIR will have another advantage over ground-based observatories: it will not have to look through layers of the very gases it is searching for. Those gases include molecular markers of habitable surface conditions (e.g., water vapor, $CO_2$, $CH_4$) and biosignatures (e.g., $O_2$, $O_3$, $CH_4$). Absorption features of all these molecules lie in the NUV, optical, and NIR wavelength ranges.

*Why so big?* LUVOIR's large apertures will allow it to find and study a greater number of exoplanets, through a combination of large collecting area, high spatial resolution, and small inner working angle (all functions of aperture diameter). These traits will also provide spectral characterization from the near-UV into the near-IR. Further, they reduce the time needed to obtain an observation of a given planet, enabling the multiple, short exposures needed to map planetary surface inhomogeneity (e.g., continents and oceans) on the nearest targets.

around M dwarf stars (e.g., Meadows et al. 2018b). The key is to measure the abundances of many molecules in the whole atmosphere, and then use those abundances to estimate the production and destruction rates predicted by physics and chemistry. To calculate the production and destruction rates, a good deal of knowledge about the atmosphere's context will be needed, including the planet's orbit, mass, and the characteristics of the host star.

If the abundance of a molecule is too high, then the production rate is underestimated and an additional biological source is indicated. Differences in rates from biological and non-biological sources on Earth can differ by orders of magnitude (Kirsanssen-Totton et al. 2018). This same logic of using rates to determine biogenicity is also being incorporated into plans to search for life in the outer solar system. In sum, scientists will try to explain the character of an atmosphere with two sets of models: one where the physics and chemistry of the atmosphere are purely driven by geological and astrophysical processes; and a second where biological sources are also considered. Those planets for which only the second set of models can explain the data are those for which evidence of life may be robustly claimed.

Therefore, confident life-detection has at least three fundamental requirements: 1) assess a wide range of atmospheric molecules, 2) measure or constrain their abundances with some precision, and 3) understand the planet's context within its system. The first requirement demands direct spectra with broad wavelength coverage from the near-ultraviolet (NUV) to the near-infrared (NIR). The LUVOIR telescope and starlight suppression system can span these wavelengths of light and access the important constituents of planetary atmospheres, including water ($H_2O$), carbon dioxide ($CO_2$), molecular oxygen ($O_2$), ozone ($O_3$), and methane ($CH_4$). Important indicators of false positive biosignatures (e.g., CO) can be observed as well.

The second requirement demands good quality spectra to measure abundances, which drives the required telescope and instrument sensitivity. The third requirement demands collection of a range of supporting information, including planet masses and far-UV spectra of stellar activity-driven emission from the host stars. The planned steps in characterization of habitable planets are summarized in the green levels of **Figure 1-5**. LUVOIR's great sensitivity and flexibility will enable the variety of observations needed to properly understand a planet as a complete system.

The dozens of rocky worlds LUVOIR will explore will inevitably have a range of sizes, orbits, and chemical compositions, and orbit various stars with different ages. Earth has been inhabited for most of its history, but in the earlier stages, the atmosphere was very different from that of the modern Earth (e.g., low $O_2$ abundances; **Figure 1-8**). Recognizing such a planet as inhabited also demands the careful assessment of atmospheric abundances and planetary context described above. If LUVOIR confirms the habitability of rocky exoplanets, or finds signs of life, we will learn how those traits respond to different planet and stellar characteristics. This will turn the habitable zone from a theoretical concept based upon a single planet to one that is empirically derived from multiple worlds, and usher in a new era of planetary science: one of **comparative astrobiology**.

**Spectra.** The coronagraph contrast and inner working angle (IWA) required to discover habitable planet candidates in the notional 2-year initial survey are sufficient to also obtain direct spectra of those candidates, although the IWA will cut off the longest wavelengths for planets around more distant stars. The optimal spectral resolution (R) and signal-to-noise ratio (SNR) pairings for accurate and efficient abundance measurements of different molecules in modern Earth-like atmospheres have been determined using spectral retrieval analysis of simulated direct spectra (Feng et al. 2018; Feng et al., in preparation). With this information and our tool for simulating direct spectra[1], the exposure time required for each spectral characterization step in **Figure 1-5** can be calculated for a habitable planet candidate around any target star. The exposure times are a steeply increasing function of the distance of the star from the Sun.

**Masses.** The surface gravity of a planet is a key parameter in modeling the physical state of its atmosphere. Simulations have shown that planet radius can be constrained from the planet's reflected light brightness and assumptions about the plausible range of albedos (Feng et al. 2018). With an adopted mass-radius relationship, the mass can also be constrained. However, the resulting constraint on surface gravity is weak and uncertain. Direct measurements of planet masses would obviously be preferable.

---

1 https://asd.gsfc.nasa.gov/luvoir/tools/

Radial velocity (RV) instruments on today's ground-based telescopes cannot measure masses of small rocky planets orbiting Sun-like stars ($\lesssim 10$ cm/s precision needed). A new generation of extremely precise RV (EPRV) instruments on future extremely large ground-based telescopes (ELTs) may be able to push down to few cm/s precision—if not systematically limited by microtelluric variability (Leet et al. 2019). The current uncertainty about future performance of EPRV instruments led the LUVOIR Study Team to consider alternatives to ground-based RV for obtaining important exoplanet mass measurements.

Astrometry is a technique that has long held promise for exoplanet discovery and mass measurement. However, the astrometric signal from an Earth-mass planet orbiting 1 AU from a G2V star at 10 pc is 0.3 µas. This astrometric precision cannot be achieved from the ground or with the Gaia mission. Therefore, the LUVOIR Study Team has chosen to incorporate precise (<1 µas) astrometric capability into the HDI instrument. This capability is also highly valuable for a number of general astrophysics applications (for example, mapping of dark matter in dwarf galaxies via stellar proper motions; **Section 1.5.2**). The individual exposure times for astrometric imaging of habitable planet candidate host stars are short (tens of seconds), but multiple visits are needed. Further discussion of astrometric exoplanet mass measurements appears in **Section 1.4.2**.

***Stellar spectra.*** Accurate spectra of the host stars are also required for detailed modeling of planetary atmospheres. In particular, the balance between near-UV and far-UV flux controls the photodissociation rates for most molecules, and thus the photochemistry of the atmosphere. The UV spectra of Sun-like and low-mass main sequence stars are produced by stellar activity, making them hard to accurately model; direct observations will be needed. LUVOIR will obtain UV spectra of all the stars harboring habitable exoplanet candidates using the LUMOS instrument. The host stars are bright (V < 11 mag) and the times required for high quality spectra are short. All stars hosting the habitable planet candidates will also be monitored to assess levels of flaring activity.

***Time variations.*** The large collecting area of LUVOIR enables another kind of characterization observation focused this time on the planetary surfaces. A planet with continents and oceans, but without a thick cloud layer, will have a non-uniform surface albedo. If exposure times are short enough, the rotation period of such a planet can be retrieved from reflected light brightness changes in high cadence images or spectra, even though the planet's surface is not spatially resolved. Constraining the rotation period of a planet is highly valuable for 3D climate modeling.

**Table 1-3** shows the rotation periods retrieved from simulated time-series LUVOIR spectra (100 total hours at 1-hour cadence), assuming the modern Earth located around Sun-twin stars at different distances (Lustig-Yaeger et al. 2019, in preparation). This suggests that the rotation period can be accurately recovered (at the < 20% level) for Earth-analog planets around Sun-like stars

**Table 1-3.** *Recovered rotation period of the Earth around stars at different distances*

| Distance (pc) | LUVOIR-A | LUVOIR-B |
|---|---|---|
| 3.0 | 24±0 | 24±1 |
| 4.0 | 24±0 | 24±2 |
| 5.0 | 24±0 | 23±3 |
| 7.0 | 24±1 | 21±6 |
| 9.0 | 23±3 | 18±8 |
| 11.0 | 22±5 | 15±9 |
| 13.0 | 21±6 | 12±11 |
| 15.0 | 20±7 | 11±12 |
| 20.0 | 16±9 | 9±13 |

out to ~11 pc with LUVOIR-A (~70 FGK stars) or ~6 pc with LUVOIR-B (~20 FGK stars). Furthermore, the same observations can reveal a rough longitudinal map of the planet (e.g., Cowan et al. 2009; **Figure 1-9**). The presence of an ocean may be inferred via careful comparison to colors of expected surface materials. Most of the data required for these investigations will be obtained during long spectroscopic observations.

In cases where a habitable planet has a non-zero obliquity, seasonal variations in surface albedo may be expected, as ice caps advance and retreat. Even more excitingly, biosignature gases arising from photosynthetic life on Earth ($O_2$ and $O_3$) exhibit strong seasonal abundance variability (Olson et al. 2018). Spectroscopy of promising planets obtained over the course of their orbits could reveal such variability, and provide powerful support for any life detection claim.

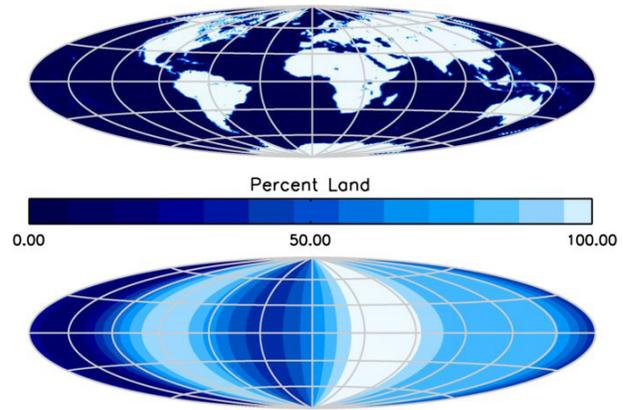

**Figure 1-9.** *Map of an ocean world. The top panel shows a view of the Earth. The bottom panel shows the surface map retrieved from imaging data obtained with the EPOXI mission, demonstrating that a longitudinal map can be recovered from spatially unresolved images. Credit: Cowan et al. (2009)*

***Characterization strategy.*** Estimating how many times to progress through the characterization steps in **Figure 1-5** (green levels) and how long that would take is complicated. At one extreme, future scientists might wish to perform detailed studies of every rocky habitable-zone planet found with LUVOIR, whether it showed atmospheric water vapor or not. More realistically in the case of a large set of habitable planet candidates, scientists could choose to progress down the pyramid only for planets that pass the test at each level. Accurately calculating how many habitable planet candidates pass each step depends on a series of unknown occurrence rates for the characteristics of an inhabited planet (culminating in $\eta_{biosphere}$, as it were). In a very real sense, LUVOIR is the experiment to begin measuring all of these unknown occurrence rates, much like the Kepler mission did for $\eta_{Earth}$ (which is really $\eta_{Earth-size}$ in the habitable zone).

In an attempt to capture these uncertainties, the LUVOIR Study Team constructed a preliminary statistical approach for estimating how many planets will be characterized in a fixed-duration program, depending on what fraction of habitable planet candidates show atmospheric water vapor. The possible target lists were drawn from the habitable planet candidates found in the initial exoEarth survey. Colors, orbits, a partial spectrum capable of detecting water vapor, UV spectra of the host stars, and masses will have been obtained for all candidates (**Section 1.3.2**).

We then carefully considered the time-consuming direct spectra, under the assumptions that we only observe candidates with water vapor detections from the initial survey and that those planets have modern Earth atmospheric abundances. Thus, these planets pass the characterization decision at every level. This provides a sense of how well LUVOIR may constrain the occurrence rate of inhabited planets among all the habitable planet candidates. **Section 3.4.2** and **Appendix B.3** contain a complete description of the analysis, along with some caveats and warnings. A simulated spectrum of a modern Earth-twin planet over the full coronagraph (ECLIPS) bandpass appears in **Figure 1-10**. This spectrum is suitable for measuring abundances of key molecules in the atmospheres of such planets.

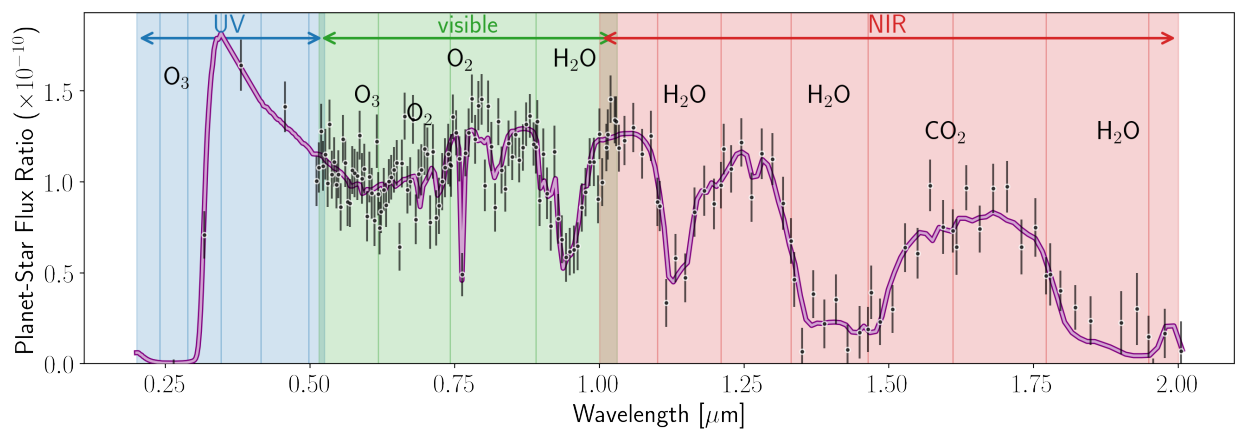

**Figure 1-10.** *Simulated LUVOIR ECLIPS spectrum of a modern Earth-twin orbiting a Sun-like star. These spectra have SNR=8.5 on the continuum across the whole bandpass, sufficient for measurement of key molecular abundances. The total time to acquire this spectrum varies dramatically depending on the distance of the system from the Sun. The shaded background regions indicate 20% bandpasses in the NUV (blue) and visible (green) channels, and 10% bandpasses in the NIR (red) channel. Credit: J. Lustig-Yaeger (UW)*

Obtaining complete spectra over the entire ECLIPS bandpass for all planets is costly, as the modern Earth is faint at the bluest and reddest wavelengths. For the present, we account for the time required to obtain complete spectra over the 290–1460 nm wavelength range. This reduces the ability to measure $O_3$ abundances, but this is not an especially useful diagnostic for the modern Earth, as atmospheric oxygen abundances are easily measured from the $O_2$ A-band at 760 nm. However, $O_3$ is a key species for the oxygen-poor Proterozoic Earth (**Figure 1-8**) and it is vital to have access to NUV wavelengths with ECLIPS. Future analysis will use additional types of inhabited planet models.

A summary of the characterization yields appears in **Table 1-4**. The time to obtain the direct planet spectra was fixed to 6 months for both LUVOIR concepts. Including time for UV spectra of the host stars brings the total program time to 6.5 months for LUVOIR-A or 6.2 months with LUVOIR-B. The time required for the mass measurements is bookkept in Signature Science Case #5 (*The formation of planetary systems*; **Section 1.4.2**).

We emphasize that this is a first attempt to estimate characterization yields in a statistical sense, and the observing program is not as efficient as it could be. The characterization yield analysis is more detailed than is typical for this stage of mission concept development, and may in truth be overly complex. However, we feel it was worthwhile to start exploring the right ways to do such an analysis, so that an optimal observing program may one day be designed. In reality, studying habitable exoplanets and searching them for biosignatures will very likely require a flexible and adaptable strategy. LUVOIR and its coronagraphs have been designed with this in mind.

**Table 1-4.** *LUVOIR's characterization of habitable planet candidates*

|  | LUVOIR-A | LUVOIR-B |
|---|---|---|
| # of hab. planet candidates | $54^{+61}_{-34}$ | $28^{+30}_{-17}$ |
| # with orbits | 54 | 28 |
| # with host star spectra | 54 | 28 |
| # with mass measurements | 54 | 28 |
| # with direct spectra from 290−1030 nm | | |
|  | 23 | 11 |
| # with direct spectra from 290−1460 nm | | |
|  | 18 | 8 |

### 1.3.3 Ocean moons in the solar system

**Section 3.5** The search for habitable environments also takes place closer to home. Scientists have discovered that several moons of the outer solar system have liquid water beneath their icy surfaces. These sub-surface oceans must be heated from below, which may also provide the energy needed for life. Plumes emanating from openings in the icy shells—such as those observed from Europa (**Figure 1-11**; Roth et al. 2014; Sparks et al. 2016) and Enceladus (e.g., Hansen et al. 2006)—may allow glimpses into the deep oceans. LUVOIR has an important role to play in determining the currently unknown strength and frequency of plume activity by providing observations over long time baselines. Such information will be a valuable support for future spacecraft visiting these other ocean worlds to search for signs of life with focused investigations.

**Signature Science Case #3:**
**The search for habitable worlds in the solar system**

#### Science Objective
Determine the strength and frequency of plume activity from Europa. Record surface changes caused by geologic activity on solar system ocean moons.

#### Description
Monitor Europa for FUV auroral emission (Lyman-α and neutral oxygen). Multi-epoch, optical/NIR imaging of 6 ocean moons.

#### Key Functional Requirements

| Telescope diameter | $\gtrsim 8\,m$ |
|---|---|
| **Total time** | |
|     LUVOIR-A | 16 days |
|     LUVOIR-B | 32 days |
| **UV spectroscopy** | |
|     Wavelength range | 1100–200 nm |
|     Spectral resolution | $R \approx 10{,}000$ |
| **Imaging** | |
|     Wavelengths | 550–1700 nm |
| **Non-sidereal tracking; Dithered mosiacs** | |
| **Multi-epoch observations** | |

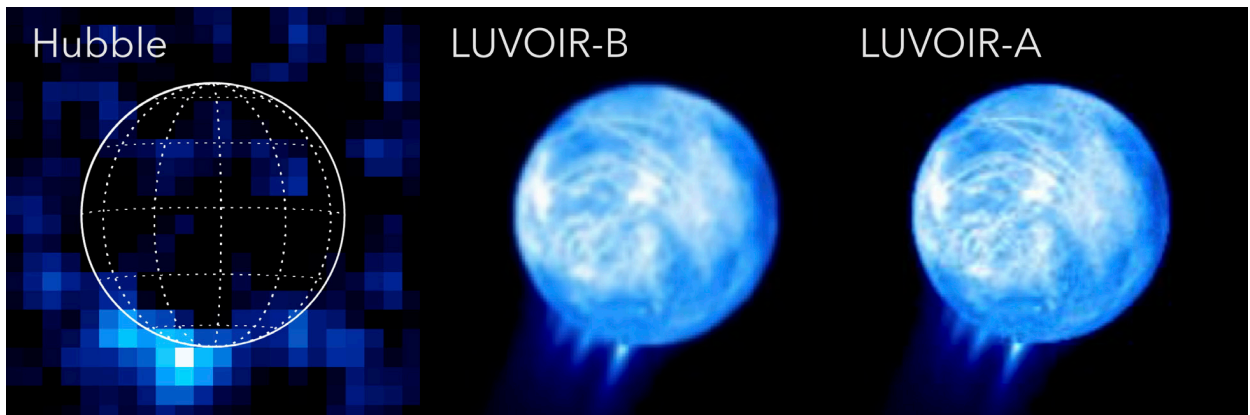

**Figure 1-11.** *LUVOIR can monitor individual plumes from solar system ocean moons. The left panel shows an aurora on Europa observed with HST (Roth et al. 2014). This UV hydrogen emission (Lyman-α) comes from dissociation of water vapor in plumes escaping through the moon's ice shell. The center and right panels show simulations of how this Ly-α emission from Europa might look observed with LUMOS on LUVOIR-B and -A. The moon's surface is bright due to reflected solar Ly-α emission, which was below the background in the HST image. Credit: G. Ballester (LPL) / R. Juanola-Parramon (NASA GSFC)*

## Solar System Science with LUVOIR

The bodies of the solar system, from smallest Kuiper Belt objects to the giant planets, represent a treasure trove of information on the formation of the solar system, atmospheric processes, and dynamical evolution. LUVOIR's observational capabilities in the solar system extend from the orbit of Venus outward. Here, we highlight a few additional solar system science cases LUVOIR could address beyond the Signature Science Cases described in this report. For some bodies, LUVOIR's imaging is comparable to flyby and orbiter quality.

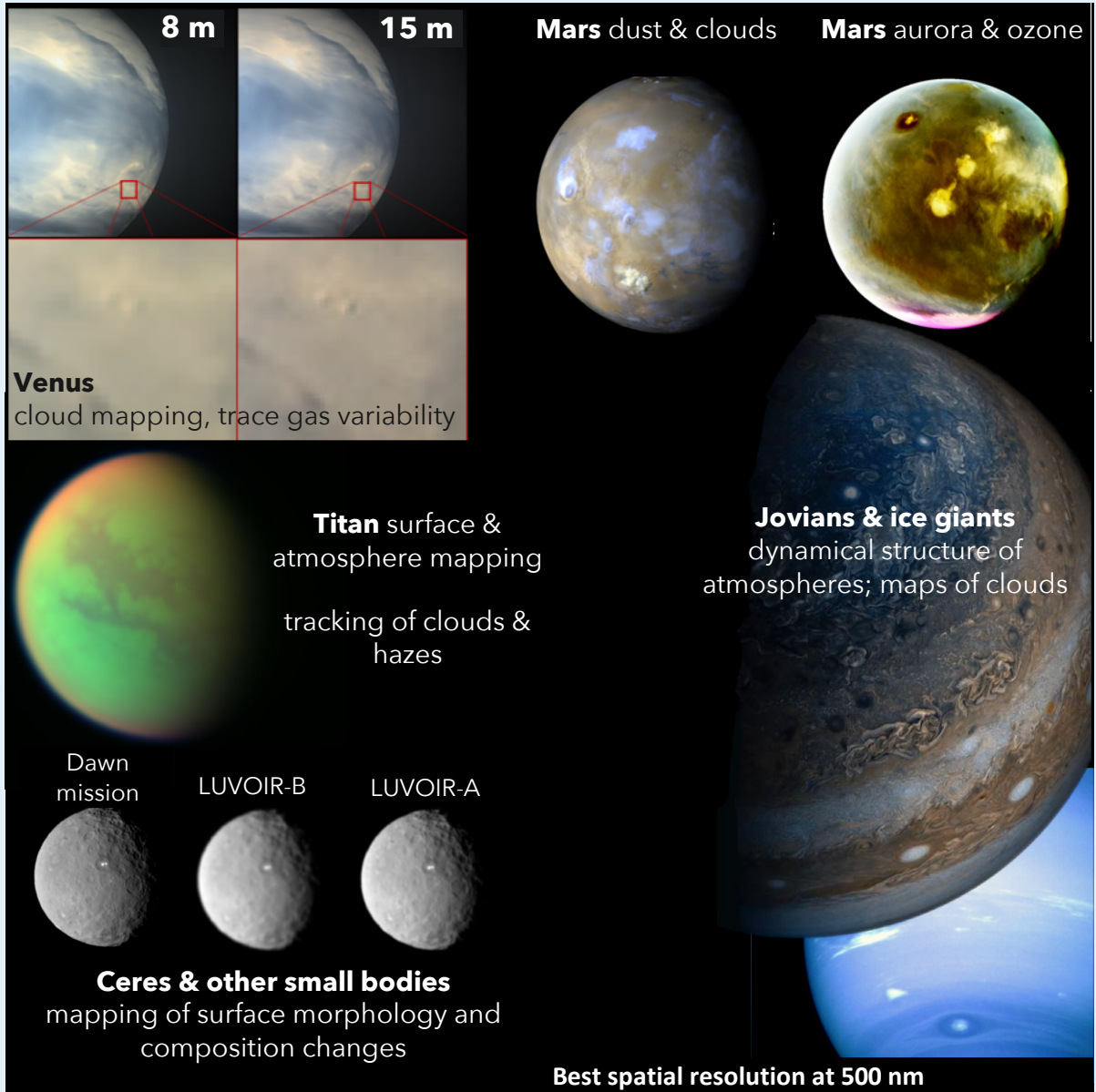

| Best spatial resolution at 500 nm | | |
|---|---|---|
| | LUVOIR-A | LUVOIR-B |
| **Venus, Mars, Ceres, Jupiter** | 7 km, 3 km, 11 km, 25 km | 14 km, 5 km, 20 km, 47 km |
| **Saturn, Uranus, Neptune, 40 AU** | 51 km, 108 km, 173 km, 232 km | 96 km, 204 km, 327 km, 438 km |

In the case of Enceladus, Cassini observations showed that the plumes are frequently (possibly continuously) active, but their strength modulates on a period that may be driven by orbital interactions with the moon Dione (Hurford et al. 2007). In the case of Europa, plumes have only been detected via Hubble FUV observations (**Figure 1-11**). The activity level is lower than that of Enceladus, but there is no information on the frequency or any possible periodicities (Roth et al. 2014). LUVOIR can provide long-term monitoring with flexible cadence, which can be challenging for in-situ spacecraft.

The FUV multi-object spectroscopy capability of LUMOS is particularly well suited to spectroscopic imaging of auroral emission from Europa's plumes in multiple atomic lines (e.g., the neutral hydrogen Lyman-$\alpha$ line at 121.5 nm and the neutral oxygen line at 130.4 nm). Furthermore, the large apertures of the LUVOIR telescopes may permit emission from individual jets to be spatially resolved. This key feature will aid disentangling plume strength vs. plume morphology.

As a proof-of-concept, the LUVOIR Study Team has designed a LUMOS monitoring campaign for Europa, consisting of a bi-monthly cadence of observations over five years (details in **Section 3.5** and **Appendix B.4.2**). We chose to omit Enceladus from the FUV program for the present, since the different magnetic environment of Saturn leads to uncertainty about the strength of any FUV auroral plume emission. The FUV spectral images of Europa will be dithered to achieve post-processed mosaics with the best possible spatial resolution. The ocean worlds observing program also includes time for two visits per year of optical/NIR imaging of 6 ocean moons (Europa, Ganymede, Callisto, Enceladus, Titan, and Triton). These high-resolution images will be taken in narrow filters placed to highlight expected surface materials (e.g., water and methane ices) and will reveal changes in surface features indicative of geologic activity (details in **Section 3.5** and **Appendix B.4.3**).

These exciting studies only scratch the surface of the solar system remote sensing that LUVOIR can do. For example, LUVOIR can provide up to about 25 km imaging resolution at Jupiter in visible light, permitting detailed monitoring of atmospheric dynamics in Jupiter, Saturn, Uranus, and Neptune over long timescales. LUVOIR's large field-of-regard will permit observations of Venus at maximum elongation and planets with orbits farther from the Sun than Earth's at opposition. Additional solar system science cases appear in **Appendix A**.

## 1.4  The exoplanet family

| Chapter 4 |

When we look at the solar system, it can appear delicately balanced to produce a living world, with our largest planet—Jupiter—playing a dominant role. However, exoplanet discoveries have shown that the planet formation process is robust and leads to a wide range of outcomes. How do we understand the solar system in this context? By exploring the character of many kinds of exoplanets (**Figure 1-12**), studying the formation and evolution of planetary systems over a wide time span, and looking for clues to the solar system's formation recorded in its minor bodies. Only then can we truly know what is typical about our own system and what is unusual.

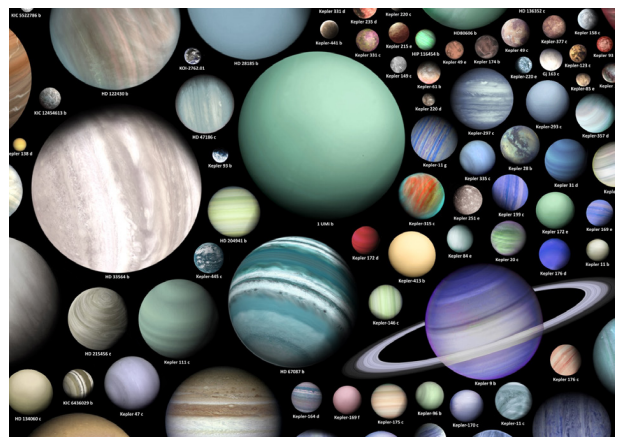

**Figure 1-12.** *Artist's conceptions of some of the 3,706 confirmed exoplanets (as of March 16, 2018). Credit: M. Vargic*

## 1.4.1 Comparative exoplanetology

**Section 4.1**  The capabilities that will allow LUVOIR to investigate dozens of potentially habitable worlds will also allow comprehensive study of a huge range of exotic exoplanets, which by and large will be easier to observe than Earth-like exoplanets. Many of these comparative planetology studies cannot be done by studying the planets of the solar system alone; for example, investigations of the effects of host star mass, stellar metallicity, system age, and a broad range of planet masses and orbits. For this diverse set of planets, the primary objectives are to 1) measure the abundances of the primary atmospheric constituents, 2) characterize hazes and clouds, and 3) assess atmospheric loss rates. These objectives will be achieved with a combination of direct and transit spectroscopy, both obtained with LUVOIR.

The initial 2-year survey for habitable planet candidates will uncover large numbers of other types of exoplanets, $648^{+251}_{-312}$ for LUVOIR-A and $576^{+166}_{-260}$ for LUVOIR-B. These planets will range from hot rocky planets to cold giant planets (**Table 1-2**). Colors

### Signature Science Case #4: Comparative Atmospheres

#### Science Objective
Investigate the properties of a wide range of exoplanets to provide fundamental data on planet formation and evolution.

#### Description
Measure abundances of key molecules (e.g., $CH_4$, $H_2O$, $CO_2$, $CO$, $NH_3$), characterize hazes, and measure atmospheric loss rates via direct and transit spectroscopy.

#### Key Functional Requirements

| Inscribed telescope diameter | $\gtrsim 6.7$ m |
|---|---|
| **Total time** | |
|     LUVOIR-A | 46 days |
|     LUVOIR-B | 49 days |
| **Inner working angle** | $\lesssim 4\,\lambda/D$ |
| **Raw contrast** | $\lesssim 1 \times 10^{-10}$ |
| **Wavelength range** | |
|     Direct | 500–1700 nm |
|     Optical/NIR transit | 200–2500 nm |
|     FUV transit | 100–140 nm |
| **Spectral resolution** | |
|     Direct | $R \gtrsim 70$, SNR $\gtrsim 15$ |
|     Optical/NIR transit | $R = 15$–$500$ |
|     FUV transit | $R \gtrsim 30{,}000$ |

for all planets and orbits for some will be obtained in the initial survey. A limited number of partial spectra capable of detecting the presence of $H_2O$ and/or $CH_4$ will also be obtained during the survey. However, comprehensive studies will require separate follow-up observing programs.

***Direct spectroscopy.*** The atmospheres of giant planets in the solar system show decreasing fractions of heavy elements to hydrogen (metallicity) with increasing planet mass, which has been interpreted as a key indicator of formation history (**Figure 1-13**). The few transiting exoplanets with atmospheric metallicity measurements appear to follow the solar system trend, but are limited to hot planets that are structurally different from cooler giant planets. LUVOIR will test whether these preliminary trends hold for a wider range of planets including ones more like those in the solar system.

For terrestrial planets, atmospheric composition is a signature of initial atmospheric formation modulated by escape, crustal recycling, and other endogenic and exogenic processes (such as volcanoes and planetesimal impacts). While scientists expect that the atmospheric

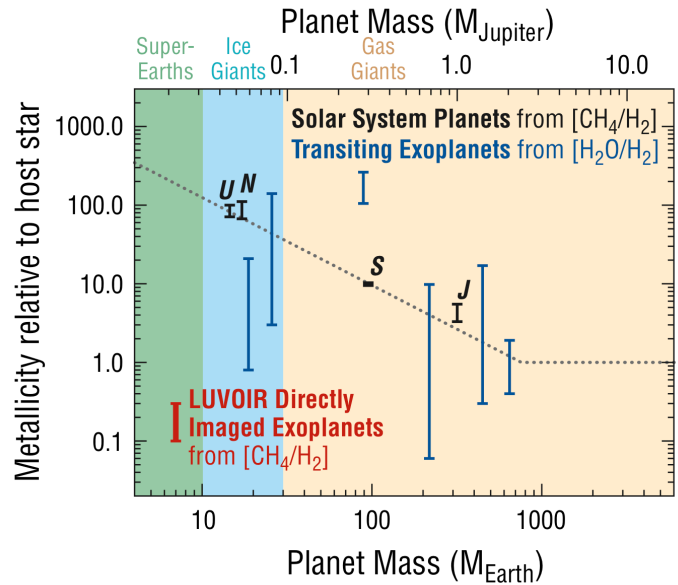

**Figure 1-13.** *LUVOIR can determine if extrasolar giant planets formed like solar system giants. The plot shows measured metallicities for solar system giant planets and a few exoplanets relative to their host stars. The dashed line shows the notional trend of decreasing metallicity with increasing planet mass predicted by the core accretion theory. The red error bar shows the expected precision of giant exoplanet metallicity measurements with LUVOIR, based on retrieval studies for WFIRST (Lupu et al. 2016). Credit: J. Bean (U of Chicago)*

chemistry of hydrogen-helium dominated giant planets can be grossly predicted, the atmospheric diversity of terrestrial planets discovered by LUVOIR will probably span a wider range. LUVOIR will assess that range, and begin to discover the classes of atmospheres and the processes that shape them.

Direct spectra of both giant and rocky exoplanet atmospheres will be needed for these investigations. **Figure 1-14** shows simulated LUVOIR ECLIPS spectra of reflected light from an assortment of exoplanets. For Neptune-size and larger planets, very high-quality spectra may be obtained in short exposure times. Atmospheric metallicity can be measured from these spectra using $CH_4$, $H_2O$, and $NH_3$ absorption bands in the optical and NIR. For terrestrial planets, the same molecular species targeted in habitable planet candidates can reveal bulk atmospheric composition and the effects of atmospheric escape processes. The short-wavelength capability of ECLIPS is critical for constraining photochemical hazes, as they typically absorb most strongly at wavelengths ≲ 500 nm.

The LUVOIR Study Team has designed a program to obtain direct optical/NIR spectra for a diverse set of currently known exoplanets with masses measured from RV (details in **Section 4.1.1** and **Appendix B.5.2**). The planets range from cold Jupiter analogs to warm sub-Neptunes. Spectra of 30 planets will be obtained with LUVOIR-A or 19 with LUVOIR-B. Inevitably, direct spectra of some non-habitable rocky planets will be obtained during the exoEarth characterization program. Together, these two programs will characterize a wide range of warm-to-cold gas-rich and rocky non-habitable exoplanets.

***Transit spectroscopy.*** In addition to direct spectroscopy, LUVOIR will also enable high quality transmission spectroscopy of transiting planets. These observations expand the range of planets LUVOIR can investigate to ones with higher temperatures, orbiting closer to their host stars. Furthermore, LUVOIR will be able to perform transmission spectroscopy of warm rocky planets transiting small red dwarf stars, which are difficult to study using direct spectroscopy due to their close proximity to their host stars.

JWST will begin transit spectroscopy studies of temperate planets around M stars. Transit spectroscopy with HDI on LUVOIR will complement those investigations by extending the transmission spectra to bluer wavelengths (200–800 nm). Between 800 nm and 2500 nm,

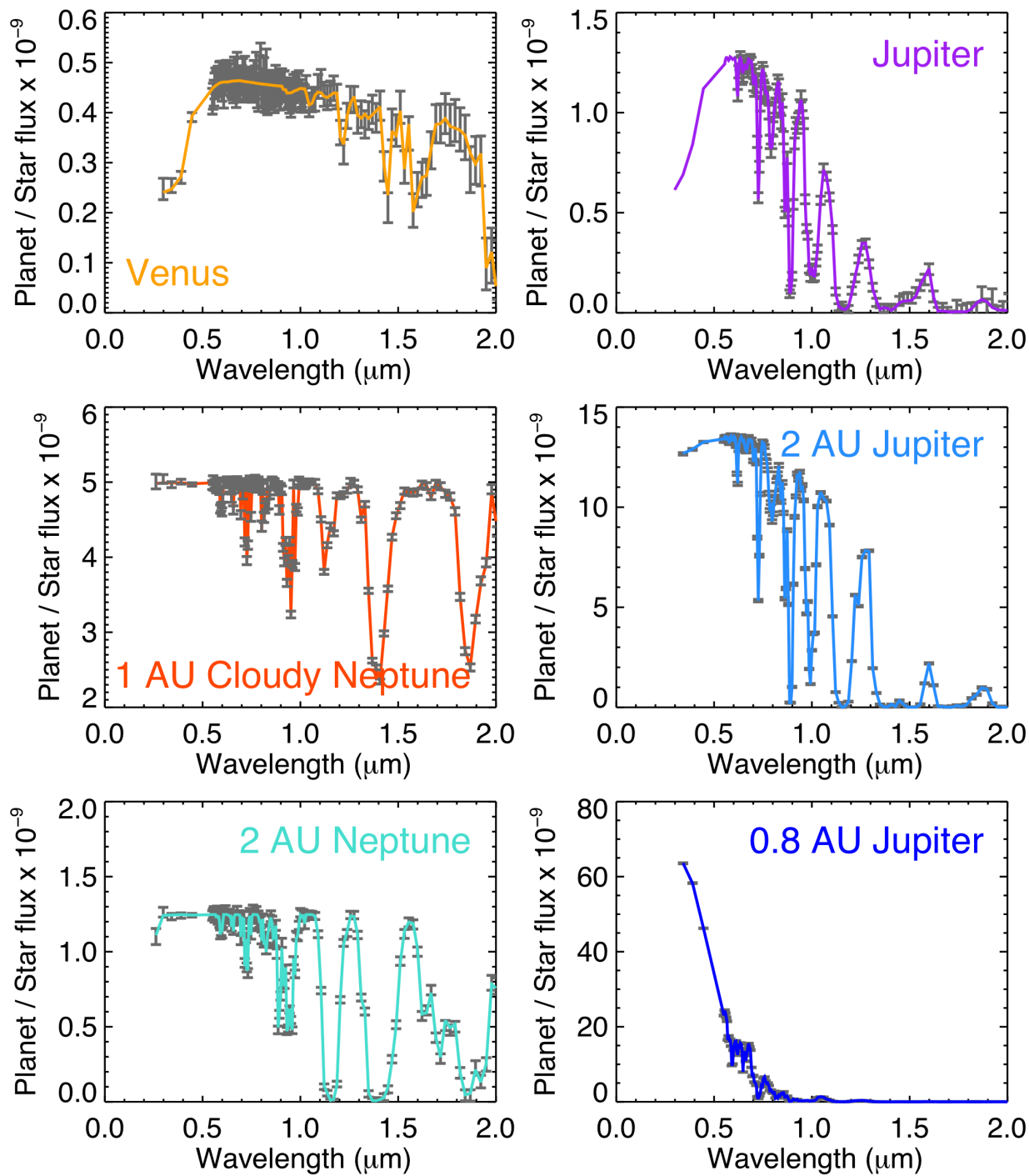

**Figure 1-14.** *LUVOIR can study a wide-range of exoplanets. The panels show simulated reflection spectra of planets orbiting a Sun-twin star at 10 pc that could be simultaneously obtained during an observation deep enough for an SNR=10 spectrum of an Earth-like planet (100/230 hours per coronagraph band for LUVOIR-A/B). Input model spectra are over-plotted with colored lines. The Venus model spectrum is based upon an atmosphere model in Arney et al. (2014). The warm Jupiter model spectra are from Cahoy et al. (2010) and the warm Neptune model spectra are from Hu & Seager (2014). Credit: LUVOIR Tools / G. Arney (NASA GSFC)*

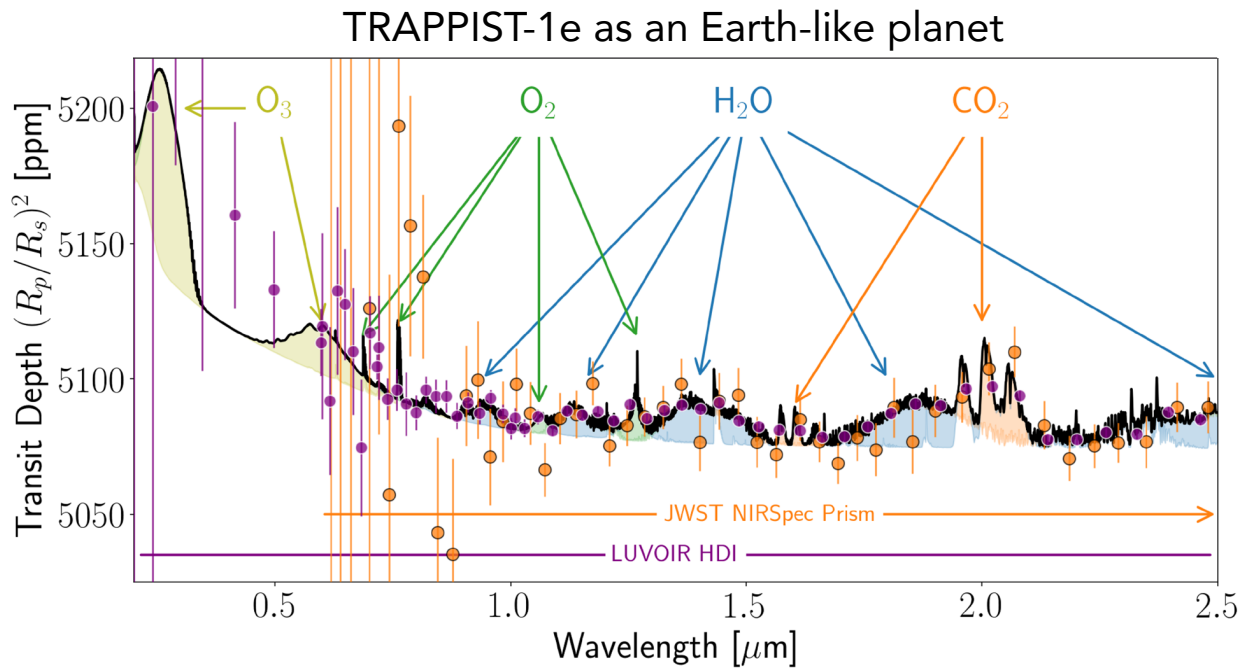

**Figure 1-15.** *Simulated spectra of TRAPPIST-1 e for 50 observed transits with JWST's NIRSpec prism (orange) and 50 transits observed with LUVOIR-A HDI (purple), assuming a water-covered planet consistent with VPL models from Lincowski et al. (2018). Signatures of multiple key atmospheric species are visible, enabling constraints on habitability. Credit: J. Lustig-Yaeger (UW)*

the greater collecting area of LUVOIR will permit improvements in sensitivity compared to JWST transit spectroscopy. At these wavelengths, thermal emission from the warm LUVOIR telescope (see **Section 1.10.1**) is negligible compared to Poisson noise from the bright stars and will not limit the photometric precision needed for transit spectroscopy. The inherent stability of the LUVOIR observatory will also likely translate into improvements in systematic limits on photometric precision.

The large apertures of the future ELTs will provide the small IWAs necessary for direct spectroscopy of warm rocky planets around nearby M dwarfs. Transit spectroscopy with the LUVOIR HDI instrument complements those studies by providing access to the 200–800 nm wavelength region that is difficult (due to limits of adaptive optics systems) or impossible (due to low transmission of Earth's atmosphere) for the ELTs. This wavelength range contains strong absorption bands of $O_2$ and $O_3$, key potential biosignature species. A simulated LUVOIR transmission spectrum of the small, warm transiting exoplanet TRAPPIST-1e appears in **Figure 1-15**.

The LUVOIR Study Team has designed a program to obtain optical/NIR transmission spectra for a set of 16 currently known transiting exoplanets. These planets have measured masses from RV and include warm-to-hot super-Earths and sub-Neptunes (details in **Section 4.1.2** and **Appendix B.5.3**). The spectra will be of sufficient quality to measure cloud layers and atmospheric molecular abundances well enough to provide meaningful constraints on bulk abundances and C/O ratios (SNR ≈ 7–250 at 1800 nm). The number of transits for each planet observed is the same for both LUVOIR-A and -B. Therefore, the total program time is the same (23 days), but with lower SNR in the LUVOIR-B spectra.

***Atmospheric escape.*** The loss of atmospheric constituents over a planet's history fundamentally alters the composition of the remaining atmosphere. A dramatic example is the dissociation of water and subsequent loss of hydrogen from the atmosphere of Venus, leading to a dense, $CO_2$-rich atmosphere. Understanding the relative roles of escape, outgassing, and accretion in a variety of exoplanet atmospheres is critical to understanding the origin and evolution of planetary atmospheres. In particular, atmospheric escape is a key factor shaping the evolution and distribution of low-mass, close-in planets (e.g., Owen & Wu 2013) and their habitability (e.g., Cockell et al. 2016). Many highly irradiated rocky planets may in fact be the remnant cores of evaporated Neptune-mass planets (e.g., Lopez et al 2012).

LUVOIR provides the key ability to measure atmospheric loss rates for exoplanets of different masses orbiting a variety of stars. Far-UV transit spectroscopy is necessary for these studies, as unbound exospheres are atomic and ionic in composition and the strong resonance absorption lines of most atoms and ions lie in the far-UV. This capability will be lost after the end of Hubble. These observations will be performed using stellar emission lines, which arise from stellar activity, as the background light source. Therefore, it is important to obtain the necessary SNR in a single transit, to mitigate the effects of varying stellar emission (Bourrier et al. 2017).

The LUVOIR Study Team has designed a UV transmission spectroscopy observing program including 10 Neptune to sub-Neptune sized exoplanets orbiting Sun-like stars and 6 sub-Neptune to super-Earth sized exoplanets orbiting M dwarf stars (details in **Section 4.1.3** and **Appendix B.5.4**). These targets were drawn from the simulated catalog for the Transiting Exoplanet Survey Satellite (TESS) mission (Barclay et al. 2018), but would certainly be modified based on future discoveries. The exposure times for each planet were set to achieve SNR $\gtrsim 8$ detections of exospheric absorption from hydrogen, oxygen, and carbon in a single transit. Given the need to obtain detections in a single transit, the LUVOIR-B program time is the same as the LUVOIR-A time, but with lower SNR.

## 1.4.2 Building planetary systems

 Beyond individual exoplanets, LUVOIR will provide unprecedented measurements to help understand the processes that give rise to the extraordinary diversity of planetary systems found over the last decades. LUVOIR will examine the formation and evolution of planetary systems over a wide range of stellar ages by 1) executing unprecedented spectroscopic studies of the most abundant materials in protoplanetary disks, 2) revealing the morphology of young planetary systems, and 3) providing the data for dynamical studies of mature planetary systems to uncover clues to their formation eons ago.

***Protoplanetary disks.*** Planet formation theories were originally built to describe the birth of the solar system. As we learned more about protoplanetary disks by observing them in star forming regions (**Figure 1-16**), even the solar system became difficult to explain. When it comes to exoplanets, our planet formation theories did not predict and cannot explain the wide range we have already discovered. We are not able to trace the planet formation process all the way from its beginning in gas and tiny dust grains to any fully formed planet, and it is abundantly obvious that important processes are missing from our theories.

Understanding the characteristics of planet-forming disks around young (1–10 million year-old) stars is furthermore critical to the search for habitable worlds. While LUVOIR's

ECLIPS coronagraph searches nearby stellar systems for inhabited planets, HDI and LUMOS will provide unique capabilities for characterizing the environments in which planets assemble and develop their nascent atmospheres. LUVOIR can quantify the evolution of material at planet-forming radii and provide new constraints on the dispersal of protoplanetary material through disk winds.

Far-UV spectroscopy is a unique tool for observing molecular gas in the inner regions of protoplanetary disks. The strongest spectral features of $H_2$ and CO, the most abundant gaseous species in protoplanetary disks, reside in the 100–170 nm wavelength range. $H_2$ is particularly important, as it makes up the bulk of the disk mass but is notoriously hard to observe at longer wavelengths. This has led some to term $H_2$ "the dark matter of planet formation." Observing cold-to-warm $H_2$ in the FUV is one of the key goals driving the 100 nm short-wavlength cutoff of LUVOIR.

The LUVOIR Study Team has designed a LUMOS spectroscopic survey of a large number of disks with different ages within five areas of the Orion Star Forming Region (details in **Section 4.2.1** and **Appendix B.6.2**). These spectra will reveal the disks' most abundant gas components and other key molecular gas species (e.g., $H_2O$, $CH_4$, and $CO_2$; **Figure 1-16**). Amazingly, a single LUMOS map of one dense area like the Orion Nebula Cluster, taking < 25 hours with LUVOIR-A, will provide a higher quality dataset than the entire medium-resolution UV spectroscopic archive of protoplanetary disks from 29 years of HST observations.

What sets the timeframe for giant planet formation? The same LUMOS spectra will also provide key measurements of the thermal and magnetohydrodynamic disk winds thought to drive disk evolution and dispersal. Winds may control the final masses of giant planets by starving gas accretion onto late-forming planet cores. Winds also affect planetary orbits by dispersing gas from preferential radial distances and thereby halting planet migration. LUVOIR's combination of high spatial resolution and UV spectroscopic coverage will provide our first maps of the launching regions of protostellar jets and disk winds,

<div style="border:1px solid">

### Signature Science Case #5:
### The formation of planetary systems

#### Science Objective
Study the formation and evolution of planetary systems over a wide age range.

#### Description
Measure the most abundant molecular species in ($H_2$, CO, $H_2O$, and OH) in Orion protoplanetary disks and trace disk winds via FUV spectroscopy. Map dust structures in young planetary systems (debris disks) via NUV/optical high-contrast imaging. Measure the masses and orbits of planets in mature systems via astrometry.

#### Key Functional Requirements

| Inscribed telescope diameter | $\gtrsim 6.7$ m |
|---|---|
| **Total time** | |
|    LUVOIR-A | 25 days |
|    LUVOIR-B | 59 days |
| **UV spectroscopy** | |
|    Wavelength range | 100–400 nm |
|    Spectral resolution | $R \approx 30,000$ |
|    Field-of-view | $\gtrsim 4$ sq. arcmin |
| **High-contrast imaging** | |
|    Wavelength range | 200–550 nm |
|    IWA | $\lesssim 4\lambda/D$ |
|    Raw contrast | $\lesssim 1 \times 10^{-9}$ |
| **Astrometry** | |
|    Wavelength | $\sim 550$ nm |
|    Final astrometric precision | $\lesssim 0.1\ \mu$as |
|    Field-of-view | $\gtrsim 6$ sq. arcmin |
| ~14 to 40 epochs per star over $\gtrsim 1$ year | |

</div>

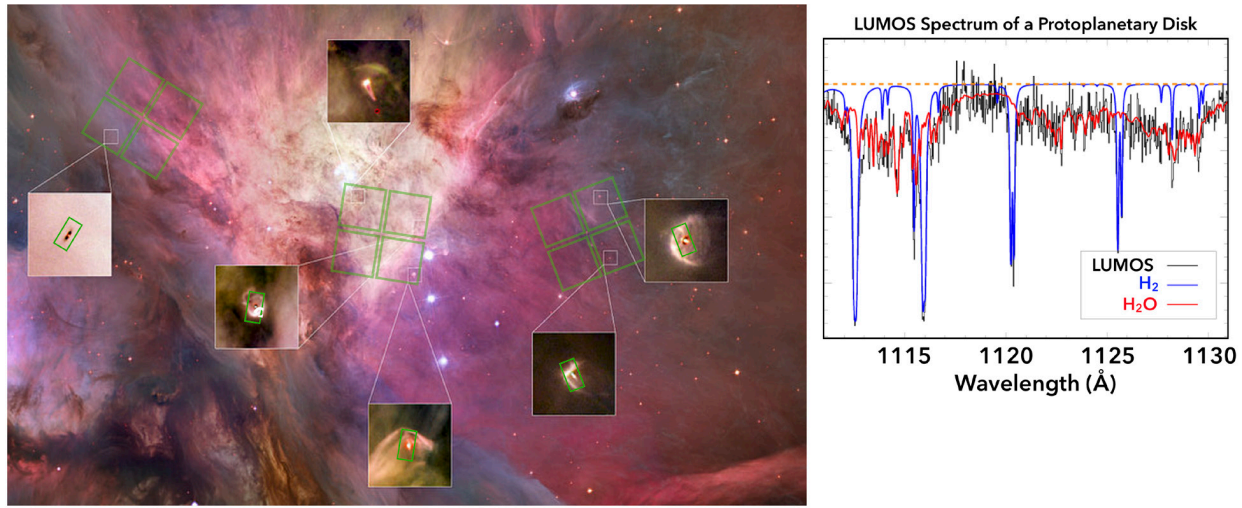

**Figure 1-16.** *LUVOIR can powerfully and efficiently measure the hidden materials of planet formation. Left: The Orion Star Forming Region with the LUMOS multi-object spectroscopic field of view overlaid three times (green squares). The blow-up panels show several planet-forming disks that fall in each field, with LUMOS spectroscopic apertures overlaid (green rectangles). UV spectra of all the disks in each field can be obtained simultaneously. Right: Simulated partial LUMOS spectrum of a protoplanetary disk showing absorption from two key molecular species ($H_2$ and $H_2O$). Credit: NASA / ESA / Hubble / K. France (CU – Boulder) / J. Tumlinson (STScI)*

providing a fundamentally new empirical framework for the relationship between star- and planet-formation.

*Young exoplanet systems.* LUVOIR can also study in real time the violent formation of rocky worlds and sculpting of planetary systems. Massive belts of asteroids and comets—the building blocks of planets—are colliding around nearby young stars, producing bright disks of dust and gas (debris disks). The ages of these systems (~10 to hundreds of millions of years) correspond to the timescale for terrestrial planet formation and early dynamical evolution of planetary systems, opening a window on processes not directly observable in mature systems.

High-contrast imaging can reveal several types of debris disk structures that shed light on planet formation processes. These structures include bright rings of dust marking the locations where Pluto-sized bodies have formed (Kenyon & Bromley 2004), as well as gaps, clumps, and warps in the dust caused by dynamical interactions between planetesimals and newly formed planets (**Figure 1-17**). Finding such structures in debris disks of different ages will place constraints on timescales for formation of various planetary bodies and allow investigation of differences with stellar type.

Many debris disks have been directly imaged with current high-contrast instruments (e.g., Hubble, Gemini Planet Imager, SPHERE). Unfortunately, these images only reveal the cold outer regions of bright debris disks around early-type stars. The small inner working angle and high spatial resolution of ECLIPS on LUVOIR will allow it to probe within the ice lines of young planetary systems, which in theory divide inner rocky planet formation regions from the cold icy regions where giant planets form.

The LUVOIR Study Team has designed a program to obtain high-contrast direct images of dust inside the ice lines of 25 debris disks (details in **Section 4.2.2** and **Appendix B.6.3**).

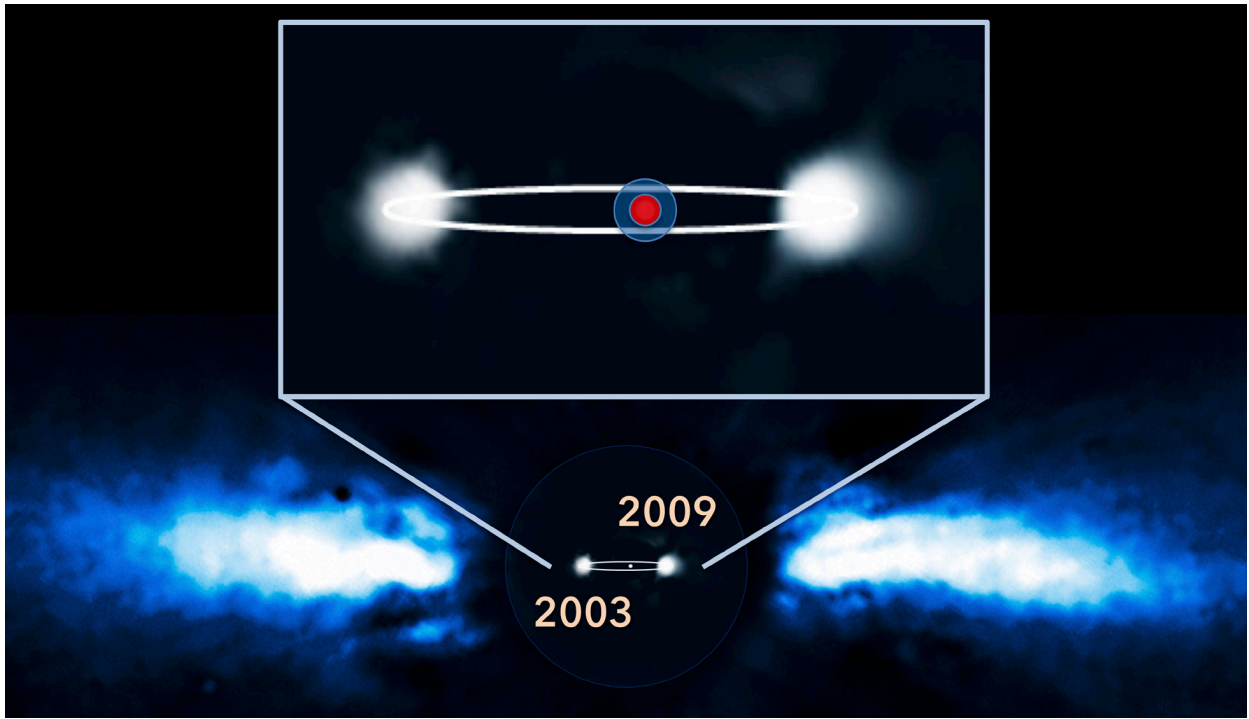

**Figure 1-17.** *LUVOIR can peer into the unseen inner regions of young planetary systems. This image shows the Beta Pictoris debris disk, with the gas giant exoplanet Beta Pic b imaged at two epochs in its orbit (Lagrange et al. 2010). This planet is responsible for the long known "warp" in the inner part of the dust disk (e.g., Heap et al. 2000). The inset panel shows a blow-up of the innermost region with the coronagraph inner working angles at 550 nm for LUVOIR-A (red circle) and LUVOIR-B (blue circle) overlaid. Credit: ESA / A.-M. Lagrange / A. Roberge (NASA GSFC)*

The host stars are in nearby young ($\lesssim$ 100 million year-old) stellar associations within 20–60 pc and will have a variety of spectral types (A, F, G, K, M). The optical wavelength imaging observations were designed to reach SNR~60 at the ansae of a dust ring with 200 times the level of the solar system zodiacal dust. For a subset of the brightest targets, high-contrast direct images at NUV wavelengths will permit a search for cometary OH emission.

***Mature exoplanet systems.*** The architecture of a whole planetary system includes the number, masses, and orbits of the planets; as well as the locations of asteroid and comet (planetesimal) belts. The arrangement of planets and belts in a system puts strong constraints on phenomena that occurred during formation and early evolution (e.g., migration). For example, say a certain system lacks an Earth-size planet in the habitable zone. Understanding why that is the case requires knowledge of whether the system has a massive planet that could have prevented formation in the first place or ejected the smaller body after formation.

In the old ($\gtrsim$ 1 Gyr) exoplanet systems LUVOIR will study, orbits and masses of planets more massive than Neptune will likely be determined via indirect means (e.g., ground-based RV, Gaia astrometry). For less massive planets, LUVOIR can measure both orbits and masses of planets using astrometry. This high-precision astrometry will be enabled by the combination of 1) a pixel geometry calibration system within the HDI, which uses a laser interferometer to measure detector distortions (Crouzier et al. 2016; more info in **Section 1.11.1**); and 2) the high sensitivity of LUVOIR, which provides enough astrophysical background sources to correct optical distortions.

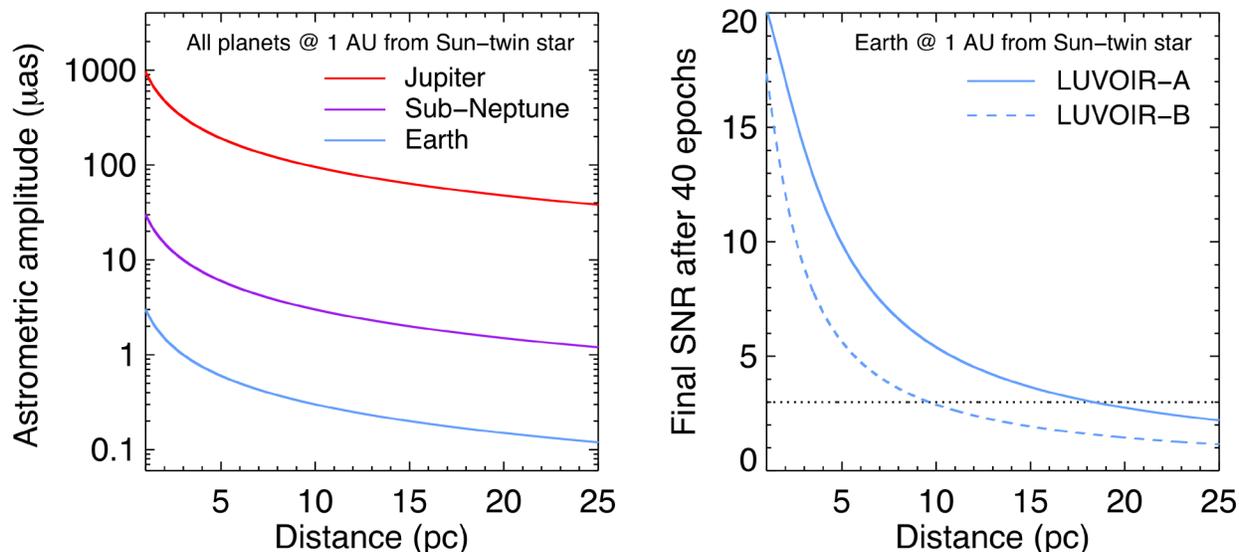

**Figure 1-18.** *LUVOIR can measure the astrometric signals of Earth-mass planets orbiting Sun-like stars and provide critical mass measurements for habitable and non-habitable planets. Left panel: Solid colored lines show the astrometric wobbles of a solar analog star due to a Jupiter-twin (red), a 10 Earth-mass sub-Neptune (purple), and an Earth-twin (blue), all with an orbital semi-major axis of 1 AU. Right panel: Predicted SNR for astrometric detections of Earth-twins around solar analogs for LUVOIR-A (solid line) and LUVOIR-B (dashed line). The curves assume 40 epochs of observations and noise from stellar jitter due to star spots, systematic noise set by detector metrology, and observational uncertainty. Using astrometry+direct imaging, a mass measurement with 25% precision requires SNR=3. Credit: D. Windemuth (UW) / A. Roberge (NASA GSFC)*

Measurement of exoplanet masses with 25% precision requires end-of-mission detection of the astrometric signal at SNR=3, if the astrometry is obtained in conjunction with coronagraphic images (e.g., Guyon et al. 2013). **Figure 1-18** indicates that ~3σ detections of the astrometric signals from Earth-mass exoplanets orbiting Sun-like stars can be obtained with a reasonable number of repeat visits for systems within at least 18 pc with LUVOIR-A and systems within about 10 pc with LUVOIR-B.

The LUVOIR Study Team has designed an exoplanet astrometry program to measure the orbits and masses of exoplanets with masses ≥1 Earth-mass orbiting the host stars of all habitable planet candidates found in Signature Science Case #1 (details in **Section 4.2.3** and **Appendix B.6.4**). That corresponds to ~50 stars for the LUVOIR-A program or ~30 stars for LUVOIR-B. The exposure times for the individual astrometric imaging observations are short; therefore, total astrometry times are dominated by overheads for repointing to the targets and calibration exposures. However, a good deal of future work must be done to more accurately determine LUVOIR's astrometric capability, including modeling of signal extraction in multi-planet systems and construction of a complete astrometric error budget.

Planets also imprint their gravitational signatures on planetesimal debris (a.k.a. exozodiacal dust), providing evidence of exoplanets too faint to be directly imaged or with periods too long for indirect detection methods (**Figure 1-19**). Although sufficiently high levels of exozodiacal dust can obscure exoplanets from view, planetesimal belts and the dusty debris they produce serve as important records of the system's early history and provide constraints

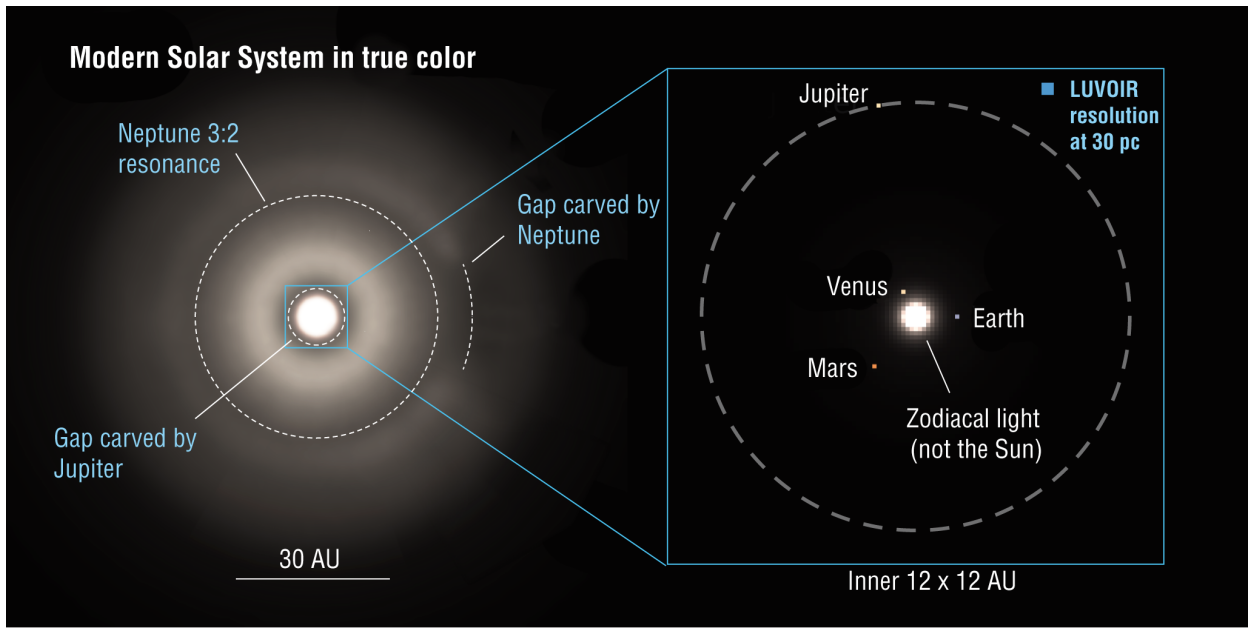

**Figure 1-19.** *The interplay between planets and planetesimal belts. The left panel shows a model of the entire solar system out to a radius of 50 AU from the Sun, while the inset panel zooms in on the inner system (Roberge et al. 2017). For ease of viewing, the Sun and possible astrophysical background sources are not included. The bright region at the center of the image is emission from warm debris dust (a.k.a. exozodiacal dust). Two circular gaps in the dust are visible, the inner one caused by Jupiter and the outer one marking the 3:2 mean motion resonance with Neptune. Neptune also creates a partial gap in its immediate vicinity. Credit: A. Roberge, T. B. Griswold (NASA GSFC) / R. Dawson (Penn State)*

on present-day orbital properties. For example, a planetesimal belt located near a planet limits how that planet's orbit has (or has not) evolved over time.

Full dynamical modeling of entire planetary systems will be a rich area of study in the LUVOIR era. Inevitably, warm inner dust distributions will also be observed during the initial 2-year exoplanet direct imaging survey. Information on dust distributions in the cold outer regions can be obtained from the ground with the Atacama Large Millimeter Array (ALMA). Therefore, no additional LUVOIR observing time is required to include planetesimal belts in dynamical studies of whole mature planetary systems.

### 1.4.3 Records of the formation of the solar system

**Section 4.3** There are many things still to be discovered and understood about the bodies within the solar system itself. Sensitive, high-resolution imaging and spectroscopy of solar system asteroids, dwarf planets, comets, and Trans-Neptunian Objects (TNOs) that will not be visited by spacecraft in the foreseeable future can provide vital information on the processes that formed the solar system ages ago. In particular, the Kuiper Belt is the only available remnant of the solar system's primordial planetesimal population accessible for direct study. The characteristics of this population, including both size and orbital distributions, are directly relevant to understanding process of planetesimal formation and the migration of the giant planets. In particular, the size distribution of the smallest bodies provides critical constraints on theories of the formation of the first planetary building blocks.

Hubble remains the most sensitive optical telescope ever built, with the ability to detect TNOs with diameters in the range of ~35 km at 40 AU. However, this limit may leave 99% of the inner TNO population invisible to observers. Deep imaging programs with LUVOIR can detect smaller TNOs than any other current or planned observatory: ~2 km bodies at 40 AU with LUVOIR-A and ~4 km with LUVOIR-B. These size limits are two orders of magnitude smaller than the limit for the Large Synoptic Survey Telescope (LSST) wide field survey and an order of magnitude smaller than the limits for proposed or potential deep drills with JWST and ground-based ELTs. The LUVOIR Study Team has designed an ambitious program of sensitive optical and NIR imaging (down to R=33 with LUVOIR-A or R=31.5 with LUVOIR-B) to search for small TNOs, with expected returns of up to ~100 objects (details in **Section 4.3** and **Appendix B.**7).

Furthermore, binary TNOs provide a highly valuable opportunity to further constrain theories of the formation and evolution of the solar system. Their separations are evidence of their original formation location, and different models of planetesimal formation make different predictions for the component sizes, mass ratios, separations, and occurrence rates of binaries. In addition, determining binary orbits allows us to measure their masses and place constraints on densities, providing key information for understanding their composition. The resolving power and point-source sensitivity of LUVOIR make it uniquely suited to find and resolve binary TNOs that are smaller and more distant than those that will be discovered with other facilities (e.g., LSST, JWST, and the ELTs).

## 1.5 The structure of the cosmos

| Chapter 5 |

The Universe began smoothly, but is now richly structured on all scales. The prevailing cold dark matter (CDM) paradigm for structure formation describes how the first seeds of structure evolved into the dynamic shape of the "cosmic web" of filaments and voids traced out by galaxies (**Figure 1-20**). The most

> ### Signature Science Case #6:
> ### Small bodies in the solar system
>
> #### Science Objective
> Investigate the formation and early evolution of the solar system.
>
> #### Description
> Execute a deep optical/NIR imaging program to detect very small Trans-Neptunian objects (2–4 km size). Measure the orbits of binary TNOs.
>
> #### Key Functional Requirements
>
> | | |
> |---|---|
> | Telescope diameter | ≳ 8 m |
> | Total time | 5 months |
> | Wavelengths | 775 nm & 1260 nm |
> | Imaging field-of-view | ≳ 6 sq. arcmin |
> | Non-sidereal tracking of moving targets | |
> | Multi-epoch observations | |

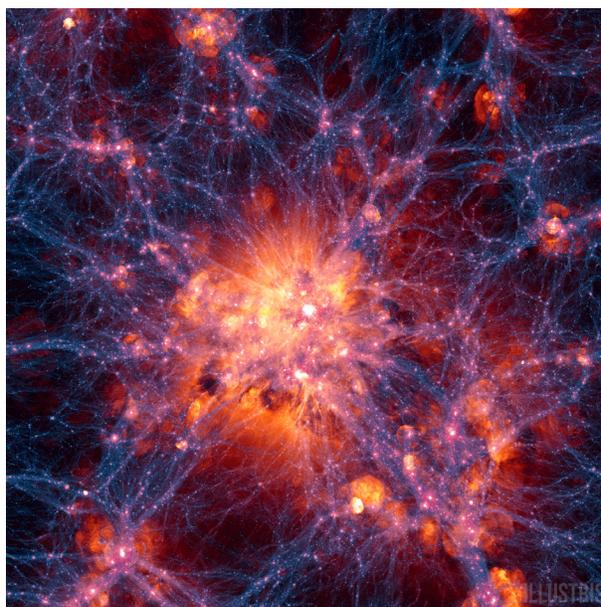

**Figure 1-20.** *Theoretical simulation of galaxy cluster formation in the presence of the dark matter web. Credit: Illustris Collaboration*

current CDM models accurately predict the numbers and properties of "dark matter halos" where galaxies and clusters of galaxies form. However, fundamental questions remain unsolved: What is the dark matter? Do dark matter-based models of structure formation correctly predict small (sub-galactic) size scales? Does dark matter decay? CDM models were built to explain large-scale structure. Whether they can explain smaller structures is a key test that will reveal whether we have captured the physics of interactions between normal matter, dark matter, and ionizing light.

To address these questions, LUVOIR will deploy an unprecedented combination of spatial resolution and photometric sensitivity at UVOIR wavelengths. With the ability to resolve 60 pc scales anywhere in the universe and detect sources down to AB=33 mag in just 10 hours, LUVOIR will open an entirely new regime in relating luminous to dark matter. LUVOIR will also be able to measure miniscule proper motions of nearby galaxies and stars to trace their dynamics under the influence of dark matter. Finally, LUVOIR's UV spectroscopic capabilities will reveal the emergence of structure and light from the cosmic dark ages.

| **Signature Science Case #7:** Connecting the smallest scales across cosmic time | |
|---|---|
| **Science Objective** | |
| Discriminate between dark matter models by measuring the shape of the matter power spectrum on <100 kpc scales. | |
| **Description** | |
| Measure the spatial distribution of extremely low-mass dwarf galaxies around four Milky Way analog galaxies within 15 Mpc. | |
| **Key Functional Requirements** | |
| Telescope diameter | $\gtrsim$ 8 m |
| Total time | |
|     LUVOIR-A | 12 days |
|     LUVOIR-B | 57 days |
| O/NIR imaging | V and J bands |
| Spatial resolution | $\lesssim$ 16 mas at 500 nm |
| Field-of-view | $\gtrsim$ 6 sq. arcmin |

## 1.5.1 Constraining dark matter with small-scale structure

**Section 5.1**    Dwarf galaxies are highly sensitive to the nature and distribution of dark matter—such as the particle mass and any self-interactions—because almost all their mass is dark matter. The space density and internal structure of dwarf galaxies are closely connected to the fundamental properties of the dark matter particle, the history of reionization, and the granular limits of the galaxy formation process. Each type of dark matter property manifests itself on astronomical scales through a gravitational signature, in the statistical description of dark matter structure. The measure of this signature is the matter power spectrum today and at earlier times.

The LUVOIR Study Team has designed an observing program to measure the matter power spectrum by mapping the regions around spiral galaxies out to ~15 Mpc to find low-mass dwarf galaxies, far more sensitively than JWST and the ELTs (details in **Section 5.1.1** and **Appendix B.8**). By reaching dwarf sub-halos of $10^7$ M$_\odot$ out to the virial radius of Milky Way-analog galaxies, LUVOIR will measure the matter power spectrum to a precision of 0.1 dex in the mass range where it is most sensitive to the dark matter particle mass and self-interaction cross-section. This program would provide the best available statistical constraints on the small-scale power spectrum over the relevant parameter space. It would also greatly advance our understanding of baryonic galaxy formation processes by characterizing large populations of dwarfs that could then be spectroscopically measured with the ELTs to obtain velocities and chemical abundances.

### 1.5.2 Constraining dark matter with astrometry

**Section 5.2** A complementary constraint on the nature of dark matter comes from using the orbits of stars within dwarf spheroidal galaxies (dSphs) to study their density profiles. These galaxies are dark matter-dominated, and their central density profiles are a sensitive probe of dark matter properties. Those density profiles can be measured using the 3D motions of individual stars within the galaxies. However, for the smallest dSphs that are most dominated by dark matter, current and planned observatories cannot precisely measure transverse velocities for sufficient numbers of stars to accurately determine the galaxies' central density profiles.

| Signature Science Case #8: Constraining dark matter using high precision astrometry | |
|---|---|
| **Science Objective** Discriminate between dark matter models by measuring the density profiles of dwarf galaxies via high-precision astrometry of individual stars. | |
| **Description** High-precision astrometric measurements of proper motions for ~100 stars in 20 dwarf spheroidal galaxies. | |
| **Key Functional Requirements** | |
| Telescope diameter | $\gtrsim$ 8 m |
| Total time | |
| LUVOIR-A | 1 month |
| LUVOIR-B | 3.5 months |
| Astrometry | |
| Systematic single-epoch position errors | $\lesssim$ 0.5 $\mu$as |
| Imaging field-of-view | $\gtrsim$ 6 sq. arcmin |
| 3 epochs per galaxy over 5 years | |

LUVOIR can provide the necessary velocity precision for enough stars, thanks to its angular resolution, sensitivity, and the pixel geometry calibration system within the HDI instrument. The LUVOIR Study Team has designed a program to measure the proper motions of stars within dSphs with ~10 km/s accuracy out to 4 Mpc with LUVOIR-A or 2 Mpc with LUVOIR-B (details in **Section 5.2** and **Appendix B.9**). Combined with radial velocities from the ELTs, these proper motion measurements will provide galaxy density profiles and dark matter maps in cases where it is not currently possible due to the difficulty of obtaining transverse velocities

### 1.5.3 Probing the impact of ionizing light on cosmic structure

**Section 5.3** The major open questions of cosmic structure formation go far beyond the nature of dark matter. The opacity and temperature of the intergalactic medium (IGM) influences where and how gas can enter dark matter halos to form stars, making the history of "reionization" a major unsolved problem. The problem is complex because dwarf galaxies are the predominant sources of ionizing photons and are also most greatly impacted by being ionized and heated.

LUVOIR will push beyond JWST and the ground-based ELTs to constrain the number and distribution of the faintest galaxies at high redshift—those that feel the effects of reionization most directly. **Figure 1-21** shows the redshift to which telescopes of different sizes can detect dwarf galaxies. With deep imaging in the I, J, and H bands, LUVOIR can measure

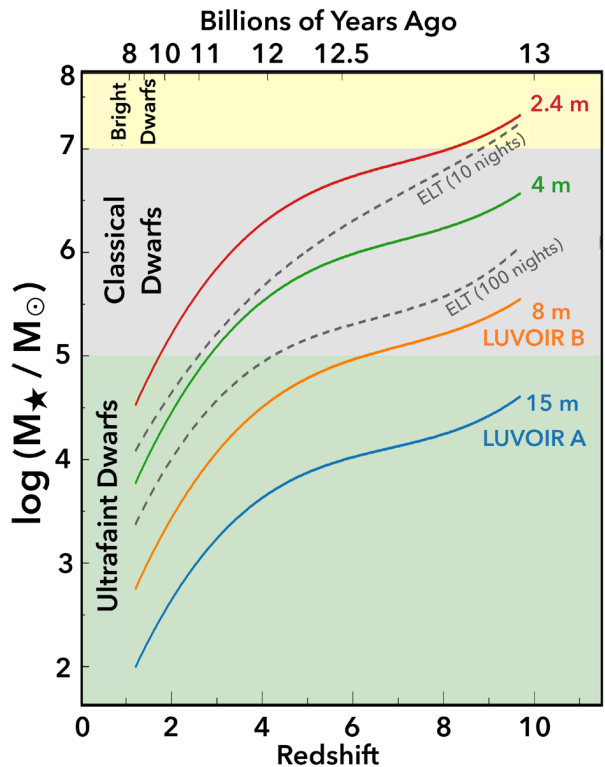

**Figure 1-21.** *LUVOIR can detect smaller galaxies than any other current or planned observatory. The curves show sensitivity to detecting galaxies of a particular total stellar mass (y-axes) as a function of redshift (and lookback time) for different telescopes: a 2.4-m space telescope, a 4-m space telescope, the LUVOIR concepts, and the 39-m ground-based ELT. The limits for the space-based telescopes are for a fiducial 500 ksec (139 hour) observation that returns a SNR=5 detection of a 200-parsec diameter source. The limits for the ELT are for a 10-night (80 hour) integration and a 100-night (800 hour) integration. Results for JWST are similar to those for the ELT 100-night exposure. Credit: M. Postman (STScI)*

### Signature Science Case #9: Tracing ionizing light over cosmic time

#### Science Objective
Study the faint end of the galaxy luminosity function to reveal the degree to which dwarf galaxies powered cosmic reionization. Quantify evolution and escape of ionizing radiation from low-redshift galaxies.

#### Description
Optical/NIR images of 12 blank sky fields to measure the number of high-redshift (z=7) dwarf galaxies per sq. arcmin down to AB≈33 mag (not feasible with LUVOIR-B). UV spectroscopy of ionizing radiation from 500 low-redshift galaxies. Multi-object UV spectroscopy of individual star-forming regions within 100 low-redshift galaxies.

#### Key Functional Requirements

| | |
|---|---|
| Telescope diameter | $\gtrsim$ 15 m |
| **Total time** | |
| LUVOIR-A | 135 days (128 in parallel) |
| LUVOIR-B | 40 days (15 in parallel) |
| **Wavelength range** | |
| Imaging | I, J, and H bands |
| Spectroscopy | 100–200 nm |
| Spatial resolution | $\lesssim$16 mas at 500 nm |
| Imaging field-of-view | $\gtrsim$ 6 sq. arcmin |
| Spectral resolution | R ≈ 500 & 10,000 |

photometric redshifts for galaxies as faint as AB≈33 mag, corresponding to the mass of the Milky Way's Fornax dwarf spheroidal galaxy at redshift z=7 (details in **Section 5.3.1** and **Appendix B.10.2**). This unprecedented depth—two magnitudes deeper than JWST—will definitively constrain the faint end slope of the galaxy luminosity function, which in turn constrains the critical influence of reionization on early galaxy formation.

For LUVOIR-A, these ultra-deep fields can be obtained in parallel with the exoplanet direct observations planned in Signature Science Case #1. While LUVOIR-B could reach the required depths given a very large investment of observing time, it may not be possible to execute the required deep imaging in parallel with exoplanet observations. Thus, the program appears prohibitively costly on LUVOIR-B. For the time being, the LUVOIR Study Team assumes that these ultra-deep fields are only feasible with LUVOIR-A.

In another investigation of reionization, LUVOIR can use its unprecedented UV sensitivity to fully account for the escape of ionizing Lyman continuum (LyC) radiation from star-forming galaxies, active galactic nuclei (AGN), and quasars of all types (details in **Section 5.3.2** and **Appendix B.10.3**). Moreover, LUMOS will enable spatial mapping of LyC escape across the face of low-redshift galaxies, allowing in-depth characterization of the environmental factors that influence escape (details in **Section 5.3.3** and **Appendix B.10.4**). The spatial resolution of LUMOS corresponds to about 1 kiloparsec or less at z < 2, the last 10 billion years of cosmic history, where measuring the LyC directly still requires UV wavelengths. In fact, while JWST will characterize the high-redshift galaxies that might be responsible for the initial reionization of the intergalactic medium (IGM), it cannot see their rest-frame LyC radiation directly because of the cumulative opacity of the intervening IGM all the way from z = 6 to now. LUVOIR will uniquely constrain ionizing radiation escape for hundreds of z < 1.2 galaxies in several deep pointings of LUMOS spectroscopy.

## 1.6 The lives of galaxies

Chapter 6

Galaxies are complex arrangements of stars, gas, dust, radiation, dark matter, and black holes with a web of interactions among them. Every individual galaxy evolves under these diverse influences, resulting in the set of observed galaxies spanning over eight orders of magnitude in stellar mass—from massive giant elliptical galaxies to tiny dwarf galaxies with only a few hundred stars. Our understanding of the physics driving these emergent patterns lags behind our ability to characterize them. We do not know what sets the minimum scale for galaxy formation, or why some massive galaxies are dominated by their bulges while others have none. We do not know how galaxies sustain star formation for much longer than their present gas supply allows, or why the most massive galaxies cease forming stars and become "quenched."

LUVOIR will apply transformational spatial resolution and sensitivity to the problem of how galaxies evolve. The uniquely powerful UV multiplexing of LUMOS will transform our ability to map the gas flows that drive galaxy evolution from the disk to the circumgalactic medium (CGM) and back. In its deep fields, HDI will reach the smallest building blocks of galaxies. And optical imaging with

### Signature Science Case #10: The cycles of galactic matter

**Science Objective**

Determine the abundances of baryons in the IGM and CGM across cosmic time. Characterize the halo around a nearby face-on spiral galaxy in detail. Examine the morphology of gas flows in nearby galaxies.

**Description**

UV absorption spectroscopy towards 100 high-redshift quasars. UV absorption spectroscopy towards 30 quasars behind a face-on spiral galaxy. UV spectroscopy of emission from ~1000 individual stellar clusters in nearby spiral galaxies.

**Key Functional Requirements**

| Telescope diameter | $\gtrsim 8$ m |
|---|---|
| Total time | |
|     LUVOIR-A | 24 days |
|     LUVOIR-B | 81 days |
| Wavelength range | 100–400 nm |
| Spectroscopic field-of-view | $\gtrsim 4$ sq. arcmin |
| Spectral resolution | $R \approx 30{,}000$ |

LUVOIR-A will be able to resolve the interiors and satellites of galaxies at 60 pc scales at any cosmic epoch.

### 1.6.1 Matter flows in and out of galaxies

**Section 6.1** Galaxies are built from gas that surrounds them and fills up their disks. Gas flows from the space between the galaxies (the intergalactic medium; IGM), into the region around a galaxy (CGM), then into the galaxy itself where it can become the fuel for new stars (**Figure 1-22**). Eventually, massive stars, supernovae, and AGN drive material back out. Much of this cycling of matter is currently unobserved, since the gas flows are hot and tenuous.

LUVOIR's uniquely powerful multi-object UV spectroscopy will reach an entirely new discovery space in the exploration of galactic gas flows. LUMOS has an array of 840 x 420 individually configurable microshutters, each of which covers 0.07" x 0.14". This angular resolution corresponds to < 100 pc out to z = 0.1 (400 Mpc) and < 1 kpc at all redshifts. The ability to probe the powerful UV diagnostics of diffuse gas, hot stars, and dust at such small spatial scales over wide fields will permit the very complex interactions of these processes to be studied as a function of local conditions (**Figure 1-23**).

LUVOIR can map the inflows of accretion and the outflows of feedback at the scale of individual star clusters. The LUVOIR Study Team has designed programs of UV absorption spectroscopy towards 100 high-redshift (z > 1) quasars and 30 quasars behind a nearby

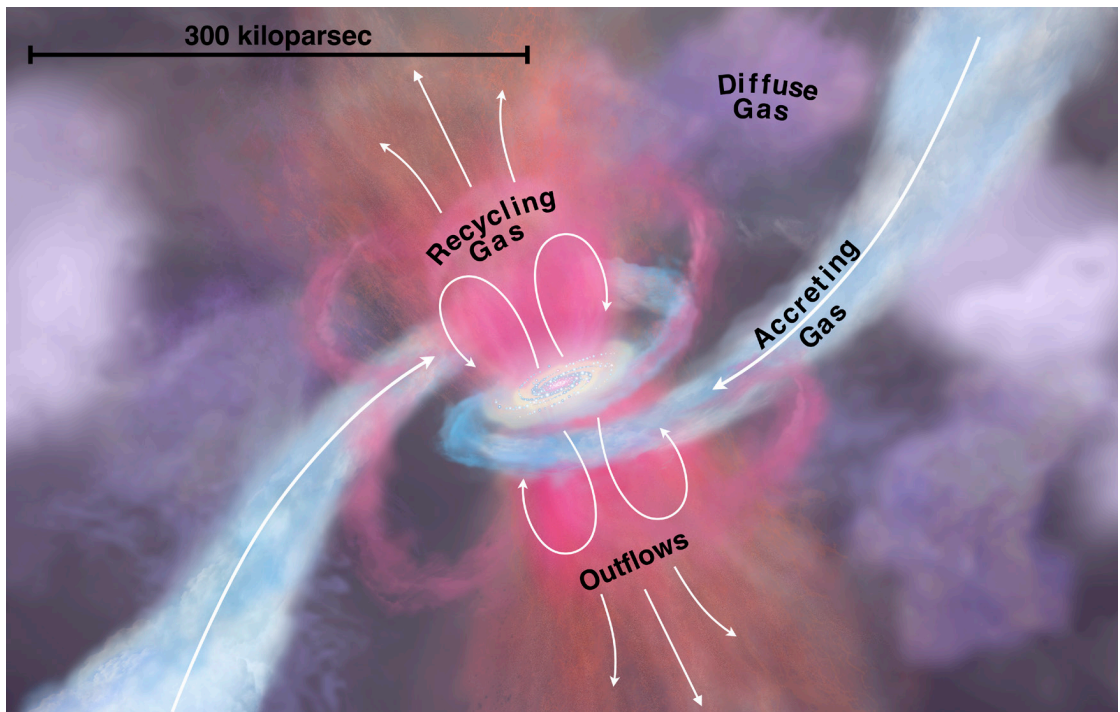

**Figure 1-22.** *Galaxies are much more than they appear to be in starlight. They are surrounded by a massive reservoir of diffuse gas that acts as their fuel tank, waste dump, and recycling center: the circumgalactic medium. It is fed by accretion out of the cosmic web and galaxy outflows, and is a critical factor in galaxy evolution. Unraveling these flows is a major goal for LUVOIR. Credit: Tumlinson, Peeples, & Werk (2017)*

face-on spiral galaxy (details in **Section 6.1.2**, **Appendix B.11.2**, and **Appendix B.11.3**). These programs will probe the temperature, density, and metallicity evolution of IGM / CGM gas and a galaxy halo in enormous detail. In another program, emission from the gas itself as it is ejected in supernova-driven outflows and AGN-powered winds will be directly mapped (details in **Section 6.1.3**, **Section 6.1.4**, and **Appendix B.11.4**). The key diagnostic lines for these z < 2 studies appear in the UV, with the far-UV playing a critical role, making a UV space telescope essential for probing these galaxy formation processes anytime over the last 10 billion years.

### 1.6.2 Galaxy assembly on multiple size scales

**Section 6.2** Since the original Hubble Deep Field in 1996, tracing the growth of galaxies has pushed all the way back to about 400 million years after the Big Bang (z ~ 11). JWST will press further back and farther down the galaxy mass function with rest-frame optical light observed in the near- and mid-IR. Because of its large size and optimization for optical wavelengths, LUVOIR with HDI will push up to 3 mag fainter. Reaching AB=33–34 mag at the diffraction limit, LUVOIR will exquisitely image the detailed assembly of galaxies from their smallest parts.

Telescope aperture is the critical factor enabling this capability, since it sets the spatial resolution and the photometric depth that can be achieved in reasonable times. **Figure 1-21** shows that LUVOIR can detect analogs of the Milky Way's smallest satellite galaxies at the time they were forming stars (z > 4). The LUVOIR-A deep fields discussed in Signature Science Case #9 (*Tracing ionizing light over cosmic time*) will capture the high-redshift building blocks of larger galaxies, thus achieving another science goal with the same dataset (details in **Section 6.2.1** and **Appendix B.12.2**). However, this program requires unfeasibly long times with LUVOIR-B.

Supplementing the LUVOIR deep fields with additional optical images at shorter wavelengths achieves yet another goal: resolving individual star-forming regions within moderate-redshift (z ~ 1–2) galaxies (details in **Section 6.2.2** and **Appendix B.12.3**). These images will have a spatial resolution that Hubble only achieves for the very limited set of gravitationally lensed galaxies (**Figure 1-24**). Finally, LUVOIR can reveal the story of star formation in individual galaxies by detecting the main-sequence turnoff in color-magnitude diagrams of resolved stellar populations, providing their ages and metallicities (details in **Section 6.2.4** and **Appendix B.12.4**). Previous observing programs could only achieve this for very nearby galaxies, limiting studies to dwarf galaxies and a couple of giant spirals; JWST will

---

**Signature Science Case #11:**
**The multi-scale assembly of galaxies**

**Science Objective**
Study galaxy formation and evolution on spatial scales of 100 pc across cosmic time.

**Description**
Deep survey for the building blocks of giant galaxies: ultra-faint dwarfs at z~7 (not feasible with LUVOIR-B). High resolution imaging of the interiors of galaxies at z~1–2. Imaging of resolved stellar populations in low-redshift galaxies to measure ages.

**Key Functional Requirements**

| Telescope diameter | $\gtrsim$ 15 m |
|---|---|
| Total time | |
|    LUVOIR-A | 139 days (122 in parallel) |
|    LUVOIR-B | 35 days (15 in parallel) |
| Wavelength range | B, V, I, J, H bands |
| Imaging field-of-view | $\gtrsim$ 6 sq. arcmin |
| Spatial resolution | $\lesssim$ 16 mas at 500 nm |

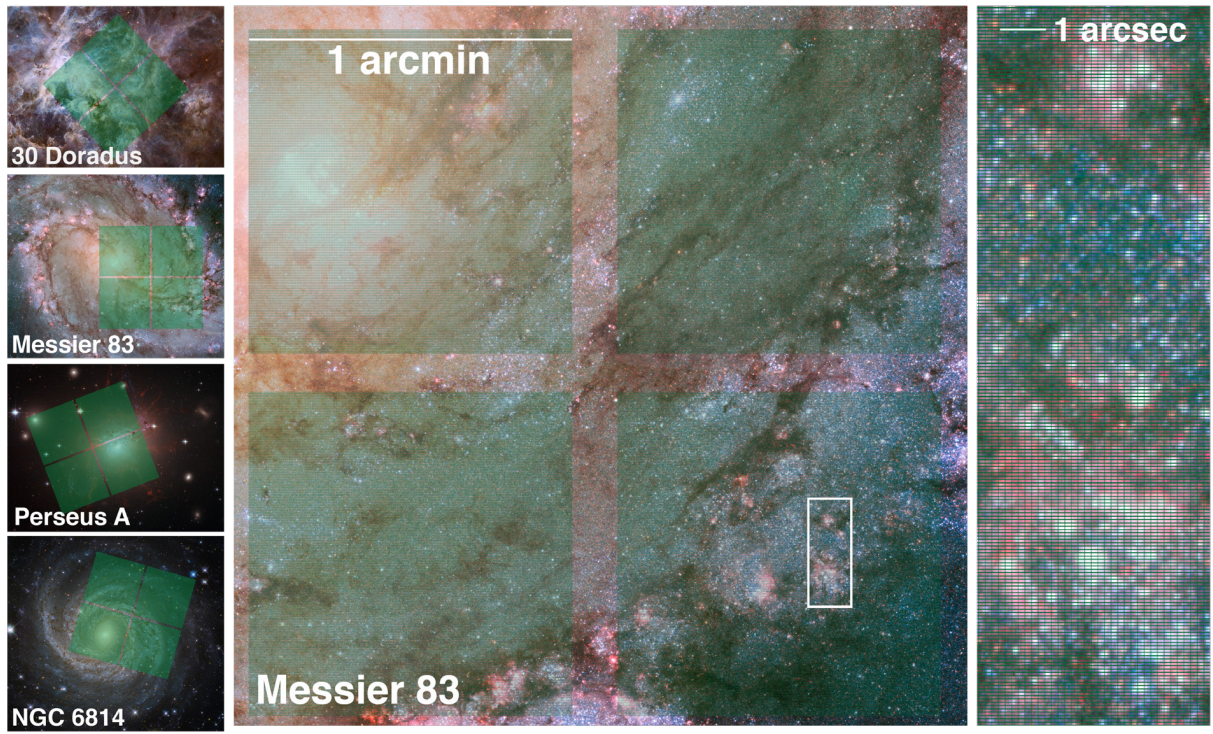

**Figure 1-23.** *LUVOIR's UV sensitivity, spatial resolution, and multiplexing capability will enable powerful probes of gas, stars, and dust at < 1 kpc scales. The 4 square panels at left show the 30 Doradus star-forming region in the Large Magellanic Cloud, the face-on starburst galaxy M83, the center of the Perseus cluster, and the NGC 6814 Seyfert galaxy. The LUMOS spectroscopic field-of-view is overlaid in each panel. The center and right panels show smaller regions within M83, comparing the 0.07 x 0.14 arcsec apertures of LUMOS to a stellar cluster and its environment. LUMOS can study the gas flows emerging from this cluster on nearly a star-by-star basis. Credit: J. Tumlinson (STScI).*

be similarly limited (**Figure 1-25**). LUVOIR will enable these studies out to the distances needed to reach the nearest giant elliptical galaxies.

### 1.6.3 The impact of stars on galaxies

Stars themselves are the most numerous sources of feedback on galaxies, returning energetic radiation to their environments over their lives and substantial amounts of kinetic energy and heavy elements when they die. The most massive stars in particular have strong effects on their host galaxies. With its high spatial resolution and uniquely powerful UV spectroscopy, LUVOIR will make fundamental contributions to our understanding of the stellar initial mass function (IMF) in two key areas: very massive stars (VMSs) and stellar multiplicity.

Stars are the product of the physical and chemical conditions of their natal gas and the early dynamical evolution that takes place before they evolve significantly as individual systems. VMSs (stars with $M_* > 150\ M_\odot$) may result from either birth conditions or early dynamical evolution, but we have scant evidence for their existence and no information on their numbers, frequency, or characteristics. Yet they can heavily influence their surrounding environment, providing 25–50% of the ionizing flux from the host cluster. Diagnostics for VMSs lie in the UV, through emission and absorption lines in the 120–200 nm range. Characterizing VMSs and other massive stars will require resolving young (< 2 million year-old) and massive (> $10^5\ M_\odot$) star clusters out to ~100 Mpc and obtaining multiplexed UV

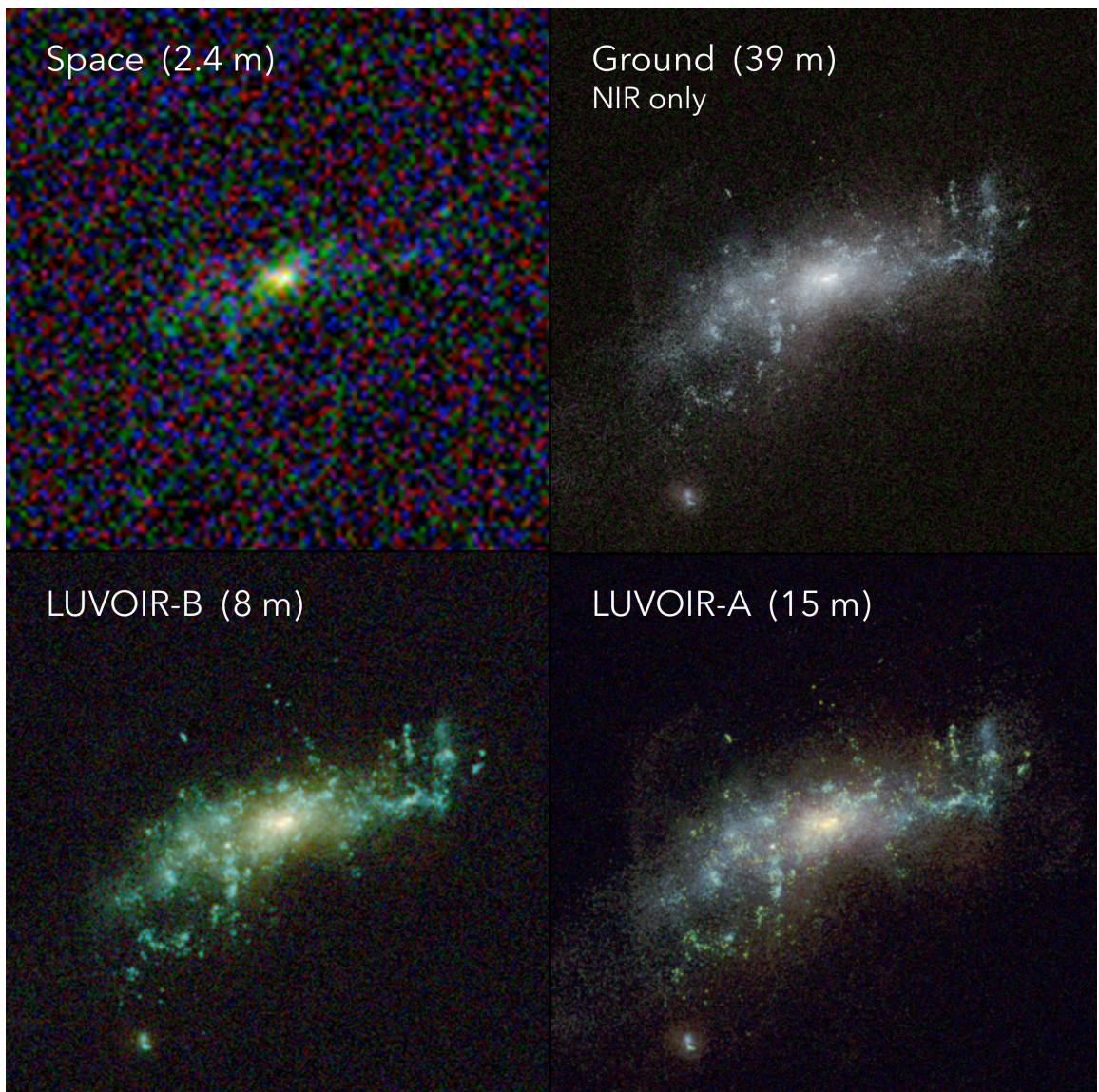

**Figure 1-24.** *LUVOIR can routinely reveal the internal details of galaxies. Simulations of a low mass (10⁹ $M_\odot$) galaxy at z=2 imaged with a 2.4-m space telescope (top left), a 39-m ground-based telescope (top right), and LUVOIR (bottom panels). The space-based simulations are a combination of B, I, and J band optical images. The ground-based simulation uses longer wavelengths (Y, J, and H bands), where extreme adaptive optics systems may provide the best spatial resolution. These simulations all assume the same total exposure time (500 ksec, corresponding to 17 nights for the ground-based simulation). Credits: M. Postman, G. Snyder (STScI)*

spectra with R > 10,000 spectral resolution to measure stellar and wind lines (details in **Section 6.3.1** and **Appendix B.13.2**).

Most stars, including massive stars, are in binary and multiple systems. Knowledge of stellar multiplicity and system properties (number of companions, distribution of short vs. long periods, distributions of mass ratios and eccentricities, and their dependence on the stellar mass) provides tight constraints on models of star formation, the IMF, and star cluster evolution. While O-star multiplicity and environmental dependency are fairly well studied, establishing the impact of local and global conditions (density, metallicity) on the

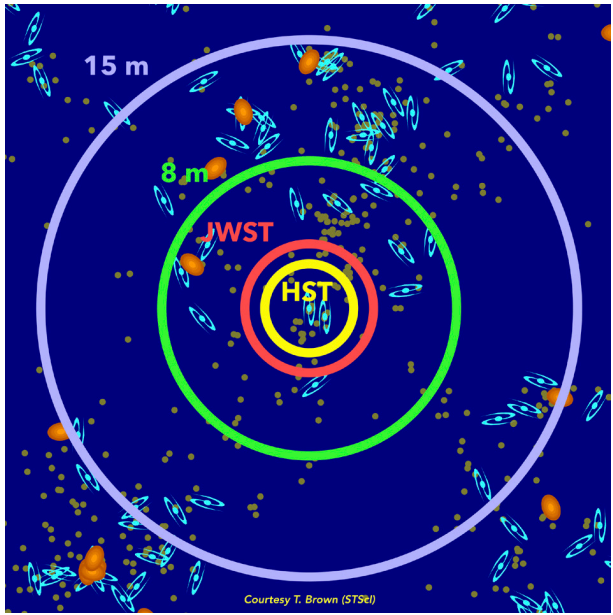

**Figure 1-25.** *Map of the local universe (24 Mpc across) with the distances out to which HST, JWST, and LUVOIR can detect the main-sequence turnoff in resolved stellar populations. Giant spirals are indicated by the blue galaxy symbols, giant ellipticals as orange blobs, and dwarf galaxies as small dots. Credit: T. Brown (STScI)*

**Signature Science Case #12:
Stars as the engines of
galactic feedback**

**Science Objective**
Study the stellar initial mass function (IMF) and the impacts of stars on galaxy evolution.

**Description**
Image very massive stars in luminous and ultra-luminous infrared galaxies and obtain spectra of key diagnostic UV emission lines. Obtain a census of binary stars in galaxies beyond the Milky Way and the Magellanic Clouds.

**Key Functional Requirements**

| Telescope diameter | $\gtrsim$ 8 m |
|---|---|
| Total time | |
|     LUVOIR-A | 25 days |
|     LUVOIR-B | 42 days |
| Wavelength range | |
|     Imaging | 200–2200 nm |
|     Spectroscopy | 100–400 nm |
| Spectroscopic field-of-view | $\gtrsim$ 4 sq. arcmin |
| Spectral resolution | R $\approx$ 10,000 |

multiplicity of B type and lower-mass stars will require a census of both short-period and long-period binaries in massive star clusters and in galaxies other than the Milky Way and the Magellanic Clouds.

Short-period spectroscopic binaries will be studied with adaptive optics-assisted spectrographs on the ELTs. However, studies of long-period binaries require high angular resolution, a very stable point-spread function, and photometry across entire ~arcmin fields-of-view, as well as UV multi-object spectroscopy to characterize the stars in the system. The LUVOIR Study Team has designed a multi-epoch observing program to obtain such observations for binary star systems in nearby galaxies (details in **Section 6.3.2** and **Appendix B.13.3**). Thus, LUVOIR will provide key measurements for formulation of a predictive theory of star formation, by characterizing massive stars both as individual entities and members of multiple systems.

## 1.7  LUVOIR's total prime mission science

The Signature Science Cases introduced here and presented in detail in **Chapters 3–6** represent a set of programs that push both LUVOIR architectures to their limits, roughly analogous to the Treasury programs on Hubble today. While the actual science the community chooses to do with LUVOIR can and will change after it launches, it remains important to show that the key science outlined in this report can be performed with the designed hardware in a notional five-year prime mission. Included with this report are scientific Design

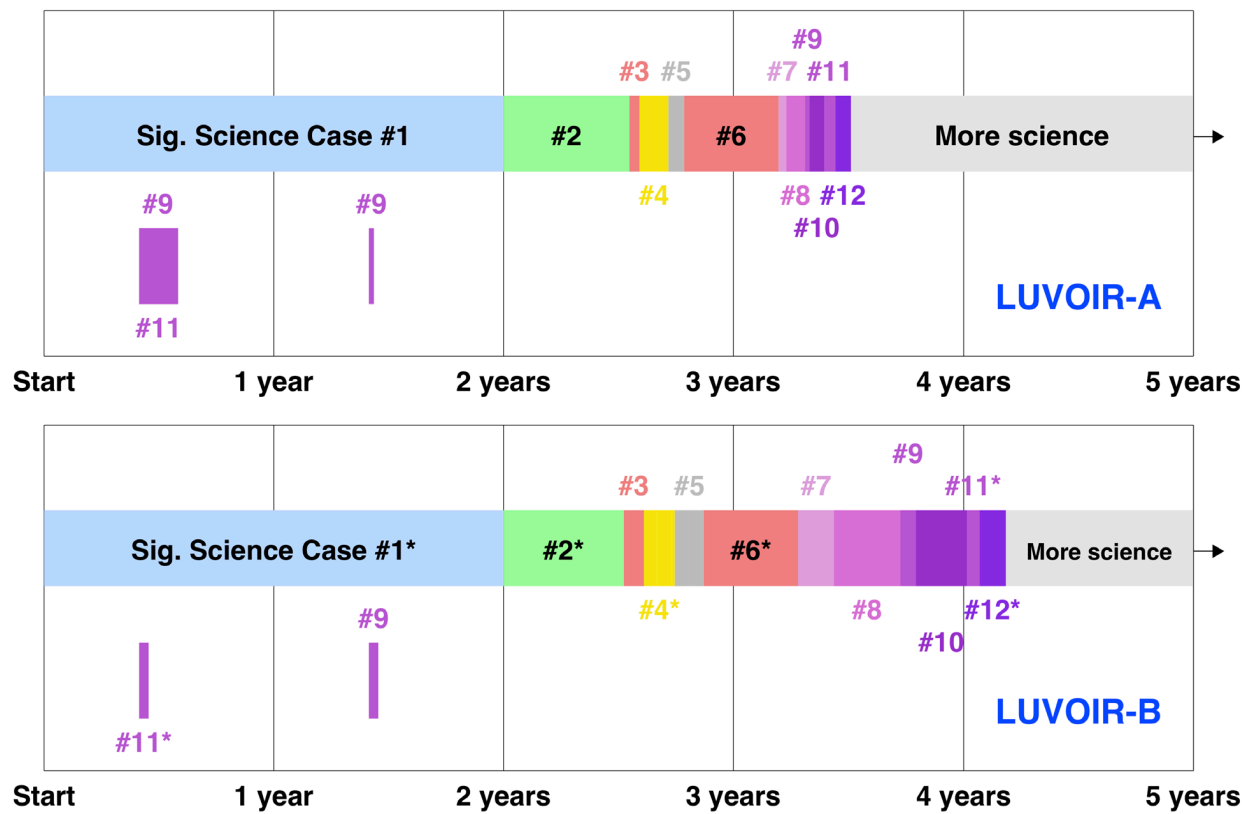

**Figure 1-26.** *LUVOIR can execute all the Signature Science Case (SSC) observing programs with time to spare in the 5-year prime mission lifetime. Programs that can be executed in parallel with other observations are shown offset below the main bars and do not count against the cumulative time. For LUVOIR-A, one program in SSC #11 uses the same data as one of the SSC #9 programs to achieve a different science objective. For LUVOIR-B, an asterisk indicates that the program provides reduced science return (e.g., fewer targets observed). In the most extreme case, the joint SSC #9 / #11 dataset mentioned above is deemed unfeasible with LUVOIR-B and therefore omitted. In reality, the observations for these programs will not be scheduled in this order, or all in single continuous time blocks, but will be interleaved. The SSCs take a total of 3.5 years with LUVOIR-A and 4.2 years with LUVOIR-B, leaving time for additional science programs (examples in* **Appendix A**). *Credit: A. Roberge (NASA GSFC)*

Reference Mission (DRM) documents that correspond to the Signature Science programs from **Chapters 3–6** (**Appendix B**). Target and observation descriptions, as well as exposure times and estimates for overheads, are provided for both LUVOIR-A and B. **Figure 1-26** shows a cumulative representation of these DRMs.

A number of the science cases (e.g., many of the exoplanet-focused programs) are given the same time allocation between LUVOIR-A and B, yielding different quantitative science returns (e.g., number of exoEarth candidates). Others demand the same science return from LUVOIR-A and -B and thus require longer exposure times with LUVOIR-B. Two programs require a deep survey for ultra-faint dwarf galaxies (in Signature Science Cases #9 and #11) that was deemed to need an unfeasibly long time with LUVOIR-B. These programs were therefore excluded from the B versions of those Signature Science Cases. Some Signature Science programs can be executed in parallel with ECLIPS exoplanet observations.

## LUVOIR and the Transient Sky

New objects on the sky often herald astrophysical phenomena new to our understanding. Pulsars, supernovae, and gamma-ray bursts have all fascinated astronomers for decades, and the 2010s have brought two new classes of transient—the still-mysterious fast radio bursts (FRBs) and the groundbreaking detection of gravitational waves (GW) from merging compact objects. The signals from these events themselves, and the information gleaned about their evolution and environment, teach us a great deal about the many way stars end their lives and generate the heavy elements in the cosmos.

Alongside the transient surveys of the 2030s, LUVOIR will bring its unique sensitivity and resolution to the transient sky. In the 2020s, the LSST will begin its nightly all-sky campaign to detect everything on the sky that changes with time. The LIGO facilities will continue and be upgraded, providing a veritable zoo of new GW sources. With its great sensitivity, high angular resolution, large field-of-regard, and fast slew speed, LUVOIR is extremely well suited for targeted, long-term monitoring of the electromagnetic counterparts of GW events and other transients.

The key result from the total DRM analysis is that both LUVOIR-A and LUVOIR-B can perform their ambitious Signature Science Case programs within five years. In fact, both observatories allow significant remaining time for other science programs like those found in **Appendix A**. The community will choose LUVOIR's science program. But even the most ambitious science programs the LUVOIR Study Team can imagine today, compressed into the first five years of LUVOIR's lifetime, leave significant opportunity for additional transformational science.

### 1.8  Magnetic fields everywhere: the science of POLLUX

Chapter 13  POLLUX is a high-resolution, point-source UV spectropolarimeter studied and designed by a consortium of European institutions, with leadership and support from the Centre National d'Etudes Spatiales (CNES). The instrument was designed to occupy the fourth instrument bay on LUVOIR-A, and represents a potential future international contribution to the LUVOIR mission. As the POLLUX study is separate from the overall NASA-led LUVOIR study, POLLUX's science goals supplement the main LUVOIR science goals and are discussed along with the instrument's technical details in **Chapter 13**.

The most innovative characteristic of POLLUX is its spectropolarimetric capability, which enables studies of a broad range of important astrophysical environments that are out of reach for standard high-resolution spectroscopy. Spectropolarimetry enables the detection and characterization of magnetic fields and local environments of astrophysical objects (**Figure 1-27**). POLLUX will address a variety of science cases related to LUVOIR's science portfolio, thereby complementing the other instruments:

- **Exoplanets:** chemical and physical properties of exoplanetary atmospheres; tidal and magnetic star-planet interactions

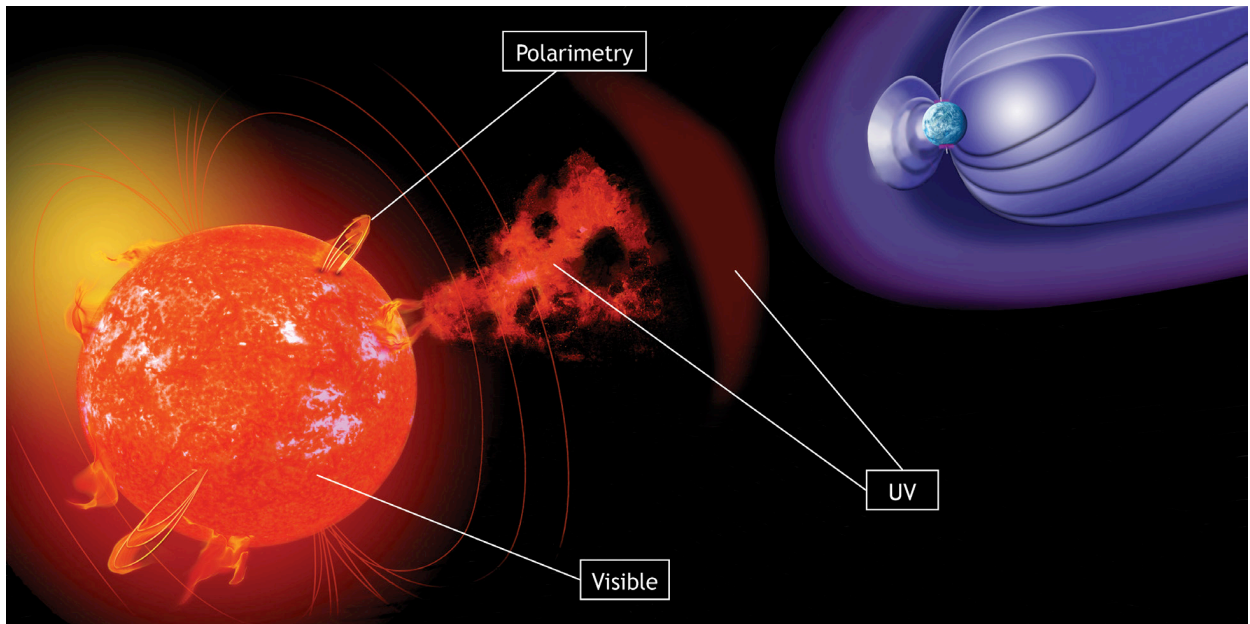

**Figure 1-27.** *Sketch of a cool star, with its dynamo magnetic field, surface faculaes and plages, wind, a coronal mass ejection, and a bow shock between the star and its planet. Credit: S. Cnudde*

- **Fundamental physics and cosmology:** the deuterium / hydrogen ratio; evolution of the cosmic microwave background temperature; variation of fundamental constants

- **Supermassive black holes:** AGN; outflows and jets; impact of AGN on galactic evolution

- **Interstellar, circumgalactic, and intergalactic media:** the various phases of the interstellar medium; the influence of magnetic fields at the Galactic scale and in IGM; the role of CGM in galaxy formation and evolution

- **Stars:** magnetic fields in star formation and evolution; magnetism in galactic massive stars; extremely metal-poor stars; novae & supernovae; white dwarfs

- **Solar system:** magnetospheres; surfaces of icy moons and small bodies; comet comae

## 1.9  Big observatories for big goals

**Chapter 7**  Each of LUVOIR's revolutionary science investigations demands distinct observations, but they share a common need for a large space telescope that can collect light from the FUV to the NIR. Further, that observatory must have a diverse and powerful toolkit of instruments. The Hubble Space Telescope, which has been one of the most productive scientific tools ever built, shares these characteristics. However, our transformative goals demand a great leap in capability over Hubble and every other current or planned observatory.

The design of LUVOIR is driven by our need for higher contrast, greater sensitivity, and finer resolution. At the same time, the design draws on a decades-long wealth of expertise and technology development garnered from several current and near-future space missions. The greatest technological challenge for LUVOIR is achieving the stable wavefront needed for high contrast direct observations of potentially habitable exoplanets. Realizing this goal requires careful attention to the integrated design of the whole observatory. The LUVOIR Study Team implemented a three-tiered system design philosophy: Wavefront Stability through Design, Wavefront Stability through Control, and Wavefront Stability through Tolerance (**Section 8.1.6**). Thus, the LUVOIR concepts minimize all mechanical and thermal disturbances, and incorporate several active control systems that work in concert.

**Table 1-5.** *Characteristics of the LUVOIR observatories*

|  | Description |
|---|---|
| Telescope diameter | LUVOIR-A: 15 meters; LUVOIR-B: 8 meters |
| Collecting area | LUVOIR-A: 155 m²; LUVOIR-B: 43.8 m² |
| Orbit | Quasi-halo orbit around Sun-Earth Lagrange 2 point |
| Diffraction limit | 500 nm |
| Telescope temperature | 270 K |
| Lifetimes | 5-year prime mission; 10 years consumables; 25-year goal for non-servicable components |
| Proposed launch date | 2039 |
| Field-of-regard | Sun-Telescope-Target angles ≥45° (3π steradians) |
| Max. slew speed | 3° / min (2x JWST speed) |
| Max. tracking speed | 60 milliarcsec / sec (2x JWST speed) |
| Total wavelength range | 100 nm–2500 nm |

The LUVOIR Study Team designed two distinct observatory concepts: the 15-m LUVOIR-A and the 8-m LUVOIR-B (**Figure 1-3**). By studying two concepts, we gain better understanding of a complex trade space, reveal how science return scales with different technical choices, and establish robustness to uncertainties such as future launch vehicle capabilities and budget constraints. It is important to recognize that LUVOIR-A and -B represent proof-of-concept point designs within a family of UV/optical/NIR observatories, demonstrating feasibility and providing information for future concept development. A summary of LUVOIR's observatory-level characteristics appears in **Table 1-5**.

### 1.9.1 Mission overview

**Section 8.1** Each LUVOIR observatory contains payload and spacecraft elements (**Figure 1-28**). The payload consists of the optical telescope assembly (OTA), the instrument modules, and the payload articulation system. The spacecraft consists of the spacecraft bus and the sunshade. The spacecraft bus provides standard functions such as orbit maintenance, attitude control, command and data handling, communications, and power. Some of the key observatory components and features are discussed below; complete details appear in **Chapters 7–12** and appendices.

Both LUVOIR observatories are designed to operate at the second Sun-Earth Lagrange point (SEL2), which provides a stable operating environment. The telescopes provide a large field-of-regard (the entire anti-Sun hemisphere plus half of the sunward hemisphere; **Figure 1-29**), facilitating time-critical observations and improving observation scheduling efficiency in general. The maximum tracking speed is 2x faster than the JWST speed, enabling observations of solar system bodies closer to the Sun. The observatories can slew 90° within 30 min (2x faster than JWST), which also aids time-critical observations and observation scheduling efficiency. The prime mission lifetime is 5 years, with 10 years of on-board consumables (e.g., propellant) and a 25-year lifetime goal for non-serviceable components (e.g., mirror segments).

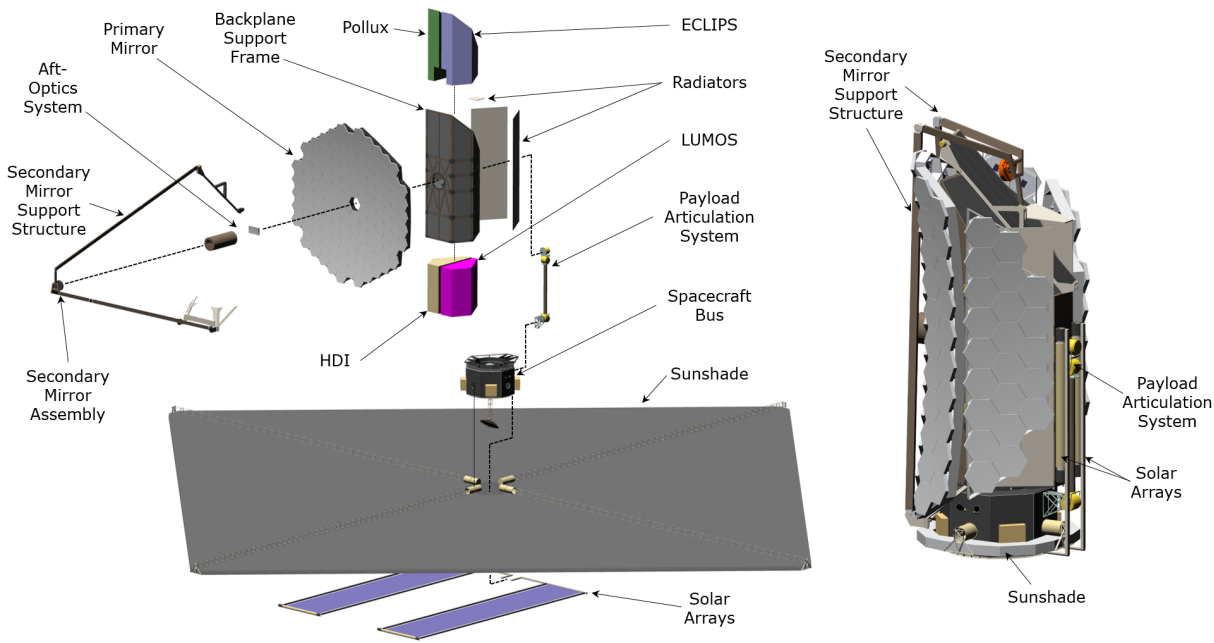

**Figure 1-28.** *Major elements of the LUVOIR-A payload and spacecraft. Similar elements appear in LUVOIR-B. Credit: A. Jones (NASA GSFC)*

### 1.9.2 Project management and mission development schedules

Chapter 12   The LUVOIR development schedule draws extensively on lessons learned from other NASA strategic missions, including Hubble, JWST, and WFIRST. At a high level, four major things that cause schedule slips and their attendant cost overruns are: immature science requirements and technologies at the beginning of the design phase, insufficiently detailed system-level designs early in development, late changes to requirements, and unstable project funding. The first three lead to re-designs that force parts of the "marching army" to march in place for a time. The last one forces projects to defer work that should be done in accordance with a master project schedule optimized for efficiency. The LUVOIR Study Team has put extensive thought into how to avoid these pitfalls.

**Table 1-6.** *NASA mission life-cycle phases*

|  | Description |
|---|---|
| Pre-Phase A | Concept development |
| Phase A | Concept and technology development. Begins with approval for formulation at Key Decision Point A (KDP A). |
| Phase B | Preliminary design and technology completion. Preliminary Design Review (PDR) occurs near end of phase. |
| Phase C | Final design and fabrication. Begins with approval for implementation at KDP C. |
| Phase D | System assembly, integration & test, launch, and checkout |
| Phase E | Operations and sustainment |
| Phase F | Closeout |

Any discussion of a mission development schedule must begin with a clear definition of terms. The NASA mission life-cycle phases and their brief descriptions are given in **Table 1-6**. The current requirement is for all technologies to be at Technology Readiness Level (TRL) ≥5 at Phase A start and TRL ≥6 at PDR near the end of Phase B. While this structure functions well for missions of low or moderate complexity, the LUVOIR Study Team believes that the process must evolve in order to manage the increased complexity of the next generation of large strategic missions.

First, key technologies should be matured to a higher level *before* the start of the mission (during Pre-Phase A); the LUVOIR Study Team recommends TRL ≥6 by Phase A start. However, the technology development should still be focused upon a specific architecture. The current Large Mission Concept Studies will be very useful for directing Pre-Phase A technology and architecture de-velopment. Second, for complex observatories, it is critical to begin work on all major components (including the spacecraft) simultaneously and develop them in parallel. Third, requirements must be completely defined by the end of Phase A, rather than left open until PDR as is currently allowed.

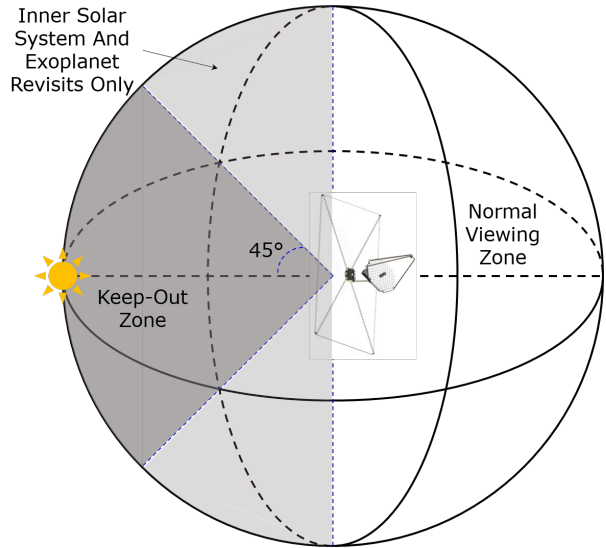

**Figure 1-29.** *The LUVOIR field-of-regard. Most observations occur in the anti-Sun hemisphere, with the sunshade in a fixed orientation normal to the Sun for thermal stability. For viewing inner solar system objects or for time-critical observations, LUVOIR can pitch up to 45° towards the Sun. Credit: M. Bolcar (NASA GSFC)*

Keeping these principles in mind, the LUVOIR Study Team has created development schedules for both LUVOIR-A and -B. High-level summaries appear in **Figure 1-30**; a discussion of the schedules appears in **Section 12.6** and the detailed schedules appear in **Appendix G**. Both architectures are built up from the sub-systems to the elements to the segments to the full missions. For example, the OTA, instrument, and payload articulation sub-systems become the payload element. The payload element plus the spacecraft element becomes the observatory segment. Then the observatory, ground, and launch vehicle segments combine into a complete LUVOIR mission. Each part of the observatory begins development at the start of Phase A, proceeds in parallel, and has its own nested set of life-cycle phases. An instrument completes its Phase D (assembly, integration & test) before the payload enters Phase D. The mission-level Phase D involves integrating the payload and spacecraft elements, launching the completed observatory, deployment, and commissioning in the final orbit.

The Large Mission Concept Study teams were instructed by NASA to assume a Phase A start on Jan 1, 2025 and create schedules extending to the end of Phase D. The launch dates are Nov 2039 for LUVOIR-A and July 2039 for LUVOIR-B. Including commissioning, the total schedule durations are 15.6 years for LUVOIR-A and 15.3 years for LUVOIR-B. The fact that the schedules for both LUVOIR concepts are so similar may be surprising. However,

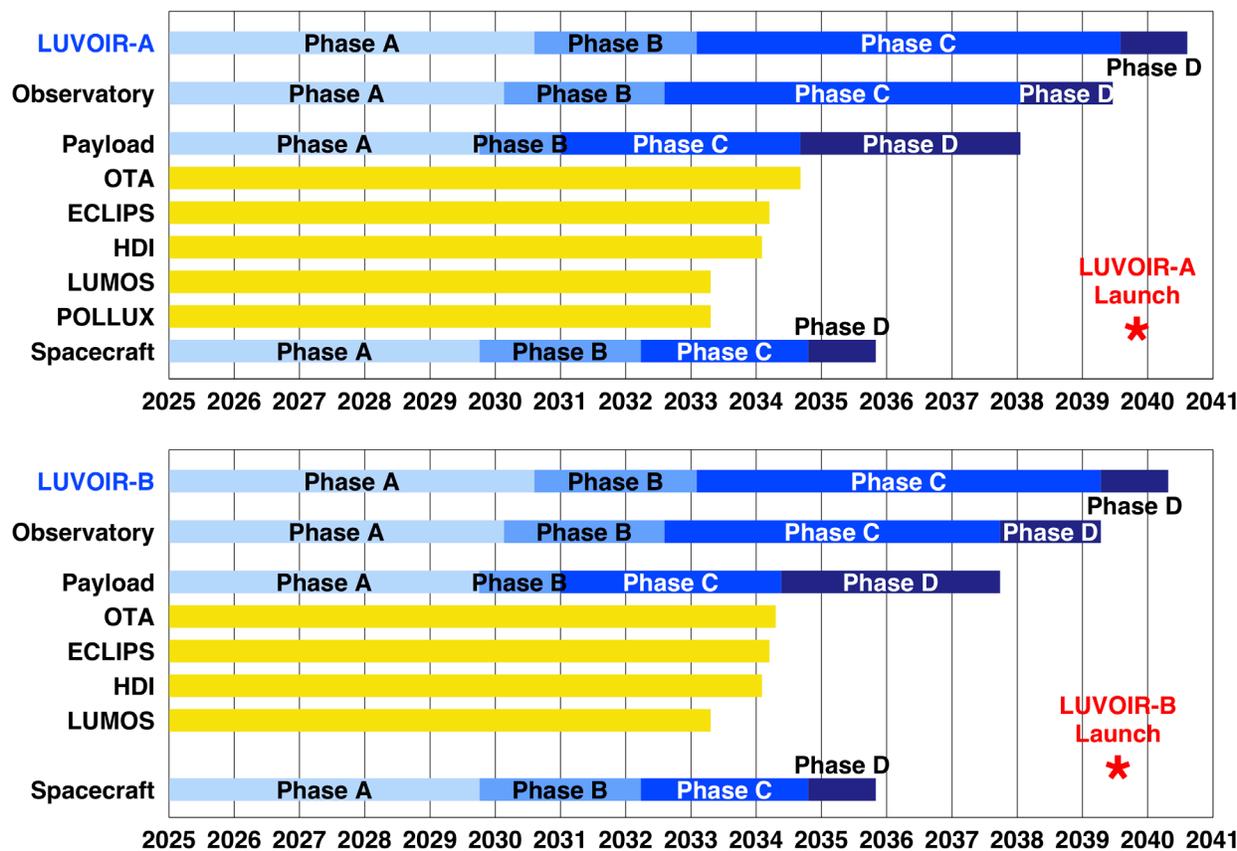

**Figure 1-30.** *High-level mission development schedules for LUVOIR-A (top) and LUVOIR-B (bottom). Including commissioning, the total development schedule durations are 15.6 years for LUVOIR-A and 15.3 years for LUVOIR-B. Assuming approval for formulation in Jan 2025 (start of Phase A), the launch dates are Nov 2039 for LUVOIR-A and July 2039 for LUVOIR-B. Credit: J. Hylan, A. Roberge (NASA GSFC)*

many of the components of LUVOIR-A and -B are nearly identical in size and complexity, and therefore have basically the same required development time.

### 1.9.3 Serviceability

**Section 8.1.2** The LUVOIR science programs described in this report can be achieved within a 5-year prime mission using the as-designed instruments. LUVOIR should be judged on the strength of that science, without relying on an extended mission. However, Congress has mandated that all future large space telescopes be serviceable to the extent practicable. The LUVOIR STDT has embraced this requirement, since gaining the greatest scientific return from a strategic mission with an extended development time and significant cost demands long operational lifetimes. One of the key features of Hubble that has allowed it to be so scientifically productive over such a long time is the fact that it has been serviced and upgraded.

The GSFC Satellite Servicing Projects Division provided guidance over the course of the LUVOIR study. The observatory designs incorporate standardized mechanical components (e.g., valves, latches, rails) and allow replacement of many spacecraft elements (e.g., control moment gyroscopes, star-trackers, avionics). Consumables (i.e., propellant) can be replenished and science instruments readily removed for replacement. Incorporating modular sub-systems and interfaces in the LUVOIR designs from the outset will also yield advantages in integration and test procedures on the ground before launch, advantages that will be realized regardless of whether LUVOIR is actually serviced.

We have considered the question "how will LUVOIR be serviced?" First, a design commitment to serviceability is not a commitment to servicing. Whether or not LUVOIR would be serviced will depend on unpredictable political and budgetary realities. If history is a guide, it is doubtful that NASA's Science Mission Directorate on its own will have the resources to create the infrastructure for servicing LUVOIR. Even so, as an illustration of how LUVOIR might be serviced, NASA's Human Exploration and Operations Mission Directorate is considering a Deep Space Gateway in cis-lunar space. The energy required to move LUVOIR from Sun-Earth L2 to Earth-Moon L1 or L2 and back is small ($\Delta v \lesssim 0.4$ km s$^{-1}$) compared to that required for launch ($\Delta v \lesssim 11$ km s$^{-1}$). Returning the fully deployed LUVOIR to cis-lunar space for servicing at the Gateway by some combination of astronauts and/or robots is a technical challenge, but not an impossible task.

### 1.9.4 Pointing and stability

**Section 8.2.6**    Both telescopes are connected to their spacecraft using a two-axis gimbal system and a Vibration Isolation and Precision Pointing System (VIPPS). The gimbal system allows the telescope to point anywhere in the anti-Sun hemisphere while keeping the sunshade normal to the Sun-observatory line (**Figure 1-29**). Pointing into the sunward hemisphere requires a tilt of the sunshade. The gimbal system also allows the telescope to roll about the boresight while keeping the rest of the observatory fixed, permitting targets to be aligned with instrument-specific apertures and enabling differential subtraction of observations at multiple roll angles.

The payload and spacecraft are physically separated from each other using the VIPPS, a non-contact interface that is controlled by voicecoil actuators over six degrees-of-freedom. The VIPPS "floats" the telescope over the spacecraft, ensuring that mechanical disturbances in the spacecraft, which would negatively affect wavefront stability, are not transmitted to the payload. Cables bridge the gap between the payload and spacecraft to transmit power and data signals. The VIPPS provides fine pointing control during science observations.

### 1.9.5 Sunshade

**Section 8.3.2**    Both observatory designs feature a deployable sunshade, which keeps sunlight out of the telescope and assists in thermal control of the payload. The LUVOIR sunshade is similar in some aspects to the sunshield developed for JWST, but does not have to provide the extreme thermal control required by JWST. Therefore, the LUVOIR sunshade is considerably simpler than the JWST sunshield. Instead of the JWST sunshield's five layers, the LUVOIR sunshade has three. These layers do not require high-precision deployment with respect to angle and separation between the layers, mitigating the need for fine positioning mechanisms. The LUVOIR sunshade is also packaged more simply, using high-TRL deployable booms to pull the sunshade from its stowed compartment at the base of the spacecraft bus. The LUVOIR sunshades are, however, larger than the JWST sunshield (55.9 m × 55.9 m for LUVOIR-A and 48.1 m × 48.1 m for LUVOIR-B).

Nominally, the LUVOIR sunshade remains in a fixed position normal to the Sun-observatory axis (**Figure 1-29**). Thus, the payload always experiences a constant thermal environment regardless of its pointing attitude, which improves thermal stability. This is a key part of achieving the extreme wavefront stability required for high contrast coronagraphic observations of exoEarths. When science observations require pointing into the sunward hemisphere (e.g., inner solar system and exoplanet revisit observations), the entire observatory will pitch towards the Sun while keeping the payload in shadow.

## 1.10  The LUVOIR telescopes

**Chapter 8** The LUVOIR Study Team's mandate was to consider UV/O/NIR space telescopes in the 8-to-16-m diameter range. LUVOIR-A features a segmented primary mirror with a 15-m diameter and four instrument bays. This observatory is designed to launch in NASA's planned Space Launch System (SLS) Block 2 vehicle. No other anticipated future launch vehicle can accommodate both the mass and volume demands of LUVOIR-A, with the possible exception of SpaceX's Starship. This risk was one of the driving considerations leading to development of the LUVOIR-B concept.

LUVOIR-B has a segmented aperture primary with an 8-m diameter and 3 instrument bays. It demonstrates the scalability of the basic LUVOIR architecture and its compatibility with a range of future launch vehicles. The fleet of available heavy-lift launch vehicles is currently in flux. Existing vehicles like the Delta IV Heavy are being phased out in favor of new ones like Space X's Falcon Heavy and Starship, Blue Origin's New Glenn, and NASA's own SLS (**Chapter 10**). At this time, the future launch vehicles that best fit LUVOIR-B's needs are the SLS Block 1B, the SpaceX Starship, and the Blue Origin New Glenn. However, LUVOIR-B is designed to fit within an industry-standard 5-m diameter launch fairing, such that the observatory can be readily adapted for other future vehicles.

### 1.10.1  Optical telescope assemblies (OTAs)

**Section 8.2.1** The LUVOIR telescopes are three-mirror anastigmats (TMAs), chosen to provide high optical quality over large fields-of-view. Both OTAs include a fast steering mirror and are diffraction limited at 500 nm, providing the best possible point-spread function (PSF) at all longer wavelengths. The PSFs at shorter wavelengths will be limited by the end-to-end wavefront error (< 35 nm RMS), but remain of sufficient quality for the science cases described in this report. The secondary-to-primary mirror separation distances were set to keep all angles of incidence across the pupils below 12°. This is needed to minimize polarization-dependent aberrations that would negatively affect coronagraph performance. If future coronagraph designs prove more tolerant to polarization effects, the primary-to-secondary separation could be reduced, easing constraints on the OTA optical design and packaging.

LUVOIR-A is an on-axis telescope with a central obscuration from the secondary mirror. While unobscured telescopes perform better with current coronagraph designs (**Figure 1-7**), the constraint on angles of incidence given above and the launch fairing volume constraint dictated that this telescope must be on-axis. To improve coronagraph performance, the central obscuration was kept as small as possible (~10% of diameter) while still allowing the light ray bundle through to the aft-optics system. Future polarization-tolerant coronagraph designs might allow an off-axis LUVOIR-A, improving its coronagraph performance. The LUVOIR-B telescope is an off-axis telescope with an unobscured primary aperture, specifically chosen to maximize performance of the coronagraph instrument for the smaller aperture diameter.

***Stability and active control.*** One of the major design drivers on LUVOIR is the need for extreme stability of the wavefront entering the coronagraph. While the design of the LUVOIR OTAs draws heavily upon that of the JWST telescope, there are several key changes to improve wavefront stability. Each LUVOIR primary mirror segment assembly (PMSA) is a single, stiff mirror segment mounted on a hexapod for 6 degree-of-freedom rigid body positioning of the segment, like the JWST PMSAs. The LUVOIR PMSAs depart from those of JWST in the areas of mechanical control systems (this sub-section), thermal control systems, and materials (next sub-section).

A critical new component of the LUVOIR PMSAs is the introduction of an edge sensor and piezoelectric (PZT) actuator control system. Each mirror segment is fitted with one edge sensor per edge (i.e., two edge sensors total per shared edge), currently baselined as capacitive edge sensors. Each sensor continuously measures the local rigid body position of each segment relative to its neighbor. Incorporating these measurements from all sensors allows for a global solution to be found for the six degree-of-freedom position of each segment.

The positions of each segment are then fed back to the hexapod fine-stage PZT actuators to rapidly (at ~100 Hz) control the segment position with picometer resolution (Saif et al. 2017). Edge sensor systems have been in use on segmented ground-based telescopes for many decades (e.g., the Keck telescopes) and are included in the designs for the ELTs. Capacitive gap measurement devices have demonstrated < 11 pm precision at up to 60 Hz in open-loop operation; even better precision is achieved in closed-loop operation (Coyle et al. 2019b).

The second innovation in the LUVOIR telescope design is the introduction of a laser metrology system to measure the rigid body position of the secondary mirror relative to both the primary mirror and the aft-optics system. By measuring the alignment of all of the major optical elements in the OTA, the actuators on the segments and secondary mirror can effectively "rigidize" the entire telescope in inertial space. Laser metrology ultimately draws its heritage from the Space Interferometry Mission (SIM) project, which demonstrated picometer precision using large and heavy beam launchers and corner cubes on ground-based testbeds. Subsequent development achieved much smaller devices suitable for attachment to lightweight mirror segments. The Laser Interferometer Space Antenna (LISA) Pathfinder mission has also further refined laser metrology technology.

Analysis shows that this hybrid approach—using a laser metrology truss on a few reference segments plus the less complex edge sensors on every segment—can provide better wavefront stability than either system on its own, provided both systems have picometer-level sensitivity. The optimal blend of measurements will depend on the ultimate precision of both systems' practical hardware and is a focus of current and future technology studies.

***Stability and thermal control.*** To minimize thermal expansion and contraction of optical elements that would negatively affect wavefront stability, a stable thermal environment must be created and the telescope materials must be carefully chosen. Precise control of the telescope operating temperature is most effectively done by actively heating the OTA. In another departure from JWST, the individual segment, secondary, and tertiary mirror assemblies incorporate a thermal control system. A heater plate immediately behind the mirror substrates radiatively heats the mirrors to 270 K ± 1 mK. This particular operating temperature is warm enough to control contamination of optical elements and ease integration & testing, yet cold enough to allow NIR observations (**Section 1.12.2**). The primary mirror backplane and secondary mirror support structures are also actively heated at key control points in order to maintain the global thermal stability of the OTA (Eisenhower et al. 2015; Park et al. 2017).

Unlike the JWST beryllium mirror segments, the baseline substrate material for the LUVOIR mirrors is Corning's Ultra-Low Expansion (ULE®) glass. At the warm operating temperature of LUVOIR, this material has been demonstrated to be very stable against any thermal changes. These glass segments, with their closed-back design, will be stiffer than the JWST beryllium segments. The OTA structure is made of composite materials with zero coefficient of thermal expansion near 270 K.

### 1.10.2 Mirrors coatings

Both observatories span a total wavelength range of 100–2500 nm. All telescope mirrors are coated with a protected, "enhanced" Al+LiF coating to enable observations at wavelengths below the effective cutoff of Hubble (~115 nm). A thin overcoat of $MgF_2$ or $AlF_3$ reduces the hygroscopic sensitivity of the LiF layer (Quijada et al. 2014). The Al+LiF coating has been used on previous FUV space telescopes (e.g., FUSE). The overcoats will be demonstrated on an approved NASA CubeSat (ARTEMIS; PI: B. Fleming), planned for launch in 2023.

### 1.11 The LUVOIR instruments

Four instrument concepts have been designed in the course of the LUVOIR study. These are the tools needed to execute the high-priority science cases described in this report. However, there are many other instrument concepts that would be compatible with the LUVOIR telescopes; a few are described in **Appendix I**. Such instruments could be chosen as first-generation instruments for a LUVOIR-like observatory or fly later as second-generation instruments.

### 1.11.1 High-Definition Imager (HDI)

**Section 8.2.2** This instrument is a wide-field imaging camera with two channels: a NUV-visible channel (UVIS) covering 200–1000 nm and an NIR channel covering 1000–2500 nm. The channels can image the same 3' x 2' sky field simultaneously for maximum efficiency or in series for maximum sensitivity. Like all the LUVOIR instruments, HDI can be operated in parallel with any other instrument. A summary of HDI's specifications appears in **Table 1-7**.

*Detectors.* The respective focal plane detector arrays provide Nyquist sampling of the PSF at 500 nm in the UVIS channel and 1000 nm in the NIR channel. The choice of Nyquist sampling at the cost of a greater number of detector pixels is driven by two goals. The first is to maximize extraction of information from the images. The second is to enable excellent image quality when HDI is operating in parallel with other instruments and the primary observation cannot accept dithering maneuvers by the observatory pointing system.

On LUVOIR-A, the UVIS focal plane of HDI is a 1.6 Gigapixel imaging array composed of twenty-four 8k x 8k CMOS-based detectors. On LUVOIR-B, the UVIS focal plane has 0.4

**Table 1-7.** *Characteristics of the High-Definition Imager (HDI)*

| HDI Characteristics | |
|---|---|
| Channels | UVIS & NIR; can be operated simultaneously |
| Total wavelength range | 200–2500 nm |
| Field-of-view | 3′ x 2′ |
| UVIS detector | CMOS; Nyquist sampled at 500 nm<br>HDI-A: 3.43 mas/pix; HDI-B: 6.45 mas/pix |
| NIR detector | HgCdTe; Nyquist sampled at 1000 nm<br>HDI-A: 6.88 mas/pix; HDI-B: 12.89 mas/pix |
| Filters | 67 science filters + low-resolution grisms |
| Special modes | High-precision astrometry; Fine guidance for observatory; Phase retrieval and wavefront sensing |

Gigapixels, comprising six of the same CMOS detectors. The 8k x 8k format has not yet been produced in flight-qualified scientific systems but is within a realistic technology trajectory from current devices. The assumed read noise and dark current are consistent with current state-of-the-art devices. A final detector decision would not need to be made for many years, allowing ample time for development. The NIR focal plane of HDI consists of a 4 x 6 array of 4k x 4k HgCdTe-based detectors on LUVOIR-A and a 2 x 3 array of the same detectors on LUVOIR-B. The 4k x 4k format has been extensively developed for the WFIRST mission (Content et al. 2013). The assumed read noise is about 2x lower than current state-of-the-art; however, it is believed this improvement can be achieved with optimization of the read-out electronics and operating temperature.

**Thermal design.** The bulk of the instrument is held at an operating temperature of 270 K. However, a passively cooled shroud around the NIR channel maintains the NIR optics at 170 K to reduce thermal background. The visible focal plane array is also passively cooled to 170 K, while the NIR focal plane array is passively cooled to 100 K.

**Filters.** The UVIS channel contains a filter selection mechanism with four wheels capable of holding 13 elements each. Thus, the mechanism allows a total of 41 science spectral elements, 4 clear slots (to allow access to each set of filters on each wheel), 1 dark slot, 4 defocus lenses, and 2 dispersed Hartman sensor elements. The defocus lenses and dispersed Hartman sensor elements are needed to align and phase the primary mirror segments after deployment. The NIR channel filter select mechanism contains three wheels capable of 10 elements each, for a total of 26 science spectral elements, 3 clear slots, and 1 dark slot.

**Astrometry.** HDI includes high-precision astrometry capability. This mode enables a range of exciting science not feasible on existing telescopes including measuring the masses of Earth-mass exoplanets orbiting Sun-like stars and the proper motions of extragalactic sources in the Local Group and beyond. To achieve the astrometric precision required for these applications (≤1 μas), both optical and detector distortions must be calibrated so that source positions can be determined with high accuracy over timescales of months to years.

The sensitivity of LUVOIR will allow optical distortions to be accurately calibrated using the large number of astrophysical sources that will be observable. Calibration of detector distortions will be achieved with an all-fiber metrology system internal to HDI (Crouzier et al. 2016). This system can measure pixel positions in the UVIS channel to a precision of $10^{-4}$ pixels. The estimated time needed to calibrate the detector pixel positions to this level is

**Table 1-8.** *Characteristics of the Extreme Coronagraph for LIving Planetary Systems (ECLIPS)*

| ECLIPS Characteristics | |
|---|---|
| Channels | NUV, VIS, & NIR; simultaneous operation of 2 of 3 channels |
| Total wavelength range | 200–2000 nm |
| Raw contrast | $1 \times 10^{-10}$ |
| Min. IWA | NUV: 4 $\lambda$/D; VIS: 3.5 $\lambda$/D; NIR: 3.5 $\lambda$/D |
| Max. OWA | NUV: 40 $\lambda$/D; VIS: 64 $\lambda$/D; NIR: 64 $\lambda$/D |
| Instantaneous bandwidth | 10–20% |
| Imaging | NUV & VIS channels |
| Spectroscopy | Spatially resolved in VIS & NIR; point-source in NIR |
| Spectral resolution, R ($\lambda/\Delta\lambda$) | VIS: 140; NIR: 70 & 200 |

~12 hours per calibration sequence. Given LUVOIR's overall stability requirements for high contrast imaging, the LUVOIR Study Team estimates that this calibration sequence will only have to be executed once during approximately monthly astrometry observing campaigns. The estimated systematic limit on astrometric precision that can be achieved in single-epoch imaging is 0.34 $\mu$as and 0.65 $\mu$as for HDI-A and HDI-B, respectively. Greater precision can be achieved by increasing the number of epochs.

*Other special modes.* HDI also provides two other special modes of operation. First, HDI will function as LUVOIR's primary fine-guidance sensor. Both the UVIS and NIR focal planes have the capability of defining small regions-of-interest around bright foreground stars. These areas can be read-out at high speeds to provide a pointing signal to the fast steering mirror and VIPPS. This can be done without interrupting regular science operations. Second, the HDI focal plane will provide defocused image data for phase retrieval and wavefront sensing. A similar function for JWST is provided by the Near-Infrared Camera (Dean et al. 2006). This dataset will be used during commissioning of the observatory to align and phase the primary mirror segments after deployment, as well as during routine maintenance of the wavefront as needed.

### 1.11.2 Extreme Coronagraph for LIving Planetary Systems (ECLIPS)

Section 8.2.3 This instrument is an internal coronagraph with the key goal of direct exoplanet observations. It has three channels: NUV (200–525 nm), visible (515–1030 nm) and NIR (1000–2000 nm). The NUV channel is capable of high-contrast imaging only, with an effective spectral resolution of R≈6, sufficient for detection of strong, broad ozone absorption and characterization of cloud/haze scattering slopes. The optical channel contains an imaging camera and integral field spectrograph (IFS) with R=140, chosen for optimal study of the narrow $O_2$ absorption band at 760 nm. The NIR channel contains an IFS with R=70 for study of broad molecular absorption bands (e.g., $CH_4$) and a point-source spectrograph with R=200 for study of $CO_2$. A summary of ECLIPS's specifications appears in **Table 1-8**.

*Wavefront correction.* All high-performance coronagraphs incorporate optical elements to perform internal corrections of the wavefront and improve performance. The coronagraphs for both LUVOIR-A (ECLIPS-A) and LUVOIR-B (ECLIPS-B) each contain two deformable mirrors (DMs) per channel. Each channel also has a dedicated, out-of-band Zernike wavefront sensor for continuous monitoring of slow, low- and mid-order wavefront errors to be corrected by the deformable mirrors.

*Coronagraph masks.* One thing that was made abundantly clear during the LUVOIR study is that a coronagraph and a telescope must be designed together to achieve the most effective high-contrast system. The coronagraph optical elements must respond to the shape of the telescope pupil, and the shape of the telescope pupil has an impact on the maximum possible performance possible with the coronagraph. While telescope segmentation has a negligible impact on coronagraph performance, the same is not true for telescope obscurations (**Figure 1-7**). At this time, all coronagraphs perform better with un-obscured telescopes; however, some are more sensitive to obscurations than others.

The LUVOIR-A OTA is an on-axis telescope (for reasons discussed in **Section 1.10.1**). Therefore, the LUVOIR Team chose an apodized pupil Lyot coronagraph (APLC) as the primary coronagraph type for ECLIPS-A. This type of coronagraph is relatively insensitive to telescope obscurations and is fundamentally similar to the WFIRST shaped-pupil coronagraph (Kasdin et al. 2019). ECLIPS-A contains several APLC masks with different IWAs, outer working angles (OWAs), and instantaneous bandwidths. The OWA is primarily set by the number of actuators in the DMs. As for all coronagraphs, the total wavelength range of each channel is covered with several smaller wavelength bands.

ECLIPS-A also includes optical elements for a vortex coronagraph (VC). This type of coronagraph has advantages in better throughput at small IWAs, but does not perform well when the target stars are resolved and the telescope has a central obscuration. However, the LUVOIR Study Team chose to include a VC in ECLIPS-A for use on the most distant target stars, which are unresolved. Since the LUVOIR-B OTA is an off-axis telescope, the primary coronagraph type for ECLIPS-B is a DM-assisted vortex coronagraph (DMVC). This provides a factor of ~2 increase in both core throughput and instantaneous bandwidth compared to the LUVOIR-A APLC. All coronagraph masks were designed to provide contrast $\approx 1\times10^{-10}$ at the IWAs.

### 1.11.3 LUVOIR UV Multi-Object Spectrograph (LUMOS)

**Section 8.2.4** This flexible and capable instrument has the key goal of efficiently studying gas over a wide range of temperatures (10s of K to $10^5$ K) and densities. The instrument has three channels: a FUV multi-object spectrograph (MOS), a NUV/VIS MOS, and an FUV imager. LUMOS was carefully designed to maximize the instrument's end-to-end throughput. Coupling the relatively high instrumental throughput with the gains in collecting area over Hubble, LUMOS can reach much fainter sources than currently possible with Hubble's UV spectrographs, roughly 100–1000 times fainter for LUMOS-A and 30–300 times fainter for LUMOS-B. A summary of LUMOS's specifications appears in **Table 1-9**.

*Microshutter arrays.* LUMOS achieves its multi-object spectroscopic capability through the use of microshutter arrays (MSAs), which are based upon those implemented for the Near Infrared Spectrograph on JWST. The MOS field-of-view is covered by a 2 × 2 grid of MSAs, with individual shutter dimensions of 100 μm × 200 μm and a total of 840 × 420 shutters. LUMOS is capable of obtaining spectra of up to 840 individual objects simultaneously and the apertures are fully configurable. LUMOS performs bright object protection by pre-imaging the target field through a neutral density filter and the MSA, enabling autonomous shutter selection and fine pointing adjustments.

**Table 1-9.** *Characteristics of the LUVOIR UV Multi-Object Spectrograph (LUMOS)*

| LUMOS Characteristics | | | | |
|---|---|---|---|---|
| | **FUV MOS** | **NUV MOS** | **VIS MOS** | **FUV Imaging** |
| Bandpass | 100–200 nm | 200–400 nm | 400–1000 nm | 100–200 nm |
| Field-of-view | 2'x 2' | A: 1.5'x 2'; B: 2'x 2' | A: 1.5'x 2'; B: 2'x 2' | A: 1.2'x 2'; B: 2'x 2'(B) |
| Spectral resolutions[a] (λ/Δλ) | A: 583; 18,000; 47,000 B: 537; 17,000; 52,000 | A: 28,000 B: 33,000 | A: 20,000 B: 28,000 | Not applicable |
| Angular resolution[a] (mas) | A: 30 B: 32 | A: 19 B: 23 | A: 21 B: 41 | A: 42 B: 40 |
| Apertures | 840 x 420 | 840 x 420 | 840 x 420 | Not applicable |

[a] Average value in best 1'x 1' of field-of-view

*Contamination control.* Achieving high throughput with LUMOS will require careful control of contaminants on all optical surfaces. Therefore, the instrument is designed to operate at a slightly warmer temperature (280 K) than the surrounding environment (270 K). This ensures that there are no cold traps in the instrument on which contaminants would tend to condense. Additionally, the instrument incorporates heaters to allow infrequent on-orbit bake-outs to release any condensed water vapor from the optics.

*Detectors and coatings.* The FUV focal planes consist of large-format microchannel plate detectors, which have a long history of reliable performance on multiple UV space telescopes and sub-orbital missions. The NUV/VIS focal plane is an array of δ-doped CMOS detectors. LUMOS optics will be coated with the same protected "enhanced" LiF coatings as the LUVOIR telescope optics.

### 1.11.4 POLLUX

Chapter 13

This UV spectropolarimeter provides observing capabilities complementary to those of LUMOS. POLLUX was studied and designed by a consortium of European institutions, with leadership and support from the Centre National d'Etudes Spatiales (CNES). POLLUX represents a potential future international contribution to the LUVOIR mission, suitable for a European Space Agency (ESA) M-class call for proposals. The instrument offers high-resolution (R=120,000) point-source spectra over 100–400 nm (stretch goal of 90–400 nm). It includes linear+circular polarization and unpolarized observing modes. POLLUX was designed to occupy the fourth instrument bay on LUVOIR-A. The technical details for POLLUX, including a technology assessment and development plan, appear in **Chapter 13** and **Appendix H**.

### 1.12 Enabling LUVOIR

Revolutionary science goals require revolutionary technology. LUVOIR builds upon investments made by HST, JWST, WFIRST, and numerous laboratory, ground-based observatory, sub-orbital, and smallSat experiments (**Figure 1-31**). However, additional advances beyond the current state-of-the-art will be necessary to enable this new great observatory. In particular, we have identified for early development three technology systems critical to LUVOIR's science objectives:

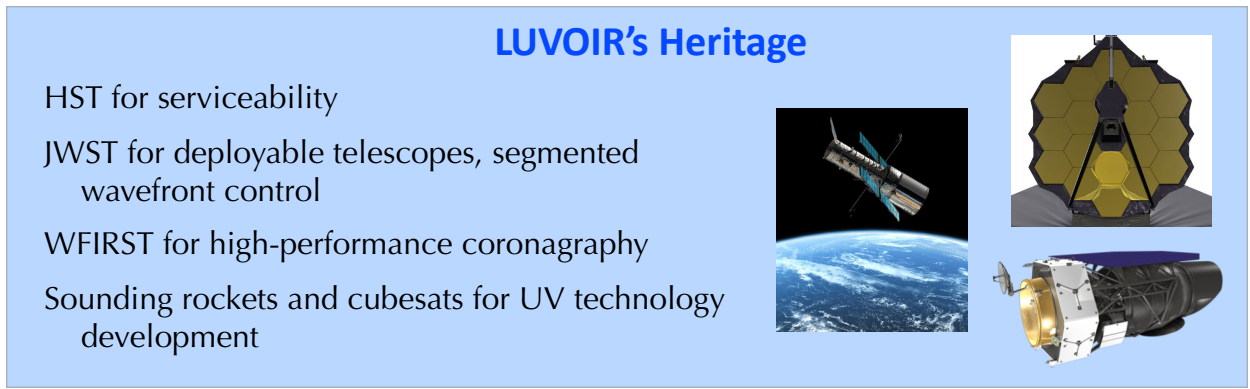

**Figure 1-31.** *The heritage of LUVOIR's design and technology. Credit: A. Roberge (NASA GSFC)*

1.  The high-contrast coronagraph instrument system

2.  The ultra-stable segmented telescope system

3.  The ultraviolet instrumentation system

## 1.12.1  Technology development

**Chapter 11**  LUVOIR's most technically challenging observational goal is the extreme starlight suppression needed to directly observe Earth-like exoplanets around Sun-like stars ($10^{-10}$ raw contrast). There are two basic instruments to do this: coronagraphs and starshades. For telescopes as large as LUVOIR, coronagraphs are the most efficient means to search for and study a large number of exoplanets (**Figure 1-7**). However, a moderately-sized starshade covering only short wavelengths may be a viable alternative to the NUV channel in ECLIPS (details in **Appendix I.1**). The highest coronagraph contrasts measured to date in broadband light are $2 \times 10^{-9}$ with an obscured aperture[2] and $6 \times 10^{-10}$ with an unobscured aperture (Trauger et al. 2012). Therefore, high-contrast performance must be improved in several ways for LUVOIR to achieve its goal of directly studying rocky exoplanets in the habitable zones of Sun-like stars.

*High-contrast coronagraph instrument system.* This technology effort covers instrument components that enable LUVOIR's exoplanet direct imaging observations. Coronagraph masks, deformable mirrors, active wavefront sensing and control, and ultra-low-noise detectors must work together as a system to suppress a parent star's glare by ten orders of magnitude to observe faint orbiting planets and measure their reflected spectra. The WFIRST Coronagraph Instrument (CGI) has made great strides in demonstrating that high-contrast with an obscured aperture is possible, developing technologies and algorithms to stabilize the wavefront coming from the telescope (low-order wavefront sensors, deformable mirrors).

In addition, the Segmented Coronagraph Design and Analysis (SCDA) Study, under the direction of NASA's Exoplanet Exploration Program, has demonstrated that high-performance coronagraphs can be designed for segmented, obscured apertures (a point of doubt a decade ago). The SCDA study showed the effect of segment gaps upon coronagraph performance, and that reasonably sized gaps have negligible impact. Central obscurations, however, do negatively impact performance for current coronagraph designs and should be

2 **https://exoplanets.nasa.gov/internal_resources/1200/**

kept as small as possible. Further work developing coronagraph designs that are less sensitive to central obscurations is underway and should continue.

*Ultra-stable segmented telescope system.* This technology effort is also critical to the high-contrast science. While the coronagraph technologies are working together to achieve high contrast, the telescope must remain stable during the integrations in order to maintain that contrast. The current requirement on the final level of wavefront stability at the coronagraph mask is ~10 picometers over ~10 minutes for the critical spatial frequencies that correspond to the high-contrast region of the coronagraph's focal plane (Coyle et al. 2019). There may be avenues to relax this requirement in future (e.g., Kasdin et al. 2019).

For now, mirrors and structures that are thermally stable and dynamically stiff are combined with an active metrology system and picometer-resolution actuators to achieve a stable optical system. A non-contact isolation system physically separates the science payload from dynamic disturbances generated on the spacecraft. All of these components combine to create a segmented telescope system that can maintain picometer-level wavefront stability during high-contrast observations.

The LUVOIR Study Team has assessed the technological maturity of each **piece** of the starlight suppression system; all are currently at TRL 3 or higher. The major remaining challenge is to show that the whole design works together as a **system** to provide the needed final coronagraph performance. In 2017, NASA funded two one-year industry studies of ultra-stable, segmented opto-mechanical systems. Results from these studies are available (Coyle et al. 2019a; Dewell et al. 2019) and have been submitted to Astro2020 as APC White Papers (Coyle et al. 2019b; East & Wells 2019; Nordt & Dewell 2019). Proposals for follow-on, two-year technology development contracts are under review. These will advance the technological maturity of stable, segmented telescopes, and will evaluate the end-to-end performance that can be achieved. In combination with further planned coronagraph development, these efforts will advance the maturity of LUVOIR's whole starlight suppression **system** to TRL 3 or higher.

*Ultraviolet instrumentation system.* Finally, a number of technologies enable LUVOIR's far-ultraviolet science observations. New high-reflectivity mirror coatings extend LUVOIR's sensitivity below 115 nm. Next-generation microshutter arrays enable multiplexed spectral observations of hundreds of objects simultaneously. And LUVOIR's sensitivity and spatial resolution goals demand new large-format detector technologies. Several sounding rocket missions have been progressively closing the gaps on these technologies, though some work still remains to meet LUVOIR's specific performance goals.

**Chapter 11** contains the LUVOIR Study Team's technology assessments, as well as a detailed development plan that matures each of these technology systems to TRL 6 through a series of component, assembly, and sub-system-level demonstrations. Engineering and manufacturing risk-reduction efforts are also identified and included in the plan. Maturing all technologies to TRL 6 prior to Phase A ensures that the technologies' performance is well understood before detailed concept development and design begins. The LUVOIR Study Team's Pre-Phase A development plan in **Chapter 11** includes a 5-year schedule with funded schedule reserves and an estimated cost with margin (total of $670M in FY20$).

## 1.12.2  Integration and test (I&T)

Section 12.4    Integration and testing of an observatory as large as LUVOIR will present challenges. The three major ones are:

- Observatory size
- Verification of required wavefront stability
- Control of contamination impacting far-UV science

Some of these challenges can leverage processes and techniques developed for previous missions, while some will require new approaches.

NASA has traditionally adhered to a "test as you fly" philosophy. This has been interpreted as a requirement to subject fully integrated flight systems to a series of tests that replicate launch and on-orbit conditions. However, LUVOIR is so large that it is impossible to exactly replicate the flight conditions in a test environment for the entire observatory (which was also true for JWST). Given that JWST has already stretched the available transport, integration, and testing infrastructure capabilities, new options must be developed for LUVOIR and other future observatories. The "test as you fly" paradigm must continue to evolve into a process combining rigorous testing of observatory units with high-fidelity, validated systems-level modeling to verify expected performance. LUVOIR's modular design, originally intended to enable on-orbit servicing, will facilitate transportation of observatory modules and unit-level testing during I&T.

LUVOIR's stability requirements will demand picometer metrology and sub-milliKelvin-level thermal stability for at least some testing. These levels of sensitivity are being demonstrated in lab facilities today over small spatial scales (e.g., Saif et al. 2019). However, they are beyond state-of-the-art for larger scales. Active design capable of correcting wavefront disturbances (discussed above) and verification by analysis may be the best ways to demonstrate that LUVOIR is capable of achieving the required wavefront stability on-orbit. The need for FUV observations means that the LUVOIR hardware will be sensitive to molecular contamination.

The general plan is to adopt the standard procedures used on previous UV and X-ray instruments and systems, and adapt them to the larger scale and complexity of LUVOIR. These include material controls, component bake-outs both on-ground and on-orbit, certifications, and routine inspections and cleaning. Both telescopes will be heated to an on-orbit operating temperature of 270 K, selected to prevent condensation of outgassed materials on optical surfaces during operations.

## 1.13  LUVOIR cost estimates

Appendix J    The LUVOIR mission represents a transformative vision for the future. Bold science goals that will revolutionize humanity's understanding of their place in the universe require ambitious capabilities and technology. Nevertheless, the LUVOIR Study Team and NASA both believe that a balanced portfolio of missions of different scales is critical for the health of the scientific community and are keenly aware that time and resources for new missions are not inexhaustible.

NASA, its government stakeholders, industry, and the scientific community must work together to find solutions that enable the capabilities needed by current and future scientists at a reasonable cost-benefit ratio. While the large amount of time and expense needed to realize a strategic mission may be warranted, long schedule delays and attendant cost over-runs are not. It is difficult to accurately cost large mission concepts at an early development stage, but the LUVOIR Study Team has put a great deal of thought and effort into enabling increased cost fidelity and mitigating cost and schedule risks. The key features of our effort are:

- Greatly increased design and engineering detail for the concepts

- A thorough technology development plan, including schedule and estimated costs, that matures technologies earlier in mission formulation

- Mitigation of launch vehicle risk via a scalable mission architecture

- Conservative programmatic and engineering margins, in some cases more conservative than required by NASA's Goddard Space Flight Center (GSFC)

- Recommendations for improvements to project management, procurement, and funding procedures (more on this below)

### 1.13.1  Methodology and results

GSFC has performed cost estimates for both LUVOIR concepts using two different internal cost estimating teams: the Resource Analysis Office (RAO) and the Cost Estimating, Modeling, and Analysis Office (CEMA). These two offices use different tools and methods, and are firewalled from each other. The RAO and CEMA cost estimate ranges appear in **Table 1-10.** These are life-cycle costs covering approval for formulation (start of Phase A) through the end of prime mission operations (end of Phase E). Estimates are presented in fiscal-year 2020 dollars (FY20$) and real-year dollars (RY$). The former are useful for comparing to the costs of other missions, while the latter are useful for future planning purposes. Cost estimates are presented as ranges of values corresponding to two different confidence intervals (50%–70%), determined from probabilistic cost analysis. Details on costing analyses appear in **Appendix J.**

**Table 1-10.** *LUVOIR's life-cycle cost estimates*

| Cost Model | FY20 ($B) | | RY ($B) | | TRL at KDP-A | TRL at PDR |
|---|---|---|---|---|---|---|
| | LUVOIR-A | LUVOIR-B | LUVOIR-A | LUVOIR-B | | |
| Phase A Ready | | | | | | |
| RAO | 13–16 | 8–10 | 19–24 | 12–15 | ≥ 5 | ≥ 6 |
| CEMA | 15–16 | 11–12 | 20–24 | 15–18 | ≥ 6 | ≥ 6 |
| Later Technology Development | | | | | | |
| RAO | 19–24 | 12–14 | 28–35 | 17–21 | 3–5 | ≥ 6 |

LUVOIR-A numbers assume a mission-enhancing contribution from ESA in the form of the POLLUX instrument; approximate costs to integrate that instrument are included. LUVOIR-B numbers do not assume any international contribution.

**Table 1-10** shows that the cost of the LUVOIR Pre-Phase A technology development plan much more than pays for itself in downstream savings. The RAO provided cost estimates for two different assumptions about technological maturity. "Phase A Ready" adopts TRL ≥ 5 for all technologies at Key Decision Point A (KDP-A), which is the start of Phase A, and TRL ≥ 6 at Preliminary Design Review (PDR), which occurs near the end of Phase B. This is the current NASA requirement. "Later Technology Development" assumes technologies are in the range TRL = 3−5 at KDP-A. The difference between the cost estimates for these two different maturity assumptions provides a rough assessment of how much money may be saved by completely executing LUVOIR's Pre-Phase A plan before starting Phase A (e.g., $6−8B in savings for LUVOIR-A in FY20$ at a price of $670M for the plan).

The CEMA "Phase A Ready" cost estimates assume the TRLs recommended by the LUVOIR Study Team (TRL ≥ 6 at KDP-A). The LUVOIR Pre-Phase A technology development plan brings all technologies to this level, more mature than the current NASA requirement. The LUVOIR Study Team looks forward to seeing the results of uniform independent cost analysis via the Astro2020 TRACE.

### 1.13.2 Improving execution of large projects

| Chapter 12 | While these cost estimates should be far more accurate than previous cost estimates for large missions at such an early stage, a word of caution is needed. The LUVOIR Study Team aimed to make the formulation and development schedules conservative in their margins; the RAO and CEMA schedule risk analysis indicated that this goal was achieved. However, these cost estimates assume that funding needed to execute the planned schedules is available year-by-year. Historically, this has often not been the case for NASA large missions.

The LUVOIR Study Team investigated funding models used by other parts of the US federal government, specifically the Department of Defense (e.g., Birkler et al. 2002). One that we believe presents significant advantages for LUVOIR (and other flagship-class observatories) involves dividing the project into multi-year blocks of work, logically constructed and each being fully funded at one time (**Section 12.1.2**). Advantages include:

- Multiple cost estimates, which increase in accuracy over time, are performed

- Congress and NASA need only commit to funding the next block of work, rather than the full mission

- Since each block is fully funded at the start, project managers can execute an optimal schedule for each block, reducing risk of delays and overruns

- The block funding model gives stakeholders multiple decision points to delay, re-scope, or cancel the project if total mission costs end up significantly higher than anticipated

While improved phasing of resources is not a panacea that solves all problems, the LUVOIR Study Team believes that such models should be further investigated and then implemented by NASA with agreement from Congress.

The LUVOIR Study Team also considered project-level avenues for further mission cost and risk reduction (**Section 12.1.1**). First, we concur with Bitten et. al. (2019) that all aspects of the design must be completely matured by the end of Phase B. Parallel manufacturing, integration, and testing operations should be used where possible. Discussions with members of the JWST Team and industry partners indicate that it is more efficient for industry to receive contracts directly from NASA instead of as a subcontract from a "prime contractor." Finally, all participants in the mission development should be incentivized for the success of the entire mission and should operate as a "badge-less" integrated team to share needed expertise across organizations.

## 1.14  Conclusion

LUVOIR is an ambitious mission concept that will provide the most science for the most scientists and has the potential to answer some of humanity's oldest questions about life in the universe. Careful attention to design and new ways of executing large missions will be needed to make LUVOIR feasible in terms of cost. As much as any of LUVOIR's technologies, improved project management is ***enabling*** for the mission. The LUVOIR Study Team believes that all this can be done, and has worked to develop plans to effectively realize our vision for a revolutionary observatory. We further believe that the mission's compelling science goals will excite the scientific community, the public, and the stakeholders in government, leading to sustained and possibly increased support for space astrophysics in the United States.

## CHAPTER 2. ROADMAP TO THE REPORT

As an aid to the reader, we provide this roadmap to the LUVOIR Final Report. It is intended to aid navigation through the document, identifying the locations of key material.

The report begins with the LUVOIR Signature Science Cases (SSCs), summarized in **Table 2-1**. In **Chapters 3–6**, we explain the motivations for these cases, identify key measurements, and set out needed observations. The telescope and instrument characteristics required for the observations are also identified. We have developed concrete observing programs for each Signature Science Case to ensure that the LUVOIR designs can execute this compelling science within the prime mission lifetime (**Appendix B**: Design Reference Missions). Additional science cases contributed by the LUVOIR STDT and the broader community appear in **Appendix A**.

**Chapter 7** presents the LUVOIR architecture and the two specific LUVOIR concepts (LUVOIR-A and LUVOIR-B) that satisfy the science objectives. The detailed technical approaches for each segment of the mission appear in **Chapters 8–10**. **Chapter 11** contains a complete discussion of LUVOIR technology development, with a proposed schedule and estimated cost for the development plan. **Chapter 12** outlines the LUVOIR management and systems engineering approach, including:

- A discussion of project-level and agency-level strategies to improve development cost and schedule performance
- An extensive set of recommended Pre-Phase A activities
- The LUVOIR Study Team's approach to integration & testing (I&T) and verification & validation (V&V)
- An overview of the project schedules. The detailed project development schedules appear in **Appendix G**

The final chapter, **Chapter 13**, contains science and technical details for an instrument (POLLUX) studied and designed by a consortium of European institutions, with leadership and support from CNES.

**Table 2-1.** *List of Signature Science Cases*

| Report Section | Signature Science Case |
| --- | --- |
| Section 3.2 | # 1: Finding habitable planet candidates |
| Section 3.3 | # 2: Searching for biosignatures and confirming habitability |
| Section 3.5 | # 3: The search for habitable worlds in the solar system |
| Section 4.1 | # 4: Comparative atmospheres |
| Section 4.2 | # 5: The formation of planetary systems |
| Section 4.3 | # 6: Small bodies in the solar system |
| Section 5.1 | # 7: Connecting the smallest scales across cosmic time |
| Section 5.2 | # 8: Constraining dark matter using high precision astrometry |
| Section 5.3 | # 9: Tracing ionizing light over cosmic time |
| Section 6.1 | # 10: The cycles of galactic matter |
| Section 6.2 | # 11: The multiscale assembly of galaxies |
| Section 6.3 | # 12: Stars as the engines of galactic feedback |

The main report chapters include high-level summaries of the tailored Concept Maturity Level 4 (CML 4*) products required for this study, as well as references to appendices where detailed descriptions and calculations are provided. **Table 2-2** provides a list of the required CML 4* products and their locations in this report.

**Table 2-2.** *Concept maturity level (CML) 4* product guide*

| Attribute | CML 4* definition | LUVOIR report section |
|---|---|---|
| Scientific Objectives and System Requirements | Working top-level scientific requirements drafted, linkages to scientific objectives identified and described.<br><br>Design reference scientific investigation defined with viable reduction options identified. | Chapters 3–6: Signature Science Cases<br>Section 7.1: Science mission profile<br>Section 7.2: Science and mission traceability<br>Appendix B: Design reference missions<br>Appendix C: Complete science traceability matrices |
| Science Data System | Design reference science data system sized to support data system flow-down requirements. | Section 9.2: Ground systems |
| Mission Development | Design Reference Mission defined, including driving requirements, initial high-level scenarios, timelines and operations modes; mass, delta-V, and power estimates; telecom and data processing approach to mission flow-down requirements. | Section 7.1: Science mission profile<br>Appendix B: Design reference missions<br>Chapter 8: Observatory segment<br>Chapter 9: Mission operations and ground segment |
| Spacecraft Systems Design | Spacecraft system architecture for design reference mission defined with mechanical configuration drawings and block diagrams to support spacecraft flow-down requirements. | Section 8.3: Spacecraft element |
| Instrument Systems Design | Instrument system architecture for design reference mission defined with mechanical configuration drawings and block diagrams to support instrument flow-down requirements and performance simulations. Instrument performance requirements traced to scientific requirements | Section 8.2: Payload element |
| Ground System / Mission Operations System Design | Mission Operations System/ Ground Data System architecture for design reference mission to support the ops scenario described. | Chapter 9: Mission operations and ground segment |
| Technical Risk Assessment and Management | Risk drivers listed. 5x5 matrix provided with relevant risk drivers (include selected mitigation / development options). | Section 12.3.2: Risk management |
| Technology | Technology options described. Baseline options selected and justified (technology roadmap). Rationale for TRL(s) explained. Risk mitigations (including fallback options, if any) for all new technologies. | Chapter 11: Technology development |
| Inheritance | Discuss all significant heritage assets used by the design reference mission. | Section 11.1: Technology items, TRL, and heritage<br>Appendix F: Heritage |
| Master Equipment Lists (MELs) | MEL documented for design reference mission to assembly level (e.g., antenna, propellant tank, star tracker, etc.). | Separate attachment: MEL spreadsheets |
| Technical Margins | Critical performance margins estimated, resource margin estimated for design reference mission (AIAA S-120 margin policies followed). | Chapter 8: Observatory segment<br>Section 12.3.3: Technical margin philosophy<br>Appendix E: Detailed technical budgets |
| Systems Engineering | Selected, high-leverage science, spacecraft, and ground system trades completed. | Section 12.3: Systems engineering<br>Appendix D: Completed technical trades |
| Launch Services | Preliminary launch vehicle(s) selection documented (NASA Launch Services used). | Chapter 10: Launch segment |

| Attribute | CML 4* definition | LUVOIR report section |
|---|---|---|
| Verification & Validation | Approach for verifying new and enabling functions of the design reference mission defined to support an acceptable risk assessment by independent reviewers. System testbeds and prototype models identified where applicable. | Section 11.2: Technology development approach<br><br>Section 12.4: Integration and test<br><br>Section 12.5: Verification and validation |
| Schedules | Top-level schedule (one page developed for design reference mission to support coarse independent cost estimates). | Section 12.6: Schedule overview<br><br>Appendix G: Detailed schedules |
| Work Breakdown Structure | NASA Standard WBS & Dictionary (down to level 2 and level 3 for spacecraft and payload used). Required for CML 3. | Section 12.1.1: Project-level strategies<br><br>Appendix J.5: LUVOIR Work Breakdown Structure (WBS) |
| Cost Estimation and Risk | Cost estimate and basis of estimate provided for design reference mission. Cost uncertainty quantified. Cost risks identified at subsystem level, with emphasis on enabling technologies. | Section 1.14: LUVOIR cost estimates<br><br>Appendix J: Astro2020 pre-Decadal cost estimation of LUVOIR<br><br>Chapter 12: Management and systems engineering |

## CHAPTER 3. IS THERE LIFE ELSEWHERE? HABITABLE EXOPLANETS AND SOLAR SYSTEM OCEAN WORLDS

After millennia of humanity wondering whether we are alone in the universe, LUVOIR's large aperture and compelling instrument suite will enable the search for habitability and life on dozens of nearby worlds. LUVOIR's exoplanet survey will be both broad and comprehensive, revolutionizing our understanding of planetary habitability, and providing the first estimates for the frequency of habitable conditions and life in the local Solar neighborhood (**Figure 3-1**). With LUVOIR, we may learn just how alone we are in the universe—or discover the dazzling prospect that that our universe is teeming with other inhabited worlds.

LUVOIR's detailed study of the orbital parameters, atmospheric compositions, surface properties, and temporal variability of many rocky planets in the habitable zones of different stars will reveal environmental conditions for a diversity of worlds at different stages of evolution, and may usher in a new era of comparative astrobiology.

LUVOIR will help to address three profound scientific questions for exoplanets:

- How common are habitable environments on worlds around other stars?
- How common is life beyond the solar system?
- How does life co-evolve with its exoplanetary environment?

LUVOIR's unequaled ability to measure fundamental surface and atmospheric properties for dozens of Earth-sized habitable zone exoplanets will address the first two questions.

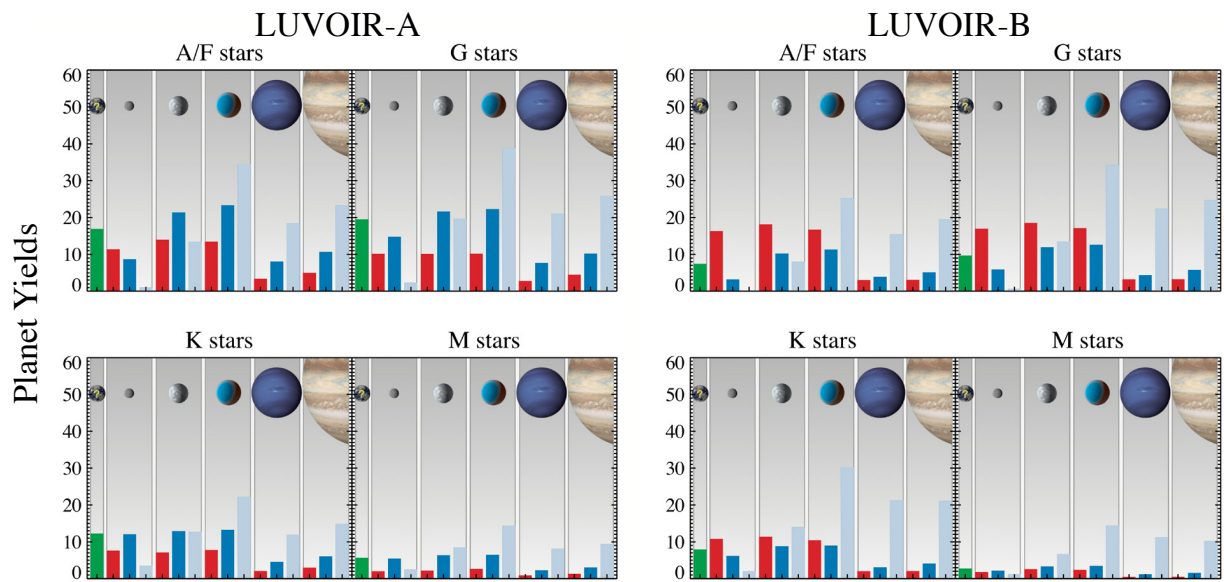

**Figure 3-1.** *Expected exoplanet detection yields for different types of stars during a 2-year exoEarth search with LUVOIR-A (left) and LUVOIR-B (right). Planet class types from left to right are: exoEarth candidates (green bar), rocky planets, super-Earths, sub-Neptunes, Neptunes, and Jupiters. Red, blue, and ice blue bars indicate hot, warm, and cold planets, respectively. Non-habitable planets shown here are detected during the exoEarth search. Color photometry and orbits are obtained for the vast majority of planets. Two partial spectra are obtained for all planets in systems with exoEarth candidates. Credit: C. Stark (STScI)*

## State of the Field in the 2030s

We anticipate significant advances in understanding solar system habitability and in small exoplanet detection and characterization between now and the mid 2030s. Exoplanet advances will necessarily focus almost exclusively on the more readily detectable planets orbiting M dwarf stars, whose worlds undergo early evolution and possibly extreme star-planet interactions that are not experienced by planets orbiting Sun-like stars. A mission like LUVOIR is essential for the study of planetary systems orbiting stars like our sun.

**The Europa Clipper:** The planned Europa Clipper (orbital insertion 2026–2031 depending on rocket type) will deliver *in situ* mass spectroscopy and directed spectroscopic measurements of this potentially habitable icy world in the outer solar system. LUVOIR can follow up on Europa Clipper discoveries with a regular cadence of observations over a longer time baseline and can monitor changes in the global plume network under different tidal heating conditions.

**Frequency of habitable planets:** On average, early M dwarfs harbor ~0.25 potentially habitable planets per star (Dressing & Charbonneau 2013, 2015). Recent detections of Proxima Cen b (Anglada-Escudé et al. 2016), LHS 1140 b (Dittmann et al. 2017), the TRAPPIST-1 planets (Gillon et al. 2017), and Ross 128 b (Bonfils et al. 2018) indicate that many rocky M dwarf habitable zone planets are nearby, and future detections are expected.

**Transiting Exoplanet Survey Satellite (TESS):** NASA's TESS (Ricker et al. 2014) launched in 2018 (Barclay et al. 2018) and will survey the entire sky for transiting habitable zone planets, with an anticipated yield of 10–20 potentially habitable M dwarf planets (Sullivan et al. 2015). These, and other planets orbiting M dwarfs discovered by ground-based telescopes, will be prime targets for future transit spectroscopy and possibly direct imaging with LUVOIR and other facilities.

**James Webb Space Telescope:** NASA's JWST is a 6.5-m infrared telescope that will launch in 2021. JWST will provide our first opportunity to probe the atmospheres of a few transiting M dwarf habitable zone planets and search for biosignatures.

**Planetary Transits and Oscillations of stars (PLATO):** ESA's PLATO mission is scheduled to launch in 2026 and will use planetary transits to search for exoplanets around stars ranging from M to G dwarfs (Rauer et al. 2014). PLATO's closest targets may be observable with LUVOIR.

**Ground-based observations:** Ground-based Extremely Large Telescopes (ELTs; 30–40-m diameter) will complement and enhance space-based observatories (Brogi et al. 2016), including LUVOIR. Ground-based observations will likely focus efforts on M dwarf stars, which have a more favorable planet-star contrast ratio ($10^{-8}$ for an M5V star with an Earth), and because adaptive optics systems perform best in the NIR (e.g., Bouchez et al. 2014). Upgrades to instrumentation (e.g., Lovis et al. 2017) may allow observations of the nearest M dwarf planets over 3–4 years with high-resolution transit spectroscopy and/or direct spectroscopy (e.g., Snellen et al. 2013; Crossfield et al. 2011). ELT direct imaging at 3–10 µm of terrestrial planets orbiting Sun-like stars may provide information on thermal emission that is complementary to LUVOIR's reflected light observations (Quanz et al. 2015).

Answering the last question will require detailed, in depth study of the most promising worlds.

LUVOIR will also revolutionize the search for habitability and life closer to home, in our outer solar system. With spatial resolution comparable to flyby missions, UV capability, monitoring on multi-year timescales, and unprecedented sensitivity, LUVOIR will provide multiple opportunities for global imaging and spectroscopy of a diverse population of potentially habitable solar system worlds, helping to tackle the question: Are there habitable environments and life elsewhere in the solar system?

## 3.1 The Earth through time as a guide

Because Earth is the best-studied planet and the only known world with surface water and life, it provides a crucial starting point to guide our studies of the remotely observable properties of habitable and inhabited worlds elsewhere. LUVOIR's search for habitable worlds and life beyond the solar system will sample habitable zone (HZ) planets orbiting stars of spectral type from F to M. The habitable zone is defined as the region around a star in which a rocky planet can support surface liquid water (Kasting et al. 1993; Kopparapu et al. 2013). These planetary systems will be at different ages, providing a window into various stages of habitable planet evolution, and possibly with dramatically different biospheres from our own planet. Studying this rich diversity of worlds requires a mission that is versatile and capable of exploring and characterizing environments unlike the modern Earth's (e.g., Kaltenegger et al. 2007; Rugheimer & Kaltenegger 2017).

To motivate and guide this search, we consider how the different environmental conditions and dominant biospheres throughout Earth's 4.6 billion-year history generated significantly different spectral features (**Figure 3-2**). LUVOIR will be capable of detecting life on a planet like Earth for all of its inhabited history. A summary of the eons of Earth's history is as follows:

The environmental conditions of the **Hadean** eon (before 4 billion years ago) are very poorly constrained due to an extremely sparse geological record, but Earth may have already have had continents during this time (Harrison et al. 2005), and life may have arisen during this eon (Bell et al. 2015; Nutman et al. 2016).

At the onset of the **Archean** (4–2.5 billion years ago), Earth likely had a flourishing anaerobic biosphere under an oxygen-poor atmosphere. The Sun was only about 75–80% as luminous as today, so a robust inventory of greenhouse gases—which likely included carbon dioxide ($CO_2$) and biogenic/geologic methane ($CH_4$)—would have been required to keep Earth hospitable, and in combination, these gases may indicate a biologically-driven atmospheric disequilibrium (Krissansen-Totton et al. 2018). Archean $CO_2$ estimates range from approximately modern levels to orders of magnitude more (e.g., Kanzaki & Murakami 2015; Rosing et al. 2010). Methane may have been between 2–3 orders of magnitude more abundant than today (Pavlov et al. 2000); methane may also have occasionally formed a photochemical organic atmospheric haze (e.g., Arney et al. 2016; Zerkle et al. 2012). Prominent spectral features from $CH_4$, $CO_2$, and possibly haze appear in the Archean Earth spectrum (**Figure 3-2**).

In the **Proterozoic** eon (2.5 billion years ago–541 million years ago), significant atmospheric oxygen accumulation occurred, irreversibly altering the chemical character of our planet's atmosphere and generating a powerful ozone UV shield that can be observed as a

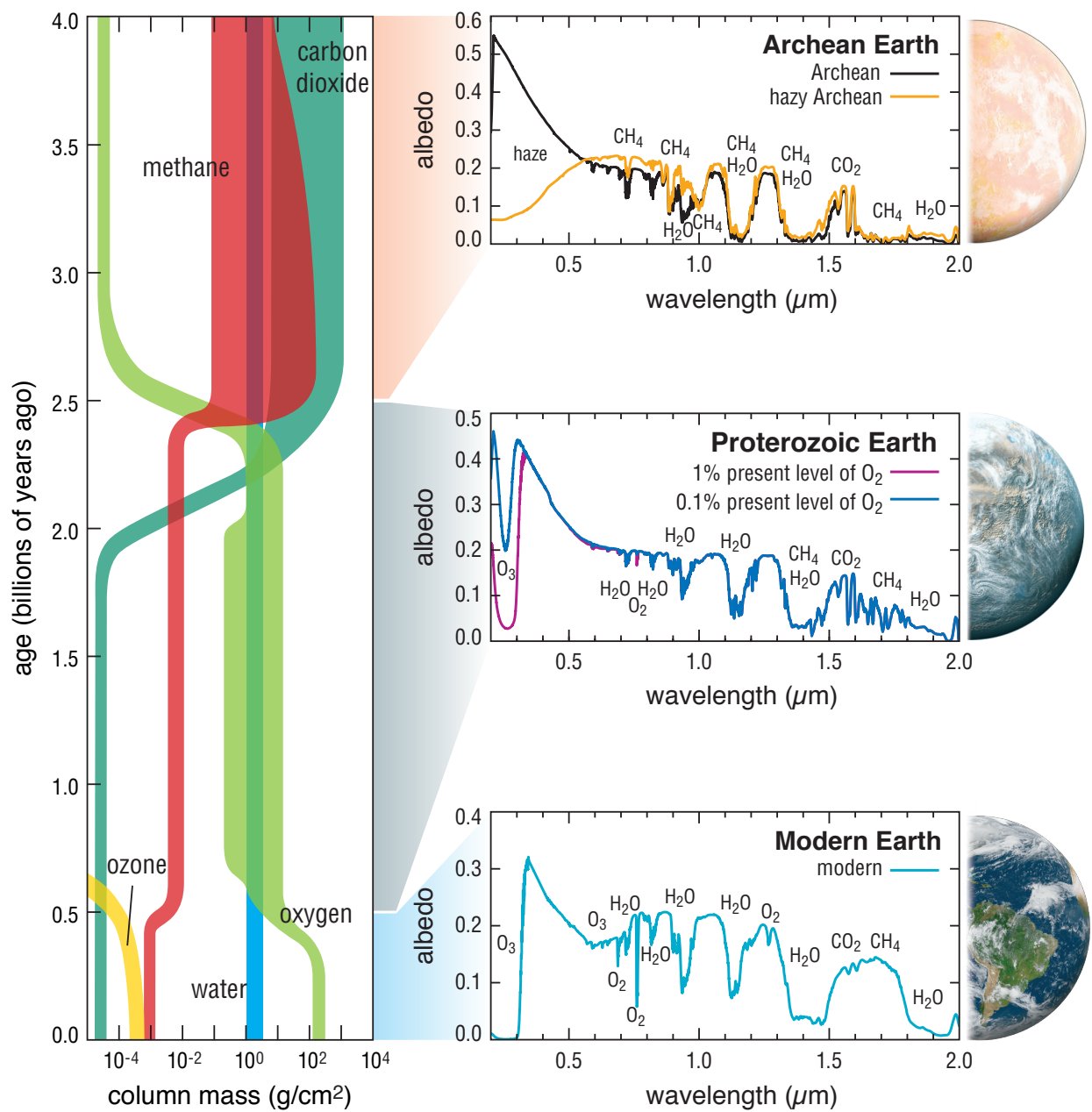

**Figure 3-2.** *The atmospheric composition of Earth has changed significantly over its history (left), as has its observable spectral features (right). For instance, the canonical biosignatures of modern Earth (e.g., $O_2$, $O_3$) cannot be observed in the Archean eon. A wide wavelength range enables higher fidelity spectral characterization by providing observations of many atmospheric species, and often multiple absorption bands of a given species. This breaks degeneracies between overlapping bands and enables the search for non-traditional biosignatures. Credit: G. Arney (NASA GSFC) / S. Domagal-Goldman (NASA GSFC) / T. B. Griswold (NASA GSFC)*

prominent spectral feature UV wavelengths. Mid-Proterozoic oxygen levels may have been only 0.1% of the present atmospheric level (PAL) (Planavsky et al. 2014), almost certainly precluding directly detectable $O_2$ spectral features for a mission like LUVOIR (Reinhard et al. 2017) and possibly leaving the $O_3$ feature in the UV as the only spectral evidence for atmospheric oxygen between the UV and the MIR (**Figure 3-3**). The atmospheric methane

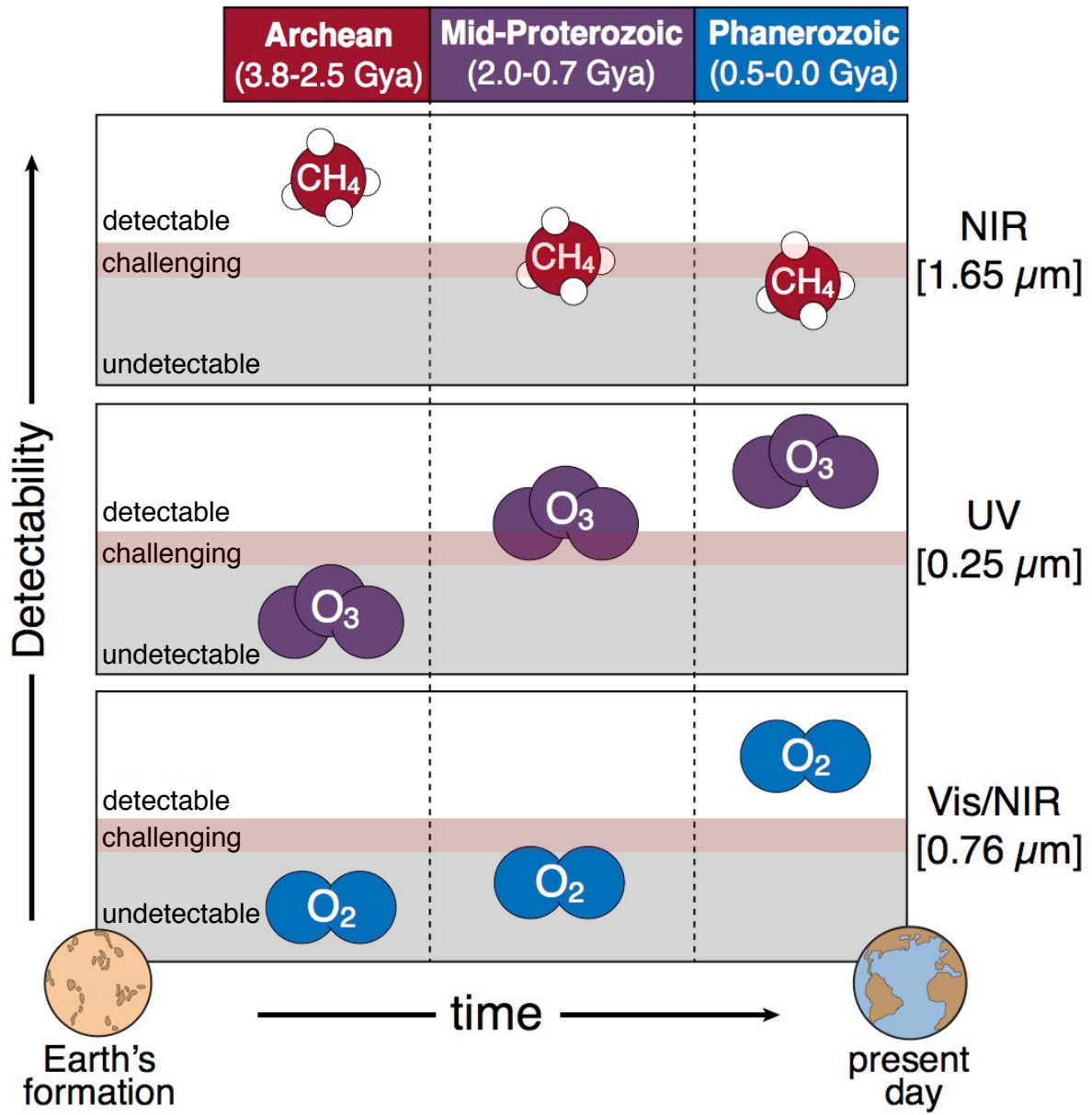

**Figure 3-3.** *The canonical biosignature of modern Earth, $O_2$, may not have been as detectable earlier in Earth's history. Other biosignatures, such as $CH_4$, may have been more easily detectable for early Earth than they are today. A wide wavelength range from the NUV to the NIR is insurance against false negative detections of life. During the mid-Proterozoic, ozone is more readily detectable than oxygen itself, while $O_2$ and $CH_4$ may have been challenging to observe. Credit: C. Reinhard (GA Tech) / E. Schwieterman, T. Lyons (UCR) / S. Olson (U. Chicago)*

abundance in the Proterozoic may also have been diminished to low (i.e., approximately modern) levels (Olson et al. 2016). However, other biogenic greenhouse gases, such as nitrous oxide ($N_2O$), may have been enhanced (Buick 2007) if they could be shielded from UV photolysis (Stanton et al. 2018).

   **Phanerozoic or modern Earth** (541 million years ago–present) has been heavily studied for signs of habitability and biosignatures in the context of remote sensing and mission development studies (e.g., Sagan et al. 1993), although it is representative of only about 12% of Earth's total history. Key gases in modern Earth's atmosphere include $O_2$, $O_3$, $CH_4$, and $CO_2$, although the latter two gases may require long observations to detect. Modern Earth can be used to validate and constrain models of the remotely observable properties of Earth as an exoplanet (e.g., Robinson et al. 2010, 2011), and its biosignatures and chemical disequilibria inform the search for life (Hitchcock & Lovelock 1967; Krissansen-Totton et al. 2016).

*The co-evolution of life and its environment.* Over Earth's history, life has co-evolved with its environment, and it is often impossible to separate the history of the two. As a result, the signs of habitability and life are occasionally the same signatures. Therefore, in this chapter, the same gases are occasionally discussed in both contexts.

Because the interactions between life and the atmosphere are complex, and can occur on diurnal, seasonal, and much longer timescales, studying them will require sampling a large number of potentially habitable exoplanets of different ages, and time-resolved high-fidelity observations over the longest possible baseline. LUVOIR is uniquely suited to pursue these goals. Its aperture enables a survey of potentially dozens of Earth-sized HZ exoplanets, offering the unique and unprecedented opportunity to witness worlds at different ages and stages of evolution, and allowing us to compare nascent worlds to ones much older.

## 3.2 Signature Science Case #1: Finding habitable planet candidates

LUVOIR will search for and characterize planets orbiting in or near the habitable zones of their host stars (Kopparapu et al. 2013), assessing the frequency of habitable worlds in our local solar neighborhood and measuring the diversity of habitable planetary environments. No currently planned mission—including TESS, JWST, and WFIRST—can carry out this census of habitable environments, or characterize habitable zone planets orbiting Sun-like (F, G, K) stars. Here, we first define what a habitable exoplanet candidate is, then we describe observations required to discover habitable exoplanet candidates, and lastly, we describe features to be sought to confirm habitability.

### 3.2.1 Defining habitable exoplanet candidates

Planetary habitability is the outcome of the complex interplay of multiple factors including the nature of the host star, the census of greenhouse gases, and planetary surface temperature and pressure. Here, we outline the processes and properties that define habitable planet candidates, also called exoEarth candidates in this report.

*Surface liquid water.* We define habitable planet/exoEarth candidates as worlds where liquid surface water is possible. While there are solar system worlds that may host habitable subsurface environments (e.g., Europa), such subsurface biospheres would not be detectable across interstellar distances. Remote detection of a global biosphere necessitates surface-atmosphere communication, so the existence of liquid water on the planetary surface is crucial.

*The habitable zone.* The habitable zone defines the region around a star where planetary surfaces can support liquid water. We adopt the conservative HZ, spanning 0.95–1.67 AU for a solar-twin star (Kopparapu et al. 2013). While a planet must orbit in the HZ to be considered an exoEarth candidate, an orbit in the HZ does not guarantee habitability or even the existence of surface liquid water, which is dependent on factors such as initial

### So Close, So Mysterious: The Planet Venus

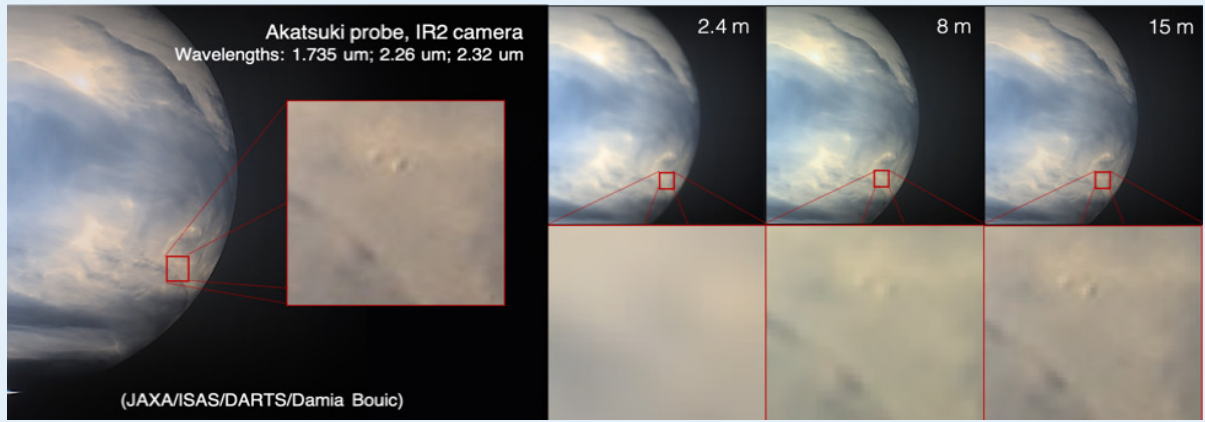

The environmental and physical conditions of Venus may occur frequently on hot exoplanets as a representation of the end state of habitability driven by high levels of stellar irradiation. LUVOIR's sun avoidance angle of 45° will allow observations of Venus at maximum solar elongation, and LUVOIR's spatial resolution can enable observations with up to 7 km per resolution element (14 km for LUVOIR-B) at 500 nm when the planet is at quadrature. By comparison, the Venus Monitoring Camera aboard ESA's Venus Express orbiter could obtain 0.2–45 km/pixel. LUVOIR can conduct observations from the UV to the NIR to shed light on processes in the Venus atmosphere. For instance, observations of $SO_2$ at 215 and 283 nm can inform studies of photochemical processes and can track large scale dynamical features in the Venus atmosphere as recently revealed by JAXA Akatsuki observations. Long temporal baseline monitoring of Venus can also inform studies of long term atmospheric dynamics: quasi-periodic variations in $SO_2$ abundance at 70 km have been shown to occur on decadal timescales. At NIR wavelengths, trace gases (HDO, $H_2O$, $SO_2$, HCl, CO, and OCS) can be observed and mapped in the sub-cloud atmosphere through spectral windows on the planetary nightside.

volatile delivery (e.g., Lissauer 2007; Raymond et al. 2006) and atmospheric/ocean loss. Atmospheric or ocean loss may be due to stellar winds (e.g., Garcia-Sage et al. 2017), luminosity evolution (Luger & Barnes 2015; Ramirez & Kaltenegger 2014; Tian 2015), and/or a small size (e.g., Mars). Consequently, although HZ planets are common, the frequency of small, temperate planets with habitable surface conditions remains unconstrained, and can only be determined via observations of exoplanetary environmental characteristics.

LUVOIR's large sample of exoplanets (approximately 50 and 30 exoEarth candidates for LUVOIR-A and -B, respectively) will be able to empirically test assumptions that underlie the HZ concept, including the prediction that atmospheric $CO_2$ abundance should increase with decreasing stellar irradiation due to temperature-dependent feedbacks in silicate weathering, a key part of the carbonate-silicate cycle that controls atmospheric $CO_2$ over long timescales (Walker et al. 1981). By measuring $CO_2$ as a function of stellar flux for a statistically significant sample (tens of planets; Checlair et al. in prep), we will be able to test this assumption (Bean et al. 2017; **Figure 3-4**).

With its large sample of planets and its wide wavelength range, LUVOIR can also explore and test alternate modes of habitability that may be possible, such as: planets with low water abundance (so-called Dune worlds; Abe et al. 2011) and planets with dense $H_2$-dominated atmospheres (Gaidos & Plerrehumbert, 2013). Finally, LUVOIR may help us understand processes that cause loss of habitability. For instance, we may discover desiccated worlds in the habitable zones of M dwarfs stars whose water was stripped away to space by the early extreme luminosity of their stellar hosts (e.g., Meadows et al. 2018).

**General atmospheric properties.** In general, Earth-like atmospheres are considered to be composed of a background of $N_2$ and greenhouse gases such as $H_2O$ and $CO_2$. The failure of these principle greenhouse gases (and others, like $CH_4$, $N_2O$, and $O_3$; **Section 3.1**) to provide a clement temperature range as a function of incident stellar radiation defines the limits of the HZ.

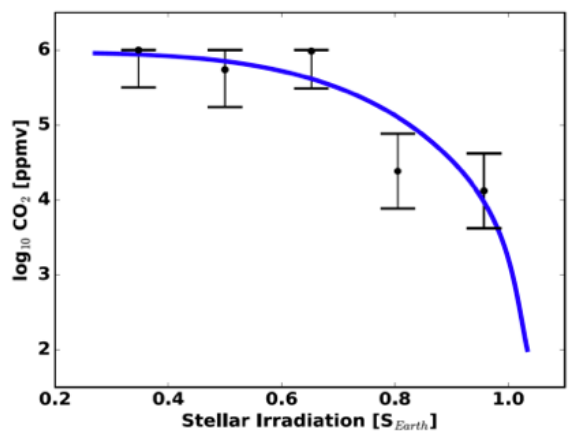

**Figure 3-4.** *Fundamental to our thinking of the habitable zone is a theory of how silicate weathering will draw down $CO_2$ as stellar irradiation increases. The blue curve shows predicted $CO_2$ needed to maintain a 290 K surface temperature, and the black points represent hypothetical planets (four planets per bin) with 1 σ error bars. A large survey of exoplanets enabled by LUVOIR would allow this theory to be empirically tested. LUVOIR will provide the largest sample of potentially habitable exoplanets of all proposed future space mission concepts, enabling statical studies such as this. Credit: Bean et al. (2017)*

Besides composition, atmospheric pressure can also affect habitability. High pressure atmospheres enhance the greenhouse effect via collisional broadening of absorption bands. Particularly low-pressure atmospheres may enable escape of water vapor due to a failure of the tropospheric 'cold trap' that would normally keep water constrained to the lowest part of the atmosphere (**Section 3.3.3**).

**Planet radii and mass.** For this report, exoEarth candidates include planets with radii greater than $0.8a^{-0.5}$ Earth radii, and less than 1.4 Earth radii, where $a$ is the HZ-normalized semi-major axis in AU. The lower radius is derived from an empirical atmospheric loss relationship for solar system bodies, where bodies smaller than this radius are less likely to retain atmospheres (Zahnle & Catling 2017). The upper limit on planet radius is a conservative interpretation of an empirically measured transition between rocky and gaseous planets at smaller semi-major axes (Rogers 2015).

The mass of a planet is a fundamental physical property. It affects a host of dynamical, interior, surface, and atmospheric processes that influence conditions for habitability and life, such as exoplanetary system architecture, plate tectonics, volcanic activity, and atmospheric structure. In the search for true habitable planets, establishing planet mass will play an important role in discriminating potentially habitable terrestrial planets from larger, low albedo gaseous planets.

In addition to mass constraints from astrometry, we will use broadband planet-star brightness ratios with reasonable limits on possible albedos to triage mini-Neptunes from high priority Earth-like targets. The geometric albedo $A_g$ relates to the planet-to-star flux ratio

## Small and Cold: Mars' Planetary Properties Have Strongly Impacted Its Habitability

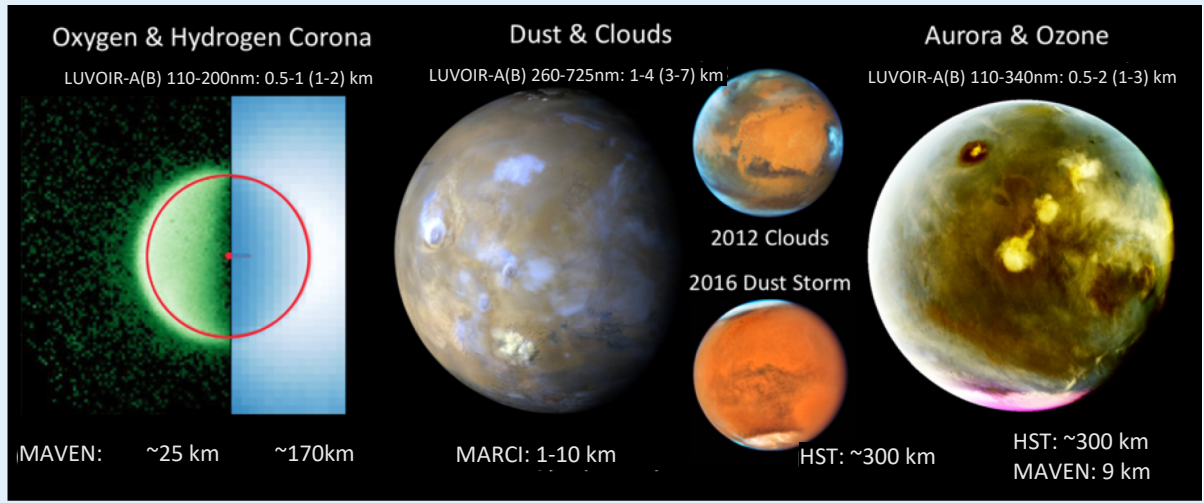

Although it may have boasted habitable conditions in the past, over geologic timescales, Mars has experienced significant atmospheric mass loss due to its small size. Yet Mars remains a compelling solar system target in our search for signs of past—and perhaps localized present—regions of habitability. LUVOIR's high resolution UV and NIR channels can reveal and characterize Martian atmospheric and photochemical processes, improving our understanding of atmospheric physics and loss processes for small, cold worlds. By combining UV/optical observations with the NIR, LUVOIR can simultaneously probe highly energetic processes (e.g., aurorae) and their molecular atmospheric precursors. With high spatial resolution, LUVOIR can provide detailed, full-disk coverage of Martian weather, including $CO_2$ ice clouds, whose climate effects may be important to understanding the outer edge of the habitable zone elsewhere. By tracking clouds and dust, we will enhance our understanding of volatile transport and deposition on Mars. Spectroscopy can map meridional transport with $O_2(a^1\Delta g)$ nightglow emission. LUVOIR will exceed the capabilities of orbiting spacecraft such as MAVEN for spatial resolution, and it provides two orders of magnitude better resolution than HST. Approximate maximum spatial resolutions of LUVOIR-A and -B together with MAVEN, MARCI, and HST resolutions are indicated in the image above.

as $F_p/F_s \propto A_g \times R_p^2$, where $R_p$ is the planet's radius. In this way, the albedo controls the shape of the planet's reflected light as a function of wavelength while the size of the planet sets the brightness level across all wavelengths. We can use these two concepts together to disambiguate mini-Neptunes from Earths (Windemuth et al., in prep).

**The stellar hosts.** Understanding the star that a given planet orbits is critical to understanding possible atmospheric evolution of an exoplanet. LUVOIR will observe planets in the HZs of F, G, K, and M dwarfs, potentially allowing a comparison of the evolution of habitability on planets orbiting stars with different luminosity evolution (Baraffe et al. 1998) and activity levels. The differing extreme UV (XUV; $\lambda$=10–124 nm) and UV ($\lambda$ < 400 nm)

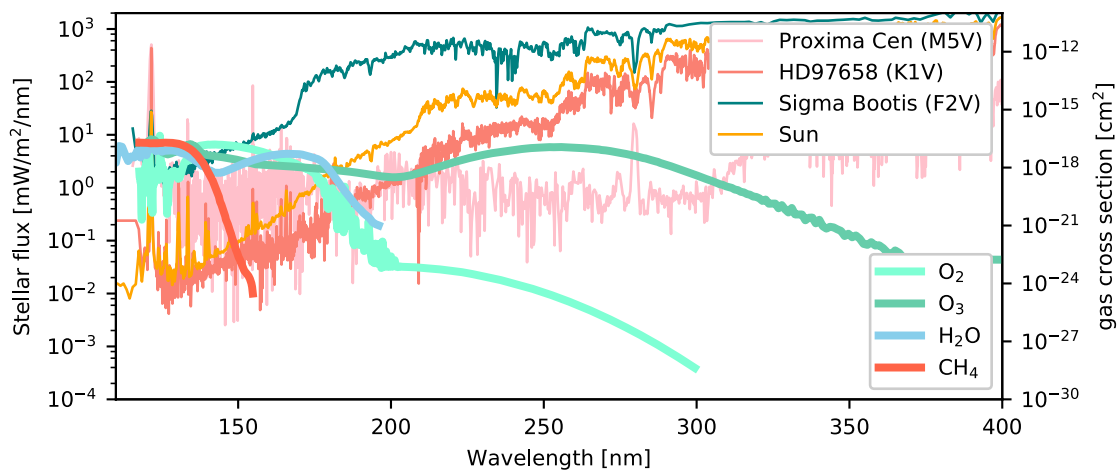

**Figure 3-5.** *LUVOIR will allow us to interpret the atmospheric compositions of planets orbiting a variety of stars, which are affected by photochemistry. To help us interpret the atmospheric compositions of potentially habitable planets, LUVOIR will obtain panchromatic spectra of all host stars. Left axis: UV spectra of the Sun, Proxima Centauri, a K1V dwarf, and a F2V dwarf. Right axis: UV absorption cross-sections for key gases. Atmospheric photochemistry is effectively controlled by the product of the stellar UV and the molecular absorption cross-sections. Thus, different UV spectra from different stars will lead to different photochemical outcomes. Credit: G. Arney (NASA GSFC) / Proxima Cen: Meadows et al. (2018); HD97658: France et al. (2016), Youngblood et al. (2016), Loyd et al. (2016); Sigma Bootis: Segura et al. (2003)*

spectra of host stars will lead to different atmospheric loss and photochemical processes (**Figure 3-5**), both of which affect planetary atmospheric composition.

*F/G/K dwarfs.* LUVOIR will primarily target F, G, and K dwarf stars, which offer the best chance of discovering planets with evolutionary histories similar to Earth's. About 90% of the LUVOIR stellar targets for the nominal exoEarth search will be F/G/K stars, which experience more moderate stellar evolution than M dwarfs. These systems represent our best chance of finding true Earth analogs to set our home world in context. K dwarfs, in particular, may be excellent targets in the search for life elsewhere (Cuntz & Guinan 2016) due to their more moderate stellar activity levels compared to M dwarfs (Richey-Yowell et al. 2019), improved planet/star contrast relative to brighter stars, and enhanced potential to detect oxygen/methane disequilibrium biosignatures relative to F and G dwarfs (Arney 2019). Observations of potentially habitable planets orbiting Sun-like stars requires a direct imaging mission because their transits are prohibitively shallow and infrequent.

*M dwarfs.* M dwarfs comprise 75% of the main sequence stars in our galaxy, so the question of whether their planets can be habitable is critical to understanding the distribution of habitable worlds and life elsewhere. LUVOIR's large aperture can allow direct observations of planets orbiting nearby M dwarfs, and they will comprise about 10% of the stars LUVOIR will examine during its nominal exoEarth survey (though LUVOIR could examine a much larger number of M dwarfs for more distant planets, and also observe transiting planets around these low mass stars).

Although F, G, and K stars may harbor habitable planets with histories and characteristics more like Earth's, M dwarf planets provide a fascinating complement from a habitability standpoint as they likely undergo a significantly different evolutionary sequence. This different evolution can include possible extreme water loss during the super-luminous pre-main sequence stellar phase (Luger & Barnes 2015), tidal locking (Heath et al. 1999), and extreme M dwarf activity (Tarter et al. 2007) that can drive severe atmospheric loss (e.g., Garcia-Sage et al. 2017). However, there are mechanisms that could allow for habitable conditions even in the face of these challenges such as late delivery of volatiles (Morbidelli et al. 2000) after the pre-main sequence phase, surface cooling provided by thick substellar cloud decks (e.g., Kopparapu et al. 2016; Yang et al. 2013; Fujii et al. 2017), and oceans that protect life from the radiation of flares (Segura et al. 2010).

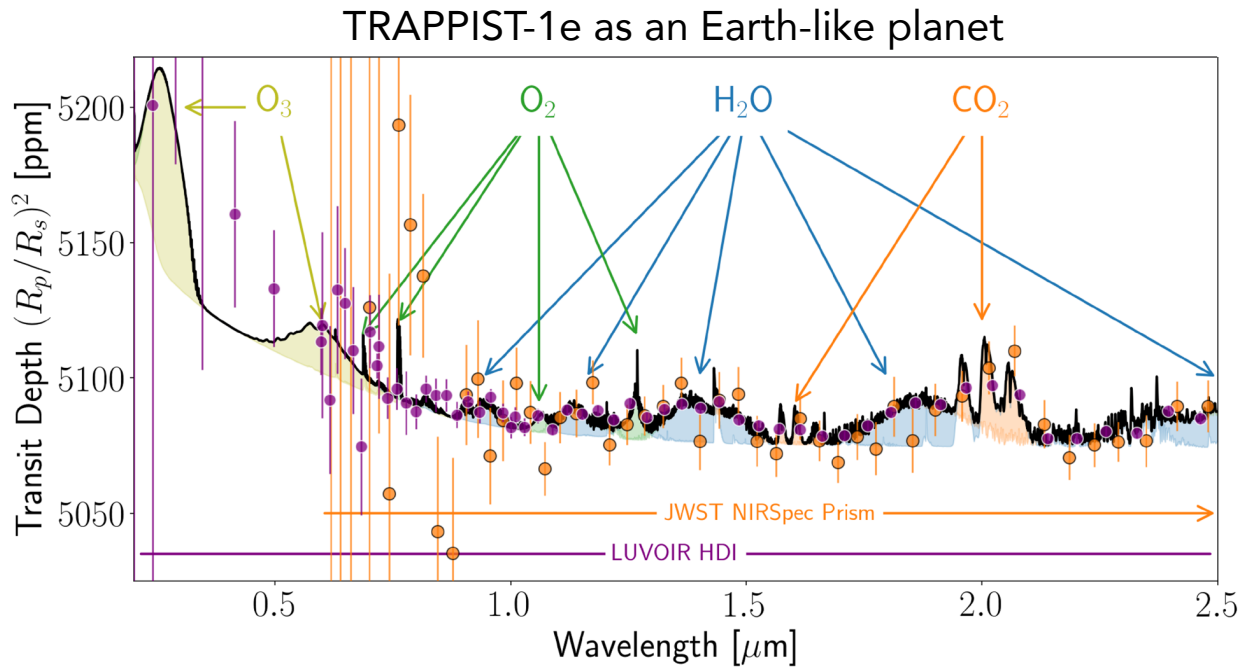

**Figure 3-6.** *LUVOIR can observe planets transiting M dwarfs at UV-Visible-NIR wavelengths, complementing observations with JWST and ground-based telescopes. Shown here is a possible transmission spectrum of TRAPPIST-1e as an ocean-covered aqua planet with a 1 bar atmosphere. Simulated observations assume 50 transits observed with LUVOIR/HDI grism and with JWST/NIRSpec prism (using the optimized readout mode from Batalha et al., 2018). JWST spectrum generated with PandExo (Batalha et al. 2017) Credit: J. Lustig-Yaeger/A. Lincowski (University of Washington)*

LUVOIR can conduct transit spectroscopy of transiting M dwarf planets in a wavelength range complementary to JWST. In particular, LUVOIR's access to the UV could test for atmospheric escape processes on transiting planets by detecting escaping H in transit (Bourrier et al. 2017; Ehrenreich et al. 2015). **Figure 3-6** shows a possible transit transmission spectrum of TRAPPIST-1e as an ocean-covered Earth-like planet with a 1 bar atmosphere from Lincowski et al. (2018). The simulated LUVOIR/HDI grism spectrum achieves 14x the SNR of the simulated JWST spectrum at 0.76 µm (the $O_2$ A band) and 4.8x the SNR of JWST at 1 µm. Larger error bars for both instruments at shorter wavelengths are driven in part by decreased luminosity of the TRAPPIST-1 star.

### 3.2.2 Discovering habitable planet candidates

Among the space-based telescopes being considered for the next decadal survey, LUVOIR has the largest collecting area and the smallest inner working angle (IWA). This will allow LUVOIR to detect and spectrally characterize a large number of planets (**Section 3.4**).

***Why direct observations?*** HZ terrestrial planets orbiting F/G/K stars are best observed through direct imaging and are inaccessible to transit observations due to the long orbital periods of these worlds and small transit depths. Space-based direct imaging will provide the first and best opportunity to image terrestrial planets in reflected light and characterize Earth-like worlds orbiting truly Sun-like stars.

Understanding the surface and near-surface environment of a potentially habitable planet is the key to discovering whether it can support a liquid water ocean. Direct imaging provides a short line of sight through the planet's atmosphere to probe the surface. Transmission spectroscopy, by contrast, observes a much longer path through the upper troposphere and stratosphere, and cannot see the planetary surface. Due to the effects of aerosols, refraction, and atmospheric opacity, transmission will have difficulty sampling even the near-surface atmosphere. Because water vapor on a habitable planet is generally confined to the troposphere, transit observations may not be able to detect water even when it is present. By contrast, LUVOIR's direct imaging spectroscopy will probe deep into planetary atmospheres, potentially all the way to the surface, providing crucial information on surface or near-surface physical conditions.

***Why space-based observations?*** While ground-based facilities with adaptive optics may be able to observe Earth-like planets around M-dwarf stars, whose contrast ratios are moderate ($\sim 10^{-7}$–$10^{-8}$), atmospheric fluctuations will prevent these facilities from reaching the $10^{-10}$ contrast required to observe Earth-sized planets around Sun-like stars (e.g. López-Morales et al. 2019; Males & Guyon 2018). In space, a stable telescope and deformable mirrors can provide the contrast needed to allow access to Earths around Sun-like stars.

***Mitigating confusion sources.*** Once LUVOIR obtains an image (**Figure 3-7**) of a potential exoplanet system, metrics are needed to distinguish planets from background stars, galaxies, brown dwarfs, and other sources of confusion like exozodiacal dust structures (**Figure 3-8**). Resolved sources like background galaxies should be straightforward to distinguish from planets, which will be unresolved point sources.

A combination of methods will be used to counter background confusion for point sources like stars and brown dwarfs. Multi-epoch observations will reveal targets with the same proper motion as the parent star, an excellent metric for target discrimination. Color photometry will reveal targets with colors consistent with planetary bodies.

***Planet orbits.*** Constraints on orbits are needed to determine whether a planet is in the habitable zone. Orbital constraints require about four detections of the planet spaced out over its orbit. Some observations will miss the planet, so we assume an average of 6 observations per star will be needed.

***Detecting atmospheric water vapor.*** LUVOIR will perform an initial habitability assessment by obtaining spectra of the water band at 0.87–1.05 µm to search for evidence of atmospheric water vapor. This band is useful for two reasons. First, it is less likely to be cut off by inner working angle constraints compared to longer wavelength water features. Second, the 1.0 µm $CH_4$ feature is accessible in this spectral range, which could provide evidence of an atmosphere rich in $CH_4$ (e.g., an

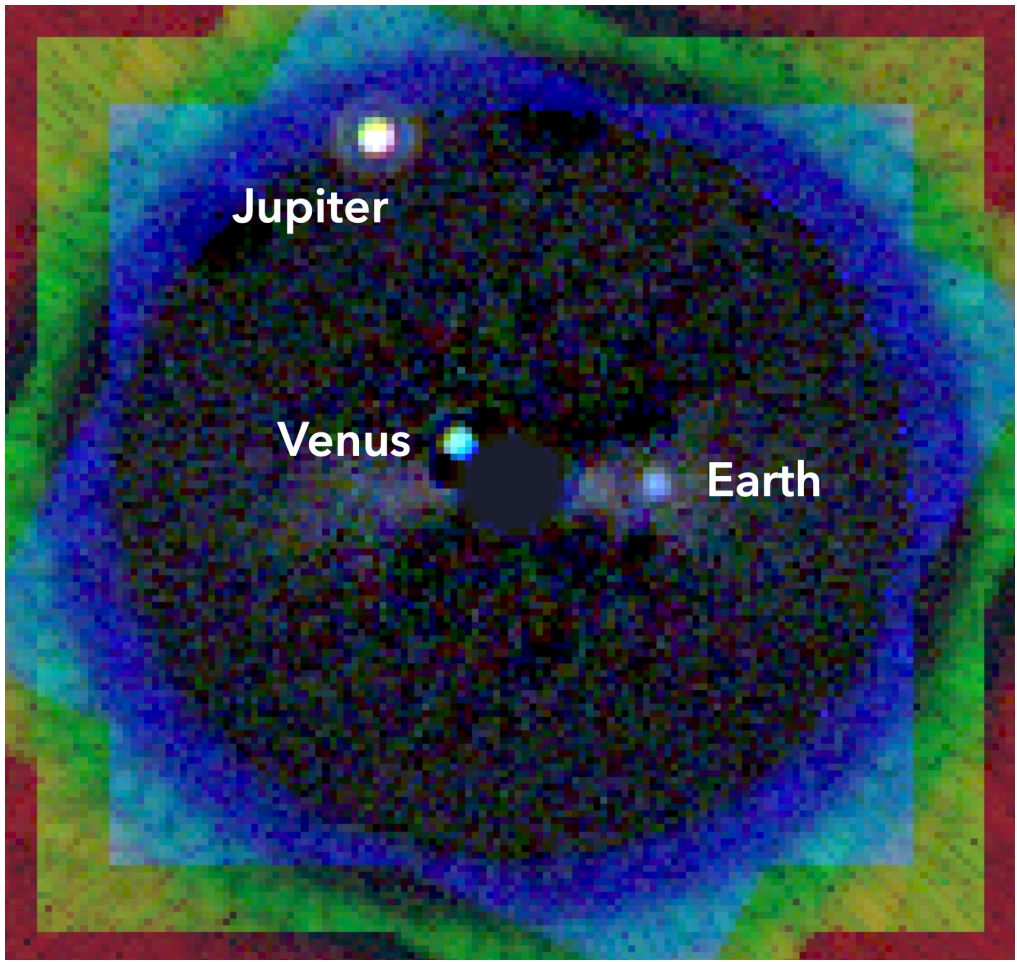

**Figure 3-7.** *Simulated image of a twin Solar System at a distance of 12.5 pc observed through the LUVOIR-A ECLIPS instrument. This RGB image is a composite of data acquired in two APLC masks (with respective working angles 3.5–12 λ/D and 7–27 λ/D) in three bandpasses (red / 800 nm; green / 700 nm; blue / 600 nm) at two observatory roll angles (27 degrees apart) over the course of 60 hours of total integration time. The coronagraph images were simulated with a diffraction model time series that includes 10 picometers of primary mirror segment jitter (random piston and tip-tilt errors applied to each mirror segment), 0.2 mas residual line-of-sight pointing jitter, and a stellar diameter of 0.75 mas. The input astrophysical scene is a model of a "modern" Solar System inclined at 60 degrees, with an exozodiacal debris disk (Roberge et al. 2017). In this scene, the Earth-like planet is observed at quadrature, appearing as a blue dot at 1 AU projected separation, to the right of the occulted star. Roll subtraction processing was used to remove starlight speckles from the "raw" co-added images. The residual structure of the exozodiacal disk—distorted by the roll subtraction— appears as a horizontally-extended diffuse cloud Credit: R. Juanola Parramon / N. Zimmerman / A. Roberge (NASA GSFC).*

Archean Earth analog or a Jupiter analog). The $H_2O$ and $CH_4$ features do not overlap significantly in this wavelength range, which will remove possible degenerate interpretations of the type of planet under observation.

We chose a detection threshold for $H_2O$ at 0.94 μm based on Feng et al. (2018) using SNR = 5, R =70 on the continuum near the band of interest. Note that an important subtlety of the Feng et al. results is that their reported SNRs values are calculated at 550 nm. We find that SNR = 10 at 550 nm from Feng et al. is equivalent to SNR = 5 on the continuum near the water band between 0.87–1.05 μm. Rocky planets in the HZ with positive water detections will be considered habitable planet/exoEarth candidates and will therefore be high priority targets for subsequent observations. Confirming habitability will be part of our detailed characterization of these worlds, discussed in the next section.

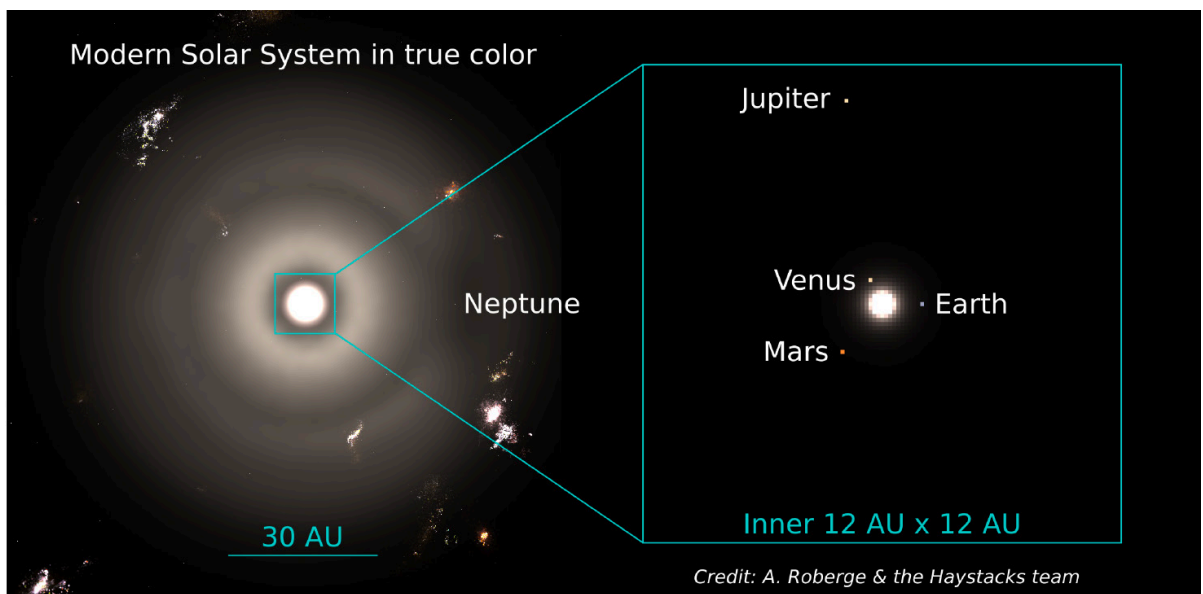

**Figure 3-8.** *Is that a planet? A high-resolution simulated view of the solar system at a distance of 10 pc including expected noise and confusion sources: scattered light from interplanetary dust (aka. exozodiacal dust), background galaxies, and background stars. For ease of viewing, the Sun is not included; the bright source in the center of the inset panel is dust. Credit: Roberge et al. (2017)*

### 3.3 Signature Science Case #2: Searching for biosignatures and confirming habitability

Detection and interpretation of biosignatures and environmental features requires high-contrast, high sensitivity, moderate resolution spectroscopy, and a broad wavelength range to detect multiple spectral features from potential biosignatures and the planetary environment (**Figure 3-2**). The environmental context is needed to help assess the likelihood that the potential biosignature has a biological, rather than planetary, source such as atmospheric loss or photochemistry (e.g., Meadows et al. 2018b). A large mirror enables detections of dozens of potentially habitable planets and access to a wider wavelength range for any given IWA. Here, we use the term exoEarth to refer to rocky planets in the habitable zones of their host stars that show positive signs of habitability. LUVOIR's survey of possibly dozens of nearby exoEarths offers the best chance in the coming decades to answer at long last whether we are alone in the universe, and to place Earth's biosphere in context.

### 3.3.1 What is a biosignature?

A planet-wide biosphere can modify its environment to leave characteristic indications of its presence known as biosignatures. These include specific gases (Des Marais et al. 2002; Sagan et al. 1993; Schwieterman et al. 2018b), atmospheric chemical disequilibria (Hitchcock & Lovelock 1967; Krissansen-Totton et al. 2018, 2016a), surface reflectance features (Gates et al. 1965; Hegde et al. 2015; Schwieterman et al. 2015; Seager et al. 2005) and time-dependent modification of environmental characteristics caused by biological processes (Meadows 2008, Olson et al. 2018).

A more generic and stronger search for biosignatures is to determine fluxes (i.e., flow rates) of gases into the environment and constrain sources and sinks for those gases. If no abiotic sources are plausible, this is strong evidence of life. This can be done by searching for environmental characteristics that are out-of-equilibrium, which implies a constant replenishment of a gas against chemical or photochemical destruction. For instance, $O_2$ is out of photochemical/geochemical steady state with $CH_4$ on Earth, implying significant surface fluxes for these gases that cannot readily be explained by abiotic processes (Hitchcock & Lovelock 1967).

In addition to detecting potential biosignatures, it is important to gather as much additional information as possible so that biosignatures can be interpreted in the context of the whole planetary and stellar environment. This is important to: 1) help rule out abiotic processes called "false positives" that can mimic biosignatures and, 2) uncover additional biogenic features to provide corroborating evidence for true biosignatures.

### 3.3.2 Potential biosignatures

In this subsection, we describe key potential biosignatures LUVOIR will seek in exoplanet atmospheres and on their surfaces. Absorption features from biosignatures and false positive discriminators are summarized in **Table 3-1**.

*Oxygen ($O_2$).* The highest priority biosignature gas to be sought is molecular oxygen (Meadows 2017). $O_2$ is the byproduct of oxygenic photosynthesis, currently the dominant metabolism on our planet and possibly the most productive metabolism for any planet (Kiang et al. 2007b, 2007a). Oxygenic photosynthesis uses cosmically abundant water, atmospheric $CO_2$, and sunlight to create biomass and power life.

Oxygen is unusual for biogenic products in having several strong features in the visible and near-infrared, including strong bands at 0.76 and 1.27 μm. Because its high abundance results in even mixing throughout the atmospheric column, it produces strong spectral features even above a planet-wide cloud deck. Oxygen (or its photochemical byproduct ozone) may have been variably present in Earth's spectrum since the Proterozoic period (2.5 billion years ago–541 million years ago; Segura et al. 2003; Kaltenegger et al. 2007), though may have been at concentrations too low to observe during the mid-Proterozoic. Because oxygen has not always been observable in our planet's spectrum, a robust search for biosignatures on exoplanets must consider a range of other gases.

**Table 3-1.** *Desired spectral features for biosignature assessment.*

| Biosignatures & False Positive Discriminants (indicated with *) | | |
|---|---|---|
| **Molecules/Feature** | **UV-VIS wavelengths (0.2–1.0 μm)** | **NIR wavelengths (1.0–2.0 μm)** |
| $O_2$ | 0.2, 0.63, 0.69, 0.76 (strong) | 1.27 |
| $O_3$ | 0.2–0.35 (strong), 0.5–0.7 | |
| $O_4$ $(O_2-O_2)$* | 0.345, 0.36, 0.38, 0.45, 0.48, 0.53, 0.57, 0.63 | 1.06, 1.27 (strong) |
| CO* | | 1.6 |
| $CO_2$* | | 1.05, 1.21, 1.44, 1.59 |
| $CH_4$ | 0.6, 0.79, 0.89, 1.0 | 1.1, 1.4, 1.7 |
| $N_2O$ | | 1.5, 1.7, 1.78, 2.0 |
| Organic haze | < 0.5 | |
| Vegetation red edge | 0.6 (halophile), 0.7 (photosynthesis) | |

**Ozone ($O_3$).** Ozone is a photochemical byproduct of $O_2$. The broad visible $O_3$ Chappuis band extends between 0.5–0.7 μm and is prominent in the spectrum of modern Earth's high $O_2$ atmosphere. The very strong $O_3$ UV Hartley-Huggins band extends between 0.2–0.35 μm and is saturated in modern Earth's spectrum. This latter band is a sensitive indicator of low $O_2$ atmospheres, as this UV band produces a strong spectral feature even when atmospheric $O_2$ levels are so low that $O_2$ itself is spectrally undetectable, which was possibly the case for mid-Proterozoic Earth (**Figure 3-2**). Over the course of a planet's orbit, changes in seasonally-dependent biological $O_2$ production might induce changes to the strength of the Hartley-Huggins UV ozone spectral band (Olson et al. 2018). This type of temporal biosignature is most easily detected at low (i.e., mid-Proterozoic-like) oxygen levels because the UV $O_3$ band is not saturated at these abundances, increasing sensitivity to modulations in its strength.

**Methane ($CH_4$).** Methane can be produced by biological and geological processes, with biological processes dominating on modern Earth at ratio of at least 65:1 (Etiope & Sherwood-Lollar 2013; IPCC 2007). Even in the Archean, most of Earth's methane production was likely biological (Kharecha et al. 2005). Because the significant abiotic production mechanisms for $CH_4$ on an Earth-like planet involve water-rock reactions (Etiope & Sherwood Lollar 2013), methane in an atmosphere can also indicate habitability by implying the existence of liquid water.

Interpretation of methane as a biosignature, even more so than oxygen, hinges on understanding its sources and sinks, which will require careful characterization of the environment. Simultaneous detection of $CH_4$ and $O_2$, which are in chemical disequilibrium with each other, is thought to be an extremely robust indicator of life (e.g., Schwieterman et al. 2018). For anoxic planets, simultaneous detection of abundant $CH_4$ and $CO_2$ (i.e., $CH_4$ atmospheric fractions > $10^{-3}$ for planets around Sun-like stars) indicates rapid $CH_4$ production rates unlikely to be explainable by abiotic processes (Krissansen-Totton et al. 2018).

Methane's 1.4 and 1.7 μm bands are strong for an Archean-type atmosphere but only weakly detectable at modern $CH_4$ concentrations. Methane also has other shorter wavelength bands that can be detectable at higher Archean-like (e.g., ~0.1% of the atmosphere) concentrations (**Figure 3-2**).

**Nitrous Oxide ($N_2O$).** Nitrous oxide is produced by life, and it may have been present in large quantities during parts of the Proterozoic eon (Buick 2007; Roberson et al. 2011). Nitrous oxide is readily photolyzed, so its accumulation in an atmosphere would likely require a robust UV shield. Known abiotic sources are minor, and include lightning, stellar activity (Airapetian et al. 2017), and reactions involving dissolved nitrates in hypersaline ponds (Samarkin et al. 2010). Nitrous oxide's strongest bands occur longward of the LUVOIR wavelength range; other bands exist between 1.4 and 2 μm, but these bands are weak and tend to overlap with other gases that are likely to be present (such as $H_2O$ and $CO_2$). However, the high abundances that have been suggested during parts of Earth's Proterozoic, or high signal-to-noise data, may render this gas detectable.

**Hazes.** Hazes are photochemically produced particles. Organic haze (i.e., carbon-containing haze), which produces a strong UV-blue absorption feature, is generated via photochemical reactions involving $CH_4$. In the presence of Earth-like $CO_2$ amounts, haze formation requires high (i.e., most likely biological) fluxes of $CH_4$ (Arney et al. 2018). In the presence of other methyl-bearing biological gases such as dimethyl sulfide ($C_2H_6S$) and methanethiol ($CH_3SH$), organic haze can form at lower $CH_4/CO_2$ ratios than $CH_4$-only photochemistry would predict, which would strengthen the interpretation of biological involvement (Arney et al. 2018).

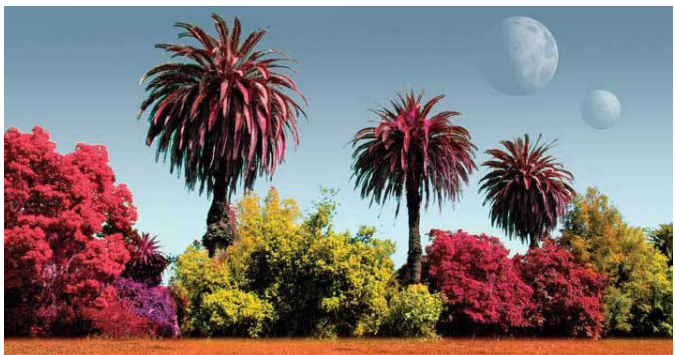

**Figure 3-9.** *LUVOIR will provide a statistical search for life on nearby exoplanets and offers the best chance of answering the question "Are we alone in the universe?" Exotic photosynthetic life may generate novel surface reflectance biosignatures, as suggested in this image, and LUVOIR's direct observing capabilities can search for these types of novel photosynthetic pigments. Credit: NASA*

**Surface reflectivity.** Leaf structure produces a steep increase in reflectivity near 0.7 μm known as the "red edge" (Gates et al. 1965; Seager et al. 2005), which produces a <10% modulation in Earth's disk-integrated brightness at quadrature (Montanes-Rodriguez et al. 2006). Other biological pigments can also produce strong reflectivity signatures (Hegde et al. 2015; Schwieterman et al. 2015a). Reflectance edges, which are discontinuities in planetary reflectance as a function of wavelength, could occur through the visible and NIR, depending on the pigmented and/or photosynthetic organism producing the signature (**Figure 3-9**). Unexpected reflectance signatures that do not match abiotic compounds could therefore suggest life, especially if present with atmospheric biosignatures.

**Temporal biosignatures.** On Earth, life produces seasonal changes in vegetation coverage and albedo, and periodic changes in gas abundances. $CO_2$ concentrations vary seasonally (~10 ppm) due to changes in temperature and sunlight intensity, which modulate photosynthetic drawdown of $CO_2$ in the spring and summer, and its release back to the atmosphere with vegetation decomposition in autumn and winter (Keeling et al. 1976). Methane also changes seasonally by ~10–20 ppm. As discussed above, at low $O_2$ levels, biological $O_2$ production might be inferred from variability via $O_3$ (Olson et al. 2018). Identifying seasonal variations like these would require observations at multiple points during the planet's orbit.

**Statistical biosignatures.** LUVOIR's large, in-depth survey of potentially habitable exoplanets provides an additional unique pathway to increase the confidence of a detection of extraterrestrial life: the integrated confidence level that biosignatures have been detected on at least one of LUVOIR's targets will rise if the same signals are detected on multiple worlds. For example, if LUVOIR detects the simultaneous presence of oxygen and methane on one rocky world in the habitable zone, this could be explained by a transient phenomenon or by a binary planet system, with one $CH_4$-rich world and one $O_2$-rich world (e.g., Rein et al. 2014). While the co-detection of $O_2$ and $CH_4$ on a single planet could be dismissed as a false positive, the simultaneous detection of these gases on multiple planets would provide stronger evidence that we are not alone in the universe.

### 3.3.3 False positive biosignature discrimination

Detecting a biosphere's gaseous byproducts is much easier than conclusively demonstrating that those gases arise from biological activity. In order to meet this higher threshold, observations must also rule out "false positives" whereby non-biological processes produce features that might in other contexts be considered biosignatures. This is true for all biosignatures, but most false positive studies have concentrated on $O_2$ and $O_3$ because they are the most well-understood and studied biosignature gases.

For all biosignatures, the top-level approach to accounting for false positives is to consider the environmental context in which a biosignature is detected. For exoplanets, this context includes the host star. Thus, any mission that attempts to detect biosignatures should pay attention to host star selection and host star characterization. The majority of work on false positives—including both a review (Meadows 2017) and an inter-model comparison (Harman et al. 2018)—highlight the importance of UV radiation. Higher amounts of radiation—specifically, radiation shortward of 200 nm and at the levels emitted by many M-type stars—dramatically increases the efficiency of photochemical false-positive generation mechanisms. Conversely, lower amounts of UV radiation—at the levels emitted by most Sun-like stars—makes most of these processes inefficient or impossible. LUVOIR's direct imaging campaign has observational biases towards F/G/K stars, for which only one known sustainable false positive mechanism exists (discussed in detail below). And LUVOIR plans to characterize the UV properties of each host star in detail, ensuring that all of its biosignatures searches (via both direct imaging and transits) are considered within the context of each planet's host star spectrum.

The only known and sustainable abiotic source of $O_2$ on planets in the habitable zones around F/G/K stars is based on theoretical studies of planets with low non-condensible gas inventories (e.g., $N_2$) (Wordsworth & Pierrehumbert 2014). This can prevent water from condensing into a cloud deck, allowing it to reach the upper atmosphere where it can be photolyzed, generating photochemical $O_2$ and $O_3$. Eventually, this can even cause oxidation of the mantle, which would allow detectable $O_2$ and $O_3$ concentrations to be sustained over geological timescales (Wordsworth et al. 2018). Planets that have undergone this process could be identified by: 1) a lack of $H_2O$ features for worlds that have experienced complete water loss (such worlds would not be regarded as habitable in the first place); 2) a lack of cloud spectral features and cloud-induced spectra variability if cloud condensation is not occurring at all at a high altitude; and/or 3) low pressure, which can be constrained from spectra (Feng et al. 2018; Schwieterman et al. 2015b). Together, these features would make a planet with this $O_2$-generation mechanism distinguishable from planets where the $O_2$ is caused by high fluxes at the surface, from biology. All of these features are detectable by LUVOIR.

There is a much larger list of non-biological mechanisms for generating $O_2$ on planets in orbit around M-type stars (Meadows 2017; Meadows et al. 2018b). These mechanisms are thought to be unlikely or impossible on the majority of LUVOIR's F/G/K target stars. However, LUVOIR would be able to identify or rule-out these processes on planets orbiting M-type stars (10% of LUVOIR's direct imaging exo-Earth survey target list, though LUVOIR can observe additional planets orbiting M dwarfs via transits), and it will be able to constrain the efficiency of these processes with host star characterization via the LUMOS instrument.

This will also serve as an assurance against the possibility that current models are underestimating the efficiency of these processes on planets orbiting Sun-like stars.

Planets around stars with high FUV ($\lambda < 200$ nm) radiation, such as M dwarfs, with $CO_2$ rich atmospheres can generate detectable abiotic photochemical $O_3$ (Domagal-Goldman et al. 2014; Harman et al. 2018). LUVOIR can identify such scenarios by detecting atmospheres with abundant $CO_2$ via near-infrared absorption features. The first three features will only become apparent in a spectrum at $CO_2$ levels even exceeding those anticipated for Archean Earth. Photolysis of $CO_2$ would also generate CO, which could be detected at 1.6 $\mu$m as a false positive indicator of efficient $CO_2$ photolysis (CO is also generally considered an "anti-biosignature" when present in large quantities since it is expected to be readily consumed by microbes, e.g., Sholes et al. (2019)). Particularly for desiccated atmospheres, abiotic photochemical $O_2$ and $O_3$ can be enhanced by a lack of $H_2O$ photolysis that would otherwise generate $O_2$- and $O_3$-destroying radicals (Gao et al. 2015). However, dry planets such as these would not be considered habitable planet candidates in the first place due to their lack of water, which otherwise would produce prominent spectral features throughout much of LUVOIR's wavelength range.

Planetary ocean loss, which can result in massive H loss, driven by high-energy radiation from the star can generate abiotic oxygen to levels where $O_2$ dominates the planet's atmospheric composition (Luger & Barnes 2015; Ramirez & Kaltenegger 2014). LUVOIR could rule this out by failing to detect $H_2O$ water vapor features or water clouds. In the case of planets that have retained some $H_2O$, massive water loss could also produce atmospheres of 10s or 100s of bars of $O_2$, which could be diagnosed through detection of collision-induced $O_4$ dimer features that occur most prominently at 1.06, and 1.27 $\mu$m, and with weaker features throughout the visible range, particularly at $O_2$ pressures higher than 3 bars (Meadows et al. 2018b; Misra et al. 2013).

Detection of $CH_4$ together with $O_2$ and $O_3$ can effectively rule out many of these abiotic oxygen scenarios even for planets orbiting M-type stars. This is because rapid destruction rates of $O_2$ and $O_3$ occur when an atmosphere contains detectable levels of $CH_4$. G dwarf photochemistry may render it difficult to accumulate large quantities of both $O_2$ and $CH_4$ in an atmosphere, but the photochemical lifetime of methane is longer in oxygenated atmospheres for planets orbiting M- and K-dwarfs (Arney 2019; Segura et al. 2005), potentially making these gases easier to simultaneously observe in these contexts.

For atmospheres with $CH_4$, but no $O_2$ (e.g., Archean Earth), detecting $CH_4$ together with $CO_2$ can help to rule out false positive $CH_4$ (Krissansen-Totton et al. 2018). Around a Sun-like star for an atmosphere with $CO_2$, $CH_4$ atmospheric fractions in excess of $10^{-3}$ can imply $CH_4$ source fluxes that are unlikely to be explainable through abiotic processes, and $CH_4$ atmospheric fractions of $> 10^{-2}$ can imply $CH_4$ source fluxes that are extremely unlikely to be abiotic (Krissansen-Totton et al. 2018). Additionally, an organic haze in the presence of detectable levels of $CO_2$ can also indicate $CH_4$ production rates consistent only with biology (Arney et al. 2018).

Finally, an important way to strengthen the interpretation of any potential biosignature is to identify independent, secondary features from life. These will be particularly strong if they are associated with the same biological process as the first biosignature—for example, surface reflectance signatures associated with $O_2$-generating photosynthesis.

**Table 3-2.** *Desired spectral features for habitability assessment.*

| Habitability Markers | | |
|---|---|---|
| Molecules/Feature | UV-VIS wavelengths (0.2–1.0 μm) | NIR wavelengths (1.0–2.0 μm) |
| $H_2O$ | 0.65, 0.72, 0.82, 0.94 | 1.12, 1.4, 1.85 |
| $H_2$ | 0.64–0.66, 0.8–0.85 | |
| $CO_2$ | | 1.05, 1.21, 1.44, 1.59 |
| $CH_4$ | 0.6, 0.79, 0.89, 1.0 | 1.1, 1.4, 1.7 |
| $S_8$ | 0.2–0.5 | |
| $H_2S$ | < 0.3 | |
| $SO_2$ | < 0.3 | |
| Ocean glint | 0.8–0.9 | 1.0–1.05, 1.3 |
| Rayleigh scattering | ≲ 0.5 | |

### 3.3.4 Confirmation of habitability

While an orbit in the habitable zone and the presence of atmospheric water are necessary to discover habitable planet *candidates*, additional information is required to confirm habitability, since not all candidates will truly be habitable. **Table 3-2** summarizes the spectral features to be sought when searching for habitable exoplanets and assessing their potential for habitability.

*Direct detection of liquid surface water.* While gas phase water may be detected through its atmospheric spectral features, direct detection of surface liquid water may also be possible for a subset of the most observable planets through detection of specular reflectance from liquid water oceans (the "glint" effect).

Ocean glint could be detected with photometric observations at continuum wavelengths (i.e., between strong atmospheric absorption bands), and especially those near 0.8–0.9, 1.0-1.05, and 1.3 μm (Robinson et al. 2010) at multiple planetary phases between quadrature and crescent. Near-infrared observations also minimize glint false positive signatures from polar ice (Cowan et al. 2012) which is less reflective at longer wavelengths.

For Earth, glint is most pronounced near a star-planet-observer (phase) angle of 150°, where a glinting planet would be nearly twice as bright as a non-glinting planet. The ability to measure planetary phase functions at such close planet-star separations will depend on the IWA of the high-contrast field-of-view and cannot be achieved for planetary systems with inclinations below about 60 degrees. An IWA of 2 λ/D in the NIR channel allows glint detection in the 1.33 μm continuum out to almost 10 pc for habitable planets around G dwarfs, and nearly 5 pc for K dwarfs.

*Measuring planet masses.* Direct imaging and reflectance spectroscopy provide no constraint on planet mass. Instead, the most direct means to measure mass is to detect the reflex motion of its host star. The radial component of this motion may be detected via the radial velocity (RV) technique while astrometry traces the sky-plane motion. However, the inherent astrophysical noise floor due to stellar activity (e.g., oscillation, granulation, spots) makes astrometry more promising than RV for lower mass planets like Earth. This is because stellar RV jitter is larger than the 9 cm/s RV signal of a Sun-Earth analog by an order of magnitude or more, while stellar activity-induced systematic limits for astrometry are ~4x smaller (e.g.,

Lagrange et al. 2011). LUVOIR, with its sub-microarcsecond (sub-µas) precision astrometry, is capable of measuring masses of exoEarths within 30 pc (**Appendix B.6**).

**Other greenhouse gases.** Besides water, other greenhouse gases will have important impacts on exoplanet habitability. These gases include $CH_4$ and $CO_2$. Measurements of greenhouse gases will be important for constraining exoplanet surface temperatures with climate models. Because LUVOIR cannot access thermal wavelengths to directly measure temperature, such modeling efforts will be needed for estimating surface climatic conditions and predicting the stability of surface liquid water.

**Scattering & aerosols.** A wide spectral range permits discovery of unexpected atmospheric absorbers and provides a longer lever arm to constrain cloud particle sizes and composition through Mie scattering effects. Blue and UV wavelengths are particularly important for observing these effects. The Rayleigh scattering slope may help to constrain atmospheric pressure (Feng et al. 2018), which is an important constraint for habitability.

Observations of exoplanets at UV-visible wavelengths may also reveal the presence of hydrocarbon, sulfuric acid or water vapor hazes in terrestrial atmospheres (e.g., Arney et al. 2016; Hu et al. 2013; **Figure 3-2**). Spectral estimates of the composition, particle size, and optical depth of the planetary aerosols improves climate estimates and could also constrain the albedo of the planet (whether it is more likely to be reflective, or dark), helping to improve the inherent size-albedo degeneracy for directly imaged, non-transiting planets.

**Signs of geologic activity.** Elemental sulfur ($S_8$) particles, produced by photochemical reactions involving volcanic gases such as $H_2S$, can produce broad absorption features at $\lambda < 0.5$ µm (Hu et al. 2013). $H_2S$ and $SO_2$ could be detected in the UV at wavelengths < 0.3 µm.

Spectral features or scattering slopes of volcanically-generated species such as $CO_2$, $H_2S$, $SO_2$, or $H_2SO_4$ aerosols, especially if these vary on monthly or yearly timescales, could indicate active volcanism (Hu et al. 2013; Kaltenegger & Sasselov 2010; Misra et al. 2015). Geological activity is important for maintaining planetary habitability on Earth over long timescales through plate tectonics, the carbonate silicate cycle (Walker et al. 1981), and seafloor weathering (Krissansen-Totton & Catling 2017).

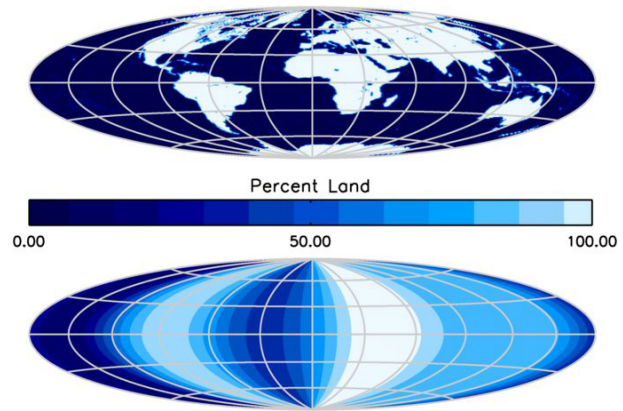

**Surface features and rotation.** A rotating planet with oceans and continents will have a non-uniform disk-integrated surface albedo, and short cadence observations over a long enough temporal baseline could determine the rotation period of a planet and even produce rough latitudinally-resolved "maps" of planetary surfaces (**Figure 3-10**; Cowan et al. 2009; Lustig-Yaeger et al. 2018). The presence of an ocean could be inferred through principal component analyses that search for blue surfaces. These types of investigations

**Figure 3-10.** *Map of an ocean world. The top panel shows a map of the Earth. The bottom panel shows the surface map retrieved from imaging data obtained with the EPOXI mission, demonstrating that a longitudinal map identifying the location and colors of oceans and continents can recovered from unresolved images. Credit: Cowan et al. (2009)*

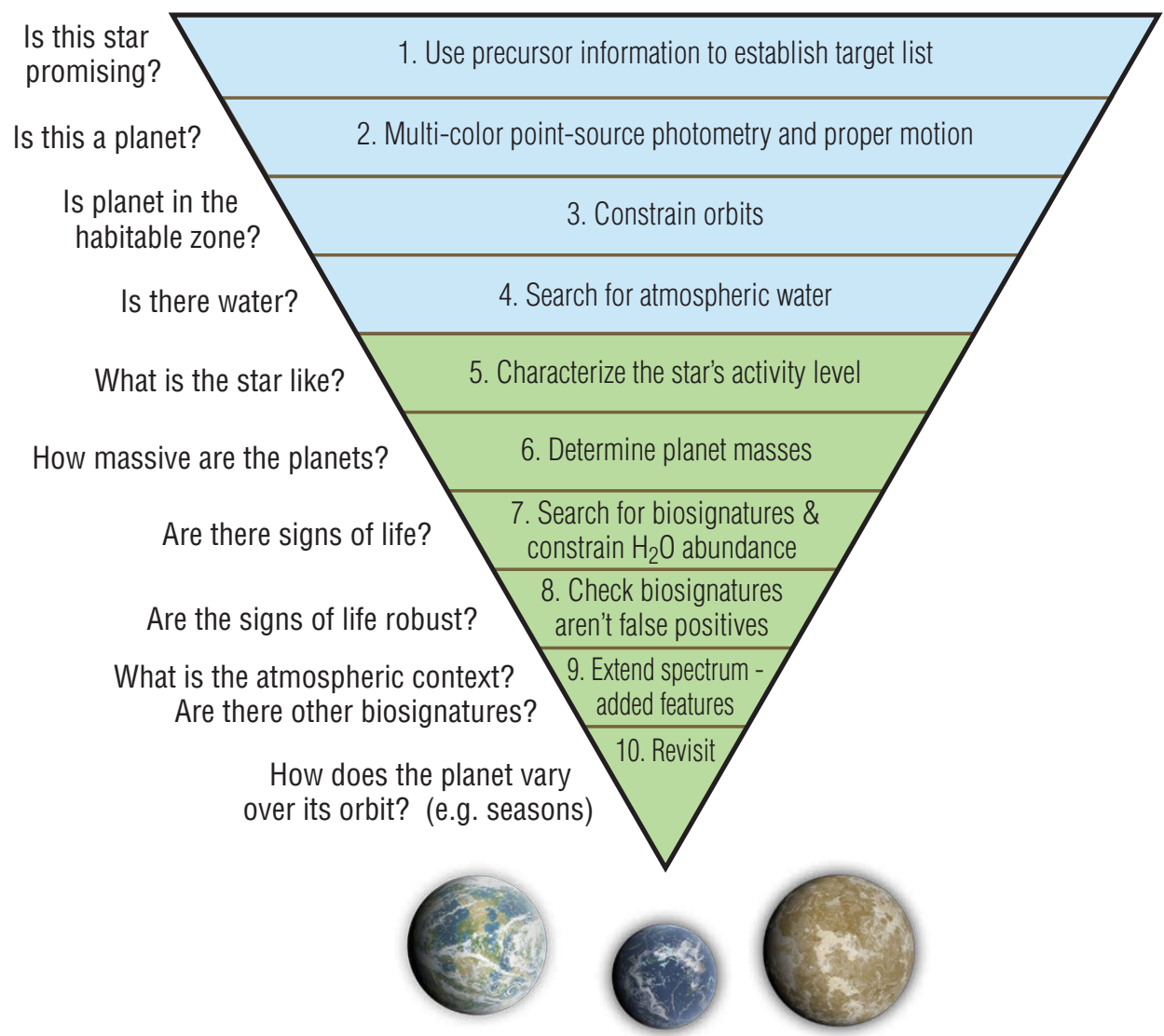

**Figure 3-11.** *Science questions and observational strategy in the search for habitable planets and life. Blue steps at the top of the figure refer to identifying habitable exoplanet candidates; green steps at the bottom of the figure refer to characterizing habitable exoplanets and searching for biosignatures. Credit: T. B. Griswold (NASA GSFC)*

can use the same data gathered during the long spectroscopy observations, with no additional exposure time needed (**Section 3.4.2**).

## 3.4 The strategy for studying habitable exoplanets

LUVOIR's strategy for studying habitable and possibly inhabited exoplanets is summarized in **Figure 3-11**. LUVOIR's exoplanet goals focus on two testable hypotheses. The first is to test whether habitable conditions are possible on planets orbiting nearby stars. The second is to determine whether any of the habitable exoplanets exhibit remotely detectable signs of life. Should we fail to detect planets in either of these categories, we will be able to place an upper limit on the frequency of planets with habitable conditions and/or life in the universe. To place useful upper limits, we are naturally driven to require large sample sizes.

In the following subsections, we first explain the strategy for exoplanet detection, followed by the strategy for exoplanet characterization, which includes the search for life and confirmation of habitability. LUVOIR's initial habitable exoplanet survey is designed as two-year program with an additional six months budgeted for characterization of exoEarth candidates.

### 3.4.1 LUVOIR's initial habitable exoplanet survey

The characteristics of habitable exoplanet candidates are described in detail in **Section 3.2.1**. Here we summarize our strategy to discover habitable exoplanet candidates, following the rationale discussed in **Section 3.2.2**. These steps follow the blue levels of the pyramid in **Figure 3-11**.

**1. Use precursor information to establish a target list.**

Since LUVOIR's search sphere is within a few tens of parsecs of the Sun, stellar targets suitable for LUVOIR's habitable exoplanet search have been known for centuries, having Bayer designations and Arabic names. Using a benefit-to-cost metric, we will optimize a list of stars to maximize the probability of detecting an exoEarth based on information available in the 2030s—such as the amount of exozodiacal dust around the star, stellar mutiplicity, and the presence of other known planets.

**2. Use multi-color point source photometry and proper motion to establish which objects in the field of view are planets.**

Background source confusion can be overcome by observing a stellar target for multiple epochs. Planets will have the same proper motion as their stellar host. Multi-color point source photometry during these same observations will also provide information to discriminate planets from background point sources (**Section 3.2.2**).

**3. Constrain planetary orbits to determine which planets are in the habitable zone.**

We require four detections of the planet spaced out over the orbit to constrain the orbit. Because some observations will miss the planet, an average of 6 visits per star may be needed for orbital determination.

**4. Search for atmospheric water to determine which planets are good candidates for habitability.**

Habitable planet candidates are those with liquid water, which we search for using the $H_2O$ feature at 0.94 μm. We require SNR = 5 on the continuum at R = 70 for water vapor detection.

*Habitable exoplanet candidate yields.* Following the steps outlined above, we use new exoplanet yield estimation methods (e.g., Stark et al. 2014) to determine the quantity and quality of exoplanet science that the LUVOIR mission concept could produce. Here we summarize the exoplanet yield that would result from surveys optimized for potentially habitable Earth-like planets with the LUVOIR concepts. This initial 2-year search represents blue steps 1–4 in **Figure 3-11**. A detailed description of the exoplanet science yield calculations, the techniques used, assumptions made, and justification for the adopted observing strategy for the initial habitable planet survey appears in **Appendix B.2**.

Potentially habitable Earth-like planets, or ExoEarth candidates, are defined according to the characteristics described in **Section 3.2.1** for rocky planets in the habitable zone. We estimate the yield of detected exoEarth candidates for the baseline LUVOIR-A 15-m mission to be $54\,^{+61}_{-34}$, where the range is set by uncertainties on $\eta_{Earth}$, exozodiacal light levels, and finite sampling uncertainties. For LUVOIR-B, this number is $28\,^{+30}_{-17}$.

The exoEarth yield (Y) is a function of a large number of astrophysical and instrumental parameters (Stark et al. 2015). The most important of these are the inscribed telescope diameter (D), the coronagraph IWA, $\eta_{Earth}$, the signal-to-noise ratio required for the broadband detection of a planet (SNR), the assumed geometric albedo of the planet (A), and the end-to-end facility throughput (T). These terms are related by the equation:

$$Y \propto D^{1.97}\left(IWA\right)^{-0.98}\left(\eta_{Earth}\right)^{0.96}\left(SNR\right)^{-0.76}A^{0.65}T^{0.35}$$

The strongest driver of yield is the inscribed telescope diameter, which is a major motivator for larger apertures to enable comparative planetology across a large population of potentially Earth-like exoplanets and maximize our chances of detecting rare phenomena and processes.

Large numbers of exoplanets can also provide constraints on the frequency of habitable conditions on worlds elsewhere. The number of exoplanet candidates ($N_{ec}$) required to constrain the fraction of planets with a given characteristic x ($\eta_x$) at a given confidence level (c) is given by:

$$N_{ec} = \frac{\log\left(1-c\right)}{\log\left(1-\eta_x\right)}$$

For LUVOIR-A, a sample of 54 planets can constrain the frequency of planets with habitable conditions to less than 5% at 95% confidence. Or, put another way, if the frequency of true habitable planets is 5% of all candidates, LUVOIR-A will detect at least one such planet at 95% confidence. For LUVOIR-B, 28 exoEarth candidates can guarantee seeing at least one true exoEarth at 95% confidence if the frequency of habitable conditions is 10% of all candidates. Thus, with LUVOIR, we have a high probability of discovering at least a small number of truly habitable "Earths." Or, in the case of a null result, we will learn how rare habitability is in our universe.

While searching for and characterizing exoEarth candidates, LUVOIR will detect dozens of other planets, from hot rocky worlds to cold gaseous planets. These other planets are detected during the same exposures required for the exoEarth search. **Figure 3-12** shows the expected exoplanet yields of all planet types when optimizing the observation plan for the detection of exoEarth candidates (green bar). Following the planet classification scheme in Kopparapu et al. (2018), the other planet class types in this figure are: rocky planets (radii of 0.5–1 $R_{Earth}$), super-Earths (radii of 1–1.75 $R_{Earth}$), sub-Neptunes (radii of 1.75–3.5 $R_{Earth}$), Neptunes (radii of 3.5–6 $R_{Earth}$), and Jupiters (radii of 6–14.3 $R_{Earth}$). As discussed previously, ExoEarth candidates are a subset of the rocky and super-Earth planets within their stars'

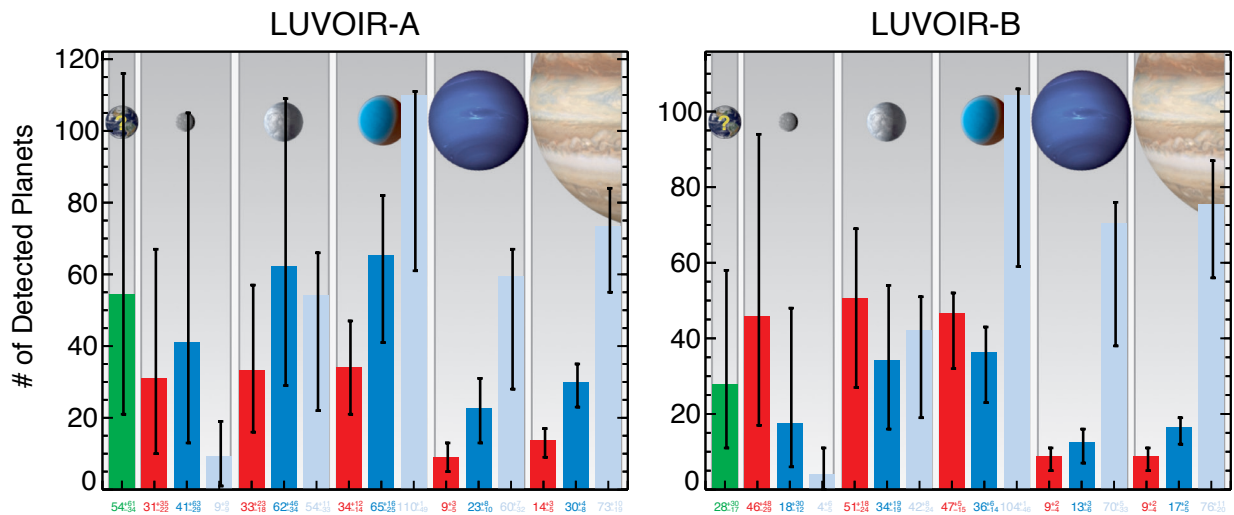

**Figure 3-12.** *Exoplanet detection yields for different classes of planets from an initial 2-year habitable planet survey with the LUVOIR-A (left) and LUVOIR-B (right) concepts. Red, blue, and ice blue bars indicate hot, warm, and cold planets, respectively. The green bar shows the expected yield of exoEarth candidates, which are a subset of the warm rocky and super-Earth planets. Planet class types from left to right are: exoEarth candidates, rocky planets, super-Earths, sub-Neptunes, Neptunes, and Jupiters. Planet types other than exoEarth candidates are detected in the same exposures required during the 2-year exoEarth search campaign. Color photometry and orbits are obtained for all planets. Two partial spectra are obtained for all planets in systems with exoEarth candidates. Credit: C. Stark (STScI)*

habitable zones and with $0.8a^{-0.5}$ < radius < 1.4 Earth radii, where $a$ is semi-major axis. "Hot" planets (red bars) receive 182 to 1x Earth's insolation (i.e., irradiation from the star); "warm" plants (blue bars) receive 1 to 0.28x Earth's insolation; "cold" planets (ice blue bars) receive 0.28 to 0.0035x Earth's insolation.

**Figure 3-13** and **Figure 3-14** summarize example target lists for LUVOIR's exoEarth search program for LUVOIR-A and -B, respectively. LUVOIR's list of target stars is already known. LUVOIR-A will observe ~290 stars (~160 for LUVOIR-B) covering a wide variety of spectral types, primarily F/G/K stars with some M dwarfs and a few A dwarfs. Target stars will be within 30 pc; most will be brighter than 7th magnitude.

In reality, yields may vary from the expected values shown in **Figure 3-12** due to astrophysical uncertainties and the actual distribution of planets around nearby stars. The yield uncertainties shown in **Figure 3-12** are estimated as the root-mean-square combination of the NASA Exoplanet Exploration Program Analysis Group SAG13[1] exoplanet occurrence rate uncertainties and the uncertainty due to the random distribution of planets around individual stars. The latter was estimated by assuming that planets are randomly assigned to stars, such that multiplicity is governed by a Poisson distribution, and that each observation represents an independent event with probability of success given by that observation's completeness. The error bars also include uncertainty in the exozodi distribution, and the Poisson noise associated with the planets and exozodi levels of individual systems. The uncertainties for the cold planet yields are likely underestimated, as the SAG13 occurrence rates are purely extrapolations in this temperature regime.

---

1 https://exoplanets.nasa.gov/exep/exopag/sag/

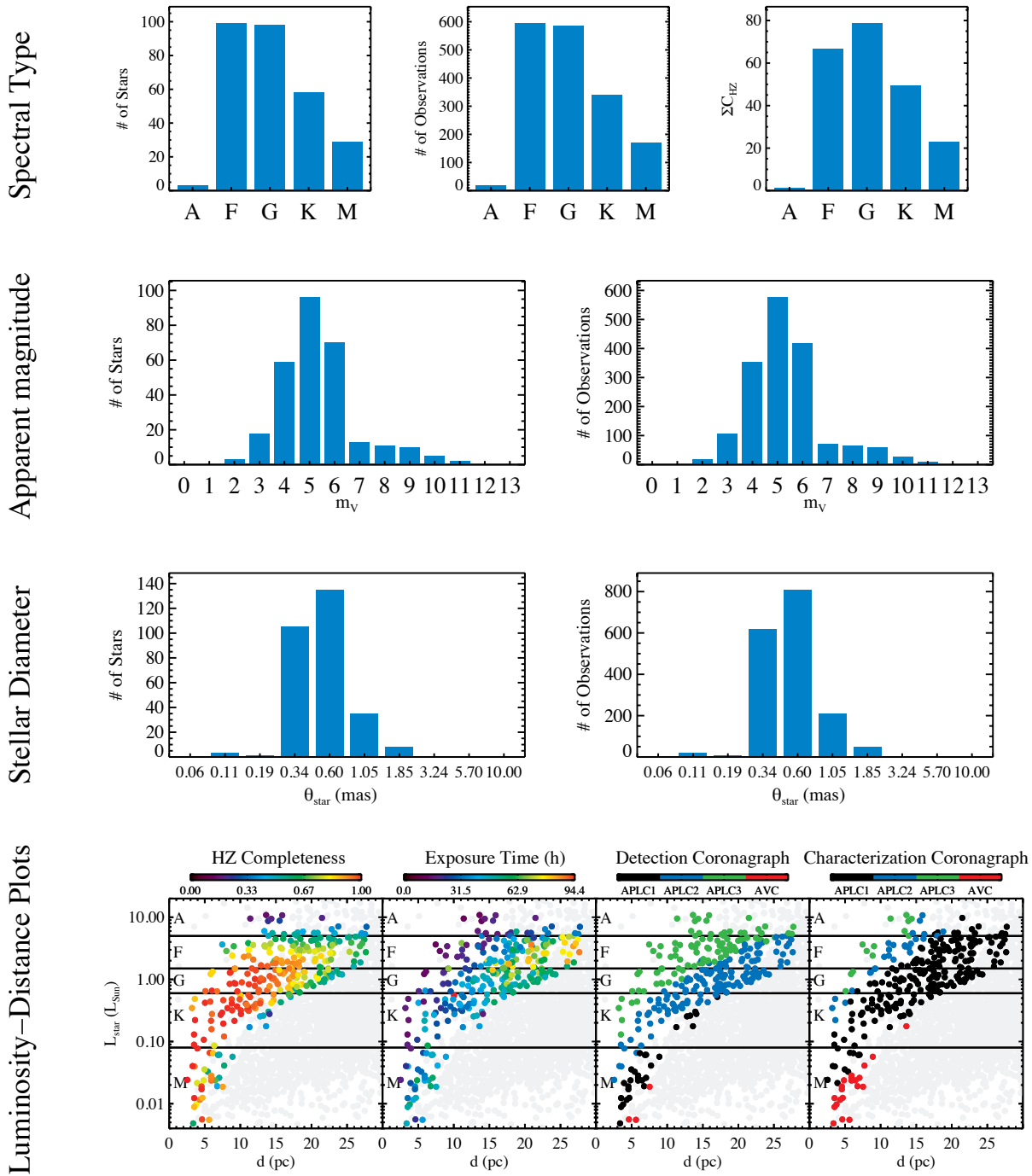

**Figure 3-13.** *Summary of targets observed in the exoEarth search campaign for the LUVOIR-A 15-m concept. Credit: C. Stark (STScI)*

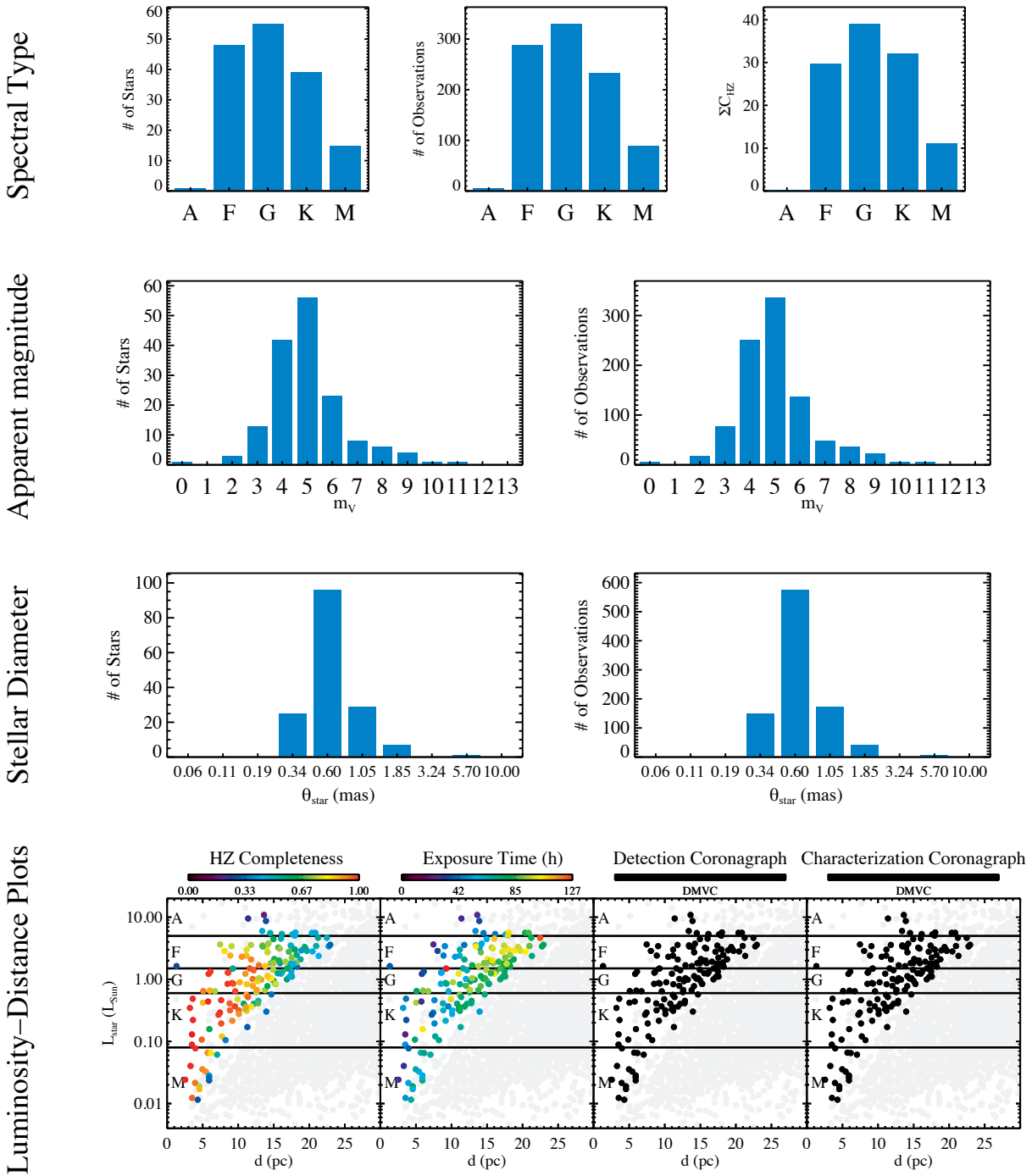

**Figure 3-14.** *Summary of targets selected for observation in the exoEarth search campaign for the LUVOIR-B 8-m concept. Credit: C. Stark (STScI)*

*The role of precursor/follow-up knowledge for exoplanet direct imaging.* Counter-intuitively, precursor knowledge of the host stars and orbits of exoEarth candidates does not greatly increase the total number of such planets discovered and characterized, in the case where a direct observation mission is target-limited rather than time-limited (i.e., where the mission can pick off the lowest hanging fruit). This is the situation for both LUVOIR concepts, A and B. However, such precursor knowledge is of great value for increasing the observing efficiency of exoplanet direct observation missions. Here we summarize the main conclusions about exoplanet precursor and supplemental studies for future missions like LUVOIR and HabEx.

1. Orbit determination is required to quantify the energy fluxes incident on an exoplanet from its host star. This is particularly necessary to show that an exoplanet is located in its star's HZ. If not done in advance of direct observations, then it should be done concurrently.

2. Mass measurements are highly desirable in order to understand an exoplanet's bulk properties and to constrain the effects of surface gravity on the observed atmospheric spectrum. This information could be obtained either through precursor, concurrent, or follow-up work.

3. Precursor observations that identify stars with detectable exoplanets (especially those in the HZ), and which develop orbital ephemerides sufficient to predict when the targets could be best observed, are desirable for mission planning. This information could reduce the mission resources (e.g., integration time) devoted to exoplanet searches and thus save them for other uses.

4. The number of HZs (or ice lines) accessible to a given imaging mission architecture is a 1:1 function of the inner working angle of that architecture. When the mission has sufficient observing resources to study all the exoplanets of interest outside its design IWA, it is target-limited and the effect of prior knowledge on mission yield is small. When there are more accessible targets outside the IWA than mission resources to study them, the effect of prior knowledge on mission yield may be substantial.

5. The primary targets of direct observation missions are stars with HZs outside the design IWA. Precursor knowledge showing the absence of a HZ planet may be used to de-prioritize certain target stars. However, a target's value to comparative planetology and planet formation studies should be considered before excluding it from observation.

The scope of the observing program(s) required to address items 1–3 depends on whether these measurements are done prior to a mission, concurrently, or afterwards. Precursor observations would need to survey all primary targets accessible to the mission design, whereas concurrent or follow-up observations would only need to study the subset of stars with planets of interest. Exoplanet orbits may be determined in advance with ground-based radial velocity instruments or concurrently from a series of exoplanet direct imaging observations with LUVOIR/HabEx. Either ground-based radial velocity instruments or a dedicated space astrometry mission could make exoplanet mass measurements. It may also be possible to make the necessary mass measurements by equipping the direct imaging mission with the appropriate astrometric or radial velocity instrument. A dedicated astrometry mission would likely be the most expensive option to fulfill items 1–3, while ground-based radial velocity would likely be the least expensive (with other options falling in between). However, a factor of 5–10 improvement in radial velocity measurement precision on Sun-like stars (to routine precision of ~ 10 cm/s or less) will be needed before this method, either from ground or space, could measure the masses of rocky exoplanets.

### 3.4.2 LUVOIR's habitable planet characterization program

Habitable planet candidates that are not filtered out by any step in the first four levels of **Figure 3-11** will be targets for subsequent observations searching for additional signs of habitability and—most compellingly—for life. These steps represent the levels 5–10 of **Figure 3-11** (shown in green). To summarize, beginning on step 5 of our characterization scheme and driven by the search for biosignatures presented in **Section 3.3.2**, we will:

**5. Characterize the star's activity level to constrain how stellar activity impacts atmospheric photochemistry and erosion.**
A planet's atmosphere can only be well characterized if its star's UV spectrum is understood because photochemistry critically impacts atmospheric composition (**Figure 3-5**). Correctly interpreting biosignatures demands careful modeling of photochemical processes. These processes can affect the gases expected to accumulate in the atmosphere and even produce biosignature false positives (**Section 3.3.3**).

**6. Constrain planet masses to separate true rocky planets from larger gas-rich imposters.**
Astrometric measurements to constrain mass will require 14 epochs per target for 54 exoEarth candidate exoplanets for LUVOIR-A and 40 epochs per star for 28 exoEarth candidate exoplanets for LUVOIR-B. The total program is dominated by overheads, with 30 days required for the LUVOIR-A program and 40.5 days required for the LUVOIR-B program. The LUVOIR astrometric program is described in detail in **Appendix B.6**.

**7. Search for biosignatures and constrain $H_2O$ abundance.**
To determine whether an exoplanet is inhabited, biosignatures such as $O_2$, $O_3$, and $CH_4$ must be searched for, and the planetary environment understood to place those biosignatures into a planetary context. During this search, additional observations of water vapor features will also be obtained to constrain water abundance in the atmosphere. These observations will require SNR ~ 10 with R = 140 in the visible channel and R = 70 in the NIR channel (Feng et al. 2018; Feng personal communication).

**8. Check that biosignatures are not false positives.**
Planets that show signs of $CH_4$, $O_2$, $O_3$, or other biosignatures must be examined closely to ensure that these spectral features are not biosignature false positives. False positive indicators are described in **Section 3.3.3**. However, most of the oxygen false positive processes described in **Section 3.3.3** are thought to only apply to planets orbiting M dwarfs, which are only 10% of LUVOIR's target stars for its nominal exoEarth search.

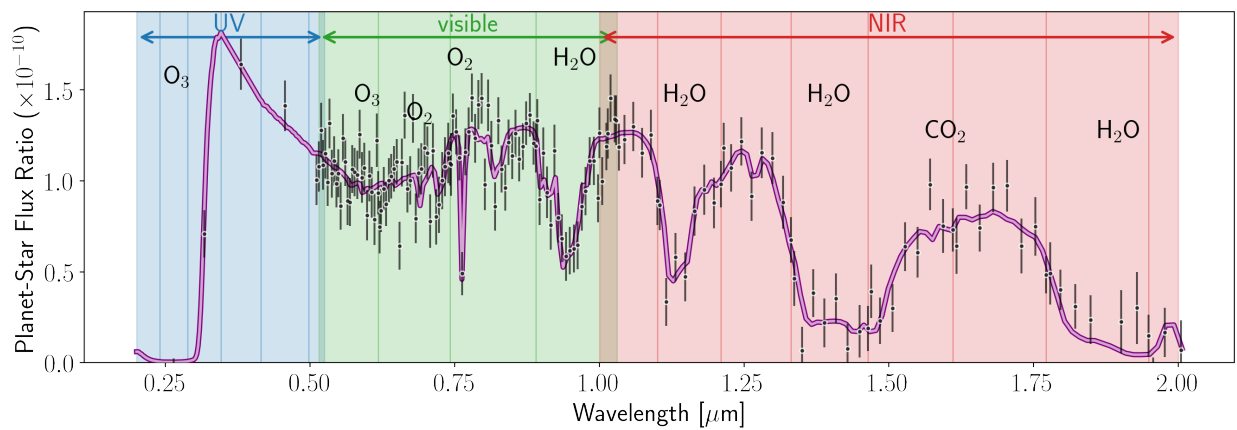

**Figure 3-15.** *Spectrum of an Earth analog exoplanet observed with LUVOIR in this planet characterization observing program (SNR = 8.5, chosen to measure spectral features) applicable to both LUVOIR-A and -B. The shaded background regions indicate 20% bandpasses in the UV (blue) and visible (green) channels, and 10% bandpasses in the NIR (red) channel. Credit: J. Lustig-Yaeger (UW)*

## 9. Extend the spectrum to find additional features.

For additional information on the atmospheric context, and for additional information on possible biosignatures, the spectrum can be extended into the UV and the longest NIR wavelengths where challenging signal-to-noise will generally necessitate longer integration times (**Appendix B.3**).

## 10. Revisit the planet to search for variations on longer time baselines.

Revisiting the planet can look for seasonal variations in atmospheric and/or surface features that could indicate biological activity (**Section 3.1**, **Section 3.3.2**).

*Spectral characterization strategy.* We anticipate detecting approximately 50 and 30 exoEarth candidate planets for LUVOIR-A and -B, respectively. Our characterization strategy involves obtaining direct spectra of sufficient quality to confirm habitable conditions, detect biosignatures, and rule out biosignature false positives on this set of potentially Earth-like exoplanets that have signs of water in their atmospheres (steps 7–10 of **Figure 3-11**). Obtaining a spectrum across the bulk of the ECLIPS wavelength range is sufficient to accomplish these goals. **Figure 3-15** shows a characteristic modern Earth spectrum with the expected noise for the observations described in this program.

As an example of notional observing programs, we estimated how many spectra we can obtain in a six-month program and in a 12-month program for all habitable exoplanet candidates.

Based on the exoEarth yields anticipated for LUVOIR-A and B (54 planets for LUVOIR A; 28 planets for LUVOIR-B), we perform a biased draw of 54 and 28 targets from the total list of stars observed in the exoEarth imaging surveys (**Appendix B.2**), weighted towards stars with higher habitable zone completeness. This becomes our biased catalog of systems with exoEarth candidates from which we obtain spectra.

All calculations use the coronagraph noise model developed in Robinson et al. (2016), which is publicly accessible from the LUVOIR website and on GitHub[2]. We consider

2 https://github.com/jlustigy/coronagraph

overheads on our observations: For each stellar target, we assume 1-hour for the combination of slew, dynamic settle, and thermal settle time. We then add 0.6 (1.25) hours for architecture A (B) to dig the dark hole for each bandpass observed. Finally, we impose a 10% overhead on the total science time for one assumed iteration of the wavefront control system. Observations assume that bandpasses in two separate channels (e.g., VIS and NIR) can be obtained simultaneously, with the third channel (e.g., UV) used for wavefront control. Planets are assumed to be of Earth-radius at quadrature and at the inner edge of their respective habitable zones.

We explored the sensitivity of the numbers of spectra obtainable for different values of $\eta_{water}$ (the fraction of exoEarth candidates with water in their atmospheres), different wavelength coverages, and 6- and 12-month program durations. This program does not represent an optimized observing strategy but rather represents a proof-of-concept study to establish feasibility. Full findings are described in detail in **Appendix B.3**. A summarized subset of findings is presented here for our nominal program that obtains spectra for all exoEarth candidates in order of lowest hanging fruit.

The strategy that obtains the largest number of planetary spectra involves observing spectra of all exoEarth candidates in order of lowest hanging fruit. If $\eta_{water}$ is small, exoEarths will on average be farther away from us and will require longer integration times to characterize; in such a scenario piece-wise spectra that strategically focus on specific molecular signatures may be a better strategy than the "full" spectral characterization strategy laid out here. Further, in a low $\eta_{water}$ scenario, some time may be shifted from the 2 year nominal exoEarth search survey to the 6 month nominal characterization program, as a 6 month reduction in the nominal search survey is expected to decrease the exoEarth candidate yield by only 10%, while it would double the time available for characterization. Maximizing the number of planets observed can enable powerful comparative planetology, so we choose a nominal program that obtains spectra of as many planets as possible.

**Figure 3-16** shows the required science exposure time per bandpass to reach SNR = 8.5, which is needed to constrain $O_2$ abundance (see **Appendix B.3.2** for details). The bluest and reddest bandpasses require longer exposure times, in some cases by orders of magnitude. Such bandpasses therefore are not considered part of the nominal survey presented here and can instead be obtained through later follow-on observations aimed at the most interesting targets. For our nominal time allocation, we choose yields for programs designed to acquire spectra from 0.29–1.46 μm, which we find strikes a balance between spectral completeness, observing time, and scientific yield.

Loss of these most time-intensive bandpasses will generally not affect the ability to detect biosignatures on targets; instead these bandpasses would enhance confidence of biosignatures on already interesting targets. The bluest channels allow access to ozone, but ozone can also be detected at shorter UV channels and at visible wavelengths for modern Earth. The reddest NIR bandpass can provide leverage for the detection of water and/or methane, but both of these species have bands at shorter wavelengths. This reddest bandpass is most valuable for detecting methane at low modern Earth-like abundances, but this type of long observation would only be sought if other biosignatures have been detected at shorter wavelengths.

**Table 3-3** shows how many spectra can be obtained for a nominal LUVOIR-A and -B program, excluding these most expensive bandpasses for the 6- and 12-month programs.

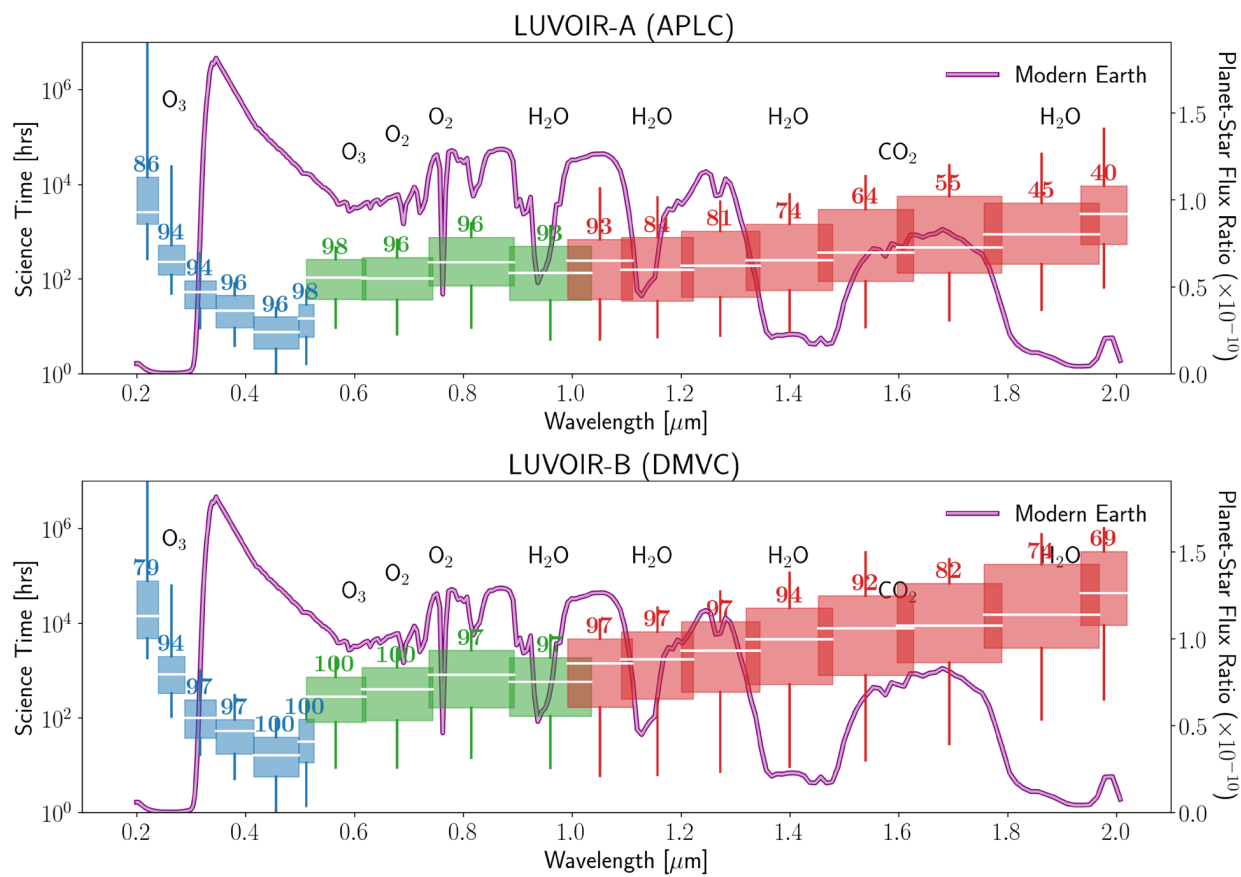

**Figure 3-16.** *Required science exposure time per bandpass (left axis) to reach SNR=8.5 for LUVOIR-A with an APLC coronagraph (top panel) and LUVOIR-B with a DMVC coronagraph (bottom panel). The blue (UV channel), green (visible channel), and red (NIR channel) "boxes and whiskers" show the 50% (box extent) and 95% (whisker extent) confidence intervals about the median (white bar) science time required to observe the spectrum shown in* **Figure 3-14***, for randomly sampled exoEarth candidates. The percentage of stars in the target catalog that can be completely observed in each bandpass, considering IWA and OWA constraints, is displayed above each box. Exposure times required to achieve a fixed precision on the spectrum can vary by orders of magnitude. Credit: J. Lustig-Yaeger (UW)*

Additional planets and/or bandpass could be sought in follow-on programs. We also show how many planets we could observe the $O_2$-A band and the UV-visible range for. If only a subset of exoEarth candidates according to $\dot{\eta}_{water}$ are observed, the average number of spectra obtained decreases according to

$$N_{spec} \approx N_0 \left(\frac{\eta_{water}}{100\%}\right)^{0.71} \left(\frac{t_{exp}}{365 \, days}\right)^{0.37}$$

for LUVOIR-A and

$$N_{spec} \approx N_0 \left(\frac{\eta_{water}}{100\%}\right)^{0.76} \left(\frac{t_{exp}}{365 \, days}\right)^{0.39},$$

where $N_0$ is the number in the second or third column.

for LUVOIR-B. The LUVOIR-B yields may increase by up to 50% if it becomes possible use the PIAA coronagraph as discussed in **Appendix B**.

Using spectroscopic observations, Earth's rotation rate can be retrieved well for planets out to 10 pc for LUVOIR-A and 5 pc for LUVOIR-B in 100-hour integrations (**Figure 3-17**). This time is comparable to the average integration time per spectral bandpass required for detection of absorption features (**Figure 3-16**), thus providing an opportunity to infer longitudinal maps and rotation rates of the planets with no needed additional exposure time.

**Table 3-3.** *Median number of exoEarth candidate spectra (SNR=8.5) observable in a 6-month and 12-month program for LUVOIR-A (top) and LUVOIR-B (bottom).*

|  | 6-month program | 12-month program |
|---|---|---|
| **LUVOIR-A (54 targets)** | | |
| $O_2$-A only (0.74–0.89 μm) | 36 | 44 |
| UV-vis only (0.29–1.03 μm) | 23 | 30 |
| # of spectra (0.29–1.46 μm) | 18 | 24 |
| **LUVOIR-B (28 targets)** | | |
| $O_2$-A only (0.74–0.89 μm) | 18 | 23 |
| UV-vis only (0.29–1.03 μm) | 11 | 15 |
| # of spectra (0.29–1.46 μm) | 8 | 11 |

## 3.5 Signature Science Case #3: The search for habitable worlds in the solar system

Within the solar system, the icy moons of the giant planets (**Figure 3-18**) remain some of the most intriguing observational targets, motivated strongly by their potential to harbor life in subsurface oceans. LUVOIR can conduct game-changing science for the solar system's ocean worlds, via multi-epoch imaging at high spatial resolution (~25 km resolution at 500 nm for Jupiter at opposition with LUVOIR-A, comparable to imaging with the Juno spacecraft) and sensitive spatially-resolved spectroscopy at UV through NIR wavelengths.

While dozens of moons have or have had oceans, a few special cases are considered the best targets in the search for life. Europa and Enceladus (moons of Jupiter and Saturn, respectively) in particular are considered the most promising places to search for evidence of past and present life, due to the fact that both of these worlds possess oceans that are in contact with their interior silicate mantles. This is a compelling scenario given the potential for active water-rock reactions that could produce the necessary conditions for the origin and maintenance of life. These moons represent end members of a different kind of habitable world whose icy surfaces, deep oceans, and interactions of endogenic and exogenic energy sources give rise to the physical and geochemical processes that might create and maintain life. Evidence for surface activity on these worlds abound; one expression of this ongoing geologic and internal activity is the detection of plumes of water erupting from Enceladus' south pole, and potential plumes of water emanating from Europa's surface. These plumes of material may allow direct access to the chemistry and composition of their ice shells and deep oceans.

*Europa.* Transient water plumes emanating from at least three different places on Europa's surface have been detected via Hubble observations of emission from hydrogen at Lyman-α (121.6 nm) and neutral O (130.4 nm) (Roth et al. 2014). Additional evidence for plumes was also found by observing changes in UV continuum absorption while Europa transited the bright surface of Jupiter (Sparks et al. 2016, 2017) and by analyzing Galileo spacecraft

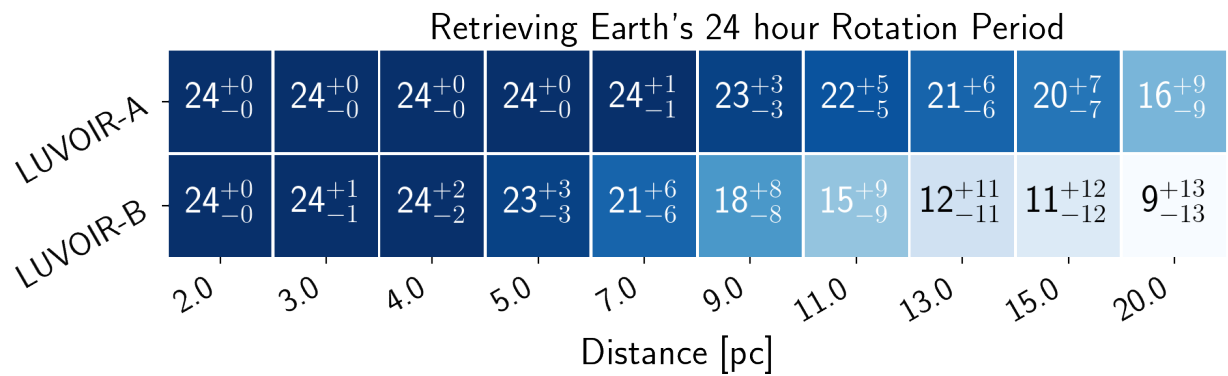

**Figure 3-17.** *Inferred rotation period for an Earth-analog exoplanet using LUVOIR-A and LUVOIR-B as a function of distance to the system. Estimated measurements and 1-sigma uncertainties were calculated for 100 hours of observations at a 1-hour cadence. The rotation period begins to decrease from the true Earth value as the periodogram begins to fit the high frequency noise. LUVOIR-A (-B) can reliably retrieve Earth's 24-hour rotation period out to about 10 pc (5 pc). Credit: J. Lustig-Yaeger (UW)*

magnetic signatures (Arnold et al. 2019; Jia et al. 2018). While one of the sources near Europa's equator has been imaged twice, thus far no other observations have been repeated and no definitive cyclic timing could be discerned. Thus, the frequency of plume activity is unknown, though it appears that any plume activity on Europa is not tidally modulated (Roth et al. 2014a). A sustained monitoring campaign of Europa with sensitive FUV spectral imaging capabilities can shed light on the frequency and locations of plumes—valuable supporting information for future *in situ* spacecraft visiting this potentially habitable world. Long time baseline monitoring of plume activity can contribute fundamental knowledge on how these plumes arise, and their frequency and dynamics, which will provide invaluable information about how surface processes and surface-subsurface exchange operates on Europa.

The plumes are of sufficient scientific interest that a UV spectroscopic capability was incorporated into the Europa Clipper mission to observe them. However, the spacecraft instrument is optimized for plume detection only during flybys and will therefore make primarily serendipitous discoveries. With a planned launch date in 2023, Europa Clipper is expected to arrive in the Jupiter system in either 2026 or 2029 (depending on launch vehicle capabilities) and operate for at least three years.

A LUVOIR campaign targeting the plumes will serve three essential supplemental roles in the effort to characterize the plumes. First, with a 100x increase in sensitivity and a 6-fold improvement in spatial resolution, LUVOIR will detect plume activity over a vastly greater dynamic range, enabling characterization of a wider range of total mass production, cycles and patterns of activity, and topographic distribution. Second, LUVOIR will be in a position to follow up Europa Clipper discoveries with a regular cadence of observations over a longer time baseline, providing supplemental measurements of regions of interest and monitoring of changes in the global plume network under different tidal heating conditions. Third, LUVOIR could provide reconnaissance for subsequent spacecraft that might be in-system in the 2040s.

***Enceladus.*** On Enceladus, plumes erupting near the south pole from the "Tiger Stripes" region were seen with the Cassini spacecraft (e.g., Hansen et al. 2006). Plumes appear to be a constant feature of Enceladus but are about four times brighter when the moon is at the apocenter of its orbit around Saturn, suggesting a connection with tidal stresses (Hurford et al. 2007). The mass of ejected material varies according to the moon's 1.37-day diurnal cycle (Hedman et al. 2013), presumably responding to openings and closings of the fissures (Nimmo et al. 2014). Enceladus' plumes may also exhibit variations on decadal timescales (Ingersoll & Ewald 2017), which could be investigated through long-term monitoring with LUVOIR if these plumes are remotely visible. However, the Enceladus plumes are tenuous (releasing a few hundred kg of $H_2O$ per second; Hansen et al. 2006) and have never been observed with remote telescopes. Given the different magnetic environment at Saturn, it is difficult to predict whether Europa-like FUV emission will be detectable with LUVOIR for Enceladus plume activity. Therefore, we chose to focus upon optical/NIR imaging of Enceladus and other ocean moons of the outer solar system, leveraging the high spatial resolution of LUVOIR to look for changes in surface morphology caused by cryo-volcanism over long timescales.

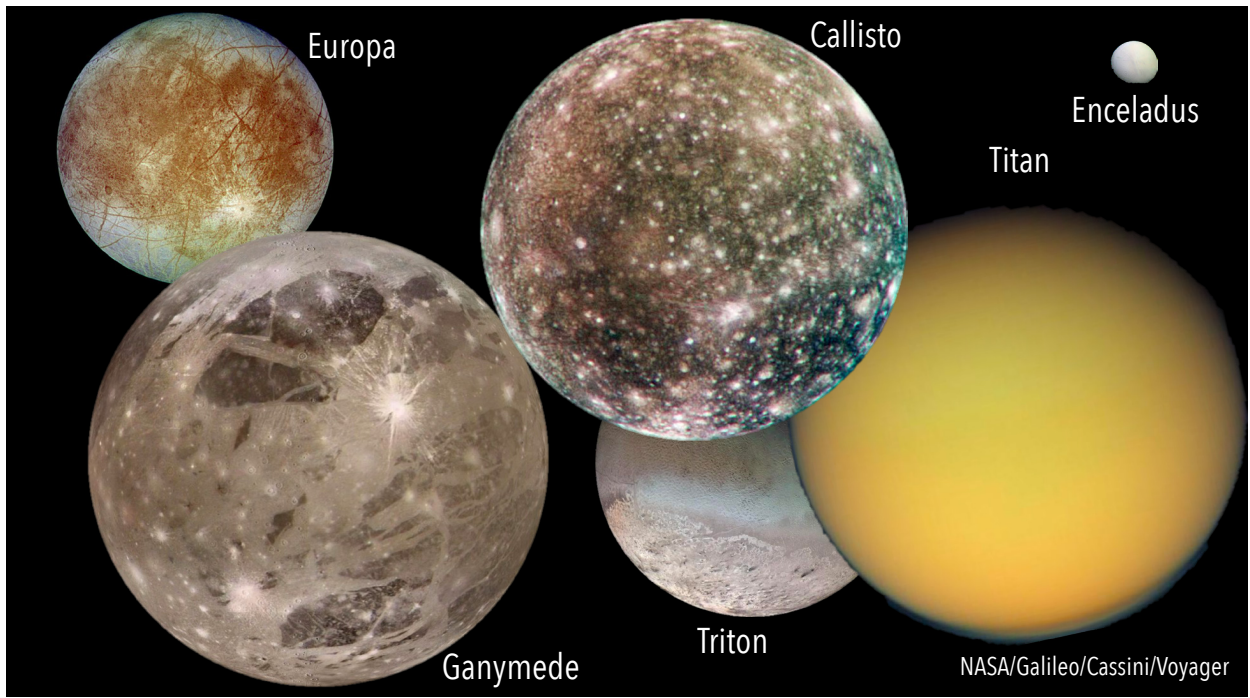

**Figure 3-18.** *Icy moons of our solar system offer incredible opportunities to find habitable conditions and possibly even life in our cosmic backyard. Europa and Enceladus in particular have subsurface watery oceans in contact with their interior silicate mantles, which may create chemical energy needed to power life. Moon sizes shown to scale relative to each other. Credit: NASA*

*Dynamics of water worlds.* A second group of ocean worlds that may be habitable includes more exotic ocean states. Ganymede has the largest known ocean in the solar system, and Titan follows closely behind. However, these two moons are massive—just larger and just smaller than the planet Mercury—which creates deep high-pressure ice layers between their oceans and deep interiors (e.g., Vance et al 2014). These high pressure ices may also exist in the interior of Callisto, Jupiter's second largest moon, although Callisto is not appreciably tidally heated the way Europa and Ganymede are. New work has shown that a lack of a water-rock interface for these moons may not prevent communication between the ocean and seafloor at least for Ganymede, where internal geologic activity could drive exotic ice-ocean-seafloor exchange (Kalousová et al. 2018).

While Titan is not known to have its own dynamo, it does possess a dense atmosphere that experiences seasonality and changing weather that has been observed by the VLT and Keck (Ádámkovics et al. 2007) and Cassini spacecraft (Turtle et al. 2011), and a surface that is covered with hydrocarbons, in the form of dunes and both transient and long-lived lakes of liquid ethane (Stofan et al. 2007). Long temporal baseline observations of Titan across LUVOIR's high-resolution VIS and NIR channels, coupled with spectroscopy, can characterize and monitor changes in Titan's atmosphere and surface. Observations of Titan's hazy nitrogen and methane atmosphere will also inform exoplanet studies, as a corollary to the hazy atmosphere conditions predicted for phases of Earth history (**Section 3.1**). Due to the composition and scattering behavior of aerosols in the Titan atmosphere, the surface is not visible at UV and optical wavelengths, but windows to the surface exist in the infrared that will be covered by LUVOIR. Moreover, with the arrival of Dragonfly at Titan in 2034 or later, LUVOIR would be poised to make global observations that couple to the local scale measurements that will be made by the *in situ* mission. Once Dragonfly has completed operations, LUVOIR can monitor atmospheric and surface phenomena on Titan discovered by Dragonfly to extend our understanding of the dynamics of this potentially habitable world over a longer temporal baseline. With LUVIOR's longevity and serviceability, observing Titan over a significant fraction of its 30-year seasonal cycle will be possible.

***Triton.*** Triton is a Europa-sized likely ocean world orbiting Neptune, and it boasts the first confirmed cryovolcanic activity in the solar system (Soderblom et al. 1990). With a hazy atmosphere similar to Pluto and a surface temperature close to the triple point of methane, Triton's atmosphere likely changes significantly over its long seasons (Hansen & Paige 1992). This process is similar to that observed on Pluto. The last spacecraft to visit the Neptune-Triton system was the Voyager spacecraft in 1989. Although Triton remains a high priority solar system ocean world target (Hendrix et al. 2019), no planned missions to this moon are yet on the horizon. With HST ACS observations, astronomers have been able to confirm that the visible surface of Triton seen by the Voyager mission has completely changed since 1989 (Bauer et al. 2010; **Figure 3-19**), but the nature of those changes remains uncertain. Triton is known to be temporally active at levels that can be observed in its light curve from small telescopes (Buratti et al. 2015). Thus there is a plethora of evidence to suggest that ongoing activity will continue on annual to decadal timescales that could be observed with unprecedented resolution with LUVOIR. Understanding the difference between seasonal and transient changes on Triton may provide key clues regarding its internal heat budget. Furthermore, a combination of mapping and spectral surveys from the UV to the NIR with LUVOIR could elucidate the nature of these changes in the absence of, and in preparation for, *in situ* spacecraft missions.

***Monitoring of ocean worlds.*** The LUVOIR team has designed monitoring campaigns for ocean world moons using the LUMOS instrument to monitor Europa for plume activity via FUV spectral images, and using the HDI instrument to obtain UV/optical/NIR images of the surfaces of icy moons with signs of subsurface oceans to search for time variable changes.

For observations of Europa, we couple long and short cadence campaigns. **Figure 3-20** shows a simulated view of how plumes on Europa might look to LUVOIR. These images assume mosaics assembled from dithered images; we adopt the strategy used for the best

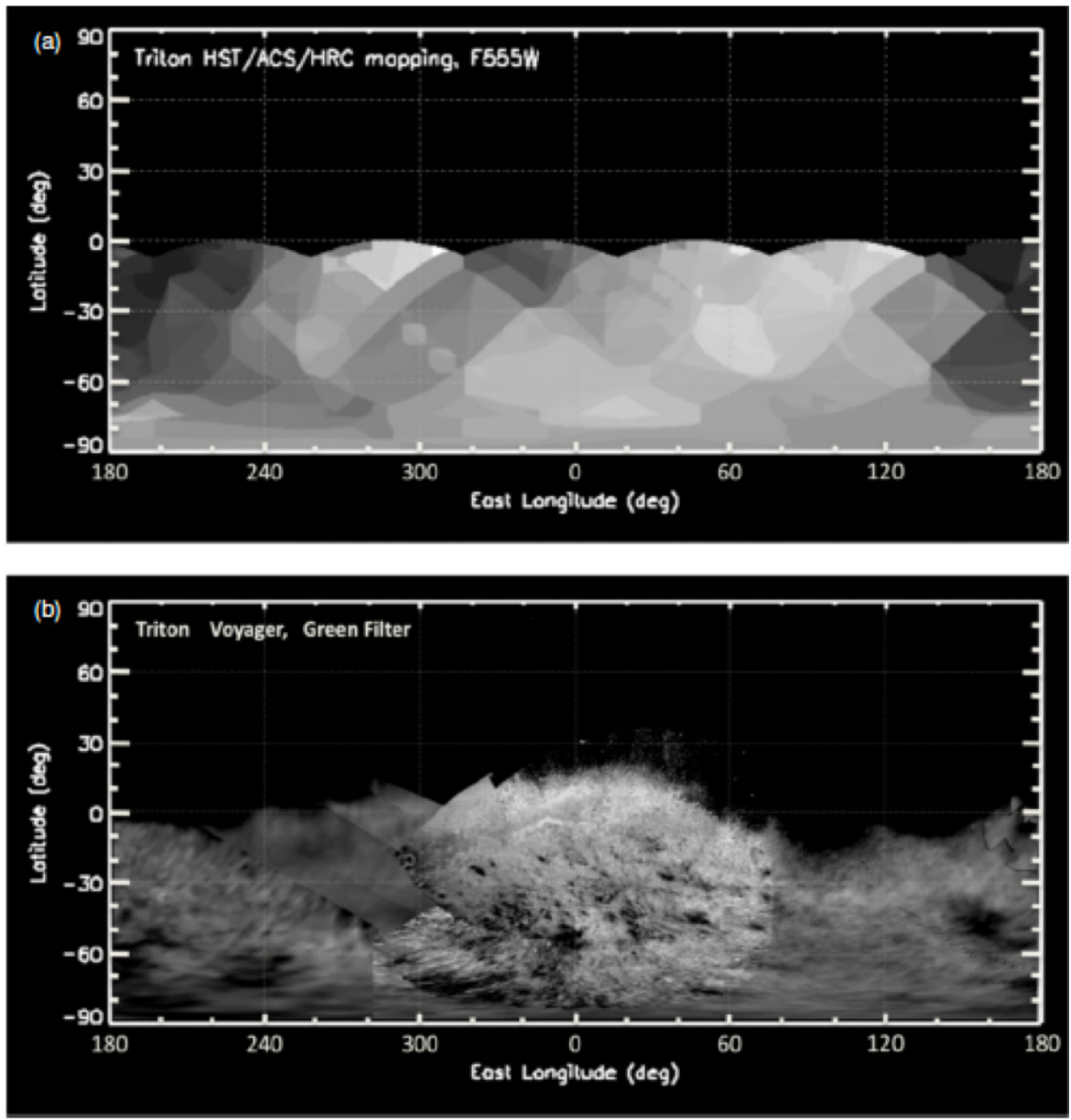

**Figure 3-19.** *Observations of Triton with the HST ACS in 2005 (a) demonstrate dramatic spatial changes in the surface albedo of Triton since Voyager's visit in 1989 (b), after Bauer et al 2010. These data, coupled with historical light curve data that show periodic reddening events and gradual seasonal changes, support an active atmosphere-surface exchange ongoing on Triton that causes dramatic surface change. Because this activity may be also related to Triton's internal heat, understanding the composition and timing of these changes is a key contribution that LUVOIR can make to understanding Triton as a compelling potentially habitable ocean world.*

HST mosaics of Pluto (Buie et al. 2010), a 16-exposure dither pattern that optimally fills a unit pixel cell. The LUMOS spectral images will simultaneously cover a wavelength range (100 nm to 200 nm) that includes the hydrogen Lyman-$\alpha$ emission line (121.6 nm) and the neutral oxygen emission line (130.4 nm). With the designed observations (detailed in **Appendix B.4**), LUVOIR-A can obtain >3-$\sigma$ detections of neutral oxygen and >5-$\sigma$ detection on Lyman-$\alpha$ in each visit for plumes comparable to those observed by Roth et al. (2014). The long cadence Europa monitoring campaign involves 6 visits/year over 5 years, which takes 7 days of total exposure time including overheads for LUVOIR-A and 23 days for LUVOIR-B. Of course, the higher spatial resolution of the LUVOIR-A observations will provide greater ability to separate and identify individual plumes.

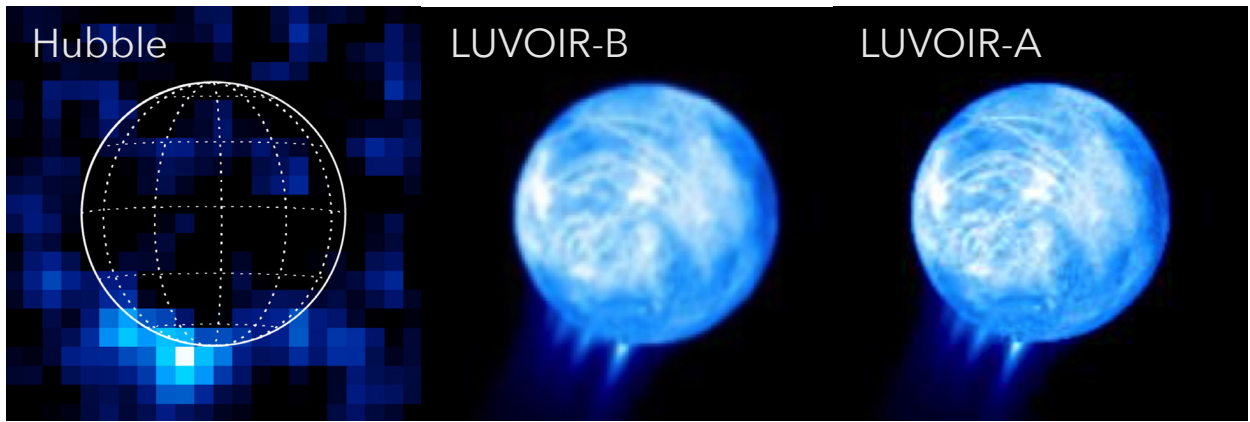

**Figure 3-20.** *Spectroscopic imaging of Europa and its water jets. The left panel shows an aurora on Europa observed with HST (Roth et al. 2014). This UV hydrogen emission (Lyman-α) comes from dissociation of water vapor in jets emanating from the surface. The center and right panels show simulations of how this Ly-α emission from Europa might look observed with LUMOS on LUVOIR-B and A using a dithering observation strategy. The moon's surface is bright due to reflected solar Ly-α emission. With LUVOIR, one could monitor the ocean worlds of the outer solar system for such activity and image the individual jets. Credit: G. Ballester (LPL) / R. Juanola-Parramon (NASA GSFC)*

To these totals, we add 4 days for a single visit of continuous short cadence observations of Europa over one whole orbit around Jupiter (3.5 day period). These data will be used to examine Europa's interaction with the Jovian magnetosphere via non-plume-related auroral emissions. About 20 mosaics can be obtained over this time period with LUVOIR-A and 5 with LUVOIR-B. This brings the total time for the Europa program to 11 days for LUVOIR-A and 27 days for LUVOIR-B. This preliminary program is intended to demonstrate feasibility of FUV monitoring of Europa with LUVOIR on various time scales and will no doubt benefit from further study and future optimization.

For imaging of the surfaces of icy moons Europa, Ganymede, Callisto, Enceladus, Titan, and Triton, we devote two visits per year to each of these six moons. At 500 nm, LUVOIR-A can resolve features as small as 25, 51, and 108 km for the Jovian, Saturnian, and Uranian systems, respectively (47, 96, and 204 km for LUVOIR-B). These targets are extremely bright for LUVOIR imaging, and we simply allot 1 hour per target per epoch, plus an hour for overheads (primarily retargeting slews). This equals 5 days of total program time for both LUVOIR-A and -B. Again, we consider this a preliminary program to demonstrate feasibility.

LUVOIR will transform our understanding of potentially habitable worlds in the solar system by characterizing geyser activity and searching for surface variability on icy moons with subsurface oceans. LUVOIR will also support *in situ* missions to these targets by providing remote, global imaging and will search for currently unknown plumes and surface morphological and compositional changes.

**Table 3-4.** *Chapter 3 Programs at a Glance*

| Chapter 3 Programs at a Glance | | | |
|---|---|---|---|
| **Goal** | **Program Details** | **Instrument + Mode** | **Key Observation Requirements** |
| **Signature Science Case #1: Finding habitable planet candidates** | | | |
| Detect habitable exoplanet candidates around a range of FGKM stars | Observations of nearby stars in target list to search for potential habitable zone rocky exoplanets, and constrain their orbits with respect to the habitable zone. | ECLIPS coronagraphic imaging | Contrast $< 10^{-10}$<br>Inner working angle $\lesssim 4\,\lambda/D$<br>Multi-color photometry near 500 nm |
| | Spectroscopic search for $H_2O$ on rocky habitable zone candidates | ECLIPS coronagraphic spectroscopy | Direct spectroscopy at R=70, SNR=5 to detect $H_2O$ at 940 nm |
| **Signature Science Case #2: Characterizing habitable planet candidates** | | | |
| Search for atmospheric biosignatures | Search for atmospheric biosignature molecules $O_2$ and $CH_4$. | ECLIPS coronagraphic spectroscopy | Direct spectroscopy at R = 140, SNR $\gtrsim$ 10 to detect $O_2$ near 760 nm<br>Direct spectroscopy at R > 70, SNR $\gtrsim$ 10 to detect $CH_4$ features near 600, 790, 890, 1000, 1100, 1400, and 1700 nm |
| Rule out false positives for life and confirm biosignatures | Measure indicators of false positive processes: stellar UV flux and presence of $O_4$, $CO_2$, and CO. | ECLIPS coronagraphic spectroscopy LUMOS FUV spectroscopy | Direct spectroscopy of various species (marked with * in **Table 3.1**) from 300–1600 nm<br>UV spectra of host stars at R~10,000 from 100–400 nm |
| | Search for secondary biosignatures from pigments and other biosignature gases. | ECLIPS coronagraphic imaging and spectroscopy | Multi-color photometry from 500–700 nm<br>Direct spectroscopy of various species (**Table 3.1**) from 200–2000 nm |
| Confirm habitability | Search for ocean glint, longitudinal ocean/continent coverage, constraints on greenhouse gases and atmospheric pressure | ECLIPS coronagraphic imaging and spectroscopy | Photometry near 500 nm, and from 900–1000 nm<br>Direct spectroscopy of various features (**Table 3.2**) from 200–2000 nm |
| **Signature Science Case #3: Potentially habitable solar system worlds** | | | |
| Characterize geyser activity from solar system ocean moons | Monitor Europa and Enceladus to determine the strength and frequency of plumes emanating from their surfaces | LUMOS FUV spatially spectroscopy | FUV spectroscopy at R $\approx$ 10,000 from 100–140 nm |

## 3.6 Conclusion

LUVOIR offers humanity's best chance of answering the ancient question "Are we alone in the universe?" If there is a global biosphere on an observable planet orbiting a star in the Sun's neighborhood, LUVOIR will find it by identifying atmospheric and surface features that can only be produced by life. If life exists within our own solar system, LUVOIR will help other missions search for it. In both cases, LUVOIR will place the search for life inside a global context, and within an expanded understanding of planetary processes.

For worlds in our solar system, LUVOIR's imaging and spectroscopic capabilities will allow global mapping of the atmospheric and mineral composition for worlds large and small. For exoplanets, LUVOIR's large aperture and wide spectral range will provide a rigorous assessment of exoplanet biosignatures as well as their global environmental contexts. This environmental context will include a search for key markers of global habitability, including signs of a hydrological cycle, and the presence of clouds and oceans. Beyond the solar system, LUVOIR will search for expressions of life and habitability on dozens of worlds orbiting a diversity of Sun-like (F/G/K) stars, representing an expansion on prior and concurrent searches for biosignatures on M-dwarf stars. Finally, LUVOIR will also complement these prior observations of potentially habitable planets around M-dwarfs with space-based observations that expand the wavelength coverage for transit spectroscopy of these worlds.

Overall, LUVOIR will conduct observations of dozens of potentially habitable worlds around a wide variety of stars. These observations will complement many other facilities, including ground-based telescopes and orbiters/landers to solar system targets. And each of these investigations will provide scientific context for the others; what we learn from LUVOIR's observations of the solar system will influence our interpretation of LUVOIR's observations of exoplanets, and vice versa. As such, LUVOIR will serve as the flagship observatory for a new area of scientific research: *comparative astrobiology*.

## CHAPTER 4.  HOW DO WE FIT IN? COMPARATIVE PLANETARY SCIENCE

While LUVOIR has been designed to characterize challenging habitable-zone planets, it will also provide superb observations of hundreds of other planets, including gas and ice giants as well as non-habitable terrestrial and sub-Neptune-sized planets around stars of diverse ages and spectral types. LUVOIR's study of such a large and varied exoplanet sample will provide a host of new targets for comparative planetary science, illuminating our understanding of planet formation, evolution, and atmospheric processes. LUVOIR will also provide unprecedented new opportunities to study and compare protoplanetary disks, planetary system architectures, and the record of the birth of our Solar System as preserved in Trans-Neptunian Objects (TNOs).

LUVOIR will contrast planets along axes of mass, size, age, distance from parent star and more. It is through such inter-comparisons that we gain insight into planetary processes. For example, the atmospheres of Venus, Earth, Mars, and Titan feature varying mixtures of atmospheric gases, clouds, and hazes. While only one of these worlds is known to be habitable, our understanding of how planetary climate is influenced by the interactions between atmospheric gasses and aerosols, incident solar flux, and gross planetary properties has been profoundly informed by comparative studies of all these worlds. By studying the planetary processes at work among exoplanets, which are already known to be stunningly diverse (**Figure 4-1**), and connecting these processes to those we understand well in the solar system, LUVOIR will markedly enhance the opportunities for such comparative exoplanet science.

The details of how protoplanetary disks evolve from primordial remnants of star formation into planetary nurseries are also critical for understanding the diversity of exoplanets and planetary systems, but key fundamental parameters remain unobserved. LUVOIR will characterize young planetary systems and sketch the architectures of mature planetary systems to provide clues to the formation and evolution processes that produce the variety of observed planets and their atmospheres. Within the solar system, the small bodies beyond Neptune contain a frozen record of the formation of the planetary system we understand most thoroughly. By characterizing a variety of planets, observing forming and evolving planetary systems, and studying the smallest trans-Neptunian objects (TNOs), LUVOIR will greatly advance our understanding of planet formation processes.

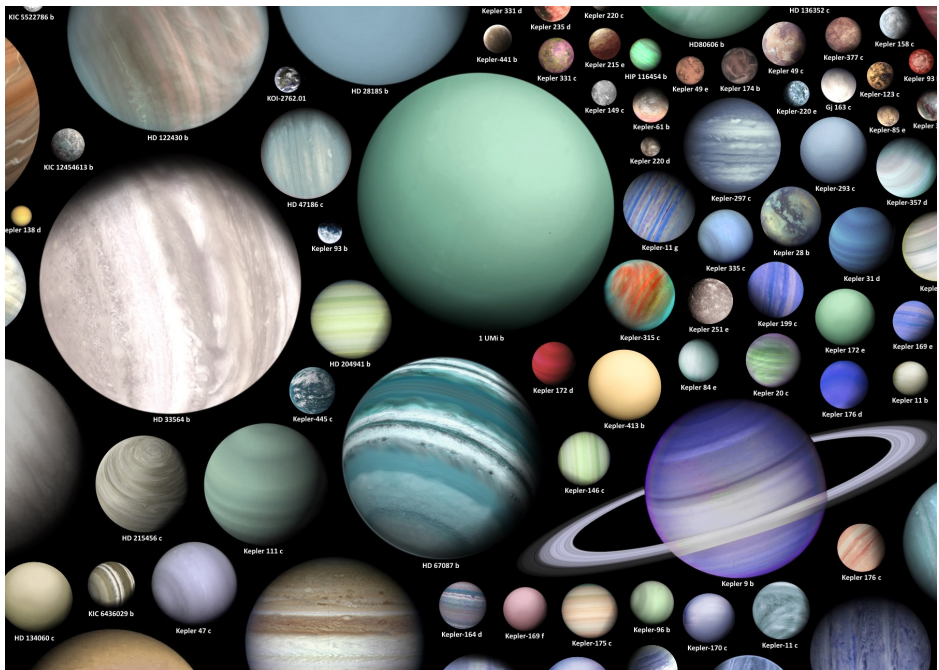

**Figure 4-1.** *Artist's conceptions of some of the 4,043 confirmed exoplanets (as of August 22, 2019). Credit: M. Vargic*

This chapter explores some of LUVOIR's capabilities for planetary origins studies and planetary science beyond the search for habitable planets and life. These "Signature Science" cases represent a few of the most compelling types of observing programs on a wide range of planets that scientists may choose to carry out with LUVOIR. Many more potential investigations are documented in the ExoPAG Science Analysis Group 15 report "Science Questions for Direct Imaging Missions" (Apai et al. 2017). Further exoplanet science cases are also described in **Appendix A**. As compelling as all of these science cases are, they should not be taken as a complete specification of LUVOIR's future potential. We fully expect that our creative community, empowered by the revolutionary capabilities of the observatory, will ask questions, acquire data, and solve problems beyond those discussed here—including those that we cannot envision today.

## 4.1 Signature Science Case #4: Comparative atmospheres

The search for exoEarth candidates will uncover hundreds of other types of planets. **Figure 3-1** illustrates the yield of detected planets of various sizes and temperatures expected to be found in the initial LUVOIR habitable planet survey. These additional exoplanets are ice and gas giants at a range of orbital distances from a variety of stars (A to M type) as well as non-habitable terrestrial and super-Earth-size planets. For example, about 150 serendipitous discoveries of planets smaller than 4 Earth radii (sub-Neptunes) around a variety of stellar types and at various temperatures are expected with LUVOIR-A. LUVOIR-B is expected to detect approximately 130 sub-Neptunes.

These planets will present exceptional opportunities for comparative studies over a much wider range of conditions than exist in the solar system, vastly improving scientists'

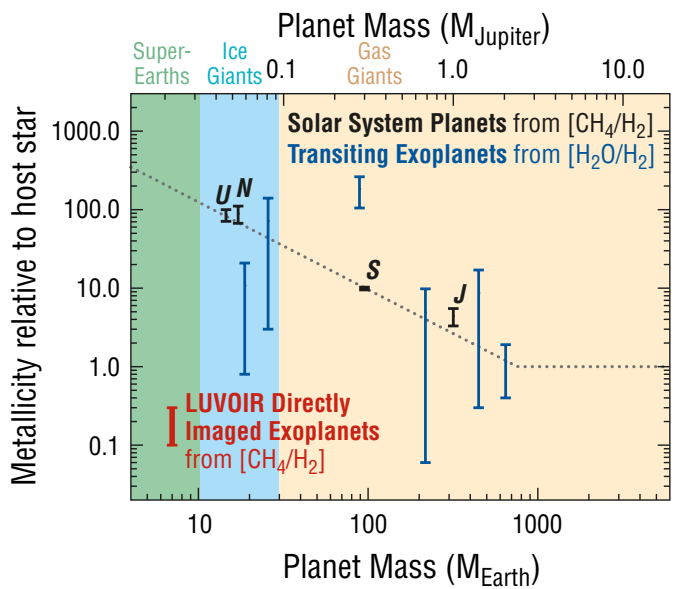

**Figure 4-2.** *Metallicity of solar system and extrasolar giant planets. Transiting planet metallicities are derived from retrievals of $H_2O$ abundance inferred from transit spectra. Solar system abundances are from remote and in situ measurement of $CH_4$. The dashed line shows the notional trend of decreasing [M/H] with increasing planet mass, a prediction of the core accretion gas giant formation mechanism. The red error bar shows expected precision of LUVOIR retrievals of gas giant $CH_4$ and $H_2O$ atmospheric abundances based on retrieval studies for WFIRST (Lupu et al. 2016). Credit: J. Bean (U of Chicago).*

# State of the Field in the 2030s

***Radial velocity***: We expect a factor of ten improvement in ground-based radial velocity (RV) measurement precision (to ~ 10 cm/s). This is roughly the size of the signal that the Earth imparts on the Sun when the solar system is observed edge-on. Depending on the effectiveness of techniques to contend with stellar jitter, instrumental noise, and signals from additional planets, precise RVs provide a possible, but not assured, path to finding and measuring the masses of terrestrial planets in the habitable zones of FGKM dwarfs in the LUVOIR era.

***Astrometry***: The ESA Gaia astrometry mission is expected to discover ~21,000 planets between 1–15 Jupiter masses in long period orbits (Perryman et al. 2014). Giant planets orbiting the nearest stars, with precise mass and orbit constraints, will be excellent targets for LUVOIR comparative planetary studies.

***Transit photometry***: NASA's all-sky Transiting Exoplanet Survey Satellite (TESS) launched in 2018 and has already found nearly 1000 planet candidates transiting bright, nearby stars. Scheduled for launch in 2019, the ESA Characterizing ExOPlanets Satellite (CHEOPS) will follow up previously detected exoplanets. The ESA PLAnetary Transits and Oscillations (PLATO) mission, planned for a 2026 launch, aims to detect transiting terrestrial planets, including potentially habitable planets orbiting Sun-like stars.

***Transit spectroscopy***: NASA's James Webb Space Telescope (JWST) will yield detailed characterization of 10–100 close-in exoplanets ranging from super-hot gas giants around all types of stars to temperate terrestrials around low-mass stars. ESA is implementing the transit spectroscopy mission ARIEL for the late 2020s, which will build on JWST by enabling a homogeneous census of the atmospheres of hundreds of close-in giant planets from the optical through NIR. However, a large sea of Earth- to Neptune-mass planets around solar-type stars will not yet be characterized by the 2030s and there are no current plans for any UV transit capability after HST.

***Microlensing***: The Wide-Field InfraRed Space Telescope (WFIRST) microlensing survey will detect thousands of exoplanets further from their host stars, leading to a statistical census of planets with masses > 0.1 $M_{Earth}$ from the outer habitable zone out to free-floating planets (Penny et al. 2019). Although WFIRST microlensing observations will probe the planet mass function, they will not directly investigate the atmospheres or habitability of these planets.

***High-contrast imaging***: WFIRST will also develop high-performance coronagraphic imaging and spectroscopic technology in space but will only be able to observe debris disks and a few large gas giant exoplanets. Second-generation instruments on the planned thirty-meter class ground-based telescopes (ELTs) will image thermal emission from young giant planets at much smaller spatial separations and around fainter stars than currently possible, will image and characterize some giants in NIR reflected light, and will detect terrestrial planets in the habitable zones of a handful of nearby low-mass stars.

***Mid-IR to millimeter disk observations***: The Atacama Large Millimeter/Submillimeter Array (ALMA) will have studied all phases of planet formation, with spectroscopy of molecular and atomic gas and high spatial resolution (but still 2 to 3 times lower than LUVOIR) mapping of cold dust in protoplanetary and debris disks. JWST will have surveyed hundreds of young stars at a resolution of > 20 AU and warm (~500 K) disk gas inside the snowline via medium-resolution ($\Delta v$~100 km/s) spectroscopy.

***Small bodies***: LSST will discover hundreds of thousands of near-Earth objects; millions of asteroids down to sub-km sizes; and tens of thousands of trans-Neptunian objects (TNOs) down to 100 km sizes. LSST will likely push to ~5 times smaller diameters on deep drill fields in a single color. JWST and ELTs (if optical wide-field imaging instruments are built for the latter) will push down to ~2× smaller bodies. JWST will have obtained spectra of intermediate size TNOs, allowing better connections between their photometric colors and their compositions.

understanding of fundamental atmospheric processes affecting habitability. For a few of these planets, characterization could be done simultaneously with long exposures aimed at habitable zone terrestrial planets. In other cases, different selections from the coronagraph's inventory of masks and additional exposure time will be needed. For all of these worlds, LUVOIR will enable a host of science investigations, ranging from studies of atmospheric dynamics in gas giants to searches for evidence of volcanic gases in the atmospheres of terrestrial planets.

Many of the large planets LUVOIR will study will have been detected by radial velocity surveys or Gaia, thereby providing masses and orbital constraints. Astrometry with HDI on LUVOIR will measure or constrain the masses of smaller planets lacking measurements (see **Section B.6.4** in **Appendix B**). Planets with known masses, incident fluxes, and measured atmospheric composition will provide a laboratory for testing theories for planet formation, atmospheric evolution, photochemistry, and cloud processes. In this section we summarize some of the highlights of such investigations made possible by LUVOIR.

### 4.1.1 Giant and terrestrial atmospheres with direct spectroscopy

***Diversity of composition.*** Atmospheric composition is of special interest as the particular species present depend on atmospheric temperature and pressure regulated by the incident stellar flux and, for giant planets, emergent planetary flux from the deep interior. The abundance of these species gives insight into atmospheric chemistry and planetary formation processes. The atmospheres of all solar system giants are enhanced over solar abundance, a fingerprint that is generally attributed to the details of planetary formation processes (**Figure 4-2**; Mordasini et al. 2016). For terrestrial planets, composition is a signature of initial volatile endowment, as modulated by escape, volcanic outgassing, impacts, crustal recycling, and other endogenic and exogenic processes.

The atmospheres of the relatively cool giant planets probed by direct imaging are directly connected to their deep interiors by a continuous convection zone and the bulk composition of their atmospheres will reflect that of the planet as a whole (Fortney et al. 2007). This is different from the atmospheres of the hot transiting giant planets, which are disconnected from their deep interiors by a radiative zone that extends to 100 to 1000 bar or more. Thus, LUVOIR direct imaging will characterize a population of planets with notably different atmospheric structure, dynamics, and composition than most of the transiting planets.

Understanding how atmospheric composition varies with planet mass and stellar properties is vital, since planet formation models predict a wide range of enrichments in elements compared to stellar abundances (e.g., Oberg et al. 2011; Fortney et al. 2013; Madhusudhan et al. 2016; Mordasini et al. 2016; Thorngren et al. 2016; Espinoza et al. 2017). Observations of transiting giant planets suggest a trend of decreasing atmospheric heavy element enrichment with increasing mass that mirrors the one seen in the solar system (**Figure 4-2**). LUVOIR users will test whether such trends hold for the cooler, more distant population of planets that more closely resemble solar system gas and ice giants. LUVOIR will also constrain the atmospheric composition of planets at the sub-Neptune/super-Earth boundary to understand if these are distinct planet types or simply a continuum of composition.

LUVOIR will achieve these goals by obtaining reflected light spectra of gas and ice giants, down to sub-Neptune/super-Earth sizes, at a variety of orbital distances around stars of varying mass and metallicity, thus balancing the population of hotter planets characterized by transit studies. Optical photometry and spectra of detected planets will readily constrain the presence or absence of atmospheres from the distinctive appearance of Rayleigh and cloud particle scattering. Atmospheric composition can be derived from optical and near-IR spectroscopy, as the wavelength ranges and spectral resolutions required for detecting gases in terrestrial atmospheres (**Chapter 3**) will also permit measurement of $H_2O$ (0.65, 0.72, 0.82, 0.94, 1.12, 1.4, 1.85 μm), $NH_3$ (1.8 μm), $CH_4$ (0.6, 0.79, 0.89, 1.0, 1.1, 1.4, 1.7 μm), Na (doublet near 0.6 μm), and K (0.77 μm) in gas giants (**Figure 4-3**) and additional species in the atmospheres of lower mass planets.

While atmospheric chemistry among the hydrogen-helium dominated giant planets can likely be approximately predicted, the atmospheric diversity of terrestrial planets discovered by LUVOIR will doubtless exceed our imagination. By carrying versatile integral field spectrographs covering a wide wavelength range, LUVOIR is well equipped to characterize the atmospheres of all types of terrestrial planets. Examples include the exoEarths or exoVenuses predicted to be the end-state of planetary evolution for some hot rocky planets (Berta-Thompson et al., 2016; Luger & Barnes, 2015; Schaefer et al., 2016). Optical and near-IR spectra will constrain $H_2O$, $O_2$, and $CO_2$ abundances for such worlds, providing insight to this critical class of planets that define the inner edge of the habitable zone.

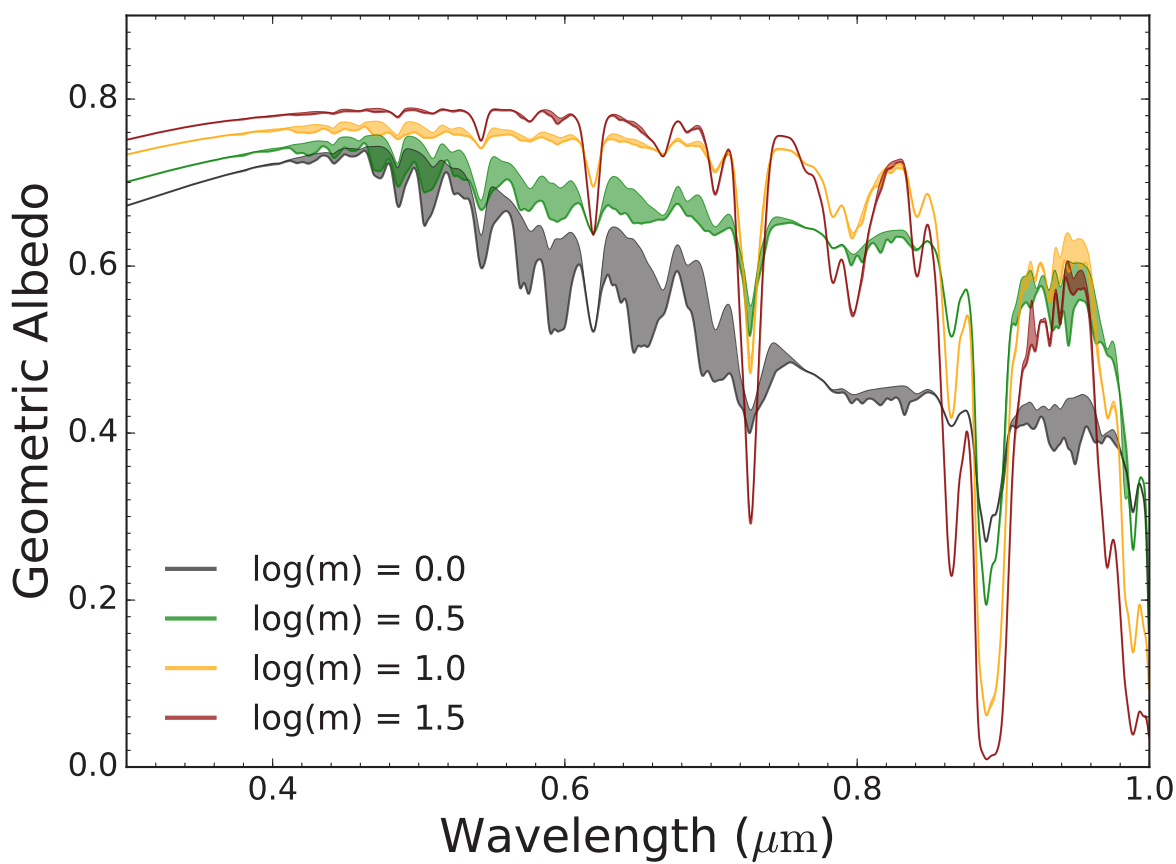

**Figure 4-3.** *Model albedo spectra of gas giant planets with atmospheric metallicity (log(m)) varying from solar to about thirty times solar. The most distinctive absorption features are those of $CH_4$. The shaded regions show the influence of gaseous $H_2O$ opacity, which, counterintuitively, is stronger at lower metallicity because of the feedback in this model between cloud height and water vapor abundance. Observations of such planets by LUVOIR users will constrain the atmospheric abundance of $CH_4$ and $H_2O$ and inform our understanding of such water vapor/cloud feedback effects. Credit: R. Macdonald (Cambridge University).*

Both very wet and very dry planets may also populate the habitable zone. Water-rich atmospheres may be found either on the path to the "exoVenus" end-state, or as a long-term stable configuration of the planet's atmosphere (Goldblatt et al., 2016). Conversely, extreme water loss can lead to "Dune-like" worlds with little surface water reservoirs but temperate climates (Abe et al., 2011). Such worlds would be identified by their Rayleigh scattering slope and lack of bright clouds or water vapor. Terrestrial-sized planets with dense envelopes that were not lost can exist beyond the habitable zone, and due to the greenhouse effect from their $H_2$ envelopes, lead to planets with stable oceans well beyond the "traditional" habitable zone (Owen & Mohanty, 2016). Pressure-induced absorption from high-pressure $H_2$ atmospheres will be detectable in the optical and near-IR by ECLIPS.

Comparing such terrestrial planets as these and others to each other and to any habitable worlds discovered by LUVOIR will place these planets in context. For example, if exoVenus-type planets turn out to be ubiquitous, the relative importance of the runaway greenhouse mechanism will be apparent. However, the diversity of such worlds (e.g., do they always display bright sulfuric acid clouds?) will emerge only through comparative studies of multiple planets. Since we cannot predict atmospheric compositions for all of the terrestrial planets LUVOIR will discover, the most important capability for characterizing these worlds will be flexibility. As discussed more fully in **Chapter 3**, LUVOIR's spectral range for coronagraphic spectroscopy provides the ability to detect and measure the abundances of a large array of possible gases, including $H_2O$, $CH_4$, $O_2$, $O_3$, and $CO_2$. **Figure 4-4** shows simulated ECLIPS spectra, of both giant and terrestrial planets, that give a flavor of LUVOIR's powerful capabilities for atmospheric characterization.

*Photochemistry.* Photochemical processes play key roles in shaping the atmospheres of all solar system planets. In Solar System giant planet stratospheres, $CH_4$ photochemistry generates hydrocarbons such as $C_2H_6$ and $C_2H_4$; these molecules can polymerize into more complex hydrocarbon species, some of which condense into aerosols. These hydrocarbons strongly absorb UV and blue light on Jupiter and Saturn. On Venus, complex sulfur chemistry transforms $SO_2$ and $H_2O$ into $H_2SO_4$ that condenses into the thick cloud deck (Yung and Demore 1982). On Earth, the UV-shielding ozone layer is the result of oxygen photochemistry, and sulfur aerosols like $H_2SO_4$ can be produced through volcanic outgassing; such aerosols can produce detectable spectral signatures that may be remotely identified as a sign of vigorous volcanism on terrestrial planets (Misra et al 2015; Hu et al 2013). The hazy disk of Titan, a consequence of $CH_4$ transport and photochemistry, is another distinctive example of the importance of hazes. No characterization of an exoplanet atmosphere—potentially habitable or not—will be complete without an understanding of the role photochemistry plays in its atmosphere.

Evidence of hazes has already been inferred from the featureless transit transmission spectra of several exoplanets (Bean et al. 2010, Kreidberg et al. 2014, Knutson et al. 2014a, 2014b, Sing et al. 2014). Exoplanet atmospheres observed by LUVOIR will sample compositional and incident UV flux conditions not found in the current solar system, yielding a diversity of haze types. Examples of the potential impact of two plausible types of hazes, sulfur hazes on a warm giant planet and hydrocarbon hazes on an Archean (4 to 2.5 billion years ago) Earth-like terrestrial planet, are shown in **Figure 4-5**. The NUV capability of ECLIPS will be crucial for constraining photochemical hazes as they often absorb strongly in the blue and UV.

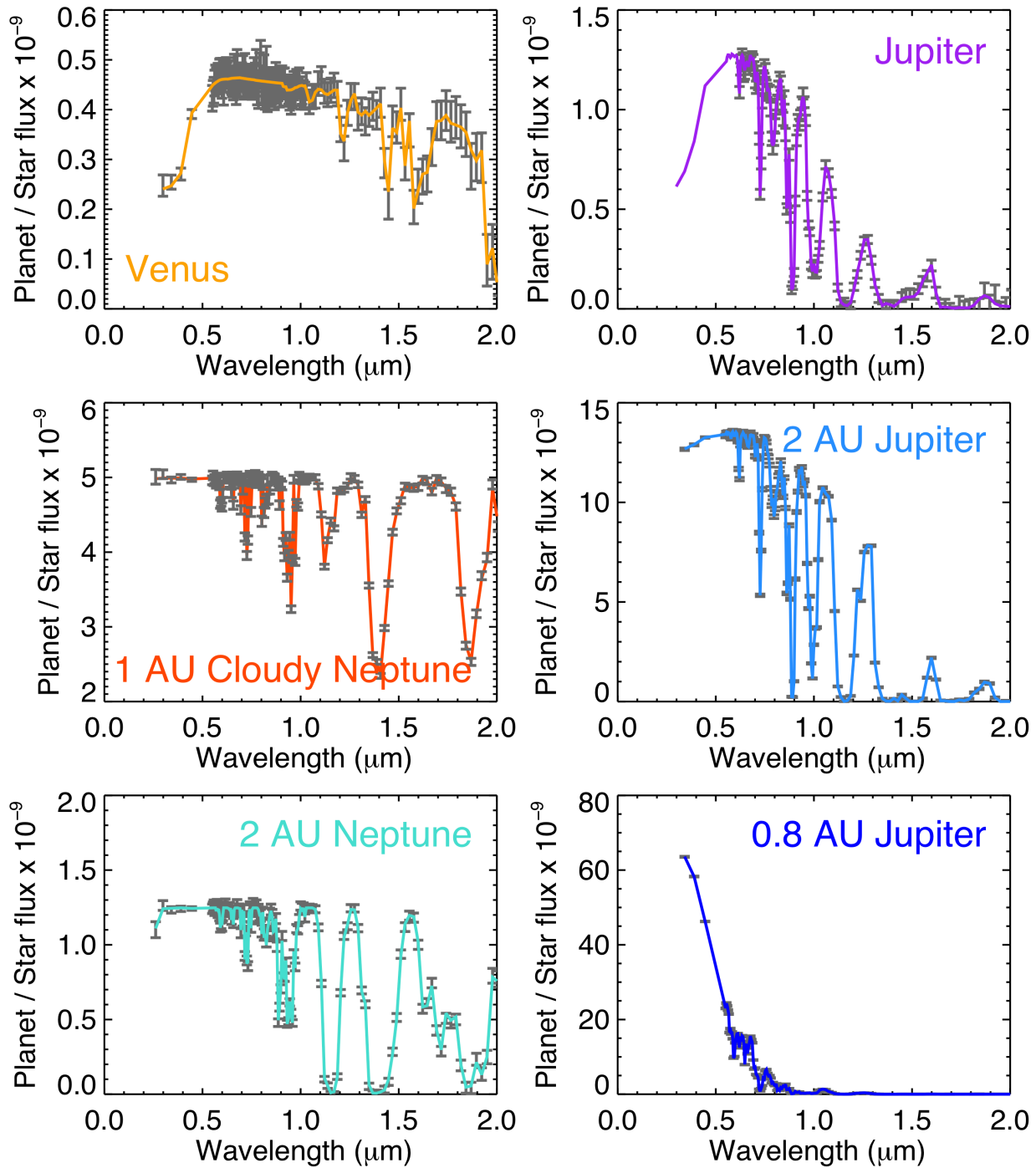

**Figure 4-4.** *Simulated LUVOIR spectra of planets orbiting a Sun-twin star at 10 pc that could be observed in the same time required to obtain a SNR = 10 spectrum of an Earth-twin (100/230 hours per coronagraph band for LUVOIR-A/B). The simulated data plotted with grey bars were generated using the online LUVOIR coronagraphic spectroscopy tool. Input model spectra are over-plotted with colored lines. The Venus model spectrum was provided by the Virtual Planet Laboratory (VPL), generated using the Spectral Mapping Atmospheric Radiative Transfer (SMART) model (Meadows & Crisp 1996; Crisp 1997). The warm Jupiter model spectra are from Cahoy et al. (2010) and the warm Neptune model spectra is from Hu & Seager (2014). Credit: LUVOIR Tools / G. Arney (NASA GSFC)*

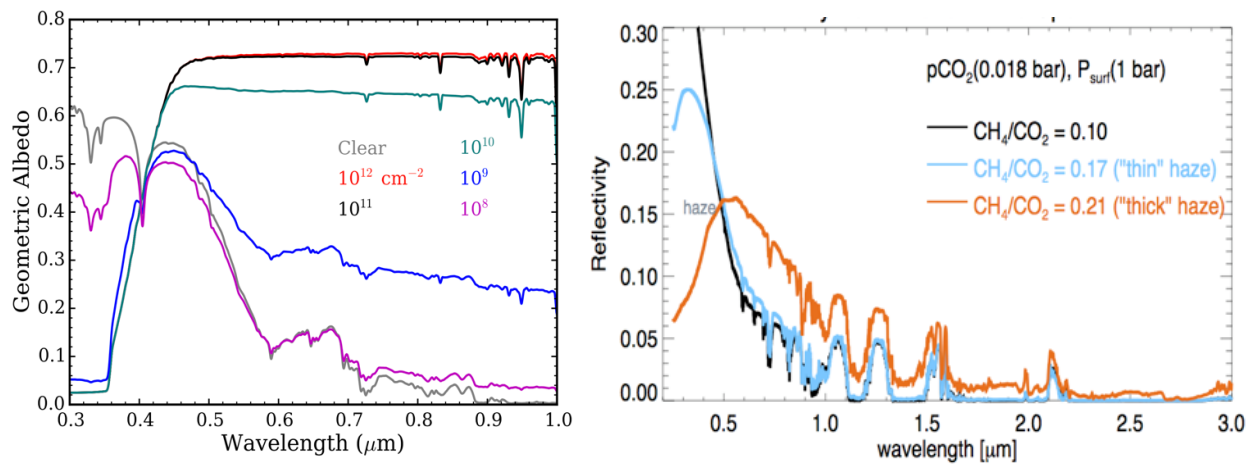

**Figure 4-5.** *Two examples of the impact of photochemical hazes on planet spectra. (Left) Model geometric albedo spectra for a cloud-free 0.8 AU Jupiter-like planet (grey) along with spectra for various column abundances of sulfur ($S_8$) haze particles. In addition to markedly raising the red continuum flux level, note the strong NUV absorption of these hazes. Credit: P. Gao (UC Berkeley). Right panel: Model geometric albedo spectra of clear and hazy Archean Earth atmospheres. Credit: G. Arney (NASA GSFC).*

Understanding the diversity of photochemical outcomes and how they relate to composition, temperature, and pressure will be a complex task, but one that will greatly inform our insight into such processes as a whole. For multi-planet systems the composition and photochemistry in one planet's atmosphere will inform our understanding of the other planets, including potentially habitable ones. Measurements of the UV spectrum of the host star will also be critical for understanding these processes and interpreting exoplanet atmospheres. LUVOIR observations will illuminate how haze opacity varies with planetary and stellar characteristics, clarify the role hazes play in potentially habitable planets, and improve our ability to model photochemical processes in all types of planetary atmospheres. For terrestrial planets, LUVOIR will also probe links between photochemistry and surface composition.

Planet surface gravity directly affects planetary spectra, so atmospheric characterization by direct imaging is greatly facilitated when limits can be placed on the planet mass (see **Section 4.2.3.1**). For planets previously detected by RV, LUVOIR images will constrain the orbital inclination and reveal the true mass of the planet instead of the minimum mass determined from RV observations. For planets without previous mass constraints, LUVOIR astrometry will directly measure planet masses as discussed in **Section 4.2**.

The Signature Science Case #4 observing programs in **Appendix B.5** provide the full details on LUVOIR's general exoplanet observing program, which begins with ECLIPS direct spectroscopy of a set of currently known warm-to-cold exoplanets spanning a range of sizes. A 424-hour program with LUVOIR-A will characterize the atmospheres of some 30 planets, including over a dozen in multi-planet systems. These planets would have masses spanning from that of Earth to ten times that of Jupiter and equilibrium temperatures from 100 to 700

K. A slightly longer 500-hour program with LUVOIR-B would characterize about 20 planets. The diverse set of planetary properties encompassed by these programs would provide outstanding opportunities for comparative planetary science, including within individual planetary systems.

## 4.1.2 Characterizing transiting planets from the NUV to NIR

LUVOIR will be an outstanding platform for studies of transiting planets of all types and will provide exceptional new science opportunities from hot Jupiters down to temperate, terrestrial mass planets. Transiting planet science is complementary to studies of directly imaged planets since the transit technique primarily uncovers planets on orbits close to their host stars that are difficult to image directly, but comparatively likely to be detectable in transit at high signal-to-noise. Moreover, the transit method reveals planet radii, which can be combined with mass measurements to link atmospheric properties to bulk composition and formation.

The Transiting Exoplanet Survey Satellite (TESS) is discovering hundreds of transiting planets orbiting the brightest stars. This sample will eventually include hundreds of hot Jupiters and hot Neptunes around Sun-like stars, along with dozens of rocky planets transiting early-to-mid M-dwarfs, which will be ideal targets for follow-up with transit spectroscopy (Ricker et al. 2015; Barclay et al. 2018). Furthermore, ground-based surveys of very late M-Dwarfs may yield additional temperate rocky planets—in fact, one of the best Earth-sized HZ targets to date is TRAPPIST-1e (Gillon et al. 2017).

JWST will thoroughly characterize the atmospheres of transiting Jupiter and Neptune-mass planets from the red optical to the near-infrared (0.6–28.5 μm), revolutionizing our understanding of hot planets orbiting close to their parent star. However, JWST lacks the coverage in the blue optical and near UV of HST and LUVOIR. These wavelengths offer unique insights into planetary atmospheres that allow us to understand cloud properties and atmospheric photochemistry. In this respect, LUVOIR is a true to successor to Hubble, however with larger aperture, higher throughput, and greater stability.

In the optical and NUV, LUVOIR will reach 9–14 times the SNR per transit compared to HST STIS, while in the NIR it will reach 3 times the SNR per transit compared to JWST NIRSpec. The primary instrument for transit spectroscopy with LUVOIR will be HDI, with its broad simultaneous wavelength coverage (200 nm–2.5 μm) and the ability to spatially scan the spectra of bright stars across the large focal plane detectors. It will have the capability of full-throughput grism observations in either short (200 nm–900 nm) or long (800 nm–2.1 μm) wavelength channels, or simultaneous observations with both bands but at half the throughput. Spectral resolution will be R ~ 500, enabling full characterization of atomic lines and molecular bands for key species. As an example, **Figure 4-6** illustrates the exceptional signal to noise LUVOIR could achieve with just a single transit for ten benchmark hot Jupiters previously observed with HST and Spitzer.

LUVOIR's high sensitivity at optical wavelengths will enable detections of atomic species such as the alkali metals for wide range of both terrestrial and non-terrestrial planets, many of which will otherwise be inaccessible due to the low throughput of the STIS instrument on HST and the wavelength limits of JWST. Likewise, LUVOIR will be able to measure Rayleigh scattering slopes, an important indicator for the total atmospheric column, for planets of all sizes at high precision.

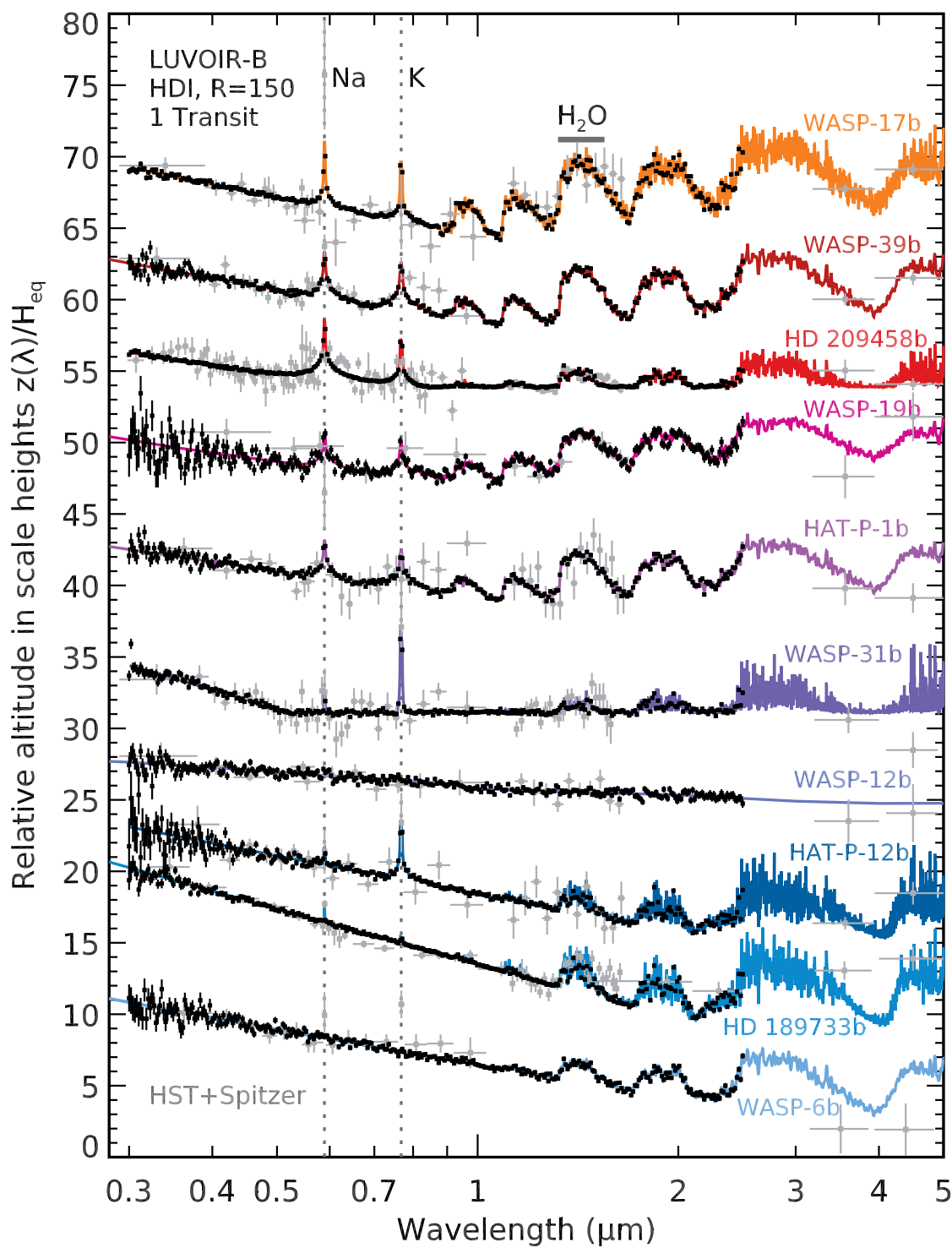

**Figure 4-6.** *Simulated LUVOIR-B transit spectra (black points with error bars) on the 10 benchmark hot Jupiters surveyed by Sing et al. (2016) with their associated model spectra (colored lines) and HST and Spitzer data (light gray). The simulated data show observations of a single transit for each planet using the beam splitter mode on HDI to simultaneously observe in both the NIR and UVIS channels at R=150. Similar observations with LUVOIR-A would be even more precise. Credit: E. Lopez (GSFC)*

Additionally, in the region of overlap with JWST, LUVOIR will still benefit from its higher SNR. Characterizing the smaller, cooler worlds will be time-intensive: JWST will need months of integration time to provide constraints on the presence of an atmosphere. The amplitude of spectral features for a temperate terrestrial planet transiting in front of a nearby M-dwarf is comparable to the single-transit photon-counting precision with JWST; therefore, in the absence of a systematic noise floor, 100 transits of such a planet could yield a $10\sigma$ detections of greenhouse gases—and this neglects the effects of cloud opacity in damping the signal of spectral absorption (Cowan et al. 2014).

Because transits last only a few hours, the one month of cumulative JWST time to characterize the atmosphere of a potentially habitable world would have to be spread out over nearly a decade for a planet in a month-long orbit; ground-based transit studies with extremely large telescopes will face the same timing challenge (Rodler & Lopez-Morales 2014) and contend with the additional hurdles of variable weather conditions, daylight, and the need to correct for contamination from the Earth's atmosphere. It is thus entirely possible that very little will be known about the atmospheres of these planets by the time JWST ends its mission—and further study will be left for a future flagship mission with equal or greater photon-gathering power. Between 0.8 and 2.1 µm transit observations with LUVOIR will distinguish between planets with Earth, Venus, or Mars-like atmospheres. At shorter wavelengths (0.2–0.8 µm), LUVOIR's capabilities will be unprecedented, providing measurements of Rayleigh scattering and possibly $O_3$ at 200–300 nm.

Barclay et al. (2018) modeled the yield of planets discovered by the TESS mission, determining that the mission would discover between 30 and 40 Earth-sized planets orbiting

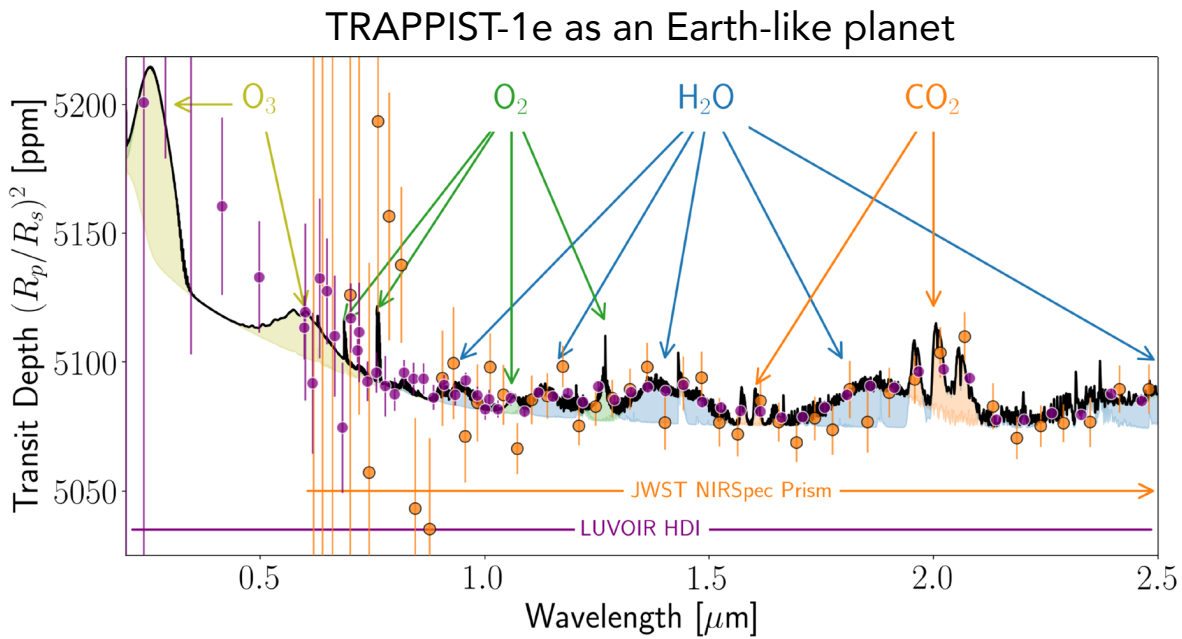

**Figure 4-7.** *Simulated transit spectrum of TRAPPIST-1e. Simulated observations assume 50 transits observed with LUVOIR-A/HDI grism (purple) and with JWST/NIRSpec prism (orange; using the optimized readout mode from Batalha et al., 2018). Spectral bands of multiple key atmospheric species are visible by eye, enabling constraints on habitability. Credit: J. Lustig-Yaeger/A. Lincowski (University of Washington). JWST spectrum generated with PandExo (Batalha et al. 2017).*

M dwarfs with K < 9, with 2 to 3 of those being on temperate orbits. A systematic survey of the most favorable transiting terrestrial planets orbiting M dwarfs will constrain atmospheric composition and structure. For example, **Figure 4-7** illustrates what could be accomplished by observing 50 transits on the potentially habitable Earth-sized planet TRAPPIST-1e. A number of molecules are immediately apparent and interesting constraints on the atmospheric chemistry and even on biomarker species such as $O_2$ will be possible.

The Signature Science Case #4 observing programs in **Appendix B.5** include HDI transit spectroscopy of 16 warm-to-hot rocky and gas giant exoplanets in about 23 days of observing time, including overheads, with either LUVOIR-A or -B as these programs were scaled by number of transits observed. The transit DRM assumes higher spectral resolution, after binning, for A than for B in this allotted time.

### 4.1.3 Atmospheric evolution and escape

Atmospheric escape is a fundamental physical process that leads atmospheric constituents to become unbound from a planet and alters the composition of the remaining atmosphere. Understanding the relative roles of escape, outgassing, and accretion in a variety of exoplanet atmospheres is critical to understanding the origin and evolution of planetary atmospheres. LUVOIR UV transit observations will both identify the escaping species and constrain the physics of escape.

The first observations of escape were obtained by Vidal-Madjar et al. (2003), who obtained STIS far-ultraviolet transmission spectra of the close-in giant planet HD209458b revealing that the planet possesses a highly extended hydrogen atmosphere due to heating by stellar X-ray and EUV photons. Subsequent HST observations have also detected various metals in the exospheres of giant planets including carbon, oxygen, and magnesium (Vidal-Madjar et al. 2004; Linsky et al. 2010), as well as escaping hydrogen from the

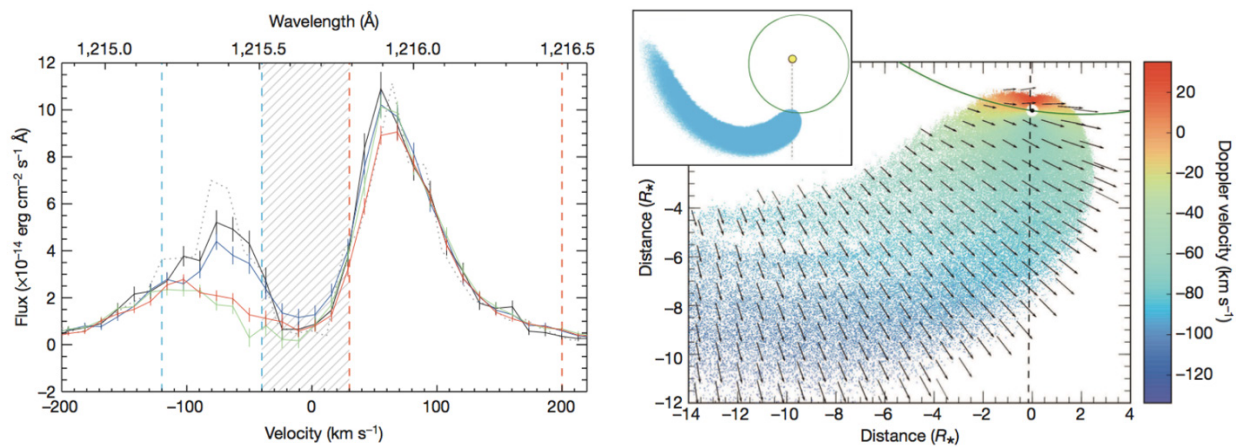

**Figure 4-8.** *FUV transit spectroscopy with LUVOIR will enable studies of atmospheric escape for a great diversity of planets. The figure on the left shows a reconstruction of escaping hydrogen observed in Lyman Alpha with HST for the hot Neptune GJ 436b (Ehrenreich et al. 2015). The simulated observation on the right shows the detection of escaping gas possible in a single transit with LUVOIR. In addition to measuring escape in Lyα for hundreds of planets, LUVOIR can also measure transits in FUV metal lines like CII at 133nm. Observing these lines at high SNR will provide key information about atmospheric composition and the structure of the escaping upper atmosphere, particularly for the low velocity material close to the planet for which Lyman alpha is completely hidden by the ISM.*

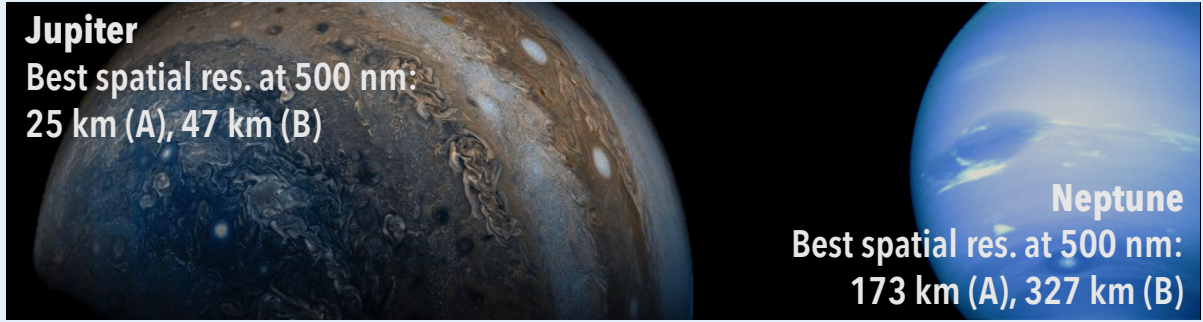

**Gas & Ice Giant Planets of our Solar System**

**Jupiter**
Best spatial res. at 500 nm:
25 km (A), 47 km (B)

**Neptune**
Best spatial res. at 500 nm:
173 km (A), 327 km (B)

Our solar system provides several nearby examples of gas giant and ice giant planets. LUVOIR can study the dynamic atmospheres of these planets and provide insights to the large exoplanets that will be studied in coming decades. At Jupiter, LUVOIR-A can deliver images with resolution comparable to Juno observations (25 km at 500 nm). Farther out in our solar system, no spacecraft has visited Uranus and Neptune for over three decades. While advances in outer planet science have been made with Hubble, the Keck 10-m telescopes, and other ground-based facilities, the extreme distances of these planets hamper high-quality imaging of their atmospheres. A 15-m aperture space telescope could capture images of Neptune that rival the quality of Voyager 2 flyby images and help prepare for future missions to the ice giants.

warm Neptune-size planet GJ436b, which showed a large tail of escaped planetary material (**Figure 4-8**; Ehrenreich et al. 2015).

Atmospheric escape is a key factor shaping the evolution and distribution of low-mass close-in planets (e.g., Owen & Wu 2013) and their habitability (e.g., Cockell et al. 2016). Indeed, many highly irradiated rocky planets (e.g., CoRoT-7b, Kepler-10b) might be the remnant cores of evaporated Neptune-mass planets (e.g., Lopez et al. 2012). As a consequence, atmospheric escape also has a major impact on our understanding of planet formation (e.g., Van Eylen et al. 2018). LUVOIR will characterize the extended atmospheres of a statistical sample of these worlds, essential for understanding the global properties of atmospheric escape and understanding how the physics of UV-driven atmospheric mass loss determines the long-term stability of all types of planetary atmospheres.

The observations will measure the velocity profile of the absorbing material along the line of sight, thereby constraining models of atmospheric escape. From the few relevant HST observations of transiting planets obtained so far it appears that the typical velocity of the escaping material is of the order of 10–20 km/s. This requires instruments with a resolving power 40,000–60,000 to resolve the velocities, readily achievable with LUMOS. Obtaining these observations simultaneously for a range of species is particularly valuable since different species probe different regions in a planet's photo-evaporative wind. For example, while Lyman-alpha can trace material beyond the planet's Hill sphere as it interacts with the stellar wind, metal lines like CII trace the inner bound portions of a planet's upper atmosphere, where the photo-evaporative wind is launched and Lyman-alpha is heavily extincted by the ISM. Additionally, detection of these metal lines provides a way to probe atmospheric compositions, which is not subject to the effects of clouds lower in the atmosphere.

The observational census of atmospheric escape requires UV transmission spectroscopy for several dozen transiting planets orbiting stars (later than spectral type A5V). A complete sample of planets is necessary to understand how atmospheric escape correlates with stellar and planetary system parameters. The observations should be carried out at both FUV and NUV wavelengths for the brightest and nearest systems (for Lyman-alpha in particular it is necessary to observe systems within 60 pc due to high ISM absorption). Moreover, because UV emission from stellar activity can be highly variable from one transit to another, it is important to obtain high SNR detections within an individual transit (Bourrier et al. 2017). The Signature Science Case #4 observing programs in **Appendix B.5** conclude with LUMOS transit spectroscopy of 16 hot exoplanets to study atmospheric escape. About five days of observing time, including overheads, with either LUVOIR-A or B, would survey these planets for escaping atmospheres.

Summarizing the observing programs in Signature Science Case #4 (**Appendix B.5**), LUVOIR-A could obtain 1) high SNR optical through NIR reflected light spectra of a diverse set of 30 warm-to-cold planets, 2) optical/NIR transit spectra of 16 warm-to-hot planets, and 3) FUV transit spectra of 16 hot planets, all in about 1.5 months of total observing time, including overheads. These planets span a range of sizes from Earth-size to Jupiter-size and orbit a variety of stars. Such a sample would directly address the atmospheric composition of giant planets, non-habitable terrestrial world characterization, and atmospheric escape studies. For LUVOIR-B the sample would be smaller for comparable observing time—about 19 directly imaged planets and 32 transiting planets at lower SNR.

## 4.2 Signature Science Case #5: The formation of planetary systems

LUVOIR will observe both young and mature planetary systems. This capability will allow fundamental, systematic investigations of the environmental influences on planet formation and evolution. These studies are enabled by:

1)  Multi-object FUV spectroscopy of protoplanetary disks with LUMOS

2)  High-contrast imaging of young exoplanet systems with ECLIPS

3)  High-precision astrometry of mature exoplanet systems with HDI

Such observations complement techniques that characterize a portion of each planetary system (e.g., transit spectroscopy) and surveys that provide planet occurrence rates (e.g., microlensing). LUVOIR also has a unique role to play in filling out our inventory of the smallest and coldest bodies in the solar system, which hold vital clues to its early history.

### 4.2.1 Protoplanetary disks

One of the main science objectives of LUVOIR will be to detect and characterize Earth-size planets in the habitable zone of nearby stars (**Chapter 3**). Understanding how and in which environments planets assemble contributes to this science objective by determining which nearby planetary systems are most likely to host life-bearing planets and by constraining planet properties that are not directly observable (e.g., bulk composition).

Recent high-resolution images of circumstellar disks around young (1–10 Myr) stars revealed complex structures (e.g., ALMA Partnership 2015; Wagner et al. 2015; Andrews et al. 2016; Perez et al. 2016; Pohl et al. 2017; Hendler et al. 2018), some of which point to advanced planet formation. A few candidate giant planets have been recently detected around Myr-old stars (e.g., Sallum et al. 2015, Keppler et al. 2018). Short timescales to assemble asteroid-size objects and giant planets are well in line with the evolution and dispersal of gas and dust in the solar nebula (Pascucci & Tachibana 2010). Thus, 1–10 Myr-old circumstellar disks provide an opportunity to study planet formation in action.

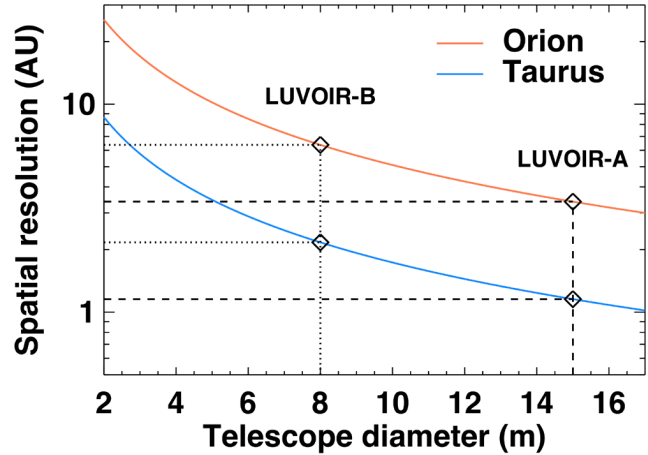

**Figure 4-9.** *LUVOIR can probe the inner regions of planet-forming disks. The plot shows the spatial resolution at 500 nm as a function of telescope aperture for two star-forming regions that have very different stellar populations. Credit: A. Roberge (NASA GSFC)*

Scaling relations in disks and exoplanets with the mass of the central star strongly suggest that sharp edges and gaps in young disks remain imprinted in the exoplanet population (e.g., Alexander & Pascucci 2012; Mulders et al. 2015). In addition, the composition of early planetary atmospheres should inform the formation history of planetary systems (e.g., Cridland et al. 2017).

At the distances of typical star-forming regions (e.g., Taurus-Auriga or Chamaeleon I), 1 AU corresponds to an angular scale of ~10 mas and emission lines from gas at ~1 AU radii will have FWHMs of ~30 km/s. Therefore, the high angular and spectral resolution of LUVOIR is a unique tool for mapping the inner disks where terrestrial planet formation occurs (**Figure 4-9**). Such high-resolution observations will link birth environments to the final architectures of exoplanetary systems.

#### 4.2.1.1 Composition and evolution of planet-forming material

UV spectroscopy is a unique tool for observing the molecular gas in the inner disk; the strongest electronic band systems of $H_2$ and CO reside in the 100–170 nm wavelength range (e.g., Herczeg et al. 2002; France et al. 2011). UV fluorescent $H_2$ spectra are sensitive to gas surface densities lower than $10^{-6}$ g cm$^{-2}$, making them an extremely useful probe of remnant gas at $r < 10$ AU. In cases where mid-IR CO spectra or traditional accretion diagnostics (e.g., H$\alpha$ equivalent widths) suggest that the inner gas disk has dissipated, far-UV $H_2$ observations can offer unambiguous evidence for the presence of a remnant molecular disk (Ingleby et al. 2011; France et al. 2012; Arulanantham et al. 2018).

LUVOIR's multi-object, high-resolution capability (R > 40,000, 4 square arcminutes per LUMOS field) enables emission-line surveys and absorption-line studies of high-inclination (i > 60°) disks. Absorption line spectroscopy through high-inclination disks, currently limited to a small number of bright stars (e.g., Roberge et al. 2000, 2001; France et al. 2014), is especially important because the large wavelength coverage of LUVOIR provides access to important molecular species in the UV and NIR, such as $H_2$, CO, OH, $H_2O$, $CO_2$, and $CH_4$. UV absorption line spectroscopy is the only direct observational technique to characterize co-spatial populations of these molecules with $H_2$, offering unique access to absolute abundance and temperature measurements without having to rely on molecular conversion factors or geometry-dependent model results as with emission-line spectroscopy.

Uniform spectral surveys of local star-forming regions are required for a systematic determination of disk abundances (a direct measurement of the initial conditions for planet-formation) and gas disk lifetimes (the timescales for gas envelope accretion and migration of planetary cores through their natal disks). For star-forming regions beyond Taurus-Auriga, the efficiency of these surveys goes up dramatically with the introduction of multi-object spectroscopy, greatly reducing the total observing time and increasing the survey efficiency in a census of star-forming regions at distances less than 1 kpc. Combining these surveys with the detailed characterization of debris disks discussed below will enable LUVOIR to trace the physical and chemical evolution of the primary disk species, constraining the composition of the gas that can be accreted onto the core of giant planets as well as that of the planetesimals that could constitute the bulk of the icy planets.

### 4.2.1.2  Mapping dispersal of protoplanetary material via disk winds

Recent theoretical work suggests that disk evolution and dispersal are driven by a combination of thermal and MHD disk winds (e.g., Alexander et al. 2014; Gorti et al. 2016; Ercolano & Pascucci 2017). Disk winds can affect all stages of planet formation: from planetesimal formation, by reducing the disk gas-to-dust mass ratio (e.g., Carrera et al. 2017); to the mass of giant planets, by starving gas accretion onto late-forming planet cores (e.g., Shu et al. 1993); to planetary orbits, by dispersing gas from preferential radial distances and halting planet migration (e.g., Alexander & Pascucci 2012; Ercolano & Rosotti 2015). However, computational challenges and poorly constrained input parameters make it difficult to predict how basic disk properties evolve and to ascertain the role of different winds in dispersing protoplanetary material.

On the observational side, high-resolution (R>30,000) optical and infrared spectroscopy yields growing evidence of slow (<40 km/s) disk winds around Myr-old stars (e.g., Pascucci & Sterzik 2009; Sacco et al. 2012; Natta et al. 2014; Simon et al. 2016). Modeling of the line profiles and line flux ratios suggest that emission arises at disk radii between 0.1 and 10 AU. On similar spatial scales, jets (ejected material moving at ~100 km/s) are also expected to be accelerated and collimated (e.g., Ray et al. 2007). Many of the brightest jet diagnostics are accessible to high-resolution imaging spectroscopy at UV and optical wavelengths (see e.g., Frank et al. 2014 for a recent review).

The high spatial resolution and wavelength coverage of LUVOIR are therefore needed to understand the origin of disk winds and the relative roles and interplay of MHD and photo-evaporative winds, as well as to clarify the physical mechanism by which jets are launched and collimated. In addition to high-resolution spectroscopy, LUVOIR slitless spectroscopy (i.e., an open micro-shutter array) and narrow-band images in UV and optical forbidden lines will map for the first time the launching region of jets and disk winds, and reveal their interaction. POLLUX spectropolarimetry will complement the science case discussed above by uniquely constraining the strength and topology of stellar magnetic fields during pre-main sequence evolution, with important implications for stellar and disk formation and evolution (e.g., Gómez de Castro et al. 2016).

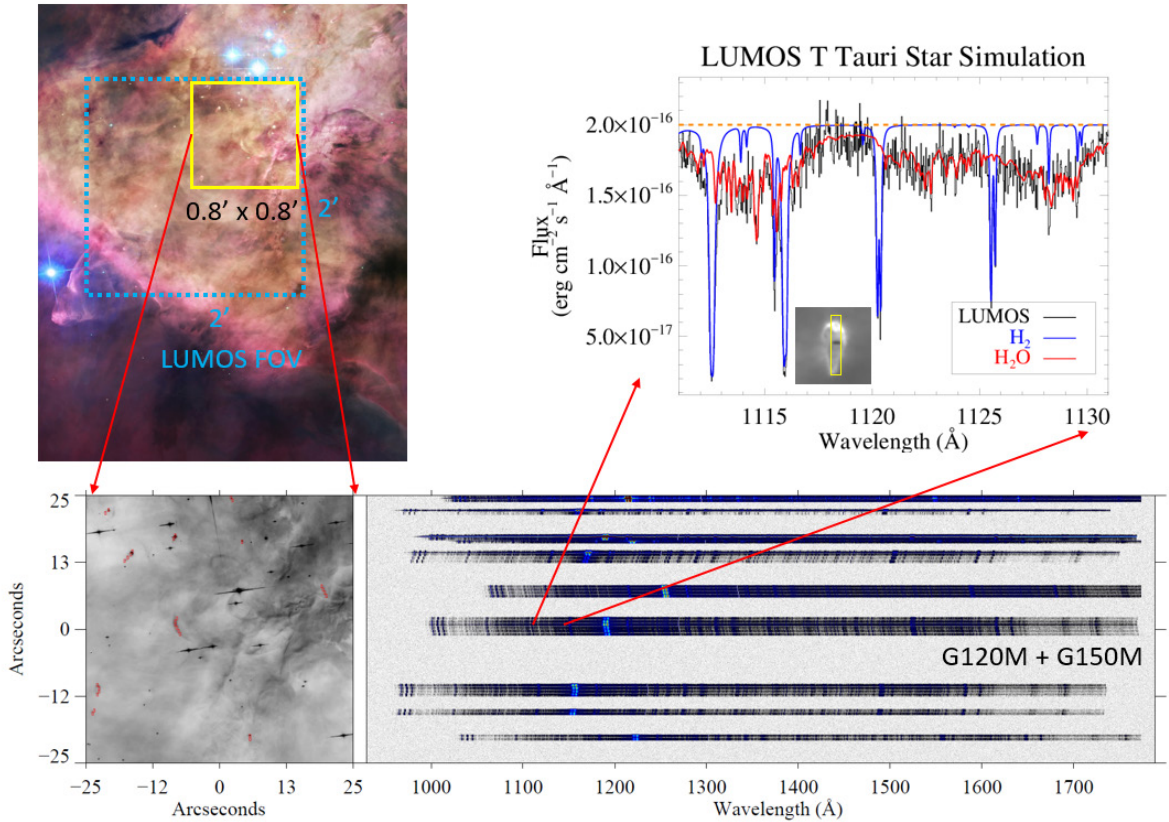

**Figure 4-10.** *Multi-object FUV spectroscopy of protoplanetary disks in the Orion Nebula. Upper left: HST-ACS image of the Orion Nebula (color credit, R. Gendler), showing the full FoV of the LUMOS spectrograph (France et al. 2017). Lower left: Blow-up of a ~0.8x0.8 arcmin region showing ~30 protostellar/protoplanetary systems (Bally et al. 1998), with the apertures of the LUMOS microshutter array overplotted (slits are oversized for display). Lower right: Two-dimensional spectra of proto-planetary disks and accreting protostars. Upper right: Blow-up of the 1111–1132 Å spectral region containing strong lines of $H_2$ and $H_2O$. The combination of spectral coverage, large collecting area, and multiplexing capability make LUVOIR ideal for surveying the composition of planet-forming environments around young stars.*

### 4.2.1.3  Disks in the Orion star forming region

The Orion complex, at a distance of ~400 pc (e.g., Kounkel et al. 2017), includes rich star-forming clusters that are more typical birth environments for stars than nearby low-density star-forming regions such as Taurus. Most stars in the Galaxy, including our own, formed in rich clusters (e.g., Lada & Lada 2003; Adams 2010) that contain massive stars whose high-energy UV and X-ray photons contribute to disperse planet-forming material around young stars (e.g., Johnstone et al. 1998; Clarke & Owen 2015). The Orion complex covers different external UV fields and the critical 1–10 Myr age range over which planet-forming disks disperse. It represents one of the best sites to study planet formation and several of its regions already have a fairly complete stellar and disk census (e.g., Fang et al., in prep.).

LUMOS is ideal for ultraviolet spectral surveys of the Orion complex (**Figure 4-10**). First, LUMOS has 50 times higher sensitivity and 2.5–3.5 times higher spectral resolution than HST/COS. Second, it carries a multi-object imaging spectroscopy mode that enables the simultaneous observation of tens to hundreds of targets (depending on source field density).

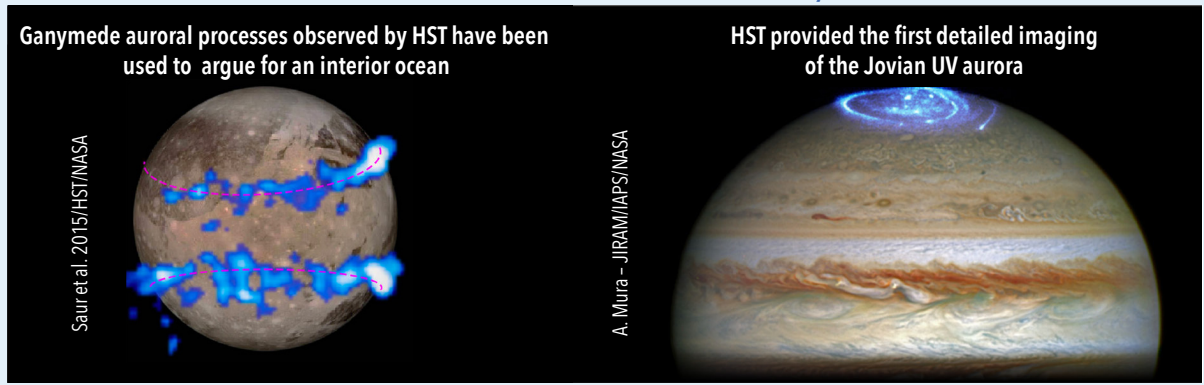

**Auroral Processes in the Solar System**

Ganymede auroral processes observed by HST have been used to argue for an interior ocean

Saur et al. 2015/HST/NASA

HST provided the first detailed imaging of the Jovian UV aurora

A. Mura – JIRAM/IAPS/NASA

Just as Hubble ultraviolet observations of jovian and saturnian aurorae complemented missions such as Juno and Cassini respectively, a future ultraviolet space telescope would complement *in situ* missions to outer planets in future decades. A large aperture space observatory provides not only increased spatial resolution, but also increased sensitivity. With increased sensitivity comes shorter exposures, reducing rotational blurring of features and permitting high-fidelity studies of the dynamics of dynamic auroral processes. For example, interactions between the magnetospheres of Ganymede and Jupiter could be studied with the resolution of a 15-m-class telescope with UV capability. Because Ganymede is the only known solar system moon with its own magnetic field (Kivelson et al. 1996), its interactions with the Jovian magnetosphere are visible to telescopes in the UV (e.g., Feldman et al. 2000; Mcgrath et al. 2013; Saur et al. 2015) and were used to argue for the presence of Ganymede's ocean. At a factor of 16 better spatial resolution than HST, LUVOIR could monitor changes in Ganymede's dipole over time, helping to understand the nature of its deep interior.

A LUMOS FUV map in dense stellar environments like the Orion Nebula Cluster would collect a higher-quality dataset (in terms of SNR, spectral resolution, angular resolution, and number of objects observed) than the entire medium-resolution UV spectroscopic archive of protoplanetary disks from the 28 years of HST observations.

In about an hour, LUMOS will reach SNR = 10 per resolution element (R = 40,000) over a 2′ x 2′ field to $F_\lambda = F(1100 \text{ Å}) = 2 \times 10^{-16}$ erg/cm$^2$/s/Å, typical of the continuum flux of young stars in Taurus scaled to the distance of the Orion complex. This means that in about ~3 hours per field LUMOS will cover the entire FUV and NUV spectral range of Orion stars enabling the detection of the most abundant molecular species in disks, e.g., $H_2$, CO, $H_2O$, and OH. With pre-defined maps of targets (Fang et al., in prep), LUMOS can cover most candidate stars in the Orion Nebula Cluster (~1 Myr, ~24′ size), and many in each of the NGC 1980 (~1–2 Myr, ~16′ size), σ Ori (~3–5 Myr, ~33′ size), λ Ori (4–8 Myr, 49′ size), 25 Ori (~7–10 Myr, 33′ size) regions in < 150 hours at FUV and NUV wavelengths. This program would allow us to trace the evolution and dispersal of the main molecular carriers of C, H, and O during planet assembly in the terrestrial and giant-planet forming regions, trace molecular and low-ionization metals from disk winds, and determine the absolute abundance patterns in the disk as a function of age. This survey will reveal how the changing disk environment affects the size, location, and composition of planets that form around other stars.

As described in the DRM for Signature Science Case #5 in **Appendix B**, with a total of 41 pointings to cover the Orion complex the LUVOIR-A program time with overheads would be 216 hours or 580 hours with LUVOIR-B. This would obtain complete FUV and NUV spectra of protostars in the Orion complex, enabling the detection of the most abundant molecular species in disks, e.g., $H_2$, CO, $H_2O$, and OH.

### 4.2.2 Young exoplanet systems

Young exoplanet systems offer valuable windows into the later stages of planet formation. Building rocky worlds and the final sculpting of planetary systems involve dynamic and sometimes violent processes. The young comets and asteroids that are the building blocks of larger planets are colliding and fragmenting, exposing dust and gas from their interiors. The orbital distribution of dusty debris from those collisions will tell us about the timescales for rocky planet formation and the properties of young planets. LUVOIR's small inner working angle will be needed to probe within the ice lines of young planetary systems, which in theory divide inner rocky planet formation regions from the cold icy regions where giant planets form.

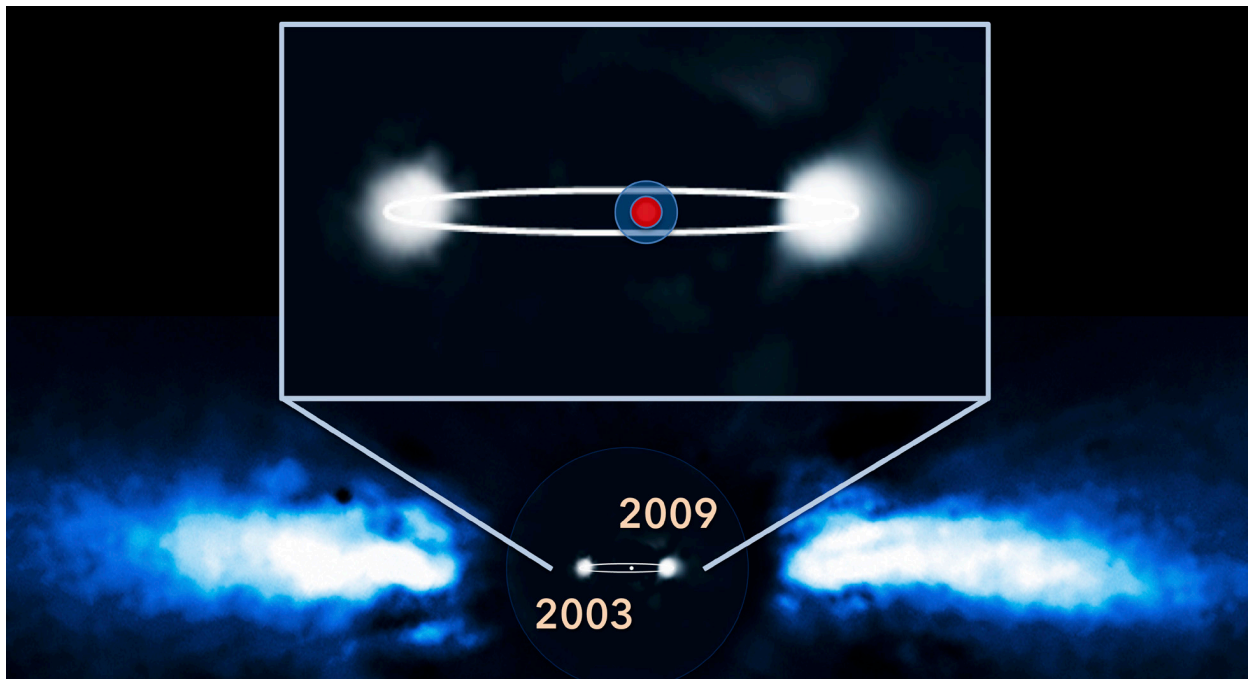

**Figure 4-11.** *LUVOIR can peer into the unseen inner regions of young planetary systems. This image shows the famous Beta Pictoris debris disk, with the gas giant exoplanet Beta Pic b imaged at two epochs in its orbit (2003 and 2009). The semi-major axis of the planet's orbit is ~ 9 AU (Lagrange et al. 2010). This planet is likely responsible for the long known "warp" in the inner portion of the dust disk (e.g., Heap et al. 2000). The inset panel shows a blow-up of the innermost region with the coronagraph inner working angles at 550 nm for LUVOIR-A (red circle) and LUVOIR-B (blue circle) overlaid. LUVOIR can search for warm dust and additional planets in the currently unseen inner region of the disk. Credit: ESA / A.-M. Lagrange*

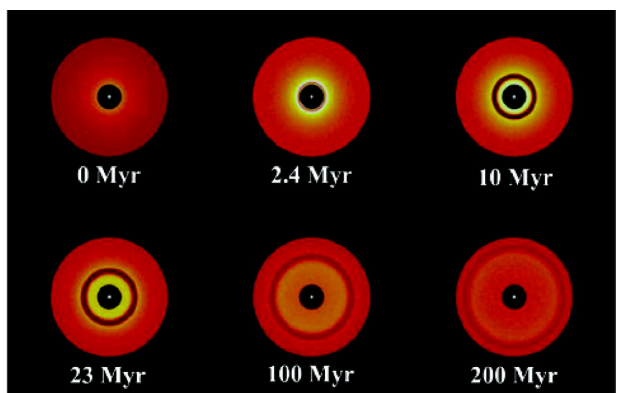

**Figure 4-12.** *Signposts of terrestrial planet formation—dust rings moving outwards with time. Through observations such as these simulations, LUVOIR will identify the epochs of terrestrial planet formation. Credit: Kenyon & Bromley (2004)*

### 4.2.2.1 Young debris belts: morphology and architecture

Young debris disks are a powerful probe of the formation and early dynamical evolution of planetary systems. On a massive scale, countless numbers of extrasolar planetesimals—the building blocks of both terrestrial and giant planets—are colliding with each other and producing bright disks of dust and gas debris that can be observed in detail (**Figure 4-11**). Since the ages of most debris disks span a plausible range of timescales for terrestrial planet formation (~10 to hundreds of Myr), the process can be observed in action, constraining when and where it occurs

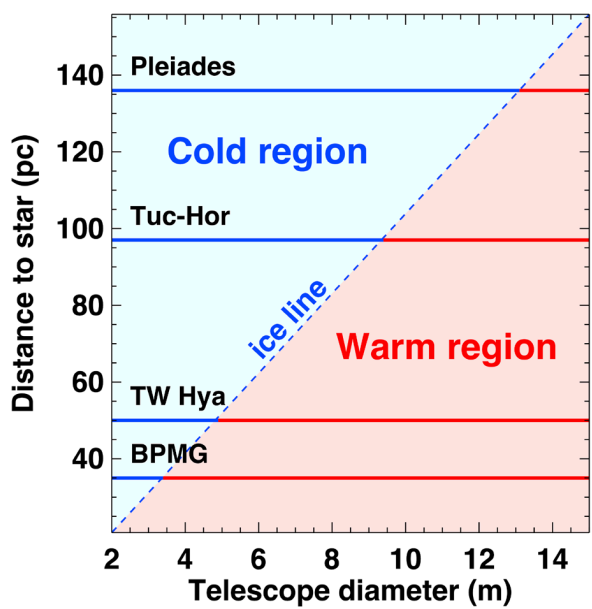

**Figure 4-13.** *LUVOIR can image the terrestrial planet forming regions of young exoplanet systems. The plot shows the distance out to which an ECLIPS-like coronagraph can observe the warm region inside the ice line of a Sun-like star (~3 AU) at 400 nm, as a function of telescope diameter. In the blue region, the ice line is inside the IWA, so only cold material can be seen. In the pink region, warm material inside the ice line can be observed as well as cold material. The mean distances to a few young (≤ 100 Myr) stellar associations are indicated with horizontal lines (BPMG = Beta Pic Moving Group, TW Hya = TW Hydrae Association, Tuc-Hor = Tucana-Horologium Association, Pleiades = Pleiades Open Cluster). Credit: A. Roberge (NASA GSFC)*

and revealing how it varies from system to system as a function of stellar properties.

Space- and ground-based instruments, including HST, ALMA, GPI, and SPHERE, have imaged debris disks containing planet-sculpted dust structures that help constrain the orbital properties and dynamical history of giant planets in young systems (e.g., HR8799; Booth et al. 2016). However, current instruments lack the spatial resolution, sensitivity, and inner working angle to image warm material in the innermost regions of debris disks. LUVOIR however can detect formation signatures as a function of separation from the host star and stellar age, mass, and metallicity.

***Dust from sites of ongoing terrestrial planet formation.*** As terrestrial planets grow in a disk of planetesimals, they dynamically stir nearby bodies, generating collisional cascades that produce copious dust concentrated in rings. The dust is subsequently removed largely by radiation pressure. The dynamical simulation in **Figure 4-12** predicts that the region of active formation—marked by bright dust rings—moves outward with time. The growth timescale is expected to depend on radial distance from the star and the local solid disk mass, which may be correlated with the host star mass and metallicity.

Planet formation theories can be tested with high-resolution, high-contrast images of young debris disks with a range of ages. Current high contrast instruments, with their large inner working angles and relatively low contrast ($> 10^{-6}$), can only obtain such images for the outer regions of bright disks around early-type stars. Millimeter and submillimeter facilities like ALMA are insensitive to warm dust in inner regions of disks. **Figure 4-13** shows that ECLIPS can obtain high-resolution images reaching within the ice lines (~3 AU for the solar system; e.g., Martin & Livio 2012) of nearby young disks within about 135 pc. LUVOIR also has the sensitivity to extend such imaging studies to fainter disks around lower mass stars than are currently possible.

***Gravitationally sculpted features associated with terrestrial planets.*** Dust structures produced by dynamical interactions between planetesimals and planets can provide constraints on young planets' masses and orbits. The Beta Pic dust disk displays an example of a planet-induced warp (**Figure 4-11**). Other types of structures include gaps cleared of planetesimals by resonances with a planet (**Figure 4-12**) and dust clumps marking planetesimals captured into resonance during a planet's migration (e.g., as invoked for Beta Pic; Dent et al. 2014). By comparing properties of planets within younger protoplanetary disks to those inferred within older debris disks, the evolution of planetary systems, planet migration, growth of planets through giant impacts after the dissipation of the primordial gas disk, and planet-planet gravitational interactions can be traced.

As described in detail in Signature Science Case #5 in **Appendix B**, LUVOIR-A ECLIPS could search for zodiacal debris disks in around 25 stars about 75 hours of observing time. For a subset of targets NUV coronagraphy will also be performed for about 6 days of total time. For LUVOIR B the comparable program would consume about 22 days.

### 4.2.3 Mature exoplanet systems

The history of a planetary system is encoded in its architecture, the system-wide configuration of planets and asteroid/comet belts. Architectural information includes the number, masses, spacing, and orbits of the planets; as well as the structure, mass, and composition of any accompanying debris from asteroids and comets. Our own solar system's architecture—including the orbits and compositions of our asteroid and Kuiper belts (e.g., DeMeo and Carry 2014), the spacing and eccentricities of the terrestrial planets (e.g., Chambers & Wetherill 1998), and the orbits of the giant planets (e.g., Raymond et al. 2004)—informs us about the formation, early environment, and habitability of the Earth. Similarly, exoplanetary system architectures provide context for exoEarths; for example, an Earth-size exoplanet in a system with a nearby, highly eccentric Jupiter would have experienced a very different environment from one tightly packed together with seven other rocky planets. By characterizing system architectures for multiple planetary systems, LUVOIR will open a window to understanding how planetary formation proceeds under a diversity of environments.

#### 4.2.3.1 Planet masses, orbits, & radii

Masses and radii constrain surface gravity and are needed to interpret spectra of planetary atmospheres. Today most planet masses are measured via the radial velocity technique, with a significant minority determined from transit timing variations. Neither of these methods

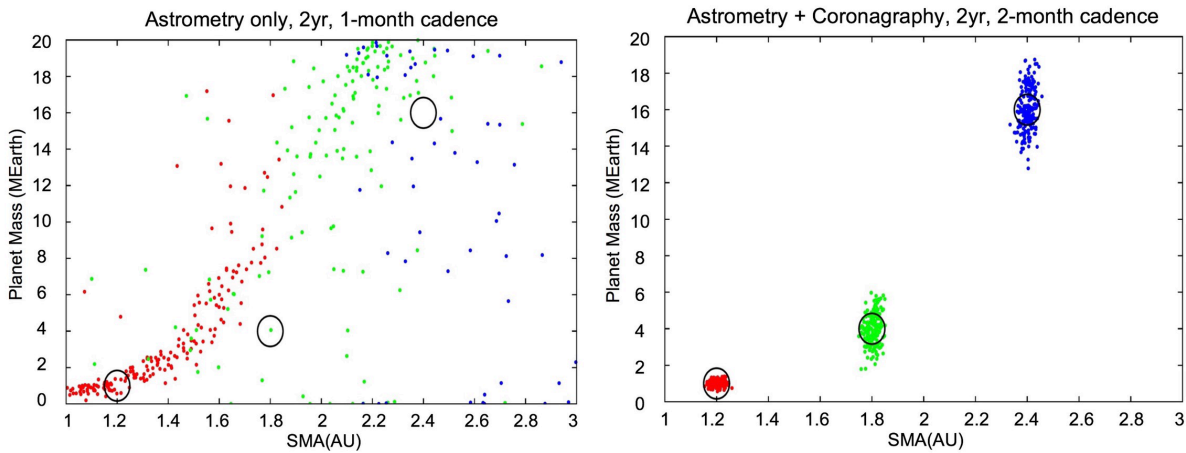

**Figure 4-14.** *The panels illustrate the efficacy of determining exoplanet mass and orbital semi-major axis (SMA) by astrometry alone (left) and astrometry combined with coronagraphy, with half the number of observations over two years (right). The black circles represent three planets for which observations were simulated. Colored points illustrate random posterior draws from Markov Chain Monte Carlo analysis of simulated data. The combination of coronagraphy and astrometry is extremely powerful in eliminating degeneracies and finding the correct orbital solutions. Figure credit: E. Bendek after Guyon et al. (2012)*

can currently provide mass measurements for true analogs of our Earth around solar-type stars. Therefore, the capability of measuring exoplanet masses with extremely precise astrometry (≤1 μas) is a key attribute of HDI. Such astrometric measurements will yield reliable orbits and masses for planets with $M ≥ 1 M_{Earth}$ orbiting ≥1 AU from G2V stars within 25 parsecs (the mass measurement limit is smaller for lower mass stars).

LUVOIR's immense light-gathering power, short integration times, large instantaneous field of regard, and rapid repointing will enable deep imaging of each exoplanet target star many times, ensuring both an accurate orbital solution as well as a high chance of planet discovery. The combination of direct imaging using ECLIPS and precision astrometry with HDI will robustly disentangle multi-planet systems much more efficiently than by astrometry alone (**Figure 4-14**; Guyon et al. 2012). The astrometry program in **Appendix B.6.4** details the capabilities and requirements for measuring planet masses with LUVOIR. While the prospects for obtaining useful planet masses with LUVOIR are very bright, more study is needed to fully understand the observatory's astrometric capabilities.

Since a planet's brightness in reflected light is a function of the planet radius, albedo, and orbital phase, determining the radius is difficult. Repeat imaging in multiple filters at different orbital phases will be key to disentangling such degeneracies (Nayak et al. 2016).

### 4.2.3.2  Orbits, compositions, and formation

Linking information on planet orbits, compositions, and dynamical history of mature exoplanet systems can provide deep insights into their past formation and evolution. For example, giant planets can affect volatile delivery to the inner solar system, leading to potential differences in composition between terrestrial planets in systems with and without giant planets. Atmospheric composition of giant planets also provides clues to their formation and evolution. Gas giant planets formed by the core accretion mechanism are expected to have atmospheric heavy element abundances in excess of those of their parent stars. Available

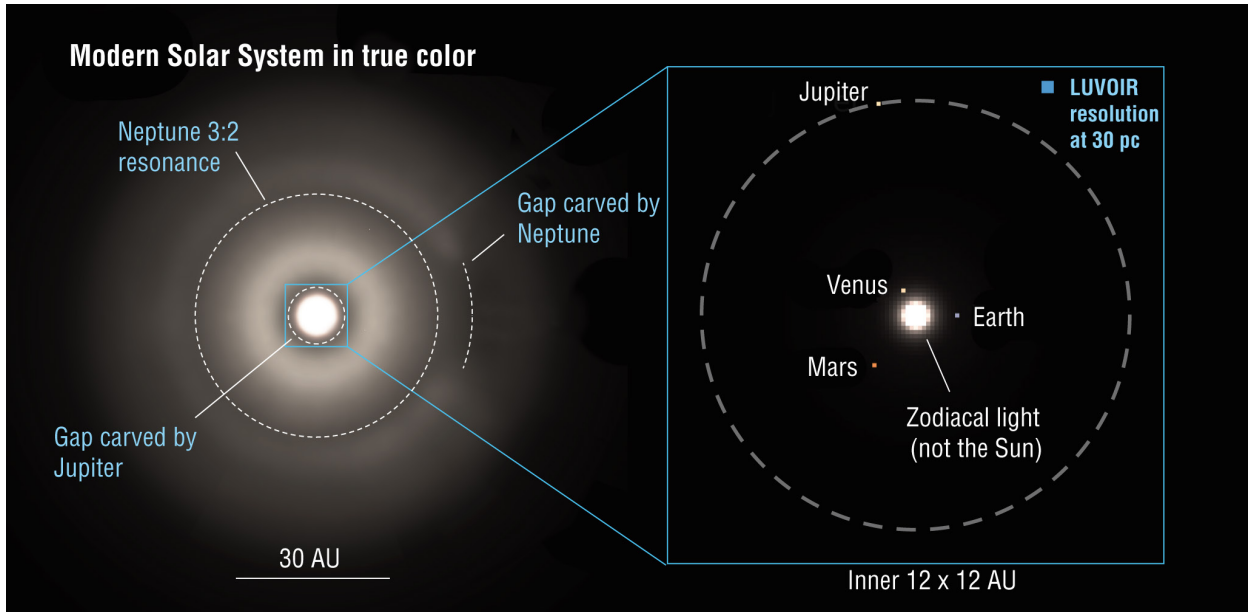

**Figure 4-15.** *Architecture of the modern solar system, showing the interplay between planets and debris dust. The left panel shows a model of the entire system out to a radius of 50 AU from the Sun, while the inset panel zooms in on the inner system. For ease of viewing, the Sun and possible astrophysical background sources are not included. The bright region at the center of the image is emission from warm debris dust (aka. exozodiacal dust). Two circular gaps in the dust are visible, the inner one caused by Jupiter and the outer one marking the 3:2 mean motion resonance with Neptune. Neptune also creates a partial gap in its immediate vicinity. Credit: Roberge et al. (2017)*

data on hot transiting planets already point to a trend of atmospheric heavy element enrichment with mass (**Figure 4-2**). LUVOIR will extend such comparisons to cooler, more solar system-like giant planets, and uncover systematic deviations from the trend that might indicate other formation mechanisms (like gravitational collapse) for subsets of planets.

Furthermore, differences in atmospheric bulk composition (recorded as C/O ratios) for planets at similar orbital distances may indicate migration has moved some of them from their birthplaces to the locations we see today. Planet migration can also establish planets in orbital resonances and alter their eccentricities. Thus, measurement of orbits, masses, and atmospheric composition for all the planets in any system together permit a much richer understanding than any of these measures alone. Integral field spectroscopy at R=70 will be able to measure volatile inventories for many gas and ice giant planets, enabling comparisons of C/O ratios over a wide range of orbital separations and identifying relationships between composition, mass, and orbit.

#### 4.2.3.3  Mature debris belts

Even in old planetary systems, asteroids, comets, and Kuiper Belt objects (KBOs) left over from the planet formation phase persist. Erosion of these planetesimals produces interplanetary dust, which can actually be the most easily observed feature of a planetary system, due to the large total surface area of the dust grains. Dust in the warm inner region of the solar system, which largely comes from comets (Nesvorný et al. 2010), is known as zodiacal dust. The spatial distribution of interplanetary dust in the solar system—with the signs of planets imprinted in it—is shown in **Figure 4-15**.

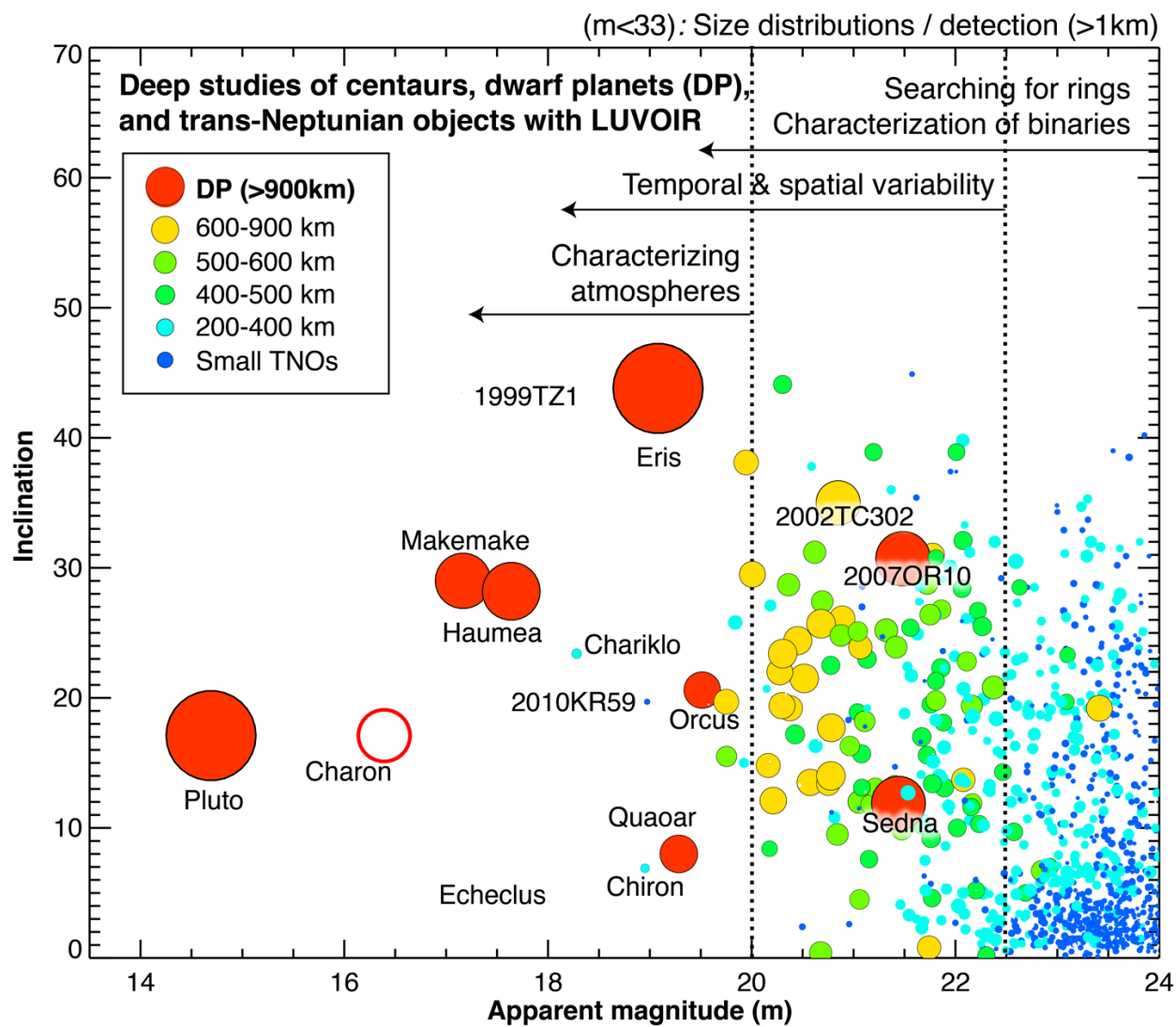

**Figure 4-16.** *Brightness of Centaurs and trans-Neptunian objects compared to the capabilities of a LUVOIR-class observatory with sensitive spectrometers in the UV/optical/NIR spectral range. LUVOIR will permit a unique characterization of the deepest regions (40–50 AU) of our solar system (objects with m<33), also permitting unprecedented studies of their atmospheres (m<20), surface composition / variability (m<22) and their origin and evolution via studies of their rings and binary configuration (m<24). Credit G. Villanueva (NASA GSFC)*

Although sufficiently high levels of exozodiacal dust can obscure exoplanets from view, planetesimal belts and the dusty debris they produce serve as records of the system's early history and provide constraints on present day orbital properties. A belt located near a planet yields constraints on its mass and orbital evolution. For example, a sufficiently massive, nearby, and/or elliptical planet would disturb the belt, which provides information on how the planet's orbit has (or has not) changed over time.

Gravitational interactions between a planet and planetesimals can trap some of the latter into dynamical resonances, which can then be revealed through clumps in the interplanetary dust. Such trapping can also serve as a record of the planet's migration history (e.g., Malhotra 1993). High-resolution, high-contrast imaging of exoplanets will simultaneously reveal interplanetary dust, providing dynamical and contextual information without additional observational demands.

As described in detail in **Section B.6.4** in **Appendix B**, the astrometry program with HDI on LUVOIR-A would consume about 10 days, with overheads, to measure the masses of habitable planets in 53 systems to ~25% precision. LUVOIR B would require about 13 days for a similar program to measure masses of planets in 29 systems.

## 4.3 Signature Science Case #6: Small bodies in the solar system

The vast collection of icy bodies in the outer solar system was witness to the primordial processes that led to the formation and final arrangement of the giant planets. This population is extremely diverse; ranging from geologically active, atmosphere-bearing dwarf planets to small undifferentiated planetesimals that are structurally reflective of primordial formation processes. Studies of the compositions, size distribution, binarity, and orbits of these objects—and links among these properties—offer our best window into the solar system's epoch of formation. Such studies will constrain the structure and composition of the solar system's protoplanetary disk, the mechanisms by which planetary building blocks grow from dust, and the dynamical history of the giant planets. Furthermore, the detailed properties of these objects are broadly applicable to understanding planetary system development around other stars.

However, the major factor limiting study of these objects is their faintness (**Figure 4-16**). They are only visible due to reflected sunlight, which experiences inverse square dilution in both the outbound and reflected directions, resulting in a $r^{-4}$ brightness function with heliocentric distance. LUVOIR would break new ground in mapping our solar system's icy reservoir at smaller sizes and greater distances, measuring the physical properties of these bodies to test planet formation theories, and characterizing the miniature worlds we call dwarf planets.

### 4.3.1 Finding the smallest planetary building blocks

It is only over the last 25 years that we have begun to map the distribution of trans-Neptunian Objects (TNOs) in the Kuiper-Edgeworth Belt (KEB), the inner of the solar system's two icy body reservoirs (Jewitt and Luu 1993). Over 2700 TNOs have been directly observed, ranging in size from ~25 km (e.g., MU69 or Ultima Thule; **Figure 4-17**) to >2300 km (e.g., Pluto, Eris). They are grouped according to their orbits into dynamical populations, which exhibit distinct physical properties and are believed to have different origins.

The collection of known TNOs will continue to grow through the efforts of current

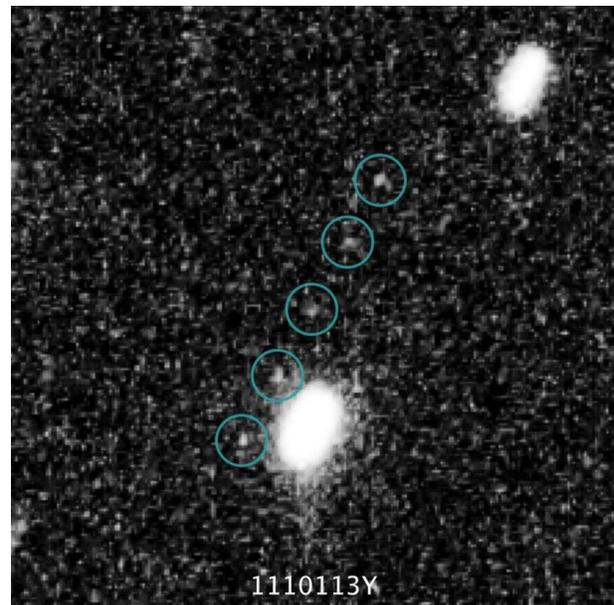

**Figure 4-17.** *The Hubble discovery images of MU69 (Ultima Thule), later visited by the New Horizons mission. This population of distant objects can be studied with LUVOIR. Credit: NASA / STScI / JHUAPL / SwRI*

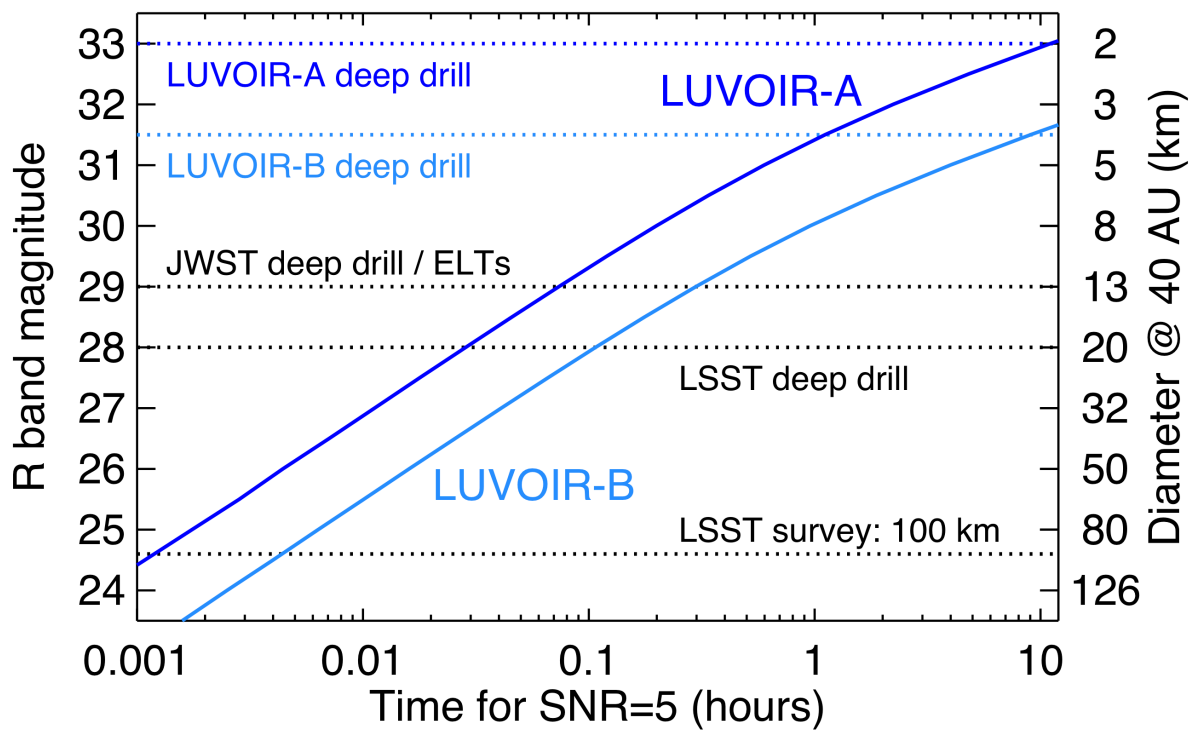

**Figure 4-18.** *LUVOIR will be sensitive to smaller trans-Neptunian objects (TNOs) than any other current or planned facility. The solid curves show the R band apparent magnitude of TNOs detectable at SNR ~ 5 as a function of exposure time for LUVOIR-A (dark blue) and LUVOIR-B (light blue). The right y-axis shows the sizes of TNOs at 40 AU corresponding to the R magnitudes. These assume a single image that is depth-optimized to the Classical Kuiper-Edgeworth Belt by non-sidereal tracking. Horizontal dotted lines show limits for other TNO surveys. Credit: R. Dawson (PSU) / A. Roberge (NASA GSFC)*

surveys like OSSOS (Bannister et al. 2016) and Pan-STARRS; and future observations with LSST, JWST, and ground-based ELTs. However, to address key unanswered questions about how planets form from disks of gas and dust, we need to explore down to smaller sizes and greater distances using a large space telescope with a wide field of view imaging instrument. Furthermore, we need sufficient tracking and revisits of these small, distant objects to characterize their orbits and place them in the context of their dynamical population.

***How do planetary building blocks grow from dust?*** To test competing theories of how the planetary building blocks known as planetesimals form and fracture, we need to probe the size distribution of small bodies down to few kilometers in size. A 2-km TNO at 40 AU corresponds to an apparent magnitude of 33, which no current or planned telescopes can reach (**Figure 4-18**). Singer et al. (2019) recently inferred a surprising deficit of 1 km TNOs from cratering on Pluto. This deficit seems inconsistent with a population sculpted by collisional equilibrium and may instead reflect the initial size distribution. Measuring this population would distinguish among different planetesimal formation theories. LUVOIR would directly probe this size range, pushing two orders of magnitude smaller than LSST's wide field survey and an order of magnitude smaller than proposed or potential deep drills with JWST or ground-based ELTs.

Furthermore, Pluto's impactors likely come from a mixture of Kuiper belt dynamical populations, which are believed to have distinct formation and dynamical histories and are

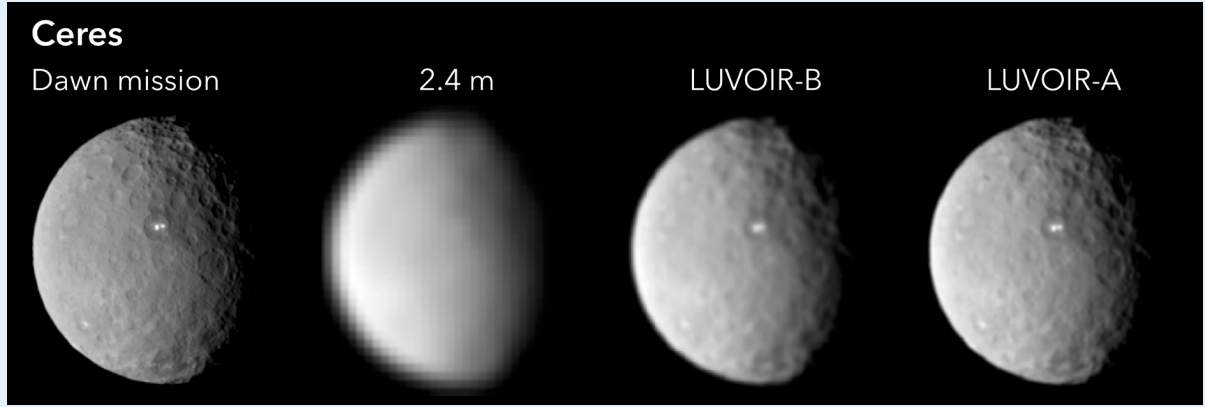

**Solar System Small Body Interiors**

LUVOIR's powerful imaging coupled with resolved spectroscopy will address critical knowledge gaps regarding the evolution of protoplanets, dwarf planets, icy satellites, and volatile sources for habitable planets. Together, these capabilities will provide fly-by quality science for most bodies in the asteroid and Kuiper belts. High-resolution imaging across the solar system will provide binary orbits and shape models that can reveal the density and internal structure of small bodies. Resolved UV/optical/NIR spectroscopy transforms albedo features in photometric data into geologic terrains and enables detection of surface composition changes.

| | Best spatial resolution at 500 nm | |
|---|---|---|
| | **LUVOIR-A** | **LUVOIR-B** |
| **Venus, Ceres, Jupiter, Saturn** | 7 km, 11 km, 25 km, 51 km | 14 km, 20 km, 47 km, 96 km |
| **Uranus, Neptune, 40 AU** | 108 km, 173 km, 232 km | 204 km, 327 km, 438 km |

known to have different size distributions. LUVOIR can distinguish these populations by tracking and revisiting small TNOs to characterize their orbits and identify to which population they belong. Additionally, directly measuring the size frequency distribution in the collisional range would also allow us to better test of our assumptions about observations of debris disks, where we probe the size distributions at much smaller mm/sub-mm sizes and extrapolate to estimate total disk masses.

The LUVOIR Study Team has designed TNO deep drill programs to reach a limiting detectable magnitude of R=33 (~ 2 km at 40 AU) with LUVOIR-A and R=31.5 (~ 4 km at 40 AU) with LUVOIR-B (details in **Appendix B.7.2**). R–J colors will be obtained for all detected bodies, and revisits performed to characterize their orbits and identify which population they belong to. These ambitious projects, which require 125 days over 5 years for LUVOIR-A and 146 days with LUVOIR-B, leverage the unique capabilities of LUVOIR. They are intended as proof-of-concept programs to establish feasibility, but no doubt will benefit from future optimization. Given an extrapolated object frequency of 10,000 per sq. degree at R=33 and 1800 per sq. degree at R=31.5, we expect to detect 117 objects in the LUVOIR-A deep drill program and 30 objects in the LUVOIR-B program (Fraser et al. 2014). Future analysis will establish what constraints may be placed on the small-end size distribution with these total samples.

***Photometric colors as tracers of TNO composition and formation history.*** Differences in photometric colors among various dynamical populations provide evidence for their distinct formation histories and the sculpting of their orbits during the early evolution of the giant planets. Moreover, assessing changes in colors with radial distance can help us understand the radial composition of the solar nebula and compare to inferences for young disks orbiting other stars. Differences in colors between small TNOs that are collisional fragments and larger, intact TNOs would aid our understanding of their material properties. LSST will provide colors of the TNOs it discovers in the main survey (~100 km size objects) but will most likely conduct the deep drill fields for faint TNOs in a single photometric bandpass only. The ability of the HDI on LUVOIR to obtain simultaneous images of the same field-of-view in both an optical and a NIR band means that the TNO deep drill programs described above will provide colors for all detected objects.

***Comets as tracers of volatiles in planet building blocks.*** Comets are the primary means through which we obtain volatile composition information from icy planetesimals. However, due to their faintness, we have previously only been able to study the compositions of comets that are very active and close to the sun. For distances beyond the orbit of Jupiter, the composition of a cometary coma has been partially sampled for only a few objects. LUVOIR will allow us to study low-activity, volatile-rich comets, inventory species that sublime further from the Sun, and explore the diversity of comet compositions. The broad spectral coverage of LUVOIR will combine observations of atomic species (O, C, S, H, and D) in the UV with the classical visible band fragment molecules (OH, CN, $C_2$, etc.) and their near-IR parents between 1–2.5 μm (e.g., $H_2O$, $CO_2$). Knowing the volatile composition of planetary building blocks is essential for understanding the composition of our solar system's nebula and how planet formation establishes the atmosphere and interior compositions of full-grown planets. While the LUVOIR Study Team has not included time for comet observations within our Signature Science Case #6 programs, a contributed supplemental science case in **Appendix A.19** provides additional information on observing active comets and asteroids with LUVOIR.

### 4.3.2 Measuring the orbits of TNO binaries

The resolving power and point-source sensitivity of LUVOIR make it well-suited to find and characterize binary TNOs by providing the sensitivity to detect the smaller component for more distant binaries than the ~50 LSST is expected to discover. The occurrence rates and properties of binary TNOs provide several critical constraints on how they form and evolve.

First, wide binaries are evidence of *in situ* formation, because they would be disrupted if they formed closer to the Sun and were scattered during the giant planets' orbital evolution. As we survey to smaller objects and wider separations, the presence or absence of binaries can trace the formation history of TNOs in these new frontiers.

Second, competing models for forming these planetary building blocks make different predictions for the component sizes, mass ratios, separations, and occurrence rates of binaries. In particular, characterizing the binary fraction of the ~2-4 km objects discovered by LUVOIR's TNO deep drill will enable us to test whether these tiny bodies are primordial—like 25 km Ultima Thule—or collisional fragments of larger bodies. The TNO population that likely formed *in situ* is observed to have a binary fraction consistent with 100% (e.g., Fraser et al. 2017), and the location of the size cut-off between these primordial bodies and smaller collisional fragments test our theories of how planetesimals form and collisionally evolve (e.g., Pan & Sari 2005, Singer et al. 2019). Finally, binary orbits allow us to measure masses and place constraints on densities, providing key information for understanding their composition and formation. Given the high scientific value of characterizing TNO binaries, this appears to be another key area for LUVOIR.

Larger and brighter TNO binaries will likely be characterized with the upcoming ELTs. Any smaller ones discovered in the LUVOIR TNO deep drill programs described above will remain beyond the reach of the ELTs. It is difficult to plan observing programs to characterize those LUVOIR-discovered binaries, as their frequency in the small TNO population is unknown. However, with component separations up to tens of thousands of km, some may be spatially resolved in the TNO deep drill images (the R band spatial resolution at 40 AU is 362 km for LUVOIR-A and 678 km for LUVOIR-B). The orbital periods will be on the order of a few to tens of days; thus the orbits of some binaries may be measured or constrained in high cadence revisits that will occur for the deep drill fields. For any other binaries that are discovered but require further observations to measure their orbits, we will plan targeted revisits to obtain additional R and J band images.

We chose to set a limit on the total time allocation for the whole notional solar system small bodies program (Signature Science Case #6) of 150 days for both LUVOIR-A and -B. Subtracting 125 days for the LUVOIR-A TNO deep drill leaves a remainder of 25 days that can be devoted to follow-up of small binary systems; thus 36 additional revisits with a limiting magnitude of R=33 are possible with LUVOIR-A. With LUVOIR-B, only 4 days remain, permitting only 7 additional revisits with a limiting magnitude of R=31.5. This appears insufficient to follow-up the faintest 4-km bodies; however, even a small increase in size results in a large decrease in needed exposure time. Further investigation of the expected frequency and characteristics of small TNO binaries is required to determine whether this follow-up is feasible within the notional LUVOIR-B time allocation. **Appendix B.7.3** provides a few additional details on the LUVOIR Study Team's program for TNO binary characterization. A contributed program appears in **Appendix A.19**.

### 4.3.3 Dwarf planets as miniature worlds

LUVOIR would provide the best characterization of distant dwarf planets short of *in situ* exploration by spacecraft like New Horizons, enabling investigation of spatial and temporal surface variations, including seasonal variations and sublimation/condensation of atmospheres. Even before New Horizons, HST images revealed strong variations in albedo over Pluto's surface (e.g., Buie et al 2010)—a preview of New Horizons' discovery of a young surface, dynamic atmosphere, and evidence for a possible sub-surface ocean (**Figure 4-19**). LUVOIR can explore far more remote (e.g., Eris) and smaller (e.g., Sedna, Makemake) dwarf planets than HST. Moreover, as searches for more distant planets like Planet 9 continue from the ground, the most fascinating targets for LUVOIR may be solar system objects yet to be discovered. While a dwarf planet characterization program is not included within Signature Science Case #6, we expect that such observations would be highly compelling guest observer programs, requiring modest amounts of observing time for these relatively bright objects (**Figure 4-16**).

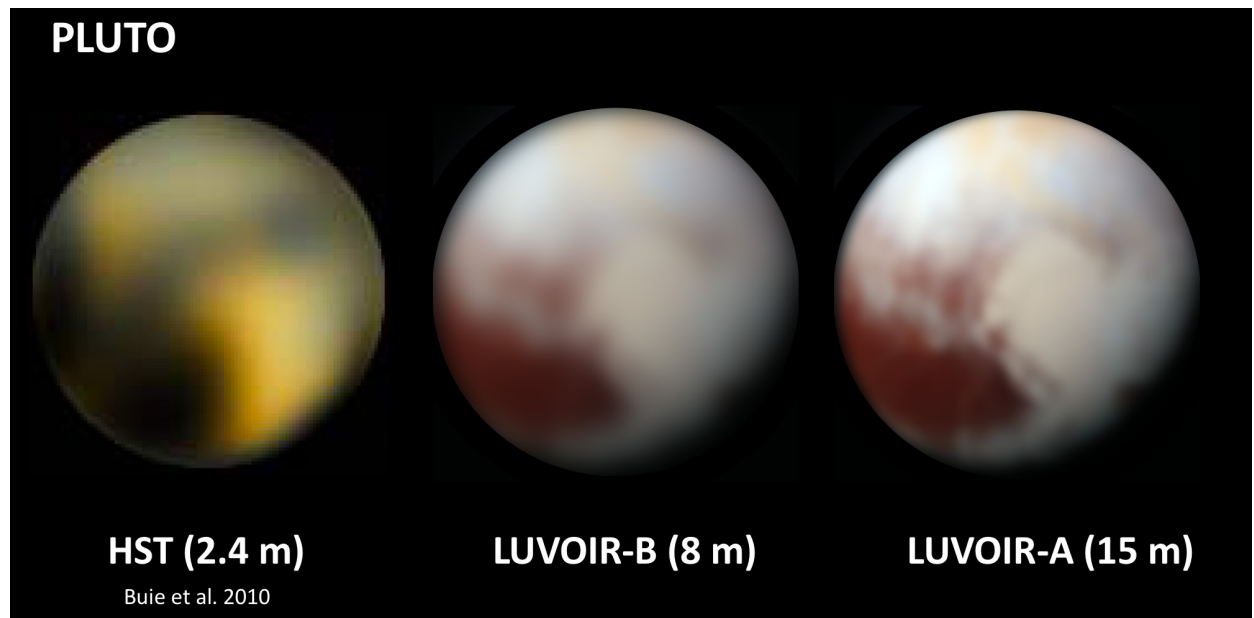

**Figure 4-19.** *LUVOIR will reveal currently unseen details of the solar system's dwarf planets.* **Left**: *The derived surface map of Pluto from Buie et al. (2010), based on high resolution imaging with HST in the range 280–550 nm. The HST image shows the 150° longitude viewing angle. The smallest resolved feature in the HST image corresponds to a projected scale of just under 570 km, assuming a diameter of 2377 km for Pluto.* **Center and Right**: *Simulated LUVOIR images of Pluto created assuming similar observing techniques as described in Buie et al. (2010). A New Horizons full-resolution color image (with a very similar longitude viewing angle as the HST image) was smoothed using Gaussians with FWHMs corresponding to projected spatial scales of 90 km for LUVOIR-A and 169 km for LUVOIR-B. These projected scales correspond to the FWHMs of the diffraction limited PSFs for LUVOIR-A and LUVOIR-B at a wavelength of 300 nm and an assumed Pluto distance of 4.5 billion km. Credit: M. Postman (STScI)*

## 4.4 Summary

Before the Kepler mission, we knew of only a few hundred exoplanets, most discovered by the RV method and thus lacking information on planetary radii. Kepler added thousands of confirmed and candidate planets to the list of known exoplanets, but perhaps the greatest contribution to exoplanet science was the discovery of exceptional diversity among planets. Super-Earth and mini-Neptune sized planets, not represented by any solar system objects, turned out to be exceptionally abundant. This diversity of planet types broadened our vision and led to a revolution in efforts to understand planet formation and evolution.

LUVOIR will enable another new era of comparative exoplanet science. Instead of comparisons of planet sizes and masses, LUVOIR will empower its users to conduct comparative studies of planetary atmospheres, including their composition, chemistry, clouds, dynamics, photochemistry, and escape processes for hundreds of planets. Likewise, by providing a new view into the process and outcomes of planet formation, LUVOIR has the potential to revolutionize our understanding of how planetary systems form and evolve. Equipped with a flexible, powerful observatory designed to characterize exoplanets of all stripes, the next generation of exoplanet and planetary astronomers will achieve a comprehensive view of the diversity of planets, the complexity of their atmospheres, and the architectures of the systems in which they reside.

**Table 4-1.** *Chapter 4 Programs at a Glance*

| Chapter 4 Programs at a Glance | | | |
|---|---|---|---|
| **Goal** | **Program Description** | **Instrument + Mode** | **Key Observation Requirements** |
| **Signature Science Case #4: Comparative Atmospheres** | | | |
| Measure atmospheric composition and haze characteristics for a wide range of exoplanets | NUV/optical/NIR direct spectra of giant through terrestrial planets at multiple separations around variety of stellar types | ECLIPS IFS high-contrast spectroscopy | Bandpass: 250–1000 nm<br>Contrast ≤ 10⁻⁹<br>IWA ≲ 4 λ/D, OWA ≳ 30 λ/D |
| | Transit spectroscopy of hot to warm mini-Neptune to rocky planets around bright stars | HDI grism spectroscopy | Bandpass: 400–2500 nm<br>R ~ 15–500<br>Stellar SNR > 100 |
| Measure atmospheric escape rates for transiting exoplanets | UV transit spectroscopy of hot exoplanets to detect outflows of multiple escaping gas species | LUMOS FUV point-source spectroscopy | Bandpass: 100–200 nm<br>R ~ 30,000<br>Stellar SNR > 20 in single transits |
| **Signature Science Case #5: The Formation of Planetary Systems** | | | |
| Study the evolution of the molecular carriers of C, H, and O in protoplanetary disks, trace disk winds, and provide absolute abundance patterns in disks as a function of age | Measure the main molecules (H₂, CO, H₂O, and OH) and low-ionization elements in Orion protoplanetary disks during planet assembly | LUMOS FUV imaging and multi-object spectroscopy | Bandpass: 100–400 nm<br>R > 30,000<br>Telescope mirror reflectivity > 60% at 105 nm<br>FOV ≳ 4 sq. arcmin |
| Map the spatial distribution of planetesimals during the late stages of planet formation | High resolution imaging of debris disks around solar-type and low-mass stars in nearby young moving groups | ECLIPS high-contrast imaging | Bandpass: 400–600 nm<br>Contrast ≤ 10⁻⁸<br>OWA ≳ 30 λ/D |
| Catalog the architectures of planetary systems around older main sequence stars | Imaging of mature exoplanet systems to determine planet and planetesimal belt locations | ECLIPS high-contrast imaging | Bandpass: 400–600 nm<br>Obtained via Sig. Sci. #1 |
| | Measure orbits and masses for all planets in systems imaged in previous row | HDI astrometry | Final astrometric precision ≲0.1 μas |
| **Signature Science Case #6: Small Bodies in the Solar System** | | | |
| Measure the sizes, colors, and orbits of the smallest outer solar system bodies | Deep multi-epoch imaging to detect 2–4 km diameter TNOs, obtain their R-J colors, and constrain their orbits | HDI broadband imaging | Bandpass: 775 and 1260 nm<br>FOV ≳6 sq. arcmin<br>Non-sidereal drift matched to orbital motion |
| Measure the orbits of the smallest TNO binaries | Follow-up imaging of binaries discovered in the TNO deep dive (previous row) | HDI broadband imaging | Bandpass: 775 nm<br>Revisits |

## CHAPTER 5. WHAT ARE THE BUILDING BLOCKS OF COSMIC STRUCTURE?

Our understanding of how the universe works is sophisticated but incomplete. The total mass-energy density of the cosmos is almost precisely equal to the critical value required for a flat universe. Most of the mass-energy density is in the dark energy component and one quarter of the total is a cold dark matter (CDM) component. There is a complex interplay between the cosmic expansion history, dark matter, radiation, and baryons, which yields many of the observable elements in the universe. Yet we are still working towards ascertaining the true nature of dark energy and dark matter. Furthermore, models remain rudimentary for how galaxies are born at very high redshift and evolve to the present day.

Ground and space-based observatories through the 2030s will significantly improve our understanding of the cosmic "dark sector" and the properties of galaxy building blocks in the early universe. These aspirations comprise some of the key goals of JWST, Euclid, LSST, WFIRST, and the planned 20-m to 40-m ground-based telescopes. Yet these telescopes will not reach into a frontier that we already recognize as critical for a comprehensive understanding of structure formation in the universe. This frontier exists at the scales of the lowest mass galaxies (with stellar masses $\leq 10^6$ $M_\odot$) that we know are there, from the first sparks of galaxy formation at z>10 to the faintest dwarf galaxies in the present day. These small masses and ultra-faint luminosities correspond to physical scales as small as a few kpc and with stellar concentrations as small as 100 parsecs. At these size and mass scales, competing scenarios for the evolution of the dark matter density field, and its associated baryonic structures, make distinguishable predictions that can be tested with observations that reach fainter than AB = 33 mag.

Peering into this ultra-faint regime requires LUVOIR, which (for the 15-m concept) will be capable of resolving 60 parsec scales at all redshifts, while reaching a 5-$\sigma$ point source limiting AB magnitude of 33 (0.23 nJy) in just 10 hours and ~35 mag (0.04 nJy) in ~10 days. Three novel investigations into the building blocks of cosmic structure that require LUVOIR are described in this chapter:

1. **Distinguishing dark matter models by probing scales below 100 kpc:** Between the universe's horizon scale and galactic scales, the structure we measure is consistent with dark matter being entirely non-relativistic and non-interacting. This does not uniquely identify a specific fundamental particle for dark matter, however. Indeed, several candidate particle classes are allowed by existing observations. This has motivated work on dark matter particle candidates that, on astronomical scales, have ensemble behaviors consistent with being "warm" (mildly relativistic) and/or "self-interacting." The interactive or radiative properties of dark matter influences the nature of structure on various scales and epochs in time. LUVOIR can measure the detectable signatures of different dark matter properties on the shapes, central densities, and statistics of structures below 100 kpc in size (corresponding to the scale of self-gravitating halos of a few million solar masses).

2. **Revealing the limits of galaxy assembly at ultra faint fluxes:** Understanding galaxy formation and evolution requires knowing how the gravity from the underlying dark matter structure interplays with baryonic cooling mechanisms, the cosmic ionizing radiation field, and feedback effects from star formation and AGN. The most massive dark matter halos have long cooling times, with stars subsequently forming late. Very low mass halos are not able to retain a steady pool of star-forming baryons due to ionizing radiation. The *source* of the ionizing photons likely comes from a great abundance of star-forming, low mass galaxies. The bulk of the physical processes that dominate the early formation and evolution of cosmic structure all take place on scales well below 1 kpc. This corresponds to miniscule volumes within massive galaxies, or in dwarf galaxies, the smallest galaxies that are able to form and retain stars. LUVOIR will observe these scales across a full range of galaxy properties from the current epoch to the end of the era of reionization owing to its sensitivity down to a few tens of picoJanskys (AB ≈ 35 mag). Such observations will be transformational as they reveal star and galaxy populations key to understanding the origins of structure but not yet seen with any current or planned telescopes.

3. **Documenting the history of ionizing light:** To understand the emergence of structure, we must characterize the influence of ionizing radiation on early galaxies. Directly observing ionizing photons at the epoch of reionization is effectively impossible due to the opacity of the intergalactic medium. LUVOIR will leverage its FUV and UV sensitivity and resolution to directly detect faint ionizing radiation escaping from z<1 galaxies in a *spatially resolved* manner for multiple objects per pointing. Such observations will reveal the environmental factors that favor the escape of radiation, providing crucial clues to determine how light escaped galaxies over the 12 billion year period from the epoch of reionization at z ≥ 7 through to the near present epoch at z ~ 0.1.

With LUVOIR, the astronomy community acquires a unique and unmatched combination of *sensitivity* and *volume* to observe, across most of cosmic time and in many different intergalactic environments, what the universe actually does on spatial scales where the confluence of the matter power spectrum, dark matter physics, and baryonic processes all interplay. LUVOIR will accomplish this by extending our census of dwarf galaxies to much fainter limits in the current epoch and extend that census, at brighter limits, across the last 13 billion years. LUVOIR will survey dwarf galaxies across cosmic time from the extended local volume in the current epoch (out to a luminosity distance of ~30 Mpc, or z=0.008), to the early universe, 13 billion years in the past (at z~7), about half-way through the reionization transition era from a visually opaque to a transparent universe. With this unparalleled sensitivity (yielding images that reach 6 times fainter than the predicted depths of even the planned extremely large ground-based telescopes of the 2020s), LUVOIR will establish the direct connection between the current-epoch matter power spectrum with its early progenitors. LUVOIR can distinguish between competing dark matter models that show significant differentiation on scales below 100 kpc. LUVOIR will also directly observe the impact of reionization on early star formation by measuring the galaxy luminosity function well below limits achievable with current or planned telescopes. Finally, LUVOIR will characterize that ionizing radiation by acquiring the spectra of UV light escaping from galaxies at wavelengths below rest-frame 912 Å, something only a large-aperture UV-sensitive telescope can accomplish.

The Signature Science Cases discussed in this chapter represent some of the most compelling types of observing programs on the building blocks of structure that scientists might

## State of the Field in the 2030s

By the year 2035, HST, Euclid, WFIRST, and JWST will have completed their missions and LSST will have achieved its 10-year survey depth. At least two 20- to 40-meter class ground-based optical-NIR observatories will be operational. SKA should be online and mapping the neutral hydrogen distribution out to high redshifts. LUVOIR will follow on the success of these facilities by bringing unparalleled sensitivity and angular resolution in the ultraviolet and optical regime.

**James Webb Space Telescope (JWST):** Deep NIRCam surveys with JWST will exist down to AB = 31.5 mag (for 0.1″ diameter source with 500 ksec exposure time). This will allow JWST users to discover many new ultra-faint sources, but will still not be deep enough to fully map the assembly of structure from small-scale components at $6 < z < 10$ that are as faint as 32–34 mag.

**LSST, Euclid and WFIRST:** These sky survey telescopes will detect many new faint dwarf galaxies in the local universe, but will not have the sub-parsec spatial resolution to precisely characterize their stellar populations nor have the stability and precision to measure their proper motions. These surveys will provide the targets that can be studied further with LUVOIR.

**Extremely Large Telescopes on the Ground:** These facilities will have similar angular resolution as LUVOIR in the NIR, owing to their planned adaptive optics systems, but will have difficulty detecting sources fainter than AB ~ 31 mag due to the brightness of the night sky. Visible band AO systems, if feasible, will be limited to extremely small fields of view (<10 arcsec). The spectroscopic capabilities of such 20- to 40-meter telescopes will be unrivaled and will be a superb complement to the imaging power of LUVOIR.

**Square Kilometer Array:** The SKA will perform wide-area deep imaging of the neutral hydrogen gas at cosmological distances (z<10) and will provide the knowledge of where to target deep imaging with LUVOIR to study how star formation was suppressed by reionization.

**LUVOIR:** In the context of studying the building blocks of galaxies, LUVOIR's unique capabilities will be its ability to study galaxy components down to stellar mass limits typical of ultra-faint dwarf galaxies at all redshifts. LUVOIR will detect sources with fluxes fainter than AB = 33 mag, where reionization suppression of star formation is predicted to be important. LUVOIR will also extend dwarf galaxy detection out to 25 Mpc in the local universe, directly detect the sources of ionizing radiation in the cosmos, and characterize the energy distribution of that ionizing radiation. LUVOIR's high precision astrometric stability and calibration will enable measurement of transverse velocities of stars within Local Group galaxies, directly measure dwarf galaxy mass density profiles, and map the gravitational fields around nearby galaxies by measuring satellite orbits within their central halos. As such, LUVOIR will constrain the kinematics and the nature of dark matter.

do with LUVOIR at the limits of its performance. As compelling as they are, they should not be taken as a complete specification of LUVOIR's future potential in these areas. We have developed concrete examples to ensure that the nominal design can do this compelling science. We fully expect that the creativity of the community, empowered by the revolutionary capabilities of the observatory, will ask questions, acquire data,and solve problems far beyond those discussed here.

## 5.1 Signature Science Case #7: Connecting the smallest scales across cosmic time

Dwarf galaxies are the smallest dark-matter dominated objects known. The least luminous dwarf galaxies contain just dozens of stars, although their host dark matter halos have masses around $10^7$ M$_\odot$. A couple dozen such faint dwarf galaxies have been discovered orbiting our Milky Way Galaxy. The SDSS Segue-1 system, at a distance of 23 kpc, has an absolute magnitude of M$_V$ = −1.5 mag (Simon et al. 2011). The Virgo-1 dwarf galaxy has M$_V$ = −0.8 mag and is 87 kpc away (Homma et al. 2016). This is only one-third of our Galaxy's virial radius (Walsh et al. 2009)—clearly the survey of the Milky Way's dwarfs is far from complete.

There is evidence that dwarf galaxies, owing to their vast numbers, were primarily responsible for the reionizing the universe and for keeping it ionized. The space density and internal structure of these galaxies can reveal fundamental properties of the dark matter particle, the level and duration of reionization, and the granular limits of the galaxy formation process. To date, limits on the thermal signature of dark matter have been placed by studies of the intergalactic medium (IGM) through Lyman-$\alpha$ forest statistics (e.g., Viel et al. 2013), strong gravitational lensing (e.g., Li et al. 2016), and dwarf-galaxy-scale statistics (e.g., Kim et al. 2017). One of the most constraining measurements places the mass of dark matter particles as greater than 2.9 keV at 95% confidence (Jethwa et al. 2018). Dark matter particles may be significantly colder (more massive) yet be interacting (either self-interacting or with a dark matter-baryon interaction mechanism), or have more nuanced properties that fall outside of the currently popular prescriptions for dark matter candidates (e.g., Buckley & Peter 2018). Each type of dark matter property manifests itself on astronomical scales through a gravitational signature, in the statistical description of dark matter structure. We can, in principle, distinguish between these signatures by measuring the matter power spectrum.

**Figure 5-1** illustrates the matter power spectrum behavior for several dark matter cases compared to CDM. At wavenumbers above k ≈ 10 h/Mpc, the distinctions between these power spectra are significant. The k ≈ 10 h/Mpc scale corresponds to total mass scales in the ~ $10^9$ range. These correspond to classical dwarf galaxies, which have stellar masses of ~ $10^7$ M$_\odot$ and are fainter than M$_V$ = −10 mag. Smaller scales, near k ≈ 100 h/Mpc, reach the regime of the ultra-faint dwarfs with total masses at or below $10^7$ M$_\odot$ and approach the lower mass limit where gravitationally bound halos can host star formation, before the necessary radiation cooling for forming stars becomes too inefficient.

By the early 2020s, the Dark Energy Survey and LSST will have completed the census of dwarf galaxies in the vicinity of the Milky Way Galaxy. Yet each massive galaxy's satellite galaxies provide just one "draw" of the underlying physics that gives rise to this population. For a robust constraint on the small scale power spectrum, the satellite dwarf population associated with other massive galaxies must be surveyed. By the late 2020s, JWST and ELTs will extend the dwarf satellite galaxy search out to several Mpc. Dwarf galaxies are identified through detection of constituent stars, as long as the membership of those stars can be confidently determined, through colors or spectroscopy over an extended field of view. LUVOIR will extend this sensitivity, using horizontal branch stars, to far beyond the Virgo Cluster's 16.5 Mpc distance. This has multiple advantages. First, a sample of massive and

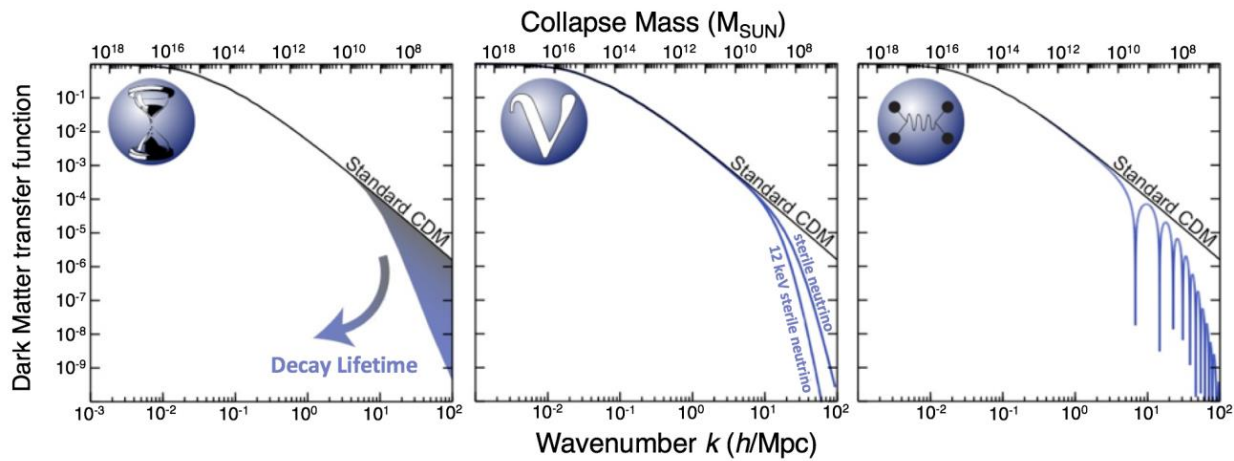

**Figure 5-1.** *The matter power spectrum as a function of spatial scale, as depicted by the "transfer function" for different dark matter particle scenarios at the current epoch (z=0). In each panel, smaller mass and physical scales are to the right. The standard cold dark matter (CDM) model prediction is shown for comparison. Left panel: Decaying dark matter model. Middle panel: A warm dark matter (WDM) model. Right panel: A self-interacting dark matter model (e.g., Buckley & Peter 2018). LUVOIR can probe the matter distribution on the small spatial and mass scales where these models differ across a large range of cosmic time. Credit: L. Moustakas (JPL), R. Massey (Durham)*

Milky Way-analog galaxies provides multiple "draws" similar to our well-studied Galaxy. Second, this much larger volume gives us access to an enormous dynamic range of environments, so that dwarf galaxies can be studied in both volatile and quiescent states.

### 5.1.1 Dwarf galaxies in the local universe: testing dark matter models

Dwarf galaxies are predicted to exist around all massive galaxies and in all environments. By reaching to several tens of Mpc from the Milky Way and well beyond the Local Group, we are identifying dwarf galaxies that are current-epoch analogs to the early galaxies responsible for re-ionizing the universe (e.g., Boylan-Kolchin et al 2015). The potential connections are complex, however. Many of the faintest local dwarf galaxies detected are chemically enriched relative to their early-universe counterparts. The stars in these low-mass neighbors of our Galaxy are created, in part, through inefficient molecular cooling while, at the same time, experience additional quenching effects from an intense ionizing radiation field[1].

To make a major breakthrough in understanding the connection between dwarf galaxies and the mass-scales of their dark matter halos will require the detection and characterization of significant samples of faint and ultra-faint dwarf galaxies in a range of environments over a wide span of cosmic time. The observed galaxies can be connected to their dark matter halo mass properties. This connection is made by combining model descriptions for abundance-matching, star formation suppression from ionizing radiation, and supernova feedback. These halo properties are, in turn, connected with the underlying matter power spectrum that they are ultimately drawn from. LUVOIR will do this both locally and at high redshift.

We calculate the predicted fidelity of the power spectrum measurements from a LUVOIR-based survey of a completeness-corrected sample of dwarf galaxies around several Milky

---

[1] By ionizing radiation, we mean photons with wavelengths shorter than the hydrogen ionization edge at 912 Å

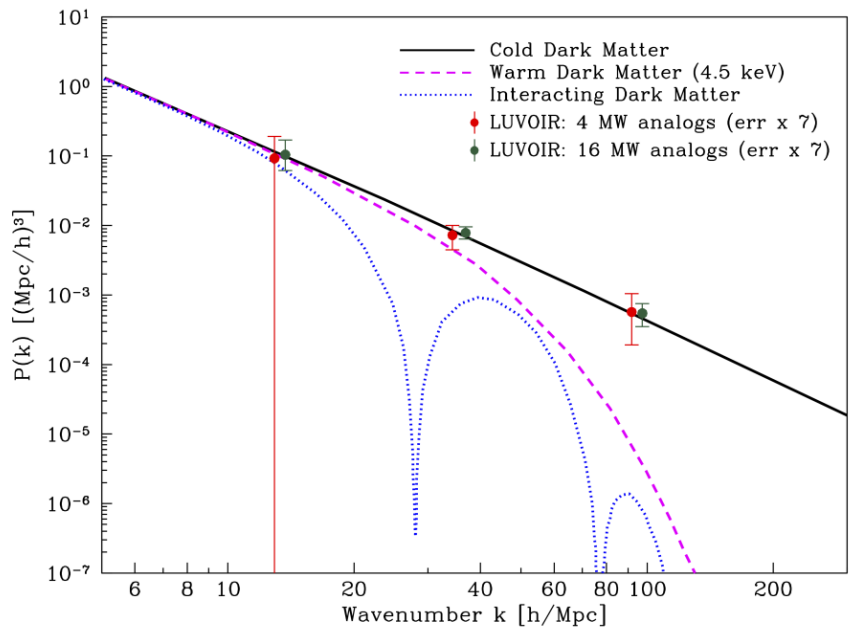

**Figure 5-2.** *Even with a sample of only 4 Milky Way analogs, a LUVOIR-based survey could distinguish between different dark matter models at high significance. The plot shows the matter power spectrum for samples of four and sixteen Milky Way analog systems. The error bars shown are magnified by a factor of 7 to make them visible in this figure. Error bars reflect only the Poisson uncertainties. For illustration, the underlying power spectra for a 4.5 keV warm dark matter model and a self-interacting dark matter model (Vogelsberger et al. 2016) are shown to illustrate the scales where these different scenarios exhibit detectable effects. Credit: L. Moustakas (JPL).*

Way Galaxy-scale systems. The results are shown in **Figure 5-2** for sample sizes of 4 and 16 Milky Way-like satellite systems. The error bars shown in **Figure 5-2** have been magnified by a factor of 7 to make them visible on the vertical scale encompassed by the plot. These calculations adopt an empirical relation between dark matter halo masses and their stellar mass (Bullock & Boylan-Kolchin 2017) that has proven robust, although how reliable this is in the regime of the true ultra-faint galaxies is the focus of ongoing work (e.g., Kim et al. 2017). At the low-mass end of the galaxy distribution, the effects of star formation suppression due to an intergalactic ionizing background flux can be significant. Barber et al (2014) develop a semi-analytic description of this effect (see also Dooley et al. 2017). For a high-level assessment of a LUVOIR-based survey of dwarf galaxies around massive systems, only the most dominant suppression signature of reionization was included.

A LUVOIR program to obtain the needed constraints on the small-scale matter power spectrum would consist of two contiguous orthogonal strips of HDI observations for each target. Each strip extends from the central galaxy out to about half the virial radius (~100 kpc); one strip is along the analog's major axis, and the other is along its minor axis. In addition, one would observe many random positions within the central 100 kpc to ensure a 50% filling fraction of the survey region (see **Figure 5-3** for a survey layout centered on NGC 3810). This observational strategy samples one quarter of the total virial-radius volume.

The dwarf galaxies discovered from these observations are then used to estimate the census of the compete population within the entire virial volume of each central galaxy targeted in the survey. The advantage of LUVOIR for this study is that the combination of its sensitivity and survey efficiency for these strips exceeds those of any current or planned observatory by orders of magnitude. These censuses of luminous dwarfs can then be converted to a mass function, using an abundance-matching framework that incorporates the relevant baryonic physics, and compared to predictions for CDM (e.g., Schneider 2015; Newton et al. 2018).

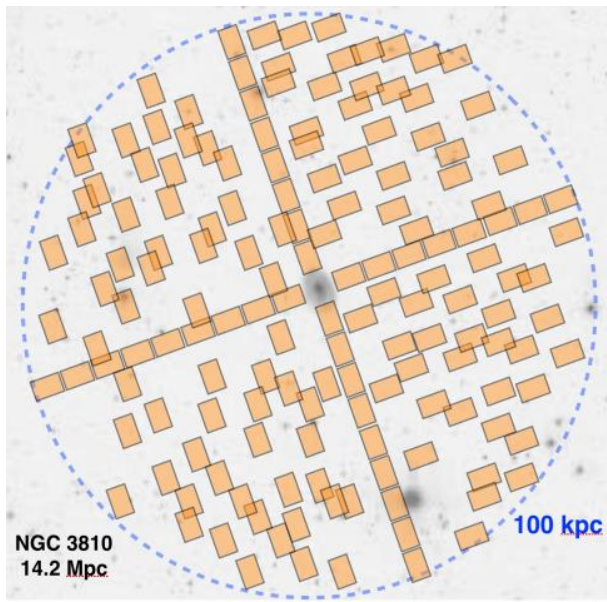

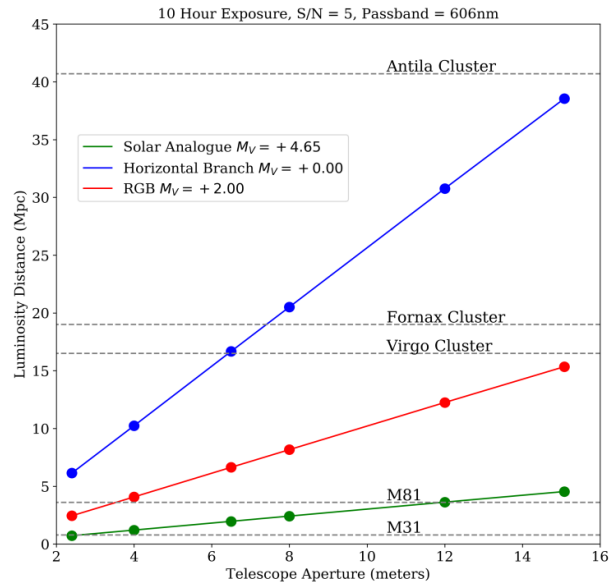

**Figure 5-3.** *The LUVOIR image tiling strategy (orange rectangles) to sample the dwarf galaxy population around the Milky Way analog NGC 3810 at a distance of ~14 Mpc. The 156 HDI pointings cover 50% of a region that extends out to half the virial radius (~100 kpc). Full sampling is performed along the major and minor axes of NGC 3810. Credit: M. Postman (STScI)*

**Figure 5-4.** *The distance to which three representative stellar types can be detected, versus aperture size of LUVOIR+HDI-like platforms. The distance reached scales linearly with telescope diameter. The distances shown here are for a SNR=5 detection with a 10-hour integration in the V-band. In such an observation, observers using LUVOIR-A could detect RGB stars out to 15 Mpc and horizontal branch stars out to 39 Mpc. With a 100-hour integration, LUVOIR-A will have the ability to detect individual RGB stars out to 27 Mpc. Credit: M. Postman (STScI).*

A LUVOIR survey to characterize the dwarf galaxy mass function around multiple Milky Way analogs will distinguish between different dark matter scenarios at >4 sigma significance as **Figure 5-2** demonstrates. A LUVOIR imaging survey around Milky Way analog systems within 15 Mpc will reach down to be to the horizontal branch ($M_V$=0 mag) in the stellar populations of each of the dwarf galaxies detected in these systems, reaching the limits of dwarf galaxy formation. LUMOS could be operated in parallel to map the absorption features from the circumgalactic medium of the central galaxy.

**Figure 5-4** demonstrates the potential of a Nyquist-sampled imager to detect and characterize stellar populations on space telescopes of various apertures. Individual horizontal branch or red giant branch stars (depending on distance) will be detectable within each candidate dwarf galaxy.

The internal structural profiles of dwarf galaxies in the local universe will also be measured as these systems will be extremely well resolved with LUVOIR. For example, the FWHM of the PSF of the LUVOIR-A HDI instrument in the r-band (650 nm) is 9 mas. An individual dwarf galaxy analogous to the Virgo-I dwarf with a half-light radius size of ~20 parsecs would, at the Virgo Cluster's distance, subtend ~0.50 arcsecond and would therefore be sampled with ~3000 distinct resolution elements. LUVOIR will fully characterize the dwarf galaxy's stellar light profile, stellar populations, and internal structure—feats not possible with any other current or upcoming telescopes.

### 5.2  Signature Science Case #8: Constraining dark matter using high precision astrometry

Dwarf spheroidal galaxies (dSph) in the local universe are extraordinary sites to explore the properties of non-baryonic dark matter. First, their mass is dominated by dark matter—they have mass-to-light ratios 10 to 100 times higher than the typical $L*$ galaxy like the Milky Way (Martin et al. 2007; Simon & Geha 2007; Strigari et al. 2008). Second, they are relatively abundant—over 40 dSph galaxies have been found in the Local Group (Simon et al. 2011; Torrealba et al. 2016) and more will be discovered in the LSST era. Third, and perhaps most striking, all well-studied dSph galaxies, covering more than four orders of magnitude in luminosity, inhabit dark matter halos with the *same* mass (~$10^7$ M$_\odot$) within their central 300 pc (Strigari et al. 2008).

The ability of dark matter to cluster in phase space is limited by intrinsic properties such as mass and kinetic temperature. Cold dark matter particles have negligible velocity dispersion and very large central phase-space density, resulting in steep central density profiles. In contrast, dark matter halos with highly relativistic particles (i.e., warm dark matter) have smaller central phase-space densities, so that density profiles saturate to form systems with flat central density profiles (constant central cores). Owing to their small masses, dSphs have the highest average phase space densities of any galaxy type, which implies that for a given dark matter model, phase-space limited cores will occupy a larger fraction of the virial radii. Hence, *the mean density profile of dSph galaxies is a fundamental constraint on the nature of dark matter*.

Current observations are unable to measure the density profile slopes within dSph galaxies because of a strong degeneracy between the inner slope of the dark matter density profile and the velocity anisotropy of the stellar orbits. Radial velocities alone cannot break this degeneracy even if the present samples of radial velocities are increased to several thousand stars (see Strigari et al. 2007). The only robust way to break the degeneracy is to combine proper motions with the radial velocities. The required measurements include proper motions for ~100 stars per galaxy with accuracies better than ~10 km/s (i.e., < 40 µas/yr at 60 kpc) and ~1000 line-of-sight velocities.

Furthermore, the orbital motions of dwarf satellite galaxies within the halo of their more massive central galaxy can reveal important properties of the dark matter distribution of galaxies like our own Milky Way and M31 (e.g., see Massari et al. 2018). While such studies have been done with existing telescopes for our Galaxy and M31, LUVOIR will be able to extend such investigations out to several Mpc and thus expand the range of galaxies whose gravitational potential is mapped in this way. Measuring orbital parameters for dwarf satellite galaxies requires proper motion measurements with similar accuracy in order to measure typical transverse orbital velocities of 10 km/s.

In the case of the brightest of these dSph galaxies such as Fornax and Sculptor, sufficient velocities and proper motions can be obtained using stellar giants. Ground-based large-aperture (~10-m) telescopes could measure the spectra, and Gaia or JWST can measure the proper motions. For the less massive dwarfs, those most dominated by dark matter, main sequence stars will have to be used to obtain the numbers of velocity and proper motion measurements needed. This will require larger (30-m class) ground-based telescopes for the velocities. The 30-m class telescopes may be able to obtain the necessary proper motions but it will be extremely challenging: it will require precisely stitching many fields together, few of which will contain sufficient astrometric reference quasars.

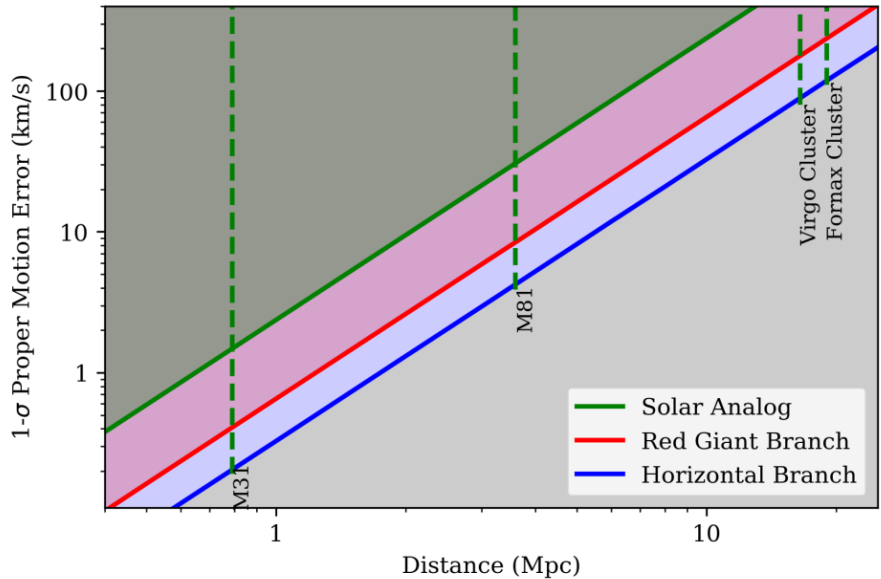

**Figure 5-5.** *The transverse velocity error from proper motion measurements achievable with the LUVOIR-A design as a function of galaxy distance. Green, red, and blue lines show the accuracies expected for solar-analog main sequence stars, red giant branch stars, and horizontal branch stars, respectively, assuming a 5-year baseline and 100 stars per epoch. The velocity errors obtained with LUVOIR-B are about 3.6 times larger. Credit: M. Postman (STScI)*

By contrast, the necessary astrometric precision can be readily achieved by LUVOIR. **Figure 5-5** shows the predicted transverse velocity error as a function of distance for LUVOIR-A. With a baseline of 5 years and 100 stars per galaxy, a transverse velocity error of 10 km/s can be achieved out to 4 Mpc (corresponding to a proper motion error of ~0.5 µas/yr) using RGB stars as tracers and out to 5.7 Mpc (~0.4 µas/yr) using horizontal branch stars. LUVOIR and ELTs working together will provide excellent constraints on the density profiles of satellite dwarf galaxies and the mass distribution around central galaxies in the nearby universe, revealing whether dark matter is kinematically hot or cold, a key characteristic that determines when and where structure forms in the universe.

### 5.3 Signature Science Case #9: Tracing ionizing light over cosmic time

#### 5.3.1 Dwarfs in the distant universe: probing the impact of reionization
The reionization era is when the universe transitioned from being opaque to UV radiation to being transparent, as it is today. The faint-end slope of the galaxy luminosity function becomes steep by z~7 (e.g., Bouwens et al. 2015; Finkelstein et al. 2015), which implies that if galaxies exist at luminosities below the *current* detection limits, then these ultra-faint dwarf galaxies should be the dominant sources for reionizing the universe. Hubble Frontier

Field data reveal high-redshift galaxies that are at least 100 times fainter in luminosity than previously observed (Livermore, Finkelstein and Lotz 2017). However, the steep faint-end slope cannot continue to indefinitely faint galaxies—there must be a turnover or cutoff. The behavior of the luminosity function at the faint end should reveal the degree to which faint galaxies powered cosmic reionization. Star formation within dwarf galaxies plays multiple roles, not the least of which is as a source of ionizing photons. While there are not yet enough identified $z \geq 7$ sources of ionizing photons to account for the universe's far UV transparency at $z < 6$ (e.g., Schenker 2015), the balance would tip in favor of high-z galaxies if there are large numbers of them that are fainter than what we presently can detect.

After reionization, the UV background should suppress star formation and gas accretion onto halos at log $(M/M_{\odot}) < 9$ (e.g., Iliev et al. 2007; Mesinger & Dijkstra 2008; Alvarez et al. 2012, Jaacks et al. 2013). Additionally, halos with virial temperatures less than $10^4$ K (log $[M/M_{\odot}] < 8$; Okamoto et al. 2008; Finlator et al. 2012) cannot cool gas via atomic line emission, and so are not expected to host efficient star formation. These processes should result in a turnover in the number density of very faint galaxies at z~7 (e.g., Jaacks et al. 2013). Existing HST observations of the Frontier Fields clusters have found no turnover down to $M_{UV} \approx -14$ mag, which corresponds to log $(M/M_{\odot}) \approx 9.5$. A direct test requires pushing fainter, down to at least M = −13.5 mag. The apparent luminosity of such faint sources at z ~ 7 corresponds to J > 33 AB mag. LUVOIR will detect extended sources at such low flux levels with a survey of a few hundred hours per pointing, which would directly test the hypothesis that the UV background suppresses star formation and produces a significant turnover in the low-mass galaxy luminosity function.

The impact of reionization on the faint end of the galaxy luminosity function is substantial (e.g., Jaacks et al. (2013). A turnover in the rest-frame UV luminosity function could be detectable starting at an absolute magnitude around M = −13.5 mag. Abundance-matching with current best estimates of the UV luminosity function implies that this corresponds to halo masses of log $(M_h/M_{\odot}) \approx 9$ at z=7, consistent with theoretical expectations for the suppression mass. Observationally, this deviation would start to be seen in NIR passbands at AB $\approx$ 33 mag. LUVOIR-enabled deep imaging surveys that reach the turnover point in galaxy LF that is predicted to occur due to UV background feedback suppression will directly test this hypothesis.

**Figure 5-6** shows the predicted number of galaxies detected in scenarios with and without a turnover in the luminosity function. The z=7 luminosity function normalization with no turnover comes from recent observational constraints (e.g., Atek et al. 2018; Livermore et al. 2017). The turnover model is from Jaacks et al. 2013. The z=7 galaxy luminosity function would deliver 30% to 40% more galaxies at an absolute magnitude of −13.5 if no turnover is present. ***LUVOIR-A could detect, at a 5-$\sigma$ confidence level, the difference in the number of observed z=7 galaxies between the no-turnover model and a reionization suppression model like those shown in* Figure 5-6 *by summing up galaxy counts in a dozen HDI pointings each with 160 hours of total exposure time***. This many deep fields is well within the expected number of independent regions targeted in the exoplanet characterization program. Running in parallel with the exoplanet program will allow HDI enough time to observe each deep galaxy field in 3 passbands (1 dropout band, 2 detection bands) with a SNR=4 source detection threshold at AB = 32.75 mag, enabling the 5-$\sigma$ discrimination between models for the evolution of the faint end of the luminosity function. We have assumed that

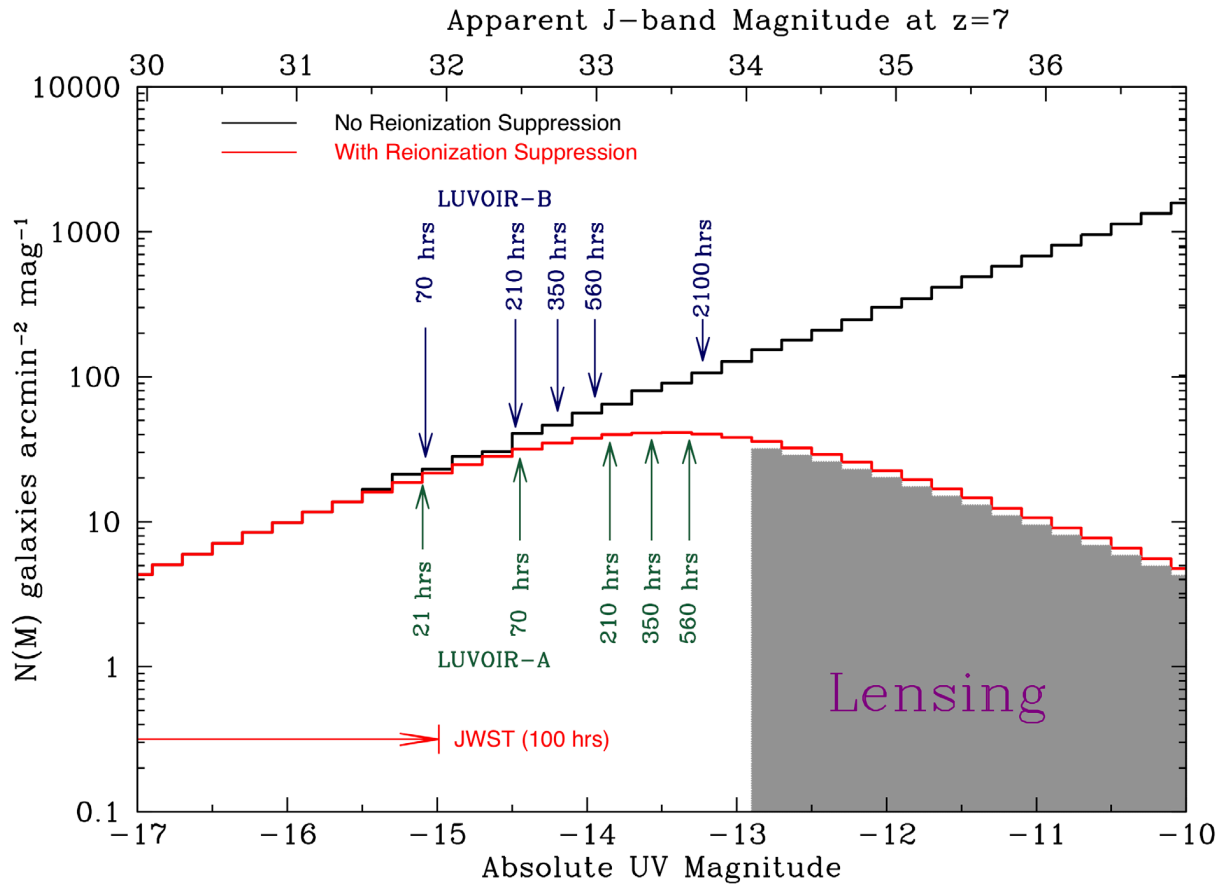

**Figure 5-6.** *Predictions for the number of z=7 galaxies per sq. arcmin. per magnitude for a model with reionization suppression (red histogram) and without reionization suppression (black histogram). Reionization suppression may introduce a detectable turnover in the luminosity function at an absolute UV magnitude around M ≈ −13.5 mag. The SNR=4 depths reached in a 3-band survey with the two LUVOIR concepts are shown, assuming simultaneous UVIS and NIR observations. This survey can be performed in parallel with the exoplanet spectroscopic characterization program. For reference, the SNR=4 depth for a 100 hour F200W JWST image is also shown. Credit: M. Postman (STScI)*

these ultra-faint z=7 galaxies are spatially resolved (~0.1 arcsec) with radii in the range of 200–300 parsecs (e.g., see Ono et al. 2013).

Another advantage of observing a dozen or more independent fields is that cosmic variance (galaxy count fluctuations due to different intervening large-scale structures) can be averaged out. LUVOIR-B would require about 520 hours per pointing to achieve the same 5-σ discrimination between reionization suppression models. The shaded region labelled "LENSING" in **Figure 5-6** shows the added depths that might be reached if powerful gravitational lensing galaxy clusters are targeted in the survey, although such fields would be unlikely to coincide with exoplanet targets. The faint z=7 galaxies would be selected using photometric redshifts. LUVOIR photometry could be augmented with infrared photometry available from future deep JWST survey fields. Even if JWST does not detect the faintest galaxies visible from the deep LUVOIR survey, the non-detection in a deep JWST survey field places an important constraint on the red-shift and can be used to reject low redshift interlopers from the sample.

Ideally, LUVOIR users would want to probe to these depths across both ionized and neutral regions (which in the 2030s, we should know of from SKA 21cm mapping, and WFIRST Lyman-α grism mapping; and/or potential NASA Probe mission Lyman-α intensity mapping). Current theories predict that we should see this turnover in ionized regions, but not in neutral regions. Models of reionization are *extremely* sensitive to this, as the galaxies near this limit dominate the ionizing emissivity. They dominate the UV luminosity density due to their numbers, but they are also more likely to have larger ionizing escape fractions than their more massive counterparts. A survey consisting of such multiple pointings would realistically only be feasible with the LUVOIR-A design.

To observe directly the build-up of the extreme low mass end of the galaxy mass function from the epoch of reionization onwards LUVOIR users will require a telescope that can detect dwarf galaxies over a broad range of redshifts. Dwarf galaxies are faint even in the nearby universe and become more so at higher redshifts. LUVOIR, however, will probe remarkably low stellar mass systems across a wide redshift range. To demonstrate this, we employ the González et al. (2012) stellar mass–luminosity (*M-L*) relation for high-redshift galaxies and derive the expected stellar mass scales that can be probed for different telescopes.

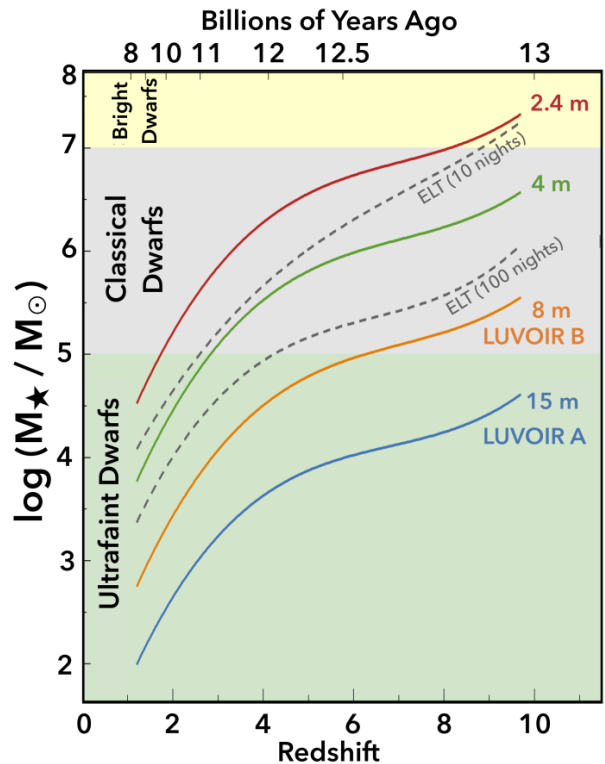

**Figure 5-7.** *The sensitivity to detecting galaxies of a particular stellar mass (y-axes) as a function of redshift (and lookback time) for different telescopes: a 2.4-m space telescope, a 4-m space telescope, the LUVOIR concepts, and the 39-m ground-based ELT. The limits for the space-based telescopes are for a fiducial 500 ksec (139 hour) observation that returns a SNR = 5 detection of a 200-parsec diameter source. The limits for the ELT are for a 10-night (80 hour) integration and a 100-night (800 hour) integration. The notional stellar mass ranges for bright, classical, and ultra-faint dwarf galaxies are indicated. Results for JWST are similar to those for the ELT 100-night exposure. These results assume that the power law slope of the UV luminosity-to-stellar mass relation remains independent of galaxy UV luminosity. Credit: M. Postman (STScI)*

**Figure 5-7** shows the results for a 500 ksec exposure with a Nyquist-sampled imager and assumes observations are made, when possible, using passbands that cover the rest-frame UV. The LUVOIR telescope concepts enable the study of the extreme low mass end ($M_* < 10^{5.5}$ M$_\odot$) of the halo mass function at many redshifts, in a single deep survey, and wider surveys of $10^6$–$10^7$ M$_\odot$ systems in much shorter exposures.

Studying the faint end of the luminosity function at high-redshift also allows users of LUVOIR to seek the progenitors of the local dwarf galaxy populations, and identify their role in the process of reionization, in the context of the underlying dark matter power spectrum

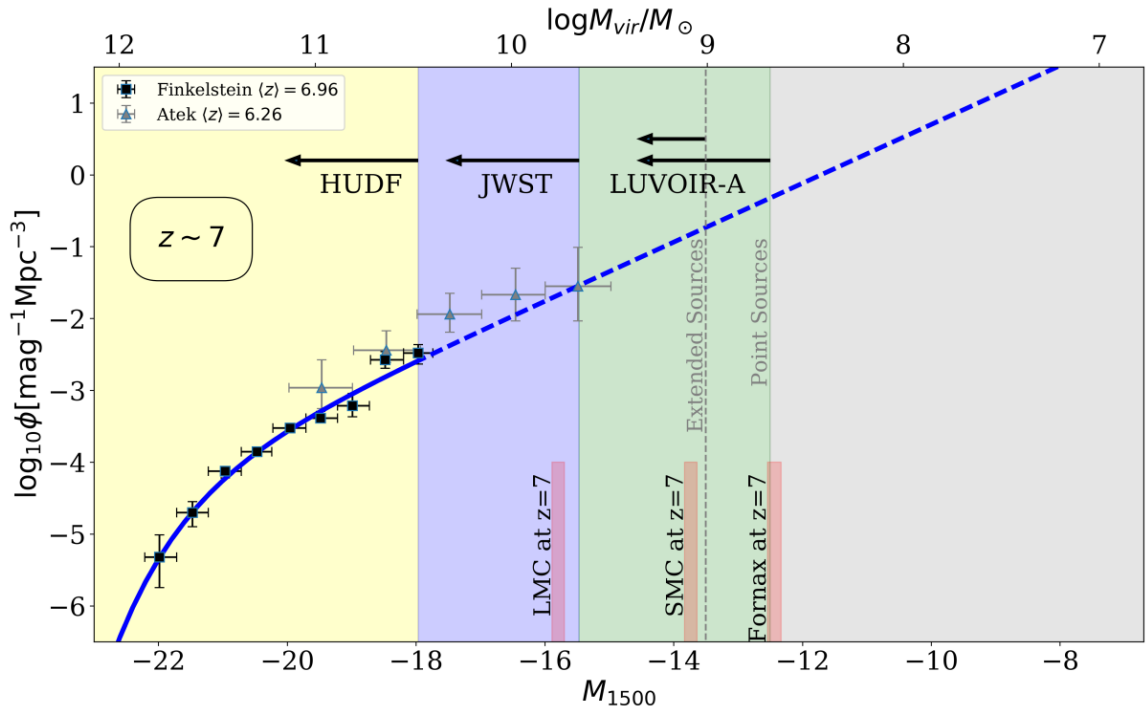

**Figure 5-8.** *The luminosities of Local Group dwarf galaxies, extrapolated to z~7, based on Fig. 4 of Boylon-Kolchin et al. (2015). The observed z~7 UV luminosity function and the best fit to the data (solid line) and an extrapolation of that fit (dashed line) are also shown. The arrows show limits that can be reached with unlensed deep HST imaging (HUDF), anticipated JWST deep fields with a limiting magnitude of AB = 31.5 mag, and with LUVOIR-A which, in 115 hours, reaches a 5-σ limiting magnitude of AB = 33.4 mag (corresponding to MUV(z~7) = −13.5 mag) for extended 0.10″ sources and AB = 34.3 mag for point sources. Credit: M. Postman (STScI)*

properties that allows them to form in the first place. There must be self-consistency between the galaxy populations expected at early times with the total galaxy populations that we detect subsequently. Part of the consistency check is against the dwarf galaxy census LUVOIR will perform in the local universe. As noted above, the faint end of the UV luminosity function can have a steep slope; a low-luminosity cutoff; or a combination of these. At z~7, the ultraviolet luminosity of local dwarf galaxy progenitors can be calculated (Boylan-Kolchin et al. 2015) and is shown in **Figure 5-8**. The most luminous local dwarf is the Large Magellanic Cloud; its progenitor at z~7 has a rest-frame absolute ultraviolet magnitude of −15.8. A LUVOIR ultra deep field (500-hour multi-band exposure) will reach AB=33.4, corresponding to $M_{UV} = −13.5$ mag at z~7 for resolved dwarf galaxies, enabling not only the detection of LMC progenitors but progenitors of the Small Magellanic Cloud (SMC) as well.

### 5.3.2 Characterizing ionizing radiation at low redshift

LUVOIR's investigation of the impact of reionization on structure formation extends to more recent periods in cosmic history as well with a characterization of the ionizing radiation that leaks from z < 1 galaxies. The emergence and sustenance of this ubiquitous background of ionizing radiation along with the accompanying evolution of large-scale structure are dependent on a key parameter, for which we have very little theoretical or observational guidance, namely the fraction of ionizing radiation escaping from these first and subsequent collapsed objects. Uncertainty in the evolution of this escape fraction, $f_{900}^{esc}$, is a major systemic unknown (Finkelstein et al. 2015; Ellis 2014; Haardt & Madau 2012), impeding our understanding of the strength of ionizing radiation feedback on the formation of structure on hierarchical and secular timescales from reionization through to the present day.

At low redshifts, the uniformity of the ionizing background radiation field appears to be tied to the low end of the neutral hydrogen column density distribution of Lyman-α forest absorbers, (12.5<log[$N_{HI}$ (cm²)]<14.5). At higher redshifts, it is an open question whether the primary source of the ionizing radiation field was the first black holes or the first stars (Madau & Haardt 2015). However, as noted earlier, low-mass dwarf galaxies could dominate the ionizing radiation budget if the steep faint end slope of the galaxy luminosity function extends to absolute UV magnitudes $M_{1500} < -13$ mag and if the escape fraction of photons below the hydrogen ionization edge in these galaxies is 5% < $f_{900}^{esc}$ < 40% (Bouwens et al. 2015; Finkelstein et al. 2015; Khaire et al. 2016, and references therein). LUVOIR can both determine the ultra faint-end shape of the luminosity function at high redshift and also characterize of the ionizing radiation itself by studying galaxies at much lower redshifts.

Although JWST is designed, in part, to discover the sources of reionization, it cannot directly observe the escape of ionizing radiation because of attenuation along the line-of-sight by intervening HI clouds in the IGM with large neutral hydrogen column densities (log[$N_{HI}$ (cm²)]>17.2, a.k.a. Lyman limit systems). The cosmic evolution of the IGM opacity is stark, leaving less than 1% transmission by the IGM at the epoch of reionization (Inoue et al. 2014; McCandliss & O'Meara 2017). Therefore, the low-z universe, accessed through the UV, has a major advantage for direct detection of escaping ionizing radiation, requiring only miniscule corrections for intergalactic attenuation. The ultimate goal is a robust determination of Lyman continuum (flux below 900 Å) luminosity function evolution across cosmic time as envisioned by Deharveng et al. (1997) and Shull et al. (2015), providing a full accounting of the ionizing radiation escape budget from star-forming galaxies, AGN, and quasars of all types.

LUMOS will spatially resolve a large number of sources in a single observation over an extended field, allowing in-depth characterization of the environmental factors that favor escape of far UV radiation. Spatial resolution is important because Lyman continuum escape depends on the relative placement of the UV emitting region with respect to intervening neutral and ionized material in surrounding disks, super-bubbles and circumgalactic streams (Dove & Shull 1994, Bland-Hawthorn & Maloney 1999, Dove et al. 2000, Shull et al. 2015). LUVOIR observations will allow an accurate determination of the sources and sinks of the ionization background radiation field. Most importantly, LUVOIR will determine what kinds of massive stellar populations generate the ionizing radiation.

GALEX showed that UV sources are intrinsically faint and rare. LUMOS, with its 4 square arcminute field-of-view, high effective area, and low background equivalent flux is the ideal instrument for surveying Lyman continuum sources. In a single pointing LUMOS will detect dozens to hundreds of galaxies across a broad range of redshift and luminosity, demonstrating the power of LUVOIR for this application. The LUMOS detection effectiveness is quantified in **Figure 5-9** and **Figure 5-10**. The top panel of **Figure 5-9** shows the cumulative number of star-forming galaxies observed to the indicated escape fraction, $f_{900}^{esc}$, at the Lyman edge (5σ detection limit) as a function of the apparent UV magnitude. The bottom panel shows the cumulative number of galaxies sampled for Lyman continuum leakage as a function of the rest frame absolute UV magnitude, $M_{1500\,Å}$. These yields are based on the estimated effective area for the G145LL grating of LUMOS ($R \approx 500$) in the LUVOIR-A concept where a peak $A_{eff}$ (1150 Å) $\approx 10^5$ cm$^2$ and low dispersion produce a background limit of AB $\approx 34$ mag (see also France et al. 2017).

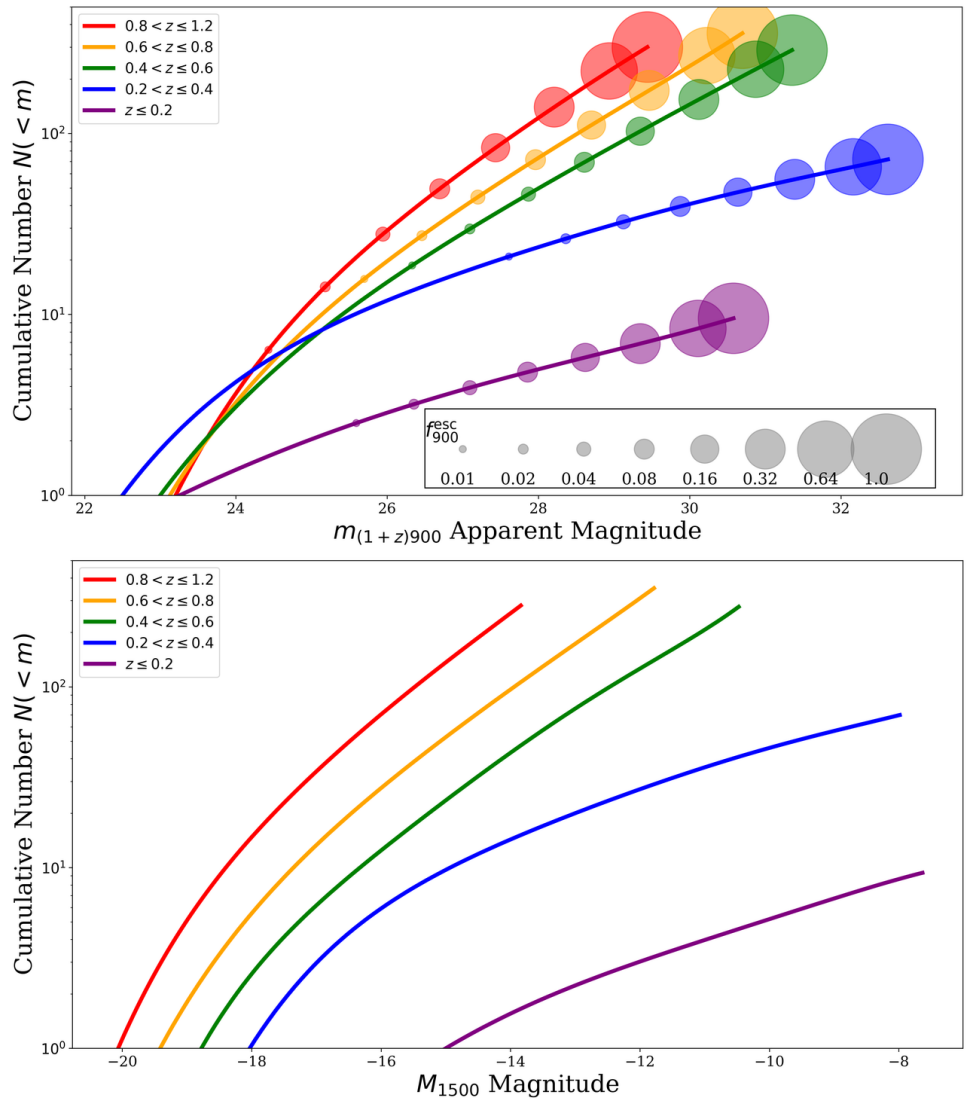

**Figure 5-9.** *LUMOS is uniquely able to explore the emission of ionizing light across cosmic time and over a broad range of galaxy environments. Top: The cumulative number of galaxies detected in a single LUMOS field of view as a function of redshift (colored lines) and escape fraction (different size circles) as a function of the 5σ limiting flux in a 10-hour observation over a 30 Å interval shortward of (1+z) \* 912 Å. Bottom: The cumulative number of galaxies detected in a single LUMOS field of view as a function of redshift (colored lines) and the absolute rest-frame 1500 Å magnitude. Credit: S. McCandliss (JHU) / J. O'Meara (Keck Observatory)*

**Figure 5-10** demonstrates that LUVOIR can observe up to 10,000 galaxies in the redshift range $0.1 < z < 1.1$ with 10 LUMOS fields observed for 10 hours each. Such a program would be a definitive exploration of Lyman continuum leakage at the faint end of the luminosity function for the $0 < z < 1$ regime as it will quantify precisely how the ionizing

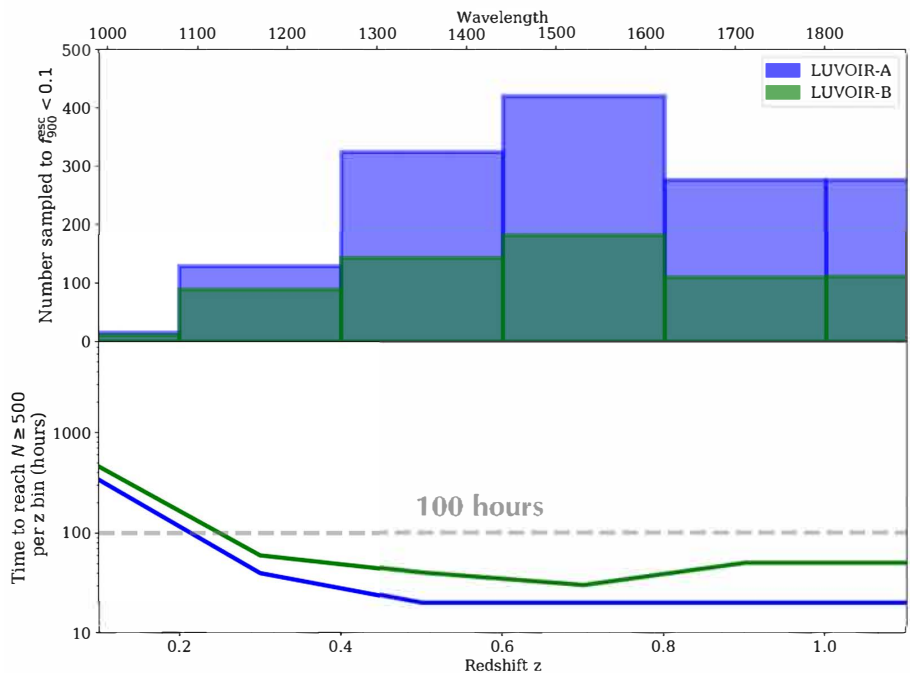

**Figure 5-10.** *The upper panel shows the number of detected objects with low (<10%) escape fraction as a function of redshift for both LUVOIR concepts. The lower panel shows the amount of time required to obtain at least 500 galaxies per redshift bin over the range 0.1 < z < 1.1. A LUMOS program of 100-hours on LUVOIR-A (150-hours on LUVOIR-B) meets this threshold for nearly all redshifts. Credit: S. McCandliss (JHU) / J. O'Meara (Keck Observatory)*

escape fraction depends on the rest-frame 1500Å luminosity. In the top panel of **Figure 5-10**, the number of galaxies sampled to $f_{900}^{esc} < 0.1$ as a function of redshift is shown for the two LUVOIR concepts. In the bottom panel, we display the total time required to populate the $\Delta z = 0.2$ redshift bins with at least 500 galaxies. A thorough quantification of the LyC escape phenomenon requires a space telescope with the capabilities provided uniquely by LUVOIR.

### 5.3.3 Mapping the escape of ionizing radiation

By studying both the ionizing radiation and the non-ionizing FUV diagnostics of star-forming regions, LUVOIR will reveal how massive stellar populations, nebular conditions, and outflows affect the escape of ionizing light.

The non-ionizing 1000–2000 Å spectra of galaxies provide key information about how galaxies form their stars, enrich their gas, and pollute the universe. This wavelength region contains unique spectral diagnostics of three distinct components: the hot stars, the ionized HII region nebulae, and the galaxy-scale outflowing winds. LUVOIR users can quickly obtain spatially-resolved diagnostics for low-redshift star-forming galaxies that probe individual star-forming regions.

Spectral diagnostics of hot stars, such as C IV and Si IV P Cygni profiles, broad He II 1640 emission, and many weak photospheric absorption lines, are just as powerful when applied to the integrated light of star-forming regions within galaxies. The ages and metallicities of star-forming regions can be determined by fitting the stellar wind features and the photospheric absorption lines (Chisholm et al. 2019). LUVOIR will therefore allow astronomers to chart how star formation proceeds within a galaxy across space and time, how the metallicity of new stars evolves, and how the gas-phase metallicity and abundance pattern responds. LUVOIR users will also be able to determine which kinds of star-forming regions permit the escape of ionizing photons.

Spectral diagnostics of the HII regions probe the gas-phase metallicity (Kewley & Ellison 2008), the abundance pattern (Garnett et al. 1995 a,b; Berg et al. 2016; Peña-Guerrero et al. 2017; Berg et al. 2019), and the ionization state and density of the gas (e.g., Byler et al. 2018). Knowledge of these conditions is critical to understanding how galaxies create and disperse the elements.

Bulk outflows of gas from galaxies, which are powered by massive stars and their explosive deaths, drive enriched material into the circumgalactic medium, and perhaps out of galaxies completely. Spectral diagnostics of the interstellar medium of star-forming galaxies reveal these ubiquitous bulk outflows (e.g., Shapley et al. 2003; Martin et al. 2005; Tremonti et al. 2007; Weiner et al. 2009; Steidel et al. 2010). The main wind diagnostics are a large number of ISM lines, and the spatial and velocity structure of Lyman-α emission (Bridge et al. 2018; Rivera-Thorsen et al. 2017; Orlitová et al. 2018; Verhamme et al. 2018).

**Table 5-1.** *Chapter 5 Programs at a Glance*

| Goal | Program Description | Instrument + Mode | Key Observation Requirements |
|---|---|---|---|
| **Chapter 5 Programs at a Glance** | | | |
| **Signature Science Case #7: Connecting the Smallest Scales Across Cosmic Time** | | | |
| Discriminate between dark matter models by measuring the shape of the matter power spectrum on <100 kpc scales | Measure the spatial distribution of extremely low-mass dwarf galaxies around 4 Milky Way analogs within 15 Mpc | HDI broadband optical and NIR imaging | Broadband imaging in V and J bands (at 550 and 1220 nm)<br><br>Imaging FOV ≥ 6 sq. arcmin |
| **Signature Science Case #8: Constraining Dark Matter via Astrometry** | | | |
| Constrain the nature of the dark matter particle by measuring the density profiles of dwarf galaxies | High precision measurements of proper motions for ~100 stars in 20 dwarf spheroidal galaxies, 10 within the Local Group (1 Mpc) and 10 beyond the Local Group (2−5 Mpc) | HDI astrometry using optical imaging | Astrometric precision < 0.5 μas/year<br><br>Three epochs per target |
| **Signature Science Case #9: Tracing Ionizing Light Over Cosmic Time** | | | |
| Study the behavior of the faint end of the galaxy luminosity function to reveal the degree to which dwarf galaxies powered cosmic reionization | Deep imaging of 12 blank sky fields in the I, J, and H bands to measure counts per sq. arcmin of dwarf galaxies at high-redshift ($z \approx 7$), with sensitivity to detect difference between rest-frame UV luminosity functions with and without reionization suppression | HDI broadband optical and NIR imaging | Broadband imaging in J and H bands (near 1220 and 1630 nm)<br><br>Simultaneous imaging in I band (near 800 nm)<br><br>Imaging FOV ≥ 6 sq. arcmin<br><br>Execute in parallel with exoplanet observations |
| Detect and quantify the evolution of ionizing radiation from low-redshift galaxies | Low-resolution FUV spectroscopy covering 12 random sky fields to characterize ionizing UV radiation from star-forming galaxies at low-redshift ($0.2 \lesssim z \lesssim 1.2$), with 500 galaxies per $\Delta z = 0.2$ redshift bin. | LUMOS multi-object FUV spectroscopy | Bandpass: 100−200 nm<br><br>$R \approx 500$<br><br>Telescope mirror reflectivity > 60% at 105 nm<br><br>Spectroscopic FOV ≳ 4 sq. arcmin<br><br>Execute in parallel with exoplanet observations |
| Map ionizing radiation escape, massive stellar populations, and outflows in low-redshift galaxies | Spectroscopy of non-ionizing and ionizing FUV from individual star-forming regions within 100 galaxies at z=0.25−0.3 | LUMOS multi- object FUV spectroscopy | Bandpass: 100−200nm<br><br>$R > 5000$ |

By acquiring both the ionizing and non-ionizing FUV spectra of galaxies on the physical scales of individual star-forming regions, LUVOIR will directly link the massive stellar populations of galaxies to the ionized nebulae and outflows they power and the ionizing radiation that may escape. A LUMOS program using the G155L grating, and targeting ~1000 star-forming regions within 100 known galaxies in the narrow redshift range z=0.25-0.3, would map where Lyman continuum escapes, and how that escape is determined by the properties of those regions (ages and metallicities of the massive stellar population, nebular conditions, and strength of the outflow). LUVOIR-A can conduct such a program in 200 hours; LUVOIR-B in 800 hours. In doing so, LUVOIR would "close the loop" between massive stellar populations, the ionizing radiation they produce, and the outflows they power, to comprehensively determine how ionizing radiation escapes from galaxies.

# CHAPTER 6. HOW DO GALAXIES EVOLVE?

This question is one of the oldest in astrophysics, dating from the very first realization a century ago that galaxies are "island universes" unto themselves. The quest to answer it has motivated some of astronomy's brightest minds to build our most ambitious telescopes. Over decades of discovery spanning the electromagnetic spectrum, astronomers have painted a rich picture of how galaxies begin as small fluctuations in the large-scale structure that aggregate into self-bound dark matter halos where gas can pile up and stars can form. These small seeds merge and grow hierarchically into larger and larger galaxies over time. Supermassive black holes lurk in the hearts of nearly every galaxy, and for brief periods might govern their evolution.

Yet our understanding of the physics behind these emergent patterns lags behind our ability to characterize them. What sets the minimum scale for galaxy formation? Why are some massive galaxies dominated by their bulges while others have none? How do galaxies sustain their star formation for much longer than their present gas supply allows? Why do the largest and smallest galaxies cease forming stars (or "quench"), and stay that way? These questions, and many more, comprise the leading edge of galaxy formation studies today (Somerville & Davé 2015; Madau & Dickinson 2014).

LUVOIR will provide astronomers with groundbreaking increases in capability for high-resolution imaging with ultra-stable image quality, highly multiplexed UV spectroscopy, and full-time access to half the sky. It will reach the bottom end of hierarchical galaxy formation, resolve the insides of galaxies to 100 parsec or better at all redshifts, and trace the flows of gas in and out of galaxies and their AGN at all cosmic times.

This chapter will focus on Signature Science—the "Cycles of Galactic Matter" and "The Hierarchical Assembly of Galaxies"—within two broad domains of galaxy evolution that encompass a wide range of unanswered questions that will remain unanswered even in the 2030s, because they require spatial resolution, sensitivity, and wavelength coverage well beyond those of other foreseeable facilities.

To understand "The Cycles of Matter," astronomers using LUVOIR will be able to deploy multi-object spectroscopy in the ultraviolet, with 30–100 times the sensitivity of Hubble and the ability to observe more than a hundred objects at once in a highly multiplexed fashion. To map "The Multiscale Assembly of Galaxies," astronomers using LUVOIR will employ extremely stable 10 milliarcsecond imaging with an unprecedented depth and wavelength coverage. These capabilities go far beyond those of any expected facility, including JWST, WFIRST, and the ground-based Extremely Large Telescopes. LUVOIR will perform breakthrough measurements at high redshift ($z > 2$) to see galaxies in fine detail (100 parsec scales) at critical phases of evolution, and in very nearby galaxies (< 10 Mpc) to see them resolved even star by star.

The Signature Science Cases here represent some of the most compelling types of observations that astronomers might do with LUVOIR at the limits of its performance. But, compelling as they are, they should not be taken as a complete specification of the LUVOIR program. We have developed concrete examples to ensure that the nominal design can do this compelling science, and so that the astronomical community has detailed examples they can adapt to their own interests.

## State of the Field in the 2030s

**Galaxy formation and evolution**: We can expect that essentially the entire sky will have been surveyed at seeing-limited resolution in the optical bands. LSST's accumulated co-add will reach ~28th magnitude in the optical over 10 years, with 0.8–1″ resolution. These maps will be joined by the massive fiber-based spectroscopic surveys that started with SDSS and 2dF and will expand deeper and wider with PFS and its contemporaries. All-sky imaging provides accurate photometry, and spectroscopy provides additional physical diagnostics of galactic dust content and stellar population age and metallicity. These vast datasets support rich multivariate analyses with statistical precision and so excel at determining galaxy population statistics and galaxy/galaxy correlations in large-scale structure and halo substructure. After these surveys, there might be very little about the large-scale distribution of galaxies that remains to be learned. These massive spectroscopic surveys also detect millions of strong gas absorbers, such Mg II and Ca II lines, which probe dense interstellar medium (ISM) and CGM gas. However, they cannot access the key UV physical diagnostics over most of cosmic time.

The other major theme of the 2020s will be high-resolution imaging in the IR, sub-millimeter, and radio. JWST and WFIRST will be the prime space observatories, bringing 50–100 mas resolution to imaging and multi-object spectroscopy. WFIRST will provide Hubble-quality imaging to a large sky area, and JWST will dissect galaxies to AB ~ 32. JWST is optimized to reach "the first galaxies," seeing the first seeds of modern galaxies at z > 15. It will also be revolutionary in its mid-IR capability, able to observe the rise of dust, ice, and molecules in galaxies over most of cosmic time. Long-wavelength facilities such as ALMA and SKA will bring their powers to bear on the neutral and molecular gas content of galaxies at high resolution. Finally, 30-m-class telescopes on the ground will, if successful with adaptive optics, bring < 10 mas imaging and spectroscopy at IR wavelengths, in a comparable resolution and depth space as JWST.

What observational parameter space will remain unexplored after all this? Space observatories are the platform of choice for high-resolution optical imaging, UV imaging and spectroscopy, highly repeatable precision photometry, high precision astrometry, and high-performance optical coronagraphy. LUVOIR's "Signature Science" motivates and uses these capabilities, while acknowledging the areas where ground-based telescopes or space facilities at other wavelengths do their best. And in the end, complementarity and collaboration between these domains drives the science forward.

**Stars, star formation, and feedback**: JWST and WFIRST will have observed stellar populations down to 0.3 solar masses for all galaxies within the local 0.5 Mpc, with lower limits for nearer galaxies. This will test theories of low-mass star formation and constrain variations of the low-mass end of the stellar initial mass function in low-density environments. ALMA will have characterized the populations of gas cores in several nearby galaxies, and will have shed light on the relation between the core mass function and the stellar initial mass function around the low mass turnover. Improved models will help interpret the complex chemistry of the pre-stellar cores. In parallel, adaptive optics-assisted 30-m-class telescopes will have assembled large samples of binary stars at a range of masses out to the Magellanic Clouds, using both radial velocity (RV) and proper motion techniques. Combinations of these facilities will have increased the number of candidate galaxy hosts of very massive stars, i.e., stars with M > 150 $M_\odot$, by at least tenfold, up from the current census of a handful.

Ultimately, we expect that the creativity of the community, empowered by the revolutionary capabilities of the observatory, will ask questions, acquire data, and solve problems that we cannot envision today. That is as it should be, as a flagship should always reach far beyond current capabilities and transcend the current limits of our imagination. With this in mind, let us ask the vital science questions that we believe will remain unsolved until LUVOIR, and consider how they drive the requirements for its aperture, resolution, and wavelength coverage. We hope these cases will stimulate the reader to design their own future with these tools.

## 6.1 Signature Science Case #10: The cycles of galactic matter

How do galaxies acquire the gas they use to form stars? How do they sustain star formation over billions of years when they appear to contain much less gas than this requires? How does feedback from star formation and active galactic nuclei (AGN) expel gas and metals, and to what extent is this feedback recycled into later star formation? What happens to a galaxy's gas when it quenches? Is it used up, ejected, or hidden? Answering these questions will go a long way toward explaining why galaxies look like they do.

Inflows and outflows of gas likely shape the evolution of star formation within a galaxy. Inflows ultimately arise from the intergalactic medium (IGM) to provide fuel for continued star formation over the lifetime of a galaxy. Outflows from the interstellar medium, driven by stellar radiation and the explosions of stars and jets from AGN, can inhibit star formation in their central regions. The balance of these flows set the rate at which galaxies accumulate heavy elements.

All these flows meet in the circumgalactic medium (CGM), a diffuse gaseous medium spanning roughly 30 times the radius and 10,000 times the volume of the visible stellar disk (**Figure 6-1**; Tumlinson, Peeples, & Werk 2017). If galaxies are the factories of creation, the CGM is their fuel tank, waste dump, and recycling center. No picture of galaxy evolution is complete without understanding the gas flows between these major components.

To organize this discussion, we will proceed into galaxies and back out again. We start with the IGM and CGM and their role as the reservoir of galactic accretion and recycling (**Section 6.1.2**). From there we will consider the flows within the CGM and how gas there is processed and recycled, and how galaxies are quenched (**Section 6.1.3**). Finally, we will consider the disk/halo interface, where accretion becomes the ISM and the ISM becomes feedback, using highly multiplexed "down-the-barrel" spectroscopy (**Section 6.1.4**). But first, we pause to reflect on why observing these processes makes LUVOIR's UV sensitivity so essential to its science case.

### 6.1.1 The essential ultraviolet

Because UV light damages cells and mutates DNA, life as we know it might not exist without Earth's UV-blocking ozone layer. It seems we can have life on Earth, or UV astronomy on the ground, but not both.

**Figure 6-2** shows why UV coverage is essential to understanding the intrinsically "multiphase" galactic gas flows. The shaded map at lower left shows the distribution of gas in a simulated Milky-Way-like galaxy from the EAGLE project (Oppenheimer et al. 2016), which spans multiple "phases" from $10^3$–$10^6$ K and eight orders of magnitude in density. In ionized gas like this, the quantum-mechanical rules of electron orbits dictate that the gas will emit

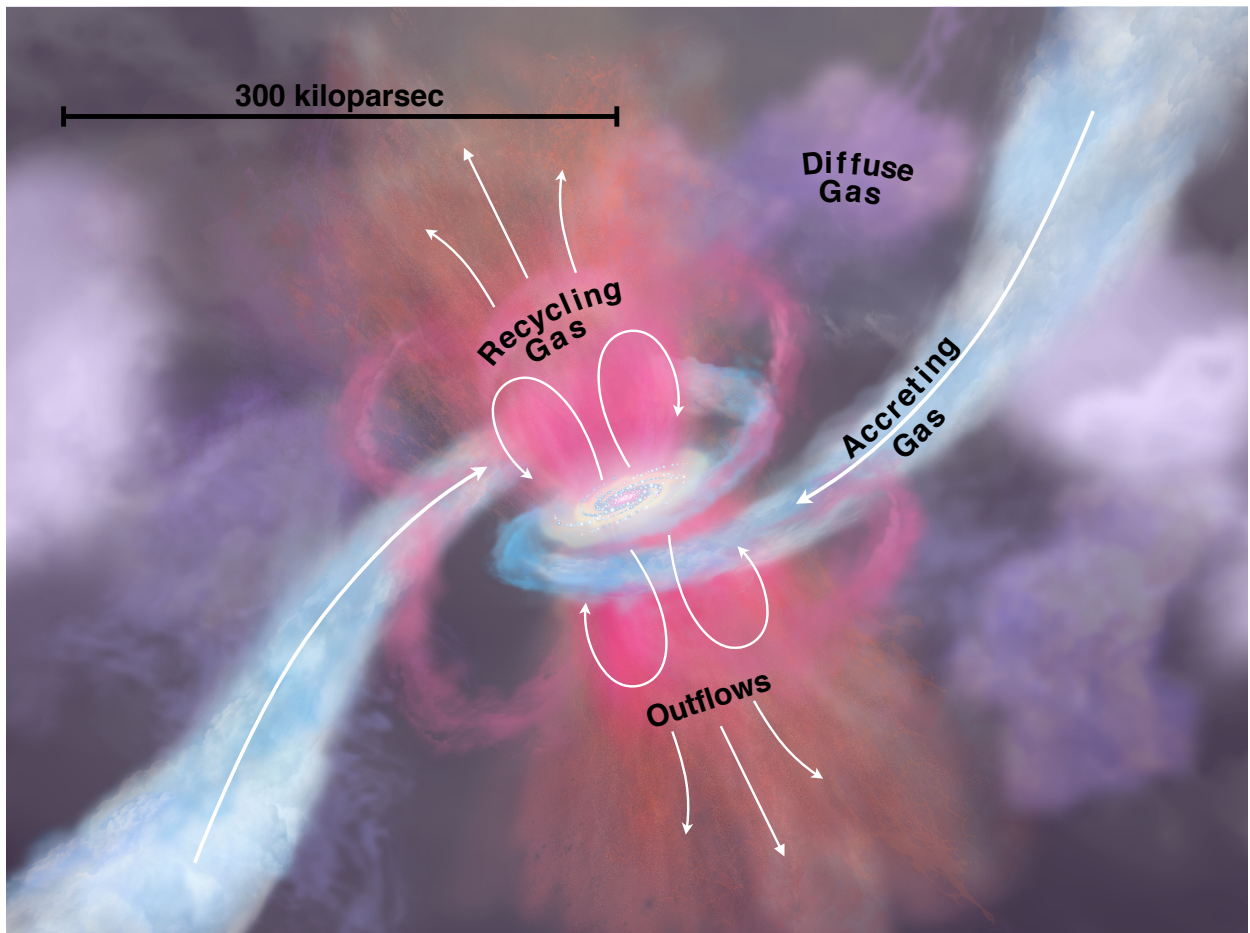

**Figure 6-1.** *"The Cycles of Matter" play out on scales far beyond the visible stellar disks. Normal galaxies are surrounded by a massive reservoir of diffuse gas that acts as their fuel tank, waste dump, and recycling center: the circumgalactic medium (CGM). It is fed by accretion out of the cosmic web and by outflows from the galaxy. Unraveling these gas flows is a major driving force for LUVOIR and its instruments. Adapted from Tumlinson, Peeples, & Werk (2017).*

and absorb energy predominantly at UV wavelengths, up to 80% according to detailed simulations (Bertone et al. 2013). Many of the transitions appear as strong UV absorption and emission lines (**Figure 6-2**). This inescapable physics means that access to UV wavelengths in space is essential if we are to resolve questions about how galaxies acquire, process, eject, and recycle their gas over the last 10 Gyr of cosmic time.

With LUVOIR, we have the opportunity to expand the reach of UV observations well past the range of Hubble's capabilities. Thanks to investment in technology development, new generations of high-reflectivity coatings with low contamination will be used to reach photons down to 1000 Å. With LUVOIR it will no longer be necessary to consider the UV as divided at the ~1150 Å boundary like with Hubble and FUSE. More details about this enabling technology is available in **Chapters 9** and **12**.

### 6.1.2  Gas flows in absorption: Accretion, feedback, and quenching

Only recently have we come to appreciate that the CGM is a major element of the mass and metal budgets of galaxies. Using Hubble's Cosmic Origins Spectrograph (COS), astronomers have found that the gaseous halos of Milky-Way-like galaxies may outweigh their disks (Werk et al. 2014), and that a large share, perhaps the majority, of all the metals ever produced by stars are outside galaxies (Peeples et al. 2014). The major strides in characterizing the CGM have left many important questions open:

1. Where are the missing baryons that are needed to fuel galaxies?

2. Where are the metals, and what do their distribution within and outside galaxies tell us about feedback?

3. How are galaxies quenched, and what happens to their CGM? Is it consumed, ejected, or heated? And how is quenching maintained?

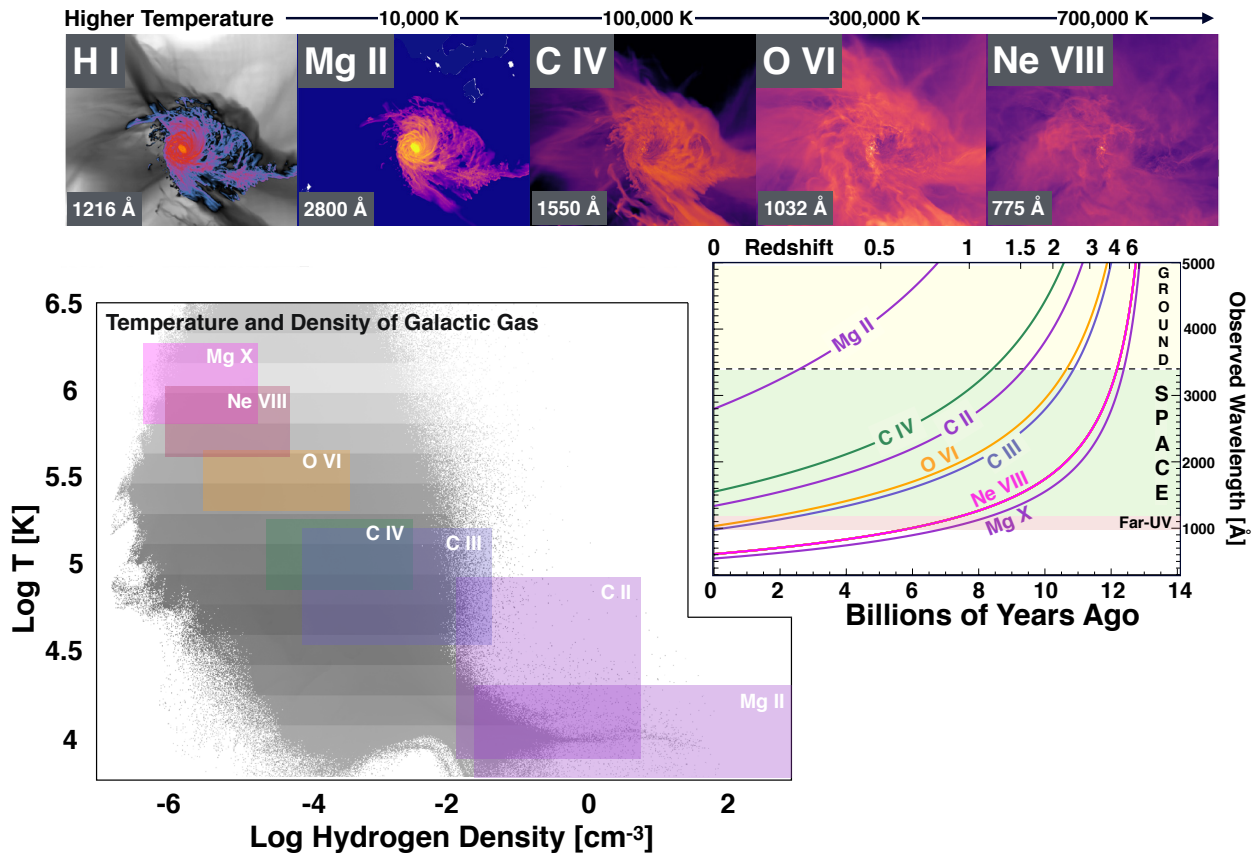

**Figure 6-2.** *Diffuse gas in and around galaxies requires UV capability for most of cosmic time. The top row shows a simulated galaxy at z = 0.7 from the FOGGIE suite (Peeples et al. 2019), rendered in some key diagnostic ions. The temperature and density regimes probed by these ions are marked in the "phase diagram" of this galaxy's gas (lower left). At upper right we show how these lines, ranging from Mg X at 680 Å to Mg II at 2800 Å, vary in observed wavelength with redshift. Even with redshift, most of this diffuse gas is visible only in the UV for the last 10 Gyr of cosmic time. X-ray lines such as O VII and O VIII (both around 20 Å) probe gas at ~1 million K but not the cooler phases where accretion and recycling occur. The 1000–1200 Å range marked "Far-UV" is critically important to capture O VI 1032 at z > 0.1 and the EUV ions Ne VIII and Mg X at z > 0.5 rather than z > 1.*

Solving these problems requires pushing the boundaries of CGM characterization far beyond the limits of today's measurements, to z = 1–2, for two major reasons. First, this period 7–10 Gyr ago encompasses the peak of cosmic star formation. Second, at z > 0.5 we gain access to a wide range of extreme UV lines, such as Ne VIII, O II to IV, and Mg X (< 800 Å in the rest frame, see **Figure 6-2**) that enable a much broader set of diagnostics of physical state and metal content that are only available with redshift.

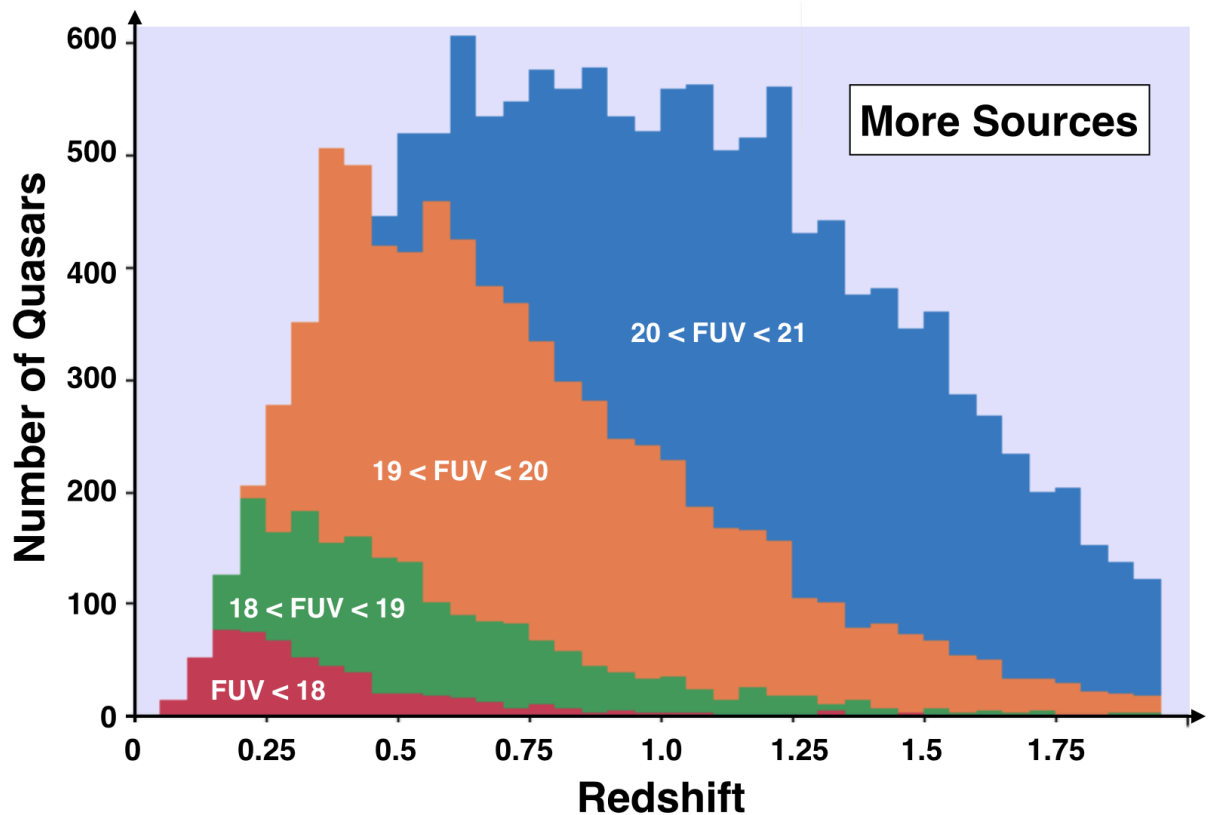

**Figure 6-3.** *Redshift distribution for SDSS quasars of various GALEX FUV magnitudes. For SNR = 20 spectroscopy, Hubble/COS is limited to the red wedge, which barely reaches past z ~ 0.7. With its 3 mag deeper grasp, LUMOS users will have many more QSOs to choose from, particularly in the vital z = 1–2 interval.*

Hubble's COS has approached the z ~ 1 frontier, but cannot advance it because it is limited to background sources of AB = 18 in the FUV (for SNR ~ 20). There are only about 1000 such QSOs on the entire sky (**Figure 6-3**), and only a handful lie at z > 1 where we can approach the cosmic star formation rate (SFR) peak with the optimal set of ions.

The LUVOIR Ultraviolet MultiObject Spectrograph (LUMOS) is designed for point-source spectroscopy 30–100 times more sensitive than Hubble/COS, at double its highest resolution (R ~ 40,000, or 7 km s⁻¹ FWHM). LUMOS users will be able to choose from any of the quasars counted in **Figure 6-3**, including thousands of choice objects at z > 1 that are too faint for a less sensitive instrument. This is a critical time in the history of the universe: star formation rates begin declining from their z ~ 2 peak, AGN have passed their epoch of maximum activity and are turning off, the "red sequence" of passive galaxies is beginning to emerge, and the first large concentrations that we call galaxy clusters are about to form. It is thus a critical time in cosmic history of which Hubble's UV spectroscopy has given us only the slightest glimpse. LUMOS will enable great strides in depth and number of sightlines, and therefore in statistical power applied to many important problems.

***Where are the missing galactic baryons?*** Normal galaxies appear to possess only a few percent of their expected budget of baryonic matter when only stars and ISM gas are taken into account (Fukugita, Hogan, & Peebles 1998; McGaugh 2005). We now know that a large mass fraction exists in the CGM, but this measurement has so far been performed only for galaxies around the mass of the Milky Way (or L*) and only at z < 0.2. (Werk et al. 2014). The large wavelength grasp of LUMOS (1000–3000 Å) provides coverage of critically important rest-frame extreme-UV ions that redshift into the FUV for z > 0.5 (Tripp 2013; **Figure 6-2**). This includes nearly every ionization state of the most abundant heavy element, oxygen, from O I (cold gas), though O VI (warm ionized gas), which LUMOS will cover simultaneously for sightlines at z ~ 1. A baryon census done with ions of a single element will eliminate the most serious systematic errors in ionization models that plague these measurements, as it is insensitive to any relative elemental abundances and only mildly sensitive to ionization corrections. The remaining oxygen ions—O VII and O VIII—absorb in the X-ray (~20 Å) and trace very low density, high temperature gas that has only been detected a few times despite investment of megaseconds of Chandra and XMM time.

Fortunately, very high ionization lines like Ne VIII (775 Å), Mg X (610 Å) and Si XII (500 Å) become available in (optically thin) IGM and CGM gas at z > 0.5, reaching a temperature regime (T > $10^6$ K) that is usually thought to be the exclusive domain of X-ray telescopes. In conjunction with a sensitive X-ray facility like Lynx, which can see O VII and O VIII, LUVOIR should be able to complete a census of CGM baryons across all its phases and more than 10 Gyr of cosmic time.

***Where are the metals that trace feedback?*** Heavy elements are "Nature's tracer particles"—the equivalent of a message in a bottle that tells us where the products of star formation have been carried over time. If we can piece together the messages from all the island universes, we could come to understand how galactic feedback works.

Hubble has extensively probed the extent of metals around galaxies, culminating in the finding that only 20% of all metals produced are still retained in the galaxies that made them (Peeples et al. 2014). The large fraction of those metals that end up in the CGM trace out a curiously bimodal metallicity distribution (Lehner et al. 2013; Wotta et al. 2016), with ~50% of the gas having metallicities around 5% solar (accretion from the IGM?), while the other half is roughly solar (metal enriched feedback?). Understanding the origins of this gas, whether it will collapse onto the galaxy or will be subsumed back into the corona, is critical to understanding how and from where galaxies replenish their fuel supply.

The same broad wavelength coverage that enables a robust baryon census will also complete a multiphase metals census over a wide range of galaxy masses, star-formation rates, and environments. By exploiting simultaneous coverage of almost every UV ion of oxygen, as shown in **Figure 6-4**, LUVOIR can avoid the systematic problems with using C, Si, O, and N variously as is done at z < 0.2. This is only possible with UV coverage and redshift that places EUV lines of O II, O III, and O IV into the space UV (Tripp 2013). Complementary X-ray measurements with Lynx will add in the hottest gas. Using this capability, LUVOIR's users can complete the low-z metals census, constrain the physics of feedback from galaxies, and assess the importance of galactic recycling with robust statistics.

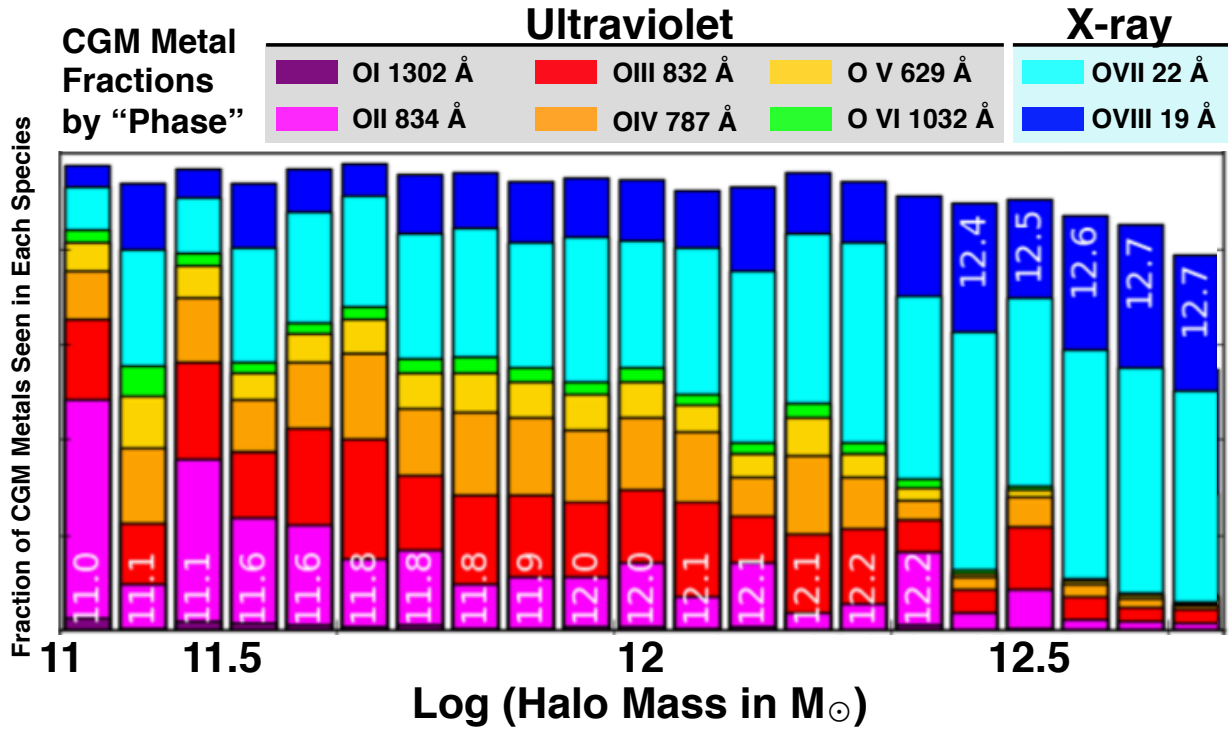

**Figure 6-4.** *The "metals budget" in the CGM of simulated galaxies from the EAGLE Zoom simulations (Oppenheimer et al. 2016). At low mass, halos are filled with low and intermediate ionization gas that can be detected with UV lines of the ions O I–O VI. At around the halo mass of the Milky Way (2x10^12 M_☉), the UV- and X-ray traced phases are comparable. Complementary UV and X-ray measurements are needed to perform a complete metals census and test the uncertain feedback physics in these models.*

**Precision cosmology with the IGM.** The IGM as traced by the Lyman alpha forest provides unique constraints on cosmological structure formation. The very high throughput of the LUMOS/G300M grating provides SNR > 30 coverage of the z ~ 1 forest in one hour. To date, less than five quasars have been observed covering the forest at SNR ~ 10 or greater at z ~ 1, at the cost of nearly 100 Hubble orbits. The baryon census and feedback studies described above obtains an unprecedented Lyman alpha forest sample with no additional exposure time. A LUVOIR program observing 100 quasar sight lines at SNR > 30 would: (1) constrain the matter power spectrum p(k) to < 5% precision on scales of 0.1 to 100 Mpc at z ~ 1; (2) measure the H I column-density distribution function f(N_HI); (3) measure the equation of state of the IGM $T=T_0(\rho/\rho_{avg})\gamma^{-1}$, to <10% precision. These measurements would constrain the formation of large-scale gaseous structure, the distribution of gas near galaxies, the radiation field produced by all sources, and the thermal history of the universe.

**A comprehensive LUMOS QSOALS campaign.** All of these studies of CGM missing baryons, metals, and the IGM can be carried out with a single, comprehensive QSO survey (**Appendix B.11.2**). For definiteness and scaling between architectures, we scope this to use the 100 brightest QSOs identified in the SDSS DR7+GALEX QSO catalog. We set the SNR goals such that 1 hour in each of the LUMOS M gratings will yield SNR ~ 20 for an FUV = 19 mag QSO. For brighter objects, we scale the exposures by FUV magnitude to maintain the same SNR. This survey of 100 QSOs will require 240 hours of exposures (and approximately

~350 hours with overheads). As an example of synergy between space and the ground, it is the future 20–30 m telescopes on the ground that will obtain the necessary galaxy redshifts to correlate with the detected gas along these LUVOIR sightlines. Complementary X-ray observations will, for the X-ray bright subset, constrain the hottest gas in these galaxy halos and complement LUVOIR's observations of Ne VIII and Mg X.

CGM characterization for thousands of galaxies at $z$~1–2 represents a fundamentally game-changing prospect for the study of the gas in, around, and between galaxies. No mission current or planned other than LUVOIR could complete such a program in a treasury scale program allocation of time. Any claim of understanding the history of baryons in the universe demands a study of this epoch of transformation in cosmic time, and LUVOIR rises to the challenge.

***How is quenching done, and maintained?*** Galaxy quenching is a prime target for LUVOIR's unique power. How galaxies quench, and remain so, is a major open question. The number density of passive galaxies has increased 10-fold over the 10 Gyr interval since z ~ 2 (Brammer et al. 2011) at the expense of the star-forming population. Galaxies undergoing quenching are the ideal laboratories to study the feedback that all galaxies experience: the galactic superwinds driven by supernovae and stellar radiation, the hot plasma ejected by jets from black holes lurking in galactic centers, and the mergers that transform galaxy shapes while triggering the consumption or ejection of pre-existing gas.

LUVOIR's first major approach to quenching will come from the QSO absorption-line treasury program from the proceeding section. The 100 sightlines in that program should pass through the halos of hundreds of galaxies that are quenched, or even underdoing active quenching through an AGN or post-starburst phase. As most of the development of the present-day red sequence occurred since z ~ 2, and the key diagnostics are rest-frame UV lines, this critical problem is a unique and compelling driver for LUVOIR's aperture and UV sensitivity.

Another unique LUVOIR application to quenching is shown in **Figure 6-5**. AGN feedback is thought to play an important role in keeping hot halos from cooling in quiescent galaxies, though it is difficult to probe the jet and radiation interactions with the surrounding CGM gas. However, increasing the density of sources to which we have access with UV spectroscopy will allow us to probe that interaction directly. The key needs for such science are high spectral resolution and sensitivity (aperture) to provide enough source density that individual AGN hosts can be studied.

In this program, the QSO absorption-line approach is adapted to examine many sightlines behind a single galaxy, the post-starburst radio galaxy Centaurus A at 4 Mpc. Using multiple sightlines to probe halo gas is currently feasible only for M31, which subtends many square degrees, large enough to encompass 20 or so QSOs at the limits of COS. LUMOS will reach dozens of QSOs within 150 kpc of the center of Cen A, or any other nearby galaxy out to 10–20 Mpc. For each galaxy, LUMOS will access to the full suite of UV diagnostic ions (including the critical O VI 1032/1038 Å doublet) and precise kinematics to probe the gas flows associated with quenching.

**Appendix B.11.3** defines a similar observing program for M51, to do the corresponding experiment for a star forming galaxy. This program observes 30 QSOs probing M51's halo out to 200 kpc, over the full wavelength range of LUMOS in a total of 117 hours of exposure time (estimated to be 146 hours with overheads). This program uses QSOs as faint as FUV

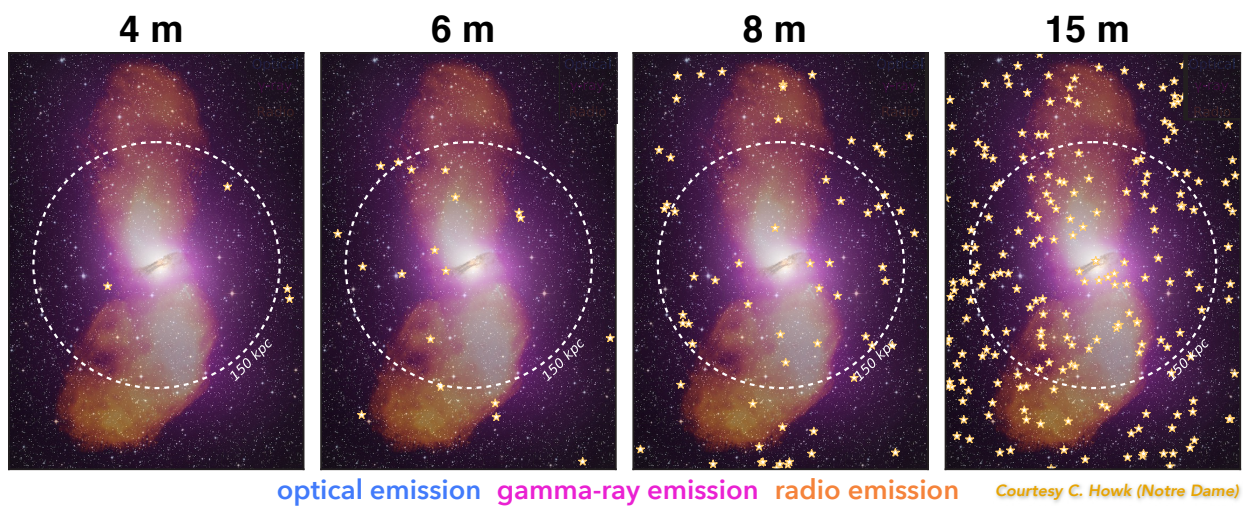

optical emission   gamma-ray emission   radio emission          Courtesy C. Howk (Notre Dame)

**Figure 6-5.** *A visualization of the sky around the radio galaxy Cen A demonstrating the density of sources available for high-resolution, high-SNR spectroscopy of background QSOs behind low-red-shift galaxies. Each star represents a QSO that could be observed to SNR > 10 in ~0.5–1 hour by a telescope of the given aperture. An 8-m telescope has a source density smaller by ~5x than that of the 15-m. The nearest QSO that is reachable by Hubble/COS in <20 orbits lies at impact parameter R = 300 kpc, two times further from the core of this galaxy than the circle representing R = 150 kpc. Background galaxies, which can also be used to map the gas, are ten times more numerous.*

~21, which are far beyond present capabilities. There are hundreds of galaxies in the local Universe for which LUVOIR could perform this same experiment, mapping the CGM in many kinds of galaxies from large to small, from star forming to quenched, and in all other phases of their evolution.

### 6.1.3 Mapping the cycles of matter in ultraviolet emission

Quasar absorption lines can detect and characterize the IGM and CGM without regard to gas density, but they reveal little about the 3D distribution of the absorbing gas. But with its millions of ~0.1″ apertures, customizable to nearly any source, LUMOS will be able to "take a picture" of a galaxy's gas flows in two dimensions, using emission from the gas itself that reveals its density, temperature, metallicity, and kinematics. This capability will enable LUVOIR's users to probe physical processes—shocks, accretion, ejecta, and recycling—at scales that even simulations today can barely reach. We will consider two applications of this capability: maps of small-scale gas dynamics in nearby galaxies, and detection of the extremely diffuse gas that fills halos at higher redshift.

***Galactic winds in high definition.*** The LUMOS microshutters map to small-scale (< 1 kpc) clouds that carry gas and metals away from the prototypical starburst galaxy, M82 (**Figure 6-6**). Tiles of the 2′ x 2′ microshutter field of view are overlaid on the galaxy and its outflow. The color zoom shows a ground-based Hα image in which small clumps appear in the flow. Did these clumps cool and form where they are? Or is this material directly eject-ed? If so, what does it imply about the mass ejection rate, and the mass rate of recycling? Only at UV wavelengths can we probe the relevant energy scales to see galactic feedback in action. Multiply this small region many-fold, and it becomes clear that we must be able to observe hundreds of such places in this complex flow to understand its true dynamics. Small scales are critical to study these flows: the LUMOS 0.14″ x 0.07″ shutters subtend parsec-scale sizes in nearby galaxies and sub-kpc sizes at z < 2 (right panel of **Figure 6-6**), where the relevant diagnostics are still in the space UV. The ability to resolve gas flows at parsec to kiloparsec scales in the key UV diagnostics lines is a unique ability of a large UV-sensitive space telescope.

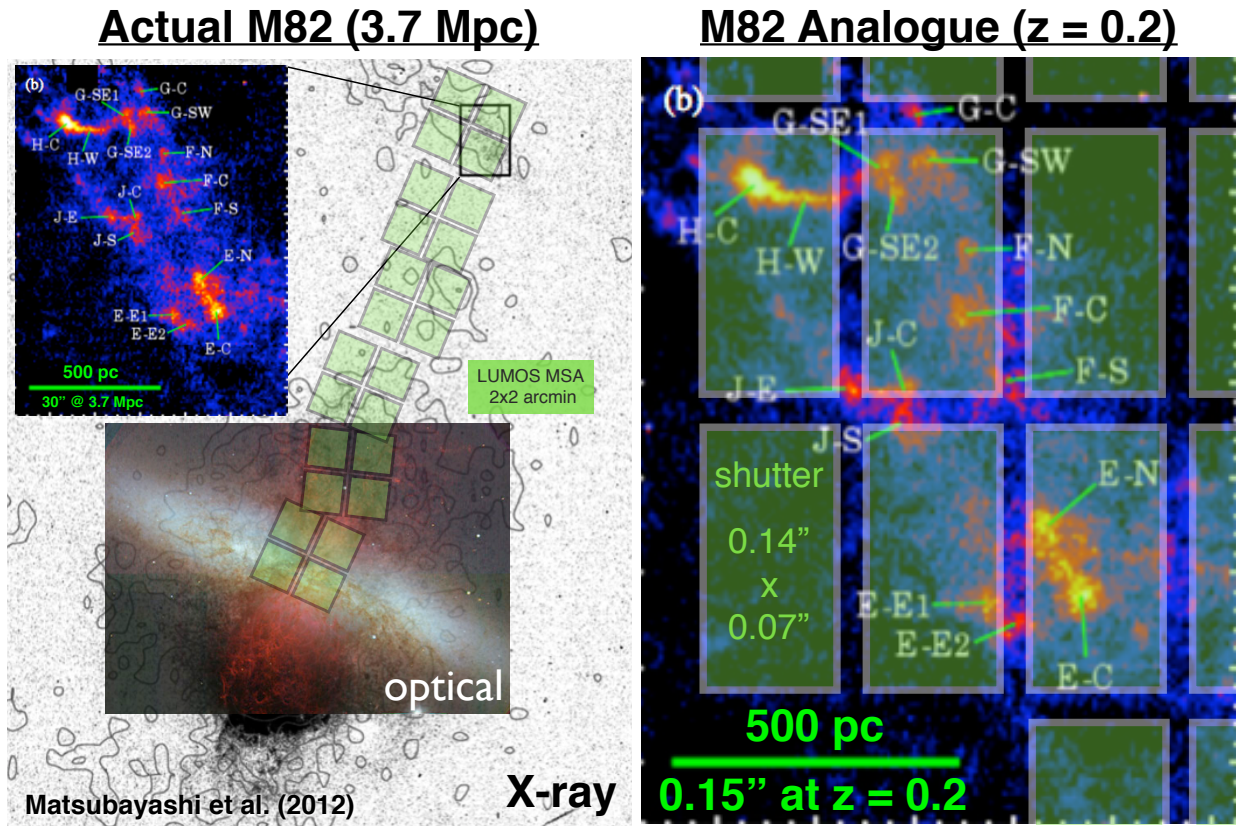

**Figure 6-6.** *Two examples of using LUMOS to examine the small-scale physics of galactic outflows. At left, X-ray and optical images of the M82 starburst are tiled by the 2′ x 2′ MSA, and one 0.14″ x 0.086″ micro shutter matches well to the bright knots of emission. At right, the zoomed region from left has been rescaled to z = 0.2 to show how these clumps of interacting gas can be mapped at the sub-kiloparsec level of individual shutters.*

The resolved gas flows observing program (**Appendix B.11.4**) uses 75 hours with LUVOIR-A to map the M82 superwind as it propagates out from the galaxy at 300 (or more) different locations in the flow spread across 6 footprints of the microshutter array. At each position, LUMOS can detect emission in Lyα, O VI, C IV, or any other diagnostic line, measure the line kinematics relative to the optical lines (such as H-α) and to other UV lines, and estimate the gas mass, metallicity, and kinetic energy. Obtaining these kinds of richly detailed physical diagnostics at ~20 parsec spatial resolution is a transformative capability for the understanding of galactic winds driven by SNe and AGN.

***Diffuse CGM, the faintest light in the universe.*** Ground-based IFU spectrographs such as VLT/MUSE and KCWI at Keck are pioneering the search for CGM gas emission at z > 2, where the relevant diagnostic lines pass into the visible bands (**Figure 6-2**). Gas reservoirs extending over hundreds of kpc appear to be illuminated by radiation from the stars and AGN in the galaxies (Cantalupo et al. 2014) and some large structures can be resolved (Martin et al. 2016). However, their spatial resolution corresponds to physical scales of > 10

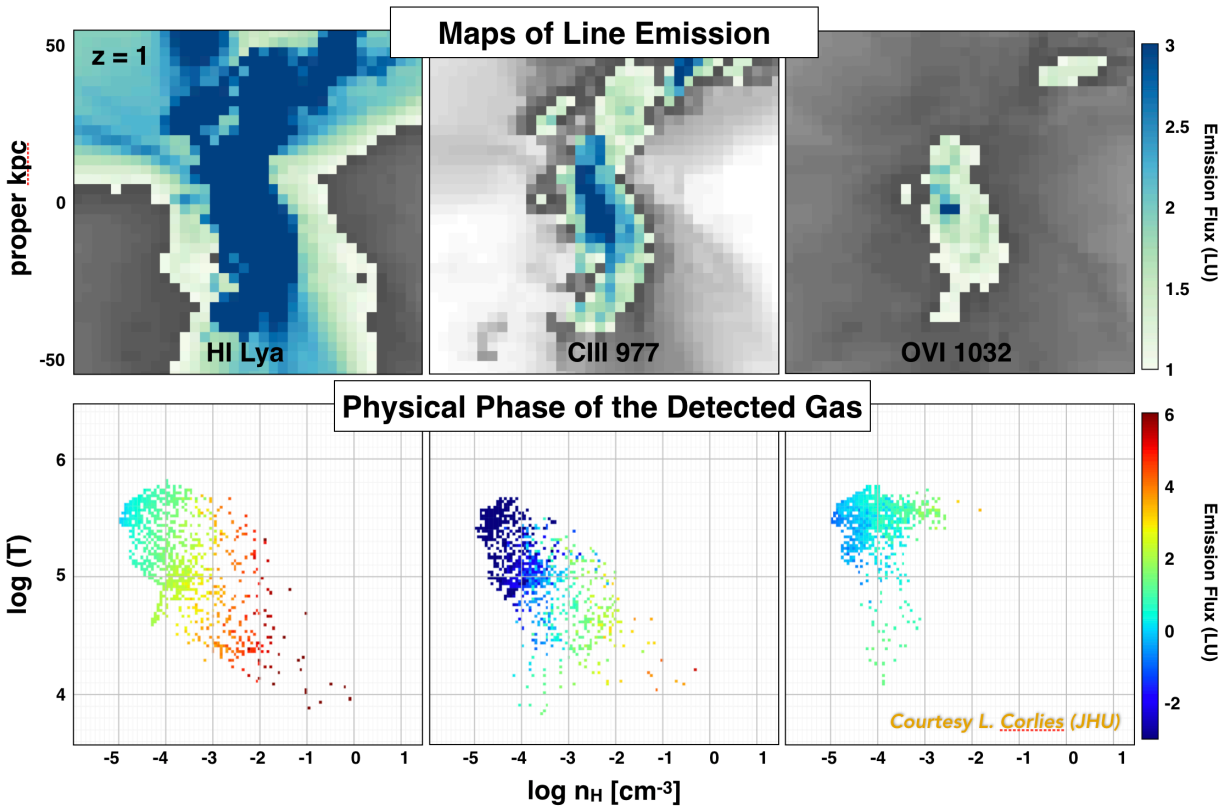

**Figure 6-7.** *A rendering of simulated CGM emission in three UV lines for a Milky-Way progenitor galaxy at z = 1. These lines range from 1000–1200Å in the rest frame, so they are uniquely visible in the UV at z < 2, or half of cosmic time. The 15-m LUVOIR will detect bright emission (blue/dark green regions) and resolve bright knots in the light green regions. Significantly smaller apertures will lack the sensitivity for all but the brightest knots, and will not resolve them even if they are detected. Credit: L. Corlies (LSST)*

kpc, and they are limited in sensitivity by bright and time-variable sky backgrounds that will make detection of the weaker metal lines difficult in individual halos. And most importantly, these instruments can access the most important UV diagnostic lines, usually Lyα, but only at z > 2, leaving most of cosmic time unexplored.

Using the same array of shutters binned into larger "virtual apertures," LUVOIR can also seek the extremely faint emission from the widely distributed diffuse CGM and structure within it. **Figure 6-7** shows such a hypothetical map from a new hydrodynamical simulation of a Milky Way progenitor galaxy at z = 1 (FOGGIE; Corlies et al. 2018). This diffuse gas will be challenging to detect even for LUVOIR, but appropriate integration times could be fitted in as parallels to the week-long exoplanet visits, for example. In deep LUMOS exposures, the structure of the CGM can be detected in multiple spectral lines, allowing observers to count up the heavy element content of this gas, to watch the flows as they are ejected and recycled, and to witness their fate when galaxies quench their star formation, all as a function of galaxy type and evolutionary state. LUVOIR could map galaxies in fields where deep imaging identifies filaments in the large-scale structure, and where ground-based ELTs have made deep redshift surveys to pinpoint the galactic structures and sources of metals to be seen in the CGM. Because this radiation is far weaker than local foreground radiation ($S_B \sim$ 100–1000 photons $cm^{-2}$ $s^{-1}$ $sr^{-1}$) ground-based telescopes seeking it at redshifts where it appears in the visible (z > 2) must perform extremely demanding sky foreground subtraction to reveal the faint underlying signal. These foregrounds are considerably lower from space (by factors of 10–100), shortening required exposure times by an equivalent factor. By binning up 0.5–1″ regions of the array (a few kpc at z < 2), LUMOS users can examine the large-scale distribution of the filaments and extended disks they feed, from the peak of cosmic star formation down to the present.

Galaxies undergoing quenching are the ideal laboratories to study the feedback that all galaxies experience: the galactic superwinds driven by supernovae and stellar radiation, the hot plasma ejected by black holes lurking in galactic centers, and the mergers that transform galaxy shapes while triggering the consumption or ejection of pre-existing gas. LUVOIR will have the collecting area to support deep, wide-field UV multi-object spectrograph (MOS) searches for CGM gas at the line emission fluxes that are expected, and with the spatial resolution to observe the transformation of star forming disks to passive spheroids at 50–100 pc spatial resolution and closely examine the influence of AGN on this process. For galaxies identified as quenching, emission maps of the surrounding CGM will determine the fate of the gas that galaxies must consume or eject and elucidate the physical mechanisms that trigger and then maintain quenching. Only a diffraction-limited space telescope with an aperture of at least 10–12 meters can achieve such spatial resolution in the optical and observe the rest-frame UV light necessary to witness the co-evolution of stars and gas in galaxies undergoing this transition.

### 6.1.4 Gas flows with down-the-barrel spectroscopy

We can now consider the final steps in our view of galactic gas flows with LUVOIR: how CGM gas turns into ISM, and how ISM gas returns to the CGM. What drives the flows that transport mass from the CGM into galaxies, and then back? The inflows are driven primarily by gravity and cooling acting on and within gas that enters the halo in filaments, strips off satellites, or becomes thermally unstable while orbiting in the halo. Gas cooling and gravity are straightforward to implement in models and simulations, even if the emergent behavior they create is hard to simulate. "Feedback"—the general term for mass, momentum, and energy from stars and AGN that influence the course of galactic evolution—is far more complex, because the underlying physical mechanisms span a huge range of density, temperature, energy, and physical scales. "Feedback" flows such as supernovae and AGN winds originate on parsec scales, but propagate to hundreds of kiloparsecs while interacting energetically, hydrodynamically, and radiatively with everything they encounter, including current inflows and past generations of outflows.

Understanding how feedback operates physically to influence galaxies as they grow is an active and abiding challenge in the astrophysics of galaxies. Theoretical models of how flows develop and propagate for various sources and physical mechanisms, including winds and radiation pressure from main-sequence OB stars, the collective effects of correlated supernovae, and jets and winds from AGN. At their best these models make specific predictions for the mass and energy transport, velocity and acceleration profiles, and time evolution of these flows as a function of the source properties (e.g., Murray et al. 2011, Thompson et al. 2011). These trends are often then implemented as "subgrid" prescriptions in large-scale numerical simulations that attempt to recover realistic galaxy populations and internal

## The LUVOIR Wide and Deep Fields

Since the original Hubble Deep Field (Williams et al. 1996), large-area surveys at the deepest limits have been a mainstay of galaxy evolution studies; the Ultra Deep Field, CANDELS, and the Frontier Fields have significantly advanced our understanding of galaxies. Much of LUVOIR's "Signature Science" will follow the same model, in which multiple scientific objectives are enabled by a single set of deep exposures over a large area.

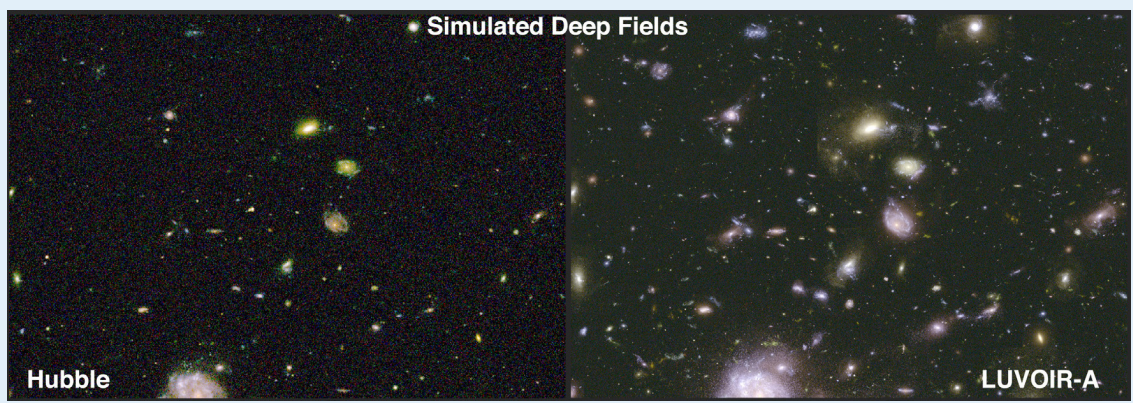

The High Definition Imager (HDI) has a field-of view of 2 x 3 arcmin, Nyquist sampled by two focal plane arrays covering NUV/VIS and NIR wavelengths that view the same portion of the sky simultaneously. Detailed calculations show that HDI on LUVOIR-A will reach 5-$\sigma$ photometric limits of AB = 33–33.5 mag for point sources in integrations of about 10 hours per band. The most basic "LUVOIR Deep Field" is a single field of view deep integration, much like Hubble's Ultra Deep Field (AB ~ 29; Beckwith et al. 2006) or Extreme Deep Field (Illingworth et al. 2013), with 10 bands taking about 100 hours of integration time to reach the limits in the table below. In the simulated view shown here, each panel (HST, LUVOIR-A) displays a field of view of 43" x 30". In terms of HDI's FOV, this scene shows only about 6% of a single HDI frame.

A LUVOIR "Wide" Field, inspired by Hubble's CANDELS program (Grogin et al. 2011), exploits LUVOIR's high mapping speed to cover 720 arcmin$^2$ in 120 tiled fields of view. If we limit this program to 1200 hours of integration time, or 10 hours per tile, 1 hour per band, the limits are AB ~ 31–32 mag, which surpasses Hubble's deepest limits by > 2 mag and matches JWST's deep limits.

These limits are unique to LUVOIR: not even 30-meter-class telescopes on the ground will reach such depths, owing to time-variable sky backgrounds. This capability enables LUVOIR to detect (1) single Sun-like stars (AB = 4.72) out to 5.5 Mpc; (2) a main sequence O star to 500 Mpc or redshift z~0.1, nearly the entire volume covered by the SDSS spectroscopic survey; and (3) a 0.001 L* galaxy at z=6, deep enough to detect the early seeds of galaxies like our own Milky Way. This is a broadly applicable capability that will advance many areas of science we contemplate as "Signature" for LUVOIR.

| The LUVOIR Deep Field: 6 arcmin$^2$ in 100 or 1000 hours | | | | | | | | | |
|---|---|---|---|---|---|---|---|---|---|
| | F225W | F275W | F336W | F475W | F606W | F775W | F850W | F125W | F160W | F220W |
| 10 hr | 33.0 | 33.2 | 33.5 | 33.6 | 33.4 | 33.0 | 32.6 | 33.2 | 33.0 | 29.7 |
| 100 hr | 34.4 | 34.6 | 34.9 | 34.9 | 34.7 | 34.3 | 33.9 | 34.5 | 34.2 | 31.0 |
| The LUVOIR Wide Fields: 720 arcmin$^2$ in 1200 hours | | | | | | | | | |
| 10 hr | 31.2 | 31.4 | 31.8 | 31.9 | 31.8 | 31.4 | 31.0 | 32.0 | 31.7 | 28.5 |

properties. These prescriptions are labeled "sub-grid" because they occur at sub-kiloparsec scales that cosmological simulations still cannot resolve.

Observers can test these physical models of feedback, but with a major limitation. Some bulk properties of galactic outflows can be revealed by spectroscopy that uses the driving sources themselves—whether AGN or star-forming regions—as background sources, in a so-called "down-the-barrel" spectrum. Inflow is detected as redshifted absorption and/or emission, while outflow is blueshifted. Often the observed profiles include absorption and emission from separate portions of the gas, which must be teased apart to reveal the details of the flow. The major limitation of current "down-the-barrel" measurements is that they typically cover most or all of a galaxy's disk, so the observed profiles average over a large number of individual sources, erasing the source-by-source variations in energy and mass that trace the key physical variables. Nevertheless, this technique has successfully demonstrated a correlation between gas velocity and, e.g., star formation rates (Martin 2005) and star formation surface density (Kornei et al. 2012), with galaxies with higher SFRs exhibiting higher outflow velocities. These signatures get stronger as the sightline approaches the galaxy's semi-minor axis, suggesting that the flows are biconical in shape and emerge up out of the disk (Bordoloi et al. 2014), as also seen in hydrodynamical simulations.

To resolve the physics at scales closer to the actual sources, astronomers using Hubble's COS instrument are making pioneering measurements of 16 individual clusters in the face-on nearby galaxy M83 (D = 4.6 Mpc). This program (Program 14681, PI Aloisi) is observing 16 UV-bright clusters across the face of M83, trying to map out the gas flows emerging from them individually. This program requires 40 orbits to execute, so doing 10 times as many clusters or a few galaxies would be a large or very large allocation of Hubble time.

**Figure 6-8** shows the potential of the LUMOS multi-object mode to transcend these limitations and fully resolve these outflows. The resolved gas flows observing program (**Appendix B.11.4**) defines observations of 100 clusters in each of 10 nearby galaxies, obtaining a sample of 1000 different star-forming regions in a wide range of galactic environments in only 100 hours of exposure time. These observations will resolve the flows at the physical scales of individual star-forming regions (~ 100 pc), allowing us to correlate outflow properties with their driving sources in detail, rather than averaged over the whole disk. Outflows can be examined as a function of the star-formation region that produced them, as a test of specific predictions for how flow velocities and mass transfer rates depend on time and energy input (Murray et al. 2011, Thompson et al. 2005). As shown in **Figure 6-8**, each of these UV spectra will contain a wealth of information about the stellar cluster (age, metallicity, stellar winds), the absorption in the flow (mass loading rates, velocities), and emission from diffuse gas (radiative transfer, LyC escape).

Not only will LUMOS enable the dissection of flows at small scales, but the multiplexing of the microshutter arrays provides a huge efficiency gain that will enable maps of resolved flows for a wide range of galaxies in the nearby Universe. It is critical to understand the wind-expulsion history of galaxies as a function of mass, as the winds are likely critical for shaping the stellar mass-halo mass and mass-metallicity relationships of galaxies, which depend strongly on galaxy mass. This is best accomplished by high-SNR, high-resolution spectroscopy with broad UV wavelength coverage. The wavelength coverage, notably to wavelengths as short as $\lambda \sim 1000$ Å, is critical. If we are to understand galaxy transformation and the role that winds may play in it, the ability to observe these flows at the relevant small scales is needed.

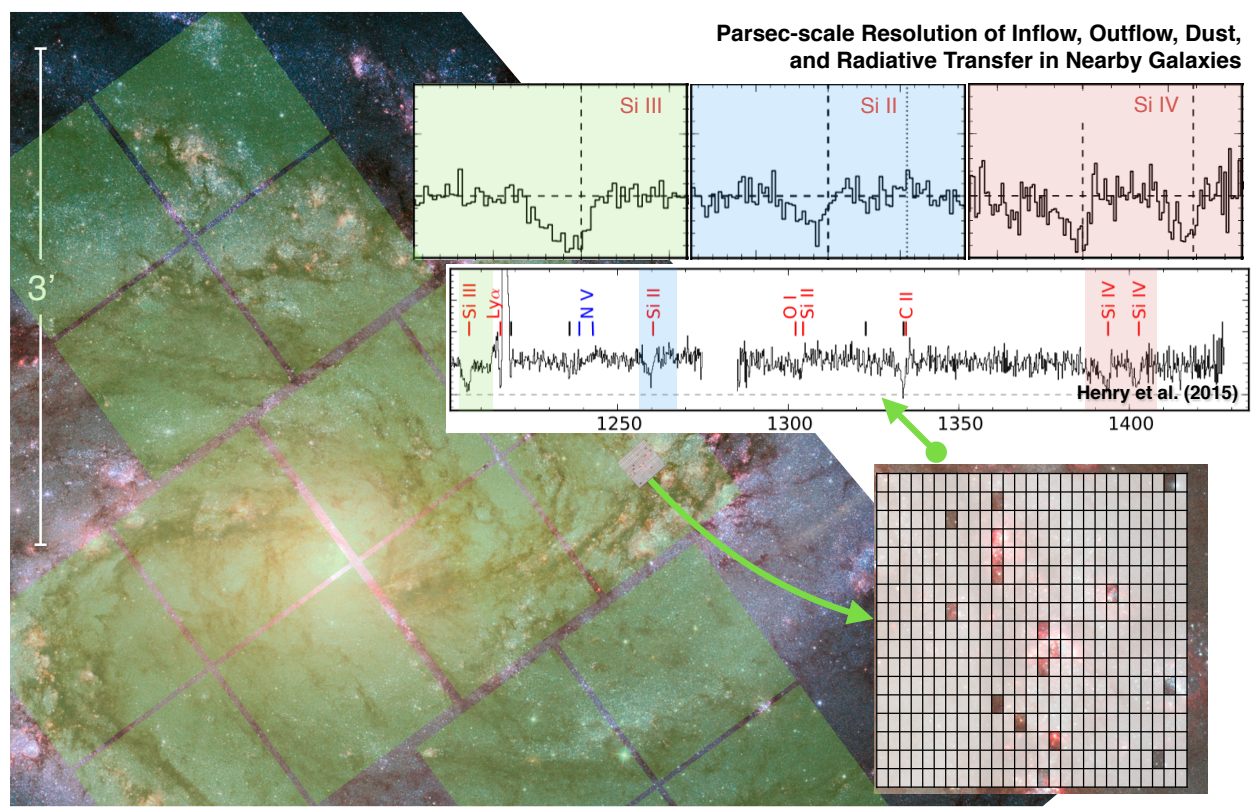

**Figure 6-8.** *LUMOS will perform intensive multi-object spectroscopy of star-forming regions, ISM gas, and galactic outflows in nearby galaxies. Here we have the footprint of the LUMOS multi-object mode overlaid on the nearby galaxy M83 at 4.6 Mpc. At this distance, the LUMOS micro shutters subtend 1.5–3 parsecs. At the top, we show three silicon lines that trace multiphase gas in outflows, as proxied by Hubble/COS spectra of low-z "green pea" galaxies by Henry et al. (2015). LUMOS users will be able to examine a wide range of ionization, metallicity, kinematics, and dust diagnostics down to 1–3 parsec scales at the positions of hundreds of individual stellar clusters and ISM simultaneously, and for many nearby galaxies in a single program.*

## 6.2 Signature Science Case #11: The multiscale assembly of galaxies

Galaxy formation is a mutiscale process spanning at least seven orders of magnitude in mass and three in size that unfolds over the 13-billion-year sweep of cosmic time, yet galaxies follow orderly scaling relations between mass, size, star formation, and metal content. This remarkable regularity challenges our current theories and even our imaginations to understand how nature does it. Part of the story is that even the largest galaxies of today began as smaller seeds at the dawn of time, and gradually built up into massive giants by acquiring gas and merging with other galaxies. Along this "merger tree" path, galaxies grow by accreting gas and merging with their neighbors large and small. Some grow enough by forming stars and merging with their neighbors to become big galaxies with prominent bulges and massive disks hosting violently active black holes. Those that grow too big, too fast can "quench," ceasing to form stars and then evolving passively as their stars age. Other galaxies remain small, perhaps eventually merging into larger ones. Nature makes galaxies of these kinds and every type in between, and no theory of origins will be complete without a full understanding of how this happened. By reaching 33–34th magnitude, LUVOIR will detect the early building blocks of galaxies like the Milky Way and fill out the merger tree that leads to galaxies at the present time. It will also see inside galaxies to unprecedented limits, resolving their internal building blocks at < 100 pc scales to unravel the processes inside galaxies that drive their evolution.

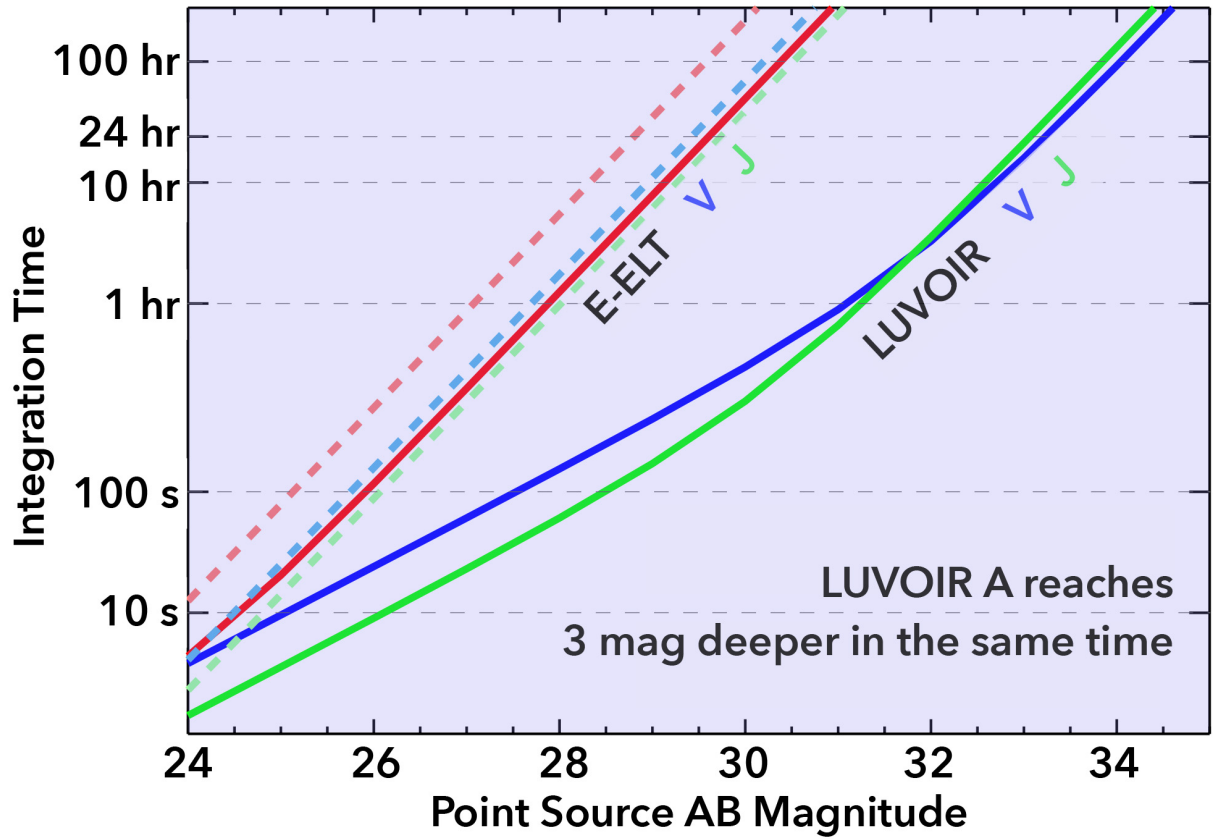

**Figure 6-9.** *Integration time required to achieve SNR = 10 for point-source photometry in three broad bands for the 15-m LUVOIR A (solid lines) and the 39-m E-ELT (dashed lines). If we take the 10-hour exposure as a nominal "large program" to which many nights or orbits would be devoted, LUVOIR reaches AB = 32.5 mag in V and J, 3.5 magnitudes deeper than E-ELT. The LUVOIR limits were derived from the HDI ETC available at* [luvoir.stsci.edu/hdi_etc](https://luvoir.stsci.edu/hdi_etc).

Why will galaxy formation remain Signature Science in the era of LUVOIR, even after JWST and the ELTs? In short, none of these highly capable facilities will be able to achieve LUVOIR's unique combination of depth, resolution, and mapping efficiency. **Depth** allows us to see the smallest building blocks (**Figure 6-9**). With enough **resolution** we can look inside galaxies at small physical scales, breaking down their formation at a level of detail that isolates the key physical processes. Finally, any observational insights must be backed by statistically significant samples collected with a high **mapping efficiency or speed**. Optimizing for these three figures of merit will enable LUVOIR's users to make revolutionary advances in the Signature Science of galaxy formation and evolution.

### 6.2.1 Galaxy assembly at the faint frontier

Pushing back the "faint frontier" has been a constant theme of galaxy formation from the earliest days, through the Hubble Deep Field, and into the present. LUVOIR will expand the faint frontier to the smallest relevant scales over nearly the whole of cosmic time. In the modern universe, galaxies occupy an enormous range of mass from "giant ellipticals" at $M^* > 10^{12}\ M_\odot$ to "ultra-faints" at $M^* < 10^4\ M_\odot$. In the prevailing hierarchal paradigm, all galaxies of any substantial size grow from the steady accumulation of gas and by the successive mergers of many smaller galactic components in a "merger tree." Every giant galaxy has many dwarfs in its past; our own Milky Way has evidently accumulated many such dwarfs over its history. Within the next few hundred million years, it will acquire its two Magellanic Clouds in a "minor merger" with an even more dramatic major merger with Andromeda to follow. Telling the full story of galaxy formation starts with being able to work backwards through this tree of galactic origins to see the roots, and this requires pushing back the faint frontier.

The brightest galaxies at any redshift will tend to be the seeds of the brightest galaxies at any later time. The galaxies we see at the dawn of time with Hubble, and soon with JWST, are not the earliest seeds of galaxies like our Milky Way. While we can probably reach Milky Way-like progenitors as far back as z = 6 with JWST (Okrochkov & Tumlinson 2010), what we see will be the main trunk of the tree, not the hundreds of other branches that existed at that time. But what is the minimum threshold for galaxy formation that LUVOIR should try to reach?

The faint frontier expanded in an unexpected direction with the discovery of "ultra-faint dwarf" (UFD) galaxies in the halo of the Milky Way. These galaxies are gravitationally bound but possess only $\sim 10^5\ M_\odot$ of stars, or even less, with characteristic radii of only 100–300 parsecs. They were detected as slight over-densities in the all-sky star map produced the Sloan Digital Sky Survey (Willman et al. 2005). The UFDs are now believed to be the ancient "fossils" of tiny galaxies that formed 80–90% of their stars before or during the epoch of reionization (Ricotti & Gnedin 2005; Brown et al. 2012; Weisz et al. 2014), like those first galaxies that preceded our own Milky Way. If so, then the ultimate "faint frontier" lies where we can detect the UFD scale over the full sweep of cosmic time.

We demonstrate the power LUVOIR will bring to such a study by deriving the plausible mass limits reached as a function of redshift in a single very deep exposure. **Figure 6-10** shows the stellar mass range that defines the "dwarf" populations of galaxies, with the UFDs at the minimum threshold (as we currently know it). The figure also shows the limits achievable by telescopes of varying size, including Hubble, JWST, and the two LUVOIR architectures. The High Definition Imager will enable LUVOIR users to reach the extreme low mass end ($M_* < 150\ M_\odot$) of the halo mass function at many redshifts in a deep survey (AB ~ 33–34) and wider surveys of $10^6$–$10^7\ M_\odot$ systems in much shorter exposures.

Only LUVOIR will be able to detect galaxies at the scale of the Ultra-Faints when they are still forming stars. This capability will reveal the entire grand sweep of galaxy formation to the earliest times, permitting full reconstruction of mass functions and merger trees. LUVOIR will also be able to trace the relationships between massive galaxies and their faintest satellites, which respond to details of the gas and star formation physics in complex ways that are still poorly understood. LUVOIR can, for instance, probe the relationship between L* galaxies and their faintest satellites, looking for evidence of when and how dwarf galaxies can be quenched by their larger, quenched neighbors ("conformal quenching") and for indications that galaxies at the UFD scale continue to form stars after reionization.

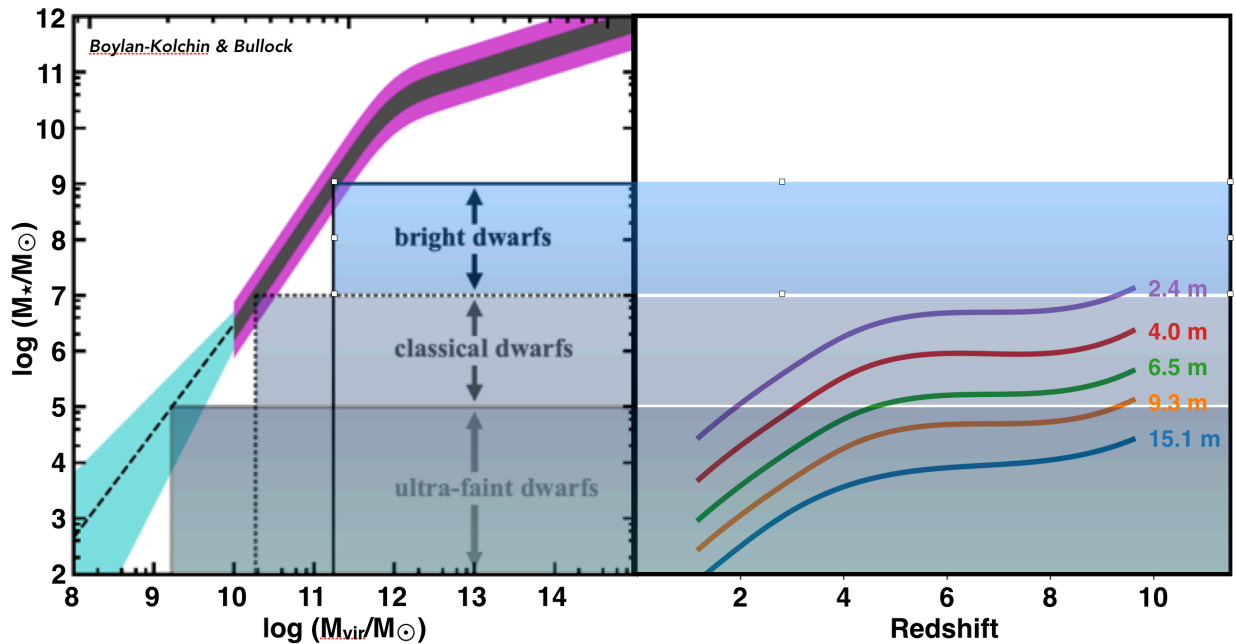

**Figure 6-10.** *LUVOIR pushes back the "faint frontier." At left, the stellar-to-dark matter ratio as a function of virial mass from Boylan-Kolchin & Bullock. Three stellar mass ranges for dwarf galaxies are shown. At right, the limits for detection by five observatories in a 500 ksec observations, assuming an extended source 200 pc in size and SNR = 5. LUVOIR's Architecture A can reach the UFD scale out to redshift z ~ 10.*

### 6.2.2 Seeing inside galaxies as they form and transform

Using Hubble, astronomers have surveyed large samples of galaxies during the rise and fall of cosmic star formation (Madau & Dickinson 2014). These studies have mapped out the relationship between stellar mass and star formation rate (Whitaker et al. 2012), and traced the rise of quenched galaxies on the "red sequence" almost back to its beginnings (Kriek et al. 2009). Perhaps the greatest surprise from this work is how small galaxies appear to be at early times. The red sequence is already identifiable at z ~ 2–3, but with most of its quenched galaxies occupying only a few kpc in size. Indeed, star forming galaxies at these redshifts average several times larger than passive galaxies at the same mass. How do these galaxies quench, and do they get smaller as they do?

Hubble itself has struggled to address what happens inside galaxies at these early times because its diffraction limit corresponds to physical scales of 300-400 pc at z = 2-3. Hubble images place only a few pixels across these small yet massive galaxies at high redshift. This is sufficient to marginally resolve galaxy disks, but not the smaller-scale processes at work within them. It is also sufficient to detect ~1 kpc star-forming clumps (Elmegreen et al. 2007, 2009; Forster Schreiber et al. 2011), but large clumps may be blended collections of smaller star-forming regions (Tamburello et al. 2017; Rigby et al. 2017; Bordoloi et al. 2016), even if the clumps are like the largest and brightest in our immediate neighborhood, 30 Doradus in the LMC (diameter d~200 pc) and the Carina Nebula (d~140 pc).

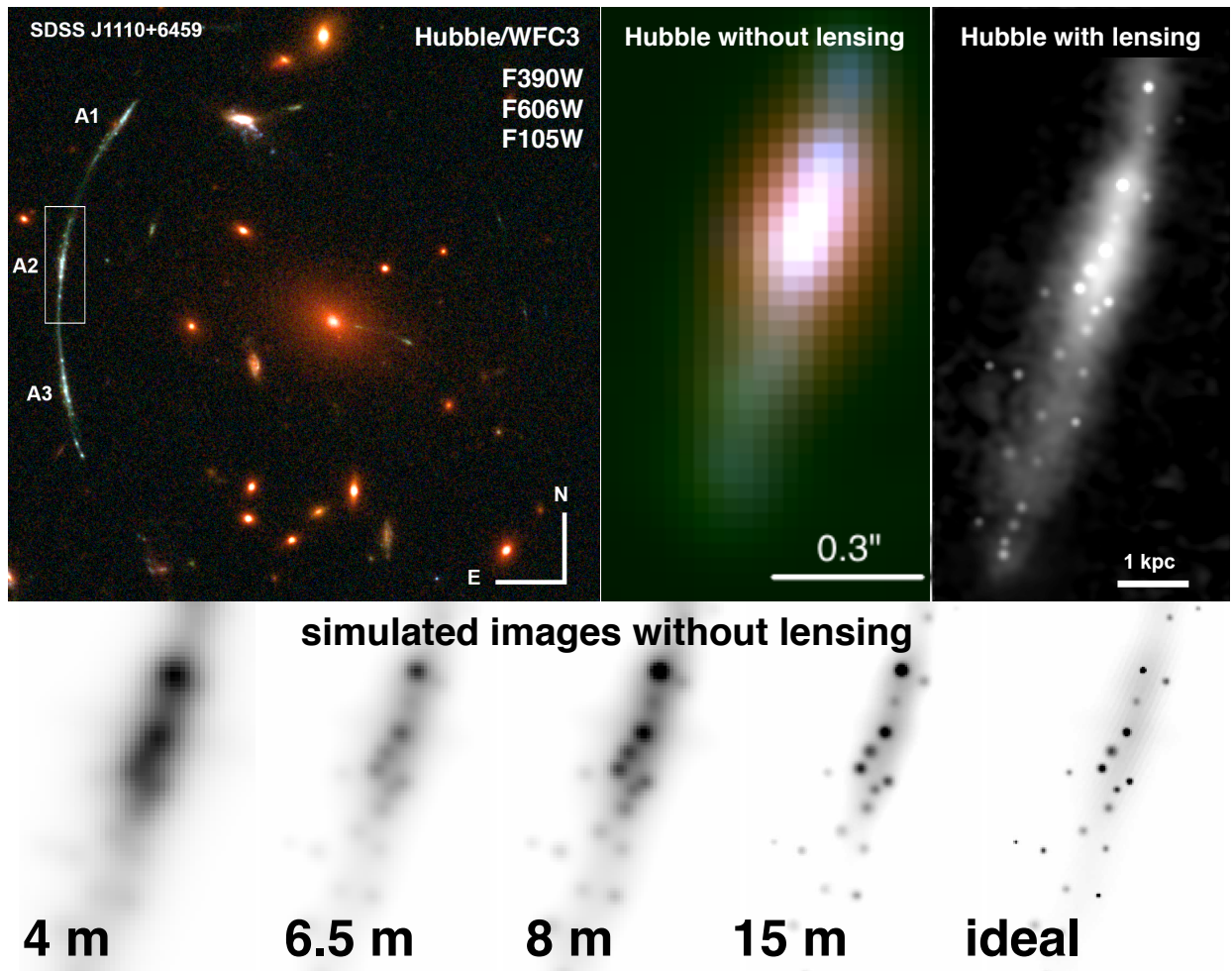

**Figure 6-11.** *LUVOIR would image distant galaxies with extremely high spatial resolution, in ways that at present are only available via fortuitous gravitational lensing. To LUVOIR, any distant galaxy appears as sharp as the best lensed galaxies with Hubble. A gravitationally lensed galaxy at z = 2.481 seen by Hubble/WFC3 (top left panel) reveals dozens of star-forming regions with radii of ~ 40 pc (top right panel). The top middle panel shows that Hubble could not resolve any of these clumps were this galaxy not lensed. The bottom panels simulate how this galaxy, were it not lensed, would appear to a large space telescope of varying size, scaling from the Hubble PSF. The simulated images show that an 8- to 15-m space telescope resolves, for an unlensed galaxy, all the structure that can be seen by Hubble with the benefits of magnification by lensing. Adapted from Johnson et al. (2017b) and Rigby et al. (2017).*

We know that galaxies have structure on 100 pc scales because of evidence from the few cases where natural gravitational telescopes are provided by foreground galaxy clusters. In these rare cases, Hubble probes individual star-forming regions in ways that illustrate the power of LUVOIR. **Figure 6-11** shows an example, where Hubble reveals two dozen star-forming clumps in a lensed galaxy, with radii of 30–50 pc. These clumps have the sizes and luminosities of the brightest star-forming regions in the nearby universe, and none would be resolvable with Hubble without lensing. The lensing reconstruction of this lensed galaxy provides a rare "truth image" of what distant star-forming galaxies actually look like. Convolving this truth image with the scaled empirical Hubble PSF shows which features

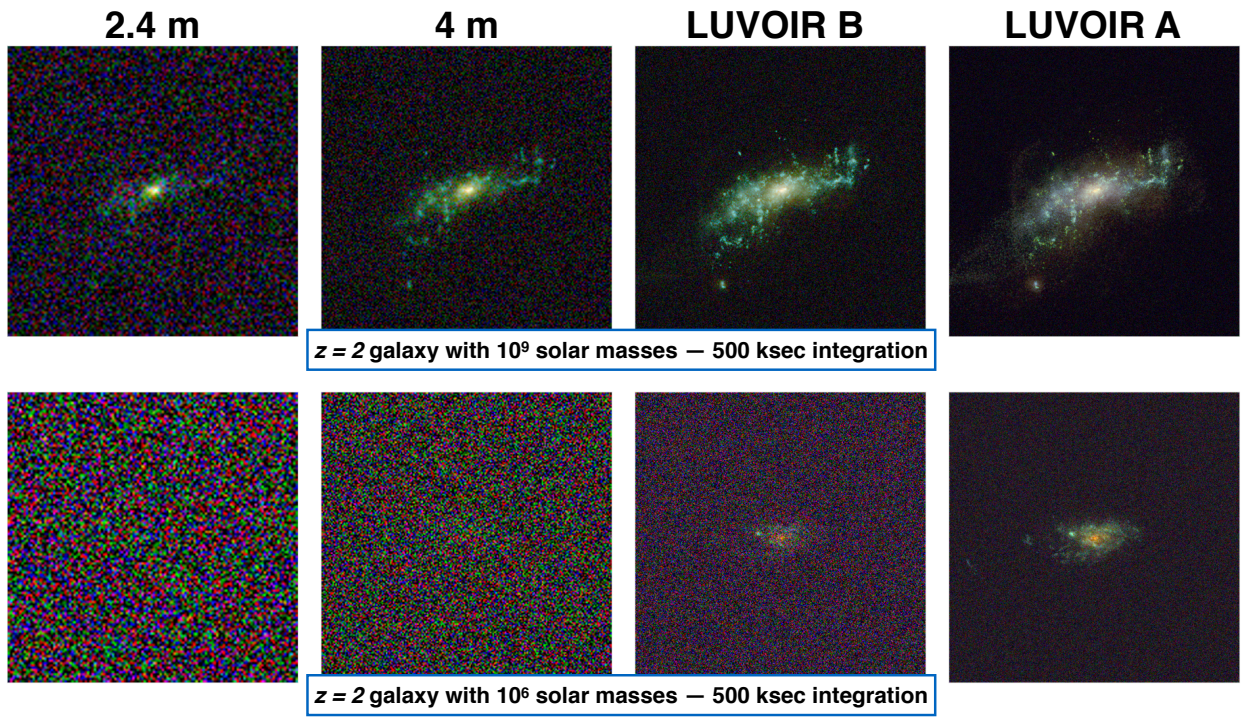

**Figure 6-12.** *Simulated galaxies showing the rich interior detail that is visible in deep exposures with telescopes of varying size. In all cases we assume extremely deep 500 ksec exposures of the $10^9$ $M_\odot$ galaxy at the top and $10^6$ $M_\odot$ galaxy at the bottom. Credit: G. Snyder and M. Postman (STScI)*

LUVOIR can recover, as a function of aperture. LUVOIR's users will be able to recover almost all the star-forming clumps in the F390W filter. This simulation shows that for a sufficiently large LUVOIR (at least ~ 8 m), any distant galaxy can be imaged with the sharpness that Hubble can now achieve for the most favorable gravitationally lensed systems, which cover only a tiny fraction of the sky. LUVOIR would therefore be able to survey star formation down to 100 pc scales for thousands of galaxies, which is terra incognita.

To illustrate the power and efficiency of LUVOIR for seeing the internal structures in galaxies at 100 pc scales, we have shown the scaling of galaxy images in LUVOIR "deep fields" as defined in **Chapter 5**. These "deep images" are seen in **Figure 6-12**, where even individual star forming regions are visible in the simulated LUVOIR-A image. These images correspond to the deep field program for Signature Science Case #9 (**Appendix B.10.2**). To illustrate the power and efficiency of LUVOIR for broader wavelength coverage, we also define a shallower addition of B and V bands to the deep I, J, H band images from Signature Science Case #9. These images will map the star-forming regions of thousands of z~2 galaxies in rest-frame UV light, for a total of only 13 additional hours with LUVOIR-A.

***Galaxy death, or quenching and compactification.*** The mystery of how galaxies "die"— or cease to form stars at any significant level—is one of the most abiding in astrophysics. Edwin Hubble's original tuning fork (1926) captured the basic truth that some galaxies are "disky," blue, and star-forming, while others are spherical, redder, and quiescent. Numerous complex and overlapping mechanisms from stellar feedback to AGN and mergers have been proposed to turn star-forming galaxies into quenched ones, but there is still no definitive physical understanding of this basic phenomenon.

The presence of fully quenched galaxies at high redshifts z > 2 is surprising enough, but even more so are the extremely compact sizes of many of these galaxies. "Ultra-compact" galaxies pack a Milky Way's worth of stars, $10^{11}$ $M_\odot$ or more, into a kiloparsec or less. These galaxies are only marginally resolved by Hubble, with > 50% of their total light falling onto just one or a few pixels. Yet these galaxies are possibly the earliest progenitors of today's massive ellipticals, and if so they are a key part of the galaxy formation puzzle.

To unravel the origins of these mysterious galaxies, we must be able to resolve their internal structures, to measure stellar content and ages of their stellar populations, to look for internal gradients in age, to follow internal dynamics, and to trace all these quantities over time as this population arises. This is Signature Science for LUVOIR owing to its exquisite spatial resolution, 100 pc or better at all redshifts, which will allow us to map the internal structures and permit age dating of stellar populations at small scales. LUVOIR's broad UV/ optical coverage and collecting area are also essential to this problem: rest-frame UV and optical colors are much more sensitive indicators of population age than are the optical / IR colors available with JWST. LUVOIR will surpass JWST in terms of both physical resolution and the diagnostics that can be applied at that resolution. The same is true of ground-based telescopes, which will exceed LUVOIR's raw spatial resolution with extreme NIR extreme AO, but will struggle to reach the same depths and cannot see rest-UV star formation diagnostics at z > 2.

### 6.2.3 Galaxies and their black holes

Supermassive black holes (SMBHs) reside at the centers of virtually all massive galaxies. The black hole masses are well-correlated with large-scale properties of their host galaxies, indicating that the SMBHs and host galaxies co-evolve. During some phases of galaxies evolution the energy feedback from the black hole might dominate the galaxy's gas accretion and star formation and ultimately quench both. Understanding the distribution of SMBH masses over cosmic time is thus a key to understanding the evolution of galaxies.

SMBH masses can be estimated from spectroscopy using the "reverberation mapping" technique, which requires high-SNR, multi-epoch spectroscopy and can be done only when the SMBH is in an actively accreting (AGN) state. Masses can also be obtained by observing the dynamical effects of the SMBH on stars within a few tens of parsecs, so called dynamical masses. This technique can work even in galaxies with quiescent nuclei, but it requires measuring stellar velocities dispersions in the small region of influence where stellar orbits are affected by the SMBH. This requirement for high spatial resolution has severely limited the number of galaxies for which the SMBH masses can measured.

**Figure 6-13** shows SMBH masses as a function of angular size distance for both quiescent galaxies (red open circles) with SMBH masses based on stellar dynamics and other techniques, and AGNs (black filled circles) with SMBH masses from reverberation mapping. The diagonal lines indicate the minimum black hole mass for a resolvable black hole radius of influence as a function of luminosity for several telescopes, including LUVOIR-A and LUVOIR-B. The goal of this project is to use LUMOS data to model the nuclear stellar dynamics to determine the SMBH masses of both quiescent and active galaxies for which the radius of influence is resolved.

Ground-based ELTs *may* be able to measure dynamical SMBH masses in quiescent galaxies, where their high spatial resolution will allow them to work close to the nucleus.

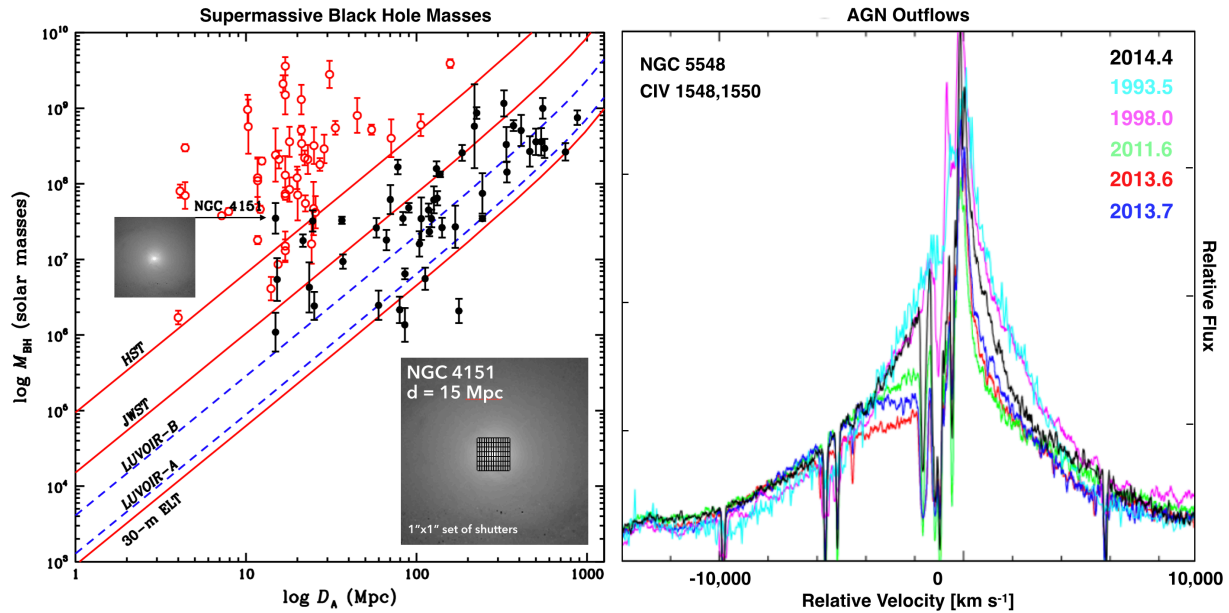

**Figure 6-13.** *AGN science with LUVOIR. At left, SMBH masses as a function of angular size distance for quiescent galaxies (red open circles) and AGNs (black filled circles). The minimum mass for which the black hole radius of influence is resolvable is shown for several telescopes, including LUVOIR-A and LUVOIR-B. Performing the experiment at the larger distances is required measure the stellar dynamical masses in the most massive AGNs (toward the right side). At right, time varying UV spectra from three different Hubble Space Telescope instruments that trace AGN outflows as a source of galaxy-wide feedback over 20 years. LUVOIR will expand this capability to fainter AGN and higher redshifts.*

However, this will require adaptive optics (AO) for the telescope to operate at the diffraction limit and will require guide stars in the field of view, which limits observations with the current largest ground based telescopes to about 20% of the available bright galaxies. However, it is unlikely that ELTs will be able to measure dynamical masses for SMBHs in luminous AGN, because at their expected Strehl ratios (even with AO) the scattered nuclear light will outshine the starlight.

The high angular resolution of LUVOIR/LUMOS enables measurements of the masses of both very large (more than a billion solar masses) and very low (less than a million solar masses) SMBHs in both active and quiescent galactic nuclei. LUVOIR's PSF is smaller and the LUMOS focal-plane microshutters on the nucleus can be closed, thus providing a crude coronagraph. As seen in the figure, it will be possible to measure stellar dynamical masses for the most massive (greater than $10^9$ solar masses) and luminous AGN.

The spectra of active galactic nuclei (AGNs) show strong, broad absorption features in the blueshifted wings of the resonance emission lines, revealing massive, high-velocity outflows of gas. It is widely believed that these outflows, by transferring energy and momentum to the host galaxy interstellar medium (ISM), are agents in quenching star-formation by heating or completely removing the ISM. This is a favored explanation for both the steep cutoff at the bright end of the galaxy luminosity function and the tight correlation between the masses of the central black holes and larger-scale properties of the host galaxies. However, the role of AGN outflows remains largely speculative without more accurate determination of the kinetic luminosity and momentum flux of the outflows. This requires repeated measurements of the physical properties of the outflows, which requires high SNR UV spectroscopy and detailed modeling.

For local AGNs, the timescales for changes in these outflows are short, weeks to months (e.g., Kaastra et al. 2014; **Figure 6-13**. Recent observations associate these outflows with accretion disk winds that are apparently triggered by increases in the AGN luminosity which is in turn driven by an increase in the accretion rate (e.g., Kriss et al. 2019). The role of these outflows in galaxy evolution remains elusive, and the key to resolving remaining ambiguity is tracing their evolution over timescales much longer than the dynamical timescales.

Just as argued above for diffuse gas around galaxies, UV wavelengths are critical to probe these AGN outflows. The most important broad absorption features—the hydrogen Lyman series, O VI 1032, 1038 Å, N V 1239, 1243 Å, Si IV 11394, 1403 Å, and C IV 1548, 1551 Å (**Figure 6-13**)—are all in the rest-frame UV, and are currently best studied in local systems where the dynamical timescales are short and it is possible to obtain high SNR spectra in a short amount of time with LUMOS. With LUVOIR, it will be possible to obtain high-quality spectra not only of nearby AGNs, but of fainter, higher-redshift AGNs as well. Over the redshift range $0.2 < z < 2.0$, highly ionized species such as Ne VII 770, 780 Å, Mg X 610, 625 Å, and Si XII 499, 521Å become observable with LUMOS, enabling studies of gas that is currently observable only in X-rays.

## 6.2.4 Dissecting galaxies one star at a time

Our ability to determine when galaxies assemble their stellar populations, and how that process depends on environment is a fundamental component to any robust theory of galaxy formation. By definition, the dwarf galaxies we see today are not the same as the dwarf galaxies and proto-galaxies that were disrupted during assembly. Our only insight into those disrupted building blocks comes from sifting through the resolved field populations of the surviving giant galaxies to reconstruct the star-formation history, chemical evolution, and kinematics of their various structures (Brown et al. 2010). Resolved stellar populations are cosmic clocks and assay meters that can assess the age and metallicity of their galaxies using well-defined relationships obeyed by stellar luminosity and color. Their most direct and accurate age diagnostic comes from resolving both the dwarf and giant stars, including the main sequence turnoff. As demonstrated by the Panchromatic Hubble Andromeda Treasury program (Dalcanton et al. 2012), fitting stellar evolution models to color-magnitude diagrams (CMDs) can reveal how episodic star formation is and how it varies across the disk. (e.g., Williams et al. 2014). This work highlights both the need for the survey depth required to obtain a CMD accurate enough to differentiate between models and the ability to map across as much of a galaxy as possible.

Unfortunately, the main sequence turnoff is too faint to detect with any existing telescope for galaxies beyond the Local Group, and beyond this the increasingly severe effects of crowding become an even more fundamental limit. Both effects greatly hinder our ability to infer much about the details of galactic assembly because the galaxies in the Local Group are not representative of the galaxy population at large. LUVOIR will transform our ability to determine stellar histories by leveraging both light gathering power and spatial resolution, extending our reach beyond the Local Group to a more diverse sample of galaxies.

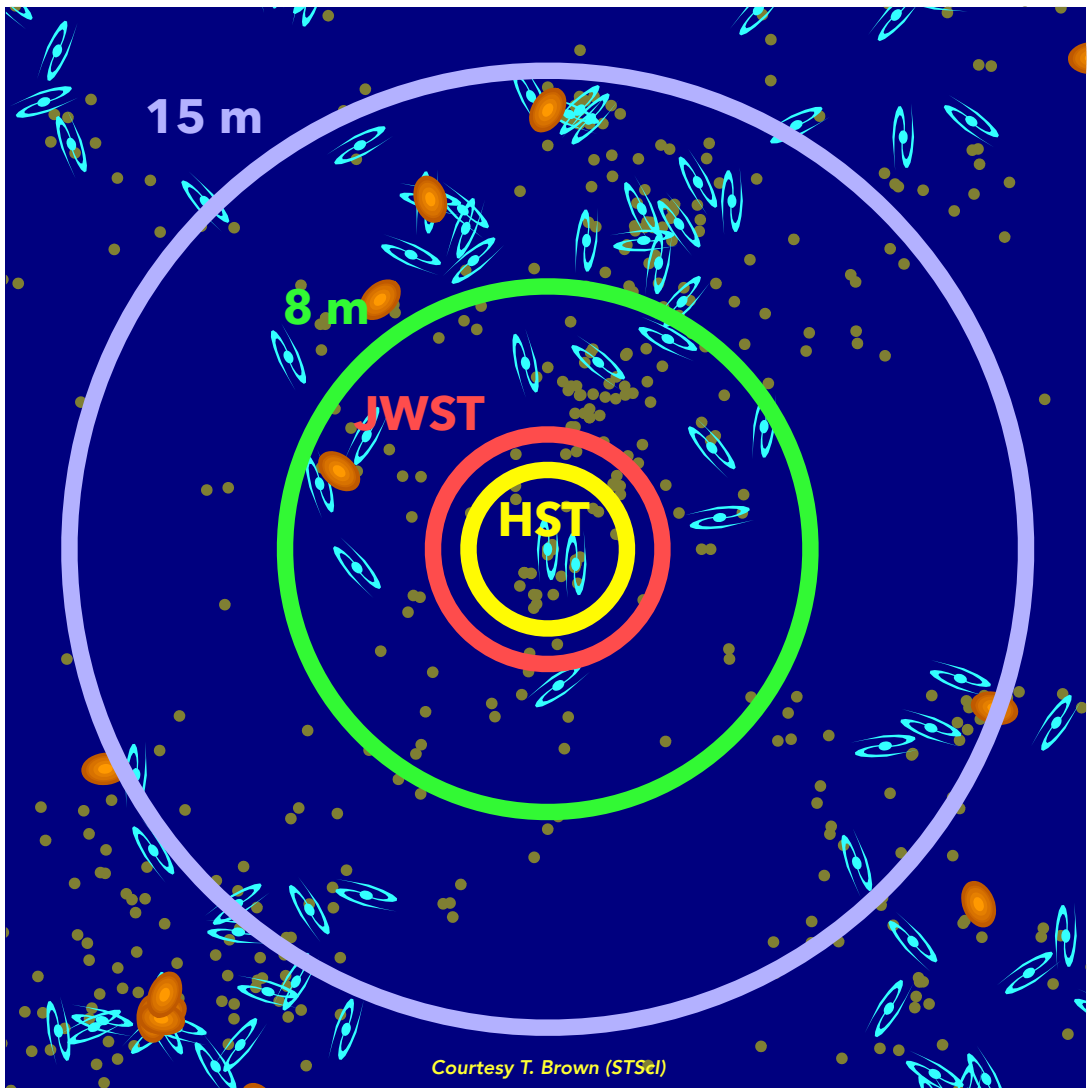

**Figure 6-14.** *Map of local universe (24 Mpc across) shown with the distances out to which HST (yellow), JWST (red), and LUVOIR (8-m, 15-m, green and violet), can detect the main sequence turnoff from a CMD in V and I passbands at SNR=5 in 100 hours. Giant spirals, like M31, are indicated by the blue galaxy symbols, giant ellipticals as orange blobs, and dwarf galaxies as small green dots. Derived using the HDI exposure time calculator at luvoir.stsci.edu/hdi_etc.*

We consider two types of surveys by way of example (**Appendix B.12.4**). First, we consider the distances that can be reached for characterization of diffuse (i.e., non-crowded) stellar populations in nearby galaxies. **Figure 6-14** shows a map of the local Universe with galaxies in their actual positions and marked by type. This observing program plans to reach stars at the main-sequence turnoff (MSTO) to age-date the star formation histories of eight ≥ L* galaxies in the local neighborhood. For LUVOIR-A, the sample includes six ~L* galaxies, two of which are early-type, for a total of 407 exposure hours. For LUVOIR-B we have defined a less ambitious sample of four L* galaxies, of which only one is early-type, for a total of 467 hours. LUVOIR-B would require more than 2000 hours to observe all eight of the galaxies in the LUVOIR-A sample. Both samples reach the MSTO at AB ~ 33–34, extremely deep observations beyond the reach of even JWST or 30m ground-based telescopes. For diffuse populations, this capability could also be used to probe faint dwarfs or age-date the outer regions of massive galaxies. At closer distances, these observations can be collected over time to build up the timeline of proper motion and work out galaxy motions in 3D (see **Chapter 5**).

For a second type of survey, we consider regions where crowding comes into play. LUVOIR's spatial resolution will enable usable photometry at much higher stellar densities than smaller telescopes. The PHAT program has empirically determined that usably accurate photometry can be obtained in UVOIR images up to a surface density of ~15 stars per square arcsecond with Hubble. Scaling this limit up by $D^{-2}$, we find that LUVOIR-A could detect the main sequence turnoff out to 5 Mpc, working with up to 400 stars per square arcsecond before crowding becomes prohibitive (**Figure 6-15**).

LUVOIR will work in concert with 30-m-class ground-based telescopes expanding our reach to other well-populated galaxy groups, with LUVOIR obtaining photometry of G dwarf stars down to V~34 mag, and the ground obtaining kinematics of much brighter giants out to the Coma Sculptor Cloud. The dwarf stars in the Coma Sculptor Cloud are effectively inaccessible from the ground, requiring gigaseconds of integration even for an isolated star. This capability for studies of resolved and semi-resolved stellar populations has a wide range of applications from mapping the history of star formation in galaxies to assessing the impact or reionization at the smallest scales.

### 6.3  Signature Science Case #12: Stars as the engines of galactic feedback

Finally we arrive at the most numerous sources of feedback on galaxies—the stars themselves. Stars, particularly massive stars, return energetic radiation to their environments throughout their lives, and then substantial amounts of kinetic energy and heavy elements when they explode as supernovae. With its high spatial resolution and uniquely powerful UV spectroscopy, LUVOIR will make fundamental contributions to the study of massive stars.

Stars form from the fragmentation of the dense parts of molecular clouds, in regions referred to as "cores." For all the simplicity of this description, significant challenges remain for a quantitative formulation of star formation. The relation between cores and stars is still matter of debate, as is the origin of the distribution of stellar birth masses (the stellar initial mass function, or IMF), the influence of the surrounding environment, and the effects of early dynamical evolution.

The reasons for these debates are the significant observational challenges faced by measurements of the relevant quantities (cores and stars at different stages of formation and early evolution), and the fact that *we do not have a predictive theory covering all stages of star formation*. Extant models are able to describe individual observational results, but not link them.

Without a theory, we cannot unequivocally interpret the observations of resolved or unresolved stars; and without unambiguous measurements, we cannot formulate a theory. In order to break this circular ambiguity, we need to perform unambiguous measurements of the early stages of star formation and the stellar IMF in a vast range of environments and parent galaxy properties, to guide models and finally achieve one of the key goals of modern astrophysics: a theory of "how stars form."

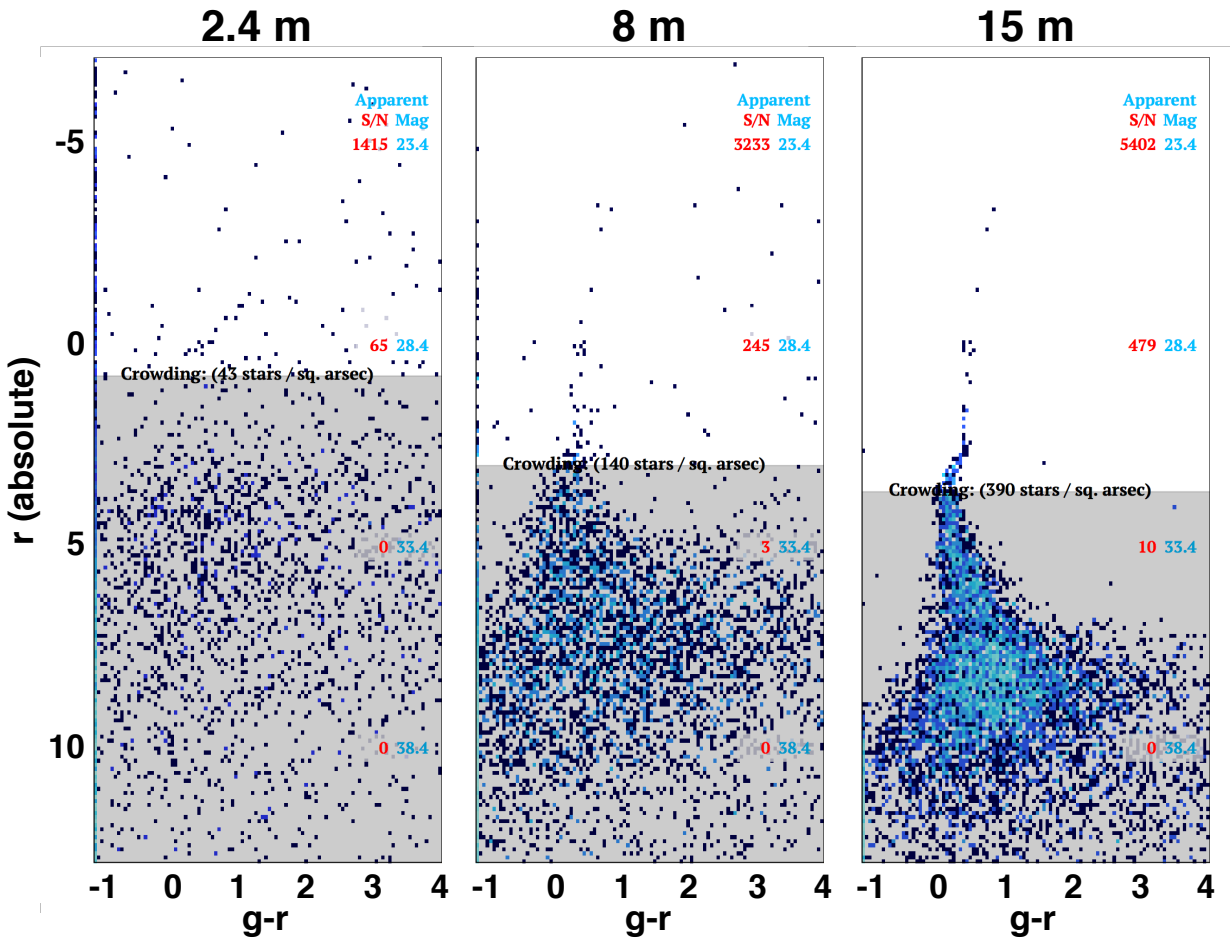

**Figure 6-15.** *Simulated CMDs for a single-age solar-metallicity population. The simulation uses a 10 Gyr-old isochrone from FSPS (Conroy et al.), placed at 5 Mpc and "observed" in 50 hours each in the g and r bands. To compute the crowding limit, we assume AB = 23 / arcsec². The Hubble / WFIRST analogue (2.4 meters) cannot overcome crowding and poor photometry. The 8-m LUVOIR can resolve the giant branch before crowding becomes too severe at the sub-giants. The 15-m LUVOIR can reach to the top of the main sequence to apply age and metallicity diagnostics. These figures were made with the LUVOIR CMD simulation tool at luvoir.stsci.edu/cmd.*

Accomplishing the characterization and quantification of star formation and the stellar IMF will require the synergistic combination of observations from a broad wavelength range, UV to millimeter. This combination is partially offered by current (Hubble, ALMA, VLT, etc.), upcoming (JWST), and planned (30m-class ground-based telescopes) facilities. Among the capabilities that are neither available nor planned, and which will be required to accomplish the overarching goal of formulating a theory of star formation, are:

1. Angular resolution for imaging: 0.007" at 500 nm (**Figure 6-16** and **Figure 6-17**), with stable point spread function (PSF) and stable astrometry across the entire field of view (FoV).

2. UV and optical imaging spectroscopy over 1'–3' FoV, with 0.01"–0.02" angular resolution.

High-

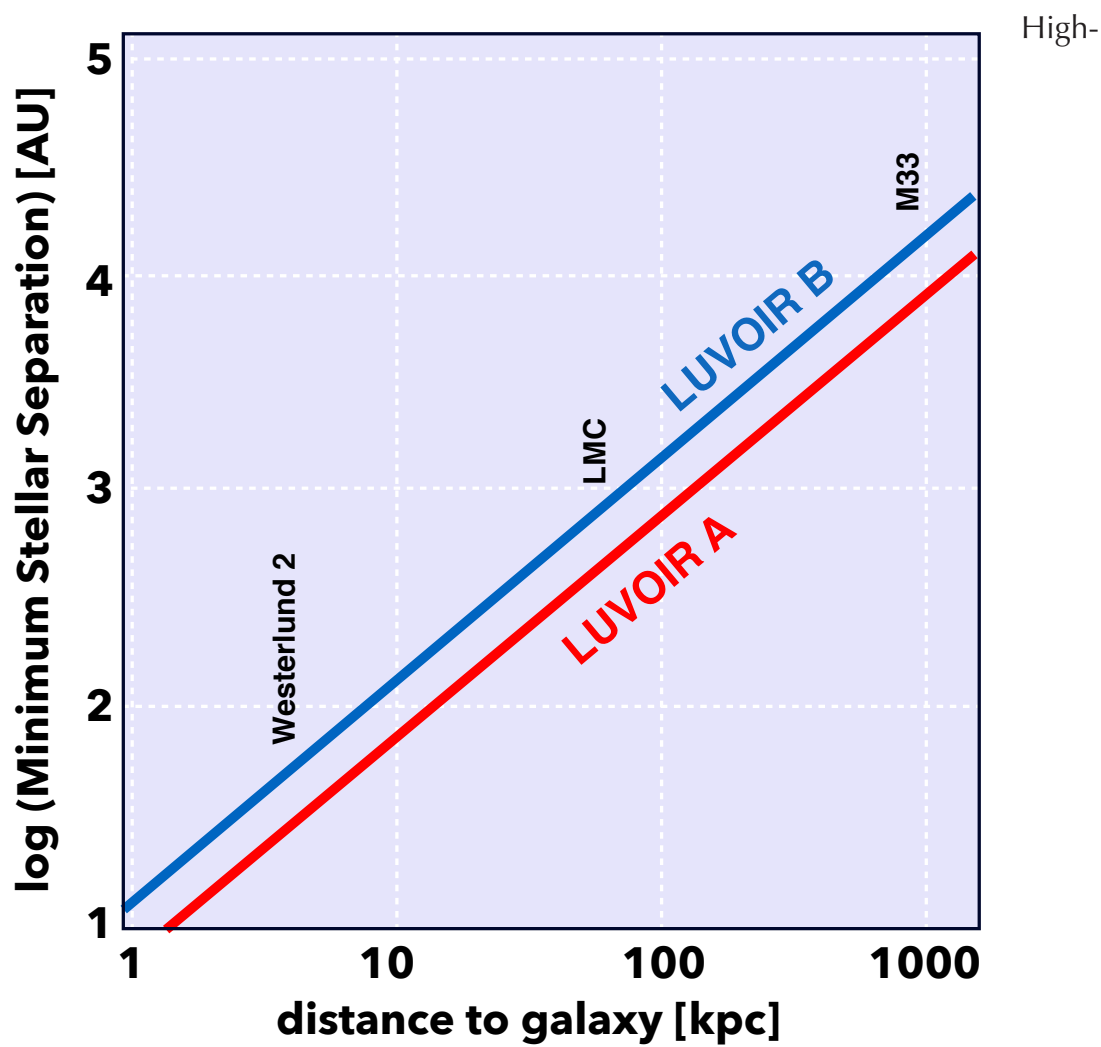

**Figure 6-16.** *The resolvable separation, in AU, between two stars as a function of distance of the stars from us, for both LUVOIR-A (red line) and LUVOIR-B (blue line). The distances of three representative systems: the galactic star-forming region Westerlund 2, the Large Magellanic Cloud and the galaxy M33, are marked. Binary stars at different stages of evolution will be resolvable by LUVOIR throughout most of the galaxies in the Local Group. Massive star signatures will be detectable in young star clusters resolved in galaxies out to about 150 Mpc.*

sensitivity UV spectroscopy with resolving power R > 3000, to resolve atomic and molecular lines from winds and photospheres in stellar spectra.

Determining the nature, average properties and relevant physical parameters, and potential environmental dependencies of star formation (and its corollary: the stellar IMF) across the full stellar mass range, from the high to low mass ends, will require synergy across multiple wavelength regions, from the UV to the mm. Important characteristics are poorly constrained, including merger rates, binary fractions below O stars, multiplicity, fraction of runaway stars, and whether very massive stars (VMSs, i.e., stars with masses > 150 $M_\odot$) exist and are common. In addition to its primary and unique role in achieving a full understanding of star formation, LUVOIR will play a complementary role in characterizing the shocks,

**LUVOIR 15m**                    **4m**

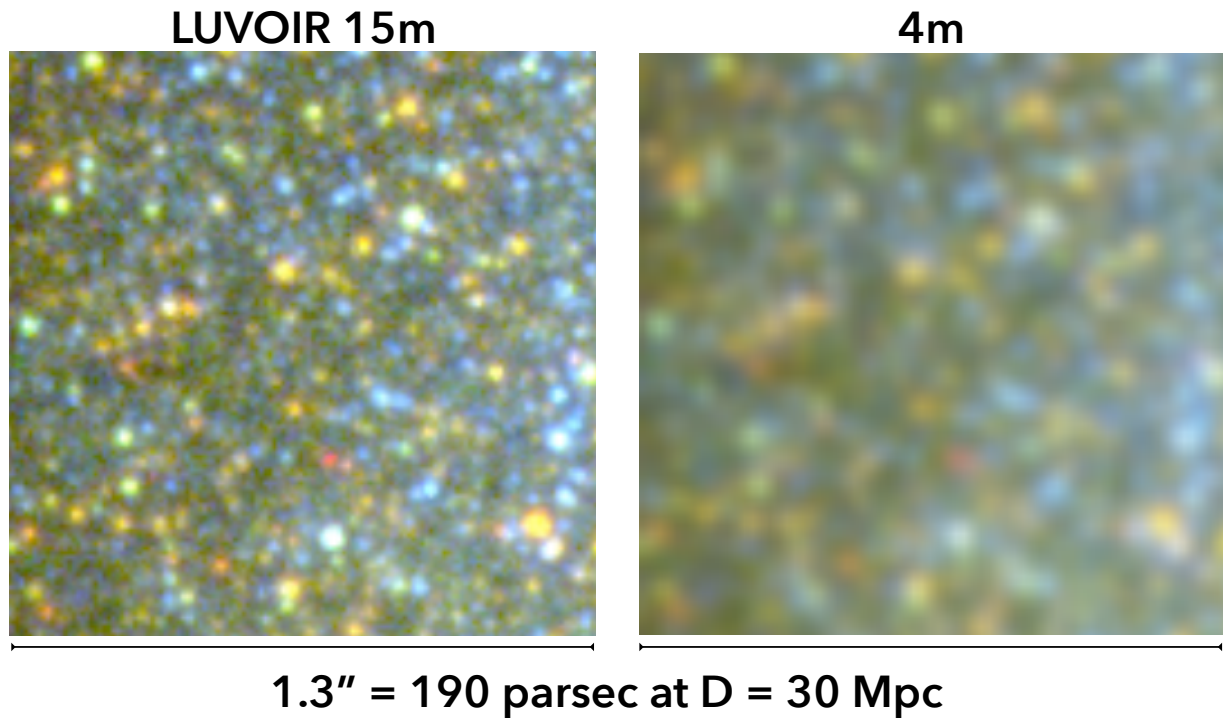

**1.3″ = 190 parsec at D = 30 Mpc**

**Figure 6-17.** *The inner region of a dwarf galaxy as viewed by LUVOIR-A and a 4-m telescope at a distance of ~31 Mpc. Each box is 1.3″ on a side, corresponding to 31 pc. The 4-m image would need ~16 times the exposure time as the LUVOIR image. The images are obtained by degrading the HST images of a region of the galaxy IC 4247, located at 4 Mpc distance (Calzetti et al. 2015).*

jets, and outflows from the individual protostars and protoclusters, and their natal cores, identified by ALMA.

Below are two example science cases that push to the limit of LUVOIR capabilities, and provide a benchmark for the large science output that LUVOIR promises in this area.

### 6.3.1 Very massive stars

VMSs are stars that exceed the standard limit of 150 $M_\odot$, and one needs to observe very young ($\lesssim 1.5$ Myr), massive star clusters ($> 10^5$ $M_\odot$) in order to detect their presence, due to small number statistics at the high end of the IMF, and rapid evolutionary timescales. Yet these stars can heavily influence their surrounding environment; for instance, they can provide between 25% and 50% of the ionizing photon flux from the host cluster.

A solid case for the presence of VMSs in R136, the central cluster in 30 Dor in the Large Magellanic Cloud, has been recently made by Crowther et al. (2010, 2016). UV spectroscopy has provided evidence for the potential presence of VMSs in an additional two unresolved star clusters in NGC 3125 (Wofford et al. 2014) and NGC 5253 (**Figure 6-18**; Smith et al. 2016). Detection of supernovae that have characteristics typical of pair-instability SNe provide indirect evidence for the presence of VMSs. Whether the VMSs are the result of birth conditions or of mergers is still a matter of debate, but models can successfully reproduce many observational properties using a rotating single VMS (Yusof et al. 2013, Kohler et al. 2015). Data on the VMS numbers and frequency in young star-forming regions are lacking, due to observational limitations: they can only be recognized in the UV.

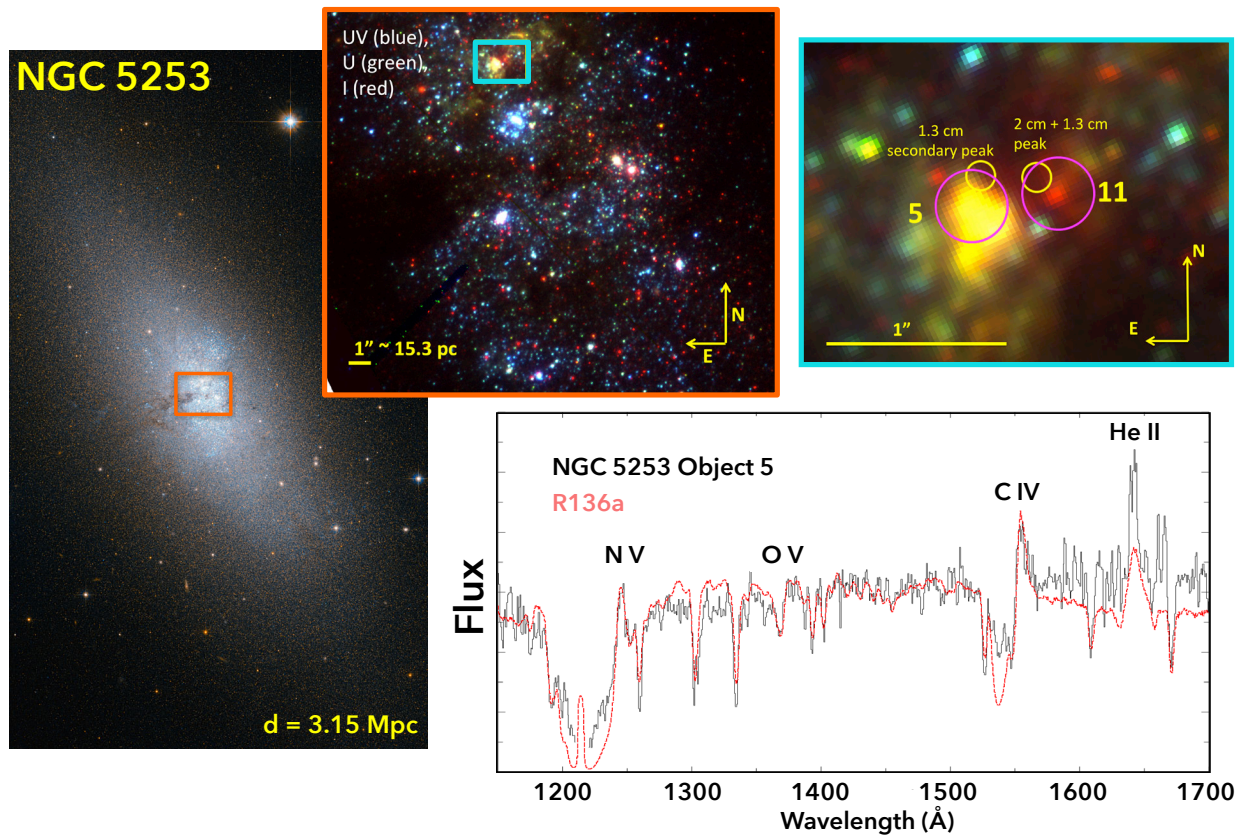

**Figure 6-18.** *LUVOIR will increase by at least tenfold the parameter space in distance and galactic properties (morphology, dust content, star formation rate, etc.) of the systems whose VMS content can be identified and characterized. Identifying very massive stars (VMSs) in star clusters. NGC 5253 is an amorphous dwarf hosting a starburst in its central ~250 pc. Most of the dust and molecular gas in this galaxy are concentrated in a ~20 pc region located at the north end of the starburst (right-top panel), where also the two main radio peaks are located (yellow circles in the top-right panel; Turner et al. 2000). The two radio peaks coincide with two clusters identified in the HST images (magenta circles in the top-right panel; called 5 and 11 by Calzetti et al. 2015). The two clusters are ~1 Myr old, with masses 1–3 × 10^5 $M_\odot$, and with AV ~ 2–50 mag of dust. UV spectroscopy of cluster 5 further reveals the presence of VMS signatures: P Cygni N V (1240 Å) and C IV (1550 Å) profiles, broad He II (1640 Å) emission, blue-shifted O V (1371 Å) wind absorption, and absence of Si IV (1400 Å) P Cygni emission/absorption (bottom panel; Smith et al. 2016), as confirmed from the comparison with the UV spectrum of the LMC cluster R136a, also containing VMSs (Crowther et al. 2016). The HST UV spectrum of cluster 5 required 3 hours of exposure. The same amount of time with LUMOS on a 15-m will detect with SNR ~ 10 a $10^6$ $M_\odot$ cluster with similar dust properties as cluster 5 out to a distance of 80 Mpc (25 times further away than NGC 5253). Credit: D. Calzetti (U Mass)*

A census of the frequency of VMSs across the full range of environments present in galaxies is required to gauge their overall impact on the evolution of galaxies, and to constrain models for their formation and evolution. The presence of VMSs can be inferred using unique UV spectral signatures: P Cygni N V (1240 Å) and C IV (1550 Å) profiles, broad He II (1640 Å) emission, blue-shifted O V (1371 Å) wind absorption, and the absence of Si IV (1400 Å) P Cygni profiles. In the optical, a VMS would be easily confused with WR emission from a lower mass O-star. Thus, UV spectroscopy is a key requirement for obtaining a census of VMSs and other massive stars in the nearby Universe, together with quantitative measurements of their properties.

UV multiplexing (multi-object or integral field spectrograph) to obtain spectra of resolved massive stars in clusters out to 0.7–1 Mpc and spectra of unresolved young, massive star clusters out to 150 Mpc (Arp 220, the prototype ULIRG, is at 77 Mpc distance) is a minimum requirement, and will secure about 200 LIRGs and ~20 ULIRGs, thus sampling the full range of galaxy properties. Large detector formats increase efficiency by covering entire star clusters/star-forming regions and/or entire star cluster populations in a single pointing. LUVOIR-A will ensure both enough sensitivity and angular resolution (0.01" at 80 Mpc corresponds to 4 pc, the size of star clusters, Ryon et al. 2017) for recovering individual star clusters and the signature of the massive stars they contain. As such, LUVOIR+LUMOS represents an ideal combination for VMS studies, and will provide more than a 1,000 increase in the cosmic volume that can be probed with Hubble in the UV.

### 6.3.2 Stellar multiplicity

Stellar multiplicity and binary frequency constrain models of star formation and the IMF (Offner et al. 2014). Hydro-dynamical simulations show that massive stars require dense and massive accretion disks to form, as these are needed to overcome the radiation pressure barrier. The disks tend to be unstable and break into complex systems. The resulting properties of the multiple systems (number of companions, distribution of separations [i.e., short vs. long orbital periods], distributions of mass ratios, and their dependence on the stellar mass) are model-dependent. Furthermore, the dynamics of binaries drive the evolution of star clusters while, simultaneously, the combination of cluster dynamics and internal stellar processes determine the internal evolution of each binary (Portegies Zwart et al. 2010), but, again, there is enormous dependency on uncertain parameters.

Short-period (spectroscopic) binaries will remain the domain of ground-based telescopes, especially the upcoming integral field spectrographs on adaptive optics-assisted (AO-assisted) 30-m+ class telescopes. Long-period binaries require high angular resolution, a very stable PSF, and high precision photometry across the entire FoV, which needs to be a few arcminutes on a side, in order to increase efficiency by targeting each cluster with as few pointings as possible. LUVOIR-A will be able to achieve accuracies as high as 0.5 μas/ yr over 5 years (**Appendix B.8**), which will increase the cosmic volume over which proper motions can be measured by 8,000 times over the AO-assisted ELTs.

In addition, the physical properties of the stars need to be characterized in order to constrain models. This demands measurements of the ***resolved*** massive stars' winds and photospheric parameters (including bolometric luminosities and masses), which can be accomplished only with spectroscopy of individual stars in the 1000–2000 Angstrom range (Wofford et al. 2012). Binaries at different stages of evolution within a star cluster have mean separations that change between a few tens and a few thousand AUs (Portegies Zwart et al. 2010). Resolving binaries with UV spectroscopy *at all stages of evolution* requires resolving mean separations across a wide range from tens to hundreds of AU; probing a range of environments requires reaching at least the distance of the Magellanic Clouds. LUMOS on LUVOIR-A will resolve binary stars with separations down to ~40 AU in the iconic high-mass star cluster Westerlund 2 (~4 kpc from the Sun) and ~500 AU in the LMC.

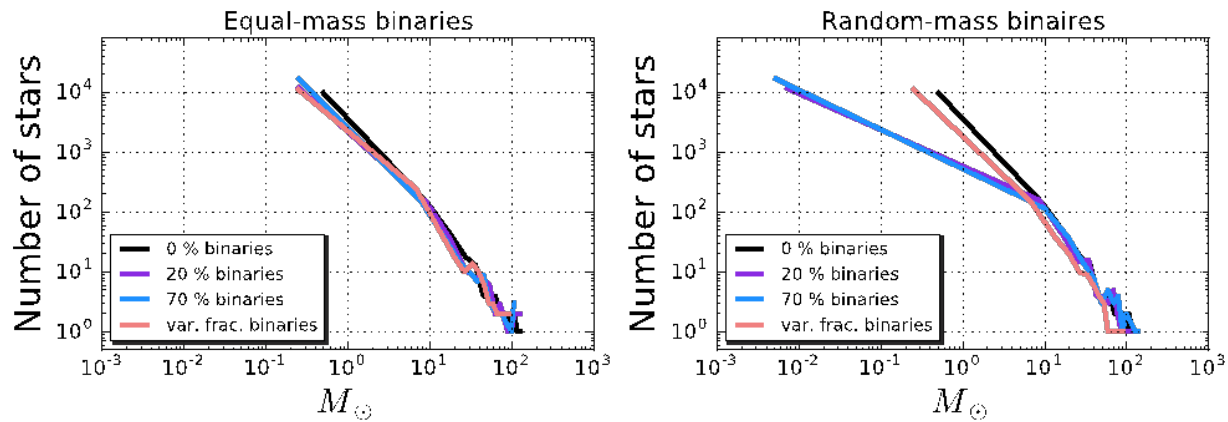

**Figure 6-19.** *Simulations indicate that very non-standard IMFs can nevertheless appear to be a common Salpeter IMF if binary stars are not properly resolved. These two panels show how different from Salpeter the IMFs can be and still be mistaken for this standard IMF. The effect is particularly large in the case of random-mass binaries (at right) where the high mass end can vary widely and still be "observed" as Salpeter. LUVOIR will be able to directly resolve these binary systems and measure the high-mass departures from standard IMFs.*

The diffraction-limited angular resolution of HDI on LUVOIR-A (0.007") will resolve and image all binary stars down to ~30 AU separation in Westerlund 2 (**Figure 6-19**), and down to smaller separations in the many dozens of intervening star-forming regions spanning a wide range of densities and other properties. A 30 AU separation is found for a system of two equal-mass late-type B stars (~3 $M_\odot$ each) with a 25-year orbital period. The tens-AU regime is interesting for the investigation of the progenitors of peculiar objects and Gamma ray bursts (GRBs), as models suggest massive stars may be interacting in the post-MS phase. This will also address numerous additional related questions, including the IMF of binary stars, and the role of binary stars in determining the upper end of the IMF. As shown in **Figure 6-19** the Salpeter IMF, which is often observed in resolved star clusters, could be the result of a significantly different actual (input) IMF, if the stars in the cluster are unresolved binary stars with random mass ratios (right panel). Overall, LUVOIR will provide both the sensitivity and angular resolution to probe more widely separated binary systems throughout most of the Local Group galaxies, thus constraining models of star formation.

**Table 6-1.** *Chapter 6 Programs at a Glance*

| Goal | Program Description | Instrument + Mode | Key Observation Requirements |
|---|---|---|---|
| **Chapter 6 Programs at a Glance** | | | |
| *Signature Science Case #10: The Cycles of Galactic Matter* | | | |
| Measure mass, metals, energetics, and fate of CGM gas that feeds galaxies | UV spectroscopy of 100 quasars (z > 1) to study absorption column densities of atoms and ions over a range of temperatures ($10^4$–$10^7$ K) and densities ($10^{-6}$–$10^2$ cm$^{-3}$) | LUMOS NUV and FUV point-source spectroscopy | Telescope diameter $\gtrsim$ 8 m<br>Bandpass: 100–400 nm<br>$R \gtrsim 30,000$<br>Telescope mirror reflectivity > 60% at 105 nm |
| Map the outflow from the starburst galaxy M82 at sub-parsec scales | Measure content and kinematics of M82 superwind at > 300 positions in flow, using emission in lines of Lyα, O VI, C IV, and Mg II | LUMOS FUV and NUV multi-object spectroscopy | Telescope diameter $\gtrsim$ 8 m<br>Bandpass: 100–400 nm<br>$R \gtrsim 30,000$<br>Telescope mirror reflectivity > 60% at 105 nm<br>MOS FOV $\gtrsim$ 4 sq. arcmin |
| Determine the spatial distribution, kinematics, metal content, and large-scale structures of the halo around a nearby galaxy | Measurements of absorption column densities and velocities for hot halo gas (Lyman-β, O VI, C IV, and Mg II) out to 200 kpc around M51 via examination of sight lines to 30 quasars through the CGM | LUMOS FUV and NUV point-source spectroscopy | Telescope diameter $\gtrsim$ 8 m<br>Bandpass: 100–400 nm<br>$R \gtrsim 30,000$<br>Telescope mirror reflectivity > 60% at 105 nm |
| Resolve galactic accretion and feedback at sub-parsec scales | Measure gas absorption and emission (Lyman-β, O VI, C IV, and Mg II) from ~1000 stellar clusters in nearby spiral galaxies (< 10 Mpc) | LUMOS FUV and NUV multi-object spectroscopy | Telescope diameter $\gtrsim$ 8 m<br>Bandpass: 100–400 nm<br>$R \gtrsim 30,000$<br>Telescope mirror reflectivity > 60% at 105 nm<br>FOV $\gtrsim$ 4 sq. arcmin |
| *Signature Science Case #11: The Multiscale Assembly of Galaxies* | | | |
| Determine how galaxies form at early times by surveying star formation in young galaxies | Study star formation at 100 pc scales within thousands of galaxies at a range of redshifts | HDI Imaging | Telescope diameter $\gtrsim$ 8 m<br>Photometric sensitivity sufficient to reach AB = 30.5 in 1 hour |
| Characterize the age and metallicity of stellar populations in different types of large galaxies | Detect main sequence turnoff in galaxies beyond the Local Group, to reach the nearest giant ellipticals (8 galaxies total) | HDI Imaging | Telescope diameter $\gtrsim$ 8 m<br>Photometric sensitivity sufficient to reach AB ~ 32 in 1 hour |
| *Signature Science Case #12: Stars as the Engines of Galactic Feedback* | | | |
| Study the impact of very massive stars (VMSs) on star forming environments | Find VMSs (M > 150 M$_\odot$, ages < 2 Myr) in luminous and ultra-luminous infrared galaxies within 150 Mpc | HDI multi-band imaging | 9 broadband filters between 200–2190 nm and 3 narrow band filters (656, 1282, and 1876 nm)<br>FOV $\gtrsim$ 6 sq. arcmin |
| | Measure diagnostics of these VMSs including: C IV, N V P Cygni profiles; broad He II emission; blue shifted O V wind absorption; absence of Si IV P Cygni emission / absorption | LUMOS multi-object spectroscopy | Bandpass: 120–170 nm<br>$R > 10,000$<br>Telescope mirror reflectivity > 60% at 105 nm<br>FOV $\gtrsim$ 4 sq. arcmin |
| Constrain models of star formation via knowledge of stellar multiplicity in galaxies beyond the Milky Way and Magellanic Clouds | Find long-period binaries in 7 regions of recent star formation in 5 galaxies within 1 Mpc. Measure proper motions over 5 years | HDI multi-band imaging | U, B, V, I broadband filters<br>FOV $\gtrsim$ 6 sq. arcmin<br>Observe each region once per year over 5 years |
| | Determine spectral types of these binaries with a single epoch of UV spectroscopy | LUMOS multi-object spectroscopy | Bandpass: 100–400 nm<br>Telescope mirror reflectivity > 60% at 105 nm<br>FOV $\gtrsim$ 4 sq. arcmin |

## CHAPTER 7. SCIENCE MISSION TRACEABILITY

In **Chapters 3–6**, the LUVOIR Study Team identified a wide range of compelling science objectives that appeal to a broad section of the astrophysics community. In this chapter, we introduce the scalable mission architecture that enables these ambitious objectives, tracing the Signature Science Cases to a set of essential, high-level capabilities. We present two mission concepts (LUVOIR-A and LUVOIR-B) that bookend the aperture size range of this architecture.

### 7.1 Science mission profile

The Signature Science Cases presented in detail in **Chapters 3–6** represent a set of programs that use the full potential of LUVOIR, and are roughly analogous to present-day Hubble Space Telescope Treasury programs. When LUVOIR launches, the actual science program that it executes will be determined by the community; however, it remains important to show that the notional science objectives outlined in this report can be performed with the designed hardware in the notional five-year prime mission.

**Appendix B** includes detailed science Design Reference Mission (DRM) documents that correspond to the observing programs within each of the Signature Science Cases from **Chapters 3–6**. The DRM documents provide target and observation descriptions, as well as exposure times including estimates for overheads, for both LUVOIR-A and -B. **Figure 7-1** shows the cumulative representation of these DRMs.

A number of the science cases (e.g., many of the exoplanet-focused programs) are given the same time allocation between LUVOIR-A and -B, yielding different quantitative science returns (e.g., number of exoEarth candidates). Others demand the same science return from LUVOIR-A and -B and thus require longer exposure times with LUVOIR-B. One LUVOIR-B program (jointly used in Signature Science Cases #9 and #11) was deemed to need an unfeasibly long time to produce the same science return as LUVOIR-A and was therefore excluded from the LUVOIR-B versions of those Signature Science Cases. Some Signature Science programs (specifically deep field galaxy observations) can be executed in parallel with ECLIPS exoplanet observations.

The key result from the DRM analysis is that both LUVOIR-A and LUVOIR-B can perform these expansive Signature Science programs within five years. In fact, both observatories allow significant remaining time for other science programs like those found in **Appendix A**. The community will choose LUVOIR's science, if they get the opportunity. Even the ambitious portfolio of science cases assembled by the LUVOIR Team leaves significant opportunity for additional transformational science.

### 7.2 Science and mission traceability

**Figure 7-2** shows the normal design process. The science objectives define an architecture that can achieve them. This architecture, combined with available technologies (or technologies identified for development) yield a design to be implemented. While a complete science traceability matrix for LUVOIR is available in **Appendix C**, it is possible to summarize the mapping from each Signature Science Case to a set of high-level mission and instrument capabilities. This mapping is shown in **Figure 7-3**.

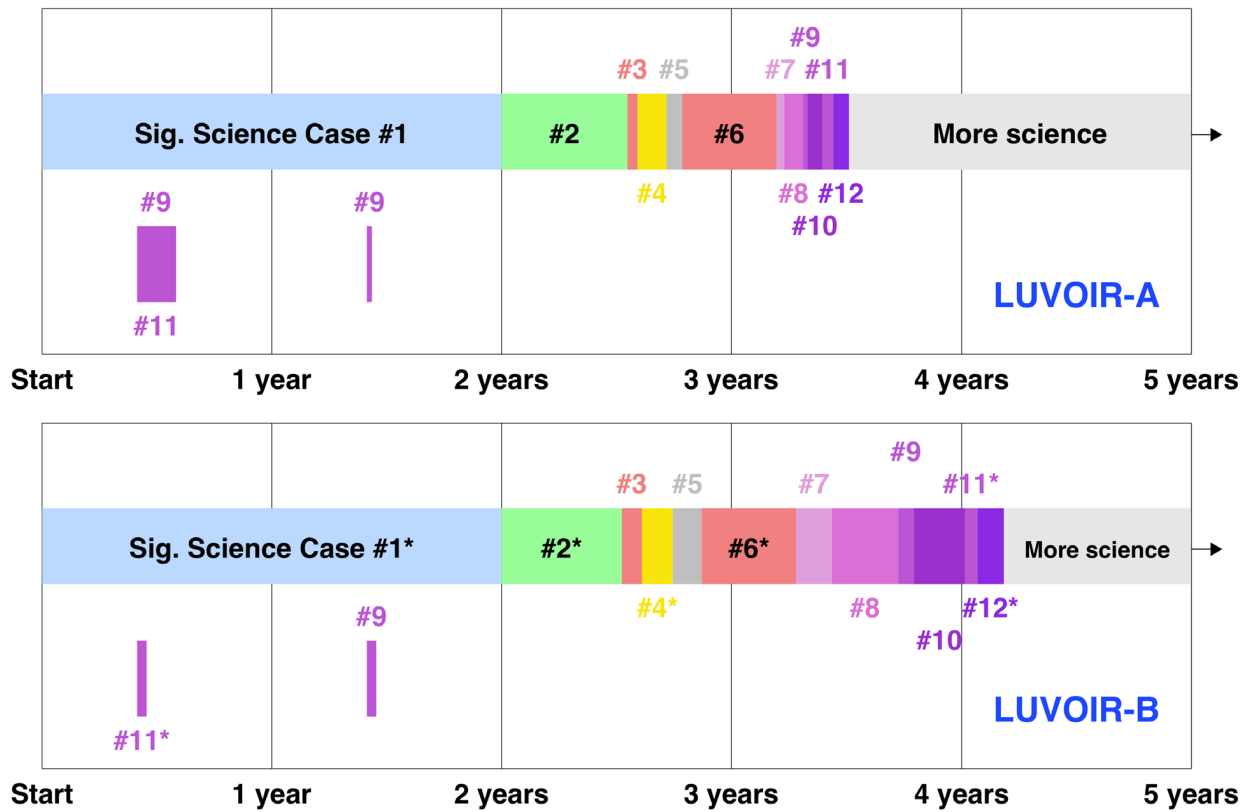

**Figure 7-1.** *LUVOIR can execute all the Signature Science Case (SSC) observing programs with time to spare in the 5-year prime mission lifetime. Details on all the individual observing programs within these SSCs appear in* **Appendix B**. *All program times include estimates for overheads. Programs that can be executed in parallel with other observations are shown offset below the main bars and do not count against the cumulative time. For LUVOIR-A, one program in SSC #11 uses the same data as one of the SSC #9 programs to achieve a different science objective. For LUVOIR-B, an asterisk indicates that the program provides reduced science return (e.g., fewer targets observed). In the most extreme case, the joint SSC #9 / #11 dataset mentioned above is deemed unfeasible with LUVOIR-B and therefore omitted. In reality, the observations for these programs will not be scheduled in this order, or all in single continuous time blocks, but will be interleaved. The SSCs take a total of 3.5 years with LUVOIR-A and 4.2 years with LUVOIR-B, leaving time for additional science programs (examples in* **Appendix A**). *Credit: A. Roberge (NASA GSFC)*

Decomposing these high-level capabilities into functional and performance requirements identifies a mission architecture and defines the technical challenges of the LUVOIR mission.

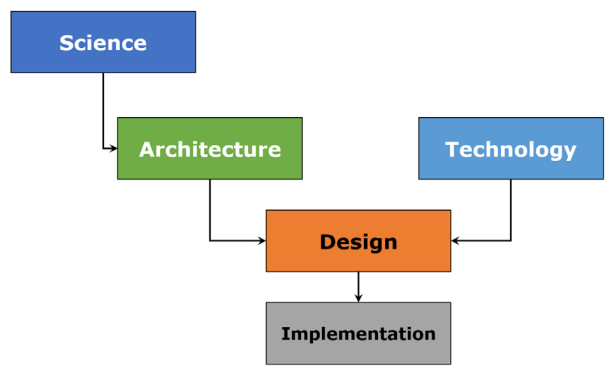

A large aperture requires many mass-efficient segments that can be compactly packaged in a rocket fairing, and autonomously deployed and aligned on orbit. Agility of the payload and a large field-of-regard require the ability to quickly repoint the system over a large range of angles, without introducing thermal or dynamic disturbances that would impact optical performance. Broad wavelength sensitivity requires high-performance optical coatings and low-noise detectors. And each of the instrument capabilities will be enabled by state-of-the-art components such as deformable mirrors, microshutters, and detectors.

**Figure 7-2.** *The normal design process. Science objectives define an architecture which, when combined with new or mature technologies, results in an implementable design.*

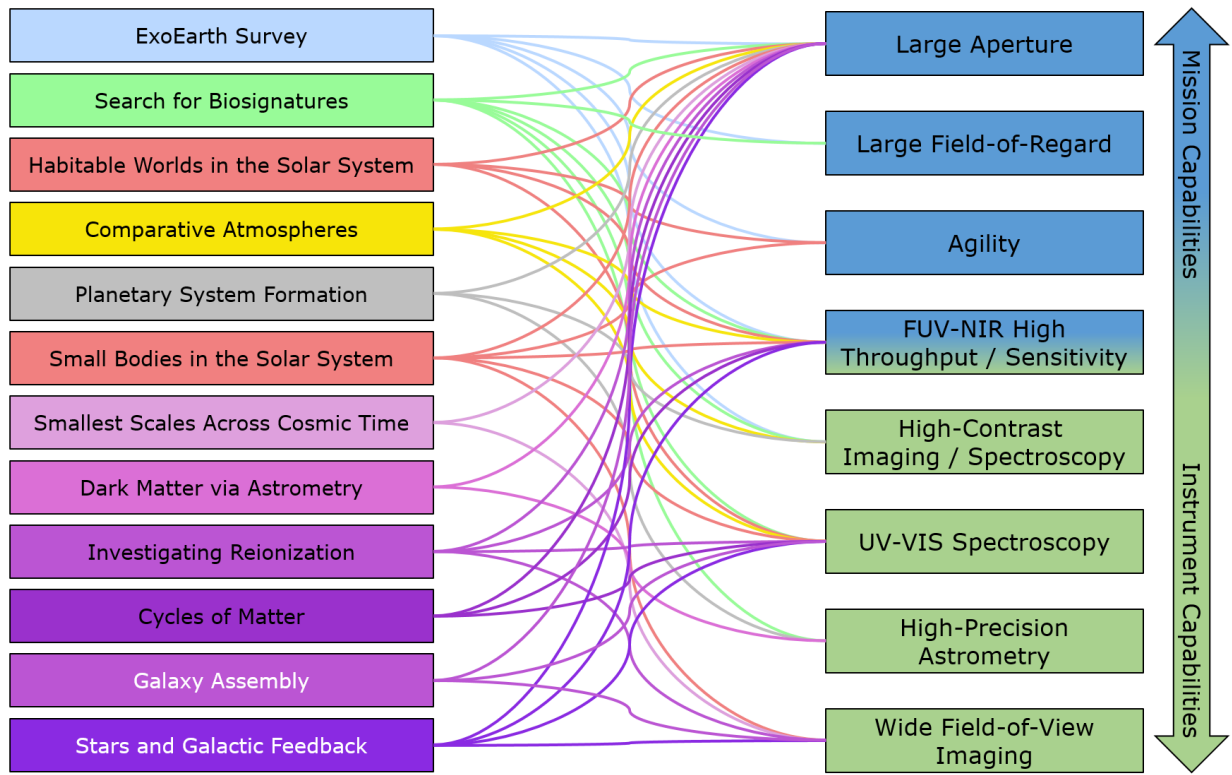

**Figure 7-3.** *Mapping between the Signature Science Cases and high-level mission and instrument capabilities. These capabilities were flowed to mission and functional performance requirements to guide the design of the LUVOIR concepts. A full accounting of the science traceability is available in* **Appendix C**.

A detailed accounting of these technical challenges, and a path to overcoming them, is provided in **Chapter 11**, but here, we highlight the single science objective that drives almost every aspect of the LUVOIR architecture: to directly image and characterize Earth-like planets around Sun-like stars. This requires either a high-contrast coronagraph instrument or a starshade, both capable of suppressing the light from a host star so that the faint reflected light from orbiting planets might be seen. The flux ratio between a Sun-like star and Earth-like planet sets the required contrast ratio to ~$10^{-10}$. Achieving such contrast with a starshade is possible, and currently being studied by other groups[1]. 

However, LUVOIR's exoplanet science goals require visiting many hundreds of stars, each several times (over 1,000 total observations), which is currently infeasible for starshades from a fuel usage perspective. LUVOIR therefore relies on an internal coronagraph instrument to achieve the ~$10^{-10}$ contrast, requiring an end-to-end optical wavefront stability that is measured in picometers. This stability requirement defines the foundation of the LUVOIR concepts and has led to a three-tiered approach to achieving stability. First, the system is designed be as passively stable as possible. Second, nested control systems sense and correct wavefront instabilities over critical spatial and temporal frequencies corresponding to the coronagraph images. Finally, new technologies continue to be explored to improve the tolerance of the system to wavefront instability, potentially reducing the requirements on the first two tiers.

## 7.3 Architecture and mission concept description

### 7.3.1 Architecture definition

Consider again the high-level capabilities identified on the right of **Figure 7-3**. We can begin to organize these mission and instrument capabilities into a hierarchy, shown in **Figure 7-4**. The mission is first divided into segments based on operational context. The observatory segment (**Chapter 8**) provides the necessary space platform from which the science observations can be made. The launch segment (**Chapter 10**) places the observatory segment into orbit, and the ground segment (**Chapter 9**) communicates with the observatory, and receives, processes, and distributes the science data for analysis.

Each segment is decomposed into elements that further define the functional roles. The observatory segment consists of two elements: the science payload element and the supporting spacecraft element. **Figure 7-4** only shows the decomposition of the observatory segment, but similar decompositions can be done for the ground and launch segments. For example, the ground segment will have a mission operations element, a science operations element, and a ground station element. The launch segment will have the launch vehicle and launch site elements.

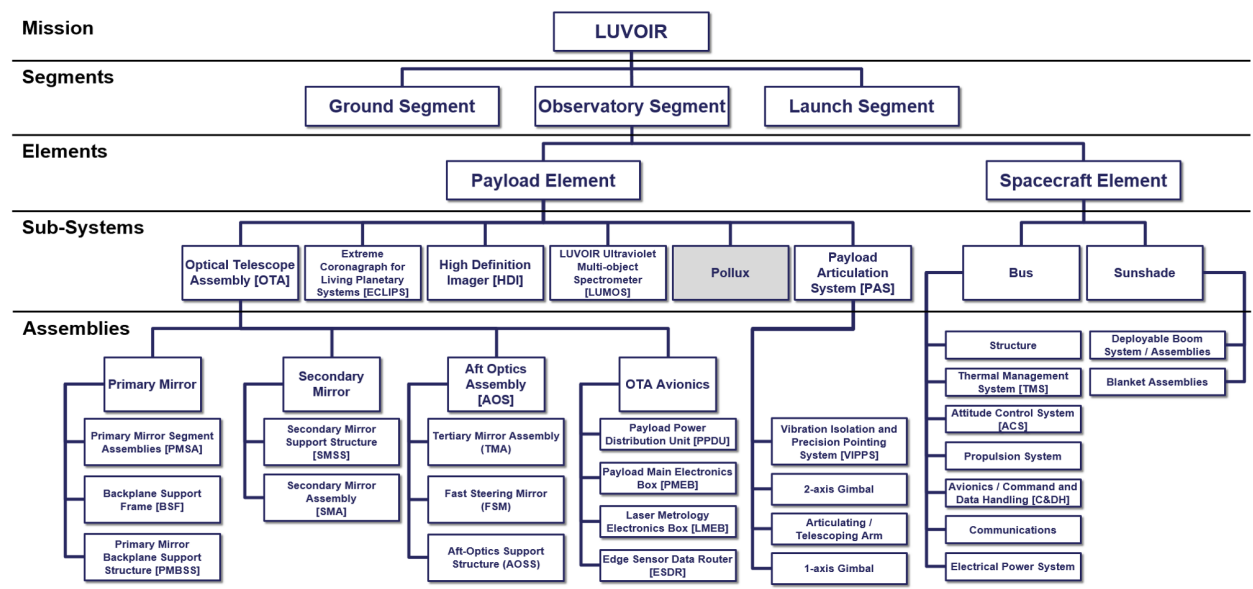

**Figure 7-4.** *The LUVOIR Mission Architecture. Both LUVOIR concepts follow this decomposition of segments, elements, sub-systems, and assemblies, with the exception that the POLLUX instrument only appears on LUVOIR-A. In this figure, only the optical telescope assembly, payload articulation system, bus, and sunshade sub-systems are decomposed into assemblies. Each of the science instruments can be similarly decomposed. The Ground Segment and Launch Segment also have their own Element/Sub-System/Assembly breakdowns as well.*

Each element is further decomposed into sub-systems, each sub-system into assemblies, each assembly into sub-assemblies, and so on. Through this decomposition process, we define the LUVOIR architecture. This architecture is product-centric, not discipline-centric, and enables us to understand how the entire system will progress from individual components up to the full set of mission assemblies needed to meet science goals. Flagship missions are, by definition, first-of-a-kind with the complexities and technologies needed for science worthy of the investment. Defining an architecture and understanding its features and challenges has been critical for planning technology development (**Chapter 11**), and future design, mission management, implementation schedules, and integration and test efforts (**Chapter 12**).

### 7.3.2 Architecture vs. concept

The Decadal Survey will recommend NASA's priorities in the next decade based on the inputs it receives from the scientific community. However, when doing so, it will be faced with a number of uncertainties, such as future budgets, the availability and capability of launch vehicles, and the as-yet-unknown discoveries of the James Webb Space Telescope (JWST), the Wide-field Infrared Survey Telescope (WFIRST), and the ground-based extremely large telescopes (ELTs).

The LUVOIR Study Team considered a range of concepts to deliver unprecedented scientific capability, while also accommodating these uncertainties. This report describes two concepts (**Figure 7-5**) that bracket a range of scientific capability, cost, and risk, allowing the Decadal Survey to evaluate LUVOIR's science yield as a function of these parameters. While we have two distinct concepts, there is a single, scalable architecture that defines those concepts—consistent with the philosophy that a continuum of options that can deliver LUVOIR science exists in between our two distinct concepts. Neither variant should be considered a baseline or descope. Rather, they are end-points in a range of science-delivery alternatives.

LUVOIR-A is a 15-m segmented, on-axis telescope, while LUVOIR-B is an 8-m segmented, off-axis (or unobscured) telescope. Throughout this report, we describe the common aspects of both the LUVOIR-A and LUVOIR-B concepts, followed by discussion of the differences between the two. Versions of the instruments associated with each concept are noted by "-A" and "-B" suffices, i.e., HDI-A is the version of HDI designed for LUVOIR-A and HDI-B is the version designed for LUVOIR-B. The common architecture is referred to simply as "LUVOIR."

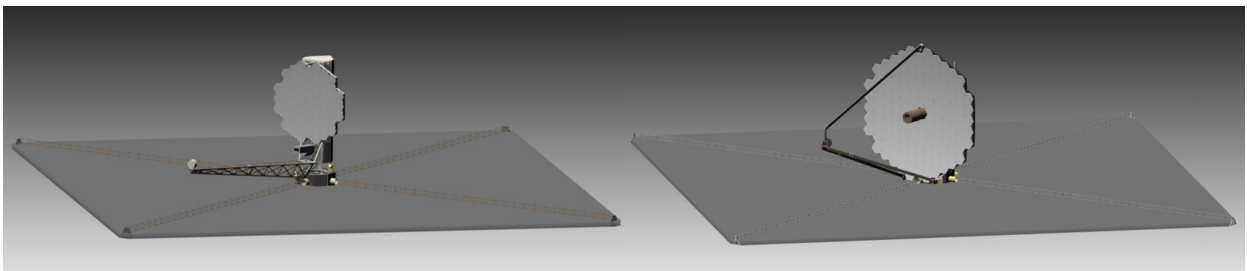

**Figure 7-5.** *LUVOIR-B (left) and LUVOIR-A (right) concepts. While there are some fundamental design differences between the two concepts, they both follow the same system architecture.*

### 7.3.3 Concept development and systems engineering

LUVOIR will be a one-of-a-kind, first-of-its-kind flagship mission. Like any mission of this class, it will be a highly complex, nested system of systems. And like any mission of this class, it will encounter challenges to its design and implementation. The LUVOIR Study Team has researched numerous past and current missions, including JWST, WFIRST, MAVEN, OSIRIS-REx, and Chandra, and has developed a project management approach to apply lessons learned from those missions to LUVOIR. Based on this research, **Chapter 12** describes a series of recommendations, from the way missions are funded and technology is developed, to how teams are organized and contracts are structured. These recommendations are driven by the goal of reducing cost and schedule risk for LUVOIR, and represent an attempt to honestly report on the real costs associated with a mission as ambitious as LUVOIR.

Another key component of the successful and efficient implementation of LUVOIR is an early emphasis on systems engineering and concept development. The scale and complexity of LUVOIR require that all aspects of its design and implementation must be considered in detail early in the project lifecycle. The goal is to uncover as many of the inevitable "unknown unknowns" before a large "marching army" of engineers, technicians, scientists, and managers is engaged. The verification and validation approach must be thoroughly understood in order to inform the number and types of models that will be necessary, as well as identify what facilities might need to be upgraded or constructed. A plan for sub-system or assembly pathfinders must be developed, so that they can be completed early enough to inform designs and test procedures. A servicing concept must be adopted so that modular interfaces can be identified and sufficiently incorporated into the designs and test plan, as well as guide development of servicing infrastructure.

These are just a few of the topics that must be addressed while the architecture and concepts continue to be developed through Pre-Phase A and Phase A, and are described in more detail in **Chapter 12**.

## CHAPTER 8. OBSERVATORY SEGMENT

This chapter describes in detail the technical design of the LUVOIR observatory segment. **Section 8.1** provides an overview of the mission properties that are common to both LUVOIR concepts, such as orbit, serviceability approach, and the overall system design. **Section 8.2** describes the sub-systems of the payload element, including the optical telescope assembly (**Section 8.2.1**), the High Definition Imager (**Section 8.2.2**), the Extreme Coronagraph for Living Planetary Systems (**Section 8.2.3**), the LUVOIR Ultraviolet Multi-object Spectrograph (**Section 8.2.4**), POLLUX (**Section 8.2.5**), and the payload articulation system (**Section 8.2.6**). **Section 8.3** describes the sub-systems of the spacecraft element, including the spacecraft bus (**Section 8.3.1**) and the sunshade (**Section 8.3.2**).

### 8.1 Overview

The LUVOIR observatory is a deployable space telescope, similar in concept and design to JWST. Whereas JWST is a cryogenic (~40 K), near- and mid-infrared observatory, LUVOIR is an ultraviolet-optical-near infrared observatory that is actively controlled to an operating temperature of 270 K. **Figure 8-1** shows an exploded view of the LUVOIR-A observatory segment, highlighting major assemblies. **Figure 8-2** shows a similar view of LUVOIR-B.

The observatory segment consists of the payload element and the spacecraft element. Later sections of this chapter discuss these elements in more detail. Here, we briefly discuss the common, observatory-level design features of LUVOIR.

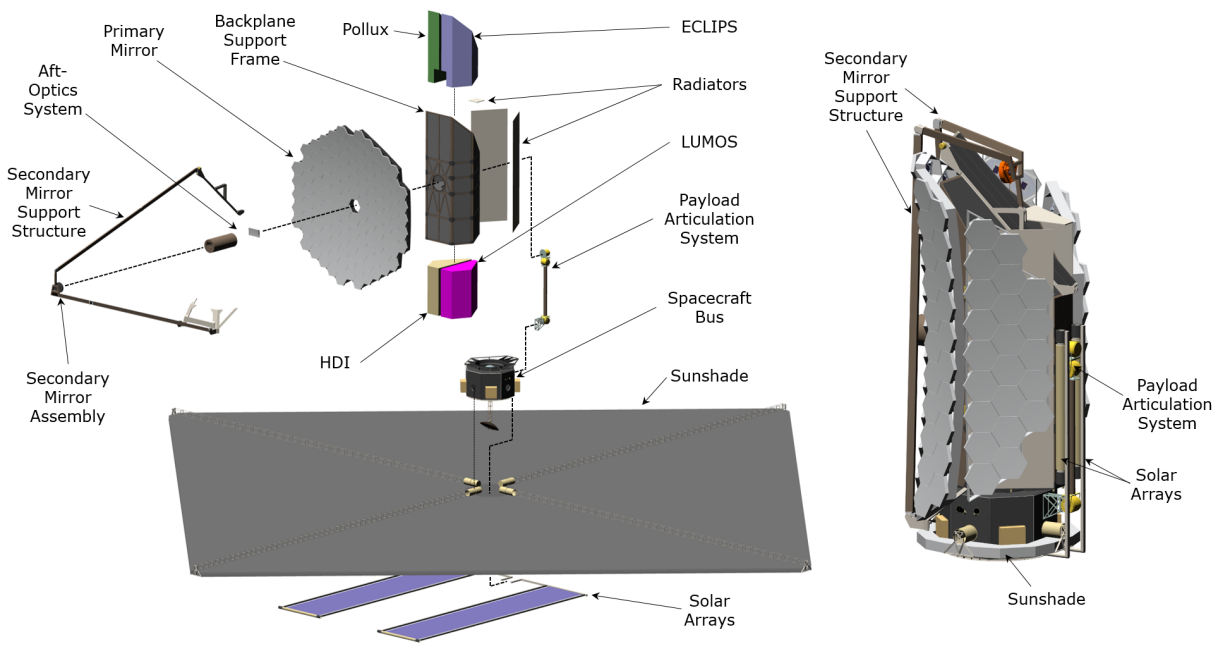

**Figure 8-1.** *Exploded view of the deployed LUVOIR-A observatory segment (left). Note that the instruments represent allocated volumes; the detailed instrument designs were omitted for clarity. The stowed configuration (right) shows how several key assemblies such as the solar arrays, sunshade, and secondary mirror support structure are stowed to enable LUVOIR to fit in the launch vehicle fairing.*

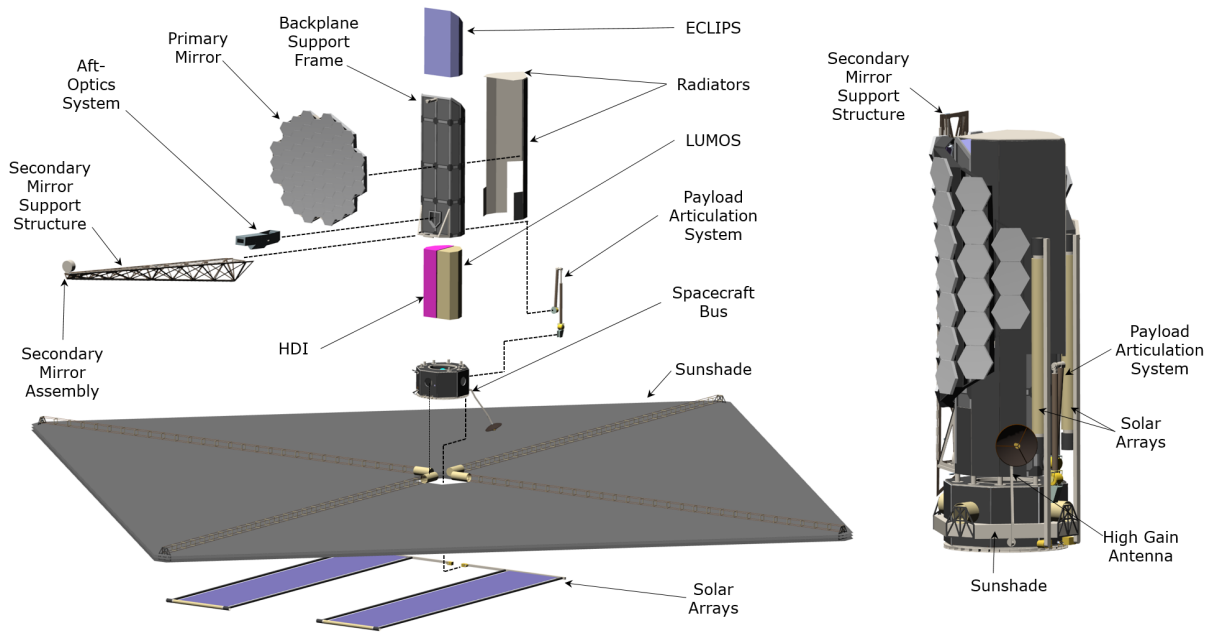

**Figure 8-2.** *Exploded view of the LUVOIR-B observatory segment.*

### 8.1.1 Orbit and radiation

The observatory will occupy a quasi-halo orbit about the Sun-Earth second Lagrange point (SEL2). From this orbit, LUVOIR will have access to the entire anti-sun hemisphere, as well as the sunward portion of the sky up to 45° from the sun-earth axis for time-critical observations. **Figure 8-3** shows LUVOIR's field-of-regard.

The total radiation dose in this orbit was evaluated over LUVOIR's expected operational periods. For the required 5-year mission lifetime, the total ionizing dose is estimated to be 20 krads. For the 10 year goal lifetime, the total ionizing dose is 27 krads. In the event LUVOIR is serviced for an extended mission (see next section), the total ionizing dose over a 25 year lifetime is estimated to be 54 krads.

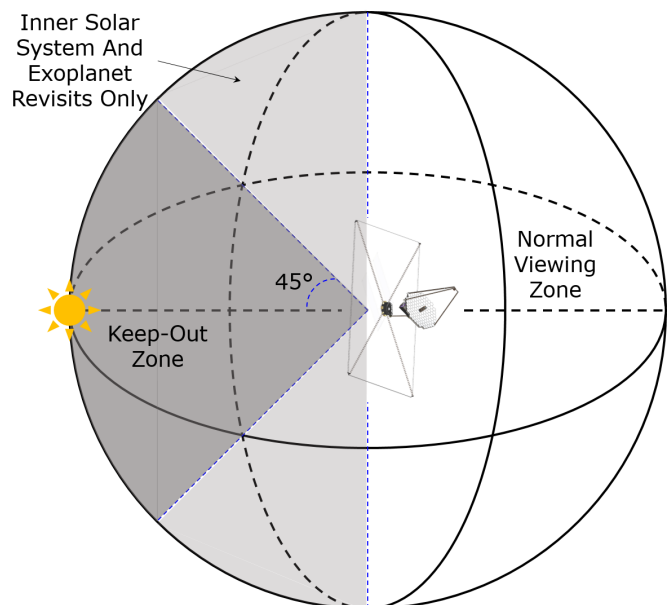

**Figure 8-3.** *The LUVOIR field-of-regard. Most observations occur in the anti-sun hemisphere, with the sunshade remaining in a fixed orientation to the sun for thermal stability. For viewing inner solar system objects, or for time-critical exoplanet revisit observations, LUVOIR can pitch up to 45° towards the sun.*

### 8.1.2 Lifetime and serviceability

The LUVOIR science programs described in this report can be achieved within a 5-year prime mission using the observatory designs described in this chapter. However, Congress has mandated that all future large space telescopes be serviceable to the extent practicable.

We have considered the question "how will LUVOIR be serviced?" We first note that a design commitment to serviceability is not a commitment to servicing. Whether or not LUVOIR will be serviced depends on political and budgetary realities that are unpredictable. If history is a guide, it is doubtful that NASA's Science Mission Directorate on its own will have the resources to create the infrastructure for servicing LUVOIR. Such an endeavor would require a commitment from the entire agency, and possibly other organizations within the federal government and industry.

We have engaged NASA Goddard's Satellite Servicing Project Division, as well as other NASA center and industry partners that are developing on-orbit servicing technologies, to ensure that the LUVOIR concepts incorporate best practices to be "servicing friendly." To this end, we imagine a tiered approach to servicing LUVOIR that can be tailored depending on the existence of future capabilities and resources. **Table 8-1** summarizes these tiers.

**Table 8-1.** *The five tiers of serviceability considered by the LUVOIR Study Team. "Accommodated by Existing Design" implies that the LUVOIR Study Team has attempted to account for the hardware, mass, volume, and power associated with a particular serviceable interface. However, at this early stage of development, we are not able to fully account for all aspects of the serviceable interface, hence many are listed as "partially" accommodated by the existing design.*

| Tier | Sub-system / Assembly | Should Be Serviceable? | Accommodated by Existing Design? | Required Lifetime [years] | Goal Lifetime [years] |
|---|---|---|---|---|---|
| 1 | Propellant | Yes | Yes | | |
| 2 | Instrument Modules | Yes | Partially | | |
| 2 | Control Moment Gyroscopes | Yes | Partially | | |
| 2 | Star Trackers | Yes | Partially | | |
| 2 | Command & Data Handling Electronics | Yes | Partially | | 10 |
| 2 | Communications Components | Yes | Partially | | |
| 2 | Electrical Power System Components | Yes | Partially | | |
| 2 | Solar Arrays | Yes | Partially | | |
| 3 | Payload Main Electronics Box | Yes | No | | |
| 3 | Payload Power Distribution Unit | Yes | No | 5 | |
| 3 | Laser Metrology Electronics Box | Yes | No | | |
| 3 | Radiators | Yes | No | | |
| 4 | Payload Articulation Gimbals | Maybe | No | | |
| 4 | VIPPS | Maybe | No | | |
| 4 | Thrusters | Maybe | No | | |
| 4 | Sunshade | Maybe | No | | TBD (10 if decided to be serviceable; 25 otherwise) |
| 4 | Mirror Segments | Maybe | No | | |
| 4 | Secondary Mirror | Maybe | No | | |
| 4 | Aft-optics | Maybe | No | | |
| 4 | Upgrade to Optical Communications | Maybe | No | | |
| 5 | Payload Element Structure | No | - | | 25 |
| 5 | Spacecraft Element Structure | No | - | | |

We consider Tier 1 to be the bare minimum requirement for serviceability: refueling. In 2013, the Robotic Refueling Mission completed an on-orbit demonstration at the International Space Station of the technologies necessary to tele-robotically refuel a client spacecraft[1]. And in 2022, the Restore-L mission will demonstrate the end-to-end process of rendezvous, capturing, refueling, and relocating a client spacecraft in low-Earth orbit[2]. While development will still be necessary to accommodate a client beyond low-Earth orbit, robotically refueling an on-orbit asset appears to be feasible and will likely be routine by the time LUVOIR is operating.

Tier 2 includes the elements of the LUVOIR observatory that we have actively designed to be serviceable, namely the science instrument modules and major spacecraft subsystems. As demonstrated by Hubble, the ability to upgrade instrumentation has allowed the observatory to far surpass the capability originally imagined for it. And replacing critical spacecraft systems such as gyros and star trackers has allowed it to continue operating for more than 29 years.

While the servicing of Hubble was reliant on the use of astronauts, many of the interfaces are similar to those that would be used by robotic servicers. Blind mate electrical and power interfaces, kinematic latching systems, and guide rails will still be used to remove and replace instrument modules or spacecraft bus orbital replacement units.

While we attempted to build this tier of serviceability into the design, we could not account for every aspect of these interfaces within the constraints of this concept study. **Table 8-1** therefore states that Tier 2 is "partially" accommodated by the design. The optical telescope assembly and instruments include mass and volume allocations for guide rails, grapples, harnessing, and blind-mate power, data, and thermal connectors. Hardware clearances have been designed to allow each of the instruments to be removed from either end of the backplane support frame, as shown in **Figure 8-4**. On the spacecraft, the avionics, communications, electrical power, and attitude control system components are located on each of the eight external panels of the bus (**Figure 8-42**) for easy access. The solar arrays can also be accessed from below the sunshade and replaced.

Tier 3 includes elements of the LUVOIR observatory that should be serviceable, but that we did not explicitly make accommodations for in the current concept designs. These elements include the payload electronics boxes within the backplane support frame, the radiator panels on the outside of the backplane support frame, and the thrusters on the spacecraft bus. The payload electronics boxes and radiators may need to be replaced as the instruments are upgraded to accommodate new performance requirements.

Tier 4 includes elements of the observatory that would be considerably more challenging to service, but would potentially enable a mission lifetime greater than 25 years. These elements include the telescope mirrors (e.g. an entire "wing" of the primary mirror could be replaced at one time), the sunshade, and the payload articulation system. Another goal might be to upgrade the currently planned Ka/S-band communication system with an optical communications terminal to accommodate greater science data bandwidth.

Finally, Tier 5 includes the payload and spacecraft element structures, heaters, and harnessing which we do not believe can, or should, be serviced.

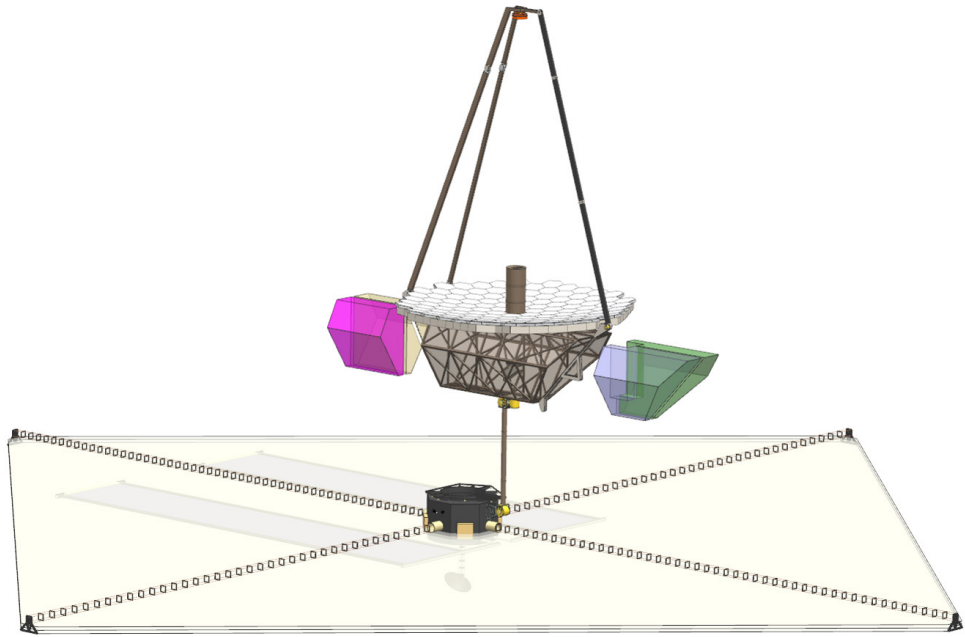

**Figure 8-4.** *Instrument hardware accommodations have been designed to allow for each of the instrument modules to be robotically removed from the payload and replaced or upgraded with new instrument capabilities.*

### 8.1.3 Mass

**Table 8-2** provides a summary of the LUVOIR-A and LUVOIR-B mass estimates. Current best estimate (CBE) masses are based on the engineering designs, engineering judgement, and actual masses for hardware that is "off the shelf". Maximum expected value (MEV) masses account for a 30% mass growth allowance[3] owing to the very preliminary nature of these concept designs. Finally, the maximum permissible value (MPV) masses account for additional mass margin and reserve against the baseline launch vehicle performance. The total mass contingency, i.e. the difference between the CBE and MPV masses as a percent of the launch vehicle capacity, is 37.2% and 59.1% for LUVOIR-A and LUVOIR-B, respectively. **Section 12.3** provides more detail on the margin philosophy adopted by the LUVOIR Study Team.

LUVOIR-A is designed to use the full capacity of a Space Launch System (SLS) Block 2 vehicle. The latest projected performance for this vehicle is a lift capacity of 44,300 kg for a characteristic energy (C3) of 0 km²/s² (NASA 2018). LUVOIR requires a C3 between -0.55 and -0.75 km²/s², so the expected lift capacity will be slightly higher than this, though we use the more conservative value of 44,300 kg in our allocation.

The LUVOIR Study Team initially required that LUVOIR-B should fit into a "conventional" heavy-lift launch vehicle, in order to reduce overall mission risk should the SLS Block 2 vehicle be developed too late or not achieve the current expected performance. By "conventional," the team implied a launch vehicle similar to the existing United Launch Alliance

---

3 A uniform 30% mass growth allowance is used on all hardware except for the primary mirror segment assemblies and secondary mirror assembly, which use a 20% mass growth allowance. These particular assemblies are based heavily on mirror assembly designs from JWST and numerous early technology development efforts, so there is higher confidence in their mass estimates

**Table 8-2.** *Estimated mass summary for both LUVOIR concepts. Both concepts have substantial mass contingency against the baseline launch vehicle's expected performance.*

| Element | Sub-System | LUVOIR-A | | LUVOIR-B | |
|---|---|---|---|---|---|
| | | Current Best Estimate (CBE) [kg] | Maximum Expected Value (MEV) [kg] | Current Best Estimate (CBE) [kg] | Maximum Expected Value (MEV) [kg] |
| Payload | Optical Telescope Assembly | 17,478 | 23,599 | 7,218 | 9,840 |
| | Payload Articulation System | 481 | 687 | 306 | 438 |
| | HDI | 671 | 958 | 430 | 615 |
| | ECLIPS | 807 | 1,153 | 780 | 1,114 |
| | LUMOS | 874 | 1,249 | 544 | 778 |
| | POLLUX | 375 | 536 | 0 | 0 |
| | Misc. | 899 | 1,284 | 376 | 537 |
| | Payload Total: | 21,585 | 29,466 | 9,654 | 13,322 |
| Spacecraft | Bus | 3,215 | 4,593 | 3,446 | 4,923 |
| | Sunshade | 670 | 957 | 586 | 837 |
| | Misc. | 194 | 278 | 174 | 248 |
| | Propellant | 2,137 | 2,137 | 1,272 | 1,272 |
| | Spacecraft Total: | 6,216 | 7,965 | 5,478 | 7,280 |
| | Observatory Total: | 27,801 | 37,431 | 15,132 | 20,602 |
| | Launch Vehicle Capacity (MPV) [kg]: | 44,300 | | 37,000 | |
| | Margin & Reserve [kg]: | 6,869 | | 16,398 | |
| | Margin & Reserve [%]: | 15.5% | | 44.3% | |
| | Total Mass Contingency (CBE to MPV): | 37.2% | | 59.1% | |

Delta IV-Heavy vehicle with a 5 m × 19.8 m fairing, and a lift capability to SEL2 of ~10,000 kg. However, after additional engineering design work, it was clear that LUVOIR-B would require a lift capacity much greater than this. A decision was made to maintain the 5-meter-fairing volume constraint on LUVOIR-B and determine a launch capacity requirement based on the system design. The latest projected minimum performance for an SLS Block 1B vehicle is a lift capacity of 37,000 kg for a C3 = 0 km²/s² (NASA 2018), which provides substantial margin against the LUVOIR-B predicted mass.

For both LUVOIR concepts, there is some risk associated with benchmarking the observatory mass against the estimated performance of launch vehicles that are still in development. We mitigate this risk through two methods. First, multiple launch vehicles exist for each LUVOIR concept, and **Chapter 10** discusses these options in more detail. Second, the scalable architecture that we have adopted ensures that some version of LUVOIR exists between LUVOIR-A and LUVOIR-B that can be designed to fit within the constraints of whichever of these launch vehicles is available. Once an actual launch vehicle is selected, a more complete mass allocation can be performed consistent with the LUVOIR concept that is developed.

### 8.1.4 Power

**Table 8-3** shows the estimated power requirements for both LUVOIR concepts. The required power is driven largely by the need to actively heat and control the large mirror and structures in the cold environment created by the sunshade at SEL2. While these power requirements

**Table 8-3.** *Estimated power summary for both LUVOIR concepts*

| Element | Sub-System | LUVOIR-A | | | | LUVOIR-B | | | |
|---|---|---|---|---|---|---|---|---|---|
| | | Current Best Estimate (CBE) [W] | | Maximum Expected Value (MEV) [W] | | Current Best Estimate (CBE) [W] | | Maximum Expected Value (MEV) [W] | |
| | | Average | Peak | Average | Peak | Average | Peak | Average | Peak |
| Payload | Optical Telescope Assembly | 9,690 | 11,881 | 16,150 | 19,801 | 5,091 | 8,305 | 8,485 | 13,841 |
| | Payload Articulation System | 86 | 215 | 143 | 358 | 86 | 215 | 143 | 358 |
| | HDI | 162 | 255 | 270 | 424 | 142 | 192 | 237 | 320 |
| | ECLIPS | 1,036 | 1,060 | 1,726 | 1,766 | 1,032 | 1,056 | 1,720 | 1,760 |
| | LUMOS | 438 | 575 | 729 | 958 | 365 | 503 | 608 | 838 |
| | POLLUX | 130 | 130 | 217 | 217 | 0 | 0 | 0 | 0 |
| | Payload Total: | 11,542 | 14,116 | 19,235 | 23,524 | 6,716 | 10,271 | 11,193 | 17,117 |
| Spacecraft | Bus | 2,824 | 3,238 | 4,706 | 5,397 | 3,092 | 3,595 | 5,154 | 5,992 |
| | Sunshade | 0 | 40 | 0 | 67 | 0 | 40 | 0 | 67 |
| | Spacecraft Total: | 2,824 | 3,278 | 4,706 | 5,464 | 3,092 | 3,635 | 5,154 | 6,059 |
| | Observatory Total: | 14,366 | 17,394 | 23,941 | 28,988 | 9,808 | 13,906 | 16,347 | 23,176 |
| | Margin & Reserve [%]: | | | 25.0% | 25.0% | | | 25.0% | 25.0% |
| | Maximum Permissible Value [W]: | | | 31,921 | 38,651 | | | 21,796 | 30,901 |
| | Total Power Contingency (CBE to MPV) [%]: | | | 55.0% | 55.0% | | | 55.0% | 55.0% |

may appear high at first glance, it is important to note that they are consistent with another large space facility: the International Space Station (ISS). The ISS requires ~90 kW of power to operate, and its four solar array wings are capable of generating up to 120 kW.

The CBE power is based on engineering designs, engineering judgement, and actual power dissipations for "off the shelf" hardware. The MEV power accounts for a 40% power growth allowance from the CBE value, due to the early nature of these concepts. Finally, the MPV power includes an additional 25% margin and reserve.

### 8.1.5 Optical performance

The optical performance requirements for LUVOIR-A and LUVOIR-B are virtually the same, so a single, comprehensive wavefront error and line-of-sight pointing error budget was generated and included in **Appendix E**. The only significant difference in wavefront error allocation between LUVOIR-A and LUVOIR-B pertains to the image motion allocation, which is described in more detail below. The values included in the budget are based on the current design concepts, with estimated allocations to some sub-systems that have yet to be designed or analyzed in detail. Therefore the values in the budget should not be considered as hard requirements, but rather to illustrate a logical set of allocations, and provide a basis for technology, engineering, and manufacturing development plans.

The wavefront error budget starts with the top-level system performance, which then branches down to the optical telescope assembly (OTA) and each of the science instruments. Contained in each of those branches is the configured wavefront error, which includes corrections from the wavefront sensing and control sub-system, the stability changes in the wavefront error, and the effective change in wavefront error caused by image motion. There is a separate system budget for each instrument since (1) the required wavefront error performance for each instrument is different, and (2) the design residual wavefront error from the OTA is different over the field-of-view for each instrument. The wavefront error performance requirement for each instrument is:

- HDI must be diffraction-limited at 500 nm in the UV-Visible channel, and at 1000 nm in the NIR channel.
- ECLIPS must be diffraction-limited at the shortest wavelength in each channel (200 nm, 515 nm, 1000 nm, respectively).
- LUMOS requires a geometric spot size of 30 mas at the microshutter entrance aperture plane. This corresponds to diffraction-limited performance of the OTA at ~500 nm.

### 8.1.6 System design philosophy

The wavefront stability required to achieve the high-contrast imaging for exoplanet science drives almost every aspect of LUVOIR system design, and leads to a three-tiered approach to achieving stability.

**Wavefront stability through design.** To achieve picometer-level wavefront stability, start by designing an overall system that is as stable as possible. This may seem like a tautology, but an alternative approach would design an unstable system that emphasizes control of instability. However, it is not clear that sufficient control is possible to maintain picometer stability at all required spatial and temporal scales. Thus a passively stable system provides the bedrock on which additional layers of control can be built as needed.

Thermal stability is achieved using materials with near-zero coefficient of thermal expansion (CTE) at the desired operating temperature (in this case, 270 K) such as Corning ULE or Schott Zerodur glass for mirror segments, and tuned composites for structures. Milli-Kelvin-level thermal sensing and control of the mirrors, structures, and interfaces ensures a stable operating temperature.

Dynamic stability is achieved via stiff structures and mirrors, passive isolation at disturbance sources, and active vibration isolation between the spacecraft (where most disturbance sources exist) and the payload (where wavefront stability is critical). The Vibration Isolation and Precision Pointing System (VIPPS; see **Section 8.2.6.3**) uses non-contact voice coil actuators to physically decouple the payload from the spacecraft and control relative attitude and translation degrees of freedom between the two elements.

**Wavefront stability through control.** After maximizing system stability through design, it is necessary to sense and control any remaining instability at the critical spatial and temporal frequencies.

Current methods of wavefront sensing and control, such as low-order wavefront sensing and phase retrieval, typically use photons from the target star that have traversed the optical system to estimate the wavefront. However, when observing dim target stars, it can take 10s to 100s of minutes to collect enough photons to generate a wavefront estimate of adequate signal to noise ratio. Techniques such as low-order wavefront sensing are also limited in the spatial frequency content that can be estimated. Thus these techniques are best suited to controlling slow, global wavefront error drifts.

Onboard metrology systems, such as segment edge sensors and/or laser metrology systems do not depend on stellar photon rates, and so can operate much faster than low-order or image-based wavefront sensing techniques. They are also better suited at measuring optic motions directly at the critical interfaces, i.e. segment-to-segment edges, or primary-to-secondary mirror displacements.

***Wavefront stability through tolerance.*** The final tier involves continuing to develop new coronagraph architectures and data processing techniques that are designed to be more robust in the presence of wavefront instability. Spatial filtering (Gong et al 2016) and post-processing techniques (Pogorelyuk & Kasdin 2019) may be able to relax sensitivities to specific wavefront error terms from 10s of picometers to 100s of picometers or even nanometers. While several of these approaches show promise, they are still very early in development and require further study.

### 8.1.7 Mechanical design and structural performance

The three overarching mechanical design considerations are the complexity, size, and stability of the design.

The complexity drives a need to triage the design efforts to focus on those engineering challenges that seem the most difficult to solve. Although some details are added in the structural design to produce higher fidelity mass properties, those design details that are considered achievable with enough skilled effort are left to be worked by future teams.

LUVOIR requires a large structure. For conceptual work, sizing rules-of-thumb are usually considered adequate as they are typically very conservative. The size of LUVOIR is closer to architectural or civil engineering scales, making rough sizing of structural elements non-intuitive. As a result, the conceptual design requires a level of structural analysis not normally performed at this phase of study. The structural analysis performed on the overall observatory only looked at primary load paths. Secondary components (for example, instrument mirror support tripods) are sized using hand calculations.

The stability requirement of ~10 picometers RMS wavefront error over a timescale of ~10 minutes drives the system to rely on near zero coefficient of thermal expansion materials, e.g., Zerodur, ULE, tailored composite materials, and state of the art thermal management systems. The LUVOIR mechanical and thermal designs incorporate this consideration to exact more robust and comprehensive sub-system designs throughout the study cycle. The stability requirement drives the thermal design far more than the mechanical design. Aside from maximizing stiffness wherever possible, the mechanical stability design focuses upon thermal isolation to make developing a thermal control system more manageable.

We also note that LUVOIR borrows many design cues from JWST. While this improves fidelity and confidence in the design, some caution must be taken given the difference in size between the two observatories. In many cases, a direct application of the JWST design was not sufficient, but rather provided a starting point from which the LUVOIR concept was refined.

### 8.1.8 Observatory thermal architecture

The observatory-level thermal design on LUVOIR leverages a passive and robust thermal system to efficiently transport and reject all internal heat dissipations to space via heat pipes and radiators. Actively-heated components are insulated to conserve heater power.

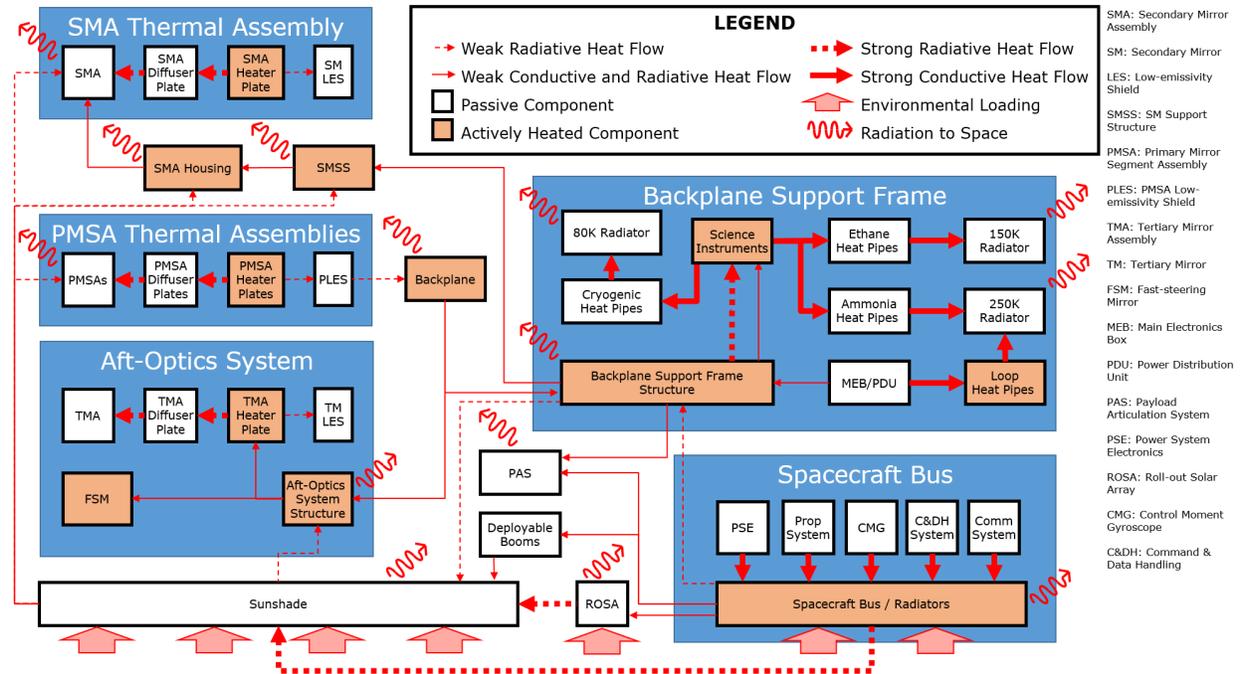

**Figure 8-5.** *The LUVOIR observatory segment thermal architecture. Active heating and passive cooling is used to maintain a stable operating temperature of 270 K. Specific assemblies that require colder operation are passively cooled via an array of radiators mounted to the external faces of the backplane support frame. Heat straps transfer heat from within each instrument to a thermal interface junction; heat pipes then transfer dissipated heat to the radiators.*

The thermal design is also modular, partitioning each separate sub-system into its own thermal zone and minimizing cross-talk between components.

The observatory-level heat map for both architectures is shown in **Figure 8-5**, with each major sub-system or assembly highlighted in blue. The only components to receive direct solar loading are the sunshade, the roll-out solar array (ROSA), and the surfaces of the spacecraft bus below the lowest sunshade layer.

### 8.1.9 Control system

A nested control system that incorporates wavefront sensing, metrology, and line-of-sight sensing is used to help achieve the picometer-level wavefront stability required by the high-contrast imaging science. Each control loop addresses a particular range of spatial and temporal frequencies. **Figure 8-6** shows a block diagram of the control system, with approximate order-of-magnitude signal rates and control bandwidths for each system.

At the lowest bandwidth, the low-order or out-of-band wavefront sensor within the ECLIPS instrument uses photons from the target star to estimate the end-to-end system wavefront error, and control slow drifts with the deformable mirrors within the instrument. This level of control is only necessary for coronagraph exoplanet observations.

At the next level, a laser metrology system is used to monitor alignment drifts between the primary mirror, secondary mirror, and the aft-optics system. These global alignment errors can be corrected with rigid-body motions of the primary mirror segments and secondary mirror, or by the deformable mirrors within ECLIPS for exoplanet observations.

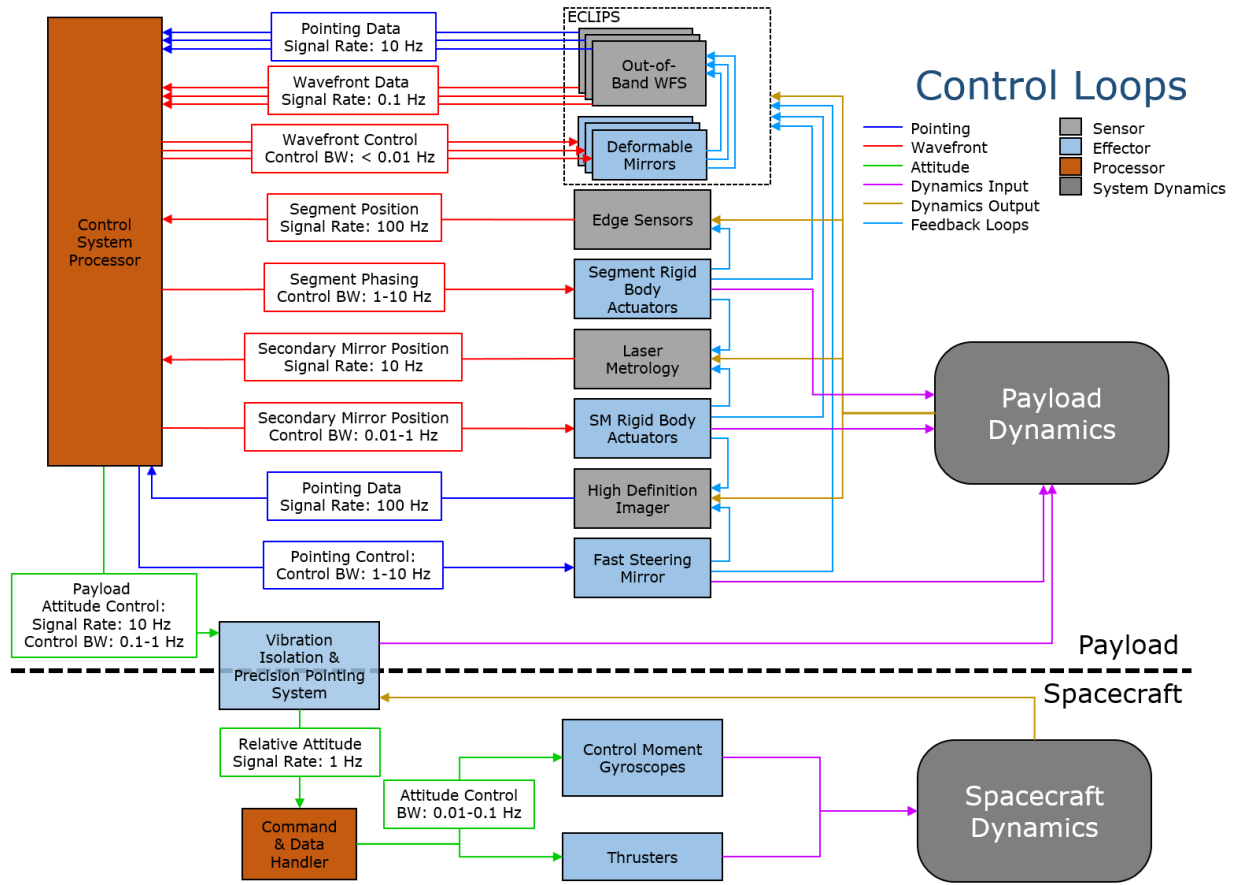

**Figure 8-6.** *Pointing and wavefront control loops baselined for the LUVOIR architecture. Payload line-of-site pointing is sensed at the HDI focal plane via fast centroiding on selected guide stars. Pointing is then controlled by three actuators with different bandwidths: the fast steering mirror controls the highest bandwidth errors, the VIPPS controls lower bandwidth errors, and the spacecraft ACS controls the lowest. The out-of-band wavefront sensor inside of the coronagraph uses the deformable mirrors to correct slow drifts in both pointing and wavefront during exoplanet observations. A segment edge sensor and laser metrology system maintains phasing and alignment of the primary and secondary mirrors. The quoted signal rates and bandwidths are approximate order-of-magnitude values.*

The fastest wavefront control loop occurs at the primary mirror segment level. Edge sensors on each segment measure rigid body motions between segments and correct those motions with rigid body actuators behind each segment.

Line of sight pointing errors are also corrected through two control systems. Within ECLIPS, pointing errors are sensed with the low-order or out-of-band wavefront sensor. Telescope pointing errors are also sensed by the HDI's fine guidance mode. Fast pointing errors are corrected by the fast steering mirror, while slower pointing errors are corrected by the Vibration Isolation and Precision Pointing System (VIPPS).

The control system processor, located within the ECLIPS instrument and discussed in more detail in **Section 8.2.3.5**, is the single processor responsible for coordinating all of these control loops. Pointing, wavefront, and metrology data are all synthesized into the appropriate commands to each effector, at the appropriate bandwidth.

### 8.1.10  Contamination control

The performance of the optical system in the far-UV is critically dependent on the cleanliness of the optical surfaces. Water and non-volatile residuals (NVR) strongly attenuate wavelengths around Lyman-$\alpha$ and shorter. Thus contamination control during the integration, test, and on-orbit operation of LUVOIR will be paramount. Several strategies will be employed to manage contamination.

First, the aluminum-coated mirrors of the OTA and LUMOS instrument will use new protected LiF coatings. LiF is known to be hygroscopic; absorbing water vapor from the air will cause the coating reflectivity to degrade. However, new coatings use a thin, protective layer of $MgF_2$ or $AlF_3$ to protect the LiF from exposure (Witt et al. 2018, Balasubramanian et al. 2017). These coatings are under development and are included in the technology development plan described in **Chapter 11**.

All mirrors and detector windows are designed with heaters that can be occasionally driven higher than the nominal 270 K setpoint. These heaters can be used on-orbit to "bake out" critical optical surfaces, and drive any deposited water vapor or non-polymerized NVR from the surface. Molecular absorber coating (MAC, a form of zeolite) will also be used on internal instrument surfaces to help capture any contaminants released by the instrument structure.

Additionally, strict contamination control processes and certifications will be used during integration and test of the payload. Whenever possible, components and assemblies will be kept under dry purge. Exposed optical surfaces will be inspected and cleaned at regular intervals, using techniques developed and demonstrated on JWST.

## 8.2  Payload element

### 8.2.1  Optical telescope assembly (OTA)

#### 8.2.1.1  Introduction

The optical telescope assembly (OTA) is the optical path every photon follows at the end of its long journey from the astronomical object to the instrument suite and detectors. The OTA must be designed to achieve all science objectives described earlier in this report, and yet be capable enough to respond to an ever-changing scientific landscape.

Some of the design drivers that shape the implementation of the OTA include:

- High-contrast imaging
- Science instrument fields-of-view
- Wavelength range
- Wavefront error
- Thermal management
- Launch vehicle mass capacity
- Launch vehicle volume
- Contamination control

As discussed in **Section 8.1.6**, the wavefront error stability needed to achieve the high-contrast exoplanet science is a critical requirement that drives the OTA design. The material choices, thermal control system, segment-level architecture, segment phasing concept, and secondary mirror support structure design have all been driven by the need to maintain wavefront stability at the 10s-of-picometer level.

**Figure 8-7** and **Figure 8-8** show block diagrams of the LUVOIR-A and LUVOIR-B OTAs, respectively, and identify all major assemblies.

### 8.2.1.2 Optical design

The LUVOIR-A telescope is a modified three-mirror anastigmat (TMA) design similar to JWST. The primary mirror and secondary mirror have conic surfaces while the tertiary mirror is an x-y polynomial surface. A flat fast-steering mirror is located at the real exit pupil after the tertiary mirror and used for fine pointing control.

The LUVOIR-B OTA is also a TMA design that uses an off-axis section of the conic primary mirror to enable an unobscured design. LUVOIR-B has a smaller 8 m diameter segmented primary mirror compared to the 15-m diameter LUVOIR-A, however both designs have a total field-of-view of 10 × 8 arcmin to accommodate similar instrument science capabilities. The smaller aperture diameter in LUVOIR-B results in less photon collecting area. To maximize the exoplanet yield from a smaller aperture, an off-axis unobscured design was selected. Not only does this help increase the collecting area, but by removing the central obscuration the diffraction point-spread function enables a more efficient coronagraph mask design, increasing exoplanet yield.

**Table 8-4** summarizes the design requirements and predicted performance for both LUVOIR OTAs, and **Figure 8-9** shows a ray-trace of each design.

**Table 8-4.** *LUVOIR OTA design requirements and predicted performance*

| Parameter | Units | Requirement | | Value | | Notes |
|---|---|---|---|---|---|---|
| | | **LUVOIR-A** | **LUVOIR-B** | **LUVOIR-A** | **LUVOIR-B** | |
| Field-of-View | arcmin | 10 x 8 | | 10 x 8 | 10 x 8 | Accommodates all instrument fields-of-view. |
| Plate Scale | arcsec / mm | 0.4–1.5 | | 0.694 | 0.699 | Driven by LUMOS single-microshutter field-of-view. |
| F/# | – | 9.1–20.0 | 17.2– 37.4 | 19.8 | 36.88 | Larger F/# allows more clearance between instruments, but larger optics within the instruments. |
| Max. Angle of Incidence for ECLIPS FOV | degrees | 12.0 | | 11.77 | 12.00 | Limits polarization aberration and contrast leakage in ECLIPS. |
| Average RMS Wavefront Error | nm | < 10.0 | | 6.25 | 3.33 | Image quality allocation. |
| Max. 100% Spot Diameter | μm | 80.0 | | 15.7 | 4.9 | Blur spot at LUMOS microshutter must be smaller than a shutter. |
| Max. Marginal Ray Height at SM | mm | 510.0 | – | 486 | – | Limit secondary mirror obscuration. |
| Max. Marginal Ray Height at PM Hole | mm | 520.0 | – | 446.9 | – | Limit primary mirror hole obscuration. |
| Max. Distance Between PM and SM Vertex | m | 20.0 | | 18.3 | 20.0 | Packaging constraint for folding inside of fairing. |
| Max. Distance Between PM and TM Vertex | m | 3.8 | 2.0 | 3.06 | 2.00 | Packaging constraint for keeping tertiary mirror inside fairing. |

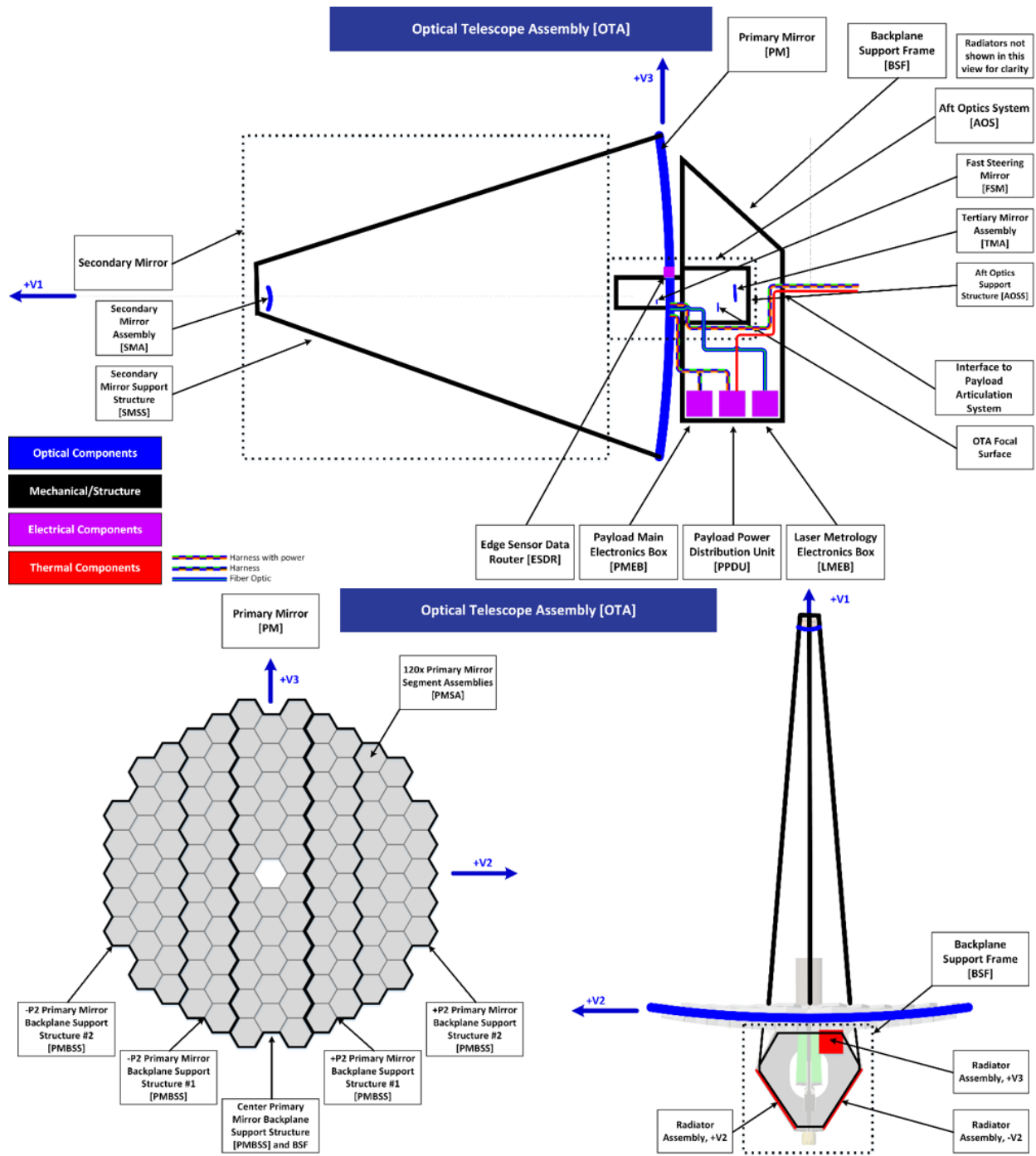

**Figure 8-7.** *LUVOIR-A OTA block diagram.*

Several correlations between requirements drove the design optimization. A requirement on the maximum angle of incidence at any optical surface is imposed by the high-contrast exoplanet science. This requirement flows from limiting the amount of polarization aberration and cross-polarization leakage in the ECLIPS instrument. Requiring the angle of incidence at any optical surface to be less than 12° directly influences the distance between the primary and secondary mirrors and the F/# of the primary mirror. The distance between the primary and secondary mirrors also influences the size of the sunshade as the secondary mirror must remain in the sunshade's shadow while the OTA is articulated to point at different objects in the sky. While it is advantageous to have a shorter separation distance for packaging, deployment, stability, and reducing the sunshade size, these are all second in priority to the maximum angle of incidence requirement, which directly impacts the exo-Earth yield science objective.

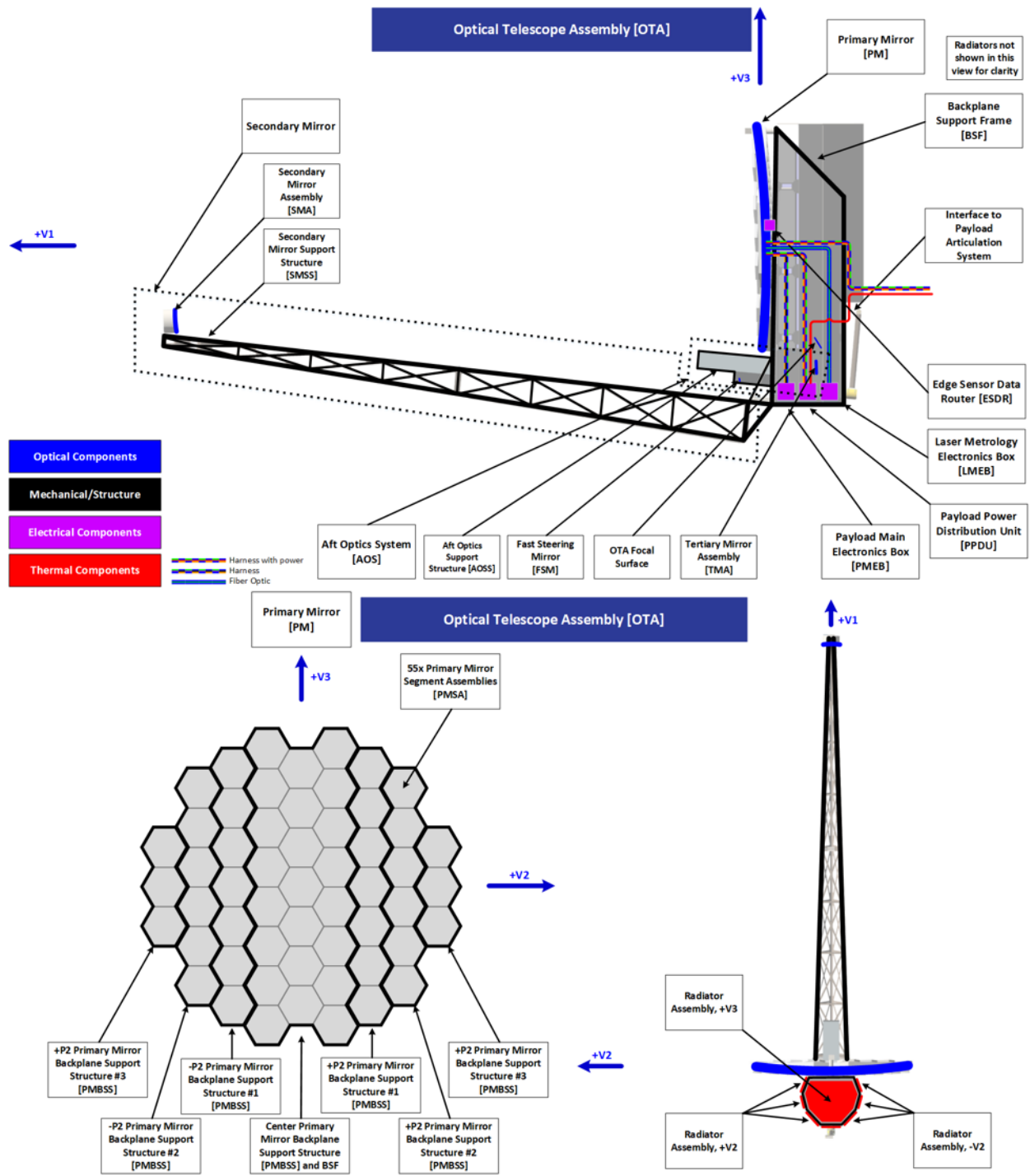

**Figure 8-8.** *LUVOIR-B OTA block diagram.*

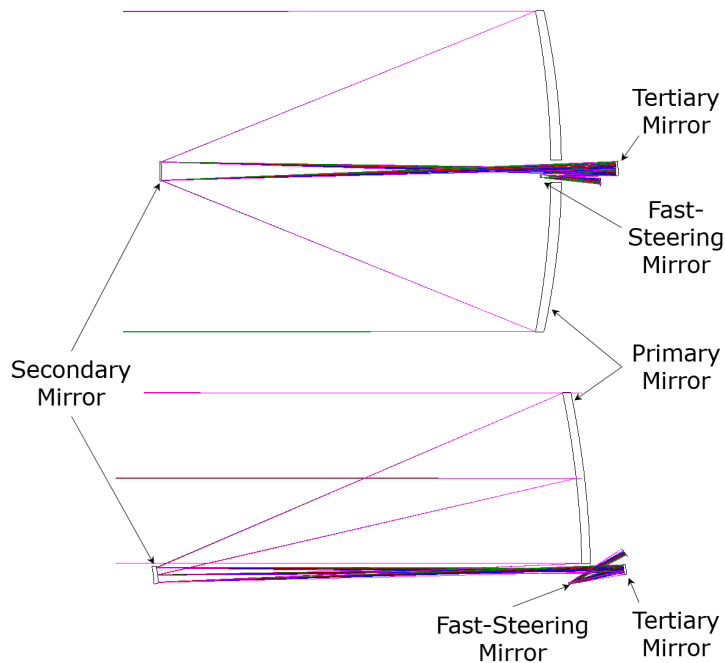

**Figure 8-9.** *Ray trace of the LUVOIR-A on-axis OTA (top) and LUVOIR-B off-axis OTA (bottom). The two images are approximately to scale.*

Another important constraint imposed on LUVOIR-A is for all rays to pass within a one-segment-sized hole in the primary mirror. A smaller hole in the primary mirror reduces the size of the entrance pupil obscuration, which has a significant impact on the coronagraph throughput, and therefore exoplanet yield. In this TMA design-form the ray bundles pass through the central hole three times before arriving at the OTA focal surface. Since the primary mirror hole is only one segment, the size of the ray bundle needs to be carefully constrained at all three passes, which degrades the wavefront error as the positions of the secondary, tertiary, and fast-steering mirrors become limited. To improve the wavefront error performance, the tertiary mirror is an x-y polynomial surface which has additional degrees of freedom to correct the induced aberrations.

The first three mirrors in LUVOIR-B are all off-axis conic sections, followed by a flat fast-steering mirror. Since this design is unobscured, it does not require the same packaging constraints for rays to fit through a central hole, thus eliminating the need for an x-y polynomial tertiary mirror surface.

An important constraint placed on the design of the LUVOIR-B OTA is the location of the OTA focal surface above the tertiary mirror. This is driven by the desire to have all science instruments located behind the primary mirror to simplify packaging of the system in the stowed configuration (i.e. with the primary mirror wings wrapped around the instrument column). Tilt of the secondary, tertiary, and fast steering mirrors positions the OTA focal surface behind the primary mirror, but drives the F/# of the OTA to be relatively slow. The longer focal length increases the field-of-view dimensions at the OTA focal surface, requiring larger optics in the instruments to capture their respective fields. There is a delicate optimization balance between packaging the instruments behind the PM and reducing the size of the instruments by shortening the focal length of the OTA.

The LUVOIR-A and LUVOIR-B hexagonal mirror segments have a projected flat-to-flat distance of 1.223 m and 0.955 m, respectively, with a 6 mm gap between segments.

On LUVOIR-A there are a total of 120 mirror segments. Symmetry reduces the number of unique mirror shapes to 20, providing a non-recurring engineering cost savings in the number of mandrels and test articles needed to manufacture the mirrors. On LUVOIR-B, there are a total of 55 mirror segments. There are no identical mirrors due to the off-axis nature of the primary mirror, however there are 24 mirror-symmetric pairs of segments. In both cases, parallel production lines will be used to efficiently manufacture, integrate, and test all of the mirror segment assemblies. **Figure 8-10** shows the pupil function for both LUVOIR concepts.

All OTA optics use a protected aluminum + "enhanced" lithium fluoride (Al+eLiF) coating. Unprotected eLiF coatings have demonstrated >80% normal incidence between 103 and 130 nm, as well as high broadband performance across the entire LUVOIR operational wavelength range consistent with standard protected-Al coatings (Quijada et al. 2014). A thin protective layer of $MgF_2$ or $AlF_3$ is added on top of the LiF layer to reduce the hygroscopic sensitivity of the LiF, without degrading the far-UV performance of the coating (Witt et al 2018, Balasubramanian et al 2017).

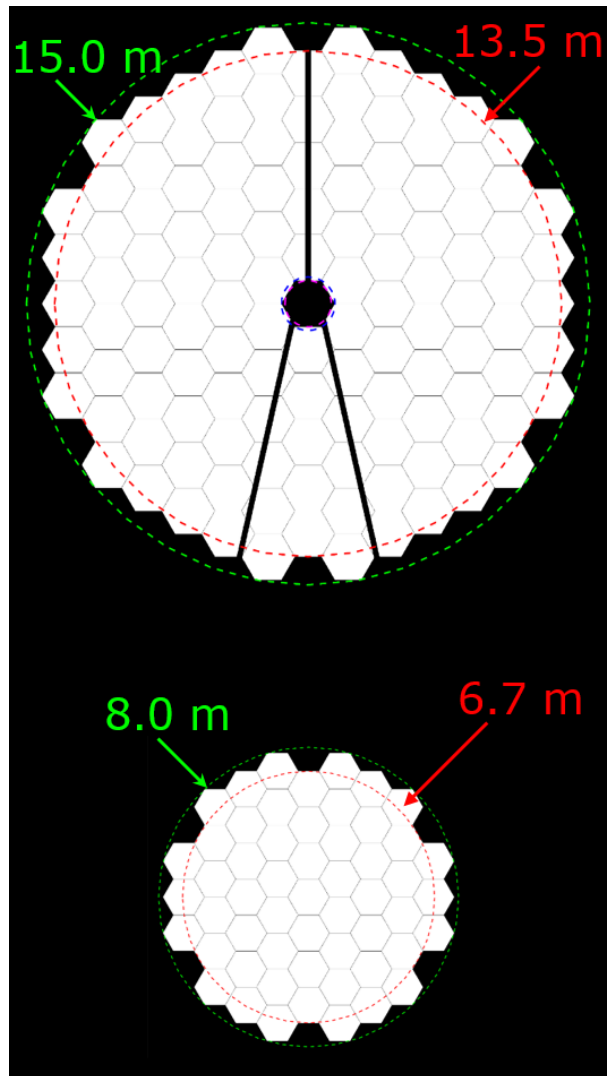

**Figure 8-10.** *LUVOIR-A (top) and LUVOIR-B (bottom) aperture functions. LUVOIR-A has 120 segments that are 1.223 m flat-to-flat. LUVOIR-B has 55 segments that are 0.955 m flat-to-flat. Both designs use 6 mm gaps. The two images are approximately to scale.*

### 8.2.1.3  Mechanical design

#### 8.2.1.3.1  Primary mirror

The packaging of both LUVOIR concepts into their respective fairings is designed to maximize the available volume for the science instruments, while at the same time trying to minimize deployment complexity. In each case, the primary mirror is folded to approximate the circumference of the fairing. The LUVOIR-A primary mirror folds with only two wings on each side in the SLS 8.4 meter-diameter fairing (see **Figure 8-11** top) striking a reasonable balance between complexity and maximizing instrument volume. The LUVOIR-B primary mirror requires three wings per side to fit into the smaller 5-meter-diameter fairing (see **Figure 8-11** bottom).

**LUVOIR-A**

**LUVOIR-B**

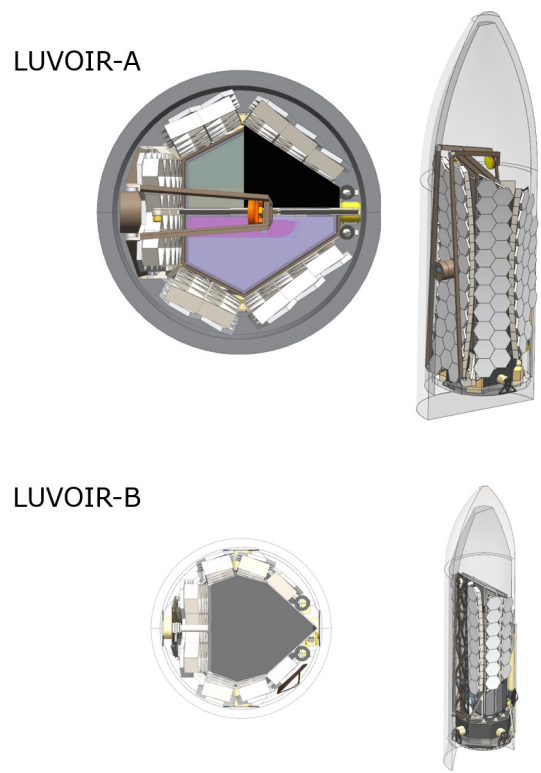

**Figure 8-11.** *LUVOIR-A (top) uses two wing folds on either side of the primary mirror to fit within the 8.4-m diameter fairing, and leave sufficient volume for instruments at the center. LUVOIR-B (bottom) uses three wing folds on either side to fit within a smaller 5.0 m fairing.*

The backplane design uses interlocking, bonded, composite plates to form a coarse iso-grid structure for each wing assembly. The plate panels form an I-beam cross-section that is 450 mm deep on LUVOIR-A, and 385 mm deep on LUVOIR-B. This design was chosen for both fabrication and design simplicity. It is likely that additional optimization of the design can reduce mass while maintaining (or increasing) stiffness.

The design of the primary mirror segment assembly also leverages the JWST design, shown in **Figure 8-12**. A single, stiff mirror segment is mounted on a hexapod for 6 de-gree-of-freedom rigid-body positioning of the segment. A delta frame interfaces the mirror segment substrate to the hexapod actuators and to the primary mirror backplane support structure.

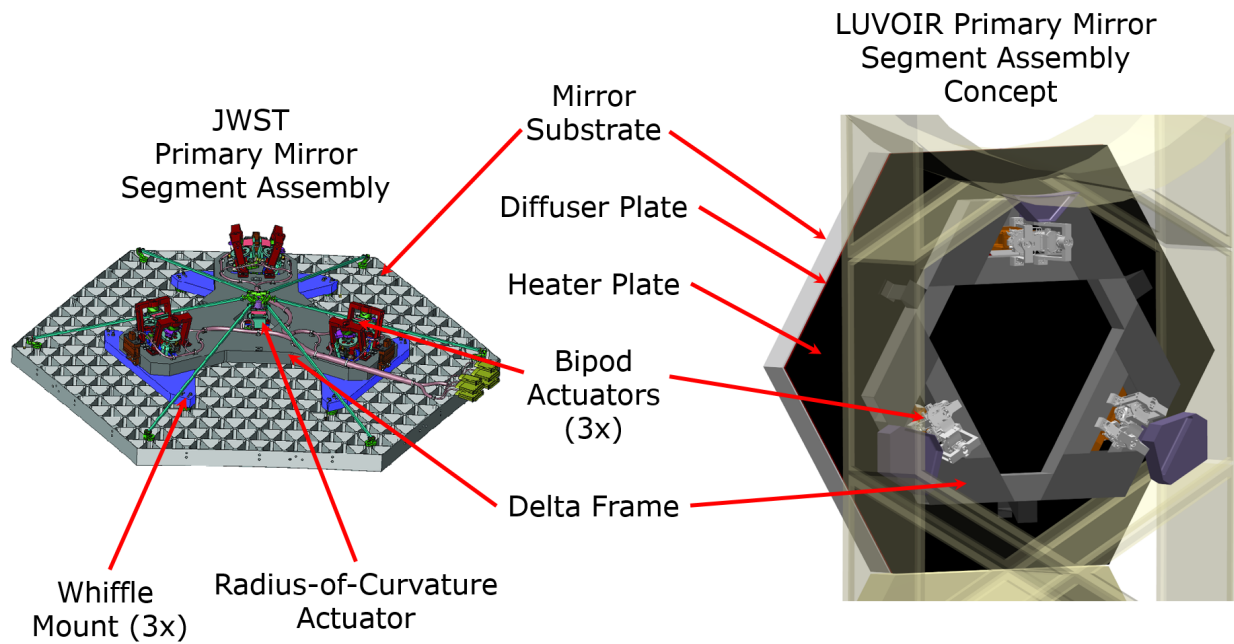

**Figure 8-12.** *Left: JWST primary mirror segment assembly, viewed from the back (Credit NASA/GSFC). Right: LUVOIR's primary mirror segment assembly concept, showing many shared components. While JWST uses a radius-of-curvature actuator, LUVOIR does not. However, LUVOIR uses heater and diffuser plates to control the segment temperature, and co-locates the mirror segment control electronics with each segment (not shown).*

The LUVOIR segment assembly architecture departs from that of JWST in the materials and the operating temperature, both optimized for stability. Instead of beryllium segments, LUVOIR has baselined Ultra-low Expansion (ULE) glass for the mirror substrate, although other low coefficient of thermal expansion (CTE) materials such as Zerodur may also be used. Additionally, LUVOIR's delta frame will be low-CTE composite material.

JWST required radius-of-curvature actuators on each segment because the mirrors deformed in a 1-g environment during integration and test, and therefore needed correctability built in. For LUVOIR, preliminary analysis indicates that the ULE mirrors, with their closed back design, will be substantially stiffer than JWST's open-backed beryllium mirrors, and thus able to hold their prescribed shape in 1-g. Therefore, LUVOIR is not baselining the use of a radius-of-curvature actuator at this time, although this will be re-evaluated as the design matures.

Another critical new component to the segment assembly architecture is the edge sensor and piezoelectric actuator control system. Each mirror segment is fitted with one edge sensor per edge (i.e., two edge sensors total per shared edge). While the current design baselines capacitive-based edge sensors, optical or inductive sensors could also be used if sufficiently mature and capable. Each sensor measures the local displacement between the segment edges. The positions of each segment are then corrected with the hexapod fine-stage piezoelectric actuators with picometer resolution. **Chapter 11** describes the technology development path for this architecture.

### 8.2.1.3.2  Secondary mirror

The distance between the primary mirror and the secondary mirror (over 18 m for LUVOIR-A and 20 m for LUVOIR-B) precludes a direct application of the folding secondary mirror support structure on JWST. On LUVOIR-A the JWST concept was modified to include an additional joint in the upper and lower arms, but otherwise the same hinges, latches, and drive motors that were used on JWST can be used on LUVOIR-A.

For LUVOIR-B the design requires no obstruction from the secondary mirror support structure, so a completely new concept was developed. The structure is composed of two sets of three folding panels that form an inner and outer boom. Both booms deploy in the same way with z-folded panels that unfold into a stiff, triangular cross section assembly (see **Figure 8-13**, and the deployment sequence video[4] for more details). The secondary mirror itself is then deployed from the end of the outer boom. All of these motions are single-axis, simple rotational motions using hinge, drive, and latch mechanisms with extensive flight heritage.

For both LUVOIR concepts, the secondary mirror assembly, including actuators and delta frame, uses a similar design as the primary mirror segment assemblies, excepting the size and shape of the optical elements.

---

4 See https://asd.gsfc.nasa.gov/luvoir/design/

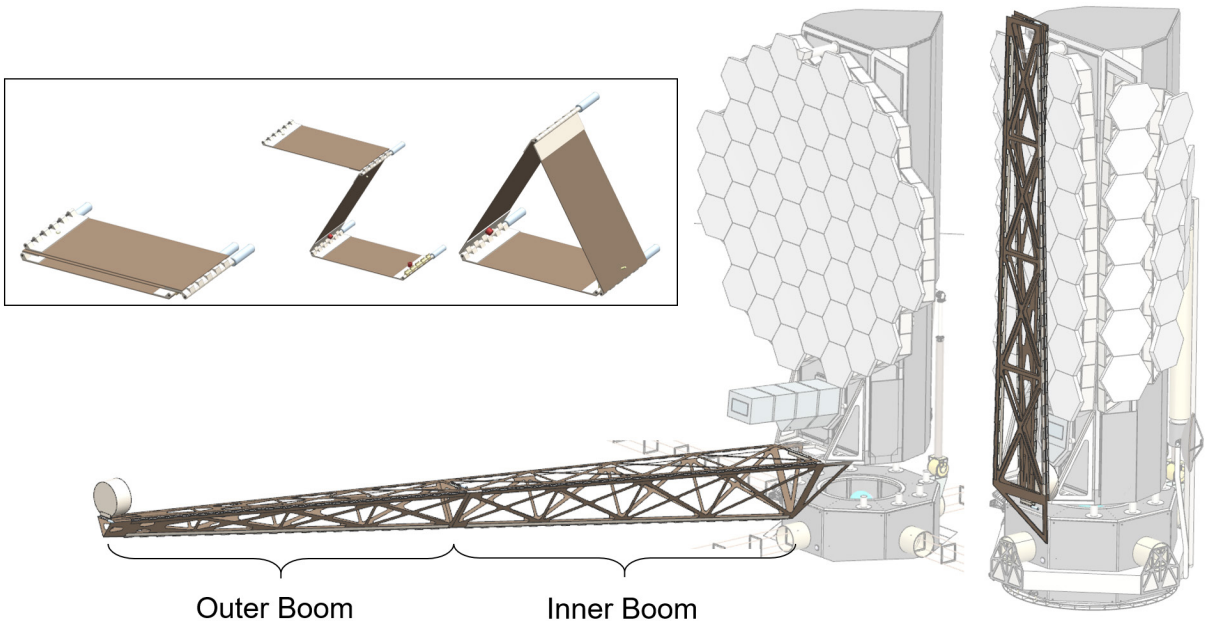

Outer Boom          Inner Boom

**Figure 8-13.** *LUVOIR-B secondary mirror support structure concept. The structure is stowed (right) as two sets of three panels in a z-fold configuration. The inset shows how the three panels unfold and lock to form triangular cross-section beam (bottom-left).*

### 8.2.1.3.3  Backplane support frame (BSF)

The BSF supports the primary and secondary mirror assemblies, all instruments, aft optics system, radiators, and directly interfaces with the spacecraft for launch. During launch, it provides a load path for all of the OTA and instrument loads through the spacecraft structure and into the launch vehicle interface. Once the payload element is deployed, the BSF interfaces to the spacecraft through the Payload Articulation System.

The BSF truss is a composite box beam assembly with bonded corner gussets and clips, using fabrication concepts similar to the JWST BSF. Honeycomb shear panels have been added within the truss to increase stiffness (see **Figure 8-14**). The primary interface between the BSF and the instruments is a centrally located bulkhead. The instrument mounting interface at the central bulkhead reduces the load path complexity and consolidates support for both the primary mirror and instruments into a central location.

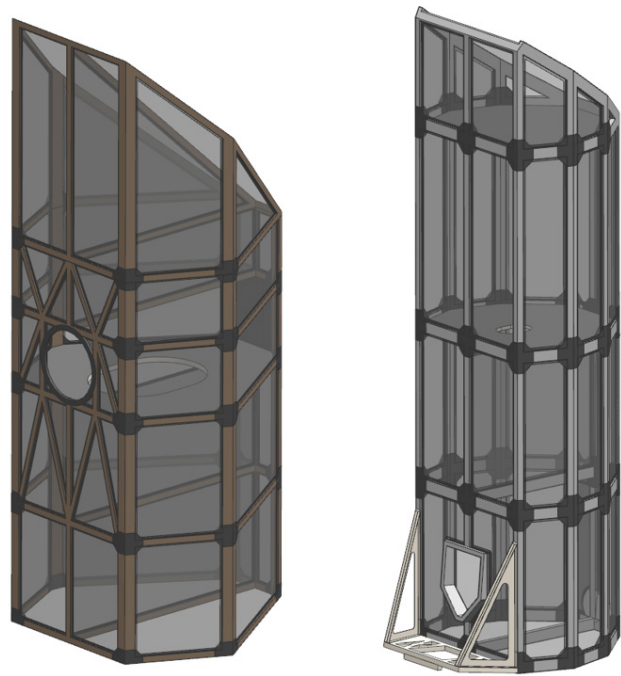

**Figure 8-14.** *The backplane support frame for LUVOIR-A (left) and LUVOIR-B (right). The main structure uses a composite box-beam assembly with honeycomb shear panels (rendered partially transparent here to show internal structure). Note the images are not to scale.*

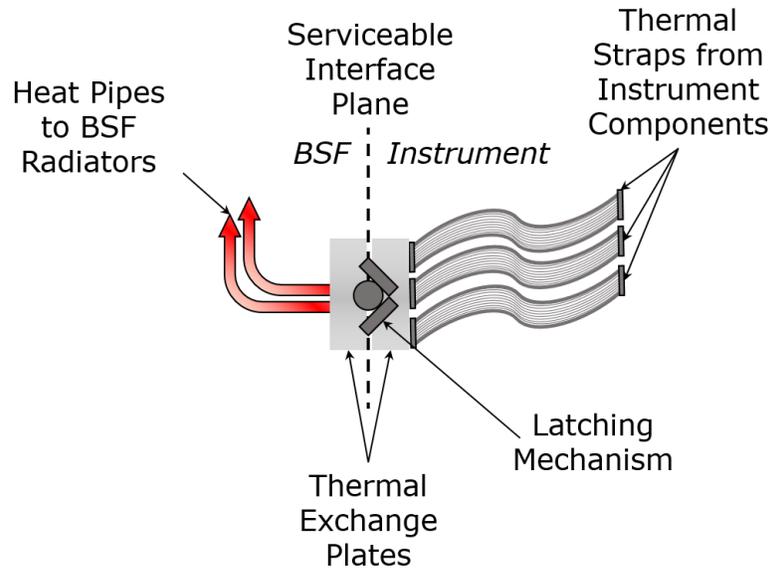

**Figure 8-15.** *A serviceable thermal interface allows for heat transfer from different instrument thermal zones to specific radiators mounted on the BSF. Within the instrument (right), thermal straps conduct heat from components to the thermal exchange plates. Within the BSF (left), heat pipes transfer heat to the respective radiators. The thermal exchange plates can be connected or separated at the interface plane via a remotely operated mechanical latch. One thermal interface is provided for each instrument thermal zone/radiator pair.*

The BSF also provides the necessary infrastructure for servicing and replacing the instrument modules. Guide rails and kinematic latches, similar to those used on Hubble, provide a repeatable mechanical interface for each instrument. Blind-mate power and data interfaces are located at central BSF bulkhead. A serviceable thermal interface is shown in **Figure 8-15**. Thermal straps within each instrument conduct heat to one side of a thermal exchange plates. Heat pipes on the BSF side of the interface then transport that heat to the specific radiator panel mounted on the external faces of the BSF. Finally, the BSF also provides standard grapple interfaces for potential robotic servicers.

#### 8.2.1.4 Thermal design
In the OTA architecture, the BSF structure represents a critical thermal and mechanical hub for the payload element. As shown in **Figure 8-1** and **Figure 8-2**, the payload element radiators are mounted to the sides of the BSF. The BSF composite beams are covered on all external sides with vapor-deposited aluminum (VDA) multi-layer insulation (MLI), and the shear panels are covered with VDA MLI on the space-facing sides and black Kapton on the internal-facing surfaces. The entire structure is actively heated to 270 K. The top and bottom of the BSF are capped with VDA MLI to prevent heat loss to space.

The primary mirror backplane support structure is composed of a lattice of composite I-beams. Only the space-facing portions of the I-beams are covered with VDA outer layer MLI, while the rest of the surfaces remain bare composite. The back surface of the backplane structure has all cavities closed out with VDA MLI, while the front surface has black Kapton outer layer MLI closeouts to prevent stray light on the optical assembly.

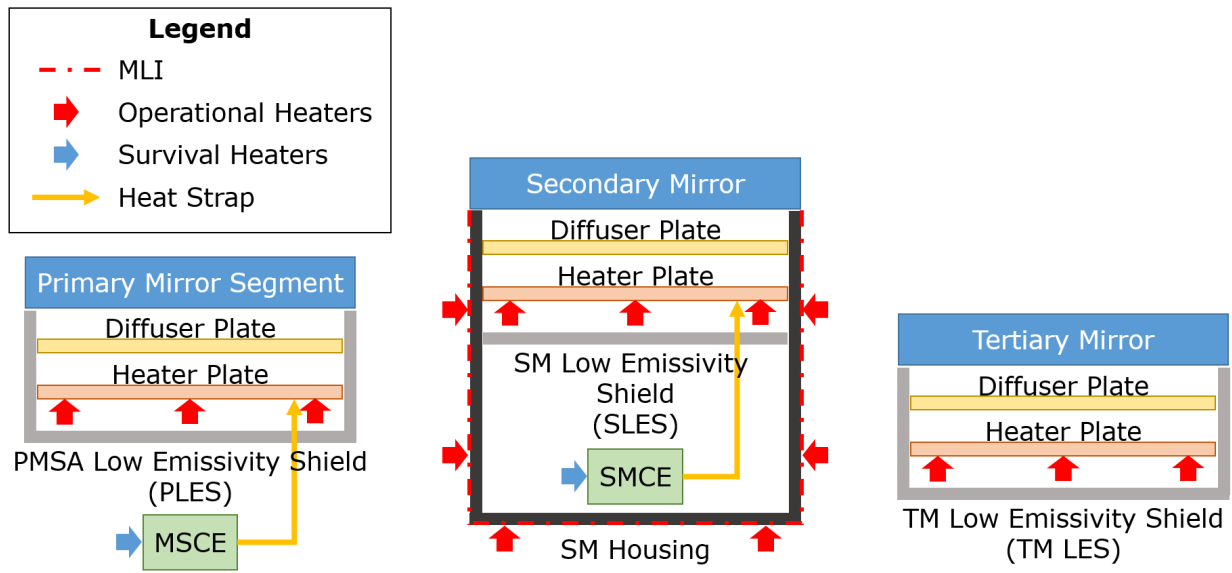

**Figure 8-16.** *Thermal block diagrams of the primary mirror segments (left), secondary mirror (center), and tertiary mirror (right), showing the thermal architecture for each mirror system. MSCE: Mirror Segment Control Electronics; SMCE: Secondary Mirror Control Electronics; MLI: Multi-layer Insulation*

The thermal assemblies of the primary mirror segments, secondary mirror, and tertiary mirror, depicted in **Figure 8-16**, are designed to allow for active heating of the mirror without direct contact between the heater and the mirrors, so that the mirror optical and thermal stabilities are not directly influenced by the heater control. In each mirror assembly, immediately behind the mirror substrates on the non-reflective side there is a diffuser plate, a heater plate, and a low-emissivity shield, each with the same surface area as the mirror face. The heater plate is actively controlled to radiatively drive the temperature on the mirror substrate to 270 K, while the diffuser plate smooths the spatial and temporal gradients from the heater plate. The low-emissivity shield is a pan-like rigid aluminum structure, designed to reduce radiative losses from the thermal assembly back towards the segment actuators and delta frame.

In contrast to the other mirrors, the fast-steering mirror has a heater directly mounted on the underside of the mirror substrate due to its large range of motion and small size, maintaining a relatively isothermal mirror surface despite not using a heater plate assembly.

A heat strap from the mirror control electronics that are co-located with each mirror segment and the secondary mirror transports excess dissipated heat from the electronics boxes to the heater plate. This reduces the amount of heat that needs to be generated at the heater plate, thus conserving total power needed for the observatory.

The secondary mirror support structure is also actively controlled to 270 K to limit alignment distortions. The support structure and non-optical surfaces of the secondary mirror are covered in black Kapton outer-layer MLI to reduce the amount of heater power lost to the environment, as well as to reduce the amount of stray light into the optical path.

### 8.2.1.5   Electrical design

The electrical architecture for the OTA is shown in **Figure 8-17**. The only significant difference between LUVOIR-A and LUVOIR-B is in the number of mirror segment electronics boxes (120 versus 55, respectively), number of edge sensor data routers (one per primary mirror wing section: 5 on LUVOIR-A, 7 on LUVOIR-B), and in the inclusion of the POLLUX instrument on LUVOIR-A. A mirror segment control electronics box is co-located with each primary mirror segment, and with the secondary mirror. A block diagram of the mirror segment control electronics is shown in **Figure 8-18**. The high data rate of the edge sensor read-out exceeds the capability of the 1553 bus, so a Low-Voltage Differential Signaling (LVDS) interface and data router is employed to transport the data to the control system processor, located inside the ECLIPS instrument.

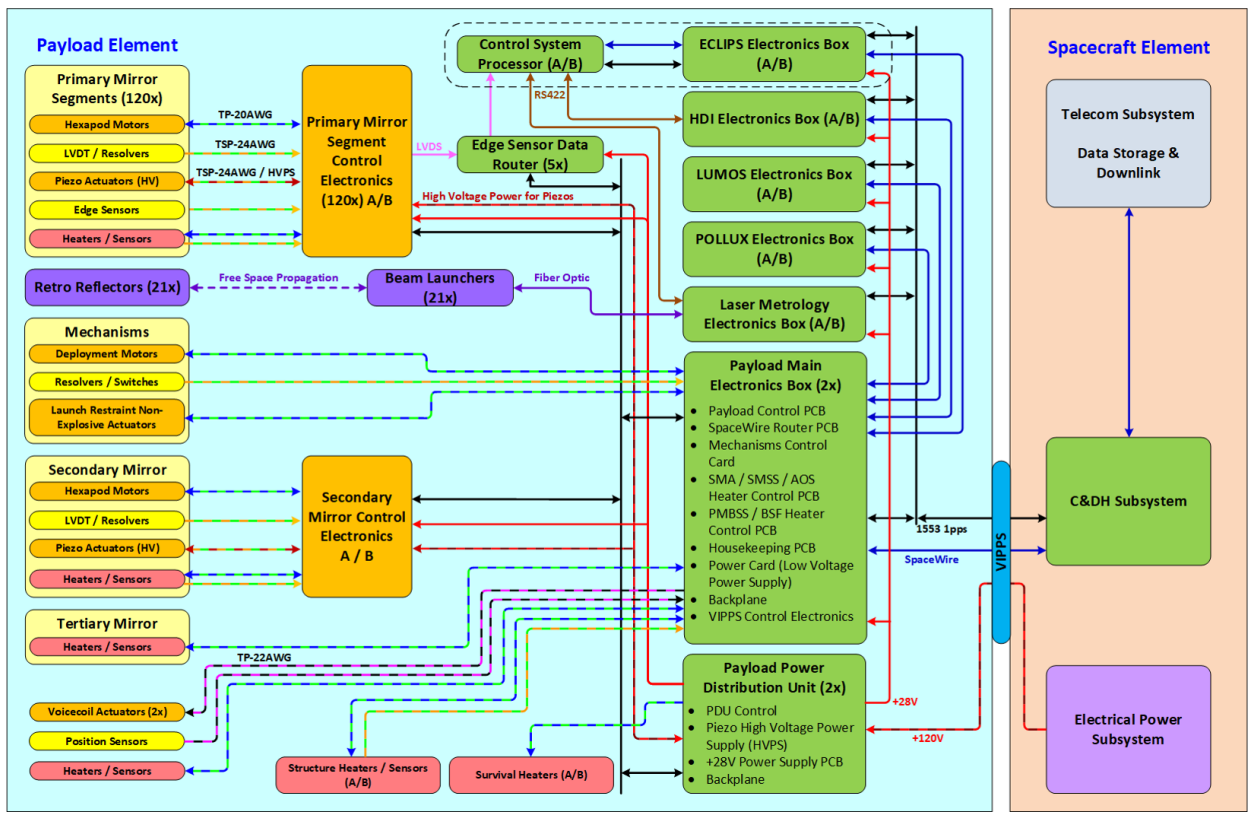

**Figure 8-17.** *The LUVOIR-A payload element electrical architecture. The system is the same for LUVOIR-B, except there are only 55x primary mirror segments, 7x edge sensor data routers, and the POLLUX instrument is not accommodated on LUVOIR-B.*

The control system processor is responsible for coordinating all of the metrology, wavefront sensing, and pointing control loops on the LUVOIR payload. Since much of its performance is dictated by the high-contrast imaging operations, the control system processor is located within the ECLIPS instrument, facilitating upgrades to the processing capability whenever the coronagraph instrument itself might be upgraded or replaced.

Aside from the control system processor, each instrument contains its own front-end electronics, main electronics box, and data storage. These elements perform all data processing, such as co-adding, windowing, and compression, and store science and telemetry data locally within the instrument. Performing this processing locally reduces the data harnessing required between the payload and spacecraft, reducing the amount of dynamic disturbance that can be transmitted from the spacecraft through the harness.

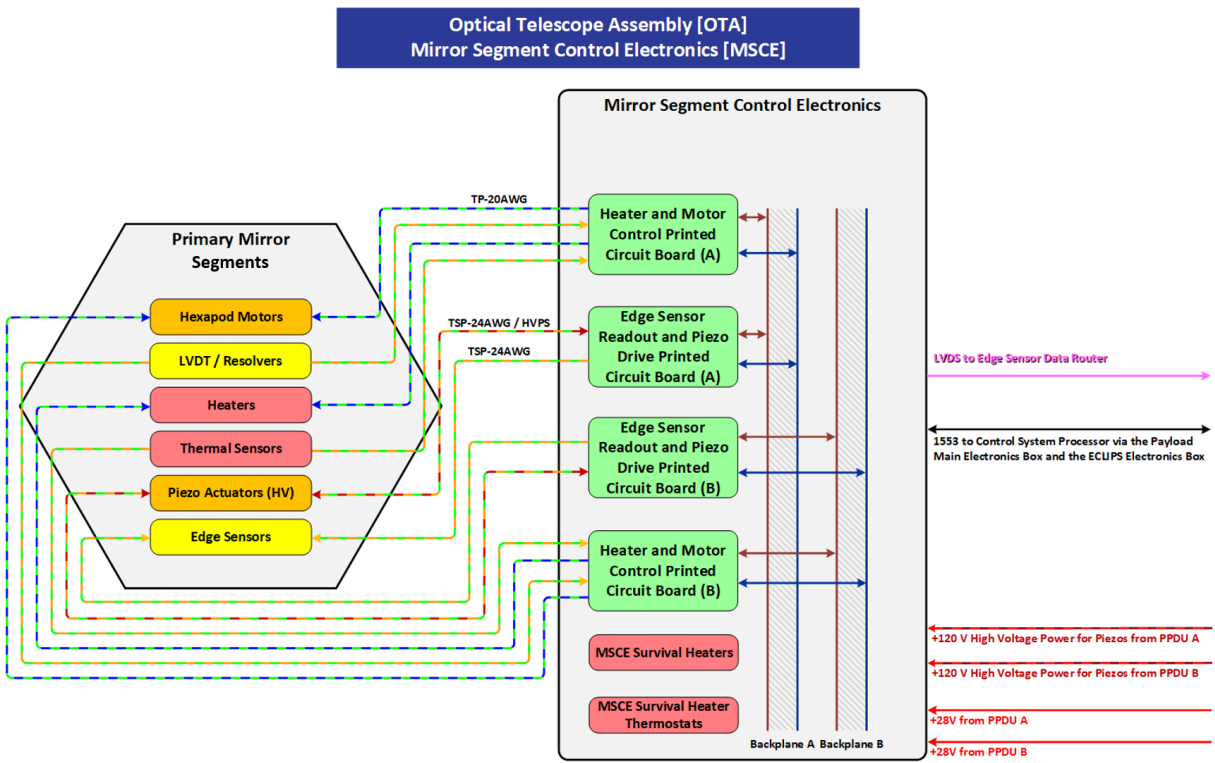

**Figure 8-18.** *Mirror segment control electronics block diagram. A single, board-redundant box is located with each mirror segment. A similar electronics box is also located with the secondary mirror, however without edge sensors or the edge sensor readout function.*

### 8.2.1.6 Metrology system

In addition to the edge sensor metrology system at the primary mirror, a laser metrology system is included to measure the relative positions in six degrees of freedom of the primary mirror, secondary mirror, and aft-optics system with a precision of 10-100 pm. The laser metrology system uses distance-measuring laser gauges mounted on a reference plane in the aft-optics system, and aimed at retroreflectors mounted to the edge of the secondary mirror. Measurements from this truss are fed back to the secondary mirror rigid body actuators in real time, effectively anchoring the secondary mirror to the aft-optics system.

Six other sets of beam launchers are mounted to a select subset of primary mirror segments, and are also aimed at retroreflectors mounted to the secondary mirror. The motion measured by these gauges is corrected at the segment actuators. Neighboring segments are then aligned to these "anchor" segments via the edge sensor system. Thus, the entire primary mirror, secondary mirror, and aft-optics system are effectively rigidized to within the measurement error of the metrology systems.

The laser gauges are all connected to a seed/injection-locked fiber laser via a fiber-optic multiplexer. The return signals from the retroreflectors are directed to phasemeter electronics for heterodyne detection of the distance changes. The measured displacements are then passed to the control system process for integration with other metrology and wavefront sensing data. **Figure 8-19** shows a block diagram of the laser metrology system.

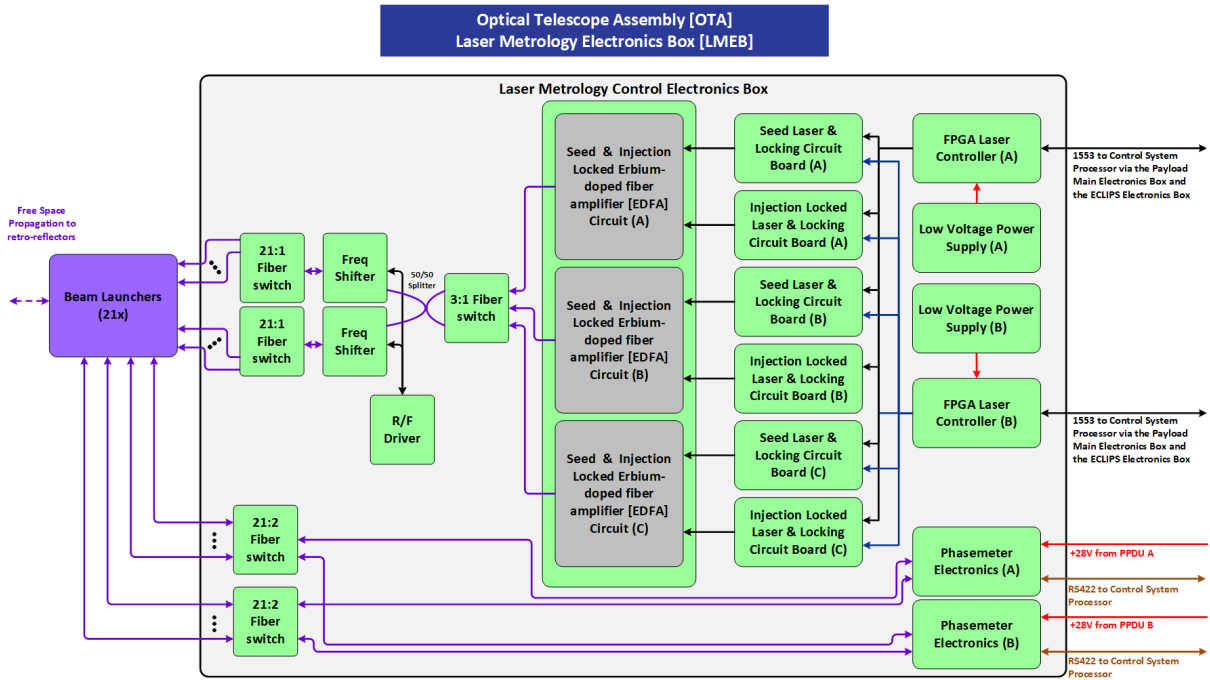

**Figure 8-19.** *Block diagram of the laser metrology system. For redundancy, three seed/injection-locked laser pairs are included in the electronics box (only one is needed to run the system at any time). There are three beam launchers mounted to each of six segments, plus an additional three launchers mounted to the aft-optics system, for a total of 21 beam launchers.*

## 8.2.2 High Definition Imager (HDI)

### 8.2.2.1 Introduction

The LUVOIR observatory will revolutionize the study of the formation and evolution of planets, stars, and galaxies through a combination of very high sensitivity, high angular resolution, and a highly stable and well-calibrated point-spread function. A key instrumental capability for LUVOIR is the High Definition Imager (HDI) instrument—the primary astronomical imaging instrument for observations in the near-UV through the near-IR.

The scientific investigations discussed in this report call for an instrument that can instantaneously observe a field-of-view that spans 6 square arcminutes and provides Nyquist sampling that takes full advantage of the angular resolution provided by the telescope. The diverse nature of the science cases also demand an instrument with an ample range of spectral elements including standard broad, medium, and narrow band filters as well as several dispersing elements to enable low-resolution slitless spectroscopy.

### 8.2.2.2 Optical design

The optical and detector specifications for HDI-A and HDI-B are summarized in **Table 8-5**. HDI is designed to be Nyquist sampled and diffraction-limited at 0.5 μm and 1.0 μm in the UVIS and NIR channels, respectively.

**Figure 8-20** shows a block diagram for HDI, and **Figure 8-21** shows the corresponding optical ray trace for HDI-A. A fold mirror is located after the focal surface of the optical telescope assembly (OTA) to direct the light to a pair of pupil relay mirrors. From there, the light passes through a channel select mechanism that determines which channel(s) is/are

**Table 8-5.** *HDI optical and detector specifications*

| Parameter | Units | HDI-A | | HDI-B | |
|-----------|-------|-------|-----|-------|-----|
| | | UVIS | NIR | UVIS | NIR |
| Bandpass | μm | 0.2–1.0 | 0.8–2.5 | 0.2–1.0 | 0.8–2.5 |
| Aperture Diameter | m | 15 | 15 | 8 | 8 |
| F/# | – | 26 | 20 | 26 | 20 |
| Focal Length | m | 390 | 300 | 208 | 160 |
| Field-of-View | arcmin | 2.93 x 1.94 | 2.97 x 1.96 | 2.73 x 1.80 | 2.75 x 1.81 |
| Plate Scale | mas / pixel | 3.43 | 6.88 | 6.45 | 12.89 |
| Diffraction-limited Spot Size | μm | 31.72 | 48.80 | 31.72 | 48.80 |
| RMS Pointing Stability | 1-s mas | 0.43 | 0.86 | 0.81 | 1.61 |
| RMS Wavefront Error | nm | < 35 | < 71 | < 35 | < 71 |
| Detector Type | – | CMOS | HgCdTe | CMOS | HgCdTe |
| Pixel Size | μm | 6.5 | 10.0 | 6.5 | 10.0 |
| Detector Format | pixels | 8192 x 8192 | 4096 x 4096 | 8192 x 8192 | 4096 x 4096 |
| Array Tiling | – | 6 x 4 | 6 x 4 | 3 x 2 | 3 x 2 |
| Total Number of Pixels | Gpix | 1.61 | 0.40 | 0.40 | 0.10 |
| Detector Temperature | K | 170 | 100 | 170 | 100 |
| Read Noise | e- | ~2.5 | ~2.5 | ~2.5 | ~2.5 |
| Dark Current | e-/pix/s | ~0.002 | ~0.002 | ~0.002 | ~0.002 |
| System Quantum Efficiency | – | 0.21 (V-band) | 0.34 (J-band) | 0.21 (V-band) | 0.34 (J-band) |

operating at any given time. The channel select mechanism has a total of five optical elements which gives HDI five operational modes:

1. **NIR transmissive:** Sends all light from the OTA to the NIR channel. This position also allows light from an internal calibration source to be directed into the UVIS channel.

2. **UVIS reflective:** Sends all light from the optical telescope to the UVIS channel. This position also allows light from an internal calibration source to be directed into the NIR channel.

3. **50/50 beamsplitter:** Allows for simultaneous operation of both channels over the entire instrument bandpass (200–1000 nm in UVIS; 1000–2500 nm in NIR), but at a reduced throughput.

4. **Dichroic beamsplitter:** Allows for simultaneous operation of the UVIS channel between 400 nm and 800 nm, and the NIR channel between 800 nm and 1.6 μm.

5. **Optimized UV reflective:** Sends all of the light from the optical telescope into the UVIS channel, but emphasizes UV throughput over broadband operation. For this element, the intention is to provide maximum reflectance in the band between 200 and 400 nm.

After the channel select mechanism, a real pupil image is formed farther downstream in the optical trace. An optical filter wheel is placed in each channel at the location of the pupil as shown in **Figure 8-20**. The UVIS channel accommodates 39 spectral filters, two grisms (R~500), and an assortment of weak lenses and dispersed Hartmann sensors to support image-based wavefront sensing operations. The NIR channel accommodates 24 spectral filters and two grisms (R~500). The mirrors in each channel are coated according to their bandpass to provide the best reflectivity: the pupil relay and UVIS channel mirrors are coated in protected aluminum while the NIR channel mirrors are coated in protected gold.

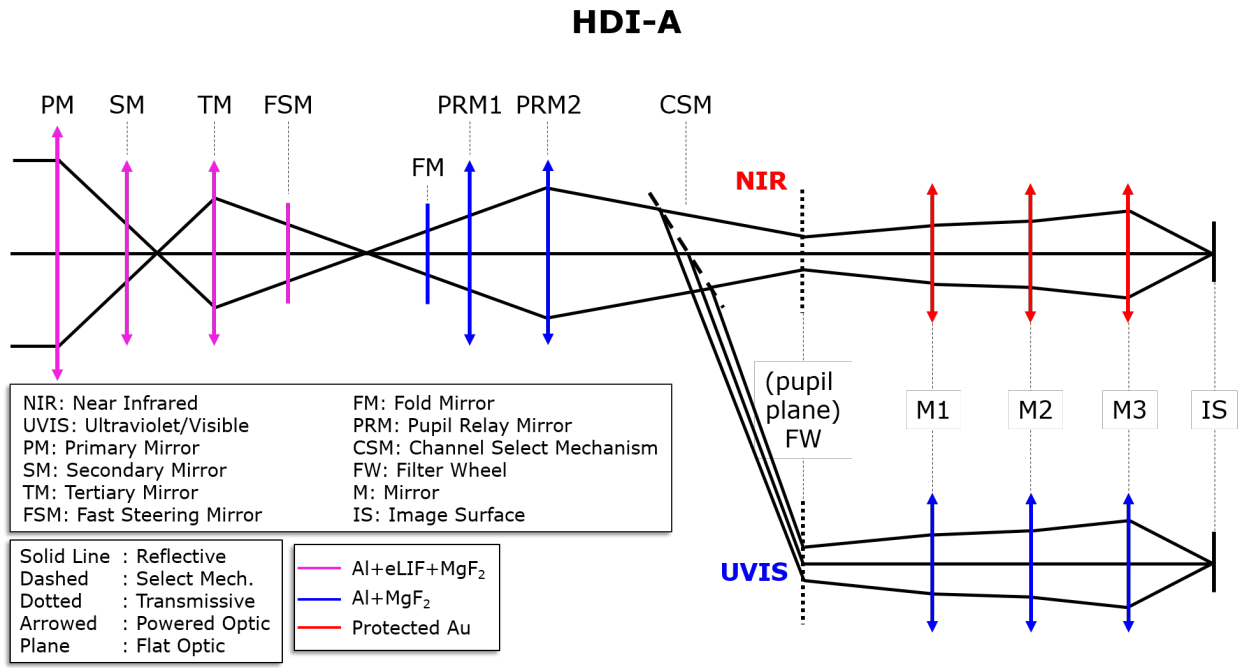

**Figure 8-20.** *HDI-A optical layout. The channel select mechanism can direct light to either the NIR channel, UVIS channel, or both simultaneously. HDI-B has an identical layout with the exception of the fold mirror (FM) prior to PRM1, which is not necessary in that design.*

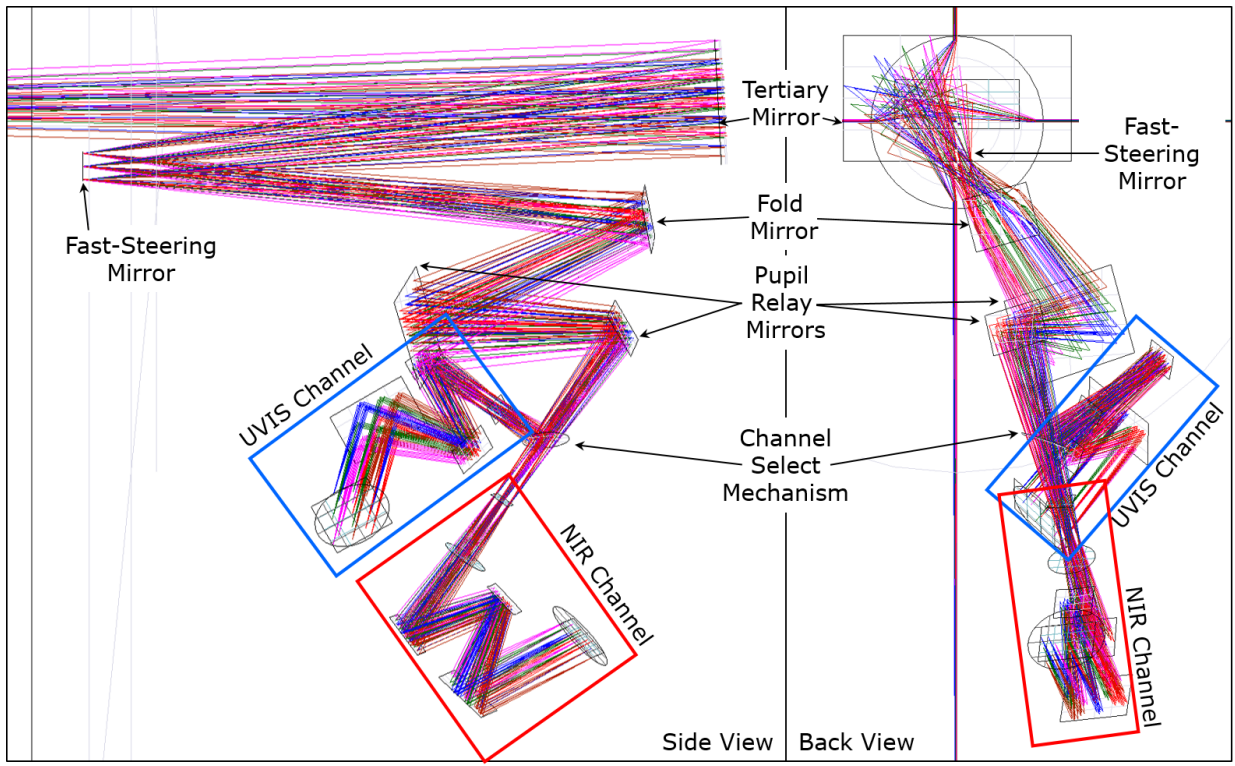

**Figure 8-21.** *HDI-A optical ray-trace*

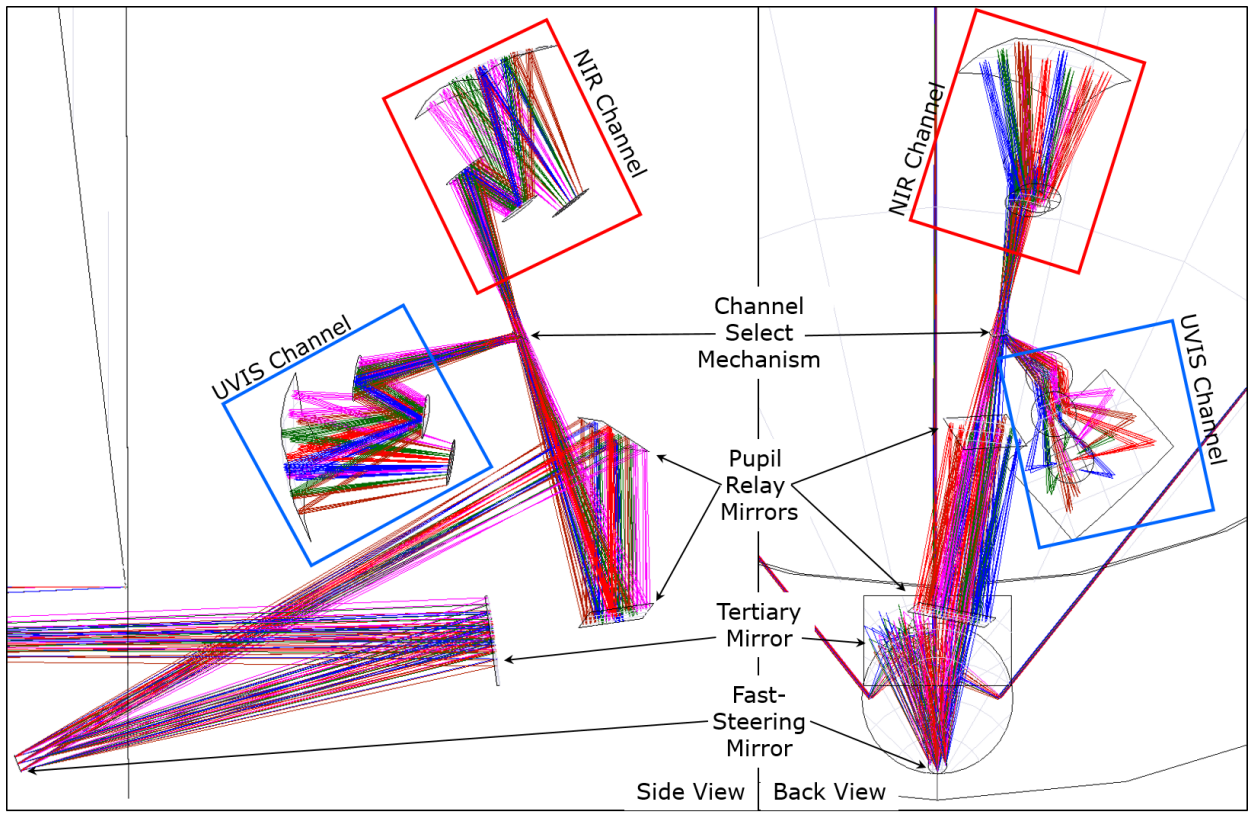

**Figure 8-22.** *HDI-B optical ray trace*

After the filter wheel in each channel, the light enters a three-mirror system that images the observed field-of-view to the detector plane.

HDI-B is very similar in concept and layout to HDI-A. The only difference between the two designs (other than variations in the surface prescriptions of the individual optical elements) is that HDI-B lacks the flat fold mirror that HDI-A has to pickoff the light behind the OTA focal surface. Instead, the first of HDI-B's two pupil relay mirrors serves this purpose (in addition to helping form the pupils for the filter wheels just as in HDI-A). This reduction in degrees of freedom for the optical design does make the design of HDI-B more challenging than HDI-A from an aberration correction point of view, requiring more complex optical surface figures. But the lack of fold mirror also decreases the required size of the instrument. **Figure 8-22** shows the ray-trace for HDI-B.

### 8.2.2.3  Mechanical design
**Figure 8-23** shows an opto-mechanical model of HDI-A and HDI-B. The benches and fixed optical supports are aluminum honeycomb panels with composite face sheets. Additional structural support is provided by composite trusses. HDI also has flexure stands that offset

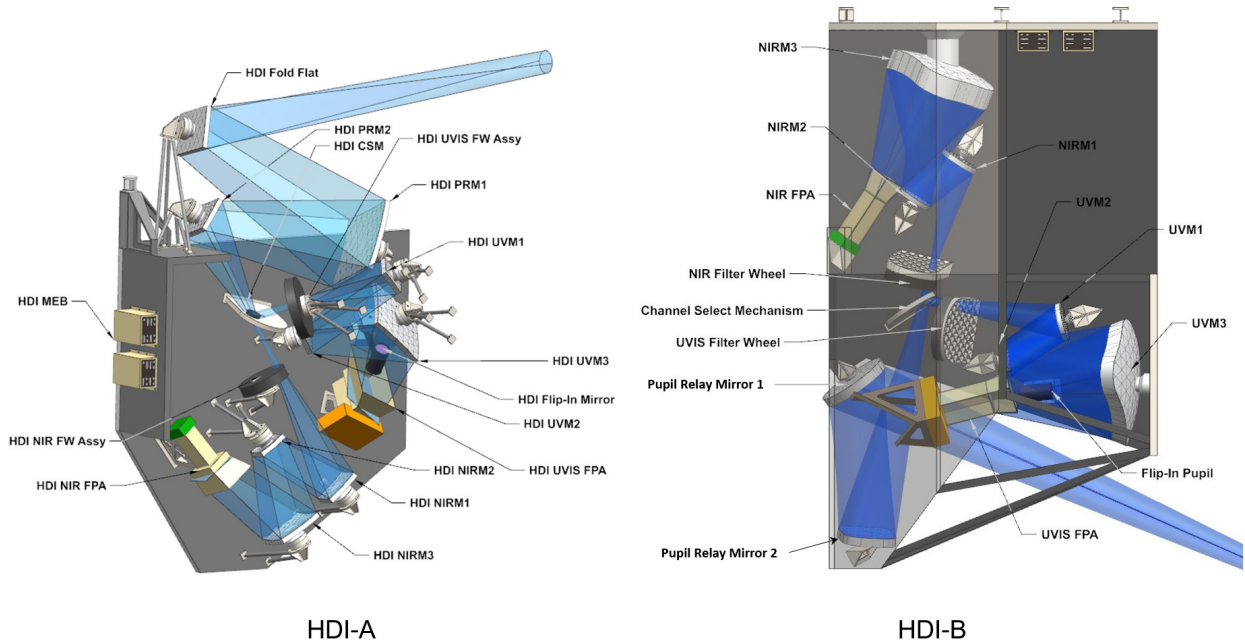

HDI-A                                    HDI-B

**Figure 8-23.** *Opto-mechanical models of the HDI instruments. HDI-A (left) picks off light from the OTA at the top of the instrument (HDI Fold Flat). HDI-B (right) picks off light from the OTA at the bottom, with the first pupil relay mirror. (Note one bench on HDI-A and the NIR channel thermal enclosure or both instruments is not shown for clarity).*

the structure from the BSF bulkhead to allow clearance for mounting of some of the upper mirrors.

The mirrors are lightweighted ULE glass with integral mushroom mount interfaces on the back. The fixed mirrors are mounted on tripods with composite legs and titanium fittings. The tripod interfaces to the mushroom mount with an adjustable 3-point interface ring.

HDI has a NIR channel requiring thermal isolation, provided by an enclosure over the channel elements. Thermal analysis indicates that the bench within the thermal enclosure did not require isolation and that optical element stand-offs made of thermal isolating material such as composite are adequate. This greatly simplified the optical bench design.

### 8.2.2.4 Thermal design

**Figure 8-24** shows a block diagram of the HDI thermal architecture, which is the same for both versions of the instrument. The bulk of the instrument optical bench is heated to 270 K. All components that require colder operating temperatures are heat strapped to a serviceable interface between the instrument and the OTA. On the OTA side of the interface, heat pipes (either ammonia, ethane, or nitrogen, depending on the temperature region) transfer the dissipated heat to the appropriate radiator on the external faces of the BSF.

An MLI blanket enclosure thermally isolates the NIR channel elements, which are passively cooled to 170 K. The UVIS focal plane, UVIS front-end electronics, and NIR channel focal plane front-end electronics are also passively cooled to 170 K. The NIR focal plane itself is passively cooled to 100 K.

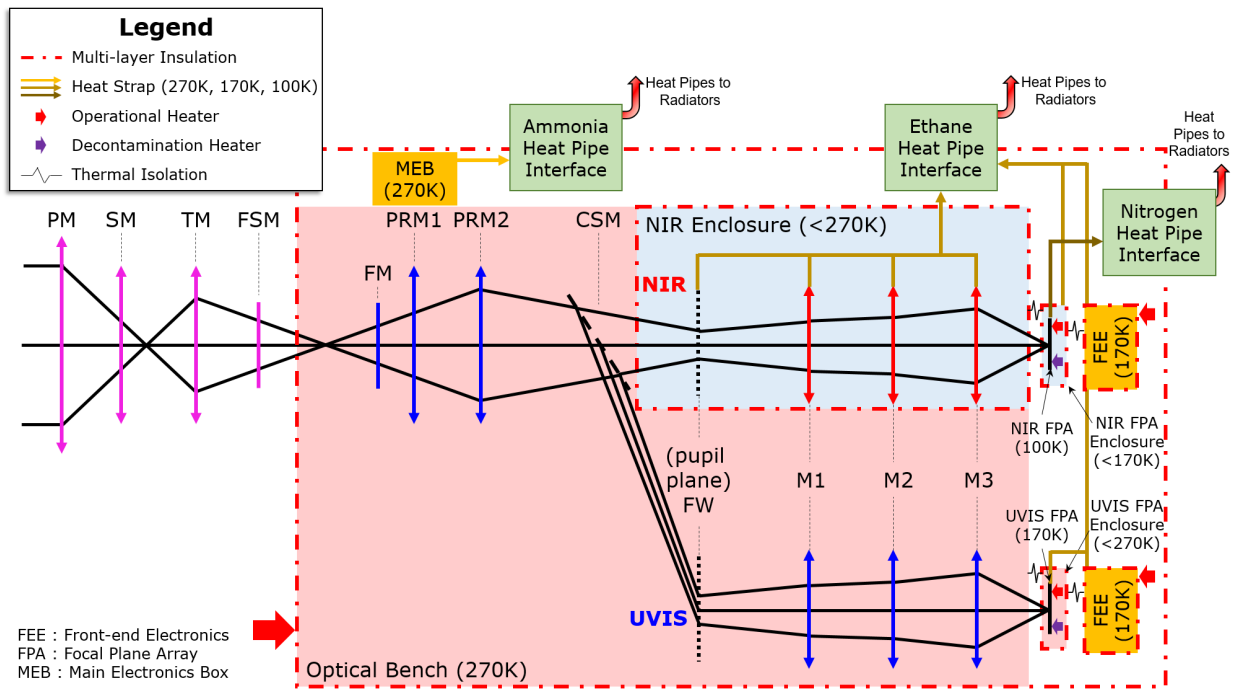

**Figure 8-24.** *HDI thermal block diagram. The thermal architecture is identical for both instruments. Passively cooled components are heat-strapped to serviceable interfaces that connect to heat pipes within the BSF, to transfer heat to the external radiators.*

### 8.2.2.5 Electrical design

**Figure 8-25** shows a block diagram of the HDI electrical architecture, which is identical for each version of the instrument excepting the number of detector ASICs (application-specific integrated circuits, 24 for each channel of HDI-A, 6 for each channel of HDI-B). A serviceable interface provides power and data connections from the OTA. The HDI Main Electronics Box controls all instrument functions, and has internal solid state memory storage for up to 48 hours of science data collection.

### 8.2.2.6 Detectors

The HDI-A UVIS channel focal plane comprises twenty-four 8k × 8k detectors arranged in a 6 × 4 tiled array. For HDI-B, the UVIS focal plane comprises six 8k × 8k detectors in a 3 × 2 array. The UVIS detectors are currently baselined to be CMOS-based devices with 6.5 μm pixels. The 8k × 8k format for each sensor included in the current design has not yet been produced in flight-qualified scientific systems but is within a realistic technology trajectory from current devices. The assumed read noise is 2.5 e-/pixel and the assumed dark current is 0.002 e-/sec/pixel, consistent with current state of the art devices.

The NIR detectors are baselined to be a 6 × 4 array of HgCdTe-based devises with 10 μm pixels on HDI-A, and 3 × 2 array of the same devices on HDI-B. The 4k × 4k format for each sensor included in the current design is TRL 5, due to the development work on H4RG-10 detectors for the WFIRST mission. A read noise of 2.5 e- per pixel and dark current levels of ~0.002 e-/sec/pixel are assumed. Current state-of-the-art read noise values are closer to 5 e-, however it is believed a ~2× improvement can be achieved with optimal operating conditions currently be investigated by WFIRST. More discussion regarding the technology development of these detector systems can be found in **Chapter 11**.

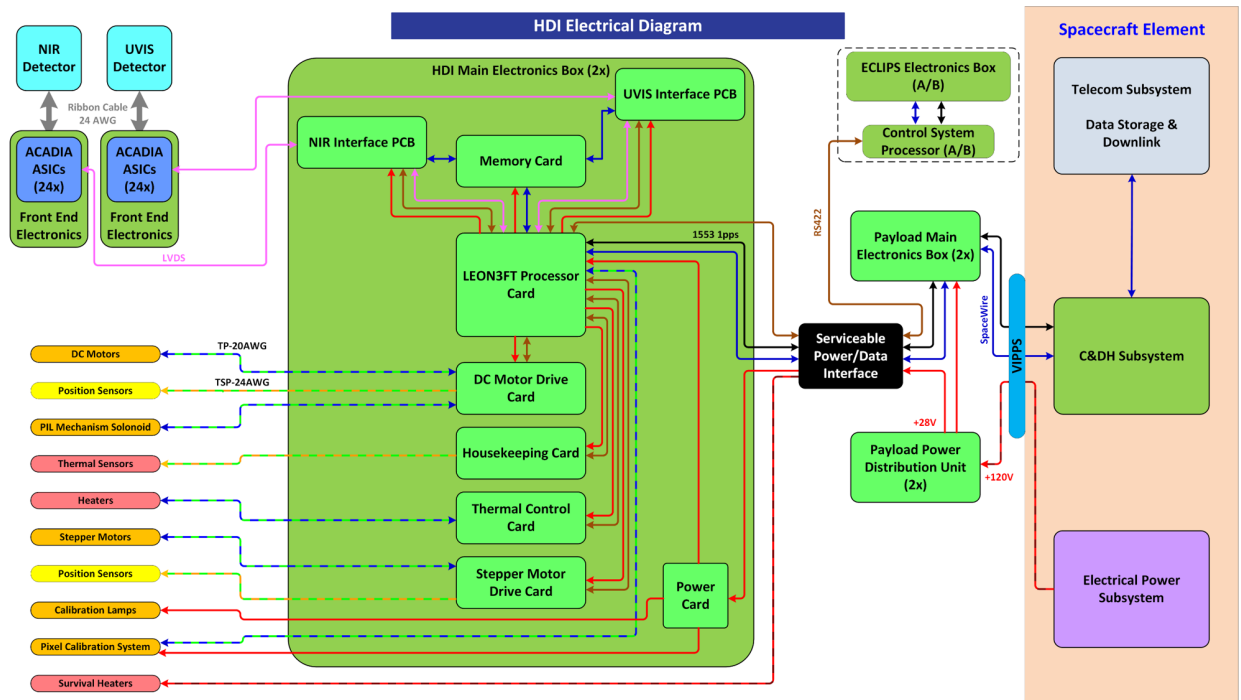

**Figure 8-25.** *HDI electrical system block diagram. The design is the same for both versions of the instrument except for the number of ACADIAs (24 in each channel of HDI-A, 6 in each channel of HDI-B).*

Both detector systems use the Application-specific integrated circuit for Control and Digitization of Images for Astronomy (ACADIA) front-end electronics to read-out and digitize the focal plane signals.

### 8.2.2.7   Special instrument modes

In addition to simultaneous UVIS and NIR imaging, HDI provides three special modes of operation. HDI's precision astrometry mode enables a range of exciting science not feasible on existing telescopes including astrometric detection of exoplanets and measuring proper motions of extragalactic sources in the Local Group and beyond. To achieve the astrometric accuracy required (better than ±1 micro-arcsecond), the position of every pixel in the UVIS detector array must be calibrated. Such a metrology calibration system is needed if one wishes to measure galaxy proper motions out at the 10–15 Mpc distance scale or to detect the stellar wobble induced by Earth-mass exoplanets orbiting their host main sequence stars. An all-fiber metrology system is included in HDI to allow for the calibration of the UVIS focal plane pixel geometry to a precision of $10^{-4}$ pixels (Crouzier et al 2016).

A second special mode of HDI is to function as LUVOIR's primary fine-guidance sensor. Both the UVIS and NIR focal planes have the capability of defining small regions-of-interest around bright foreground stars within each of the focal plane sensor chips. These regions-of-interest can be read-out at high speeds (up to 500 kHz) to provide a pointing signal to the fast-steering mirror and VIPPS. This can be done without interrupting regular science operations.

**Table 8-6.** *HDI data volume summary. Integration times and frame counts assume typical values, and are not strict requirements for every observation. The onboard instrument memory is sized to store 48 hours of nominal science data, including 20% overhead and 30% margin.*

| Parameter | Units | HDI-A | | HDI-B | |
|---|---|---|---|---|---|
| | | UVIS | NIR | UVIS | NIR |
| Detector Array Size | - | 6 x 4 | 6 x 4 | 3 x 2 | 3 x 2 |
| Number of Detectors | - | 24 | 24 | 6 | 6 |
| Detector Format | pixels | 8192 x 8192 | 4096 x 4096 | 8192 x 8192 | 4096 x 4096 |
| Pixels / Detector | Mpixels | 67.1 | 16.8 | 67.1 | 16.8 |
| Outputs / Detector | - | 32 | 32 | 32 | 32 |
| Bits / Pixel | bits | 16 | 16 | 16 | 16 |
| **Nominal Science Mode** | | | | | |
| Integration Time / Frame | s | 20 | 10 | 20 | 10 |
| Number of Co-added Frames | frames | 50 | 50 | 50 | 50 |
| Total Integration Time / Image | s | 1000 | 500 | 1000 | 500 |
| Average Data Rate | Mbps | 26 | 13 | 6 | 3 |
| Readout Rate / Detector | kHz | 105 | 52 | 105 | 52 |
| Total Uncompressed Data Rate | Mbps | 39 | | 10 | |
| Total Compressed Data Rate (1.6:1) | Mbps | 24 | | 6 | |
| 48-hour Storage Requirement (includes 20% overhead, 30% margin) | Tbits | 6.5 | | 1.6 | |
| **High-Speed Science Mode** | | | | | |
| Maximum Pixel Read Rate | kHz | 500 | 500 | 500 | 500 |
| Integration Time / Frame | s | 0.05 | 0.05 | 0.05 | 0.05 |
| Number of Co-added Frames | frames | 1 | 1 | 1 | 1 |
| Total Integration Time / Image | s | 0.05 | 0.05 | 0.05 | 0.05 |
| Maximum Pixels Read / Output | pixels | 25,000 | 25,000 | 25,000 | 25,000 |
| Maximum Pixels Read / Detector | pixels | 800,000 | 800,000 | 800,000 | 800,000 |
| High-Speed Window Size | pixels / side | 894 | 894 | 894 | 894 |
| Data Rate | Mbps | 256 | 256 | 256 | 256 |
| Total Uncompressed Data Rate | Mbps | 512 | | 512 | |
| Total Compressed Data Rate (1.6:1) | Mbps | 320 | | 320 | |
| Max. High-speed Observation Time | hours | 3.6 | | 0.9 | |

Finally, the HDI UVIS focal plane will serve a function similar to that of the Near Infrared Camera (NIRCam) on JWST, and provide defocused image data for phase retrieval and wavefront sensing. This dataset will be used during commissioning of the observatory to align and phase the primary mirror segments after deployment, as well as during routine maintenance of the wavefront as needed. Six elements in the UVIS filter wheel are dedicated to supporting this mode of operation: 4 weak lenses for generating defocused point-spread function data, and 2 dispersed Hartmann sensors for performing coarse phasing of the primary mirror segments.

### 8.2.2.8   Data volume

**Table 8-6** summarizes the HDI data volume for each version of the instrument. HDI's internal storage is sized for 48 hours of nominal science data and overhead, plus 30% margin (6.5 Tbits on HDI-A, 1.6 Tbits on HDI-B).

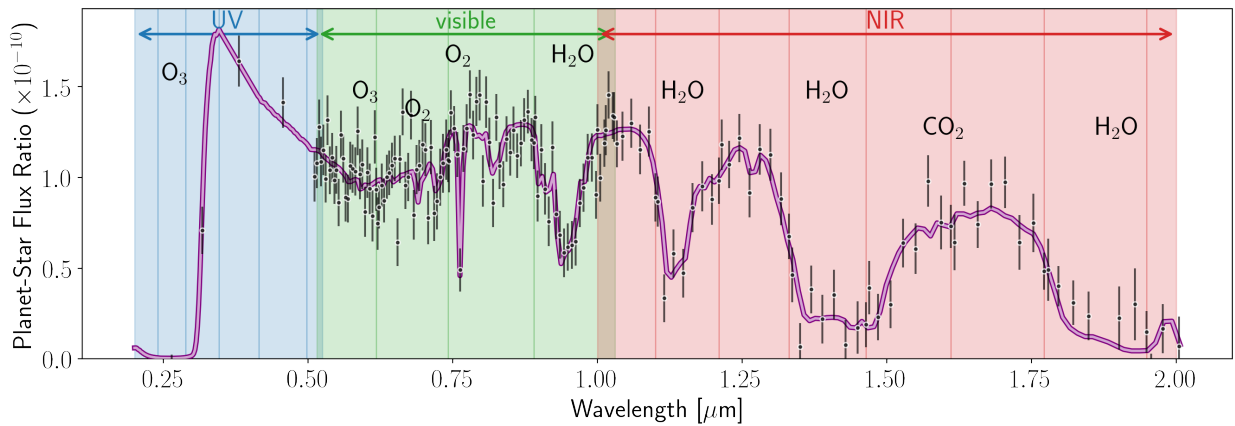

**Figure 8-26.** *Simulated LUVOIR ECLIPS spectrum of a modern Earth-twin orbiting a Sun-like star. These spectra have SNR=8.5 on the continuum across the whole bandpass, sufficient for measurement of key molecular abundances. The total time to acquire this spectrum varies dramatically depending the distance of the system from the Sun. The shaded background regions indicate 20% bandpasses in the NUV (blue) and visible (green) channels, and 10% bandpasses in the NIR (red) channel. Credit: J. Lustig-Yaeger (UW)*

### 8.2.3 Extreme Coronagraph for Living Planetary Systems (ECLIPS)

#### 8.2.3.1 Introduction

The scientific goals of ECLIPS are commensurate with two key science themes:

(1) measuring the occurrence rate of biomarkers in the atmospheres of rocky planets orbiting in the habitable zone of their host stars, and

(2) studying the diversity of exoplanet systems.

The former science theme is significantly more demanding on the instrument and drives the ECLIPS design. Any mission aimed at measuring the occurrence rate of biomarkers in the atmosphere of nearby habitable-zone rocky planets, ought to first be capable of detecting a statistically significant ensemble of exo-earth candidates. The detectability of exoplanets in long coronagraph exposures depends on both the level of contrast achieved in the high-contrast region or "dark hole" of the focal plane, and on the throughput of the coronagraph at the apparent separation of the planets, expressed using the Inner and Outer Working Angle (IWA and OWA, respectively) scalar metrics.

The characterization of identified exo-earth candidates is equally as important as their detection. **Figure 8-26** shows a simulated spectrum of a mature earth analog as detected by LUVOIR-A with the ECLIPS instrument. This example illustrates the most salient characteristics of the atmosphere of earth analogs which we seek to characterize with great precision using ECLIPS, and translates into the following three requirements on the instrument:

(1) Continuous spectral coverage from 200 nm to 2.0 µm in order to capture the spectral features associated with carbon and oxygen based molecules, which help discriminate between different atmospheric compositions,

(2) Spectral resolution of R = 140 in the visible, and

(3) Spectral resolutions of R =70 or 200 in the near-IR.

Spectroscopy of faint exo-earths beyond ~1.6 μm will be limited to the closest and brightest targets due to the thermal background from the 270 K telescope. Nevertheless, redder spectral coverage will be invaluable to studying the details of the atmospheres of our nearest neighbors, as well as characterizing larger planets.

There is currently no flight heritage for a coronagraph such as the one proposed for LUVOIR. There are several ground based systems that use coronagraphs, however their overall requirements are far less stringent than those imposed on ECLIPS, and their contrast ratios are several orders of magnitude less. The WFIRST Coronagraph Instrument (CGI) is the most similar coronagraph to ECLIPS, but it is still under development.

The CGI development has and will continue to help advance the state-of-the-art in terms of how to successfully implement a coronagraph instrument on a space telescope, despite WFIRST and LUVOIR being very different observatories. The experience gained from integrating, testing, and operating a high-contrast coronagraph on-orbit will be invaluable. Some of the most critical pieces of information that will be learned by the operation of WFIRST CGI will be how to post-process high-contrast data with positively verifiable targets. Performing transit spectroscopy with HST is an analogous example: this was a new field of data analysis, and the experience from working with Hubble data informed the Kepler mission and will feed directly into JWST. Similarly, working with real high-contrast data from WFIRST will inform the design of LUVOIR and ECLIPS. We note, however, that even though LUVOIR/ECLIPS will benefit a great deal from the successful operation of WFIRST/CGI, it should not be considered a pre-requisite for LUVOIR/ECLIPS.

Separate from the WFIRST CGI efforts, the development of segmented-aperture coronagraphs has rapidly progressed over the past half-decade. There are currently many promising candidate architectures, including apodized pupil Lyot coronagraphs (APLC), vortex coronagraphs (VC), phase-induced amplitude apodization (PIAA) coronagraphs, hybrid Lyot coronagraphs (HLC), and nulling coronagraphs (NC). The landscape of coronagraph design may look very different at the time that an observatory such as LUVOIR is actually built, and it is recommended that each of these candidates continue to be developed and evaluated to maximize the possible science yield (see **Chapter 11** for more detail on the coronagraph technology development plan).

Based on the two LUVOIR telescope designs (obscured vs. unobscured), and the best existing analysis of coronagraph performance to date, ECLIPS will use two coronagraph architectures: an APLC and a VC. The APLC is the primary coronagraph instrument for LUVOIR-A, as it is the most compatible with the obscured, on-axis aperture. However, a secondary VC coronagraph is included for observations where stellar diameters are not resolved. On LUVOIR-B, the reverse is true. The VC is the primary coronagraph as it is most compatible with the unobscured, off-axis aperture. An APLC coronagraph is included as a secondary coronagraph for a few optimal targets. The instrument can be switched between the two types of coronagraphs by selecting the appropriate set of masks to be inserted at the specified optical planes.

### 8.2.3.2  Optical design

**Figure 8-27** shows a block diagram for the ECLIPS instrument, which is identical for both versions of the instrument excepting the order of the pupil relay optics that pick-off the beam from the OTA. **Table 8-7** summarizes the optical design parameters of the ECLIPS instrument.

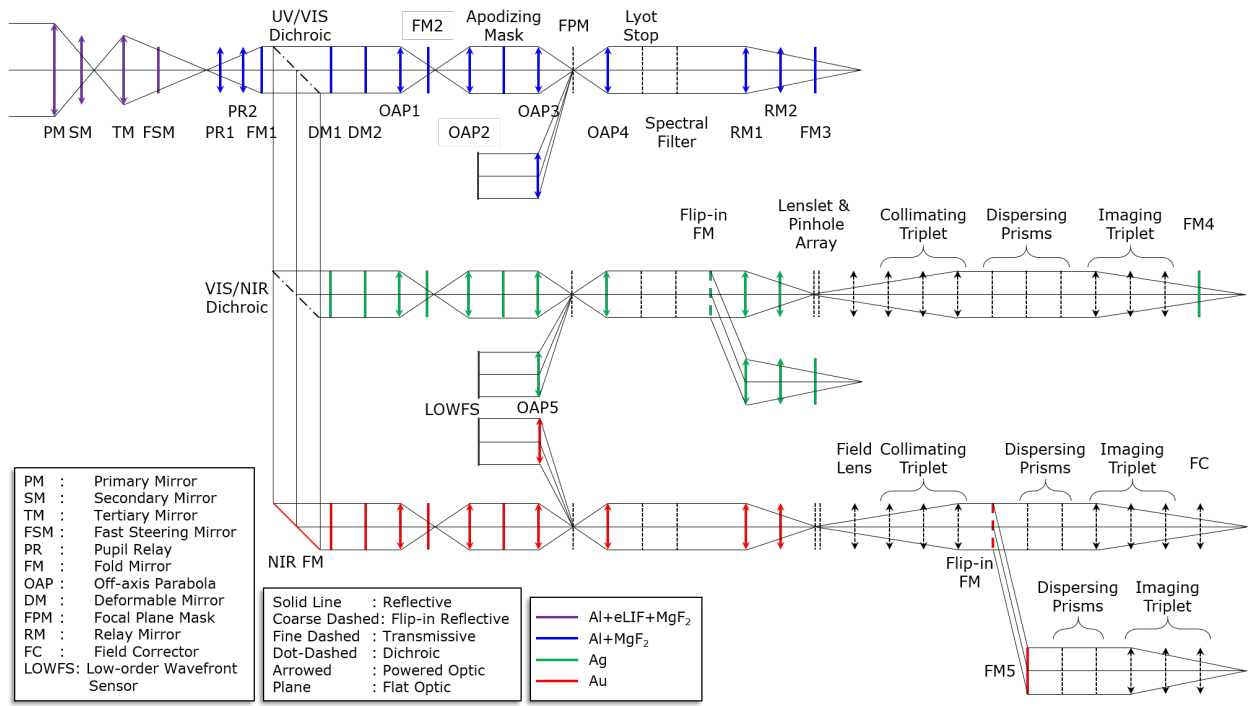

**Figure 8-27.** *Optical layout of the ECLIPS-A instrument. The optical train is divided into three spectral channels (UV, Visible, NIR) to make optimal use of coating and detector technologies. The UV channel (top) has only an imaging camera. The visible channel (middle) has both an imaging camera and an integral field spectrograph that can be selected via a flip-in mirror after the spectral filters. The NIR channel (bottom) has an integral field spectrograph and a single point source spectrograph that can be selected via a flip-in mirror. The inset at the upper right shows the optical throughput at the center of each filter bandpass for each channel. All three channels also have a low-order / out-of-band wavefront sensor that uses light reflected from the focal plane mask to sense wavefront drifts. ECLIPS-B has a similar layout with the exception that the dichroic channel splitting occurs before fold mirror 1, and there is one additional fold mirror in each channel prior to the deformable mirrors.*

Three mirrors are used to pick-off the light from the OTA focal plane, collimate it, and re-image the telescope pupil onto the first deformable mirror (DM) in each channel. Once the light is collimated by these pre-optics, it is separated into each channel via a series of dichroic beamsplitters. The mirrors in each channel are coated according to their bandpass to provide the best reflectivity: the pre-optics and UV channel mirrors are coated in protected aluminum, the visible channel mirrors in protected silver, and the NIR channel mirrors in protected gold. To minimize polarization aberration effects within the instrument, all 90° fold mirrors are used in crossed-pairs, such that one mirror compensates the polarization effects of the other. There is one exception in ECLIPS-B, where a single, uncompensated compound fold mirror was needed to orient the optical system in the available volume. This mirror resulted from a late design change with instrument volume; with additional design iterations this mirror can either be removed entirely, or compensated by an additional fold mirror.

After the first DM, the beam propagates to a second DM that is not in a pupil plane. A pair of relay mirrors then re-image the pupil from the first DM onto the apodizing mask. A second pair of relay mirrors re-images the pupil again onto the Lyot stop, with the co-ronagraph focal plane mask is located at the intermediate focus. After passing through the

**Table 8-7.** *Optical design parameters of the ECLIPS instrument. Unlike the HDI and LUMOS instruments, both versions of the ECLIPS instrument have the same specifications. This is due to the fact that parameters such as pupil size and imaging F/# are driven not by the telescope aperture, but by manufacturing tolerances on components such as the deformable mirrors and coronagraphic masks.*

| Parameter | Units | Value | | |
|---|---|---|---|---|
| | | **UV** | **VIS** | **NIR** |
| Raw Contrast | – | $1\times10^{-10}$ | | |
| Total Bandpass | nm | 200–525 | 515–1030 | 1000–2000 |
| Inner Working Angle | – | 4 λ/D | 3.5 λ/D | 3.5 λ/D |
| Outer Working Angle | – | 40 λ/D | 64 λ/D | 64 λ/D |
| RMS Wavefront Error | nm | 14 | 37 | 71 |
| | | **Spectrometer Lenslet Array Parameters** | | |
| Sampling at Lenslet Array | – | – | Nyquist at 515 nm | Nyquist at 1000 nm |
| Lenslet Diameter | μm | – | 124 | 120 |
| Pinhole Diameter | μm | – | 30 | 40 |
| Lenslet Packing | – | – | Square | Hexagonal |
| Total Lenslet Count | – | – | 72,300 | 128,000 |
| F/# at Lenslet Array | – | – | 481.5 | 248.0 |
| | | **Integral Field Spectrometer Detector Parameters** | | |
| Sampling at Detector | – | – | Nyquist at 515 nm | Nyquist at 1000 nm |
| Average Optical Throughput | % | | 23.8 | 32.5 |
| Pixel Size | μm | – | 12 | 10 |
| Magnification | – | – | 1 : 1.47 | 1 : 1 |
| F/# at Detector | – | – | 11.8 | 8.0 |
| Total Spectral Length* | pixels | – | 34 | 20 |
| Resolving Power | | – | 140 | 70 |
| | | **Single Point Source Spectrometer** | | |
| Pixel Size | μm | – | – | 10 |
| F/# at Detector | – | – | – | 8.0 |
| Average Optical Throughput | % | | – | 32.6 |
| Resolving Power | – | – | – | 200.0 |
| | | **Imaging Camera Parameters** | | |
| Sampling at Detector | – | Nyquist at 200 nm | Nyquist at 515 nm | - |
| F/# at Detector | – | 130.0 | 50.5 | - |
| Average Optical Throughput | % | 9.9 | 27.7 | – |
| Effective Focal Length | mm | 2,628 | 1,020 | - |
| Pixel Size | μm | 13 | 13 | - |

*Includes isolation pixels.

Lyot stop, the optical bandpass is further reduced to 10–20% by spectral filters for science observations.

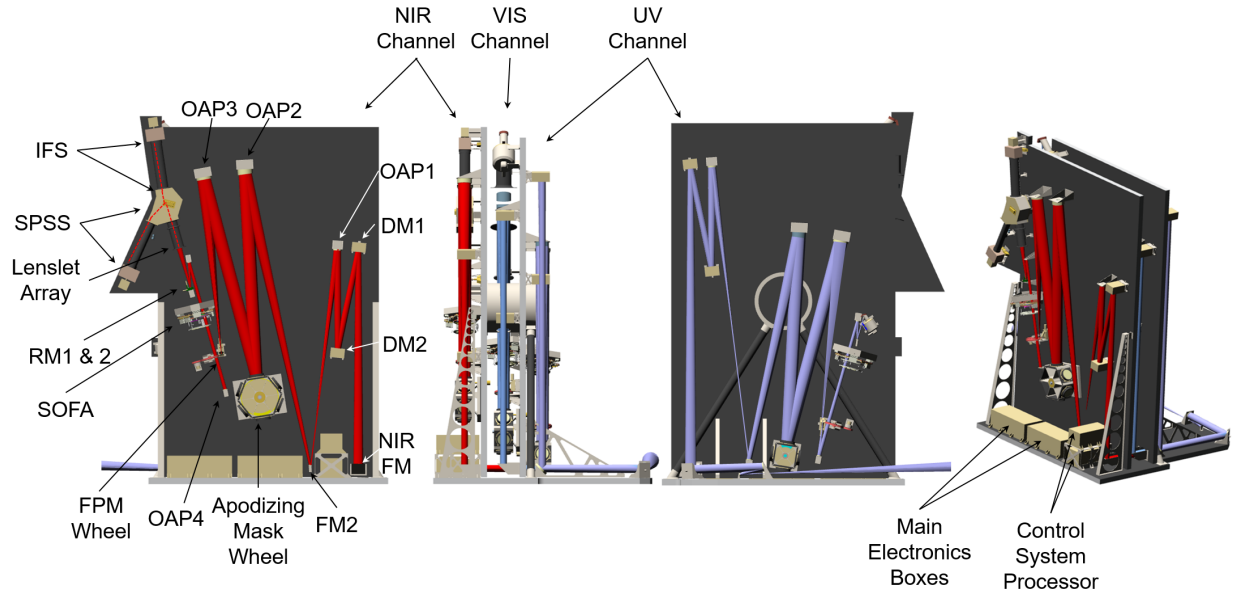

**Figure 8-28.** *ECLIPS-A opto-mechanical design. The three channels are separated by two optical benches that are mounted to a third interface bench. Elements in the NIR channel (left) are identified using the same nomenclature as* **Figure 8-27**. *The UV channel layout is similar, except after the selectable optical filter assembly (SOFA), the beam is relayed directly to an imaging camera. The Visible channel has a similar layout to the NIR channel, except there is no SPSS, and the beam can instead be relayed to an imaging camera after the SOFA.*

Depending on the channel, after passing through the spectral filters, the light can be sent to one of several back-end instruments. In the UV channel, the only back-end instrument is an imaging camera. In the visible channel, a flip-in mirror can be used to direct the light to either an imaging camera or an R=140 integral field spectrograph (IFS). In the NIR channel, a flip-in mirror can be used to direct the light to either an R=70 IFS, or an R=200 single point source spectrograph (SPSS).

In addition to the primary science beam path, each channel also supports a low-order / out-of-band wavefront sensor (LOWFS / OBWFS). The focal plane mask doubles as a Zernike wavefront sensing mask. The core of the focal plane mask is reflected into the LOWFS/OBWFS camera. A divot in the center of the focal plane mask provides the phase reference for the Zernike wavefront sensor (Shi et al. 2015).

Although not explicitly included in the opto-mechanical design at this stage, internal photometric and spectral calibration sources have been accounted for in the total estimated mass and power for the instrument. The calibration sources would be unique to each channel.

### 8.2.3.3  Mechanical design

**Figure 8-28** shows an opto-mechanical model of ECLIPS-A. Two optical benches hold the three channels of the instrument. An interface bench at the base of the instrument holds the two optical benches and interfaces to the BSF bulkhead. The benches are aluminum honeycomb panels with composite face sheets. Additional structural support is provided by composite gussets and a central, cylindrical hub that braces the two benches near the

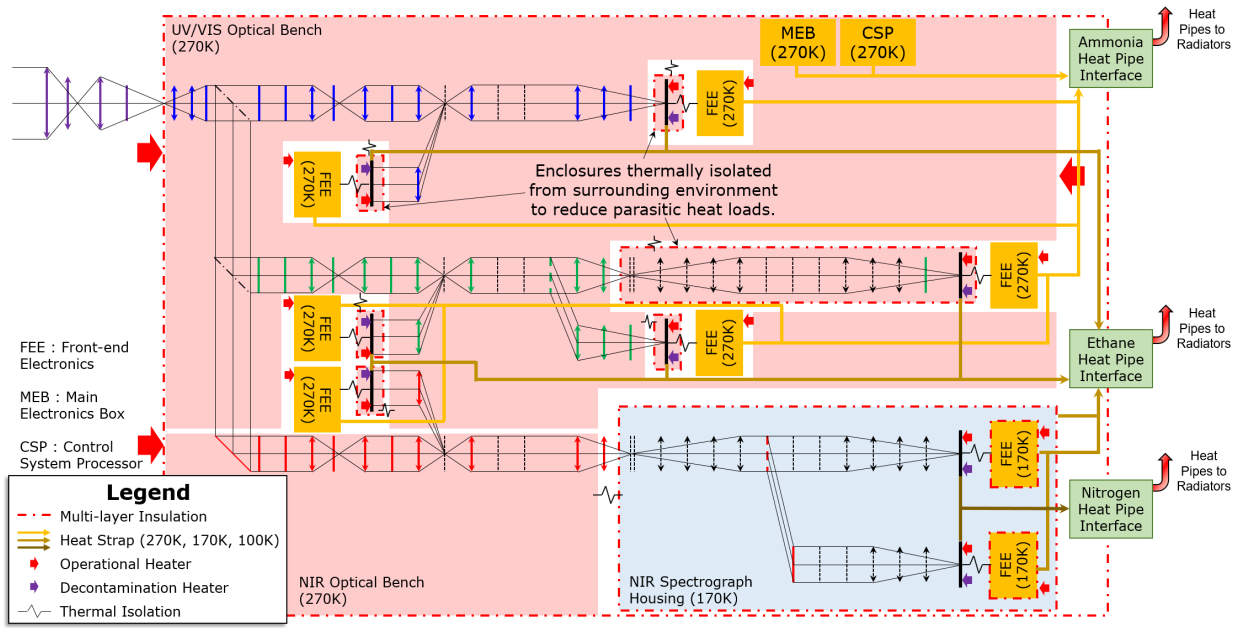

**Figure 8-29.** *ECLIPS thermal block diagram. The thermal architecture is identical for both instruments. Passively cooled components are heat-strapped to serviceable interfaces that connect to heat pipes within the BSF, to transfer heat to the external radiators. Each UV- and VIS-channel detector housing is thermally isolated from the surrounding environment with MLI and flexures to minimize the parasitic heat-load from the actively heated optical benches to the passively cooled detectors.*

instrument center-of-mass. The ECLIPS-B design is nearly identical, save small modifications to accommodate packaging within the LUVOIR-B instrument volume.

#### 8.2.3.4 Thermal design
**Figure 8-29** shows a block diagram of the ECLIPS thermal architecture, which is the same for both versions of the instrument. The instrument optical benches are held at 270 K. All components that require colder operating temperatures are heat strapped to a serviceable interface between the instrument and the OTA. On the OTA side of the interface, heat pipes (either ammonia, ethane, or nitrogen, depending on the temperature region) transfer the dissipated heat to the appropriate radiator on the external faces of the BSF.

An enclosure thermally isolates the NIR IFS and SPSS optical elements, which are passively cooled to 170 K. All detectors in the UV and VIS channels are also passively cooled to 170K, as well as both of the NIR detector front-end electronics. Both NIR focal planes are passively cooled to 100 K.

#### 8.2.3.5 Electrical design
**Figure 8-30** shows a block diagram of the ECLIPS electrical architecture, which is identical for both versions of the instrument. A serviceable interface provides power and data connections from the OTA. The ECLIPS main electronics box controls all instrument functions, and has internal solid state memory storage for up to 48 hours data collection.

In addition to the main electronics box, ECLIPS also houses the control system processor. This is the primary processor that operates all of the control systems on LUVOIR. It processes focal plane data and LOWFS/OBWFS data from ECLIPS to control the deformable mirrors.

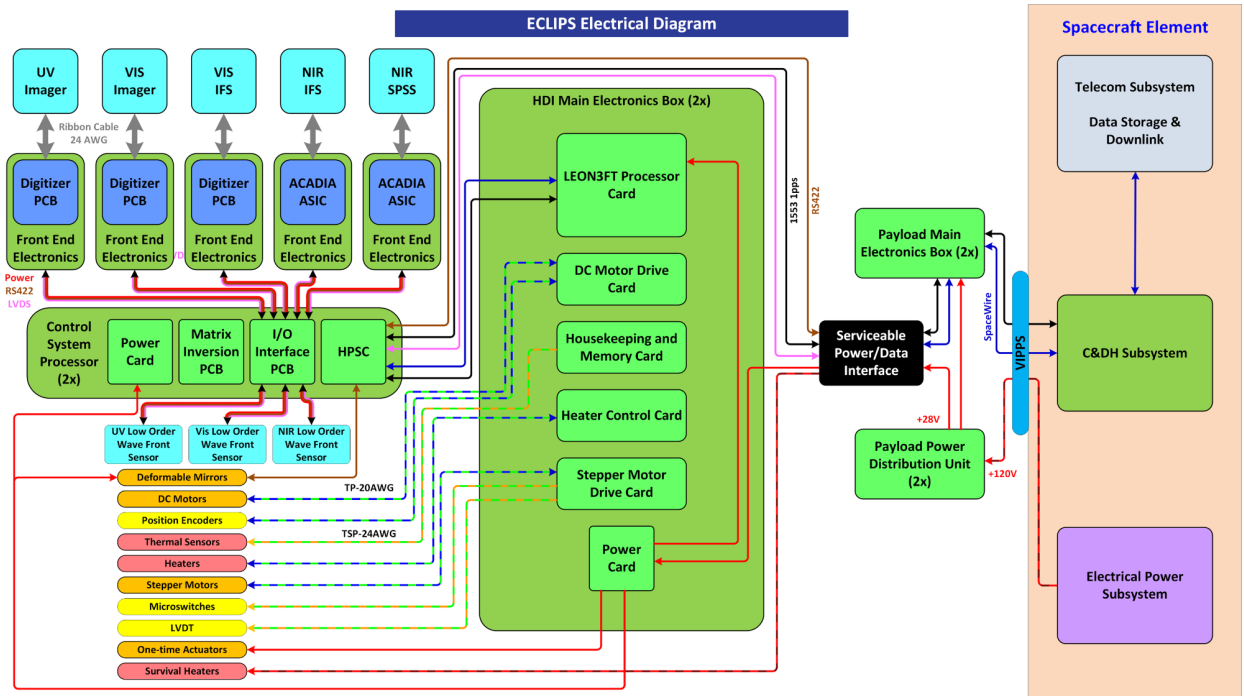

**Figure 8-30.** *ECLIPS Electrical Block Diagram. The electrical architecture is the same for both versions of the instrument.*

It processes data from the primary mirror edges sensors and the laser metrology system to control primary mirror segment and secondary mirror rigid body positions. And it processes fine-guidance data from HDI to control the fast steering mirror and Vibration Isolation and Precision Pointing System (VIPPS) for precision line-of-sight pointing.

The control system processor uses a hybrid architecture consisting of a central processing unit to handle overall algorithm execution and data I/O, and field programmable gate arrays to perform dedicated, computationally intensive tasks such as matrix inversions or fast Fourier transforms. The control system processor performance requirements are largely driven by the need to do focal-plane high-contrast imaging and low-order/out-of-band wavefront sensing and control within the coronagraph. Therefore, the control system processor was located within the ECLIPS instrument volume, such that should ECLIPS ever be replaced with a higher performance coronagraph, a new processor capable of supporting the instrument could be included within the same module.

### 8.2.3.6  Detectors

As with the coronagraph architectures, the technology development of detectors for exoplanet science is proceeding rapidly. By the time LUVOIR is to be developed, it is likely that detectors with better sensitivity, less noise, and higher radiation tolerance will exist. However, to benchmark system performance and science yields now, the Study Team has selected a suite of detectors that exist today and only require a moderate amount of engineering development to be capable of achieving the science objectives described in earlier chapters. **Chapter 11** provides more detail on other detector technologies that may prove superior with sufficient technology development. The detector technologies are specific to

**Table 8-8.** *ECLIPS data volume summary. Integration times and frame counts assume typical values, and are not strict requirements for every observation. The onboard instrument memory is sized to store 48 hours of nominal science data, including 20% overhead and 30% margin.*

| ECLIPS | | | | | | |
|---|---|---|---|---|---|---|
| **Parameter** | **Units** | **UV** | **VIS (Imager)** | **VIS (IFS)** | **NIR (IFS)** | **NIR (SPSS)** |
| Array Format | – | 1 x 1 | 1 x 1 | 1 x 1 | 1 x 1 | 1 x 1 |
| Number of Detectors | – | 1 | 1 | 1 | 1 | 1 |
| Detector Format | pixels | 1024 x 1024 | 1024 x 1024 | 4096 x 4096 | 4096 x 4096 | 4096 x 4096 |
| Pixels / Detector | Mpixels | 1.0 | 1.0 | 17.8 | 17.8 | 17.8 |
| Number of Outputs / Detector | – | 1 | 1 | 8 | 1 | 1 |
| Bits / pixel | bits | 8 | 8 | 8 | 16 | 16 |
| **Nominal Science Mode** | | | | | | |
| Integration Time / Frame | s | 1 | 1 | 1 | 100 | 100 |
| Number of Co-added Frames | – | 100 | 100 | 100 | 1 | 1 |
| Total Integration Time / Image | s | 100 | 100 | 100 | 100 | 100 |
| Average Data Rate | Mbps | 0.084 | 0.084 | 1.342 | 2.684 | 2.684 |
| Readout Rate / Detector | kHz | 1049 | 1049 | 2097 | 168 | 168 |
| Total Uncompressed Data Rate** | Mpbs | 4.110 | | | | |
| 48-hour Storage Requirement (includes 20% overhead, 30% margin) | Gbits | 1,108 | | | | |

**Note that the VIS Imager and VIS IFS cannot be used in parallel, and the NIR IFS and NIR SPSS cannot be used in parallel. Thus only one of each detector can contribute to the data volume at a time. We choose one of the NIR detectors and the larger of the two VIS detectors to determine the maximum 48-hour data volume.

each channel and our baseline choices rely heavily on technologies that have been used on previous missions and/or are planned in the WFIRST baseline instruments.

In the UV and visible channels, electron multiplying CCDs (EMCCDs) will be used, with the UV channel devices being δ-doped for better shortwave performance. Current devices are being developed and tested for the WFIRST coronagraph mission with 1k × 1k, 13 μm pixels. These 1k × 1k devices will be sufficient for the UV and visible channel imagers, however to cover the field-of-view and spectral resolution of the visible channel IFS, a 4k × 4k device will be required. In photon-counting mode, these devices produce <1 e- read noise, and $1 \times 10^{-4}$ e-/pixel/second dark current (Harding et al. 2016).

The baseline technology for the NIR channel is the Teledyne HAWAII 4RG (H4RG) HgCdTe detector. Current devices have a format of 4k × 4k with 10 μm pixels, and are three-side buttable for building larger arrays. H4RGs are the baseline detector for the WFIRST wide-field instrument and are an evolution of H2RG detectors used on the JWST NIRCam instrument. Current devices have a median read noise of ~5 e-/pix, and a median dark current: $2 \times 10^{-3}$ e-/pix/sec. It is believed that the read noise can be further reduced to 1–2 e-/pix.

### 8.2.3.7  Data volume

**Table 8-8** summarizes the ECLIPS data volume, which is the same for both versions of the instrument. The non-photon-counting NIR detectors require a higher bit-depth (16-bits) than the photon-counting UV and VIS-channel EMCCDs (8-bits each).

As with HDI, ECLIPS has enough internal storage for 48 hours of continuous science data collection plus 30% margin, or ~923 Gbits (same for both instrument versions).

#### 8.2.3.8  Coronagraph masks

Each of the three channels is equipped with a series of element select mechanisms that accommodate the various apodizing, focal plane, and Lyot masks, as well as spectral filters. The number of elements in these mechanisms is driven by two considerations. First, each channel can only operate over an instantaneous 10–20% spectral bandpass at a time, due to limitations on wavefront control techniques and mask designs, requiring between six and eight spectral filters per channel. The second consideration is that each combination of spectral bandpass, IWA, and OWA, requires a specific set of apodizing, focal plane, and Lyot stop masks. Each channel has between 8 and 11 mask combinations to enable different optimized high-contrast regions on the focal plane.

#### 8.2.3.9  Deformable mirrors

Our choice of deformable mirror (DM) technology is based on the need for high-density devices to achieve large outer-working angles. For instance achieving high contrast for the outer region of the habitable zone around the most nearby stars requires an outer-working angle of ~48 $\lambda$/D. Providing wavefront control over a dark hole of that size requires ~96 actuators across the pupil. Imaging of outer giant planets drives the actuator count higher, to 128 actuators across the pupil to achieve a ~64 $\lambda$/D outer working angle. In order to keep the optical design compact a 128 ×128 actuator micro-electro-mechanical systems (MEMS) DM device is baselined. A secondary technical driver is the fact that such compact DMs can achieve small Fresnel numbers, and thus are more amenable to DM-based correction of amplitude errors. While this technology will not be matured by WFIRST at the component level, there are avenues to do so using smaller satellites in the near future. More details on DM technology development are provided in **Chapter 11**.

### 8.2.4  LUVOIR Ultraviolet Multi-object Spectrograph (LUMOS)

#### 8.2.4.1  Introduction

A fundamentally important science driver for LUVOIR is its ability to study gas in the cosmos, its relationship to, and evolution with, star and galaxy formation, and how this gas is transferred from one site to another. Understanding the flow of matter and energy from the intergalactic medium  to the circumgalactic media, and ultimately into galaxies where it can serve as a reservoir for future generations of star and planet formation, is essentially a challenge in characterizing the ionic, atomic, and molecular gas at each phase in this cycle. LUVOIR will be capable of characterizing the composition and temperature of this material in unprecedented scope and detail; on scales as large as the cosmic web and as small as the atmospheres of planets around other stars. A key characteristic of this science case is that the strongest emission and absorption lines in the gasses to be studied—therefore the highest information content for understanding the physical conditions in these objects—reside at ultraviolet wavelengths, roughly 100–400 nm. The LUVOIR Ultra-violet Multi-object Spectrograph (LUMOS) instrument is designed to make revolutionary observational contributions to all of the disciplines that call for high-resolution spectroscopy, multi-object spectroscopy, and imaging in the ultraviolet bandpass.

### 8.2.4.2  Optical design

LUMOS is the only instrument on LUVOIR where the two versions of the instrument for LUVOIR-A and LUVOIR-B significantly differ in their opto-mechanical design. Tighter packaging constraints on LUVOIR-B drove LUMOS-B to have a more compact design. Each version of the instrument is therefore discussed separately here. **Table 8-9** summarizes the optical performance and first order specifications for both instruments.

#### 8.2.4.2.1  LUMOS-A

LUMOS-A contains three channels: a Far-UV (FUV) multi-object spectrograph (MOS), a near-UV/visible (NUV/VIS) MOS, and an FUV imager. **Figure 8-31** shows a block diagram of the LUMOS-A layout, while **Figure 8-32** shows the corresponding ray-traces of each channel separately. After light reflects off of the fast-steering mirror in the OTA into LUMOS, it passes through a 2 x 2 grid of microshutter arrays (MSA) which defines the field of view for the MOS. Light passing through the MSA is folded into the spectrograph by a convex biconic optic. A second fixed aberration-correcting toroidal mirror directs the beam to the FUV or NUV grating wheel. The NUV grating wheel picks off the beam enroute to the FUV grating wheel if those modes are selected.

**Table 8-9.** *LUMOS optical performance specifications*

| Instrument Parameter | Units | G120M | G150M | G180M | G155L | G145LL | G165LL | G300M | G700L | FUV Imager |
|---|---|---|---|---|---|---|---|---|---|---|
| Optimized Spectral Bandpass | nm | 100–140 | 130–170 | 160–200 | 100–200 | 100–200 | 110–200 | 200–400 | 400–1000 | 100–200 |
| Actual Spectral Bandpass | nm | 93–159 | 111–189 | 141–219 | 93–267 | 93–210 | 110–270 | 193–460 | 340–1000 | 100–200 |
| Field of View | arcmin | 2 x 2 | 2 x 2 | 2 x 2 | 2 x 2 | 2 x 2 | 2 x 2 | 1.5 x 2 (A) 2 x 2 (B) | 1.5 x 2 (A) 2 x 2 (B) | 1.2 x 2 (A) 2 x 2 (B) |
| Spectral Resolving Power ($\lambda/\Delta\lambda$) [Required] | – | $3.0 \times 10^4$ | $3.0 \times 10^4$ | $3.0 \times 10^4$ | 8,000 | 500 | 500 | $2.0 \times 10^4$ | $1.5 \times 10^4$ | – |
| Imaging Resolution [Required] | mas | 50 | 50 | 50 | 50 | 50 | 50 | 50 | 50 | 50 |
| **LUMOS-A** | | | | | | | | | | |
| Effective Area, Peak | cm² | $3.1 \times 10^5$ | $1.4 \times 10^5$ | $1.6 \times 10^5$ | $3.0 \times 10^5$ | $1.8 \times 10^5$ | $3.1 \times 10^5$ | $3.2 \times 10^5$ | $3.0 \times 10^5$ | $\sim 1.0 \times 10^5$ |
| Effective Area, Central Wavelength | cm² | $2.9 \times 10^5$ | $1.0 \times 10^5$ | $1.6 \times 10^5$ | $9.1 \times 10^4$ | $3.1 \times 10^4$ | $9.1 \times 10^4$ | $2.5 \times 10^5$ | $2.7 \times 10^5$ | $\sim 1.0 \times 10^5$ |
| Grating Ruling Density | lines/mm | 1,950 | 2,050 | 2,020 | 815 | 21 | 43 | 450 | 150 | – |
| Average Resolving Power (Full Field-of-View) | – | $3.0 \times 10^4$ | $3.6 \times 10^4$ | $4.1 \times 10^4$ | $1.4 \times 10^4$ | 408 | 820 | $2.1 \times 10^4$ | $1.4 \times 10^4$ | – |
| Average Resolving Power (Best 1' x 1') | – | $3.9 \times 10^4$ | $4.7 \times 10^4$ | $5.4 \times 10^4$ | $1.8 \times 10^4$ | 583 | 1,170 | $2.8 \times 10^4$ | $2.0 \times 10^4$ | – |
| Average Angular Resolution (Full Field-of-View) | mas | 37 | 43 | 39 | 42 | 38 | 38 | 31 | 33 | 50 |
| Average Angular Resolution (Best 1' x 1') | mas | 28 | 32 | 30 | 34 | 26 | 26 | 19 | 21 | 42 |
| **LUMOS-B** | | | | | | | | | | |
| Effective Area, Peak | cm² | $8.6 \times 10^4$ | $3.8 \times 10^4$ | $4.4 \times 10^4$ | $8.4 \times 10^4$ | $5.1 \times 10^4$ | – | $9.0 \times 10^4$ | $8.4 \times 10^4$ | $\sim 2.8 \times 10^4$ |
| Effective Area, Central Wavelength | cm² | $8.1 \times 10^4$ | $2.8 \times 10^4$ | $4.4 \times 10^4$ | $2.5 \times 104$ | 8,700 | – | $6.8 \times 10^4$ | $7.4 \times 10^4$ | $\sim 2.8 \times 10^4$ |
| Grating Ruling Density | lines/mm | 2,555 | 2,630 | 2,610 | 1,060 | 50 | – | 865 | 349 | – |
| Average Resolving Power (Full Field-of-View) | – | $2.9 \times 10^4$ | $3.7 \times 10^4$ | $5.0 \times 10^4$ | $1.4 \times 10^4$ | 455 | – | $2.0 \times 10^4$ | $1.8 \times 10^4$ | – |
| Average Resolving Power (Best 1' x 1') | – | $4.0 \times 10^4$ | $5.2 \times 10^4$ | $5.9 \times 10^4$ | $1.7 \times 10^4$ | 537 | – | $3.3 \times 10^4$ | $2.8 \times 10^4$ | – |
| Average Angular Resolution (Full Field-of-View) | mas | 41 | 42 | 43 | 48 | 28 | – | 25 | 48 | 48 |
| Average Angular Resolution (Best 1' x 1') | mas | 31 | 32 | 33 | 39 | 23 | – | 23 | 41 | 40 |

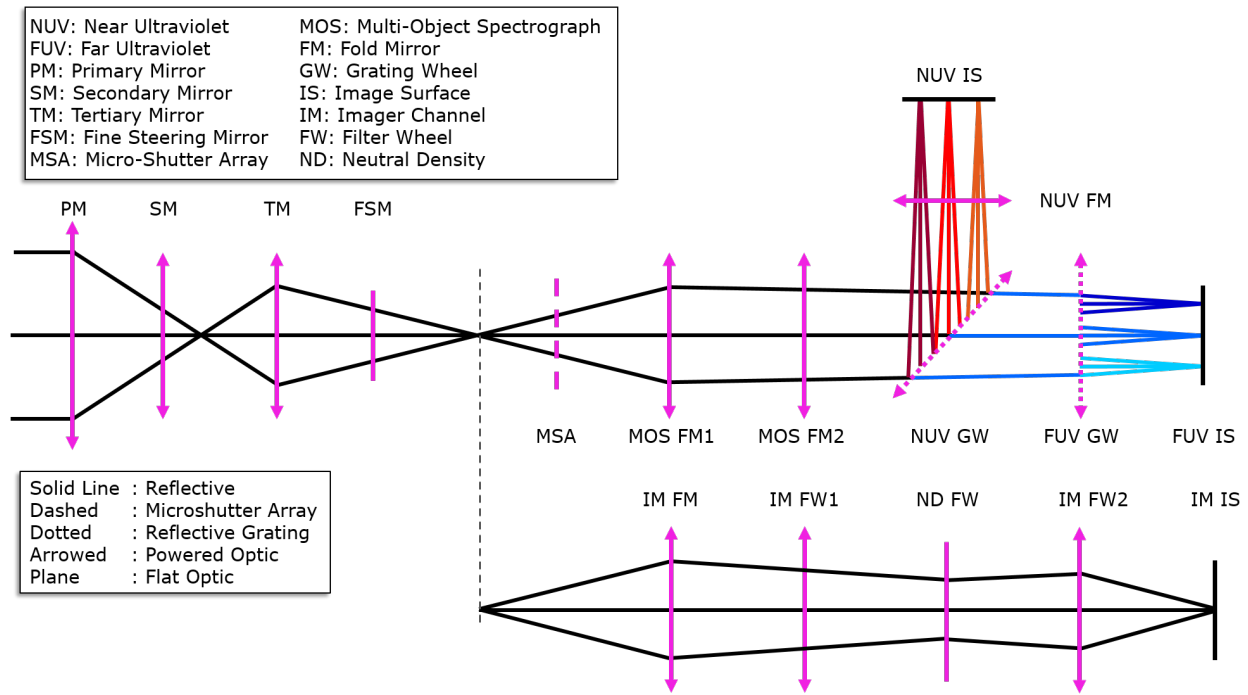

**Figure 8-31.** *Optical layout of LUMOS-A instrument. There are two independent channels: a FUV – NUV/VIS multi-object spectrograph (top), and a FUV imager (bottom).*

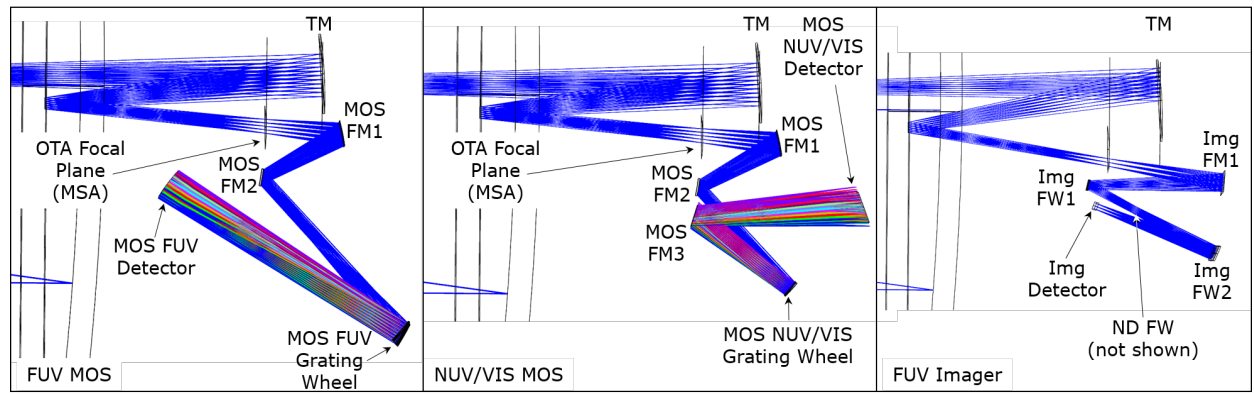

**Figure 8-32.** *Ray-trace of each channel in the LUMOS-A instrument. The FUV multi-object spectrograph channel (left) and NUV/VIS multi-object spectrograph channel (center) share the same field-of-view. The appropriate detector is selected via a flip-in grating mechanism (NUV GW). The FUV imaging channel (right) has an independent field-of-view.*

The FUV MOS modes include a set of medium, low, and very low spectral resolution settings contained within the FUV grating wheel that all maintain imaging performance that is less than a shutter in the MSA, essentially creating an array of long-slits that can be used for point-source spectroscopy or extended source imaging spectroscopy. The gratings are set within a wheel mechanism and selected by wheel position. Grating spectral resolutions are summarized in **Table 8-9**.

The ~250 mm diameter medium resolution gratings are toroidal figured and holographically ruled, building on the heritage of HST Cosmic Origins Spectrograph for aberration-controlling, focusing diffraction gratings. The observing modes are designed to have roughly 10 nm of optimized overlap with the neighboring setting to allow for robust spectral stitching and allow the user to select the most sensitive or highest angular resolution mode, depending on their science requirements.

The G155L mode serves as a good comparison for the only previous medium resolution FUV imaging spectrograph, the HST Space Telescope Imaging Spectrometer (STIS) G140M mode. LUMOS G155L delivers 20 times the simultaneous bandpass (100nm vs 5nm) and four times the one-dimensional field-of-view (120" vs 28"). The G145LL and G165LL modes are designed to minimize detector background and detector footprint, enabling the maximum number of faint objects to be observed simultaneously for FUV spectroscopic deep fields.

In order to access the NUV/VIS MOS modes, the NUV grating mechanism rotates the desired grating into place, preventing the light from reaching the FUV grating mechanism. After reflecting off of the NUV grating, the light is directed towards the NUV fold mirror which in turn sends the light towards the channel's focal plane array. The NUV/VIS grating modes are G300M and G700L, the details of which are included in **Table 8-9**.

The majority of the LUVOIR imaging science is addressed through the HDI instrument (200nm–2.5μm), and LUMOS provides a complimentary FUV imaging capability from 100–200nm. The LUMOS-A FUV imaging aperture subtends a 1.2′ x 2′ FOV, and is physically offset from the MOS MSA array. Light from the OTA enters this channel through an unobstructed open aperture with separate pickoff mirrors and subsequent beam paths from the FUV MOS. This is not the case for LUMOS-B which is discussed below. The incident light is folded into the imaging channel off of a biconic convex fold optic and then through two identical reflective filter wheel assemblies that serve to define the imaging bandpass in this mode. A neutral-density filter wheel is inserted between the two filter wheels to accommodate FUV bright object protection. The images are then recorded on a single 170mm × 110mm CsI photocathode MCP detector.

### 8.2.4.2.2  LUMOS-B

**Figure 8-33** shows a block diagram of the LUMOS-B layout, and **Figure 8-34** shows the corresponding ray-trace of each mode separately. As mentioned before, there are not separate pickoffs for the MOS and imaging channels in LUMOS-B. All the light entering LUMOS from the OTA does so by first passing through the MSA before being reflected by a fold mirror. While the second optic after the MSA in LUMOS-A was a single mirror, **Figure 8-33** shows that in LUMOS-B, it is a mechanism that switches between several elements depending on the mode of operation. For MOS modes, a simple broadband reflective mirror rotates into place, while a narrowband filter is used for imaging. After reflecting off of an element in

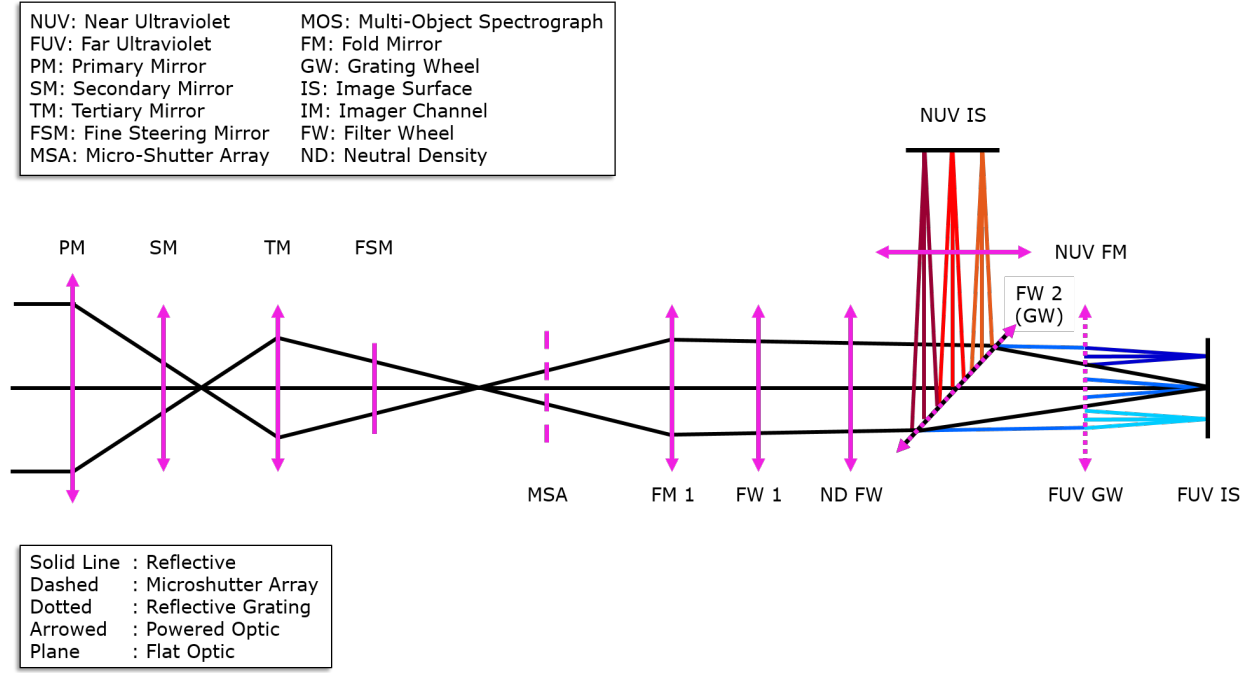

**Figure 8-33.** *Optical layout of LUMOS-B instrument. In this version, there is a single channel that can be switched between FUV MOS, NUV/VIS MOS, and FUV imaging modes, depending on which element is selected in filterwheel (FW) 2. The imaging is done through the microshutter array aperture of the MOS.*

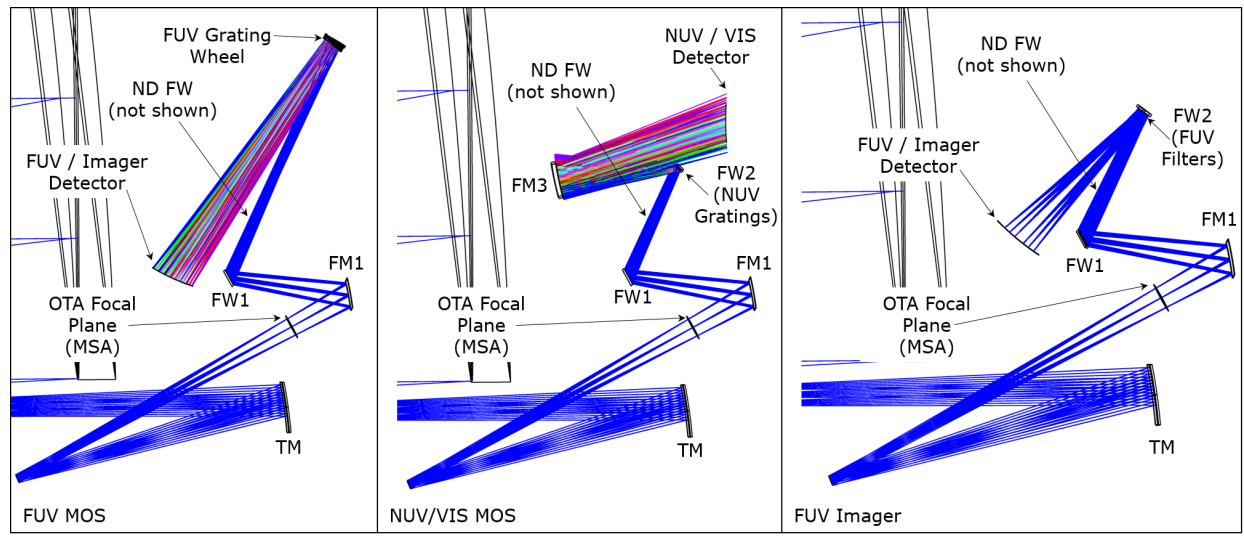

**Figure 8-34.** *Ray trace of each mode of the LUMOS-B instrument. All three modes share the same field-of-view defined by the microshutter array, and can be selected by inserting the appropriate combination of filters, mirror, and gratings at the FW1, NUV GW/FW2, and FUV GW planes.*

FW1 (and passing through the neutral density filter wheel), the light follows one of the following paths depending on the mode of operation:

1. **FUV MOS:** the light reflects off of the FUV grating mechanism before reaching the FUV focal plane array.

2. **NUV/VIS MOS:** the light reflects off of an element within a mechanism that holds elements for both the NUV/VIS MOS and imaging modes. For NUV/VIS MOS modes, the light then reflects off of the NUV fold mirror before reaching the NUV FPA.

3. **Imaging:** the light reflects off an element within the NUV/VIS grating mechanism. For imaging, the light is directed towards the FUV FPA which tilts into one of two positions depending on if the imaging or FUV MOS mode is being used.

As with ECLIPS, although not explicitly included in the opto-mechanical design for either LUMOS-A or LUMOS-B, a photometric and spectral calibration source has been accounted for in each instrument's estimated mass and power. A mechanism allows the calibration source beam to be scanned across the micro-shutter array so that the response at each shutter location can be calibrated.

### 8.2.4.3  Mechanical design

#### 8.2.4.3.1  LUMOS-A
The left side of **Figure 8-35** shows an opto-mechanical model of LUMOS-A. The structure is designed with aluminum honeycomb panels with composite face sheets. Additional structural support is provided by composite trusses. The large mirrors and gratings are lightweighted ULE substrates with an integral mushroom mount interface on the back. The fixed mirrors are mounted on tripods with composite legs and titanium fittings. The tripod interfaces to the mushroom mount with an adjustable 3-point interface ring. The tripod design combined with the three point mount ring provides a wide range of motion options for optical alignment.

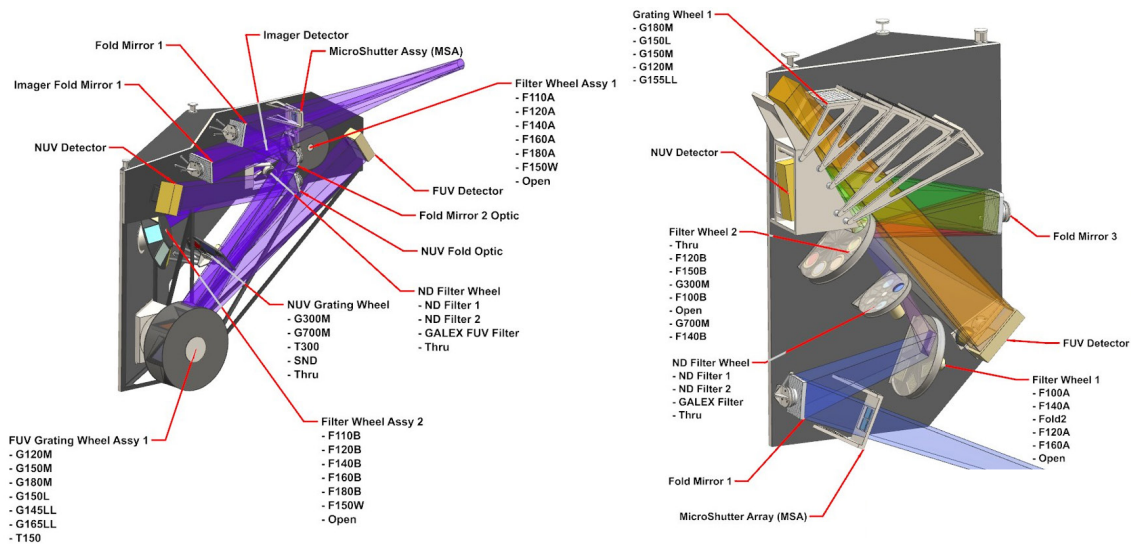

**Figure 8-35.** *LUMOS-A (left) and LUMOS-B (right) opto-mechanical designs. For LUMOS-A, light enters from the OTA at the top of the instrument. On LUMOS-B, light enters from the OTA at the bottom of the instrument. The images are not to scale.*

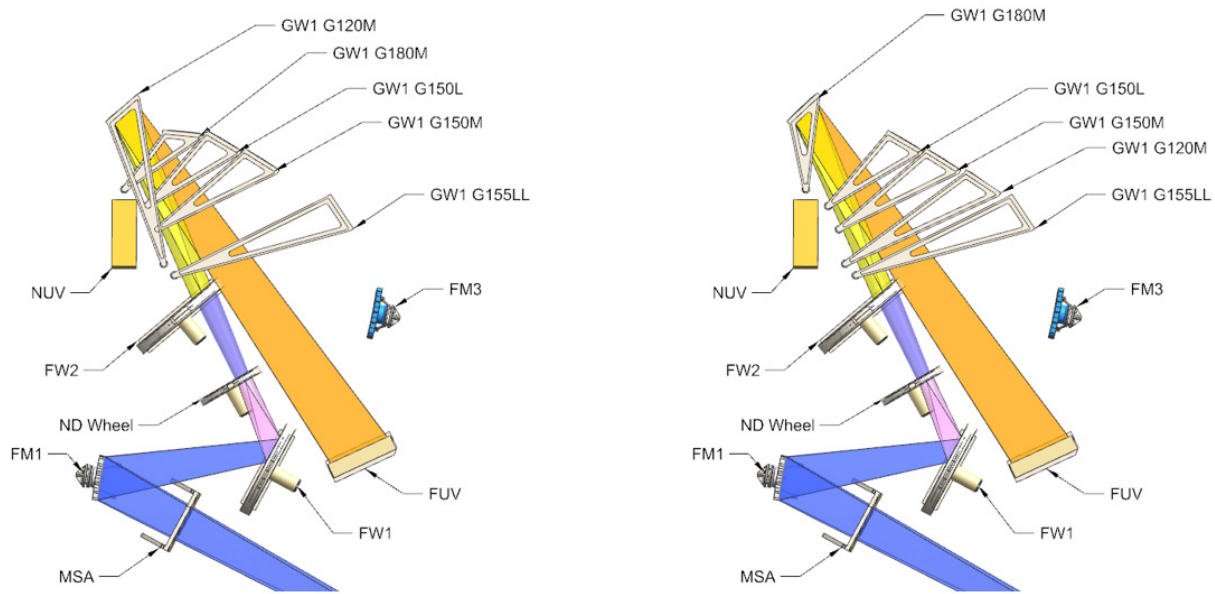

**Figure 8-36.** *Illustration of the operation of the FUV grating mechanism in LUVOIR-B. Each grating is held in a separate arm that can be moved into the appropriate position in the beam path when needed. Each arm is independent and does not interfere with the motion of any other arm.*

### 8.2.4.3.2 LUMOS-B

The right side of **Figure 8-35** shows an opto-mechanical model of LUMOS-B, which uses the same material and mechanical properties as LUMOS-A. One unique feature of LUMOS-B is the FUV grating mechanism. The grating mechanism assembly has a kinematically keyed structure that each of the arms swing and latch into when in the engaged positon. The location of the arm pivot points and the varying depth of the arms allows each of the arms to pass over or under one another, providing a clear path to the engaged position. This is demonstrated in **Figure 8-36**, which shows a comparison of the G120M and G180M modes engaged, as an example.

### 8.2.4.4 Thermal design

**Figure 8-37** shows a block diagram of the LUMOS-A thermal architecture, which is similar to that of LUMOS-B save the differences in the optical layout. Contamination control is a primary concern for the LUMOS instrument, and thus it is designed to operate at a slightly warmer temperature than the surrounding structure and instruments (280 K vs. 270 K). Heaters actively control the components to 280 K to force contaminants to deposit on the colder surfaces surrounding the instrument and not the warmer mirrors in LUMOS itself. However, the inclusion of the near-UV MOS detector requires a 170 K environment to operate, and is therefore passively cooled and isolated from the 280 K components. Additionally,

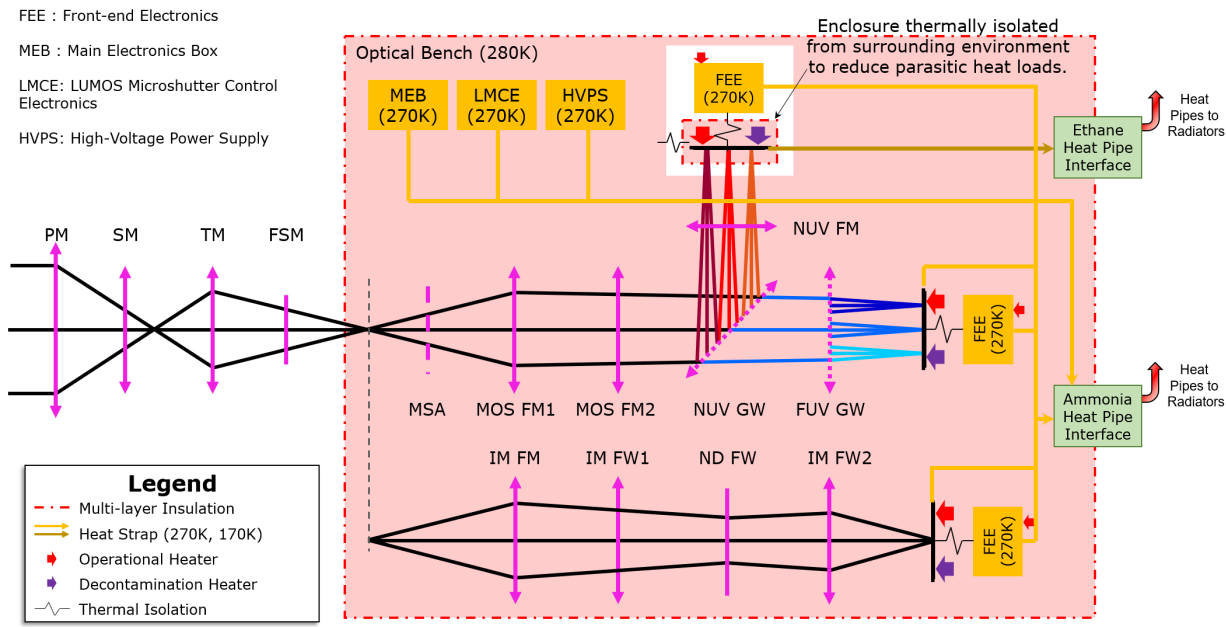

**Figure 8-37.** *LUMOS-A thermal architecture. LUMOS-B has a similar design.*

contamination control heaters on all mirrors, gratings, and detectors are sized to allow for infrequent on-orbit bake-outs to help evaporate any accumulated water from the optics.

### 8.2.4.5  Electrical design

**Figure 8-38** shows a block diagram of the LUMOS electrical architecture, which is identical for each version of the instrument excepting (1) the number of detector ASICs for the NUV focal plane (28 for LUMOS-A versus 14 for LUMOS-B), and (2) the presence of the FUV imaging detector in LUMOS-A, which is not present in LUMOS-B (where the FUV MOS detector is also used for imaging). A serviceable interface provides power and data connections from the OTA. The LUMOS main electronics box controls all instrument functions, and has internal solid state memory storage for up to 48 hours data collection.

### 8.2.4.6  Detectors

The far-UV MOS and imager focal planes consist of large-format microchannel plate (MCP) detectors, which leverage a long heritage of reliable performance on multiple NASA space and sub-orbital missions. All of the FUV MOS modes (except G145LL) are focused onto a 2 x 1 array of large-format microchannel plate detectors, with the full spectral bandpass spanning two detector faces. The shorter wavelengths of each mode fall onto a short-wavelength optimized MCP with a CsI photocathode (like HST-COS), while the longer wavelengths fall onto a GaN photocathode MCP optimized for longer wavelengths, but not sensitive to wavelengths shorter than 110 nm. The G145LL focuses the entire 2' x 2' FOV onto the short wavelength optimized MCP, while the G165LL focuses the FOV onto the longer wavelength optimized MCP.

The far-UV imaging channel detector in LUMOS-A is a single CsI photocathode MCP.

The near-UV MOS focal plane consists of an array of δ-doped CMOS detectors with 8,192 × 8,192, 6.5-μm pixels per tile. For LUMOS-A, there are 4 x 7 detector tiles, while on LUMOS-B there are 2 x 7 detector tiles.

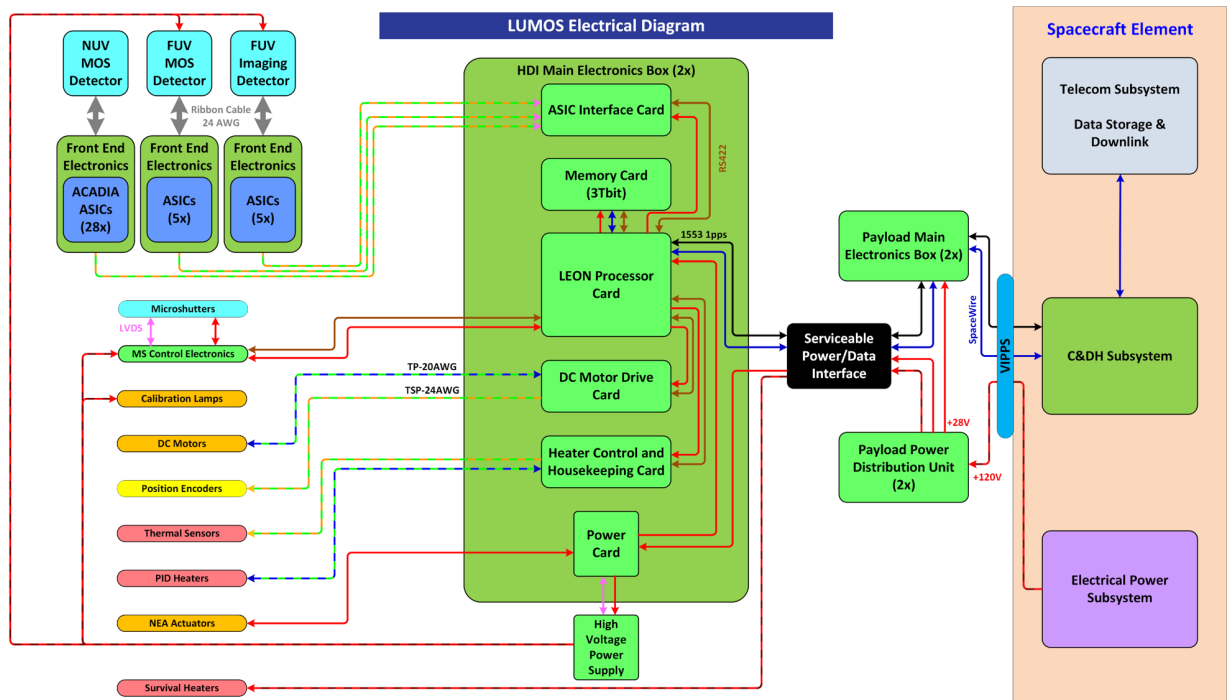

**Figure 8-38.** *LUMOS electrical block diagram. Both instruments have the same architecture except-ing the differences in the number of focal planes.*

### 8.2.4.7 Data volume

**Table 8-10** summarizes the LUMOS data volume for each version of the instrument. Since the NUV/VIS MOS nominal science mode and the FUV MOS bright object science mode

**Table 8-10.** *Data volume summary for the LUMOS instruments.*

| Parameter | Units | LUMOS-A | | LUMOS-B |
|---|---|---|---|---|
| | | **FUV MOS** | **FUV Imager** | **FUV MOS/Imager** |
| Array Format | – | 2 x 1 | 1 x 1 | 2 x 1 |
| Number of MCPs | – | 2 | 1 | 2 |
| Bits / x-coordinate / read | bits | 15 | 15 | 15 |
| Bits / y-coordinate / read | bits | 15 | 15 | 15 |
| Bits / energy / read | bits | 6 | 6 | 6 |
| Total bits / read | bits | 36 | 36 | 36 |
| **Nominal Science Mode** | | | | |
| Read Rate | Hz | 1,000 | 1,000 | 1,000 |
| Average Data Rate / MCP | Kbps | 36 | 36 | 36 |
| Total Data Rate | Kbps | 72 | 36 | 72 |
| Total Data Volume | Gbits | 12.44 | 6.22 | 12.44 |
| Total 48-hour Storage Requirement* | Gbits | 19 | 10 | 19 |
| **Bright Object Science Mode** | | | | |
| Read Rate | MHz | 3.0 | 3.0 | 3.0 |
| Average Data Rate / MCP | Mbps | 108 | 108 | 108 |
| Total Observation Time | s | 3,600 | 3,600 | 3,600 |
| Total Data Volume | Gbits | 389 | 389 | 389 |
| Total 1-hour Storage Requirement* | Gbits | 607 | 607 | 607 |

| | | LUMOS-A | LUMOS-B |
|---|---|---|---|
| | | **NUV/VIS MOS** | **NUV/VIS MOS** |
| Array Format | – | 4 x 7 | 2 x 7 |
| Number of Detectors | – | 28 | 14 |
| Detector Format | pixels | 8192 x 8192 | 8192 x 8192 |
| Pixels / Detector | Mpixels | 67.1 | 67.1 |
| Outputs / Detector | – | 32 | 32 |
| Bits / Pixel | bits | 16 | 16 |
| **Nominal Science Mode** | | | |
| Integration Time / Frame | s | 12,600 | 12,600 |
| Number of Co-added Frames | frames | 1 | 1 |
| Total Integration / Image | s | 12,600 | 12,600 |
| Average Data Rate | Mbps | 2.39 | 1.19 |
| Readout Rate | kHz | 0.166 | 0.166 |
| Total 48-hour Storage Requirement* | Gbits | 643 | 322 |
| **Bright Object Science Mode** | | | |
| Maximum Pixel Read Rate | kHz | 400 | 400 |
| Integration Time / Frame | s | 0.1 | 0.1 |
| Number of Co-added Frames | – | 1 | 1 |
| Total Integration / Image | s | 0.1 | 0.1 |
| Time to Readout 1 Frame | s | 0.00 | 0.00 |
| Average Data Rate | Gbps | 5.73 | 2.87 |
| Max. Bright Object Observation Time | s | 71.9 | 71.9 |

*Includes 20% overhead and 30% margin.

dominate the data volumes, the data storage is sized to accommodate 48-hours of continuous NUV/VIS MOS nominal science data collection, plus one hour of bright object science data collection with any one of the FUV detectors, including 20% overhead and 30% margin. Thus the total internal storage is 1.3 Tbits on LUMOS-A, and 929 Gbits on LUMOS-B.

## 8.2.5 POLLUX

The POLLUX instrument was studied by a consortium of European partners, led by the Centre National d'Etudes Spatiales (CNES). POLLUX is a proof-of-concept demonstration of a mission-enhancing instrument for the fourth instrument bay on LUVOIR-A. This instrument is an UV spectropolarimeter that complements the LUMOS instrument in both capability and scientific objectives. It combines high-resolution (R > 120,000) spectroscopy in the far- and near-UV (~100–400 nm) with polarimetry. The POLLUX science objectives, instrument design, and technology needs are discussed in more detail in **Chapter 13**.

## 8.2.6 Payload articulation system (PAS)

### 8.2.6.1 Introduction

The payload articulation system has two key components: the gimbal arm, and the Vibration Isolation and Precision Pointing System (VIPPS). **Figure 8-39** shows the PAS for each LUVOIR concept[5]. The PAS can repoint the LUVOIR payload to any position on sky within 45 minutes (30 minute goal), and enable tracking of targets at a rate of 60 mas/s.

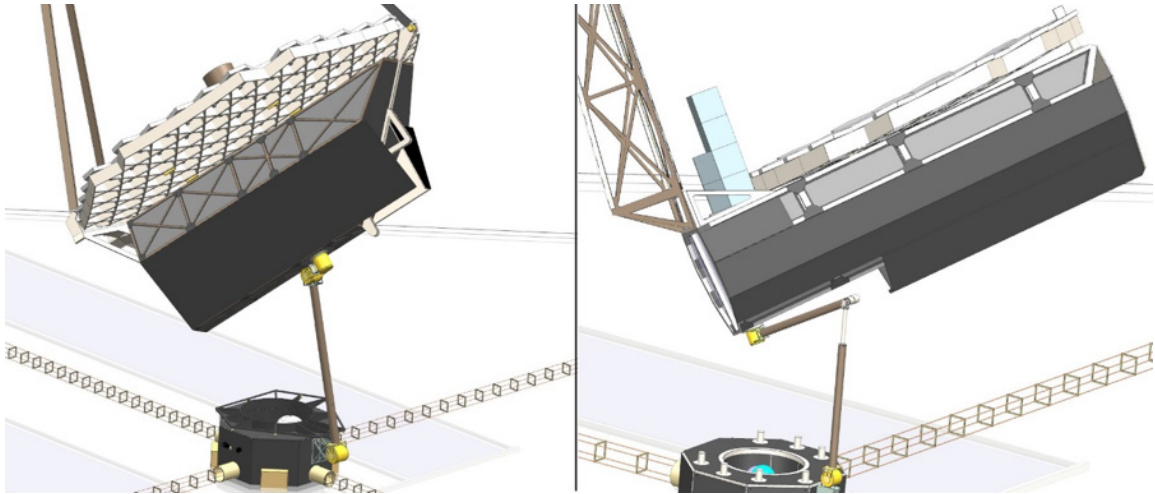

**Figure 8-39.** *Payload articulation system for LUVOIR-A (left) and LUVOIR-B (right).*

### 8.2.6.2  Gimbal arm

The gimbal arm is responsible for coarse pointing of the payload relative to the spacecraft. The purpose for pointing the payload independently of the spacecraft and sunshade is two-fold. First, by keeping the sunshade in a fixed orientation relative to the sun, the change in the thermal load on the sunshade is minimized. This creates a very stable thermal environment on the dark side of the sunshade. As the payload repoints, it will experience different view factors to the sunshade, but by keeping the dark side of the sunshade cold and stable, the resulting thermal instability can be reduced.

The second purpose of the gimbal arm is to maintain alignment of the payload center of gravity with the center of solar radiation pressure on the sunshade. This minimizes momentum build-up in the spacecraft attitude control system, reducing the operational overhead and fuel consumption from momentum unloading.

The specific implementation of the gimbal arm and mechanisms was not studied in detail, however the Space Station Remote Manipulator System (SSRMS) aboard the International Space Station is larger and more capable than LUVOIR's gimbal arm would need to be. SSRMS is 17 meters long (compared to LUVOIR's ~7 meters), has seven degrees of freedom (compared to LUVOIR's required 3), and can manipulate payloads up to 116,000 kg (compared to LUVOIR's less than 44,000 kg). The SSRMS demonstrates feasibility of the LUVOIR gimbal arm system.

### 8.2.6.3  Vibration Isolation and Precision Pointing System (VIPPS)

Dynamic stability is achieved with stiff mirrors and structures, passive isolation at the disturbance sources, and the non-contact VIPPS that actively isolates the payload from the spacecraft. The VIPPS "floats" the telescope and controls the payload attitude relative to the interface plane via six non-contact voicecoil actuators. The VIPPS effectively isolates any dynamic disturbances from the spacecraft attitude control system from transmitting to the payload and exciting resonances that contribute to wavefront instability. The VIPPS also provides fine pointing control of the payload during science observations. **Figure 8-40** shows a notional block diagram of the VIPPS control loops, while **Figure 8-41** shows a model of the VIPPS system.

---

5 See also the deployment video of each concept at <u>https://asd.gsfc.nasa.gov/luvoir/design/</u>

**Figure 8-40.** *VIPPS control architecture. PL: Payload, SC: Spacecraft, WFE: wavefront error, LOS: line-of-sight. Credit: Lockheed Martin*

**Figure 8-41.** *Model of the VIPPS interface. Six non-contact voicecoil actuators control the relative position and attitude of the two plates. Credit: Lockheed Martin*

**Table 8-11.** *Harness summary between the payload and spacecraft elements. This cable bundle must bridge the non-contact interface of the VIPPS.*

| Qty. | Type | Purpose |
|------|------|---------|
| 8 | Spacewire | Science Data |
| 2 | 1553 Bus | Telemetry, Housekeeping |
| 24 | 10 AWG | 120-v Power |

While the VIPPS provides ideal mechanical isolation, power and data signals must still be transmitted between the payload and spacecraft. Using a 120-V bus for the electrical power system allows for fewer, smaller-gauge conductors to be used, minimizing the number of cables and cable stiffness that must bridge the non-contact gap. Similarly, by performing as much of the data-processing on the payload side as possible, the cabling requirements for data transfer between the payload and spacecraft are minimized. **Table 8-11** summarizes the harness requirements between the payload and spacecraft elements. **Chapter 11** describes a technology development effort for the VIPPS that includes characterizing and mitigating the effects of cable stiffness on the isolation system. Additional details about the VIPPS capabilities and technology status can be found in Dewell et al. 2019.

## 8.3 Spacecraft element

### 8.3.1 Bus
The following sections discuss each of the spacecraft bus sub-systems in detail. All of these sub-systems use high-TRL components with extensive flight heritage.

#### 8.3.1.1 Mechanical design
The spacecraft bus assembly, shown in **Figure 8-42**, is an octagonal structure with a central cylinder, upper and lower decks, outer equipment panels and radials. It interfaces with the payload element through the BSF launch locks during launch, and through the PAS while on orbit. It also interfaces with the launch vehicle through the payload launch vehicle interface ring. The outer equipment panels provide mounting and support structure for the solar arrays, electronics boxes, control moment gyroscopes, sunshade membrane, and the sunshade deployment masts. These panels are easily accessible from the exterior of the bus for servicing the internal components. An entire panel, with associated components, can be removed from the bus and replaced with new or upgraded components. **Figure 8-43** shows a block diagram of the LUVOIR-A spacecraft bus; LUVOIR-B's configuration is similar.

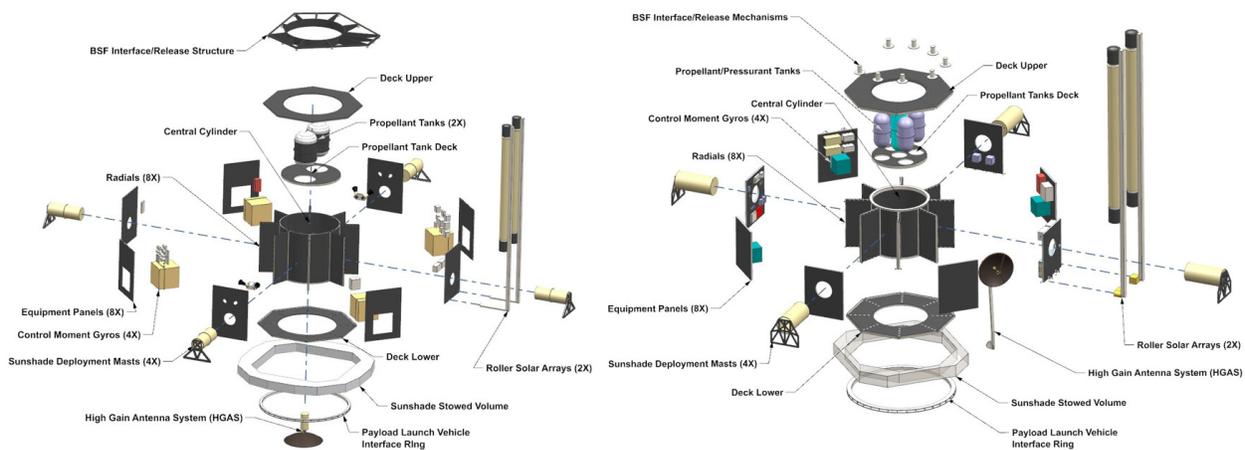

**Figure 8-42.** *Exploded view of the Spacecraft Bus for LUVOIR-A (left) and LUVOIR-B (right). Images are not to scale.*

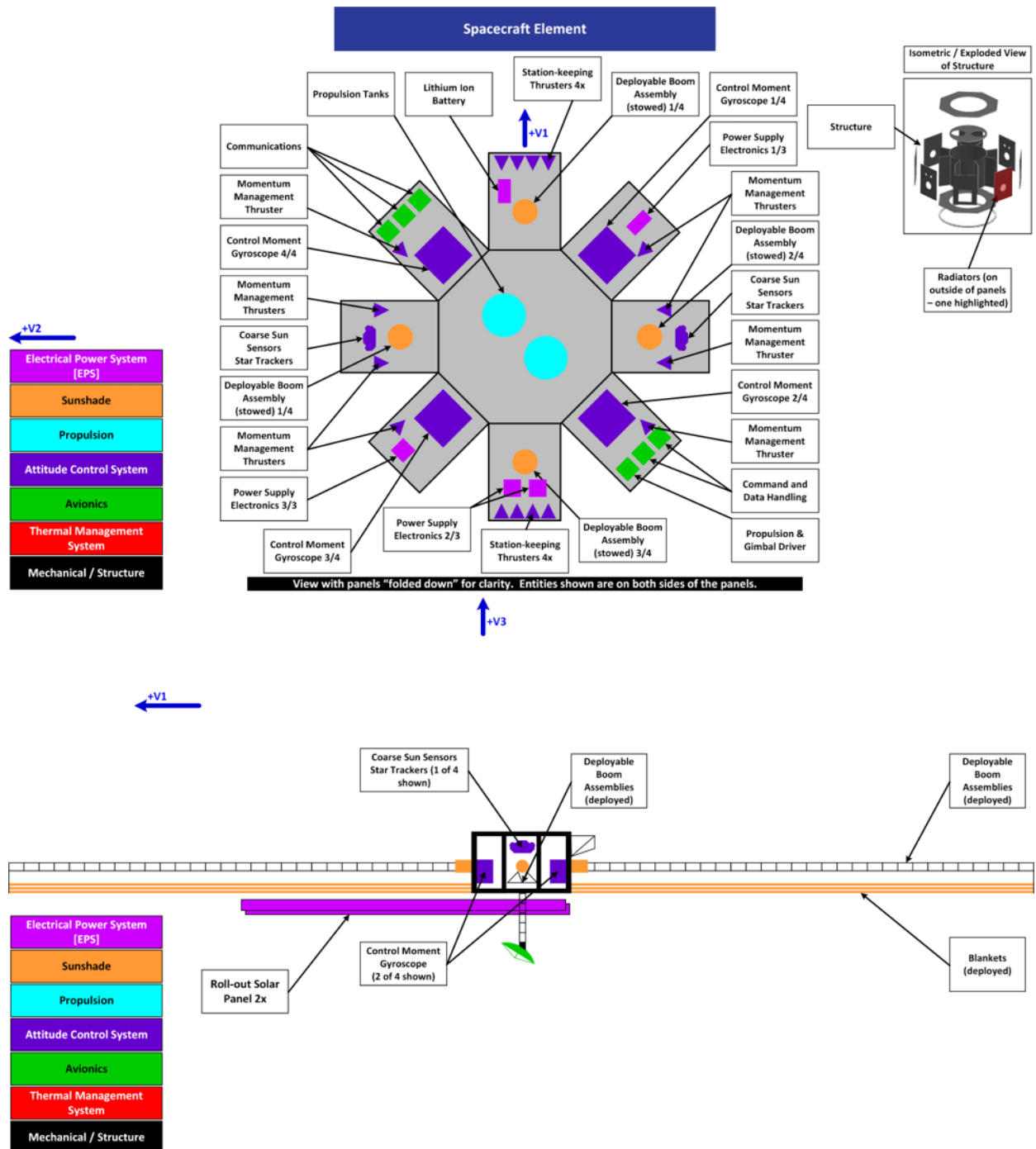

**Figure 8-43.** *LUVOIR-A spacecraft bus block diagram, showing the location of each internal subsystem. For clarity, the 8 sides of the bus are shown "folded down."*

The spacecraft bus geometry is driven by the need to maximize the primary mirror area while providing a robust load path for the BSF to the launch vehicle. The components in the bus need to fit into a low profile structure to maximize the available fairing volume for the payload element. While the load path needs to be minimized to reduce the mass of the spacecraft, analysis shows that the footprint stance of the load path needs to be as wide as possible due to lateral modes in the very tall payload element. As a result, a wide central cylinder is the primary load path and the bus structure that carries the other spacecraft components are secondary structure. This provides a simple design with reduced mass and a low profile. The primary structure uses composite face sheets with aluminum honeycomb core panels, interfacing flanges and rings, and a clip and bushing joinery method similar to Lunar Reconnaissance Orbiter and other composite panel missions.

### 8.3.1.2  Thermal management system (TMS)

The power system electronics boxes, four control moment gyroscopes of the attitude control system, and the boxes for the command and data handling and communication sub-systems all directly mount to the internal-faces of the spacecraft bus outer panels with silicone thermal interface material to facilitate heat conduction from their baseplates. The external faces of these panels are covered with Z93 white paint where heat rejection is desired. Germanium black Kapton outer layer multi-layer insulation blankets with $\varepsilon^* = 0.03$ cover all non-radiator surfaces. These external-facing panels are also actively heated to maintain a $270 \pm 3$ K temperature. Additionally, ammonia heat pipes are embedded in certain outer panels that have high heat-dissipating components mounted on the internal face. These components require both through-thickness conduction to the external-facing radiator as well as lateral spreading to reduce the bus panel in-plane temperature gradients. For all of the internal-facing surfaces inside the bus enclosure, the panels are left as bare composite and the boxes are covered with high-emissivity black Kapton to isothermalize the bus via radiative heat exchange.

### 8.3.1.3  Attitude control system (ACS)

The spacecraft is three-axis stabilized and inertially fixed, with the spacecraft axial vector parallel to the sun-spacecraft axis. The ACS is responsible for counteracting external disturbances (primarily solar pressure torques), as well as changing the yaw axis of the observatory and reacting against gimbal pitch changes during payload retargeting maneuvers. The ACS is also responsible for pitching the entire observatory sunward for specific observations.

The ACS sensor suite serves two functions: star trackers, gyroscopes, accelerometers, and coarse sun sensors determine the attitude and inertial position of the spacecraft relative to the sun, while proximity sensors on the VIPPS determine the relative attitude between the spacecraft and the payload. ACS actuators, consisting of four control moment gyroscopes (CMGs) and twelve 5-lb. thrusters respond to the sensor suite to maintain the absolute attitude and inertial position of the spacecraft. The four CMGs are arranged to provide full three-axis stabilization with redundancy. The thrusters are arranged to accommodate simultaneous momentum desaturation and orbit maintenance stationkeeping maneuvers.

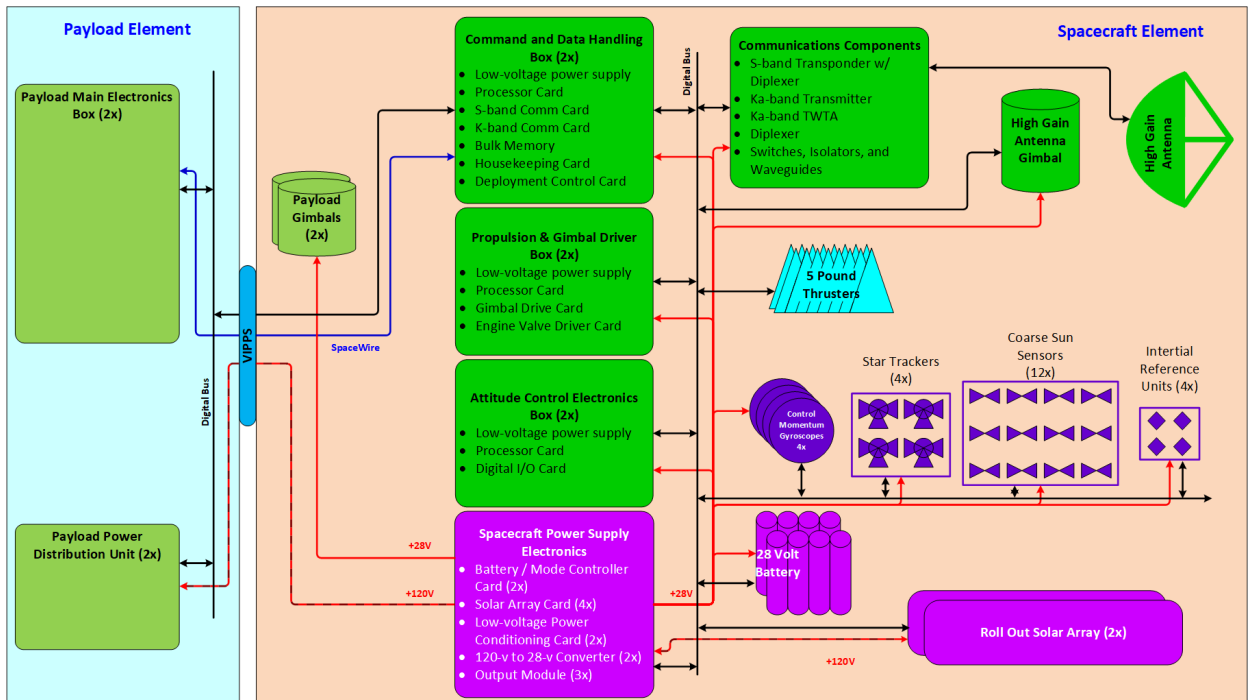

**Figure 8-44.** *Electrical block diagram of the LUVOIR spacecraft avionics system. The overall design is the same for both LUVOIR concepts.*

### 8.3.1.4  Propulsion system

A set of four 5-lb. long-duration thrusters are used for all mid-course correction maneuvers, as well as for the orbit insertion burn to enter the SEL2 halo orbit. An additional twelve 5-lb. thrusters are arranged around the spacecraft bus to support attitude control and orbit maintenance maneuvers.

Both concepts use a bi-propellant system. LUVOIR-A uses a pressure-regulated system, while LUVOIR-B uses a simpler blowdown system. Aside from the additional pressure-regulation hardware on LUVOIR-A, the two propulsion systems are identical for both concepts. Propellant masses are derived from the required maneuver list discussed in **Chapter 9**, and are based on the MEV mass of the system. Margin and reserve is then applied to the propellant. Propulsion system schematics are provided in **Appendix E**.

### 8.3.1.5  Avionics / command & data handling (C&DH)

The C&DH system consists of a solid state recorder, a board-level computer for command and control, and dedicated controller boards for spacecraft deployment and gimbal mechanisms. To ensure the payload never sees direct sunlight, a separate ACS safe-hold processor takes over in the event of a C&DH system failure to maintain a safe attitude. Each instrument contains enough internal storage for two day's-worth of data. Therefore, the solid state recorder on the spacecraft only needs to act as a buffer for each downlink, and is sized to hold only a single downlink's worth of data (~8.6 Gbits for LUVOIR-A, 4.3 Gbits for LUVOIR-B). **Figure 8-44** shows a block diagram of the spacecraft avionics and their interface to spacecraft subsystems and payload.

**Table 8-12.** *Solar array sizing based on each concept's required peak power.*

| Parameter | Value | | Units | Comment |
|---|---|---|---|---|
| | **LUVOIR-A** | **LUVOIR-B** | | |
| Peak MPV Observatory Power Needed | 38,650 | 30,902 | W | |
| Peak Power Required from Solar Array | 48,313 | 38,627 | W | Accounts for inefficiencies in load path |
| Solar Constant | 1353 | | W/m$^2$ | At SEL2 |
| Solar Array Performance / Unit Area | 457 | | W/m$^2$ | |
| Solar Inclination Angle | 45 | | degrees | Required for special sunward observations |
| Beginning-of-Life Performance / Unit Area | 233 | | W/m$^2$ | Includes inherent degradation and solar angle |
| End-of-Life Performance / Unit Area | 176 | | W/m$^2$ | Assumes 10 year life |
| Required Solar Array Area | 274 | 219 | m$^2$ | |

The C&DH system is also responsible for failure detection and correction. Following an anomaly, the C&DH will command the observatory to enter safe mode, in which several critical precautions are taken:

1. Each instrument is issued a "safe" command. Shutters are closed or introduced into the beam paths to prevent exposure of the focal planes to unexpected sources. In LUMOS, high-voltage power to the detectors and microshutters is cut to prevent damage to the electronics.

2. The VIPPS and gimbal systems are locked to prevent any changes in payload pointing.

3. In the event of a loss of the spacecraft primary C&DH processor, the independent attitude control system safe-hold processor takes over to ensure that the sunshade orientation relative to the sun remains fixed, to avoid direct exposure of the payload to sunlight.

Once the anomaly has been resolved, the system will enter standby mode. In this mode, instrument beam paths are still shuttered, however power is restored to all focal planes and mechanisms. The VIPPS and gimbal systems are unlocked and re-engaged. Depending on the nature and duration of the anomaly, it may be necessary to repeat portions of the commissioning process (see **Section 9.1.3**) if the alignment of the OTA drifted beyond the capture range of the metrology system.

### 8.3.1.6 Communications

The communications system uses Ka-band for data downlink and S-band for telemetry and housekeeping. On LUVOIR-A, a 1.8 m high gain antenna is used to connect with ground stations. On LUVOIR-B the antenna is 1.2 m in diameter to accommodate packaging constraints in the smaller launch fairing. The smaller antenna on size on LUVOIR-B is offset by the fact that the data volume on LUVOIR-B is also smaller, allowing for similar data rates, transmitter power, and contact times between the two concepts. Commands and telemetry during launch and ascent will use S-band omni antennas on the spacecraft and payload. Link budgets for both LUVOIR concepts are available in **Section 9.2**. For both LUVOIR concepts, 48-hours' worth of data can be downlinked in 4 hours of contact time via the Ka-band downlink.

### 8.3.1.7  Electrical power system (EPS)

Roll-Out Solar Arrays (ROSAs) are baselined for both LUVOIR concepts. The geometry of the solar array system was provided by Deployable Space Systems (DSS)[6], although other vendors make similar roll-out systems that can also be used here. **Table 8-12** summarizes the total solar array area required for each concept, based on the maximum possible value (MPV) power requirements provided in **Table 8-3**. It should be noted that while these MPV peak power numbers appear high, they are commensurate with a space facility of similar scale: the International Space Station requires ~75-90 kW to operate, and the solar arrays are capable of providing up to 120 kW at beginning of life.

The arrays are mounted vertically for launch and must be supported by the payload element in this position. To deploy, the arrays pivot down, then roll out using extension booms[7]. After the arrays are deployed they are stationary. The arrays are oversized so as to provide sufficient power to the observatory even when the observatory is tilted by up to 45° for special sunward observations. This design choice was made to reduce dynamic instabilities from a solar array drive mechanism. The ROSA system has flight heritage on the International Space Station and provides a very efficient way to package a very large solar array.

A 24 amp-hour battery is included for providing power during launch and ascent. The power system provides 28-V to spacecraft subsystems, and 120-V directly to the payload. A 120-v payload bus was chosen primarily to minimize the power harness across the VIPPS interface. The higher voltage allows for fewer, thinner conductors, and therefore more flexible cables, to bridge the non-contact interface.

LUVOIR adopts a Class-A mission risk posture throughout the observatory (both the spacecraft and the payload). All electrical system boards are side-A / side-B or box-level redundant, all mechanisms have dual-windings, and all electrical power and data interfaces are either cross-strapped, redundant, or both.

### 8.3.2 Sunshade

A central component of the LUVOIR thermal management architecture is the deployable sunshade. The sunshade isolates the payload from solar thermal loads and provides a cold environment in which the payload can be thermally stabilized with active heaters, and in which system radiators have sufficient cold-sink temperatures.

Although the sunshade appears similar in concept and design to the sunshield of JWST, it differs in two important aspects. First, it is larger. While JWST's sunshield is ~21 x 14 m (point-to-point in each dimension), the LUVOIR-A sunshade is ~78 x 78 m (67 x 67 m for LUVOIR-B), as shown in **Figure 8-45**. The size of the sunshade is determined by the need to keep the entire payload element in shadow as it articulates through its range of pointing orientations.

The second difference pertains to the complexity and thermal performance requirements of each system. The terminology sunshade vs sunshield is meant to emphasize this distinction. A sunshield is a precision deployable thermal system; a sunshade is not much more than a blanket. LUVOIR's sunshade does not need to achieve the extremely cold temperatures of JWST's sunshield. The temperature requirement is driven by the need for an adequate cold sink for the instrument radiators mounted to the BSF; an average temperature of 95 K is sufficient to passively cool the instrument components to 170 K[8]. Whereas JWST's sunshield is an extremely complex system of 5 layers that need to be precision deployed to tight tolerances on the angle and separation between the layers, LUVOIR's sunshade is simpler. A minimum of two sheets of single-layer insulation are needed to meet the thermal performance requirements, although three layers are included in the design for redundancy against micrometeroid strikes. The spacing and angle between these layers is not critical, so long as they do not touch and create a thermal short.

---

6 https://www.dss-space.com/
7 See deployment videos at https://asd.gsfc.nasa.gov/luvoir/design

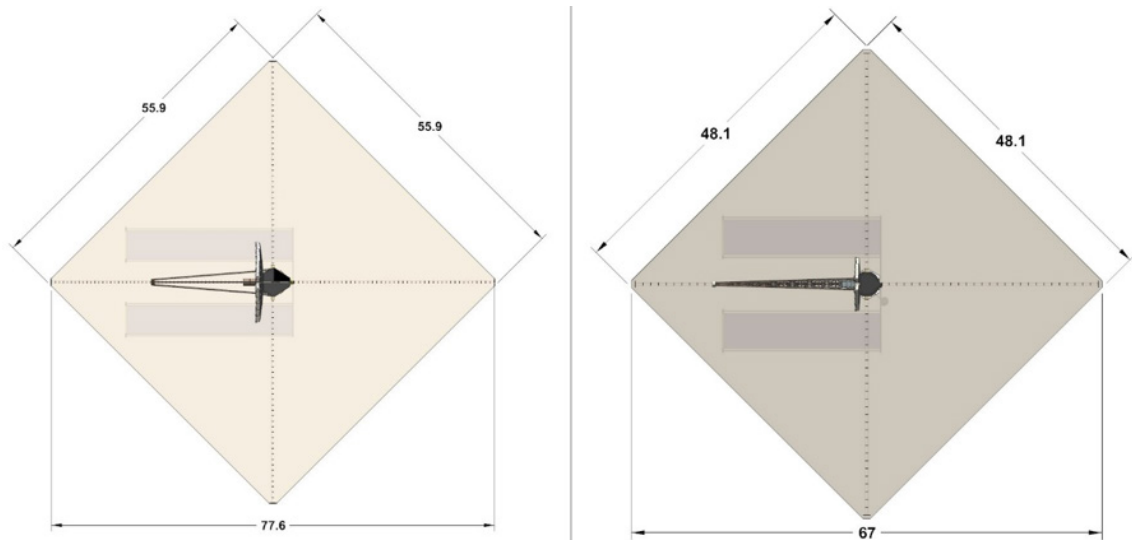

**Figure 8-45.** *LUVOIR sunshade dimensions, in units of meters. LUVOIR-A is shown on the left, while LUVOIR-B is shown on the right. The size of the sunshade is largely driven by the primary-to-secondary mirror separation, and the need to keep the secondary mirror in shadow for all possible pointings of the observatory.*

The manner in which the sunshade is deployed also differs from JWST, which uses two rigid pallets to hold the sunshield system. These pallets are stowed up around the telescope and instruments, cocooning the payload during launch and ascent. After launch, the pallets fold down, and port and starboard deployable booms pull the sunshield layers out from the pallet, before spreaders and tensioners fine-position the five sunshield layers. In contrast, LUVOIR packages the entire sunshade at the base of the spacecraft. Four deployable booms pull the folded sunshade out in each of four directions: fore, aft, port, and starboard. This method has several advantages over JWST, as well as several new challenges. Advantages include a smaller relative mass and volume for the stowed sunshade system, the use of high TRL deployment mechanisms, such as the coilable boom systems, and overall lower risk to the deployment due to fewer mechanisms. New challenges include venting of the stowed sunshade during ascent, and the fact that the payload will be exposed once the fairing is jettisoned. Preliminary thermal modeling indicates that a roll about the long axis of the stowed observatory is sufficient to keep most surfaces below critical heating temperatures, although more detailed study will be needed as the designs mature.

---

8 Note that the 100 K NIR detectors are cooled with a radiator that does not have a direct view of the sunshade for any pointing orientation.

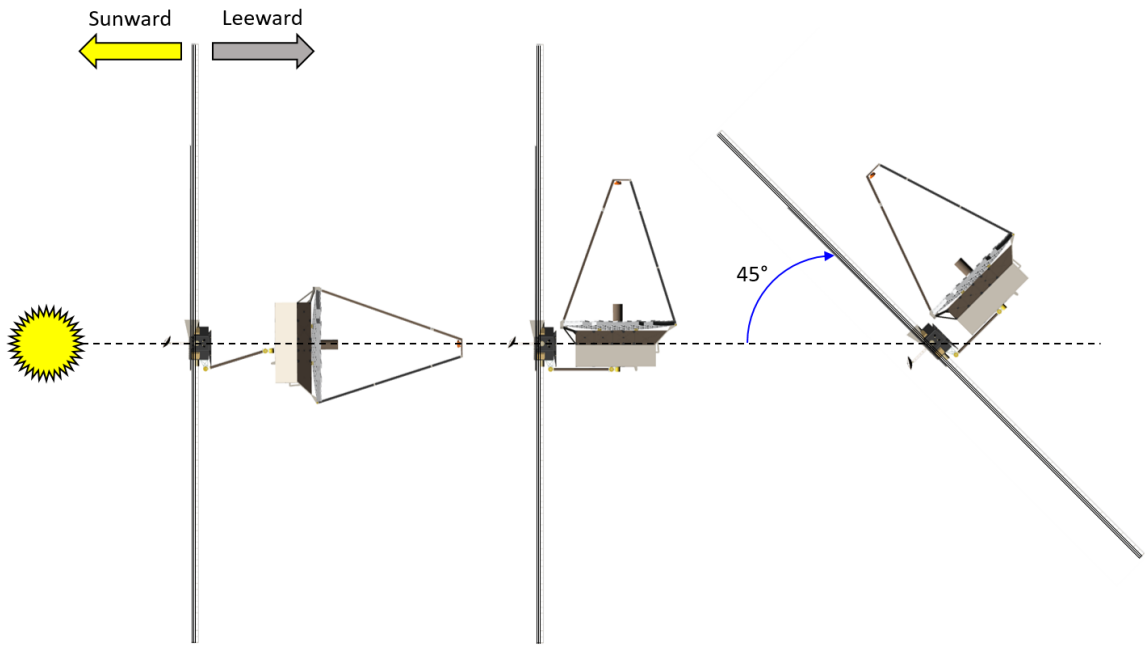

**Figure 8-46.** *Different pointing scenarios of the LUVOIR observatory. Most science observations maintain the sunshade normal to the Sun-spacecraft axis (left and center). The thermal load on the sunward side of the sunshade remains constant generating a thermally-stable environment on the leeward side, regardless of payload pointing. Exoplanet revisit observations and solar system targets of opportunity may require the entire observatory to pitch towards the sun to a 45° angle with the Sun-spacecraft axis (right).*

Nominally, the sunshade remains in a fixed attitude normal to the Sun-observatory axis, maintaining a constant thermal load on the sunward side of the sunshade as the payload slews to different targets on the leeward side. The payload always sees a constant thermal environment, regardless of its pointing attitude. Two science observations necessitate exceptions to this nominal attitude and require the entire observatory to pitch toward the sun (while still keeping the payload in shadow). Exoplanet revisit observations must be scheduled at critical time intervals to maximize observational completeness and may require access to sunward portions of the sky. Also, observations of specific solar system bodies, such as Venus and comets, require pointing the observatory up to 45° from the Sun-Earth axis. **Figure 8-46** shows these different pointing scenarios for the observatory.

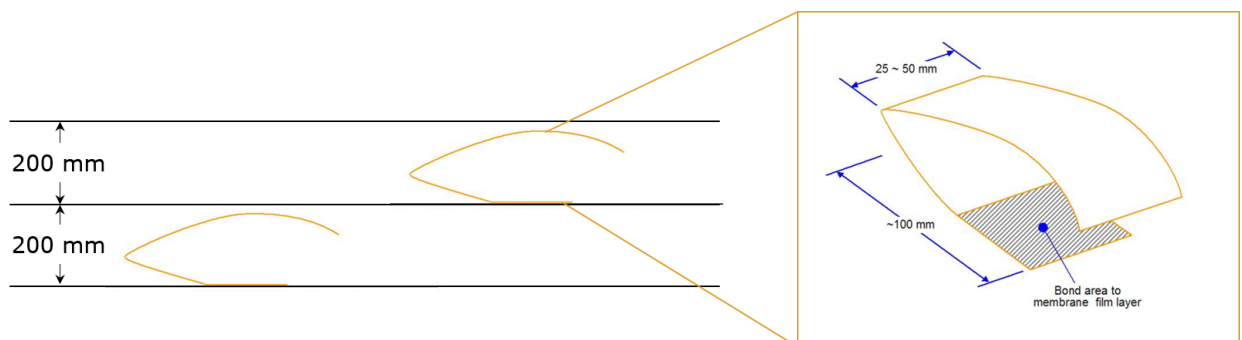

**Figure 8-47.** *Notional sunshade separator. Heat-formed Kapton springs are bonded at regular intervals between the sunshade layers, and can be compressed flat when the sunshade is folded and stowed.*

### 8.3.2.1  Blanket assemblies

The LUVOIR sunshade does not require the level of precision and thermal control the JWST sunshield requires since LUVOIR is operating at far warmer temperatures. The LUVOIR sunshade blankets are a 3 membrane design with sufficient separation and minimal tensioning to ensure the layers do not touch. One concept for ensuring layer separation is the inclusion of kapton "springs" between the layers, shown in **Figure 8-47**. The extendable booms have been sized to provide deployment and some light tensioning of the membranes, but the deployed surface finish of the membranes does not require a highly tensioned design.

The sunshade size is defined by the requirement to keep the payload element in shadow at all times. Since the PAS allows a range of pointing options for the payload element, the sunshade was sized for the worst case orientation.

The sunshade membranes use 0.0254 mm-thick Kapton for all 3 layers with a layer spacing of 200 mm. The sun-facing side of the bottom-most layer uses a silicon-doped vapor-deposited aluminum coating. All inner layers use a vapor-deposited aluminum coating, and the payload-facing side of the top-most layer is coated in black Kapton to reduce the straylight impact on the optical system. This top layer achieves an average temperature of 95 K on LUVOIR-A and 75 K on LUVOIR-B.

### 8.3.2.2  Deployable boom system

Four Northrop Grumman Coilable Boom Systems are baselined to pull the sunshade membranes away from the spacecraft bus and provide final tensioning. These mechanisms have high TRL, with 44.5 m versions having flown on the Naval Research Lab Low-Power Atmospheric Compensation Experiment (LACE) mission in 1990; shorter versions have flown more recently on numerous missions, including the Nuclear Spectroscopic Telescope Array (NuSTAR). Alternative deployable booms are also available from Roccor, and have been used to deploy light-sail units in ground demonstrations. LUVOIR-A and LUVOIR-B would require boom lengths of ~40 m and ~35 m, respectively. The booms were sized based on a 130N pull-out force. This drives the size of the canisters that house the stowed booms, the cross section of the deployed boom, and the overall mass of the boom sub-system. Although no sunshade this large has deployed in orbit, there has been some ground testing for this concept in solar sail applications.

# CHAPTER 9. MISSION OPERATIONS AND GROUND SEGMENT

## 9.1 Mission operations
LUVOIR mission operations are divided into six mission phases:

1. Launch & ascent

2. Cruise

3. Commissioning

4. Stationkeeping

5. Slew & settle

6. Science operations

**Figure 9-1** shows a high-level timeline of the launch & ascent, cruise, and commissioning operations. These first three modes are only ever performed once in LUVOIR's nominal 5–10 year lifetime. Modes 4–6 will be used routinely during normal operations. In the following sections, we discuss each mode in more detail.

### 9.1.1 Launch & ascent
Both LUVOIR-A and LUVOIR-B are baselined to launch aboard a Space Launch System vehicle from Kennedy Space Center, Cape Canaveral, FL. Alternative launch vehicles such as SpaceX's Starship and Blue Origin's New Glenn are being considered, however detailed launch timelines have not been evaluated for these vehicles yet.

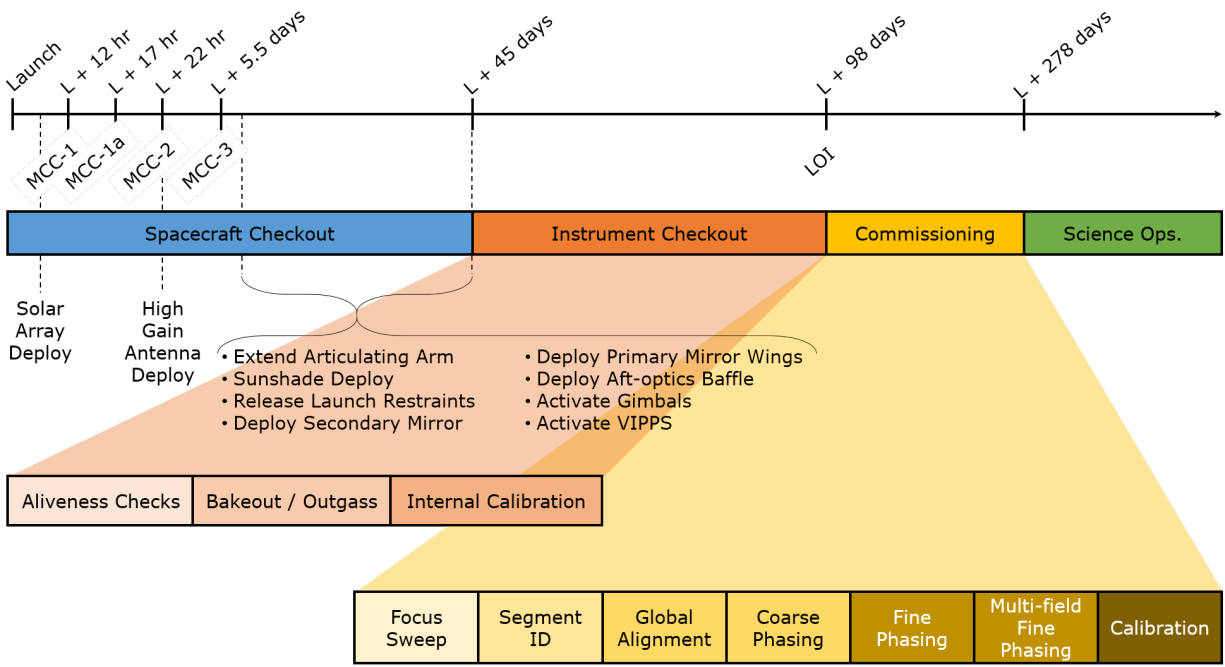

**Figure 9-1.** *Launch, ascent, cruise, and commissioning phase timeline. MCC: Mid-course correction; LOI: L2 orbit insertion; VIPPS: Vibration Isolation and Precision Pointing System*

After initial separation of the solid rocket boosters and core stage, there are two second stage engine burns. Observatory separation from the second stage occurs ~53 minutes after launch.

### 9.1.2 Cruise to second Sun-Earth Lagrange point (SEL2)

Following launch vehicle separation, the solar array is deployed to power LUVOIR's internal systems. Mid-course correction (MCC) burns are used to account for launch vehicle performance variability. The first three MCC burns are executed 12 hours, 17 hours, and 22 hours after launch, respectively, using a bank of 5-lb. thrusters on the spacecraft. Up to this point, communication with the observatory is achieved via S-band omni-directional antennas. Following MCC-2, the high-gain antenna is deployed for remaining communications. A final MCC burn occurs ~5.5 days after launch, after which the remaining observatory deployments can occur in the following order:

1. Release payload-to-spacecraft launch restraints; extend payload articulation system telescoping arm.

2. Deploy sunshade.

3. Release launch restraint mechanisms on payload element.

4. Deploy secondary mirror.

5. Deploy primary mirror wings.

6. Deploy aft-optics baffle.

7. Release and activate payload articulation system gimbals.

8. Activate Vibration Isolation and Precision Pointing System (VIPPS).

**Table 9-1** lists each of the cruise phase maneuvers and provides a Δv budget and listing of burn times for both LUVOIR concepts.

**Table 9-1.** *Cruise-phase maneuver list and Δv budget for both LUVOIR concepts. Values are based on each concept's maximum expected value (MEV) mass, and propellant mass is derived from these values. Margin and reserve are then applied to the system and propellant masses. In keeping with the serviceability approach discussed in* **Chapter 8**, *propellant for 10 years of stationkeeping is included, and assumes the on-orbit refueling would be performed to extend the mission beyond this limit. MCC: Mid-course correction; LOI: L2 orbit insertion; L: Launch*

| Maneuver | Δv (m/s) | Mission Time | Burn Time (min) | | Notes |
|---|---|---|---|---|---|
| | | | A | B | |
| Launch | – | – | – | – | Requires a C3 = -0.55 to -0.75 km²/s² |
| MCC-1 | 45 | L + 12 hr | 143 | 85 | Dependent on launch vehicle performance reliability |
| MCC-1a | 10 | L + 17 hr | 31 | 19 | Clean up for MCC-1 efficiency |
| MCC-2 | 20 | L + 22 hr | 62 | 37 | |
| MCC-3 | 5 | L + 5.5 days | 31 | 19 | |
| LOI | 30 | L + 98 days | 185 | 110 | |
| Stationkeeping | 80 | – | < 1 | < 1 | Allocation assumes 8 m/s per year for 10 years |

Deployments are expected to complete ~45 days after launch. Once fully deployed, a series of aliveness checks will be completed on all payload systems. A "bake-out" will also be performed as a contamination control measure. During this operation, heaters are driven to a maximum set-point to facilitate residual outgassing and drying of composite structure materials. Critical optical surfaces will be held at warmer temperatures to prevent the settling of contamination on these surfaces. Following this bake-out, instrument calibrations using internal sources are completed.

A final orbit-insertion burn occurs ~98 days after launch, marking the end of the cruise phase of mission operations.

### 9.1.3 Commissioning

Once the observatory enters its final orbit at Sun-Earth L2, commissioning can begin to align the optical telescope assembly (OTA) and set final internal instrument alignments. A high-level sequence of the alignment steps is shown in **Figure 9-1**, Each step is designed to position the primary mirror segment and secondary mirror within the capture range of the subsequent step. Fine phasing is accomplished using phase-diverse phase retrieval on defocused images collected by the HDI instrument. Multi-field fine phasing is accomplished using phase-retrieval images captured in each of the instruments' separate fields-of-view. Every step in the commissioning flow up to this point uses the same processes and algorithms that will be used for JWST commissioning. Once fine phasing is completed, the edge sensor and laser truss active metrology system, which JWST does not have, will maintain final alignment of the OTA throughout science operations.

### 9.1.4 Stationkeeping

The quasi-halo orbit about the Sun-Earth L2 point is unstable, and requires periodic orbit maintenance maneuvers. Additionally, over time the control moment gyroscopes will saturate their ability to provide angular momentum changes to control the spacecraft attitude. "Momentum dump" maneuvers are necessary to desaturate the gyroscopes. These two maneuvers—orbit maintenance and momentum management—are standard procedures for inertially-stabilized spacecraft, and are combined into a single stationkeeping maneuver that occurs approximately every three weeks. **Figure 9-2** shows a block diagram of the stationkeeping process. These stationkeeping maneuvers will require, on average, a $\Delta v$ of 8 m/s per year.

### 9.1.5 Slew & settle

Following a science observation or stationkeeping maneuver, LUVOIR will be ready to slew to a new target for observation, based on previously uploaded coordinates. Slews will incorporate pitch-angle changes using the gimbal system, and azimuth angle (yaw) adjustments using the spacecraft attitude control system. For special sunward observations, the spacecraft attitude control system may also need to pitch the entire observatory. The VIPPS, gimbal system, and attitude control system were designed to accommodate a required slew rate of 2°/min (the designed system is capable of achieving a maximum slew rate of 3°/min, however settle times have not yet been analyzed for this scenario), allowing the observatory to repoint anywhere in the anti-sun hemisphere within 45 minutes (or 30 minutes assuming the faster 3°/min. slew rate goal).

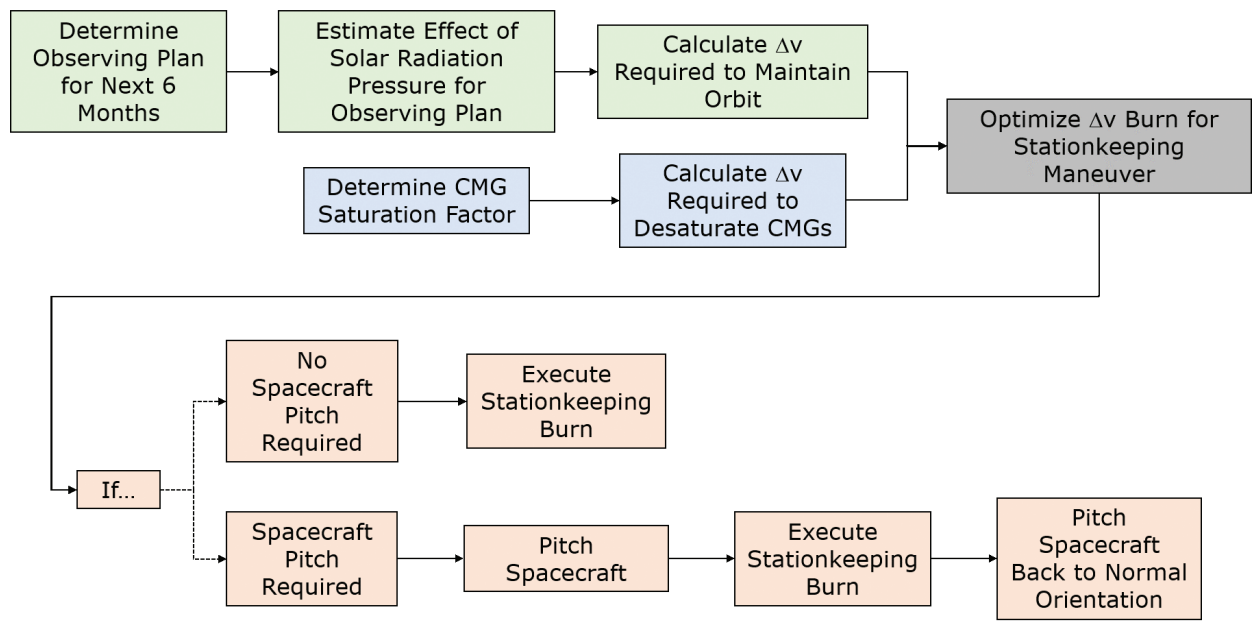

**Figure 9-2.** *Stationkeeping flow diagram. Orbit maintenance and momentum management maneuvers are combined to minimize the impact on science observing time. Depending on the direction of the required ∆v, an additional pitch of the spacecraft may be necessary. CMG: control moment gyroscope*

During a slew, the VIPPS will be actively isolating the payload from the spacecraft, reducing the dynamic inputs to the payload and thus reducing the amount of settling time that will be required once the final attitude is achieved. **Figure 9-3** shows a block diagram of the slew control system.

Star trackers will be used to coarsely align the observatory to the target field, after which fine guiding will be done by the HDI instrument. Acquisition of suitable guide stars known to be in the field-of-view will be performed autonomously using the control system processor (see **Section 8.2.3.5**). Once fine guidance has begun, the observatory will enter a prescribed "settle" period. The settle period is intended to allow residual thermal and dynamic drifts that are not compensated by the active heaters and vibration isolation system to dampen out. The length of the settle period will depend on the type of science observation to be carried out. Based on integrated modeling of payload dynamics, with the VIPPS system active during slews, most general astrophysics and solar system observations will achieve the necessary diffraction-limited wavefront error and line-of-sight pointing in less than 10 minutes (Dewell et al. 2019). For most exoplanet observations, the settle time will be a few 10s of minutes to achieve the necessary picometer-level wavefront error stability (Dewell et al. 2019).

### 9.1.6 Science operations

The science observation operations depend on the type of science observation being conducted.

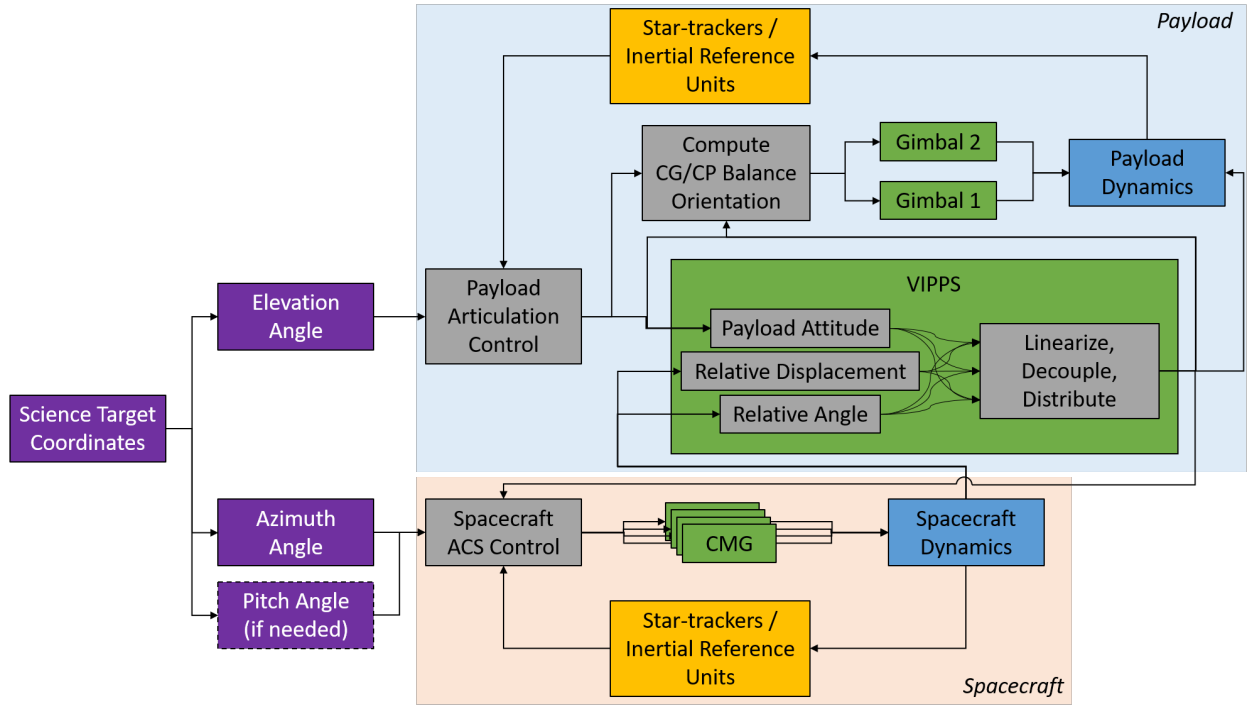

**Figure 9-3.** *Slew control system block diagram. Science coordinates are decomposed into elevation angles and azimuth angles. Azimuth angles are achieved by rolling the spacecraft about the sun-line using the attitude control system (ACS). Star-trackers and inertial reference units provide feedback on the spacecraft attitude. If required, the spacecraft ACS can also pitch the observatory sunward for critical observations. Elevation angles are achieved by pitching two gimbals in the payload articulation system, as well as using the Vibration Isolation and Precision Pointing System (VIPPS). The two gimbals are used to maintain alignment of the observatory center-of-gravity (CG) with the solar radiation center-of-pressure (CP) to reduce momentum buildup. Star-trackers and inertial reference units on the payload provide feedback to the payload articulation control system. The VIPPS provides fine payload attitude control. Relative angle and displacement sensors in the VIPPS provide commands to the spacecraft ACS to maintain stroke on the VIPPS actuators. CMG: control moment gyroscopes*

### 9.1.6.1 General astrophysics observations

General astrophysics observations using HDI, LUMOS, or POLLUX may begin shortly after a slew is complete. For LUMOS and HDI observations, it may be necessary to execute an additional roll maneuver about the telescope boresight to align the target to the instrument field-of-view, especially within the LUMOS microshutter apertures.

After this roll maneuver, an additional settle period may be necessary. While waiting for this settling to occur, internal instrument calibrations may be performed, if necessary, as well as instrument "set-up," selecting appropriate filters and channels. Once complete, science exposures may begin.

During general astrophysics observations, parallel observation with other instruments is supported. Since each instrument's data storage is self-contained within that instrument's main electronics box, it is possible to have HDI, LUMOS, and POLLUX collecting data simultaneously under the condition that secondary instrument operations do not impact those of the primary instrument. For example, if a specific LUMOS observation is being executed, HDI may collect science data on whatever field it happens to be pointed at, but HDI would not be able to request pointing dithers to fill in focal plane gaps during this time. Similarly, if HDI were operating as prime, LUMOS would not be able to request roll maneuvers to align targets to the microshutter array.

### 9.1.6.2  Solar system observations

Most solar system observations will likely require engaging LUVOIR's target tracking capability (up to 60 mas/s, approximately twice that of JWST). Following a slew to target, the target will first be acquired by HDI using in-field guide stars. A trajectory will be computed, and tracking will begin. For cases in which it is desired to observe the object with LUMOS, ECLIPS, or POLLUX, an additional offset will be computed to place the object in their respective fields-of-view. Depending on the speed and duration of the observation, tracking may be accomplished by the fast steering mirror, VIPPS, gimbal system, spacecraft ACS, or any combination thereof.

### 9.1.6.3  Exoplanet observations

High-contrast exoplanet observations will be the most demanding of LUVOIR's capabilities. Following a slew, a longer settling period on the order of 10 minutes will be required for the active control systems to help stabilize the wavefront error before acquisition of high-contrast images can begin. Once engaged, the "dark-hole" acquisition algorithm will use images from the coronagraph focal planes to generate updates to the deformable mirrors, slowly increasing the contrast in the focal plane to the desired level ($\sim 10^{-10}$ for most observations). Once the contrast is achieved, science observations will begin by taking long integrations to reveal targets of interest. During this time, onboard metrology systems and low-order and out-of-band wavefront sensors will work to maintain the wavefront error and high-contrast image. This entire process is executed autonomously by the onboard control system processor (see **Section 8.2.3.5**).

It is during the dark-hole acquisition process and science integrations that picometer-level wavefront error stability is required. If the wavefront error is drifting faster than the dark-hole wavefront control algorithm can keep up, the system will never achieve $10^{-10}$ contrast. Similarly, once the contrast is achieved, if the system drifts faster than the low-order or out-of-band wavefront sensors can keep up, then the contrast will degrade and long science integrations will not be possible.

Parallel observations are also possible during exoplanet science observations, again assuming the secondary instrument operations do not impact the high-contrast performance. Given the picometer-level stability requirement, this may preclude even changing filters after the initial instrument setup, or the actuation of mechanisms within the other instruments during the exoplanet observation.

### 9.2  Ground systems

The LUVOIR ground segment will be based on the models of other large observatories such as HST and JWST. Primary and backup mission operations centers will be responsible for overall control of the observatory, orbit determination, telemetry, and mission planning. Similarly, a science operations center will be responsible for planning the observation schedule, supporting a guest-observer program, processing science data, and distributing and archiving the data for the user base. Two Ka/S-band ground stations are also baselined to provide two downlink passes per day; a third station would provide additional margin against missed passes due to weather or other conditions but is not strictly necessary to meet mission requirements. The ground stations consist of an existing 18-m dish in White Sands, NM, and a newly constructed 18-m dish in South Africa. Since LUVOIR does not require constant contact, these ground stations could be shared with other missions to reduce operating costs. **Figure 9-4** shows a schematic of the LUVOIR ground gegment.

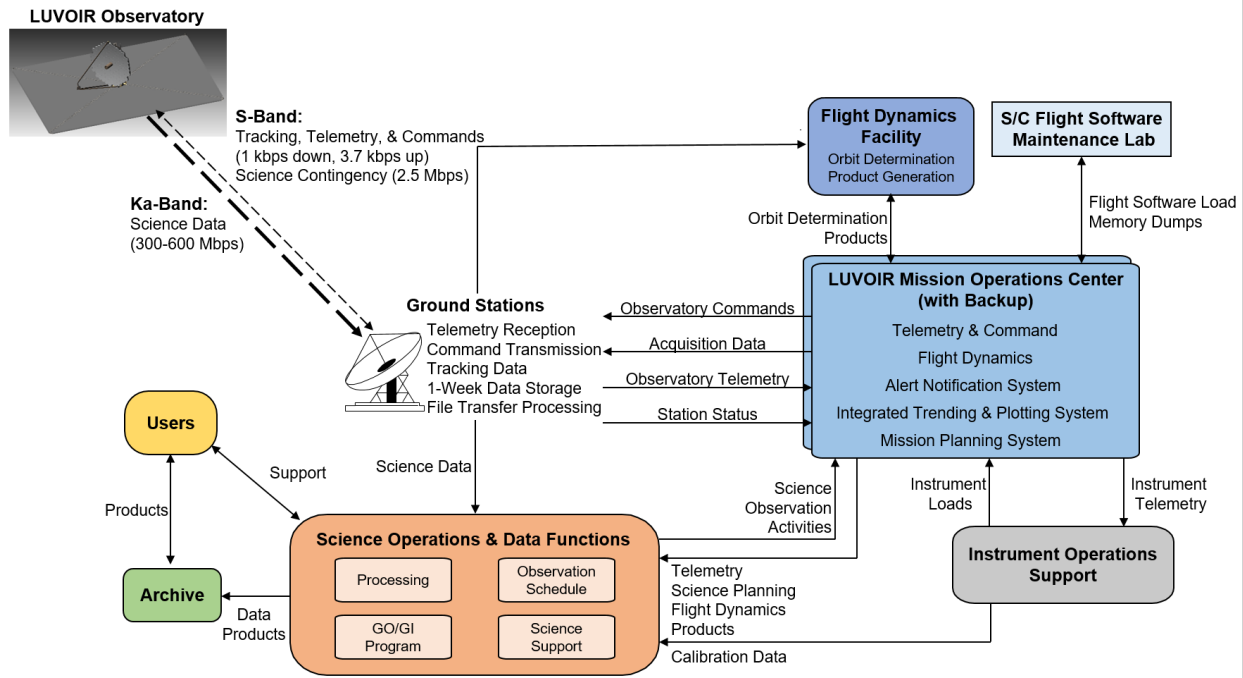

**Figure 9-4.** *LUVOIR ground segment. Primary and backup mission operations centers are responsible for overall control of the observatory. A science operations center is responsible for observation planning, science data distribution, and archiving. Two 18-m ground stations, one at White Sands, NM and one in South Africa, provide adequate coverage for two downlinks per day.*

**Table 9-2.** *Total 48-hour data storage for each instrument. All values include 20% overhead and 30% margin.*

| Instrument | LUVOIR-A [TBITS] | LUVOIR-B [TBITS] |
|---|---|---|
| HDI | 6.5 | 1.6 |
| LUMOS | 1.3 | 0.9 |
| ECLIPS | 1.1 | 1.1 |
| POLLUX | 0.1 | – |
| **Total** | 8.9 | 3.6 |

Each LUVOIR instrument is designed with enough onboard memory to store 48-hours of nominal science data, including 20% overhead and 30% margin. **Table 9-2** summarizes these data volumes. However, we anticipate that LUVOIR will use two ~4-hour contacts per day to transmit science data and telemetry and receive commands. The typical time between downlinks is 8 hours. Assuming file generation at the science operations center takes ~1 hour, the typical data latency is <13 hours (8 hours of science data collection, <4 hours of transmit, 1 hour of file generation). In the event of missed contacts, each instrument has enough onboard storage for 48 hours of data collection, and the maximum possible data latency is 53 hours (48 + 4 + 1).

**Table 9-3** details the downlink budgets for both LUVOIR concepts. With two ground stations, there is a total or 20 hours of available contact time for science data and telemetry

**Table 9-3.** *Downlink budget for both LUVOIR concepts. LUVOIR-A requires 4.2 hours (out of a 20-hour window) to downlink the maximum 48-hour science data volume of 8.9 Tbits. LUVOIR-B requires 3.5 hours (out of a 20-hour window) to downlink the maximum 48-hour science data volume of 3.6 Tbits. QPSK: quadrature phase shift keying; LDPC: low density parity check; HPBW: half power beam width; EIRP: equivalent isotropically radiated power; Eb/No: normalized signal-to-noise ratio*

| | | **LUVOIR-A** | | | **LUVOIR-B** | | |
|---|---|---|---|---|---|---|---|
| Scenario: | | Ka-band Science | S-band Science Contingency | S-band Telemetry | Ka-band Science | S-band Science Contingency | S-band Telemetry |
| Link Type: | | Space-Ground (Downlink) | | | Space-Ground (Downlink) | | |
| Modulation Scheme: | | QPSK | | | QPSK | | |
| Coding: | | LDPC Rate 1/2 | | | LDPC Rate 1/2 | | |
| Bit-Error-Rate: | | $10^{-6}$ | | | $10^{-6}$ | | |
| 1/2 Angle HPBW (deg): | | 0.22 | 2.591 | 90 | 0.22 | 2.591 | 90 |
| Polarization: | | Circular | | | Circular | | |
| Contact Time (hrs): | | 4.2 | 19 | 4.2 | 3.5 | 19 | 3.5 |
| Transmit Frequency | MHz | 26,500 | 2,250 | 2,250 | 26,500 | 2,250 | 2,250 |
| Transmitter Power | W | 100 | 20 | 20 | 100 | 20 | 20 |
| Transmitter Power - dB | dBW | 20 | 13 | 13 | 20 | 13 | 13 |
| Transmitter Antenna Diameter | m | 1.80 | 1.80 | 0.04 | 1.22 | 1.22 | 0.04 |
| Transmitter EIRP | dBWi | 67.87 | 39.86 | 3.91 | 64.46 | 36.45 | 3.91 |
| Min. Elevation Angle | deg | 20 | 10 | 10 | 20 | 10 | 10 |
| Slant Range | km | 1,800,000 | 1,800,000 | 1,800,000 | 1,800,000 | 1,800,000 | 1,800,000 |
| Free Space Loss | dB | 246.0 | 224.6 | 224.6 | 246.0 | 224.6 | 224.6 |
| Total Propagation Effects | dB | 1.879 | 0.359 | 0.359 | 1.879 | 0.359 | 0.359 |
| Power Received | dB | -180.1 | -185.2 | -221.2 | -183.5 | -188.6 | -221.2 |
| Receiver Antenna Diameter | m | 18.3 | 18.3 | 18.3 | 18.3 | 18.3 | 18.3 |
| Received Carrier to Noise Density | dB | 96.77 | 73.86 | 37.91 | 93.36 | 70.45 | 37.91 |
| Information (Data) Rate | Mbps | 600 | 2.5 | 0.001 | 300 | 1.8 | 0.001 |
| Total Data Transmitted per Contact | Gbits | 9,072 | 171 | 0.015 | 3,780 | 123 | 0.013 |
| Total Information (Data) Rate | dB-bps | 87.78 | 63.98 | 30.00 | 84.77 | 62.55 | 30.00 |
| Received Eb/No | – | 8.99 | 9.89 | 7.91 | 8.59 | 7.90 | 7.91 |
| Implementation Loss | dB | 3 | 3 | 3 | 3 | 3 | 3 |
| Required Eb/No At Decoder | dB | 1.89 | 1.89 | 1.89 | 1.89 | 1.89 | 1.89 |
| Margin | dB | 4.10 | 5.00 | 3.02 | 3.70 | 3.01 | 3.02 |

per day. The link budgets show that both LUVOIR concepts can downlink two-day's worth of science data in approximately four hours of contact time using the primary Ka-band science channel, leaving ample margin for missed contacts or future data volume growth with updated instruments. If needed, science data can also be downlinked via an S-band contingency channel. Under this contingency condition, LUVOIR-A can transmit 171 Gbits of critical science data, and LUVOIR-B can transmit 123 Gbits using 19 hours out of the 20-hour contact window. **Table 9-4** shows the S-band command uplink budget for each LUVOIR concept.

**Table 9-4.** *Uplink budget for both LUVOIR concepts. QPSK: quadrature phase shift keying; LDPC: low density parity check; HPBW: half power beam width; EIRP: equivalent isotropically radiated power; Eb/No: normalized signal-to-noise ratio*

| | | LUVOIR-A | LUVOIR-B |
|---|---|---|---|
| Scenario: | | S-band Command | S-band Command |
| Link Type: | | Ground-Space | Ground-Space |
| Modulation Scheme: | | QPSK | QPSK |
| Coding: | | LDPC Rate 1/2 | LDPC Rate 1/2 |
| Bit-Error-Rate: | | $10^{-6}$ | $10^{-6}$ |
| 1/2 Angle HPBW (deg): | | 90 | 90 |
| Polarization: | | Circular | Circular |
| Contact Time (hrs): | | 4.2 | 3.5 |
| Transmit Frequency | MHz | 2,100 | 2,100 |
| Transmitter Power | W | 310 | 310 |
| Transmitter Power - dB | dBW | 25 | 25 |
| Transmitter Antenna Diameter | m | 18.3 | 18.3 |
| Transmitter EIRP | dBWi | 72.64 | 72.64 |
| Min. Elevation Angle | deg | 10 | 10 |
| Slant Range | km | 1,800,000 | 1,800,000 |
| Free Space Loss | dB | 224.0 | 224.0 |
| Total Propagation Effects | dB | 0.350 | 0.350 |
| Power Received | dB | -151.8 | -151.8 |
| Receiver Antenna Diameter | m | 0.01 | 0.01 |
| Received Carrier to Noise Density | dB | 43.62 | 43.62 |
| Information (Data) Rate | Mbps | 0.004 | 0.004 |
| Total Data Transmitted per Contact | Gbits | 56.0 | 46.6 |
| Total Information (Data) Rate | dB-bps | 35.68 | 35.68 |
| Received Eb/No | – | 7.94 | 7.94 |
| Implementation Loss | dB | 3 | 3 |
| Required Eb/No At Decoder | dB | 1.89 | 1.89 |
| Margin | dB | 3.05 | 3.05 |

## CHAPTER 10.  LAUNCH VEHICLE

While the science priorities for NASA over the next decade and beyond will come into focus following the Astro2020 Decadal Survey, project teams will be left with uncertainty with respect to the capabilities of potential launch vehicles. There are ongoing studies to understand how large telescopes like LUVOIR might someday be assembled in space. However, until such time as facilities exist to allow for assembly in space, the most efficient and most economical method of putting a large aperture telescope into space is a traditional launch vehicle.

The selection of a launch vehicle imposes constraints on the size and the mass of the observatory and, as with all technologies, the long-term landscape for launch vehicle options is not yet mature. This is one of the reasons why our team studied two distinct conceptual designs that bookend a breadth of options that could be implemented. LUVOIR is designed to be scalable based on the capability of the launch vehicles that will be available should LUVOIR's science portfolio be prioritized.

Initially, the LUVOIR Team levied a goal that LUVOIR-B should fit into a "conventional" heavy-lift vehicle. By "conventional," the team implied a launch vehicle similar to the existing United Launch Alliance Delta IV-Heavy vehicle with a 5-m x 19.8-m fairing, and a lift capability to the second Sun-Earth Lagrange point (SEL2) of ~10,000 kg. However, it became clear during the engineering process that LUVOIR-B would require a lift capacity greater than the ~10,000 kg of the Delta IV to maintain a healthy science portfolio. Mass aside, LUVOIR-B does still fit volumetrically within the smaller, 5-m-class fairing.

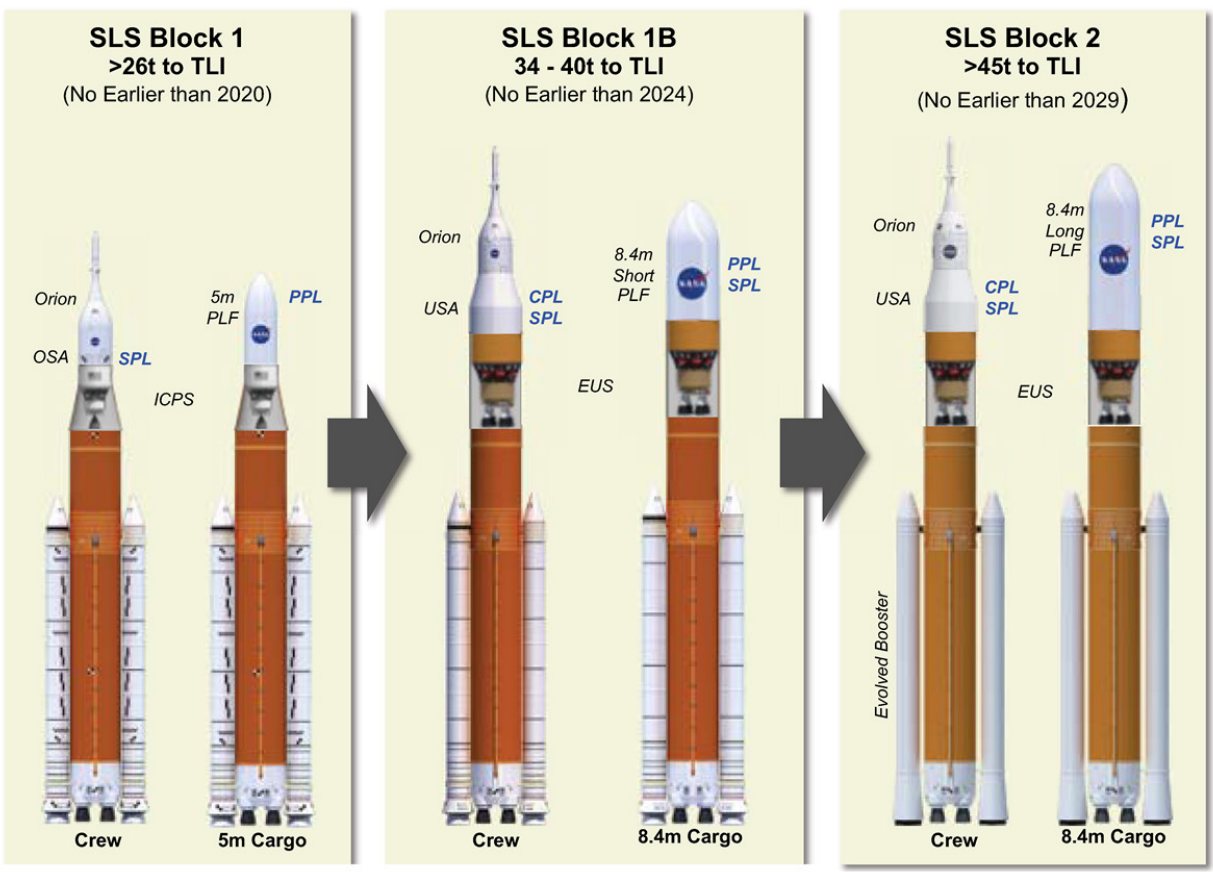

**Figure 10-1.** *Various configurations of NASA's Space Launch System. TLI: trans-lunar insertion; ICPS: interim cryogenic propulsion stage; OSA: Orion stage adapter; SPL: secondary payload; PLF: payload fairing; PPL: primary payload; CPL: co-manifested payload; EUS: exploration upper stage; USA: universal stage adapter*

## 10.1  Baseline launch vehicle

NASA's Space Launch System (SLS) is the baseline launch vehicle for LUVOIR. As shown in **Figure 10-1**, the SLS comes in a variety of configurations that are expected to come online over time; each of the LUVOIR concepts uses a different configuration, shown in **Figure 10-2.** The baseline launch vehicle for LUVOIR-A is the SLS Block 2 cargo configuration. The baseline launch vehicle for LUVOIR-B is the SLS Block 1B cargo configuration.

As the SLS continues to develop, the LUVOIR Team has received regular reports from the SLS team on their progress. As of May 2019, they indicated that all components for the Block 1B configuration were at a critical design review level of development and all components except the fairing were under procurement. The delay with getting the fairing under contract is a funding, not technical, delay. The SLS team is still determining which fairing sizes will ultimately be of the most interest to potential customers.

The SLS team expects the first launch of a Block 1 configuration to be in 2020 with the Block 1B launching and flying regularly in the late 2020s. The Block 2 configuration is expected to be operational in the early 2030s.

**Figure 10-3** through **Figure 10-6** show both LUVOIR-A and LUVOIR-B from a variety of angles in the stowed configuration.

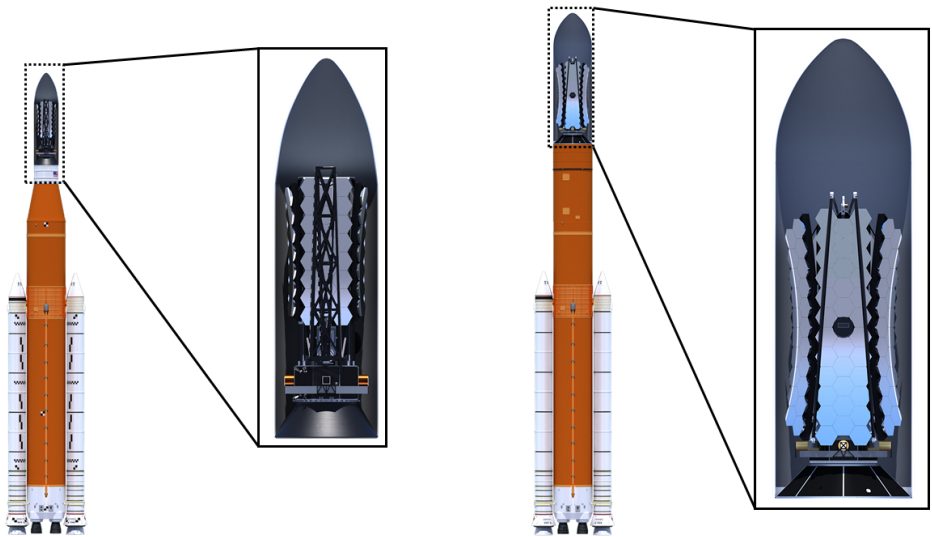

**Figure 10-2.** *LUVOIR shown in the SLS. LUVOIR-A (right) would use the SLS Block 2 cargo and LUVOIR-B (left)would use the SLS Block 1B cargo. Credit: NASA/SLS*

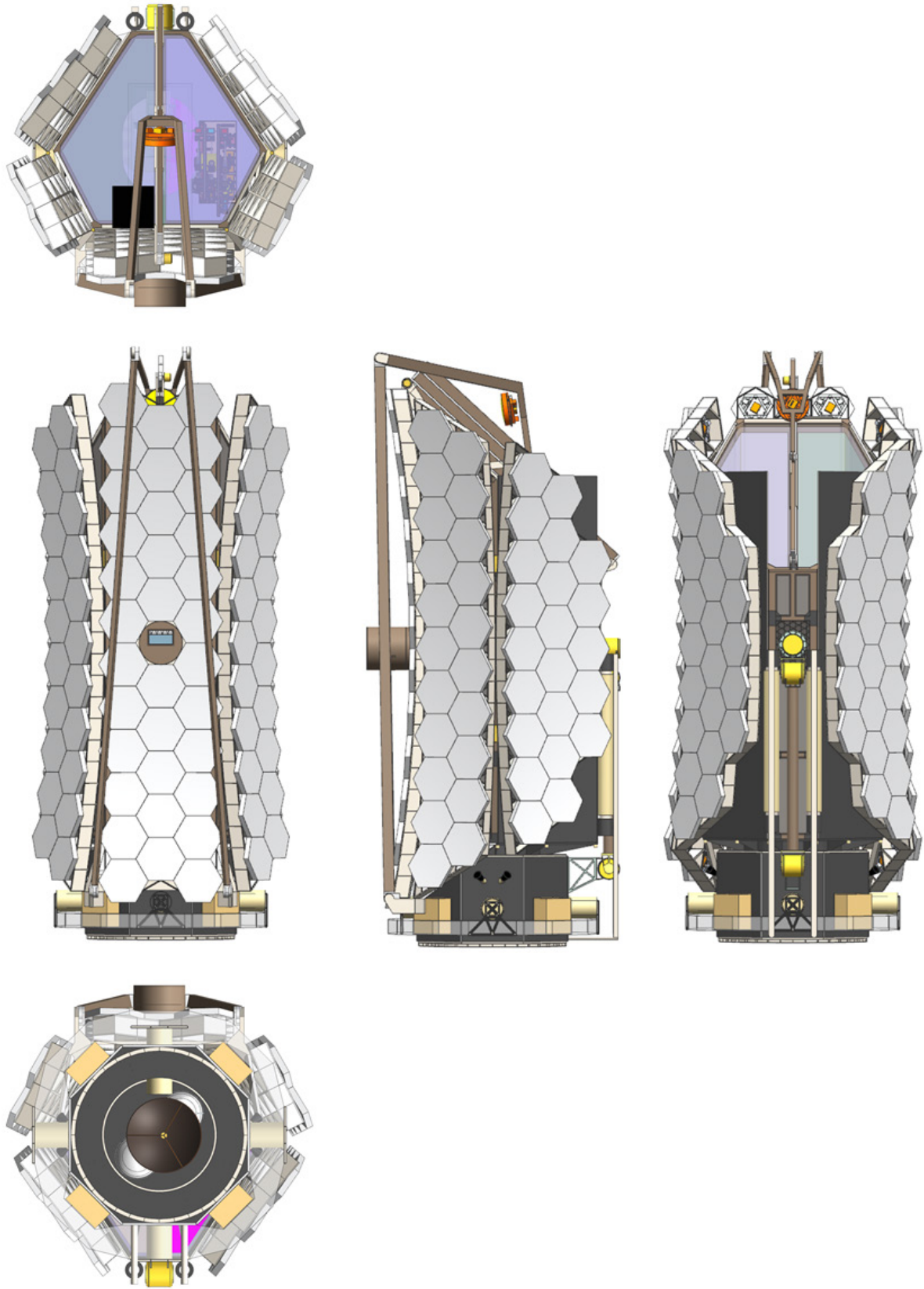

**Figure 10-3.** *Various views of LUVOIR-A in the stowed configuration*

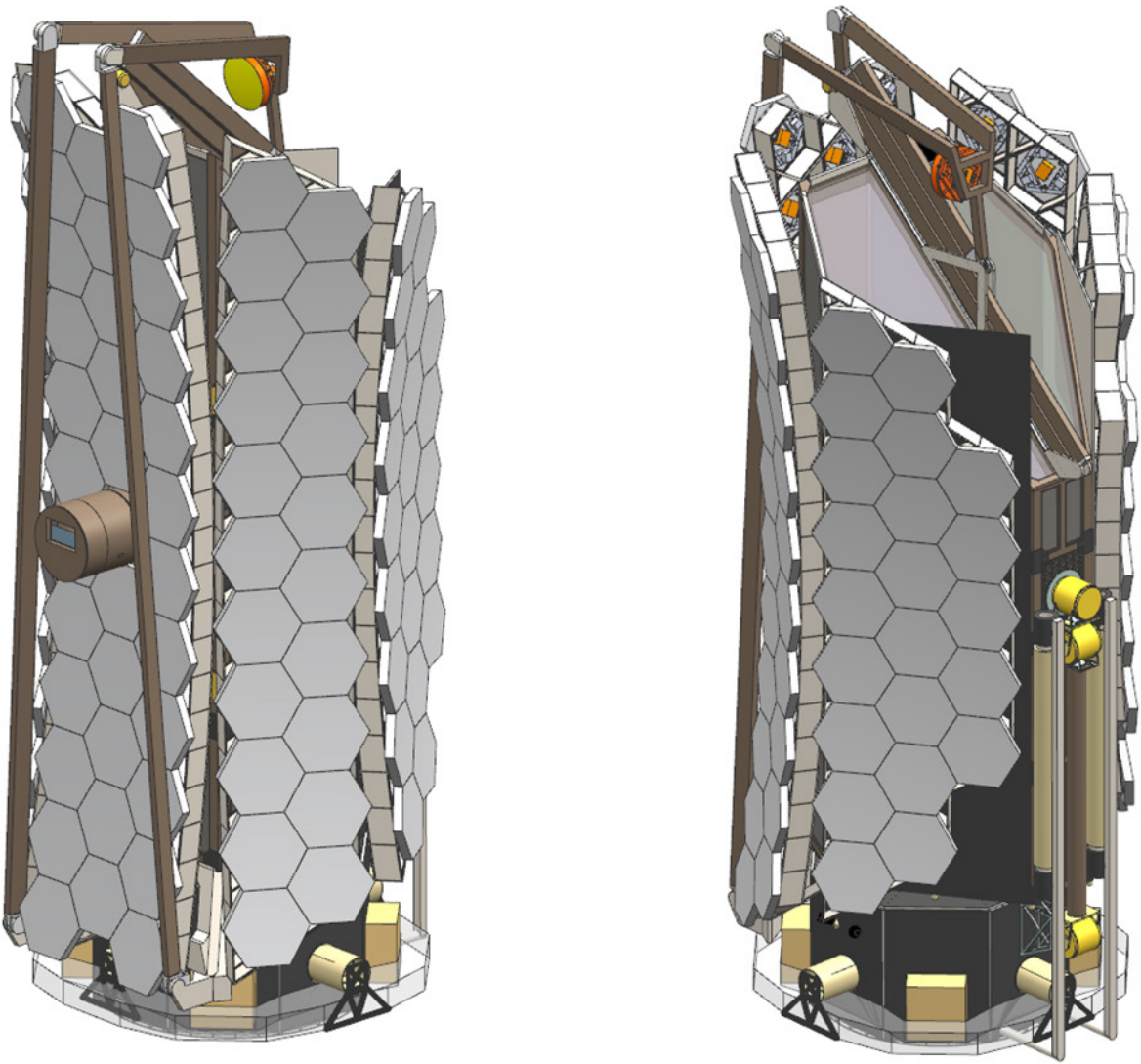

**Figure 10-4.** *Isometric views of LUVOIR-A in the stowed configuration.*

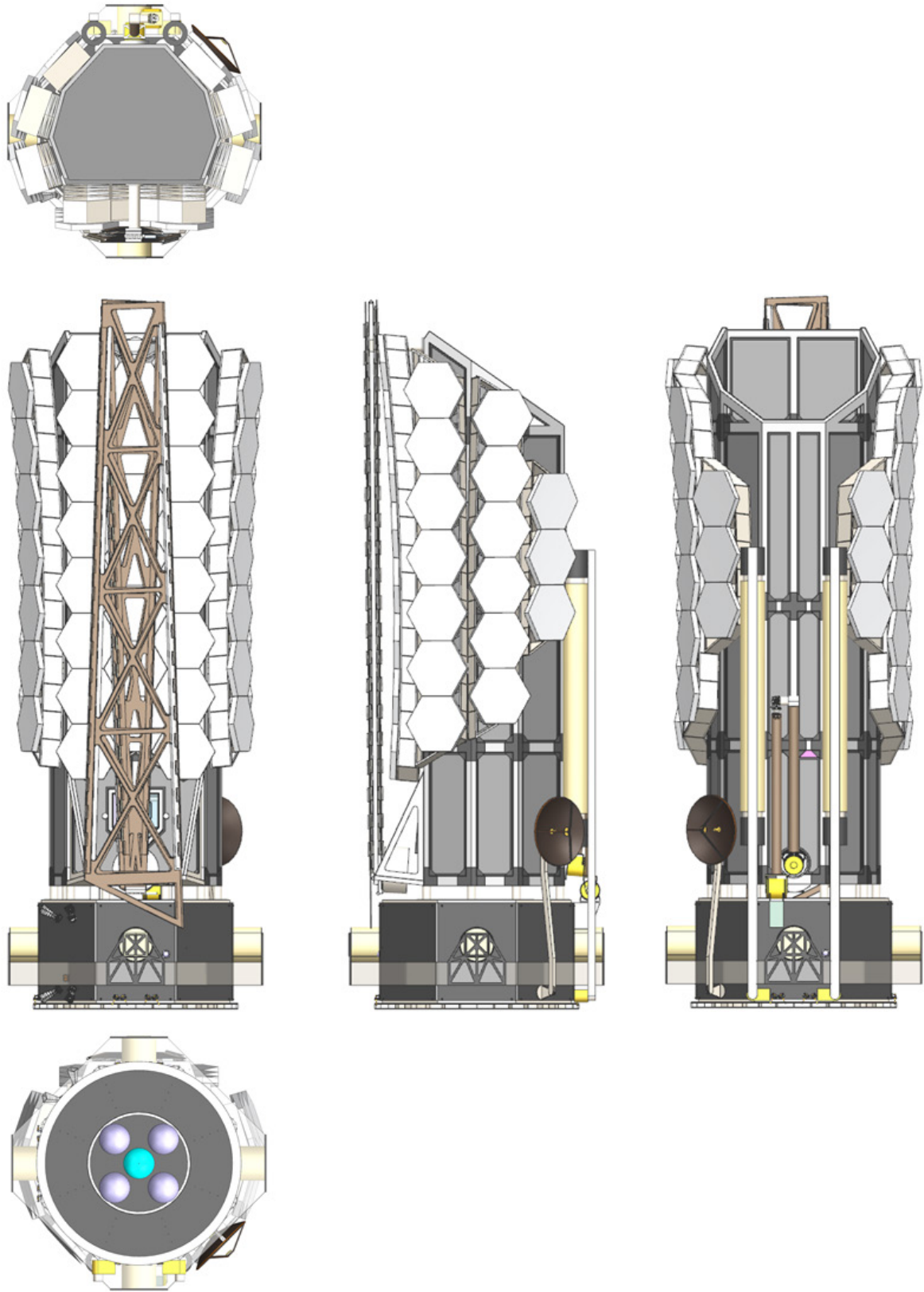

**Figure 10-5.** *Various views of LUVOIR-B in the stowed configuration*

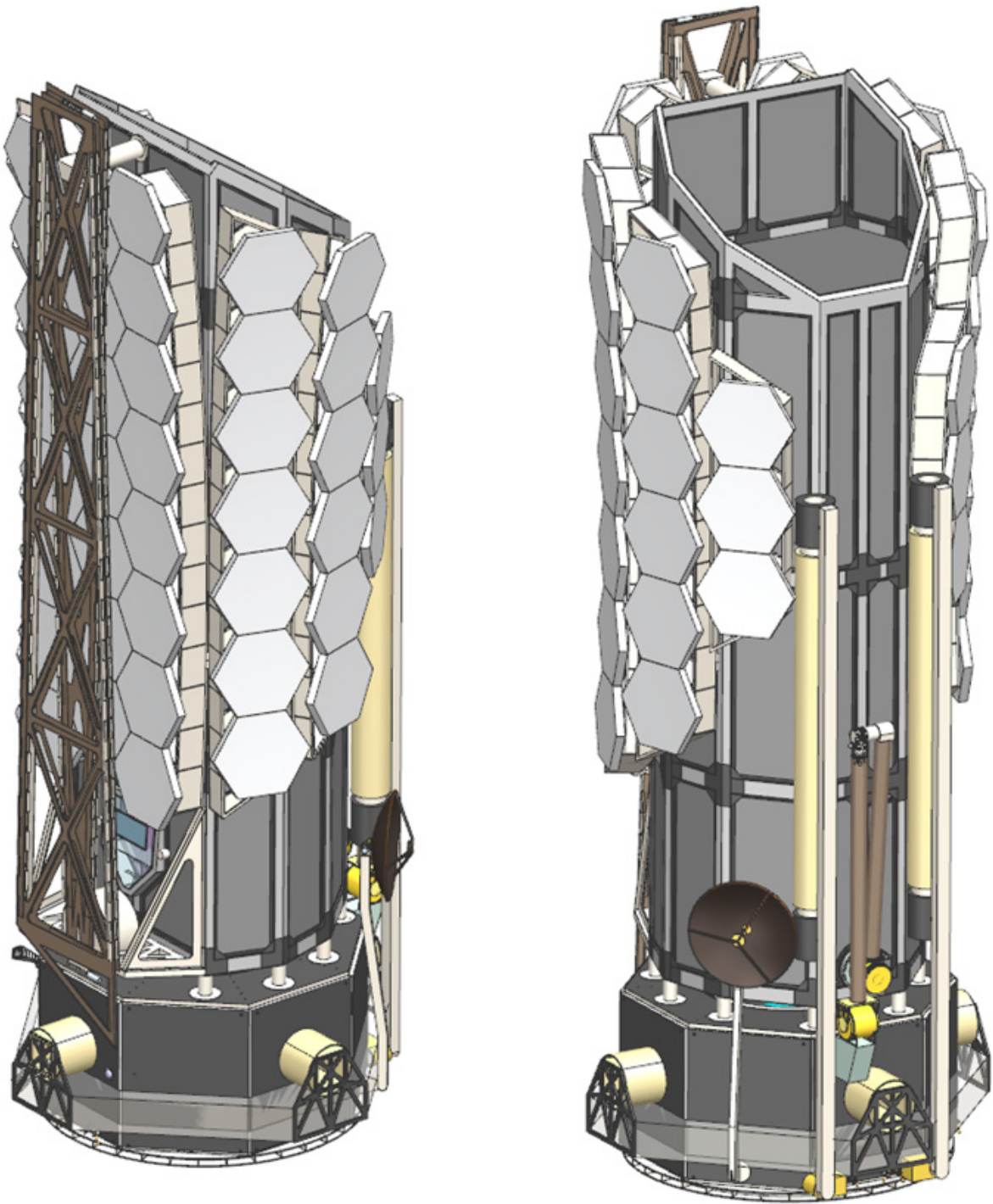

**Figure 10-6.** *Isometric views of LUVOIR-B in the stowed configuration*

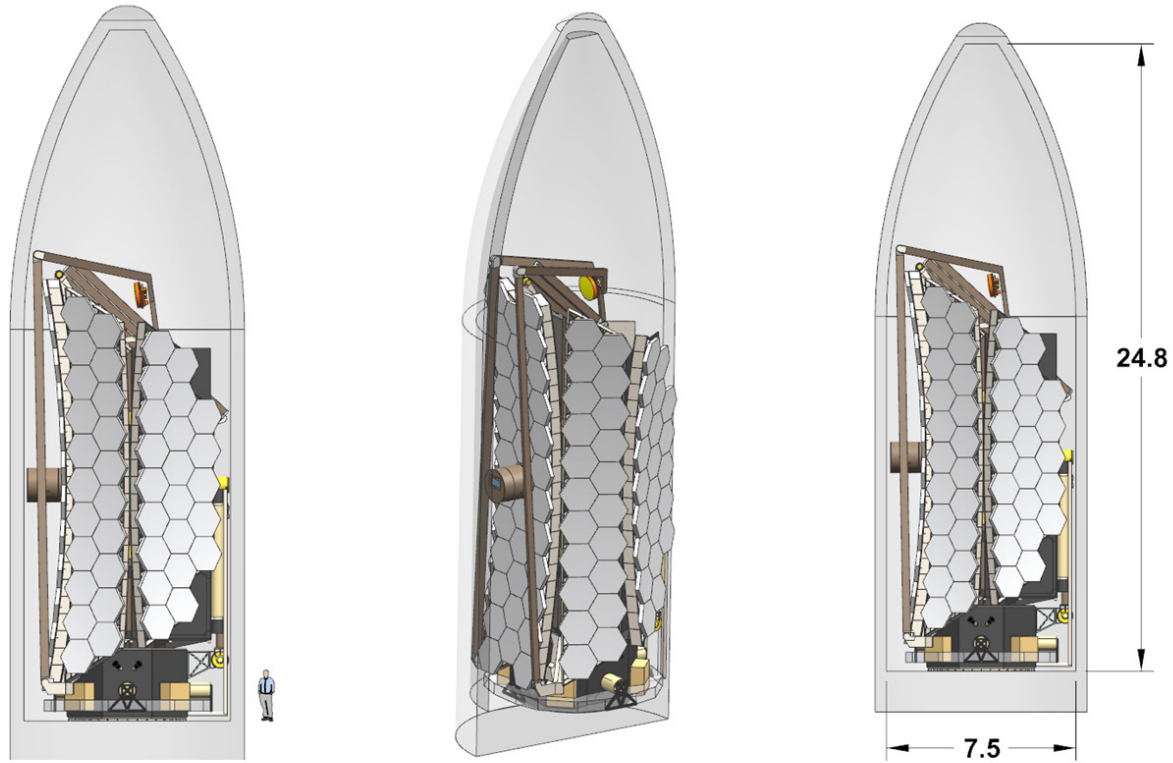

**Figure 10-7.** *LUVOIR-A in the SLS 8.4-m fairing*

**Figure 10-7** and **Figure 10-8** show LUVOIR stowed in both the SLS 8.4-m fairing as well as a "conventional" 5 m fairing.

## 10.2 Alternate launch vehicles
LUVOIR has been designed to the SLS vehicle to demonstrate an observatory and spacecraft design that closes. However, the future landscape for launch vehicles should provide more options with the advent of commercial launch vehicles.

### 10.2.1 SpaceX Starship
The SpaceX Starship is a launch vehicle in the preliminary design phase. As such, there are not yet many details publicly available. However, the LUVOIR Team has communicated with representatives from SpaceX and performed a preliminary assessment of the compatibility of LUVOIR with Starship.

SpaceX has indicated that the Starship will be able to lift as much as 150,000 kgs to SEL2. This incredible capacity is enabled by launching that mass first into low earth orbit and then refueling a booster for transfer to other orbits. The final fairing dimensions are still being determined but SpaceX did conduct a preliminary analysis of a fairing whose shape was altered to fit LUVOIR-A (based on this study's final concept models) and they reported that it was a viable option. Without modification, LUVOIR-B can fit into the currently planned Starship fairing with room to spare as shown in **Figure 10-9**.

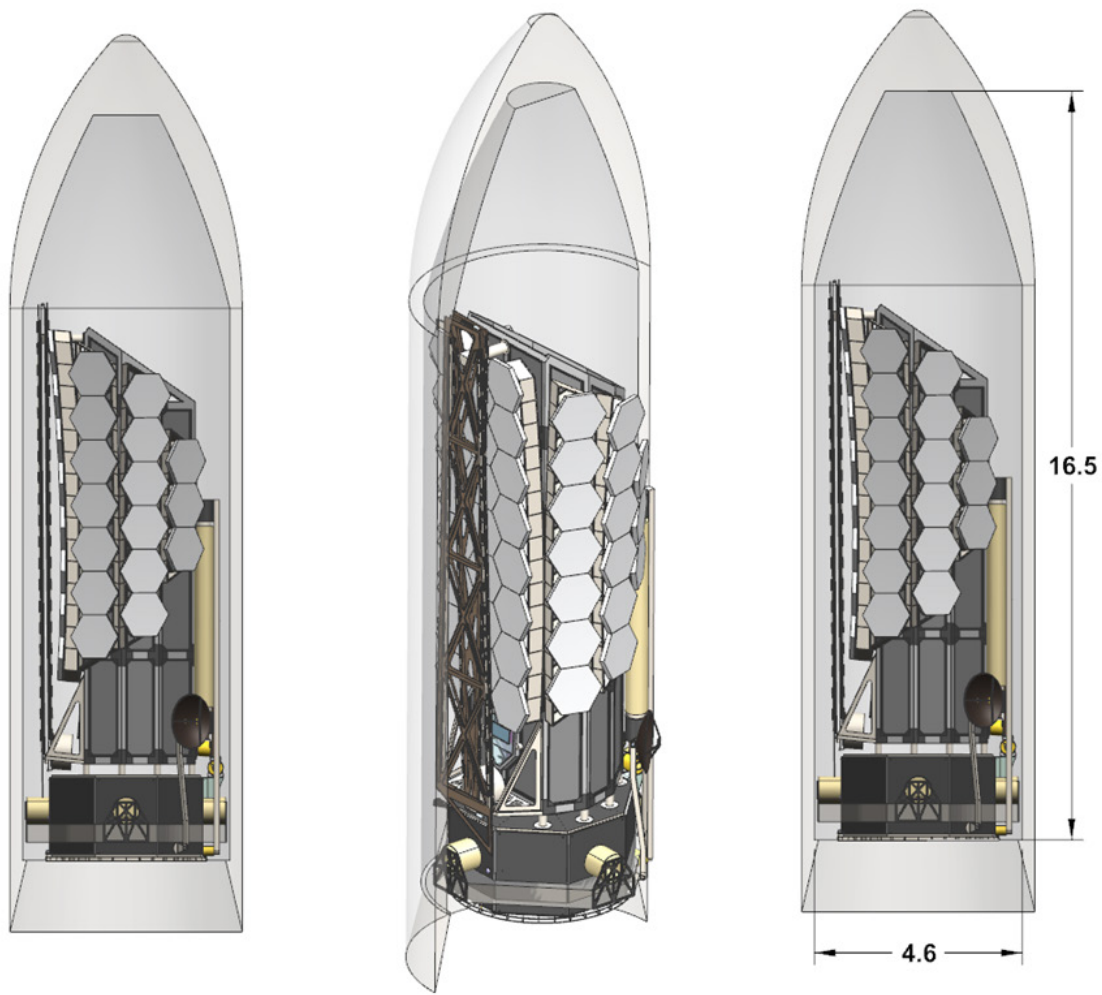

**Figure 10-8.** *LUVOIR-B in a "conventional" 5-m fairing*

## 10.2.2  Blue Origin New Glenn

The Blue Origin New Glenn launch vehicle may be another option for the smaller version of LUVOIR. While details of the New Glenn's performance are also not publically available, the LUVOIR Team has had several conversations with representatives from Blue Origin to determine compatibility of LUVOIR with the New Glenn.

The New Glenn passed its preliminary design review (PDR) in December, 2018. In 2019, Blue Origin expects to qualify the engines for the launch vehicle. They are also completing the construction of their Florida facility and expect to start to populate it with tooling late in the year. Also later in 2019, Blue Origin is scheduled to conduct the critical design review of the launch vehicle.

Blue Origins is already under contract to launch several privately-owned satellites using the New Glenn. They have also been awarded funding towards the development of more powerful engines.

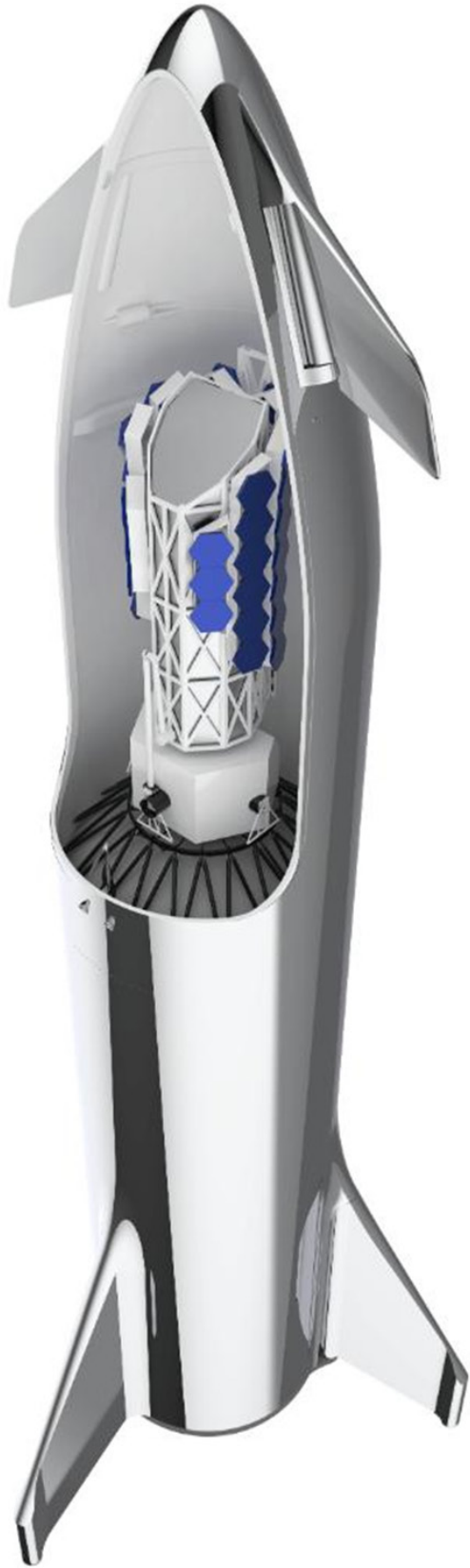

**Figure 10-9.** *LUVOIR-B shown inside the nominal starship fairing. With small changes to the fairing, LUVOIR-A will also fit inside the SpaceX Starship. Credit: SpaceX.*

**Table 10-1.** *Launch vehicle compatibility with each LUVOIR concept. Mass compatibility implies that the launch vehicle can lift the current maximum permissible value (MPV) mass for each LUVOIR concept to SEL2. Volume compatibility implies that the stowed LUVOIR concept fits within the static envelope of the launch vehicle's fairing.*

| | SLS Block 1 | | SLS Block 1B | | SLS Block 2 | | SpaceX Starship | | Blue Origin New Glenn | |
|---|---|---|---|---|---|---|---|---|---|---|
| | **Mass** | **Volume** | **Mass** | **Volume** | **Mass** | **Volume** | **Mass** | **Volume** | **Mass** | **Volume** |
| **LUVOIR-A** | No | No | Yes* | Yes | Yes | Yes | Yes | Yes** | No | No |
| **LUVOIR-B** | Yes | Yes | Yes | Yes | Yes | Yes | Yes | Yes | Yes | Yes |

*Pending improved engine performance

**Pending minor modifications to Starship cargo fairing

It is now clear that the New Glenn fairing is large enough to fit LUVOIR-B with room to spare. However, the exact lift capacity of New Glenn is unknown because it is currently being certified by the NASA Launch Services Program. Blue Origins expects the lift capacity to increase over time once it starts launching and margins can be reduced. Based on preliminary evaluations, the New Glenn may be another alternative launch vehicle for LUVOIR-B.

## 10.3 Summary and acknowledgements

**Table 10-1** summarizes the launch vehicle compatibility with each LUVOIR concept. By the time LUVOIR launches in 2039, either version of LUVOIR will have several options available to it for a launch vehicle.

Some of the launch vehicle information presented was provided to the LUVOIR team via private communications with representatives from SpaceX, Blue Origin, and NASA's Launch Services Program. We are grateful for their collaboration.

# CHAPTER 11.  TECHNOLOGY DEVELOPMENT

Three technology systems enable the LUVOIR science objectives discussed earlier in this report, and require development to technology readiness level (TRL) 6 prior to the start of Phase A. The high-contrast coronagraph instrument is the technology system that enables LUVOIR's exoplanet science. The ultra-stable segmented telescope system includes technologies that enable picometer-level wavefront and contrast stability during the exoplanet science observations. Finally, the ultraviolet instrumentation technology system includes technologies to enable LUVOIR's far- and near-UV observations as part of the general astrophysics and solar-system science objectives.

Each of the technology systems comprise the individual technology components that must be developed. **Table 11-1** through **11-3** list the individual technology components, their state-of-the-art, and the capability needed for each of the three technology systems. When concept studies like LUVOIR count the number of technologies at a given TRL, it is traditionally only the technology components, such as actuators, mirrors, or detectors that are considered. However, it is not sufficient to develop technology components only. Instead, complete technology systems must be demonstrated to achieve the necessary performance in the context of a specific architecture. These system-level demonstrations often require supporting engineering and manufacturing development activities to ensure a cost-effective implementation.

It is important to note that the high-contrast coronagraph instrument and ultra-stable segmented telescope technology systems are tightly coupled; the performance capabilities of one system directly affect the requirements of the other. For example, while all high-contrast coronagraphs require picometer-level wavefront stability, different types of coronagraphs are more or less sensitive to certain wavefront spatial frequencies. The selection of a particular coronagraph architecture will drive the ultra-stable segmented telescope design. Conversely, the limiting performance of the telescope system may require the prioritization of one coronagraph design over another.

Similarly, the ultraviolet instrument technology system is loosely coupled to the other two technology systems. Coating polarization properties, uniformity, and reflectivity across the entire wavelength band impact coronagraph contrast and throughput, and therefore exoplanet yield. Conversely, the telescope optical design will affect throughput and optical performance in the ultraviolet instruments, LUMOS and HDI.

This coupling between technology systems, shown in **Figure 11-1**, requires continuous cross-validation between the technology development paths. It is therefore critical that the technology development plan be executed in parallel with a detailed Pre-Phase A architecture study, as discussed in **Section 12.2**. The architecture study provides the necessary systems-level perspective to provide cross-validation of the technologies' performance. The architecture study can also further improve technology development efficiency by determining which development paths are no longer viable in the larger context of the system architecture, or when a targeted influx of resources may result in a technology breakthrough that enables or enhances a portion of the architecture. We recommend that a Pre-Phase A project office be formed to coordinate the architecture study and technology development activities and direct resources to each as appropriate.

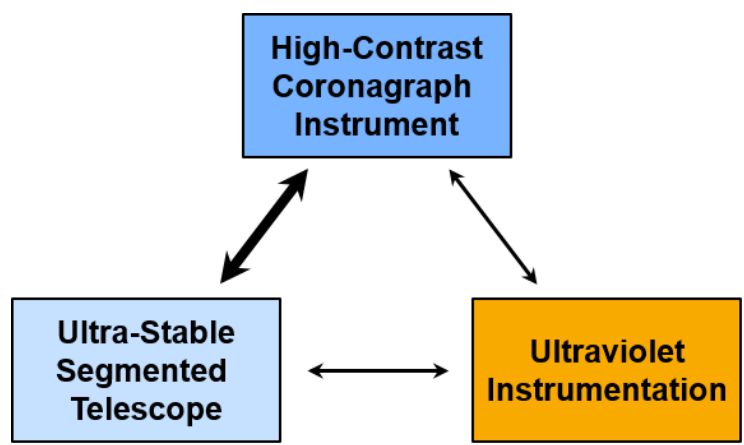

**Figure 11-1.** *The coronagraph and ultra-stable telescope systems are tightly coupled, and both are loosely coupled to the ultraviolet instrumentation system. This coupling requires continuous cross-validation to understand the impacts that any one system has on the other two.*

**Table 11-1.** *Technology components in the high-contrast coronagraph instrument technology system.*

| Section | Technology Component | Implementation Options | State of the Art | Capability Needed | FY19 TRL | In LUVOIR Baseline? |
|---------|---------------------|------------------------|------------------|-------------------|----------|---------------------|
| 12.2.1.3 | Coronagraph Architecture | Apodized Pupil Lyot Coronagraph (APLC) | $6.3 \times 10^{-6}$ over 6% bandpass in air. Validated models with WFIRST CGI SPC demonstrations | $1 \times 10^{-10}$ raw contrast<br>>10% bandpass<br><4 λ/D inner working angle<br>64 λ/D outer working angle<br>Robust to stellar diameter and jitter | 4 | ✓ |
| | | Vortex Coronagraph (VC) | $8.5 \times 10^{-9}$ contrast over 10% band with unobscured pupil. SCDA modeling for unobscured, segmented pupil | | 3 | ✓ |
| | | Phase-Induced Amplitude Apodization (PIAA) | SCDA modeling results for unobscured, segmented pupil | | 3 | |
| | | Hybrid-Lyot Coronagraph (HLC) | $3.6 \times 10^{-10}$ contrast over 10% band in DST. SCDA modeling for unosbcured segmented pupil | | 3 | |
| | | Nulling Coronagraph (NC) | $5 \times 10^{-9}$ narrowband at 2.5 λ/D | | 3 | |
| 12.2.1.4 | Deformable Mirrors | Micro-Electro-Mechanical Systems (MEMS) | Available up to 64 x 64 actuators; $8.5 \times 10^{-9}$ contrast demonstrated with 32 x 32 actuators | 128 x 128 actuators<br>Stable actuators (low creep)<br>Diffraction-limited surface quality (< 3 nm surface roughness) | 4 | ✓ |
| | | Lead-Magnesium-Niobate (PMN) Macro-scale | $<1 \times 10^{-8}$ contrast demonstrated with 48 x 48 actuator Xinetics DMs (WFIRST CGI Testbed) | | 5 | |
| 12.2.1.7 | Wavefront Sensing | Out-of-band Wavefront Sensing | Model predicting <10 pm residual error with nonlinear ZWFS, Mv = 5 source | Wavefront stabilty ~10 pm RMS<br>~1 Hz bandwidth with Mv < 9 source<br>Able to capture wavefront spatial frequencies on the order of segment-to-segment drift and DM actuators | 3 | ✓ |
| | | Low-order Wavefront Sensing | <0.36 mas RMS line-of-sight residual error; <30 pm RMS focus, Mv = 5 source (WFIRST CGI Testbed) | | 6 | ✓ |
| | | Artificial Guide Star | Concept study for guide star spacecraft and wavefront sensing control loop completed. | | 3 | |
| 12.2.1.10 | UV/VIS Low-noise Detector | Electron-Multiplying CCD | 1k x 1k WFIRST Detector:<br>$7 \times 10{-}5$ e-pix/s dark current<br>0 e- read noise<br>$2.3 \times 10{-}3$ CIC | $3 \times 10^{-5}$ e-/pix/s dark current<br>0 e- read noise<br>$1.3 \times 10^{-3}$ e-/pix CIC<br>>80% QE at all detection wavelengths<br>4k x 4k array size | 4 | ✓ |
| | | Hole-Multiplying CCD | Prototype devices fabricated with gains > 10x (>20x in at least one device) | | 3 | |
| 12.2.1.12 | NIR Low-noise Detector | HgCdTe Photodiode Array | H4RG-10 currently meets needed capability @ 170 K | $< 1 \times 10^{-3}$ e-/pix/s dark current<br>< 3e- read noise<br>4k x 4k array size | 5 | ✓ |
| | | HgCdTe Avalanche Photodiode | $1.5 \times 10^{-3}$ e-/pix/s dark current<br>< 1 e- read noise<br>320 x 256 array size<br>Requires < 100 K temperatures | | 4 | |

**Table 11-2.** *Technology components in the ultra-stable segmented telescope technology system.*

| Section | Technology Component | Implementation Options | State of the Art | Capability Needed | FY19 TRL | In LUVOIR Baseline? |
|---------|---------------------|------------------------|------------------|-------------------|----------|---------------------|
| 12.2.2.4 | Mirror Substrate | Closed-back ULE (rigid body actuated) | 7.5 nm RMS surface figure area with no actuated figure correction | ~5 nm RMS surface figure error<br>> 400 Hz first free mode<br>19 kg/m² areal density | 5 | ✓ |
| | | Closed-back ULE (surface figure actuated) | < 200 Hz first free mode<br>~10 kg/m² areal density | | 4 | |
| | | Open-back Zerodur (rigid body actuated) | Meets wavefront error requirement, but first mode and areal density are challenges | | 4 | |
| 12.2.2.6 | Actuators | Combined piezo/mechanical | JWST mechanical actuators; Off-the-shelf PZT actuator with 5 pm resolution | > 10 mm stroke<br>< 10 pm resolution<br>< 1 pm / 10 min creep<br>Long lifetime | 3 | ✓ |
| | | All-piezo | 20 mm travel with 5 nm coarse resolution and 5 pm fine resolution | | 3 | |
| 12.2.2.8 | Edge Sensors | Capacitive | 5 pm in gap dimension, 60 Hz readout | <4 pm sensitivity at 50–100 Hz rate (control bandwidth of 5–10 Hz) | 3 | ✓ |
| | | Inductive | 1 nm / sqrt(Hz) for 1–100 Hz in shear; 100 nm / sqrt(Hz) for 1–10 Hz in gap | | 3 | |
| | | Optical | 20 pm / sqrt(Hz) up to 100 Hz | | 3 | |
| | | High-speed Speckle Interferometry | < 5 pm RMS at kHz rates; requires center-of-curvature location and high-speed computing | | 3 | |
| 12.2.2.9 | Laser Metrology | Laser truss with phasemeter electronics | Planar lightwave circuit; 0.1 nm gauge error; LISA-Pathfinder heritage laser | < 100 pm sensitivity at 10 Hz rate (control bandwidth of 1 Hz) | 4 | ✓ |
| 12.2.2.11 | Vibration Isolation | Non-contact Isolation System | > 40 dB transmissiability isolation > 1 Hz; Requires electronics development and performance validation | > 40 dB transmissiability isolation > 1 Hz | 4 | ✓ |

Finally, as we discuss in **Chapter 12**, it is critical to retire technology development risk as early in the project lifecycle as possible. Current NASA guidance suggests that all technologies be at a minimum of TRL 6 prior to the mission preliminary design review (PDR). We believe this is too late for a project as large and complex as LUVOIR. By the time the mission has reached PDR, many of the sub-system assemblies and components have already been fabricated and integrated. Uncovering a technology shortfall that late would require significant cost and schedule overruns as redesigns would ripple through the subsystems. Therefore, it is a primary goal of this technology development plan to have all technologies at TRL 6 prior to the start of Phase A. This ensures that all of the technologies' performance is well understood before detailed design begins. Of course, we recognize that as designs mature, it may be necessary to re-validate some technologies during Phases A and B. Therefore, we recommend all technology models and testbeds be maintained through this time.

In the following sections, we discuss the technology development plan for each technology system. We identify the individual technology components that make up the system and distinguish them from heritage hardware (**Section 11.1**). We outline specific development activities to mature those technology components, and critical assembly-, sub-system, and system-level demonstrations (**Section 11.2**). We also identify significant engineering and manufacturing development activities that are required to enable the demonstrations.

**Table 11-3.** *Technology components in the ultraviolet instrumentation technology system.*

| Section | Technology Component | Implementation Options | State of the Art | Capability Needed | FY19 TRL | In LUVOIR Baseline? |
|---------|---------------------|------------------------|------------------|-------------------|----------|---------------------|
| 12.2.3.4 | Far-UV Broadband Coating | Al + eLiF + MgF$_2$ | Meets performance requirements, but requires demonstration on meter-class optics; requires validation of uniformity, repeatability, environmental stability | >50% reflectivity (100–115nm) | 3 | ✓ |
| | | Al + eLiF + AlF$_3$ | | >80% reflectivity (115–200nm) >88% reflectivity (200–850nm) >96% refelctivity (> 850nm) | 3 | |
| | | Al + eLiF | Meets performance requirements, but is environmentally unstable | <1% reflectance nonuniformity (over entire primary mirror) over corongraph bandpass (200–2000 nm) | 5 | |
| 12.2.3.6 | Microshutter Arrays | Next-gen Electrostatic Microshutter Arrays | 840 x 420 prototype demonstrated, but requires development to survive launch environment | 840 x 420 array format, two-side buttable | 3 | ✓ |
| 12.2.3.7 | Large-format Microchannel Plates | CsI | Meets requirements for 100–150 nm | 200 mm x 200 mm tile size >30% QE between 100–200 nm | 6 | ✓ |
| | | GaN | Meet requirements for 150–200 nm range; requries development for large tile size and integration with cross-strip readout. GaN has better solar blind performance. | | 4 | ✓ |
| | | Bi-alkali | | | 4 | |
| | | Funnel microchannels | Demonstrated 50% improved quantum efficiency with CsI photocathode | | 4 | |
| 12.2.3.8 | Large-format High-resolution Focal Plane Arrays | 8k x 8k CMOS | 4k x 4k devices exist, require development for 8k x 8k and readout optimization | 8k x 8k format, <7 micron pixels, three-side buttable | 4 | ✓ |
| | | 4k x 4k CCD | 8k x 8k devices exist with 18 micron pixels; lacks programmable high-speed region-of-interest readout for guiding capability | ~1 e- read noise ~1x10$^{-4}$ e-/pix/s dark current at 170 K | 5 | |

## 11.1  Technology items, TRL, and heritage

The LUVOIR design incorporates flight, qualification, and design heritage elements from many missions, including the Hubble Space Telescope (HST), James Webb Space Telescope (JWST), and the Wide Field Infrared Survey Telescope (WFIRST), respectively. Use of heritage components establishes confidence in the design's feasibility, as well as cost and schedule realism. Throughout the concept study, the LUVOIR team has sought to take advantage of heritage designs and components to the greatest extent possible to ensure an appropriate level of risk is constrained to truly new items or applications, i.e., technology.

Because LUVOIR is planned to launch in the late 2030s, ostensibly all of the missions we cite will have flight heritage as LUVOIR enters its implementation phase. However, at this phase of our study, we conducted an inheritance review to ensure we understand the scope of effort and magnitude of cost that is either implicit in our assumptions or explicit in our technology-development plan. We will conduct follow-up inheritance reviews in Pre-Phase A and beyond to update our assumptions and technology development plans as needed.

We define three types of heritage in order to be transparent about the risk associated with our assumptions: design, qualification, and flight. ***Design heritage*** is a form of "future heritage" that leverages hardware design efforts by projects that themselves are early in development, such as WFIRST and its Coronagraph Instrument (CGI). Most design heritage hardware has a TRL of 6. Design heritage hardware carries the greatest risk of the three categories, and we attempted to minimize its use in the LUVOIR designs.

***Qualification*** heritage refers to hardware that has been designed, built, and fully flight qualified for a particular mission's operational environment, but has not yet flown and operated in that environment. An example of qualification heritage is most of the components in the LUVOIR design that leverage JWST. These components have received significantly more analysis and testing than the design heritage components, and so carry less risk, yet still have not been fully validated in an operational flight environment. Most qualification heritage hardware is TRL 7, however if it would require significant additional engineering to be adapted to the LUVOIR operational environment, it is assessed at TRL 6 instead.

***Flight*** heritage refers to hardware that has successfully flown and operated in space. The TRL may vary between 7 and 9 depending on how much the flight environment and usage differ from LUVOIR's. For example, HST clearly has flight heritage, and yet LUVOIR will have to ensure any HST-derived hardware is designed and qualified for the LUVOIR environment and operation. Another example is harnesses. The specific types of cables themselves are all TRL 9, but some degree of engineering is required to determine length, connectors, and routing. Only flight-proven items that are truly "off-the-shelf", such as launch restraint mechanisms or thermostats, are assessed at TRL 9.

**Appendix F** contains a comprehensive listing of LUVOIR's hardware down to architecture level 5 (or lower), and identifies the heritage type, what mission(s) the heritage is derived from, and the current TRL of the items. Any item in the LUVOIR design that does not fall into one of the three heritage categories is considered part of the technology development effort of LUVOIR, and is discussed in the next section.

## 11.2  Technology development approach

In the following sub-sections we detail specific tasks associated with developing each of the three LUVOIR technology systems. A grassroots estimated cost and duration of each activity is also provided. Many of these activities are executed in parallel, enabling the entire technology development program to be completed in ~5.3 years—achieving TRL 6 for all of LUVOIR's technology systems prior to the start of Phase A in early 2025. The total cost of the technology development plan, including all engineering and manufacturing development activities, is ~$412M (FY20), phased as shown in **Figure 11-2**. This schedule and cost estimate includes funded schedule reserve applied to technology activities at a rate of 15 weeks per year. To this, we add 30% cost reserve for a total technology development cost of $536M (FY20).

Activity cost and duration estimates are based on several sources. First, we considered similar technology development efforts in the past. For example, the Advanced Mirror Segment Demonstration (Matthews et al. 2003) that led to JWST's lightweighted beryllium mirror segments serves as a model to LUVOIR's mirror segment assembly development effort. Secondly, a detailed technology development schedule, laying out specific design, fabrication, test, and analysis activities, allows for a grassroots estimate of labor and material costs for each activity. This detailed Pre-Phase A plan is available as part of the integrated master schedule shown in **Appendix G**, and a high-level schedule summary is shown in **Figure 11-3**. Finally, estimates were vetted by the LUVOIR Technology Working Group, a team of over 90 subject matter experts from NASA centers, industry, academia, and international partners. These experts have helped identify the highest-TRL solutions for each of LUVOIR's technology needs and the specific actions needed to mature those solutions to TRL 6.

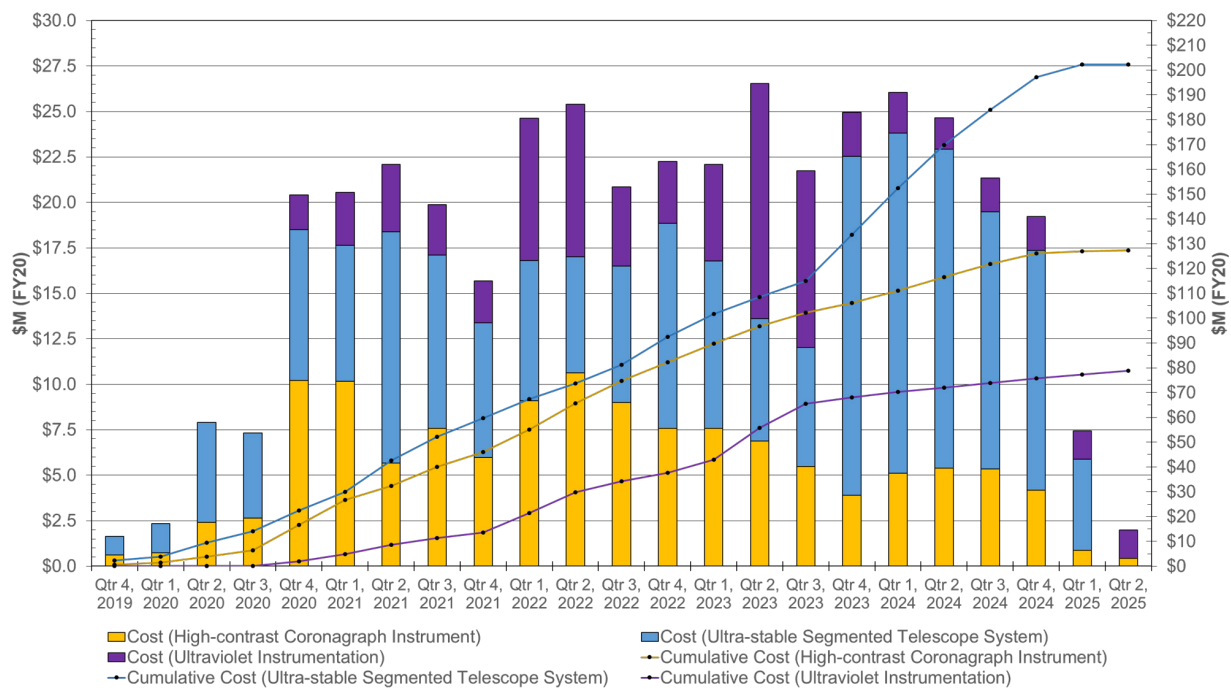

**Figure 11-2.** *Technology development costs by technology system. Quarter-by-quarter costs reference the left axis, cumulative costs reference the right axis. All costs are in FY20 $M, and include all technology, engineering, and manufacturing development activities, as well as funded schedule reserve. Note that an additional 30% cost reserve is not included in these values, but instead applied at the top-level.*

We have leveraged existing, already-funded technology development activities to the greatest extent possible. Since 2012, more than $80M[1] has been invested in LUVOIR-related technologies through the Strategic Astrophysics Technology program, Astrophysics Research and Analysis program, HQ-directed work packages to NASA centers, and the System-level Segmented Telescope Design studies. Additional investments made by government and industry internal research and development programs and Small Business Innovative Research grants push that number well above $100M, which still does not include the large investment made by WFIRST in high-contrast coronagraph technologies. Each of these existing technology development efforts are identified in the relevant sections of the full schedule in **Appendix G**, and lead directly into the LUVOIR-specific activities identified here.

---

1 We estimated this number based on private communications with Principal Investigators of all efforts that were relevant to LUVOIR.

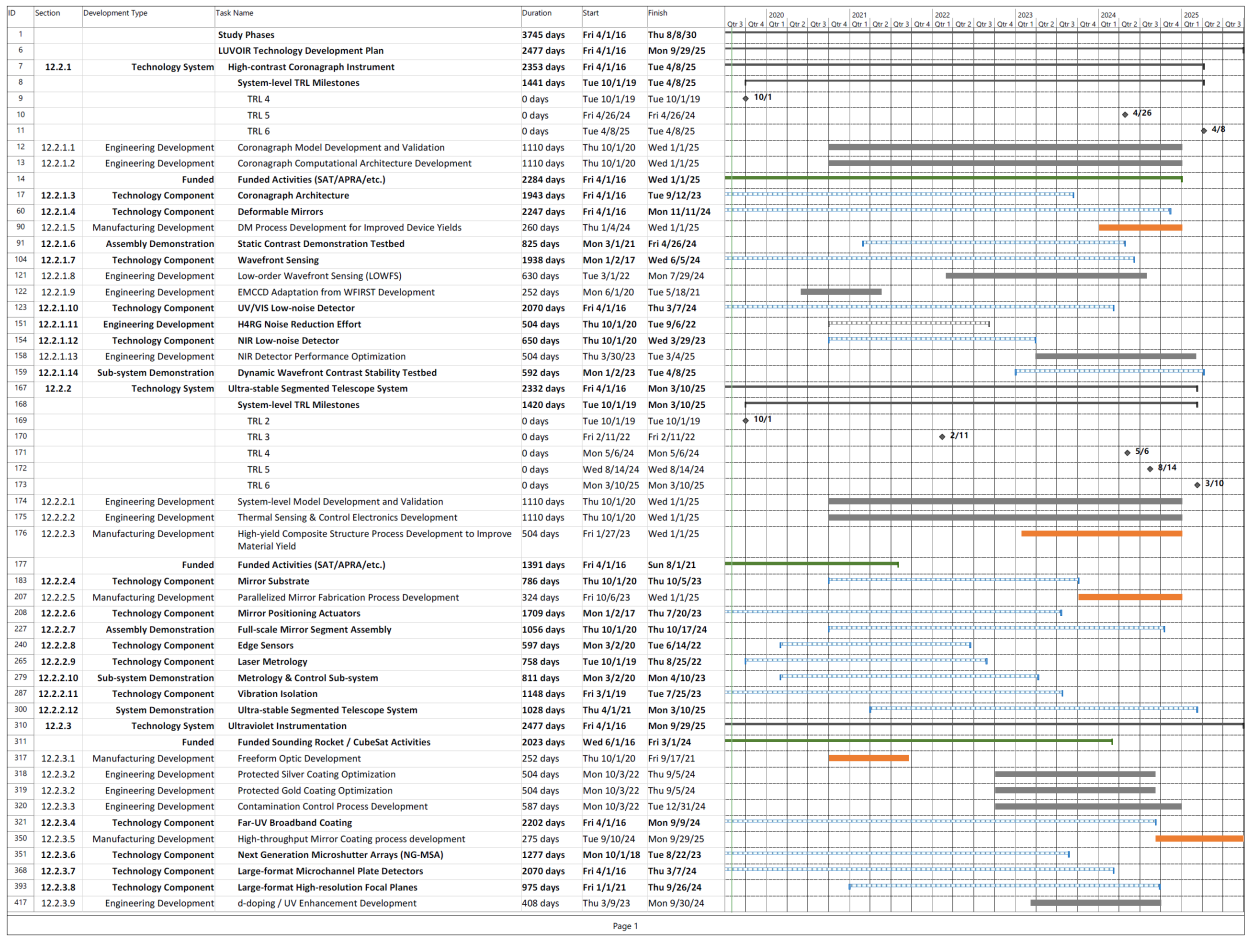

**Figure 11-3.** *Summary schedule of the technology development activities outlined in this section. Technology development efforts are shown in blue, engineering development efforts in gray, and manufacturing development efforts in orange. Currently funded technology activities that are relevant to LUVOIR are shown in green. A more detailed schedule is available in* **Appendix G**.

In each of the following sub-sections we summarize each technology development task. A technology system is matured through component, assembly, sub-system, and system-level demonstrations, identified by each sub-section heading. At the lowest level technology component, we identify the different implementation options that are available, which option(s) currently exist in the LUVOIR baseline concepts, and recommend a path to developing and down-selecting the technology components for higher-level demonstrations. Engineering and manufacturing development activities are also discussed throughout.

### 11.2.1  Technology system: high-contrast coronagraph instrument
**Figure 11-4** shows a high-level summary of the high-contrast coronagraph instrument development plan, and **Table 11-4** summarizes each activity's duration and cost.

### 11.2.1.1  Engineering development: coronagraph model development and validation

*Estimated Duration of Activity*: 5 years
*Estimated Cost of Activity*: $6.6M

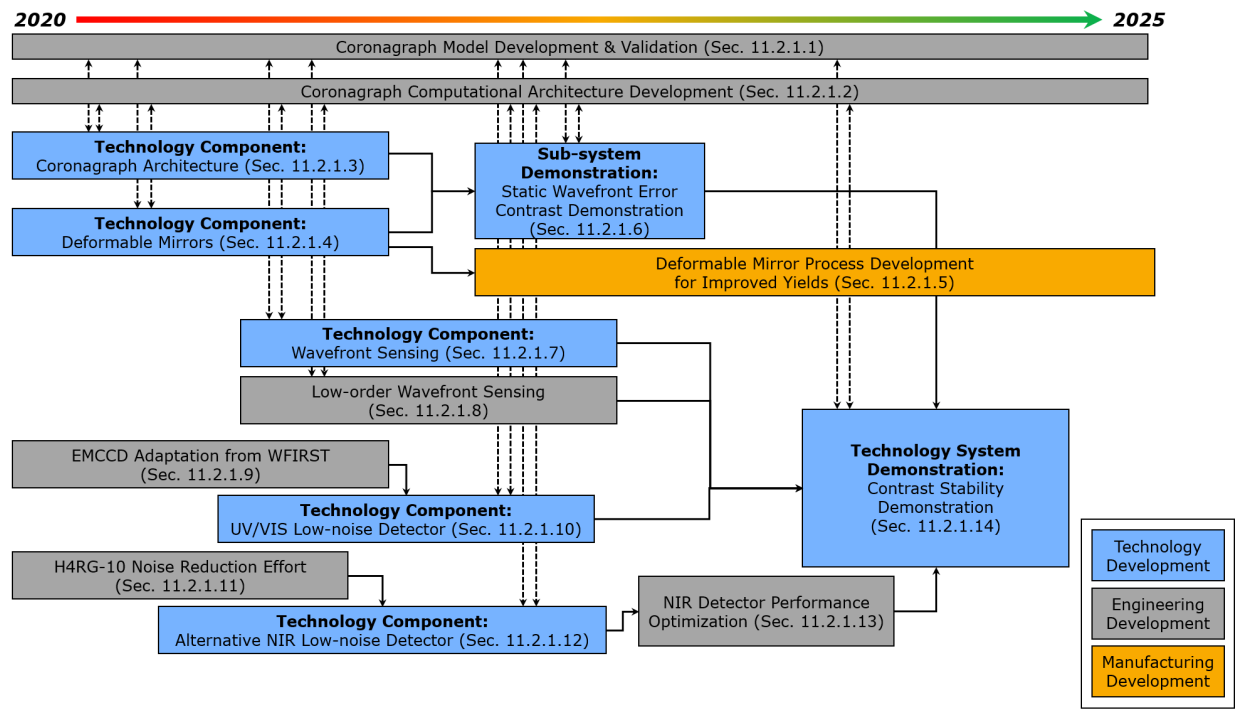

**Figure 11-4.** *High-level flow of the high-contrast coronagraph instrument technology system development. Technology items are blue, engineering development items are identified in gray, and manufacturing development items are shown in orange. The top-level model development & validation and computational architecture efforts inform, and are informed by, each of the other technology development efforts.*

High fidelity models of predicted coronagraph performance—in both laboratory testbeds and on-orbit—will be critical to validating the technology readiness of the complete high-contrast coronagraph instrument system. These models require continuous development for the duration of the entire LUVOIR program. The models also require continuous validation against the various assembly, sub-system, and system-level demonstrations that are executed during the technology development program to ensure that the end-to-end system-level performance is being accurately predicted.

**Table 11-4.** *Summary of high-contrast coronagraph instrument technology development activities.*

| Section | Name | Development Type | Duration [years] | Cost [$M FY20] |
|---|---|---|---|---|
| 11.2.1.1 | Coronagraph Model Development and Validation | Engineering | 5.0 | 6.6 |
| 11.2.1.2 | Coronagraph Computational Architecture Development | Engineering | 5.0 | 11.6 |
| 11.2.1.3 | Coronagraph Architecture | Technology Component | 2.6 | 13.0 |
| 11.2.1.4 | Deformable Mirrors | Technology Component | 4.1 | 32.3 |
| 11.2.1.5 | Deformable Mirror Process Development for Improved Yields | Manufacturing | 1.0 | 6.3 |
| 11.2.1.6 | Static Contrast Demonstration | Technology Sub-system | 3.2 | 4.9 |
| 11.2.1.7 | Wavefront Sensing | Technology Component | 3.4 | 9.3 |
| 11.2.1.8 | Low-order Wavefront Sensing | Engineering | 2.4 | 3.5 |
| 11.2.1.9 | EMCCD Adaptation from WFIRST Development | Engineering | 1.0 | 2.6 |
| 11.2.1.10 | UV/VIS Low-noise Detector | Technology Component | 4.4 | 20.0 |
| 11.2.1.11 | H4RG Noise Reduction Effort | Engineering | 2.0 | 8.8 |
| 11.2.1.12 | NIR Low-noise Detector | Technology Component | 2.5 | 1.6 |
| 11.2.1.13 | NIR Low-noise Detector Performance Optimization | Engineering | 2.0 | 2.2 |
| 11.2.1.14 | Dynamic Wavefront Error Contrast Stability Demonstration | Technology System | 2.3 | 5.5 |
| | | | Total Cost: | 128 |

It is important to note that substantial progress has been made on this task already, in large part due to the development of the WFIRST CGI instrument, and recent investments in the Segmented Coronagraph Design and Analysis program (Shaklan et al. 2019). Many of the algorithms and modeling tools exist already. The focus of this engineering development effort is specifically on curating a model of the LUVOIR architecture to ensure a detailed understanding of the system as a whole throughout the project lifecycle.

### 11.2.1.2  Engineering development: coronagraph computational architecture development

> *Estimated Duration of Activity*:  5 years
> *Estimated Cost of Activity*: $11.6M

Eventually, the high-contrast coronagraph instrument system will need to run autonomously on-orbit, coordinating all of the wavefront sensing and control routines of not just the coronagraph, but also of the entire payload element. During Pre-Phase A, it is important to begin developing computational hardware and software architectures that are capable of executing autonomous control, and have a clear path to implementation on a flight platform. Existing flight-qualified computer systems have the necessary raw capabilities; however, engineering development is needed to optimize the architecture and algorithm implementation.

### 11.2.1.3  Technology component: coronagraph architecture

> *Estimated Duration of Activity*: 2.6 years
> *Estimated Cost of Activity*: $13M (~$2.6M per team)

| Implementation Options | FY19 TRL | In LUVOIR Baseline? |
|---|---|---|
| Apodized Pupil Lyot Coronagraph (APLC) | 4 | ✓ |
| Vortex Coronagraph (VC) | 3 | ✓ |
| Phase-Induced Amplitude Apodization (PIAA) | 3 | |
| Hybrid-Lyot Coronagraph (HLC) | 3 | |
| Nulling Coronagraph (NC) | 3 | |

Apodized pupil Lyot coronagraphs (APLCs, N'Diaye et al. 2015; Zimmerman et al. 2016) and vortex coronagraphs (VCs, Ruane et al. 2016) have been baselined for the LUVOIR concepts. APLC is the primary coronagraph option on LUVOIR-A, with a secondary VC mode that is optimal for a small set of target stars. VC is the primary coronagraph on LUVOIR-B, with a secondary APLC mode that is optimal for a small set of target stars.

The APLC has heritage with the Shaped-Pupil Coronagraph (SPC) on WFIRST/CGI (Cady et al. 2015; Groff et al. 2017). The fundamental technology is identical: a binary apodization mask in a pupil plane compensates for diffraction from pupil obscurations. The APLC masks are more complex than the SPC masks, due to high-level of segmentation in LUVOIR's pupil. Current analysis shows that APLC works particularly well with the LUVOIR-A obscured-aperture design and is robust to error sources such as stellar diameter, pointing jitter, and low-order aberration drift (Stark et al. 2019).

The VC has two basic implementations: vector and scalar masks. VCs have been shown to be particularly well-suited to the LUVOIR-B unobscured aperture, with high-throughput and robustness to stellar diameter, pointing jitter, and low-order aberrations. However, the vector vortex masks rely on a liquid-crystal material that exhibits polarization leakage with current manufacturing capabilities. Development is required to reduce this polarization leakage so that dual-polarization operation can be achieved; otherwise, the instrument must operate in a single polarization at a time, and total throughput is reduced by ~50%. Scalar vortex masks do not share the polarization leakage issue, but have not been developed as much in recent years.

Phase-induced amplitude apodization (PIAA) coronagraphs have recently been shown to be a very promising option for the LUVOIR-B unobscured aperture, exhibiting very high throughput at small inner working angles. Additional evaluation is underway to explore the robustness of these PIAA designs to various noise sources, and to determine the compatibility of the PIAA optical design with the LUVOIR ECLIPS instrument (Shaklan et al. 2019, Belikov et al. 2019).

Hybrid Lyot coronagraphs (HLCs) also share heritage with WFIRST/CGI (Seo at al. 2017), and have recently been shown to be compatible with segmented apertures such as LUVOIR's. However, additional study is still required to understand the limitations of their performance.

Finally, nulling coronagraphs (NCs) are a different family of coronagraph altogether that use destructive interference to reject starlight, instead of a series of diffractive masks. NCs are inherently compatible with complex aperture geometries, and have been demonstrated at $10^{-9}$ contrast with a segmented deformable mirror with narrowband light. Additional development is necessary for broadband (10–20%) operation, and to establish sensitivities to error sources such as stellar diameter, pointing jitter, and low-order aberration drift (Lyon et al. 2008).

Another challenge for all of these coronagraph architectures is contrast degradation due to polarization aberrations. As light traverses the optical system, each polarization state accumulates different aberrations, and the coronagraph wavefront control system can only correct the average wavefront error from both polarization states. Any residual contributes to contrast leakage. Recent results (Will and Fienup 2019) show that through careful optical design and limiting the angles-of-incidence within the optical system, this issue can be mitigated.

**Recommended development path:**

Leverage the progress made by the Segmented Coronagraph Design and Analysis study (Shaklan et al. 2019) and continue to pursue all five coronagraph architectures in parallel. Design and simulate coronagraph masks and/or optical layouts that are compatible with telescope concepts being considered by the parallel Pre-Phase A architecture and concept development studies. Analyze sensitivity to various noise sources, including stellar diameter, pointing error, polarization aberration, aberration drift, detector effects, DM errors, etc., and optimize the designs accordingly. Produce at least one set of masks and/or optics for each coronagraph architecture to be demonstrated in the static wavefront error contrast sub-system-level demonstration (**Section 11.2.1.6**).

### 11.2.1.4 Technology component: deformable mirrors

*Estimated Duration of Activity*: 4.1 years
*Estimated Cost of Activity*: $32.3M

| Implementation Options | FY19 TRL | In LUVOIR Baseline? |
|---|---|---|
| Micro-Electro-Mechanical Systems (MEMS) | 4 | ✓ |
| Lead-Magnesium-Niobate (PMN) Macro-scale | 5 | |

Macro-scale deformable mirrors using lead-magnesium-niobate (PMN) based electrostrictive actuators are being developed for WFIRST/CGI in 48 x 48 actuator formats. These deformable mirrors (DMs) have been used extensively in laboratory and ground-based coronagraph demonstrations. However, the large size, weight, and power ("SWaP") properties, and substantial harness requirements do not make these DMs optimal for spaceflight applications. Additional development would be needed to reduce the SWaP requirements on the DM electronics, improve interconnect packaging and reliability, and thoroughly characterize actuator stability performance at the picometer level.

MEMS DMs have also seen broad use in coronagraph instruments in ground-based observatories, most notably the 4k Boston Micromachines devices on the Gemini Planet Imager. They are preferred for spaceflight applications; their compact size yields more favorable diffractive properties relevant to wavefront control, in addition to having reduced SWaP requirements. Development is needed to improve actuator count and yield (up to 128 x 128 actuators), as well as reduce the effects of surface defects from the fabrication process on overall coronagraph performance, and thoroughly characterize actuator stability performance at the picometer level.

**Recommended development path:**

Pursue both DM candidates in parallel early on, with a goal to fabricate at least one pair of 100%-yield, 64 x 64 actuator DMs for each implementation option, including all necessary control electronics. Evaluate the DMs on measured and predicted performance in the context of a flight-like coronagraph instrument, and select a single DM candidate for additional development. Integrate the selected pair of 64 x 64 actuator DMs into the static wavefront error contrast sub-system-level demonstration for additional evaluation (**Section 11.2.1.6**).

Following the down-selection, design and fabricate two full-scale (128 × 128 actuators) DMs, including flight-traceable control electronics. Perform complete functional, performance, and environmental qualification testing of these DMs to achieve TRL 6 at the component level.

### 11.2.1.5  Manufacturing development: deformable mirror process development for improved yields

> *Estimated Duration of Activity*: 1 year
> *Estimated Cost of Activity*: $6.3M

Once a candidate DM is established, and the basic fabrication process has been verified, investment in additional process refinement is needed. For MEMs devices in particular, wafer yield and device yield are significant challenges; it often takes multiple fabrication runs to end up with a single device at 100% actuator yield. LUVOIR will need a minimum of six flight devices, as well as spares and test units. Investment in improving manufacturing yield early will reduce schedule risk when the time comes to fabricate the flight devices.

### 11.2.1.6  Sub-system demonstration: static contrast demonstration

> *Estimated Duration of Activity*: 3.2 years
> *Estimated Cost of Activity*: $4.9M

The first assembly-level demonstration aims to achieve TRL 5 for the high-contrast coronagraph instrument system. This demonstration brings together the coronagraph architecture with a pair of DMs (minimum 64 × 64 actuators each, with 100% yield) in a thermally-controlled, vacuum environment. The performance goal is to achieve contrasts of $1\times10^{-10}$ over a minimum bandwidth of 10% at the relevant inner working angle (~3–4 λ/D). For this TRL 5 demonstration the coronagraph stimulus is a static, segmented aperture pupil that is traceable to the telescope concepts being considered by the parallel Pre-Phase A study. No dynamic wavefront disturbances, aside from ambient environmental effects, are included in this demonstration. Therefore, a low-order wavefront sensor may be used, but is not necessary if the ambient environment and DMs are stable enough.

Demonstrate each of the five coronagraph architectures that were developed at the component level in this testbed. Evaluate and compare final performance, and prioritize or down-select the coronagraph architectures, as appropriate, for additional development.

It is important to note that several facilities that would be suitable for this demonstration already exist today (Patterson et al 2019, Sayson et al. 2019, Soummer et al. 2019, Subedi et al. 2019). One of these facilities could be repurposed and dedicated to this technology development plan. However it is critical to maintaining the development schedule that the testbed be fully available to this effort, and not shared with other missions or technology development efforts. If an existing testbed cannot be repurposed in such a way, then a new testbed must be constructed. The estimated cost provide here includes the development of a brand new facility.

### 11.2.1.7  Technology component: wavefront sensing

> *Estimated Duration of Activity*: 3.4 years
> *Estimated Cost of Activity*: $9.3M

| Implementation Options | FY19 TRL | In LUVOIR Baseline? |
|---|---|---|
| Out-of-band Wavefront Sensing | 3 | ✓ |
| Low-order Wavefront Sensing | 6 | ✓ |
| Artificial Guide Star | 3 | |

In order to maintain wavefront stability during coronagraph operations, some form of concurrent wavefront sensing and control system will be required. Low-order wavefront sensing (LOWFS) is being developed and implemented on WFIRST/CGI (Shi et al. 2016), and is capable of tracking pointing errors, as well as slow drifts in low-order aberrations (global focus, astigmatism, coma, etc.). However, LOWFS systems are limited in both the speed at which they can sense aberrations, and the spatial frequency of the sensed aberrations. It is likely that additional concurrent wavefront sensing techniques will be necessary.

Out-of-band wavefront sensing (OBWFS) seeks to use light that is out-of-band (either spatially or spectrally) from the science band to determine wavefront drifts within the science band. Since OBWFS can access higher-spatial frequencies than LOWFS, it is possible to use the information from the OBWFS to monitor DM actuator drift, and even primary-mirror segment level aberrations. Furthermore, since the OBWFS would use broadband light outside of the 10-20% science band, or could use an off-axis bright guide star or even an internal light source, OBWFS would generally be able to sense the full range of wavefront drifts much faster than LOWFS, allowing for shorter control loop times and relaxed stability requirements on the optical system.

One approach to improving the speed of the wavefront sensing system is to borrow a technique from ground-based observatories and use an artificial guide star (AGS). Early studies have evaluated the use of a laser source on a CubeSat or SmallSat platform, flying in formation with the telescope at a distance of 40,000–80,000 km (Douglas et al. 2019). Such a source, coupled with an out-of-band Zernike wavefront sensor can improve sensing loop rates from < 1 Hz to > 10 Hz. It is important to note that the AGS would likely only be needed for the dimmest of target stars.

**Recommended development path:**
Conduct a survey of OBWFS concepts via computer modeling and simulation to determine which options yield the best performance and are the most implementable. Select one or two OBWFS concepts and fabricate the necessary components. A simple testbed, without coronagraph elements, would be suitable to verify the OBWFS performance predicted by the earlier simulations. Once verified, integrate the OBWFS components into the wavefront error contrast stability system-level demonstration (**Section 11.2.1.14**).

Continue studying the artificial guide star approach to determine feasibility and impact on system performance. A decision on whether to include the AGS in the LUVOIR baseline will be dependent on whether the faster closed-loop control rates will be necessary, based on system-level integrated and coronagraph modeling efforts.

### 11.2.1.8 Engineering development: low-order wavefront sensing

*Estimated Duration of Activity*: 2.4 years
*Estimated Cost of Activity*: $3.5M

Current LOWFS technology has benefitted from years of investment on ground-based tele-scopes, and as part of the WFIRST/CGI technology development effort, and will continue to be developed as that instrument matures. While OBWFS techniques are explored as a new technology for LUVOIR, it is recommended that the WFIRST LOWFS system be adapted to LUVOIR as well—both to complement an OBWFS system, but also to serve as a fallback solution should OBWFS prove unviable.

Under this effort, the limits of LOWFS performance should be explored in the context of the coronagraph architectures being developed for LUVOIR, and expected performance of the LUVOIR system (target star brightness, system throughput, expected disturbances, etc.). Much of this development will likely also inform the OBWFS development activities, as the two techniques are closely related.

### 11.2.1.9 Engineering development: EMCCD adaptation from WFIRST development

> *Estimated Duration of Activity*: 1 year
> *Estimated Cost of Activity*: $2.6M

Electron-multiplying CCDs (EMCCDs) are also being developed for WFIRST/CGI, and can achieve the low read- and dark-noise requirements for high-contrast imaging (Nemati 2014). However, radiation exposure reduces the long-term performance of these devices (Nemati et al. 2016). An improvement in quantum efficiency at the red end of the visible spectrum (~800–1000 nm) is also desirable to enhance exoEarth detection yields.

Under this effort, EMCCD development would be continued in the context of a LUVOIR coronagraph system. Focus is needed on improving radiation tolerance through shielding design and readout electronics optimization, and on improved red-end quantum efficiency via substrate thickness, deep depletion implementation, and optical coatings.

### 11.2.1.10 Technology component: UV/VIS low-noise detector

> *Estimated Duration of Activity*: 4.4 years
> *Estimated Cost of Activity*: $20M

| Implementation Options | FY19 TRL | In LUVOIR Baseline? |
|---|---|---|
| Electron-Multiplying CCD | 4 | ✓ |
| Hole-Multiplying CCD | 3 | |

In parallel with early EMCCD engineering development, hole-multiplying CCDs (HMCCDs) should also be developed as a potential alternative. HMCCDs are inherently radiation hard, and do not suffer long-term degradation under continuous exposure. Furthermore, this ra-diation hardness allows thicker substrates to be used in the devices, improving long-wave-length quantum efficiency.

**Recommended development path:**
Building off current development activities that are already funded, design and fabricate a 1k x 1k pixel HMCCD device and evaluate its noise and sensitivity performance relative to the existing EMCCDs. Select a single candidate technology for continued development.

Incorporate this 1k x 1k candidate into the coronagraph testbeds for validation at the system level.

Following the down-select, design and fabricate a 4k x 4k device, including all necessary readout electronics. Complete functional, performance, radiation, and environmental qualification testing to achieve a component-level TRL 6.

### 11.2.1.11 Engineering development: H4RG noise reduction effort

> *Estimated Duration of Activity*: 2 years
> *Estimated Cost of Activity*: $8.8M

HgCdTe photodiode arrays are high-TRL, high-performance near-infrared (NIR) detectors. The baselined Teledyne H4RG-10 detectors have direct heritage to the H4RG detectors baselined on WFIRST, and H2RG detectors used in JWST. However, for use in a high-contrast coronagraph, it is desirable to reduce read noise and dark current further, if possible. Additional gains in NIR detector performance will also benefit the HDI instrument's NIR channel focal plane.

This engineering effort explores two paths that have been identified to potentially achieve this goal: reducing the pixel size (to smaller than 10 μm), and optimizing the readout electronics, operating temperature, and clock rates for lower-noise performance.

### 11.2.1.12 Technology component: NIR low-noise detector

> *Estimated Duration of Activity*: 2.5 years
> *Estimated Cost of Activity*: $1.6M

| Implementation Options | FY19 TRL | In LUVOIR Baseline? |
|---|---|---|
| HgCdTe Photodiode Array | 5 | ✓ |
| HgCdTe Avalanche Photodiode Array | 4 | |

In parallel with the H4RG noise reduction engineering development effort, HgCdTe avalanche photo diode (APD) detectors should also be evaluated as a potential low-noise alternative NIR detector. These detectors are photon-counting versions of the HgCdTe photodiode arrays used in H4RGs. Development is needed to improve array size (currently limited to a few hundred pixels on a side) and dark current, as well as general evaluation of detector performance for use with high-contrast imaging systems (Rauscher et al. 2016), and at operational temperatures > 100 K.

**Recommended development path:**
Invest in the development of a 1k x 1k HgCdTe APD array and evaluate its noise and sensitivity performance relative to the H4RG. Select a single candidate technology for continued development.

### 11.2.1.13 Engineering development: NIR low-noise detector performance optimization

> *Estimated Duration of Activity*: 2 years
> *Estimated Cost of Activity*: $2.2M

Following selection of a NIR detector candidate, continue investment in optimizing detector performance for use with a high-contrast imaging system. Specific attention should be made to the operational thermal environment that is required to achieve the best performance, and how that thermal environment might be enabled in the context of the overall LUVOIR system.

### 11.2.1.14 System-level demonstration: dynamic wavefront error contrast stability demonstration

*Estimated Duration of Activity*: 2.3 years
*Estimated Cost of Activity*: $5.5M

Following successful demonstration of the static wavefront error coronagraph performance, upgrade the testbed to incorporate a telescope simulator stimulus. This telescope simulator should reproduce a segmented and/or obscured telescope pupil that is similar to the concepts being considered by the parallel Pre-Phase A LUVOIR study. Most importantly, the telescope simulator should reproduce wavefront drifts that are commensurate with those that are expected to be present in the on-orbit LUVOIR system. These wavefront drifts include, but are not limited to, pointing errors; slow, low-order aberrations induced by thermal drifts; fast, high-order aberrations induced by structure dynamics; segment-to-segment phasing errors; polarization effects; etc. In short, the stimulus provided by the telescope simulator should fully represent the expected on-orbit performance of the LUVOIR optical telescope assembly.

For this demonstration, the prioritized coronagraph candidates incorporate the out-of-band wavefront sensor concept to perform concurrent wavefront sensing and control during the contrast demonstration. If the OBWFS does not fully supersede the use of a LOWFS, then a LOWFS system should also be incorporated. Successful demonstration includes maintaining a dark-hole contrast of $1 \times 10^{-10}$ over the designed image-plane region, with a minimum 10% bandpass, in the presence of the wavefront error drifts from the telescope simulator. Upon achieving this, the high-contrast coronagraph instrument system would be TRL 6.

### 11.2.2 Technology system: ultra-stable segmented telescope

**Figure 11-5** shows a high-level summary of the ultra-stable segmented telescope technology development plan, and **Table 11-5** summarizes each activity's duration and cost.

### 11.2.2.1 Engineering development: System-level model development and validation

*Estimated Duration of Activity*: 5 years
*Estimated Cost of Activity*: $15.2M

As with the coronagraph technology system, high-fidelity structural-thermal-optical-control system models will be critical to the validation of picometer-level wavefront error stability in both laboratory demonstrations and on-orbit. These models require continuous development for the duration of the LUVOIR program. The models also require continuous validation against the various assembly, sub-system, and system-level demonstrations that are executed during the technology development program to ensure that the end-to-end system-level performance is accurately predicted.

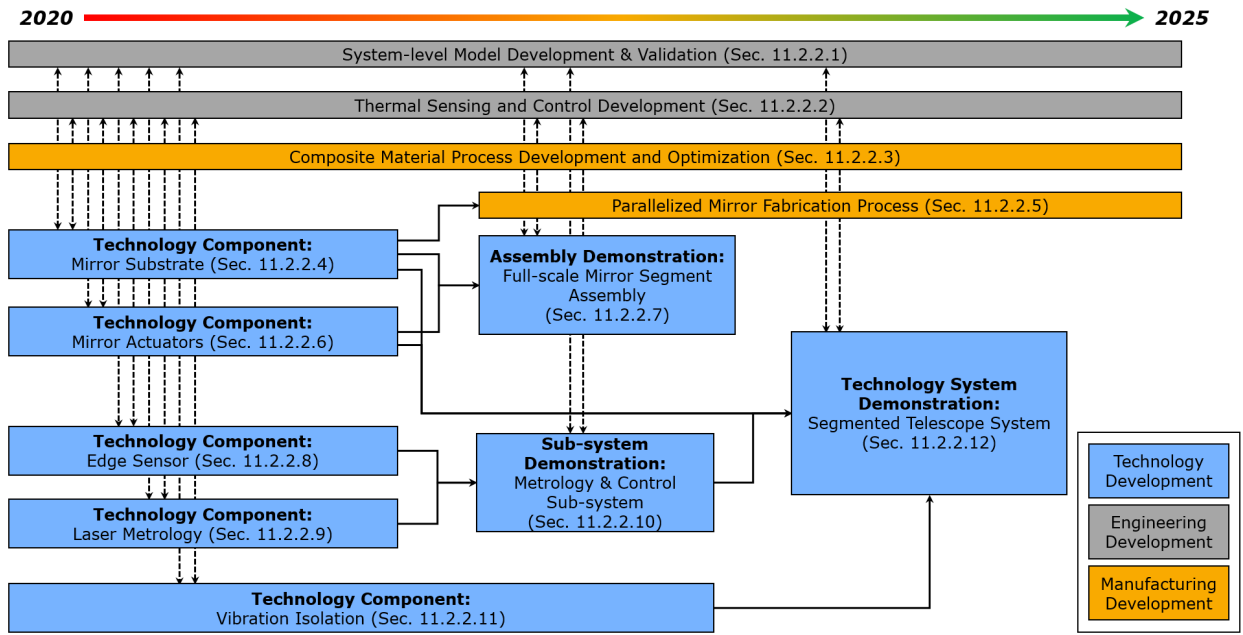

**Figure 11-5.** *High-level flow of the ultra-stable segmented telescope technology system development. Technology items are blue, engineering development items are gray, and manufacturing development items are orange. The system-level model development & validation and thermal sensing & control efforts inform, and are informed by, each of the other technology development efforts.*

**Table 11-5.** *Summary of ultra-stable segmented telescope technology development activities.*

| Section | Name | Development Type | Duration [years] | Cost [$M FY20] |
|---|---|---|---|---|
| 11.2.2.1 | System-level Model Development and Validation | Engineering | 5.0 | 15.2 |
| 11.2.2.2 | Thermal Sensing and Control Development | Engineering | 5.0 | 10.3 |
| 11.2.2.3 | Composite Material Process Development and Optimization | Manufacturing | 2.0 | 13.0 |
| 11.2.2.4 | Mirror Substrate | Technology Component | 3.0 | 15.0 |
| 11.2.2.5 | Parallelized Mirror Fabrication Process Development | Manufacturing | 1.2 | 19.0 |
| 11.2.2.6 | Mirror Positioning Actuators | Technology Component | 3.7 | 9.4 |
| 11.2.2.7 | Full-scale Mirror Segment Assembly | Technology Assembly | 4.0 | 22.3 |
| 11.2.2.8 | Edge Sensor | Technology Component | 2.3 | 9.0 |
| 11.2.2.9 | Laser Metrology | Technology Component | 2.9 | 10.8 |
| 11.2.2.10 | Metrology & Control Sub-system | Technology Sub-system | 3.1 | 5.6 |
| 11.2.2.11 | Vibration Isolation & Precision Pointing System | Technology Component | 4.4 | 18.4 |
| 11.2.2.12 | Segmented Telescope System | Technology System | 4.0 | 54.4 |
| | | | **Total Cost:** | **202** |

## 11.2.2.2  Engineering development: thermal sensing and control development

> *Estimated Duration of Activity*: 5 years
> *Estimated Cost of Activity*: $10.3M

To achieve system-level wavefront stability, LUVOIR will require active thermal control, with critical structural and optical surfaces maintained at milli-Kelvin level stability. This level of thermal sensing and control has been demonstrated over small scales in the laboratory using rack-mounted electronics. Under this engineering development effort, the electronics will be ported to flight-traceable form factors, with an emphasis on reducing size, weight, and power characteristics. Additional analysis via modeling, included as part of the system-level model development and validation activity (**Section 11.2.2.1**) and guided by the parallel Pre-Phase A LUVOIR study, will optimize heater and sensor location and control algorithms to achieve the best thermal stability.

### 11.2.2.3 Manufacturing development: composite material process development and optimization

> *Estimated Duration of Activity*: 2 years
> *Estimated Cost of Activity*: $13M

Preliminary architecture studies indicate that present-day composite structure materials can achieve the necessary material properties (coefficient of thermal expansion, coefficient of moisture expansion, etc.) for overall system stability. However, current manufacturing processes would yield a high scrap rate of structure materials that do not meet the strict selection criteria. Under this effort, material characterization and composite structure fabrication will be optimized to better select optimal structure components while simultaneously reducing scrap rate (Coyle et al. 2019).

At the same time, additional development and optimization of composite layups, structure joints, and bonding practices will further improve overall system stability. It may also be necessary to develop infrastructure to fabricate single composite pieces of the size needed by LUVOIR.

### 11.2.2.4 Technology component: mirror substrate

> *Estimated Duration of Activity*: 3 years
> *Estimated Cost of Activity*: $15M

| Implementation Options | FY19 TRL | In LUVOIR Baseline? |
|---|---|---|
| Closed-back ULE (rigid body actuated) | 5 | ✓ |
| Closed-back ULE (surface figure actuated) | 4 | |
| Open-back Zerodur (rigid body actuated) | 4 | |

Mirror segment substrates were studied in detail during early mirror development efforts for JWST. The Advanced Mirror Segment Demonstration developed not only lightweight beryllium mirrors for JWST, but also glass mirror substrates (Matthews et al. 2003). SiC mirror segment assemblies were also studied in later development programs (Hickey et al. 2010, Wellman et al. 2012). For LUVOIR, materials with near-zero coefficient of thermal expansion (CTE) at the LUVOIR operating temperature of 270 K are desired for optimal thermal stability. Thus, candidate mirror segments using closed-back ULE or open-back Zerodur are under consideration.

Preliminary studies indicate that surface figure actuation of these mirror segments is not necessary to achieve overall system performance, although such actuation could reduce fabrication tolerances at the expense of greater complexity at the segment assembly level (Coyle et al. 2019). The Pre-Phase A LUVOIR study should continue to explore this option as a trade, and select two of three implementation options for continued development.

**Recommended development path:**
Based on the Pre-Phase A trade study, pursue two mirror substrate candidates in parallel. Fabricate two subscale test-article mirrors. The test articles should be fabricated such that if they were full-scale segments they would achieve the necessary areal density, first mode frequency, and optical performance, as determined by the Pre-Phase A study. Emphasis should also be placed on achieving radius-of-curvature matching in the fabrication process and improving gravity release error performance to demonstrate that radius of curvature actuators are not needed. A representative mounting structure should also be incorporated to show that support print-through can be mitigated. Complete optical performance and environmental (thermal, vacuum, vibration, acoustic) testing of the subscale test articles and evaluate their performance. Down-select to a single mirror substrate candidate.

Following the down-select, design and fabricate a full-scale (1-meter-class) mirror segment substrate. This full-scale mirror segment should achieve all necessary performance goals (areal density, first-mode frequency, optical performance, thermal performance, etc.). Complete optical performance and environmental testing to bring the full-scale mirror segment substrate technology component to TRL 6.

### 11.2.2.5 Manufacturing development: parallelized mirror fabrication process development

> *Estimated Duration of Activity*: 1.2 years
> *Estimated Cost of Activity*: $19M

Once a viable mirror substrate candidate is demonstrated at TRL 6, it will be necessary to build the infrastructure to produce a large quantity of these mirrors in a short time to reduce schedule risk later in the project lifecycle. Under this manufacturing development effort, establish a mirror production line for high-throughput mirror processing. Wherever possible, leverage the investments made by the Thirty Meter Telescope and European Extremely Large Telescope in segment manufacturing.

### 11.2.2.6 Technology component: mirror positioning actuators

> *Estimated Duration of Activity*: 3.7 years
> *Estimated Cost of Activity*: $9.4M

| Implementation Options | FY19 TRL | In LUVOIR Baseline? |
|---|---|---|
| Combined piezo/mechanical | 3 | ✓ |
| All-piezo | 3 | |

Rigid-body positioning of the primary mirror segments will require actuators that simultaneously have high dynamic range (millimeters of travel) and ultra-fine resolution (picometer step sizes). The JWST mechanical bipod actuators currently achieve the necessary level of coarse and fine stage actuation, and could be repurposed for LUVOIR, with modest optimization of their design to meet LUVOIR's mass and volume requirements.

For the ultra-fine actuation, piezo-electric (PZT) actuators are the primary candidate. Commercial off-the-shelf PZT actuators with 5 picometer resolution are available and have been demonstrated in controlled laboratory environments (Saif et al. 2019). However, development is needed to incorporate PZT actuators with the coarse and fine stage mechanical bipod actuators. PZT launch survivability and reliability must also be studied and improved.

**Recommended development path:**
Review all-piezo and piezo-mechanical hybrid actuators, and evaluate expected performance against system requirements. Design an initial test article actuator based on evaluation. Co-develop the necessary drive electronics to manage the actuator motion. Fabricate a test article actuator and verify its performance in a laboratory setting to achieve TRL 4.

Following the TRL 4 demonstration, design and fabricate a complete, flight-like bipod actuator and electronics. Complete functional and performance testing of the actuator with a mirror segment mass simulator (TRL 5) and complete environmental qualification testing to bring the actuator technology component to TRL 6.

### 11.2.2.7  Assembly demonstration: full-scale mirror segment assembly

> *Estimated Duration of Activity*: 4 years
> *Estimated Cost of Activity*: $22.3M

Following the fabrication of a full-scale mirror segment substrate and the development of segment rigid body actuators, a full-scale primary mirror segment assembly can be demonstrated. While the earlier technology components are being developed, study and design a full mirror assembly architecture that incorporates support structure and thermal management systems. Fabricate and integrate all components into a complete mirror segment assembly, including actuator control electronics and harnessing. Functional, performance, and environmental qualification testing will bring the full-scale mirror segment assembly to TRL 6.

### 11.2.2.8  Technology component: edge sensor

> *Estimated Duration of Activity*: 2.3 years
> *Estimated Cost of Activity*: $9M

| Implementation Options | FY19 TRL | In LUVOIR Baseline? |
|---|---|---|
| Capacitive | 3 | ✓ |
| Inductive | 3 | |
| Optical | 3 | |
| High-speed Speckle Interferometry | 3 | |

Mirror segment co-phasing at the picometer level is critical to maintaining contrast stability in the high-contrast coronagraph instrument. Edge sensors can provide high-speed direct measurements of mirror segment positions relative to one another, which can be used to drive the mirror segment rigid-body actuators to maintain co-phasing. Development challenges include designing a sensing geometry that can measure edge piston, gap, shear and/or dihedral angle at the required levels.

An alternative to sensing mirror dephasing directly at the edges is to use high-speed interferometry (HSI) to view and align the entire primary mirror at once. HSI has been developed for lab verification of picometer dynamics (Saif et al. 2019) of actuators, structures, and materials. If used in specular mode, an HSI could be placed at the end of a boom at the center of curvature to view the entire primary mirror. Alternatively, if used in speckle mode, an HSI may be located elsewhere on the spacecraft to monitor segment motions. Additional study is needed to determine how such a metrology system could be incorporated into the architecture.

**Recommended development path:**

Develop two candidate edge sensors and associated electronics. Fabricate the edge sensors and electronics and verify their performance measuring the relative position of a single edge over all relevant degrees of freedom. Based on performance, down-select to a single edge sensor candidate.

Fabricate additional edge sensors necessary to establish an edge sensor network to measure the relative positions of multiple dynamic edges. Complete functional, performance, and environmental qualification testing of the edge sensor network to bring the edge sensor technology component to TRL 6.

In parallel, continue to develop the HSI concept and determine its viability as an on-orbit metrology system.

### 11.2.2.9  Technology component: laser metrology

*Estimated Duration of Activity*: 2.9 years
*Estimated Cost of Activity*: $10.8M

| Implementation Options | FY19 TRL | In LUVOIR Baseline? |
|---|---|---|
| Laser truss with phasemeter electronics | 4 | ✓ |

In addition to edge sensors, a laser metrology system will help maintain global alignment of the primary mirror, secondary mirror, and aft-optics system. A subset of mirror segments, outfitted with laser metrology beam-launchers and aimed at retroreflectors mounted to the secondary mirror connect the relative positions of the primary and secondary mirrors. Another set of beam-launchers mounted to a reference datum in the aft-optics system, and also aimed at retroreflectors on the secondary mirror, connect the relative positions of the secondary mirror and aft-optics system. Thus, the wavefront error resulting from the drift in relative positions of the primary mirror, secondary mirror, and aft-optics system can be measured to picometer resolution. These measurements are then used in the overall control system to command primary mirror segment actuators, secondary mirror actuators, and deformable mirror actuators to maintain wavefront stability. The primary development activity needed for the laser metrology system is to design compact, thermally stable beam-launchers, and to develop low-noise, flight traceable electronics. Laser reliability is also of concern and should be studied.

**Recommended development path:**

Design and fabricate a compact beam-launcher and associated phasemeter electronics. Complete functional and performance testing of the laser distance measurement system to verify its expected performance. Fabricate additional beam-launchers necessary to create a laser truss, sufficient to measure the relative position of a single surface in six degrees of freedom. Perform functional, performance, and environmental qualification testing of the laser truss to bring the laser metrology technology component to TRL 6.

### 11.2.2.10 Sub-system demonstration: metrology & control sub-system

> *Estimated Duration of Activity*: 3.1 years
> *Estimated Cost of Activity*: $5.6M

During the development of the edge sensors and laser metrology technology components, develop a nested control system architecture, including all necessary control algorithms. Incorporate an edge sensor network, laser truss, and closed-loop position control feedback of multiple optical surfaces into a testbed demonstration to bring the metrology and control sub-system to TRL 5.

### 11.2.2.11 Technology component: Vibration Isolation and Precision Pointing System (VIPPS)

> *Estimated Duration of Activity*: 4.4 years
> *Estimate Cost of Activity*: $18.4M

| Implementation Options | FY19 TRL | In LUVOIR Baseline? |
|---|---|---|
| Non-contact Isolation System | 4 | ✓ |

Active vibration isolation is necessary to isolate the optical payload from likely disturbance sources on the spacecraft (control moment gyroscopes, flexible structure modes, thruster impulses, etc.). While passive isolation systems and dampers will be used throughout the system, preliminary studies show that a non-contact isolation system that floats the payload relative to the spacecraft is critical to maintaining overall dynamic stability of the optical system (Dewell et al. 2019). Development challenges include demonstrating the fundamental isolation performance at the necessary levels, characterizing the contribution of power and data cables to dynamic isolation shorting, and developing flight-traceable control electronics.

**Recommended development path:**

Four development efforts will mature the individual pieces of the non-contact isolation system, bringing the entire system to TRL 5: (1) Demonstrate real-time control of the non-contact interface in 5 degrees of freedom in a 1-g environment; (2) Design, and test low size-weight-and-power electronics in a thermal-vacuum environment; (3) Measure and evaluate the impact of cable stiffness on the non-contact interface; and (4) Develop and validate control-structure-dynamics integrated models, anchored by the Pre-Phase A LUVOIR study. Finally, bringing all components together in a flight CubeSat demonstration will demonstrate system performance scalable to LUVOIR, demonstrating TRL 6 for the VIPPS system concept.

### 11.2.2.12  System demonstration: segmented telescope system

*Estimated Duration of Activity*: 4 years
*Estimated Cost of Activity*: $54.4M

A high-fidelity sub-scale segmented telescope system that incorporates all of the above technology components, assemblies and sub-systems will demonstrate the ultra-stable segmented telescope technology system at TRL 6. The individual mirror segments should be sub-scale versions of the full-scale mirror segment assembly (**Section 11.2.2.7**). Alignment of the subscale segment assemblies, secondary mirror, and aft-optics system will be maintained by the metrology and control sub-system (**Section 11.2.2.10**). Isolate the subscale segmented telescope in five degrees of freedom in a 1-g environment with the vibration isolation sub-system (**Section 11.2.2.11**). Inject disturbances that are traceable to those expected from the on-orbit spacecraft at the input of the isolation system, and verify performance of the end-to-end segmented telescope system with external metrology. Following functional and performance testing of this complete system, the ultra-stable segmented telescope technology system will be TRL 6.

### 11.2.3  Technology system: ultraviolet instrumentation

**Figure 11-6** shows a high-level summary of the ultraviolet instrumentation technology development plan, and **Table 11-6** summarizes each activity's duration and cost.

### 11.2.3.1  Manufacturing development: freeform optic development

*Estimated Duration of Activity*: 1 year
*Estimated Cost of Activity*: $6.2M

Large freeform optics and gratings are used in both the LUMOS and HDI instruments, as well as the tertiary mirror of the LUVOIR-A telescope. Freeform optics allow fewer optical surfaces to be used to correct wavefront aberration, enabling better performing instruments in more compact packages, and with higher throughput. Under this effort, investigate and develop fabrication and test processes to reduce the time and cost involved in manufacturing large freeform optics.

### 11.2.3.2  Engineering development: VIS and NIR coating optimization

*Estimated Duration of Activity*: 2 years
*Estimated Cost of Activity*: $4.4M

Existing protected silver and protected gold mirror coatings are sufficient for all visible and near-infrared instrument channels. However, coating prescription and process optimization is needed to fine-tune the coating performance for LUVOIR's specific science goals. Bandpass filters and dichroic beamsplitters will also be designed and optimized under this effort.

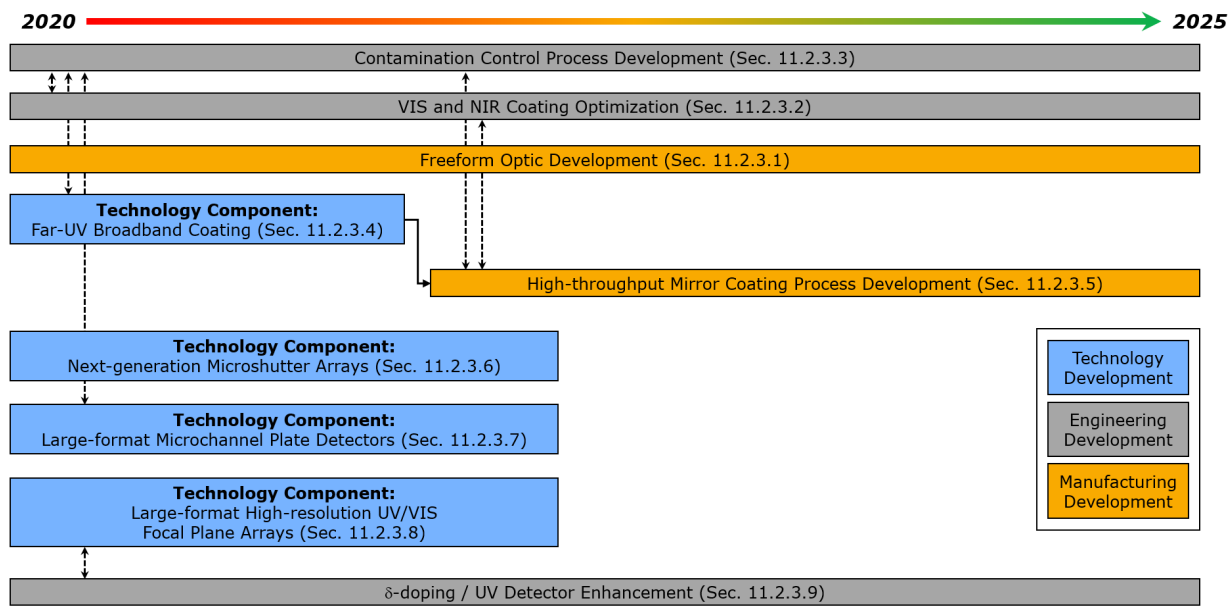

**Figure 11-6.** *High-level flow of the ultraviolet instrumentation technology system development. Technology items are blue, engineering development items are identified in gray, and manufacturing development items are shown in orange. The contamination control process development effort is coupled to the development coatings for all of the optical elements, as well as the far-UV coating technology development.*

**Table 11-6.** *Summary of ultraviolet instrumentation technology development activities.*

| Section | Name | Development Type | Duration [years] | Cost [$M FY20] |
|---------|------|------------------|------------------|----------------|
| 11.2.3.1 | Freeform Optic Development | Manufacturing | 1.0 | 6.2 |
| 11.2.3.2 | VIS and NIR Coating Optimization | Engineering | 2.0 | 4.4 |
| 11.2.3.3 | Contamination Control Process Development | Engineering | 2.3 | 2.4 |
| 11.2.3.4 | Far-UV Broadband Coating | Technology Component | 4.0 | 5.0 |
| 11.2.3.5 | High-throughput Mirror Coating Process Development | Manufacturing | 1.1 | 6.6 |
| 11.2.3.6 | Next-generation Microshutter Arrays | Technology Component | 2.0 | 7.3 |
| 11.2.3.7 | Large-format Microchannel Plate (MCP) Detectors | Technology Component | 3.2 | 8.0 |
| 11.2.3.8 | Large-format High-resolution Focal Plane Arrays | Technology Component | 3.0 | 37.5 |
| 11.2.3.9 | d-doping / UV Enhancement Development | Engineering | 1.6 | 3.4 |
| | | | **Total Cost:** | **81** |

## 11.2.3.3  Engineering development: contamination control process development

> *Estimated Duration of Activity*: 2.3 years
> *Estimated Cost of Activity*: $2.4M

LUVOIR's far-UV science objectives will require strict control of contamination at all levels of integration and test. While existing contamination control processes can meet the cleanliness requirements, the scale of LUVOIR will pose a challenge. This engineering development effort will study and develop contamination control processes to facilitate typical activities that might occur during LUVOIR's integration and test phase. This will include analysis of molecular absorber coatings as an on-orbit contamination control measure. Draft a preliminary contamination control plan, and document best practices for optic protection and cleaning at all stages of integration and test.

### 11.2.3.4  Technology component: far-UV broadband coating

> *Estimated Duration of Activity*: 4 years
> *Estimated Cost of Activity*: $5M

| Implementation Options | FY19 TRL | In LUVOIR Baseline? |
|---|---|---|
| Al + eLiF + MgF$_2$ | 3 | ✓ |
| Al + eLiF + AlF$_3$ | 3 | |
| Al + eLiF | 5 | |

LUVOIR's far-UV science objectives require a broadband mirror coating with high-reflectivity in the far-UV (>100 nm). Enhanced LiF-protected aluminum coatings have demonstrated the necessary optical performance, but are susceptible to degradation when subjected to even moderate humidity (Quijada et al. 2014). An additional protective layer of MgF$_2$ or AlF$_3$ has been shown to mitigate this degradation without significantly affecting performance (Balasubramanian et al. 2017). Development challenges include demonstrating repeatable, highly-uniform coating depositions on large-scale (1-meter-class) optics. Additional optimization of the coating prescription to maximize reflectivity at appropriate wavelengths is also desirable. It is important to note that multiple sounding-rocket and CubeSat missions are expected to fly within the next few years that will demonstrate several of these advanced coatings over 0.5-meter-class optics.

**Recommended development path:**
Develop and optimize the coating process and implement on sub-scale mirror samples. Evaluate and verify coating performance. Repeat coating deposition multiple time (minimum of 3) to verify repeatability. Begin age-testing the sub-scale samples by storing in a controlled environment and subjecting to routine measurements.

Follow the sub-scale sample demonstration, coat a full-scale (1-meter-class) mirror and verify coating performance and uniformity. Complete optical, radiation, and environmental qualification testing to bring the far-UV broadband coating technology component to TRL 6.

### 11.2.3.5  Manufacturing development: high-throughput mirror coating process development

> *Estimated Duration of Activity*: 1.1 years
> *Estimated Cost of Activity*: $6.6M

Once the mirror coating has been developed and the process has been optimized, a production line must be developed to facilitate coating a large number of mirrors in a short time. Ideally, this infrastructure would be integrated with the parallelized mirror fabrication development effort to leverage efficiencies.

### 11.2.3.6 Technology component: next-generation microshutter arrays

> *Estimated Duration of Activity*: 2 years
> *Estimated Cost of Activity*: $7.3M

| Implementation Options | FY19 TRL | In LUVOIR Baseline? |
|---|---|---|
| Next-gen Electrostatic Microshutter Arrays | 3 | ✓ |

The multi-object spectroscopy capability on LUVOIR is enabled by an array of microshutters, similar to those used on JWST/NIRSpec (Kutyrev et al. 2008). An existing technology development effort is already funded to mature the development of "next generation" microshutter arrays to TRL 4, and incorporates multiple design improvements over the JWST-style microshutters. These next-generation microshutters will also fly on the FORTIS sounding rocket in 2022. Once the existing technology development effort is completed, the remaining challenges include fabricating large-format (840 × 420) arrays and completing environmental qualification.

**Recommended development path:**

Upon completion of the existing development effort, design and fabricate a large format (840 x 420) microshutter array with accompanying control electronics. Complete functional, performance, and environmental qualification testing to bring the next-generation microshutter array technology component to TRL 6.

### 11.2.3.7 Technology component: large-format microchannel plate (MCP) detectors

> *Estimated Duration of Activity*: 3.2 years
> *Estimated Cost of Activity*: $8M

| Implementation Options | FY19 TRL | In LUVOIR Baseline? |
|---|---|---|
| CsI | 6 | ✓ |
| GaN | 4 | ✓ |
| Bi-alkali | 4 | |
| Funnel microchannels | 4 | |

Large-format, low-noise microchannel plates are being developed as part of numerous sounding rocket experiments (Fleming et al. 2011, Hoadley et al. 2016). Remaining development challenges include the demonstration of GaN microchannel plates to enable the bluest end of the far-UV detector, and demonstrating a tiled micro-channel plate focal plane array. New funnel-style micro-channels have also been shown to improve quantum efficiency by 50% (Matoba et al. 2014) and their incorporation into the baseline MCP architecture should be explored.

**Recommended development path:**

Design and fabricate a 200 mm x 200 mm CsI microchannel plate detector, 200 mm x 200 mm GaN microchannel plate detector, and all associated readout electronics. Verify performance of each detector. Integrate both microchannel plate detectors into a single focal plane array, with a maximum gap size of 15 mm. Complete functional, performance, and environmental qualification testing to bring the large-format microchannel plate detector technology component to TRL 6.

### 11.2.3.8 Technology component: large-format high-resolution focal plane arrays

> *Estimated Duration of Activity*: 3 years
> *Estimated Cost of Activity*: $37.5M

| Implementation Options | FY19 TRL | In LUVOIR Baseline? |
|---|---|---|
| 8k x 8k CMOS | 4 | ✓ |
| 4k x 4k CCD | 5 | |

The HDI UVIS-channel and LUMOS near-UV multi-object spectrograph channel focal plane arrays require a large number (>20) of large format (8k × 8k) CMOS detectors with small pixels (<7 μm). While commercial off-the-shelf CMOS arrays of this format exist, they have not been adopted for spaceflight, nor do they exhibit optimal noise performance for scientific operations. Development challenges include optimizing detector and readout electronics for low noise performance, and flight qualifying large-format detector packages. Optimizing sensor packaging to enable three-side buttability is also needed.

**Recommended development path:**
Design and fabricate an 8k × 8k × 6 μm CMOS detector and associated readout electronics in a three-side-buttable, vertically-integrated package. Verify detector sensitivity and noise performance, incorporating δ-doping for enhanced UV sensitivity as necessary. Complete functional, performance, and radiation testing to achieve TRL 5. Fabricate additional sensors and electronics and integrate into a single focal-plane array. Complete functional, performance, and environmental qualification testing to achieve TRL 6.

### 11.2.3.9  Engineering development: δ-doping / UV enhancement development

> *Estimated Duration of Activity*: 1.6 years
> *Estimated Cost of Activity*: $3.4M

Several detectors, including the HDI UVIS-channel detector, the ECLIPS UV-channel detector, and the LUMOS near-UV multi-object spectrograph detector require enhanced near-UV (200–400 nm) sensitivity. Several UV-enhancing techniques, such as δ–doping, have been developed for multiple UV sounding rocket and CubeSat missions (Nikzad et al. 2017). Under this effort, UV-enhancing processes will be developed for the large-format, buttable CMOS arrays and low-noise EM (or HM) CCD detectors baselined for LUVOIR. For HMCCD devices, additional effort may be necessary to accommodate the reversed polarity of the process.

# CHAPTER 12.  MANAGEMENT AND SYSTEMS ENGINEERING

LUVOIR is a complex, flagship-level mission concept. Any project that evolves from it brings with it unique management, systems engineering, integration and test (I&T), and verification and validation challenges compared to smaller missions. This chapter addresses key strategies that are believed to be critical to efficiently managing, formulating, and implementing any flagship mission, not just LUVOIR. The Chapter is organized as follows:

- **Section 12.1** provides an overview of the LUVOIR management approach, including specific project-level strategies (**Section 12.1.1**) and an alternative agency-level funding strategy to improve cost and schedule performance (**Section 12.1.2**)

- **Section 12.2** describes the Pre-Phase A project office that would begin implementing many of the strategies outlined in the previous section.

- **Section 12.3** addresses the systems engineering approach for LUVOIR, including mission class (**Section 12.3.1**), project risks (**Section 12.3.2**), technical margin philosophy (**Section 12.3.3**), the trade studies that were performed during this study (**Section 12.3.4**) and those that we recommend for study in Pre-Phase A (**Section 12.3.5**)

- **Sections 12.4** and **12.5** describe the integration and test plan and the verification and validation approach, respectively.

- **Section 12.6** provides an overview of the development schedule.

- **Section 12.7** concludes the chapter.

## 12.1  Management approach

The LUVOIR management approach is derived from the NASA Procedural Requirements (NPR) that govern project management:

- *NPR 7120.5 NASA Space Flight Program and Project Management Requirements*

- *NASA/SP-2014-3705 NASA Space Flight Program and Project Management Handbook*

- *NPR 8705.4 Risk Classification for NASA Payloads*

Any project that evolves from the LUVOIR mission concept study will be managed in accordance with these documents, among other relevant NPRs.

Within the constraints of these documents, there are a variety of strategies that can be employed to optimize several variables including but not limited to cost, schedule, and science yield. These strategies can be tailored to balance the interests of the Agency and the project with the availability of resources. As part of this concept study, the LUVOIR Study Team has researched lessons learned from multiple current and past NASA missions. We have also considered how other government organizations, such as the Department of Defense, manage their large projects. From this research, and using successful missions such as MAVEN, Cassini, OSIRIS-REx, and Chandra as a guide, we have identified a tailored set of strategies that may improve on the cost and schedule execution performance during the development and build of NASA's flagship missions.

For the purposes of this report, we chose to select and document strategies that are intended to minimize the overall cost and development schedule of the project. We recognize that some of these strategies require an increase to costs for certain phases or years of the project; overall, this will reduce the total cost and minimize the risks of schedule overruns. For example, we fully expect that our effort to minimize **total** cost and the Phase A through D schedule will result in higher **year-to-year** costs in some years, due to the larger workforce required at any given time to execute parallel development paths.

All of the project-level strategies presented in **Section 12.1.1** can be implemented by the project, through tailoring of the governing management requirements, as permitted by the Agency. There is one additional recommendation for an alternative funding strategy, presented in **Section 12.1.2**. Although it is not required to implement LUVOIR, it would improve any flagship's development cost and schedule performance, as indicated by its repeated prominence in many "lessons observed" documents from previous NASA flagship missions. This new funding strategy could only be implemented at the agency-level and by Congress.

### 12.1.1  Project-level strategies

#### 12.1.1.1  Early technology development

NASA requirements for instruments and missions allow a project to mature technologies to a Technology Readiness Level (TRL) of 6 by the Mission Preliminary Design Review (PDR). However, due to the nested nature of LUVOIR (and other flagship missions), it is inadequate to allow the development of technologies this late in the mission. By Mission PDR, nested entities such as assemblies and sub-systems could be through design, fabrication, integration, and test, as shown in **Figure 12-1**. Should a technology fail to mature as expected, it could alter the overall architecture of LUVOIR and subsequently alter the conceptual and detailed designs. In either case, the size of the development team grows rapidly from Phase A to Phase B, and it is imperative for a project of this magnitude to minimize changes that have a ripple effect across the entire architecture (Gady 2018; Bitten et al. 2019). On flagship missions such as LUVOIR, we believe that all technologies should achieve a TRL 6 before the start of Phase A. Our technology development plan, shown in **Chapter 11**, and Pre-Phase A Activities described in **Section 12.2**, demonstrate how a LUVOIR project would achieve that goal.

It is critical to establish technology projects early. The technology should be developed to fit an evolving architecture and conceptual design as part of Pre-Phase A. The technology development, in turn, informs the technical team as to the feasibility of the current architecture and conceptual design. It is imperative to let this cycle play out with an agile team that is still relatively small compared to the team in Phases A-D. Resources are more efficiently used with a smaller team to develop the technologies, define the science requirements, mature the architecture, and perform any final design trades before ramping up the marching army that grows once Phase A begins. The architecture and the conceptual design, informed by technology development, is what will lead to clearly defined engineering requirements for the mission segments, elements, sub-systems and so on. This process is discussed in more detail in **Section 12.2**.

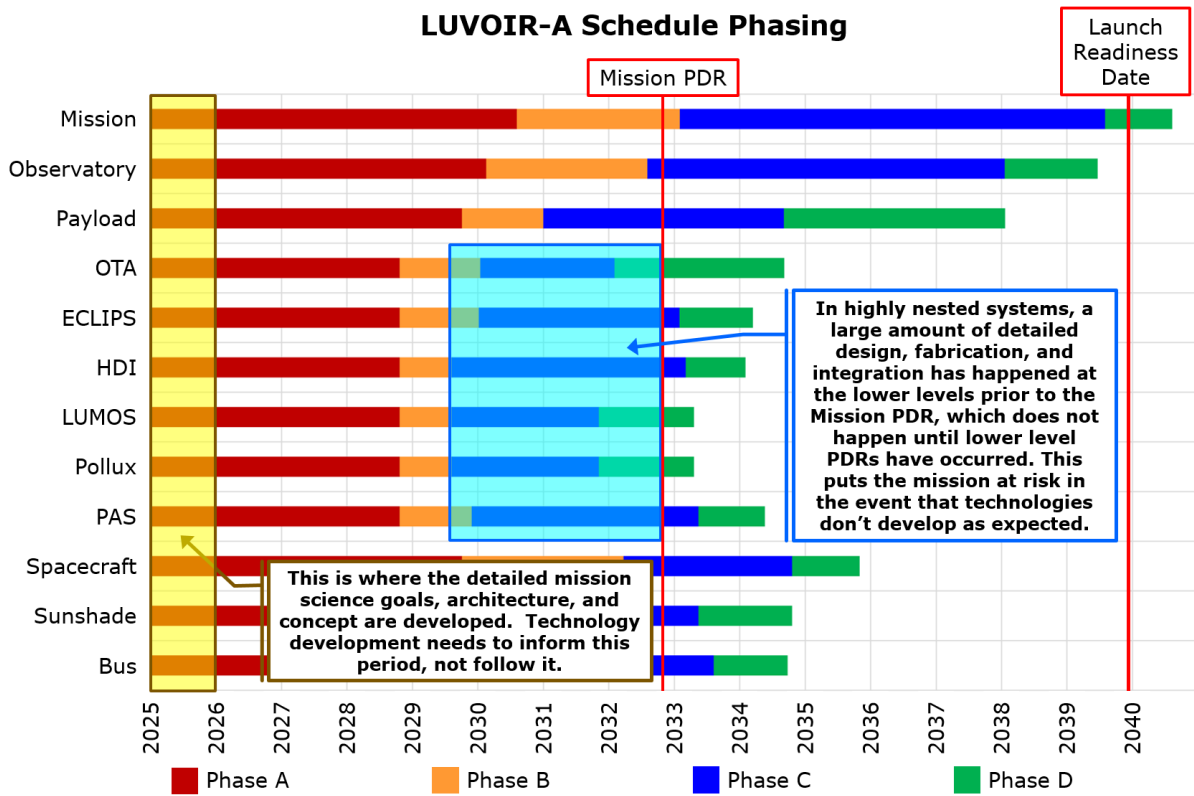

**Figure 12-1.** *LUVOIR-A phasing. This illustrates that a significant amount of the conceptual design, detailed design, and some I&T at the lower levels is complete by the time the Mission gets to PDR. The need to go back and re-engineer large portions of a mission after PDR due to failed technology development is routinely cited as a reason for NASA missions being late and over budget.*

It is also critical to understand that developing technologies that are not consistent with a specific architecture or concept design run the risk of needing to be re-engineered, potentially negating the benefits of developing technology early. The LUVOIR Pre-Phase A strategy is to mature enabling technologies—specific to a given architecture—to TRL 6 using subscale system demonstrations as described in **Chapter 11**.

While this strategy will be implemented at the project level for LUVOIR, NASA could consider requiring this strategy at the Agency level for all flagship missions. Allowing technologies to continue to develop as late as Mission PDR is an unacceptable risk posture, but individual projects (or project managers) may have varying levels of acceptance of this strategy. Making early technology development an Agency-level requirement could help improve cost and schedule performance on many of NASA's large projects.

### 12.1.1.2 Managing complexity with requirements definition

Technical complexity, defined by multiple interconnected systems, sub-systems, and assemblies, is a driver for risk, schedule, and cost; however, these risks can be minimized with appropriate considerations and plans if they are instituted early in the project's life-cycle. Understanding the boundaries and interfaces of the interconnected parts or systems can break down the complexity, simplifying ground-based verification of overall system performance.

This is particularly true for complex, highly integrated flagship-level missions. Complexity can be reduced by establishing full and clear science requirements early on, including the evaluation of the feasibility of requirement allocations. While requirements are always subject to be reviewed, modified, or waived during the design process, minimizing the number of open requirements and the length of time they are open through the detailed design phase will reduce risks to the schedule and cost (Alban 2016). Directed technology development before Phase A and early Phase A engineering test units, engineering design units, and pathfinders can be used to converge on requirements that are difficult to close without hardware design support to augment the systems engineering work.

Despite the complexity of the system and the nested nature of a large flagship mission, LUVOIR should define requirements and critical Interface Control Documents (ICDs) using a strong systems engineering discipline that enables each of the parts of the mission architecture to operate as independently as possible. It is understood that all of these systems interact; therefore product requirements evolution is inevitable. However, maturing the most challenging of these interfaces earlier in the development cycle and off the critical path will save time and reduce risk once on the critical path (discussed as part of the Pre-Phase A activities in **Section 12.2**). The more each product can exist in its own design space, the easier it is for designs to progress in parallel. Such parallel progress is critical for making the schedule as efficient as possible. These ICDs should also address the governance rules when hardware is delivered for integration into the next higher-level system as discussed further in **Section 12.1.1.7**.

### 12.1.1.3  Managing complexity with pathfinders

LUVOIR will also manage complexity through the use of pathfinders. LUVOIR's pathfinders are envisioned in two general groups: (1) those used to inform designs; and (2) those used to inform testing processes and procedures. The former will include engineering design validation units, engineering development units, and engineering test units.

An example of a "design-informing pathfinder" is one used to flesh out the modular design. For example, a pathfinder could be used to demonstrate de-integration of the OTA at the wing-fold level and then re-integration to show how repeatable the alignment is to within a certain tolerance. The pathfinder could consist of a full wing or a few PMSAs placed in one wing of the primary mirror backplane support structure (PMBSS), sufficient to allow engineers to understand how to create the modular interface. Such engineering design validation units could also be used to validate the realism of the initial alignment tolerances, error budgets, and on-orbit mechanism ranges required for opto-mechanical control. An example of a "test process and procedure pathfinder" unit is one that could be used in a thermal vacuum chamber to optimize a testing sequence and troubleshoot bugs in an initial test plan and procedure.

The I&T section (**Section 12.4.5**) explores some other possible pathfinders that LUVOIR would need. As the architecture and the design matures, other pathfinders would certainly be considered. Identifying pathfinders and other types of engineering development units is part of the Pre-Phase A activities identified and discussed in **Section 12.2**. The intent is to use these pathfinders off the critical schedule path to help mature and optimize designs, plans, procedures, test setups, and integration sequences as much as possible for a "one-of-a-kind design" such as LUVOIR.

"Carefully planned rehearsals of test equipment, procedures, and personnel was originally a lesson from Chandra. In the case of JWST, this was highly successful for the telescope, which used a pathfinder including four flight spare mirrors and three separate cryogenic tests that checked out all of the optical and thermal testing." (Feinberg et al. 2018)

### 12.1.1.4  Managing complexity with modular design

LUVOIR intends to further manage complexity by making the architecture and design modular. While not necessary to achieve the science goals described in this report, LUVOIR is being designed for serviceability to enable science programs beyond those detailed in **Chapters 3–6** via a longer operational lifetime, potential instrument upgrades, and replacement or upgrades of spacecraft parts (Launius et al. 2014; Cowen 2015; Wiseman 2015). With a large, complex observatory, designing for serviceability may add design effort, but it also forces the design to be modular and for entities to be easily accessible. This access to components, subsystems, and systems provides an additional benefit during I&T, including lowering transportation complexity during this phase of the project. In addition, incorporating modularity in the design in Phase A does not occur on the critical path or with a marching army in place. Failing to utilize a modular design will bring added integration complexities that will need to be dealt with while a marching army is in place and on the critical path.

Modularity that facilitates potential future servicing means LUVOIR is built up from well-defined modules designed for initial alignment, de-integration, and re-alignment during the build process. It also enables de-integration by modules to reduce complexity of shipping and reassembly at a test site and the launch site. This also enables clear interface requirements definitions for partners. The I&T section (**Section 12.4.2**) further examines the details and benefits of modularity.

### 12.1.1.5  Enable parallel manufacturing, integration, and test operations

The more things that can be done in parallel, the more efficient the schedule can be. Successfully initiating parallel operations requires significant upfront planning. There needs to be a master plan and vision—a storyboard—with all of those parallel paths mapped out rigorously in an integrated master schedule. This plan will be iterated multiple times as the architecture and the concept designs evolve. Project management needs to be vigilant and diligent in their handling of these relationships. Constant, honest communication is essential.

Parallel operations require significant training because pseudo-independent teams of people will be required to execute identical operations at the same time (Werneth 2001). This means there cannot be a single expert acting as a bottleneck anywhere that parallel activities occur. While an expert will be required to initiate a given plan, training of and delegation of authority to multiple teams will be required.

LUVOIR intends to enable parallel developments in many critical areas—especially those related to the fabrication, assembly, and testing of the primary mirror segment assemblies (PMSAs). The current concepts for LUVOIR-A and -B have a total number of PMSAs ranging from 55 to 120. In order to keep the development schedule implementable, a number of mirror substrates need to be manufactured in parallel (Cole 2017). Likewise, once the substrates are manufactured, the assembly and subsequent testing of the PMSAs needs to be done in parallel. LUVOIR's current development schedule requires four teams to assemble and then test PMSAs in parallel. Once the PMSAs are qualified and the PMBSS sections are assembled and qualified, PMSAs will need to be aligned and attached to each of the PMBSS sections in parallel. This requires that all PMBSS sections be assembled and tested in parallel prior to these operations.

Of course, parallel operations will rapidly grind to a halt if the project funding is inadequate or not phased correctly. More information on project funding can be found in **Sections 12.1.2**, **12.2.5**, and **Appendix J**. The integration and test section (**Section 12.4.2**) explores further justifications for and examples of parallel manufacturing and I&T operations that would be useful to LUVOIR.

### 12.1.1.6  Acquisition and partner strategy

There are several important reasons for NASA to act as the "prime contractor" (or "prime") on its large, flagship missions. Some of the primary benefits are:

1.  NASA will set the requirements for the mission in Phase A, and maintain those requirements throughout the mission lifecycle, coordinating requirements changes across all of the various partner interfaces.

2.  NASA will have control over the interfaces, will define the procurements, and will set the required pass/fail criteria for the final product deliverables. This includes setting the acceptance performance metrics, delivery schedules, governance rules for higher levels of integration, and any other aspects that are necessary for inspection.

3.  With NASA acting as prime—issuing multiple, smaller, openly competed contracts for individual products—industry partners avoid a long, costly proposal process for a single "winner take all" contract award. This enables earlier and broader involvement from partners by soliciting work on aspects of the mission that more closely align with individual partners' capabilities and expertise.

4.  By allowing NASA to procure things at the subsystem level, NASA is able to select the "best in class" providers according to specific expertise and capability.

5.  NASA retains expertise by being the system integrator. This also enables NASA to grow and maintain its systems engineering core competency.

6.  While LUVOIR is anticipated to be primarily a US mission, there is significant international interest. It is an inherently governmental role to carefully manage international contributions and relationships.

In Pre-Phase A, NASA can continue to contract with industry and academia using numerous Broad Agency Announcements (BAAs) to leverage their knowledge and expertise to help mature many parallel technologies. Once NASA sets the system-level requirements and has defined critical system/sub-system interfaces and the higher-level integration interface requirements, multiple Requests for Procurement (RFPs) can be openly competed to build individual LUVOIR products. Since the interfaces and requirements have been clearly defined prior to issuing these requests, partner organizations can follow their own internal design and development processes. Establishing this entire process upfront enables earlier team selection, broader buy-in, and supports the forward-looking, integrated team approach recommended in **Section 12.1.1.8**. This process provides the benefit of early partnerships between government and industry/academia to build the mission as one team, with all members and partners having a stake in the success of the project.

### 12.1.1.7  Managing institutional requirements

We propose that a single NASA Center be responsible for managing the mission and formally delivering the segment-level products, with the other NASA centers, industry partners, international partners, and academic partners involved with the engineering and development of LUVOIR.

The LUVOIR team embraces a multi-center and multi-partner approach to mission development, since the engineering challenge is so broad. However, with this breadth of development comes the challenge that each entity has their own set of rules governing the design, manufacturing, integration, and test of products. While all NASA requirements stem from agency-level requirements, they are all tailored and optimized for each center's capabilities and experience. Likewise, each industry, international, and academic partner has their own set of rules.

Having multiple rules does not necessarily present problems during the development of individual products. However, it can become problematic during integration and test at higher levels of assembly, where anomalies can arise when multiple products from multiple partners come together in one place. These anomalies create inefficient scenarios when there is debate as to which rules are going to be followed—the ones under which a product was developed or the ones under which that product is being integrated into a larger system.

As an example, assume that the payload is being integrated and tested at one NASA center. The lower level products will need to be informed of institutional requirements at that NASA center and how those rules will affect the handling of those products—before it arrives for integration—so that they can be accommodated in the design.

The LUVOIR team will address and communicate which rules will be in place at various levels of assembly ahead of time so that any issues or conflicts can be avoided as much as possible. These scenarios should be anticipated and their resolution process should be communicated as part of the interface, performance, specifications, delivery, and acceptance testing requirements. One default position would be that, unless there is a waiver, the "higher-level" system rules will be followed when anomalies occur. These rules would need to be established as part of Phase A and should be defined by the NASA project office. **Figure 12-2** demonstrates the institutional requirements flow.

This does not preclude agreements that allow partners to use their own internal rules for design, development and testing of products. It does acknowledge that different entities do things differently and ultimately those products are going to be handled at a higher level of assembly at a center that may impose rules in conflict with those from a lower level. The intent is to inform all stakeholders and to minimize compatibility issues that will inevitably present themselves during integration and test at various levels.

### 12.1.1.8  Integrated 'one team' environment

The LUVOIR mission architecture, shown in **Figure 12-3**, represents the mission organization, i.e., the flow of products from the bottom into higher-level products at the top. The management of those products will likewise be organized in a pyramidal fashion, creating clearly defined roles with clearly defined lines of authority and accountability (Mitchell 2015)—both being critical to the successful execution of LUVOIR.

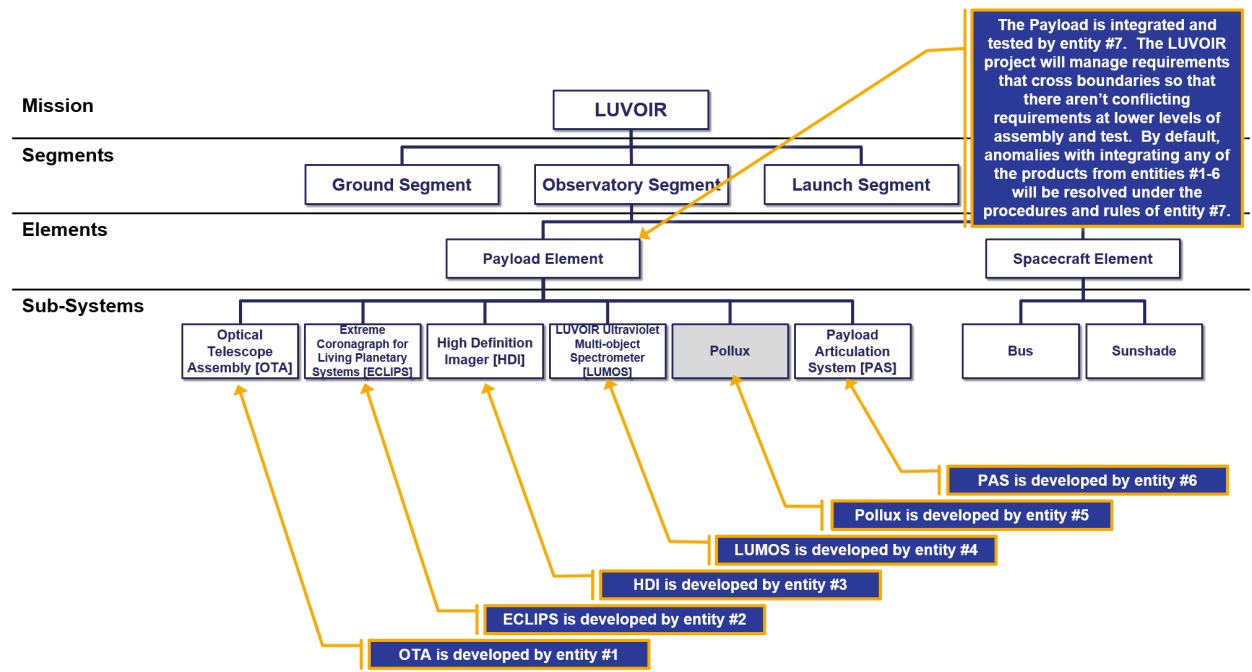

**Figure 12-2.** *A part of the LUVOIR mission architecture, with notes on how institutional requirements at a given level of assembly and testing can influence the development of products to minimize conflicts at these higher levels of assembly.*

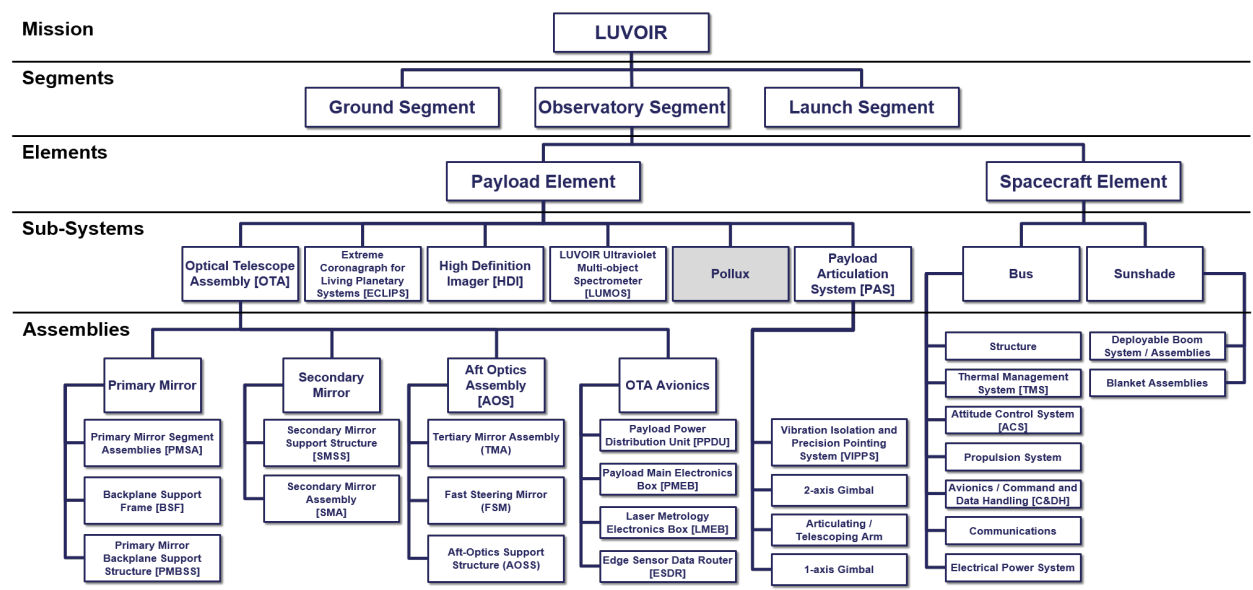

**Figure 12-3.** *The LUVOIR mission architecture serves as a guide for how the team will be structured. Management lines of authority and accountability will flow down from the top, with team members affiliated with the product they are delivering and not by the organization identified on their badge.*

While we envision that a single NASA center will be responsible for managing the project and for delivering top-level products, we acknowledge and embrace the fact that the success of the project will require the active involvement of many partners, including other NASA centers, industry, academia, and international agencies. While this brings the benefits of the diversity of expertise represented by those groups, it also brings challenges that stem from the global distribution of partners and the different cultures that exist at these different institutions. It is therefore critical that the project investigates ways to structure contracts and partner agreements in a way that encourages cooperation between entities. This could increase the likelihood that the project has the best people working on the specific products that need their attention, regardless of their affiliation; just because contractor X is responsible for product A, does not mean that someone from contractor Y should not be able to provide their expertise to solving issues with that product.

Of course, there are limits to and restrictions on this approach. There exists a need to protect each partner's intellectual property. And federal regulations governing International Traffic in Arms Regulations (ITAR) and Export Administration Regulations (EAR) must be followed. These are constraints that procurement and legal teams will need to work within. Ultimately, mission success needs to drive the makeup of the team while still rewarding partners for supplying the necessary human capital and technical know-how.

Complex missions require forward-looking, proactive, integrated teams with clearly defined roles, adequate resources, and a positive project culture. In particular, positive project culture instills trust and empowers individuals to take ownership. It supports clear communication and camaraderie within the framework of well-known and understood deliverables, schedules, and processes. And there are few experiences as rewarding as working on a challenging flight project with exciting and compelling science goals, a positive project culture, and an integrated 'One Team' attitude.

> "From all of the respondents, the single most important aspect of our ultimate success was the development of a truly integrated team." (Arenberg et al. 2014)

### 12.1.1.9  Team, experience, and depth

Another factor in project success—based on lessons learned from previous flagship missions—is having a leadership team with hands-on, relevant space flight mission development experience (Werneth 2001; Martin 2012). There is no substitute for end-to-end project lifecycle experience. A project of LUVOIR's magnitude will extend for many years and the project needs to be multiple people deep in critical positions in anticipation that not everyone will be able to finish the project. A key element to the success of LUVOIR (or any flagship) will be the training of emerging talented engineers, scientists, and managers throughout the project lifecycle.

Consider the architecture diagram in **Figure 12-3** and pick any product block: there should be at least two people who are subject matter experts (SMEs) that are cognizant of that product and who can each step in and lead should the other not be available. In this context, it is completely acceptable that one of those two people be a more junior person who is being mentored by the more experienced, senior person. In that way, both LUVOIR and the Agency benefit from training on the job.

### 12.1.2  A recommended alternative agency-level funding strategy to improve flagship cost and schedule performance

As the LUVOIR Study Team researched lessons learned from other flagship and government large projects to inform the management strategies presented above, one issue was consistently cited as a significant factor in mission cost and schedule overruns: the phasing and stability of funding appropriations. While this issue is beyond the control of any given project, or project manager, this section discusses principles that would help reduce the total costs of LUVOIR (or any flagship).

The concepts, costs, and schedules presented in this report assume an ideal funding profile—that all funds are available when they are needed (Lee et al. 1994). This assumption was made per the direction of NASA HQ, which convened the study that led to this report. We acknowledge the optimism in this assumption; LUVOIR will of course need to adapt the planned implementation based on the funding profile provided by NASA and by Congress. This is how past NASA flagships have been developed, and is likely how future missions will continue to be developed. Yet the significance of resource phasing and stability is so great, that we take this opportunity to propose an alternative approach, to be considered by NASA and by Congress.

Below, we first present the current method of funding NASA's large missions. We then present an alternative approach based on the research completed during this study. We recognize these funding profiles are beyond the direct control of any project or the Decadal Survey. We discuss them here as part of our team's holistic approach to considering how a LUVOIR-like project would be best implemented and could, perhaps, turn a lesson observed by many previous NASA flagship missions into a lesson *learned*.

### 12.1.2.1  Current NASA flagship funding strategy

There are two main aspects of funding NASA's large missions. First, the total cost of the mission is estimated. Second, the funds are disbursed according to that estimate.

**Estimating costs**

Estimating the cost of NASA's large missions is extraordinarily difficult. By the very nature of NASA's work, each new mission has never been done before. Therefore, one cannot completely rely on previous missions to determine the costs of new missions. Often, costs must be extrapolated from one or two data points that have some correlation to the mission being estimated, or from many data points that have very little to no correlation to the mission being estimated. Neither approach leads to high confidence in the final result.

Grassroots estimates attempt to circumvent this issue by breaking the work down into ever smaller pieces, until one can have relatively high confidence in estimating the amount of time, personnel, and materials needed to complete each piece of work. It is difficult to estimate the cost of a year-long effort to design a complete instrument, but it is easy to estimate the cost of a week's-worth of an optical engineer's time to design a single optical component.

This implies that the path to the most accurate, highest-confidence cost estimates is a detailed understanding of the design and *all* of the work that needs to be done. The opposite is also true: the more immature a design is, the less accurate its cost estimates will be. Historically, NASA has relied on relatively immature mission concepts to estimate the costs of the fully implemented missions. Immature technologies, evolving science requirements, and developing concepts all ensured that those cost estimates could never accurately represent all of the work that would be needed to implement the final mission.

Fortunately, this study (and other flagship studies conducted in advance of Astro2020) have reached an unprecedented level of detail and maturity for pre-decadal mission concepts (tailored Concept Maturity Level 4; Hertz 2015; NASA 2019). However, more study is required before accurate cost estimates can be provided for any of these missions. This creates a conundrum for any Decadal Survey panel: it must make decisions about missions in advance of an accurate understanding of their costs. In **Section 12.1.2.2**, we describe a potential strategy that can help resolve this conflict, by advancing the maturity of LUVOIR in order to obtain an accurate cost estimate, while providing the astrophysics community with the flexibility and authority required to make a final decision on LUVOIR once its full costs are better known.

**Fund disbursement**

NASA receives annually appropriated funding, leading NASA's flagship missions to have a limited, not-to-exceed appropriation allowance each fiscal year. Furthermore, that allowance is typically based on what funds are available to the agency, and not on what the project actually needs to accomplish that year's work. In a flat budget, this available funding, or "funding wedge," only begins to open up as previous missions near the end of their development.

Flagship missions are a complex, nested, system of systems that ideally follow an optimal, integrated master schedule to be implemented. Parallel paths of development and assembly are carefully planned and timed to be completed and integrated into the next higher-level of assembly. When funds are disbursed based on availability, and not on need, it causes the project to replan that schedule, often deferring critical work until later years. This cascades until eventually the deferred work must be completed while a large, marching army of engineers, designers, managers, and scientists are waiting for its completion. This is the most expensive time in a mission's lifecycle, and any delays in schedule execution quickly deplete available reserves.

In addition to the above limited funding "allowance" not being optimal, there is an added complexity for annual appropriations. NASA has received an **on-time** annual budget only 7 times throughout its more than 60-year history (Martin et al. 2012). Continuing resolutions (CRs) are the norm, not the exception. Under a CR, each new fiscal year's budget is set to the previous year's level of funding until an actual appropriation is passed. Thus, even when projects are able to replan their schedule to account for the sub-optimal phasing described above, a CR requires even more replanning and deferment, compounding the effect. Many documented NASA flagship lessons-learned documents (Martin 2012; Feinberg et al. 2018; Arenberg et al. 2014; Bitten et al. 2019; Crooke et al. 2019) attribute a portion of flagship cost and schedule overruns to a non-optimized funding profile and unstable funding due to CRs.

> "NASA leaders must temper the Agency's culture of optimism by demanding realistic cost and schedule estimates, well-defined and stable requirements, and mature technologies early in the development. They must also ensure that funding is phased appropriately, funding instability is identified as a risk, and project managers are appropriately rewarded and held accountable for meeting project cost and schedule goals." (Martin 2012)

These two issues, inaccurate cost estimates and unstable sub-optimal funding appropriations, continue to have a significant impact on NASA's ability to complete missions on schedule and within budget. This is not to say these are the only issues. Indeed **Section 12.1.1** identifies numerous other strategies within a project manager's control that also contribute to a project's successful implementation. But NASA has shown in the past, with the Apollo Program and with the Shuttle Return to Flight after the loss of Challenger, that fully funding and front-loading a program can lead to a successful, timely completion.

> "Funding instability (receiving less money than planned or funds are disbursed on a schedule different than planned) is cited as among the most significant challenges to project management at NASA that causes management to delay work which in turn can lead to cost increases and schedule delays. When planned funding does not materialize, project managers may defer development of critical technologies to a time when integration of those technologies may be more difficult or when the cost of material and labor may be greater" (Martin 2012)

### 12.1.2.2  Recommended future NASA flagship funding strategy

We propose that large projects be broken into discrete blocks of work, with each block of work being fully funded at one time. This is an alternative to the current status quo, where missions are incrementally funded year-to-year based on cost estimates (and caps based on those estimates) for the full project lifecycle that are made when the design is immature and the full lifecycle costs are uncertain.

**Figure 12-4** shows an example of how this new method would work, based on the LUVOIR implementation schedule presented in **Section 12.6**. Six funding blocks are defined over the course of the project lifecycle, with five of those blocks corresponding to the project formulation period (Pre-Phase A through Phase B). Each block is defined by a gate, called a Funding Decision Point (FDP), to be differentiated from NASA's standard Key Decision Points (KDPs) that define mission phases. At each FDP, four things occur:

1. The project must demonstrate that it has successfully met all of the necessary criteria that define the completion of that block of work.

2. A new mission cost is estimated, based on the latest designs, requirements, and technology readiness, including an accurate and reliable estimate for the next block of work.

3. The project generates a high-fidelity budget request for the next block of work to be completed.

4. NASA decides to commit to funding only the next block of work, based on the budget request, updated cost estimate, and project success.

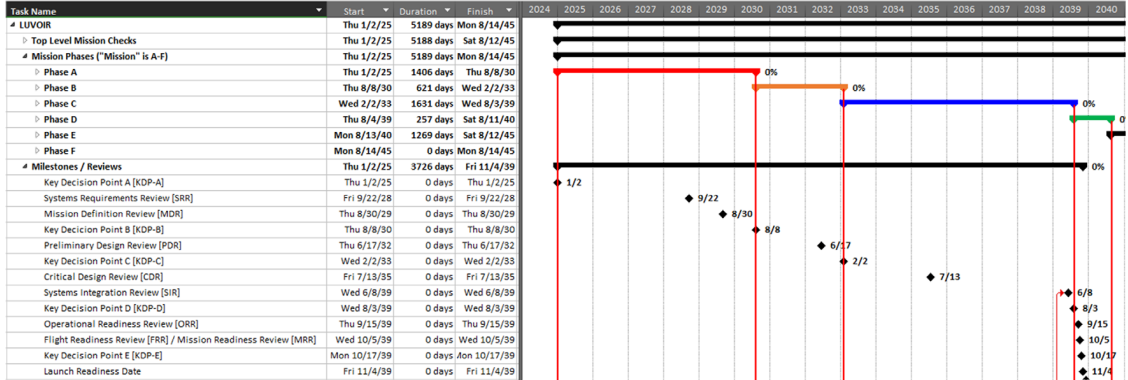

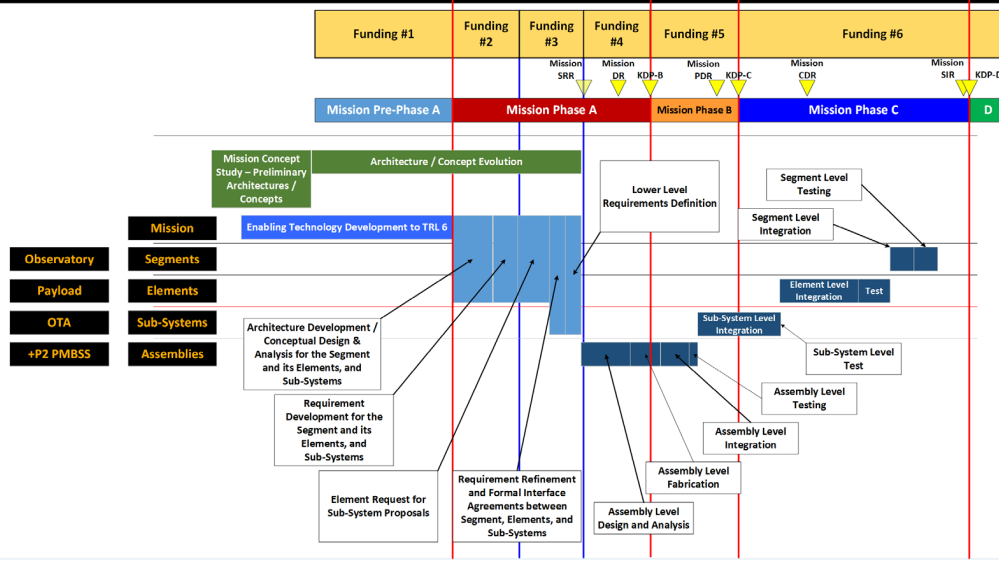

**Figure 12-4.** *Example of funding blocks, based on the LUVOIR-A implementation schedule. Each funding block covers a set amount of work, with a defined gateway that the project must pass before it receives the next block of funding.*

**Table 12-1** summarizes the notional funding blocks, FDPs, and success criteria we envision for LUVOIR.

It is important to also note that each funding block is *fully funded* at its initiation. This is at odds with NASA's current annual appropriations. But there are precedents for providing multi-year federal appropriations for large projects. The Department of Defense routinely uses funding vehicles such as No-year Funding, Incremental Funding, Multiyear Procurement, Block Buy Contracting, Economic Order Quantity Authority, and Advanced Procurement for large capital investments such as submarines, fighter jets, aircraft carriers and other warships (Birkler et al. 2002; O'Rourke 2006; 2007; 2019; O'Rourke & Schwartz 2019; Crooke et al. 2019). These different funding vehicles have different restrictions on their use. However, they represent the ability to commit multiple years'-worth of appropriated funds at one time, to ensure the optimal expenditure of those funds over the course of the program.

**Table 12-1.** *Funding blocks and Funding Decision Point (FDP) criteria, based on the LUVOIR implementation schedule. FDPs correlate with key project milestones that signal an advance in the fidelity of the system design, and therefore a more accurate estimate of the mission cost.*

| Funding Block | Funding Decision Point | Decision Point Criteria |
|---|---|---|
| 1 | Start of Pre-Phase A | Decadal Prioritization. Agency decision to proceed with mission Pre-Phase A study. |
| 2 | Start of Phase A | All technology systems demonstrated to TRL 6.. |
| 3 | Issue Requests for Proposals (RFPs) | Requirements developed to sub-system level. Ready to issue RFPs for all industry, academic, and international partners. |
| 4 | Mission System Requirements Review (SRR) | All requirements developed to lowest level. Project successfully passes Mission SRR. |
| 5 | Key Decision Point (KDP) - B | Mission satisfies all criteria for completing Phase A. |
| 6 | KDP - C | Mission completes Preliminary Design Review and satisfies all criteria for completing Phase B. |

> "A proposed approach to developing Flagship missions should help eliminate some of the historical issues in Flagship development. This approach would ensure that a programmatic baseline is established after both the technology and design have matured such that an accurate estimate could be developed. It has been shown that a policy that sets the programmatic baseline after a mature design had been developed, provides the ability to manage the program to cost." (Bitten et al. 2019)

The approach we outline here has several advantages over the current way NASA's flagships are funded and executed:

1. Since multiple cost estimates are performed on designs that become more and more detailed over the course of mission formulation, the accuracy of those cost estimates increases at each step. Runaway cost growth is signaled by the cost estimate at any FDP being outside the error bars of the previous FDP.

2. Congress and NASA need only commit to funding the next block of work, and not the full mission. Each FDP provides the opportunity to delay, augment, or deny the next funding block in response to potential cost growth or inadequate project performance. The predetermined FDPs also allow NASA and Congress to better plan budget requests and appropriations.

3. Since each block of work is fully funded up front at one time, project managers can execute an optimal schedule within the block, without worrying about the unpredictability of annual appropriations and continuing resolutions. Parallel development work can be optimally planned and executed. Cost and schedule reserves can be better defined and, most importantly, used for what they are meant for the "unknown unknowns" that will inevitably be revealed, and not to execute deferred work while the marching army waits.

4. There will be a lower risk of cost overruns on the flagship mission, and a lower risk of any overruns that do happen negatively impacting other projects and priorities.

In the short-term, the funding that is provided for any given block of work is much less likely to experience cost overruns compared to the current model of full-project funding. In the long run, the existence of FDPs gives stakeholders the opportunity to cancel or delay the project if total mission costs end up being significantly higher than what was anticipated or what can be afforded. Finally, while cost overruns in NASA's Astrophysics Division do not generally impact smaller projects, the proposed funding model would help insure smaller projects are not negatively impacted in the future, because the funding for the block of work for the flagship mission would be provided up-front and effectively be placed in a separate silo from the year-to-year funding for smaller projects.

We recognize that this proposal is a dramatic change from the way NASA flagships are currently executed, and indeed from the way NASA as an agency is currently funded. Yet accurate cost estimation and funding stability have been shown over and over again to be critical to controlling cost and schedule growth. The above proposal allows for an alternate approach—using methods already available to the federal government—to control runaway flagship costs and routine delays in schedule.

## 12.2 Pre-Phase A activities

To facilitate many of the management strategies discussed in this chapter, we recommend that a project office, depicted in **Figure 12-5**, be established. While a project manager has overall authority for the Pre-Phase A activities, a project development team composed of each activity lead provides coordination of the major Pre-phase A activities, which are described in the following sections.

### 12.2.1 Project management and project development team

The project management team is responsible for the overall logistics and execution of the Pre-Phase A study activities. The project manager has final authority on all project decisions, and is ultimately the one responsible for project success. In addition to project management

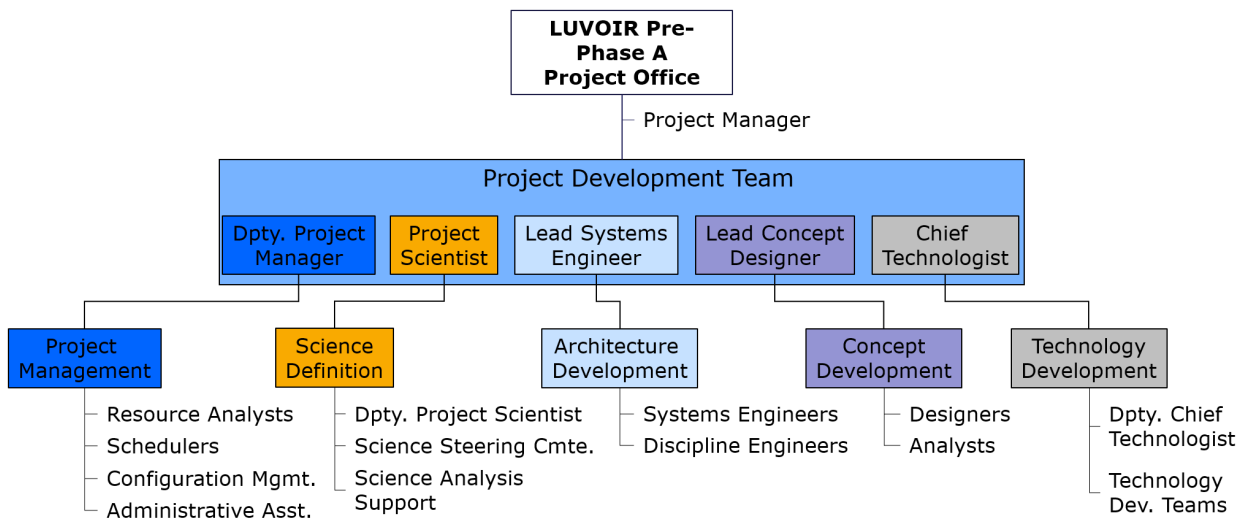

**Figure 12-5.** *Organization of a Pre-Phase A project office to coordinate project management, science definition, and architecture, concept, and technology development efforts.*

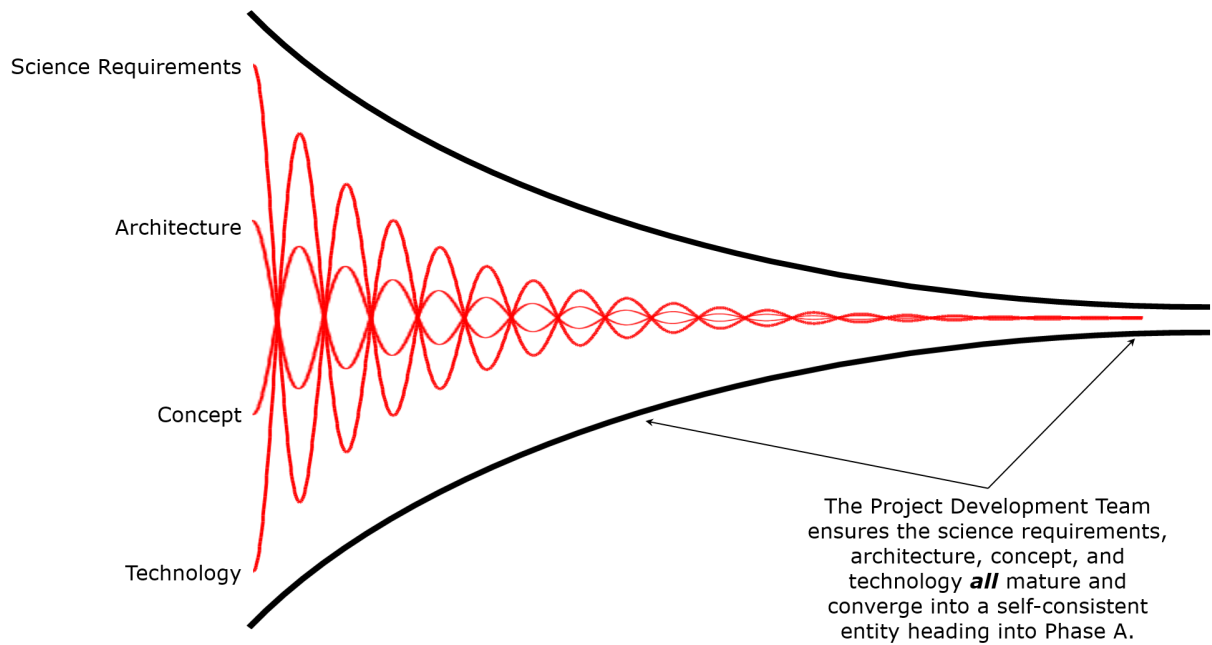

Science Requirements

Architecture

Concept

Technology

The Project Development Team
ensures the science requirements,
architecture, concept, and
technology **all** mature and
converge into a self-consistent
entity heading into Phase A.

**Figure 12-6.** *This is a conceptual image of how the project's science requirements, architecture, concepts, and technologies will converge before entering Phase A. The Project Development Team—comprising the Deputy Project Manager, Project Scientist, Lead System Engineer, Lead Concept Designer, and Chief Technologist—will be responsible for ensuring this convergence.*

skills, the project manager will need a workable understanding of the science, engineering, and technology aspects of the entire endeavor.

Of course, a "workable understanding" does not imply expertise. A Project Development Team, comprising the leads from each sub-activity, will serve as a leadership council to inform project management decisions. This council will ensure that proper coordination between each activity is occurring: that the science requirements and technology demonstration are flowing into the architecture, that the concept is being validated against the science objectives, that project milestones are being met within budget and schedule, and so on. The Project Development Team is critical to ensuring that the science, architecture, concept, and technology all converge to a single, cohesive entity by Phase A, as depicted in **Figure 12-6**.

### 12.2.2  Science definition

A science steering committee (SSC) will be responsible for defining the science objectives of the mission. The SSC, established as part of the Pre-Phase A project office, will be maintained as an organization throughout the project lifecycle. The members of the SSC will be drawn from the community, in ways that both maintain organizational knowledge and bring in new expertise on a regular basis. Throughout Pre-Phase A, the SSC will establish a process by which science objectives are proposed, reviewed, and accepted by the project. It will work with the rest of the project team to decompose science objectives into requirements to guide the architecture, concept, and technology development. Finally, the science team will perform the necessary analysis to validate the concept designs against the science objectives.

One of the project-level management recommendations discussed earlier is to manage system complexity with requirements definition. This is largely thought of as an engineering responsibility to ensure that all "To Be Determined" or "To Be Resolved" (TBD, TBR) requirements are filled in before the detailed design phases begin. However, there is an equal responsibility on the science team to assist in creating well-posed requirements while limiting scope creep. Late changes or additions to the top-level science objectives will ripple throughout the sub-systems and assemblies, and can significantly impact their design and implementation cost and schedule. The SSC will not only be responsible for defining the science requirements, but also protecting them from external pressures to change mission scope.

### 12.2.3  Architecture and concept development

The Pre-Phase A project office will be responsible for: fielding architecture and concept development teams to continue maturing the architecture and concepts that have been presented in this report; exploring additional trade studies that this study was unable to conduct; and refining the concept designs in response to the Decadal Survey recommendations.

The architecture development team will also begin the extensive systems engineering required to implement a mission as complex as LUVOIR. This includes long time-horizon planning that must be considered in conjunction with the system architecture development. Such planning would include facilities, pathfinders, servicing, verification and validation, and interface development in preparation for partner participation. The work of the Project Office on each of these areas during Pre-Phase A constitutes the next sub-sections of this report.

### 12.2.3.1  Facility development planning

The scale of the fully developed LUVOIR systems presents a challenge to integration, test, verification, and validation. While some facilities may exist for certain aspects of the LUVOIR test campaign, as discussion in **Section 12.4.4**, many will require upgrades. In some cases, whole new facilities may need to be constructed. The project office will survey existing facilities to identify these gaps and develop a plan to implement facility upgrades or new construction. Starting this process early will ensure that funding can be secured in time to begin construction efforts to reduce schedule risk later in the project lifecycle.

### 12.2.3.2  Pathfinder planning

Pathfinders are valuable tools in managing system complexity. As discussed in **Section 12.1.1.3**, pathfinders will be used to inform both designs and testing processes and procedures. As the architecture evolves, the team will identify where pathfinders can best be used to reduce risk. Some of these pathfinders may leverage testbeds that will be built as part of the technology development program. Additional potentially useful pathfinders are explored in **Section 12.4.5**.

### 12.2.3.3  Servicing approach

During the course of our study, we engaged with several groups developing on-orbit servicing infrastructure to understand how to build serviceability into the LUVOIR architecture.

However, our study was limited to how to make LUVOIR service**able**, and not how one would actually go about servicing it (i.e., service *in situ* at L2 vs cis-lunar space; robotic servicing vs. human-assisted, etc.). Whether LUVOIR gets serviced will be left for future leaders of NASA and its stakeholders to decide. *How* LUVOIR gets serviced will likely be decided by NASA's Human Exploration and Operations Mission Directorate (HEOMD). Enabling the servicing of LUVOIR will require government-wide participation, leveraging assets within NASA, the Department of Defense, and numerous commercial entities. The Pre-Phase A project office can begin laying the foundation for this effort, developing relationships with these entities, incorporating servicing into the architecture as the necessary infrastructure matures, and planning for the long-term funding that will be required for a 5- or 10-year servicing cadence. Further details of how LUVOIR accommodates servicing is discussed in **Section 8.1.2**.

### 12.2.3.4  Verification and validation approach

As with new facilities, the verification and validation (V&V) of LUVOIR will be challenged by its scale and complexity, and will likely require a heavier reliance on validation-by-analysis than previous systems. Understanding how the modularity of LUVOIR's architecture may lead to new verification and validation approaches will require study, and will need to be developed along with the architecture. Most importantly, the models that will enable verification and validation by analysis will be developed as part of the technology development plan in Pre-Phase A. These models will require continuous development, maintenance, and exercising throughout the project lifecycle. **Section 12.5** provides further details on LUVOIR's approach to V&V.

### 12.2.3.5  Interface development

As with previous flagship missions, NASA will involve partners across the federal government, industry, academia, and international agencies. As described in **Sections 12.1.1.6**, **12.1.1.7**, and **12.1.1.8**, NASA will leverage industry and academia in Pre-Phase A to help mature technologies. After all technologies have been matured, and after NASA sets and defines all requirements, it will be critical to LUVOIR's successful implementation to ensure that partnership agreements and interfaces are clearly defined. The Project Office will define the interfaces in the context of the system architecture.

### 12.2.4  Technology development

Finally, NASA will coordinate and execute the technology development plan to mature all technologies to TRL 6 as described in **Chapter 11**. As the architecture continues to mature, new technology needs may be identified, or some of the technologies we've discussed here may be retired as no longer necessary. Conversely, advancements in these (or other newer) technologies may drive the architecture in a preferred direction. An agile project office with direct control over the technology development program is better positioned to respond to this changing landscape for efficient use of architecture and technology development funds.

### 12.2.5  Pre-Phase A staffing and cost

The activities described above will require resources in addition to those described in **Chapter 11** for the technology development plan. The Project Office will be led by a project

**Table 12-2.** *Pre-Phase A project costs by year, in FY20$. These costs include all of the technology development efforts described in Chapter 11, as well as the Pre-Phase A activities described earlier in this section. This table assumes the project office is formed at the start of FY20 and runs until the start of Phase A on 1/1/2025.*

| LUVOIR Pre-Phase A Costs by Year (FY20$) | | | | | | | | | |
|---|---|---|---|---|---|---|---|---|---|
| | CY2020 | CY2021 | CY2022 | CY2023 | CY2024 | CY2025 | Baseline Cost | 30% Cost Reserve | Total Cost |
| Technology Development to TRL 6 (Chapter 11) | $40M | $80M | $93M | $97M | $92M | $10M | $412M | $124M | $536M |
| Pre-Phase A Project Office (Chapter 12) | $6M | $24M | $24M | $24M | $24M | 0* | $102M | $31M | $133M |
| Totals: | $46M | $104M | $117M | $121M | $116M | $10M | $514M | $155M | $669M |

*Phase A starts on 1/1/25, and costs associated with that effort are included in the total mission lifecycle cost, reported elsewhere.

manager and deputy project manager, and will require administrative, resource analysis, configuration management, and scheduling support staff.

For the SSC, we imagine a model similar to the Science and Technology Definition Teams (STDTs) that were commissioned for these pre-Decadal study teams. However, the membership of the science working group must receive funding, as their level of involvement and responsibility will be greater than what was expected of the Pre-Decadal STDTs. A science support analysis team that conducts detailed simulations of the performance of the observatory will also be necessary.

The architecture and concept development teams will need to be considerably larger than what was available for the pre-Decadal study teams. An emphasis on systems engineering will be needed for the long-horizon activities discussed in **Section 12.2.3**. Finally, the technology development office can be relatively small, as most of the effort is already captured in the technology development plans. A chief and deputy technologist would be sufficient to coordinate the technology development activities with the architecture development study.

Assuming this full complement of managers, scientists, engineers, and support staff, and their additional costs (materials, travel, etc.), we estimate the cost of operating this Pre-Phase A project to be ~$24M per year (FY20$). The total Pre-Phase A costs, including all of the technology development efforts described in **Chapter 11**, are summarized in **Table 12-2**. We include a 30% cost reserve, for a total Pre-Phase A cost of ~$670M over a roughly 5-year period.

## 12.3 Systems engineering

For successful execution of a mission such as LUVOIR, it is imperative that the system complexity not be underestimated. A strong, coordinated systems engineering effort early in the project lifecycle will be critical to this goal. How will the system be integrated? How will it be verified and validated? How will interfaces be designed and controlled? These are a few of the standard questions that any system engineering management plan must answer, but will be all the more challenging due to LUVOIR's scale, complexity, and long time horizon.

A model-based systems engineering approach will be necessary to efficiently represent and analyze the nested and interconnected LUVOIR system. Preliminary integrated models were developed as part of this concept study that incorporated structural dynamic response with linear optical model wavefront predictions. However, the complexity of this mission coupled with the dynamic range and precision required of the analyses—even at this early stage—caused this relatively simple set of models to stretch the capabilities of our small study team. As the LUVOIR architecture continues to evolve, emphasis on high-fidelity, efficient integrated modeling will be necessary. This is especially true because of the reliance on these models for final system verification and validation.

The foundation for this systems engineering process will be laid in Pre-Phase A, when detailed architecture and concept development will pick up where the study described in this report leaves off. Although the LUVOIR engineering team performed many trade studies, we were unable to exhaustively explore every side avenue that revealed itself. Furthermore, as the study progressed, new information became available, such as new launch vehicle capabilities and a better understanding of coronagraph performance drivers. The Pre-Phase A team will need to re-evaluate the architecture and concepts described in this report in the context of the Decadal Survey recommendations, and identify which portions of the trade space may need to be re-explored.

As these trade studies are performed, a clearly-defined decision process must be established to narrow the trade space. These decisions cannot be made solely by the systems engineering team, but must incorporate science yield analyses and demonstrated technology performance metrics. As **Section 12.2** describes, a Pre-Phase A project office that incorporates science, architecture, technology, and concept development teams will be established to facilitate this process.

### 12.3.1  Mission class and fault tolerance

LUVOIR will be executed as a Class A mission with a development approach that provides all required mission assurance protocols and preventive measures and procedures. Both LUVOIR concepts described in this report are consistent with design choices that are fault tolerant, with block, functional, and sub-system internal redundancies including all required cross-strapping, autonomous fault detection, isolation, and recovery. Payload accommodations meet all pointing, power, thermal, and data handling requirements with significant margin.

### 12.3.2  Risk management

NASA employs well-defined risk management and risk-informed decision-making processes, as documented in NASA Procedural Requirement 8000.4B. For such a large endeavor, the management approach should address known risks with the potential for high-impact consequences as cost and schedule overruns. Early risk-informed decision processes can minimize risk, which is a known historical cause of cost growth.

**Figure 12-7.** *A 5x5 matrix showing the likelihood and consequence of LUVOIR's top-five risks.*

For flagship missions, performance is always the driving metric, to ensure the project and Agency meet their commitments to the scientific community. However, it is critical that performance requirements are compatible with cost and schedule constraints. They are not independent from each other. This is why it is critical to establish and maintain firm agreement on high-level requirements. A commitment to fully understand the required performance up front and to resist changes late in the project lifecycle is required to reduce the risk that meeting performance objectives will lead to runaway costs and schedule delays.

We present LUVOIR's top five risks, summarized in **Figure 12-7**, and explain the LUVOIR Study Team's proposed approach to each of them below.

### 12.3.2.1 Integration and test facilities
**Rank:** 1
**ID:** IT1A, IT1B
**Likelihood:** 5 (LUVOIR-A), 4 (LUVOIR-B)
**Consequence:** 4

**Statement**: Given the size of LUVOIR, there is a possibility that existing integration and test facilities may not be able to accommodate the fully integrated observatory segment, resulting in the need to construct new facilities.

**Approach**: Research. A preliminary survey of existing facilities has been completed as part of this study. While it appears that large enough facilities exist for most integration and test activities, a full accounting of required optical and mechanical ground support equipment has not been completed. As the architecture continues to develop in Pre-Phase A, a thorough assessment of available facilities and their capabilities will be completed, and a facility development plan will be created. This plan will identify any facility gaps, and specify which existing facilities will need to be upgraded and which facilities may need to be constructed.

### 12.3.2.2 Technology development
**Rank:** 2
**ID:** TD1
**Likelihood:** 3
**Consequence:** 4

**Statement**: Given the complexity of the LUVOIR technology development plan (as presented in **Chapter 11**), there is a possibility that the technology development process may require more resources than are currently planned for, resulting in cost and schedule growth in Pre-Phase A.

**Approach**: Mitigate. The technology development plan is a grassroots estimate of the tasks necessary to mature all of LUVOIR's technologies to TRL 6 by Phase A start, as well as the cost and duration of those tasks. While we have high confidence that these technologies have a path to maturity, the very nature of technology development is open-ended; if everything was known about a technology's performance, it would not need to be developed. Funded schedule reserve (15 weeks / year; Bitten et al. 2014) and cost reserves (30% of total estimated cost; Hayhurst et al. 2015) are included in the baseline technology development plan to help mitigate cost and schedule growth. Regardless, the LUVOIR Team advocates that Phase A should not start until all technologies are at TRL 6 and have been demonstrated and proven. If TRL 6 is not achieved or if more resources are needed, we recommend delaying the start of Phase A as necessary.

### 12.3.2.3  Verification and validation approach
**Rank:** 3
**ID:** VV1
**Likelihood:** 3
**Consequence:** 3

**Statement**: Given the precise wavefront stability requirements for high-contrast imaging, there is a possibility that verification and validation by test may not be possible at full scale in a ground environment, resulting in a deviation from the "test as you fly" approach.

**Approach**: Mitigate. The verification and validation by analysis techniques developed on missions such as Chandra, JWST, and WFIRST will be extended to LUVOIR. The Pre-Phase A technology development plan includes two activities dedicated to the engineering development of coronagraph and ultra-stable system models. These models will be correlated with sub-scale technology demonstrations and used to verify and validate full-scale on-orbit performance. It is not yet clear at what level of assembly LUVOIR will need to deviate from the "test as you fly" approach; it is likely to occur at lower levels of assembly than on previous missions. As the models are developed in parallel with the system architecture during Pre-Phase A, a detailed verification and validation plan will be developed, and any necessary waivers will be written.

### 12.3.2.4  Contamination control
**Rank:** 4
**ID:** CC1
**Likelihood:** 3
**Consequence:** 3

**Statement**: Given strict requirements on contamination to enable LUVOIR's far-UV science, there is a possibility that existing engineering best practices, integration processes, and material certifications fail to adequately control contamination on LUVOIR's optics, resulting in a reduction in science yield.

**Approach**: Research. Existing contamination control processes that have been used on other UV missions such as FUSE and HST are adequate in theory; however, they may prove very difficult to implement on a system as large and as complex as LUVOIR. A Pre-Phase A Engineering Development activity is planned to survey contamination control processes and evaluate how to best implement them on the LUVOIR system. The results of this activity will include recommendations for new techniques, processes, facilities, or materials that will need to be in place prior to starting LUVOIR integration and test, and that can be developed during Phase A/B.

### 12.3.2.5 Launch vehicle
**Rank:** 5
**ID:** LV1A, LV1B
**Likelihood:** 3 (LUVOIR-A), 2 (LUVOIR-B)
**Consequence:** 2

**Statement**: Given that the baseline launch vehicles (SLS Block 1B and SLS Block 2) are still in development, there is a possibility that the predicted performance will not be realized in time for LUVOIR's needs, resulting in the need to descope LUVOIR's size and capabilities.

**Approach**: Mitigate. The LUVOIR architecture is designed to be scalable for exactly this scenario, allowing the final concept to be tailored to the capabilities of a specific launch vehicle. By the time LUVOIR enters Phase A in 2025, all three potential launch vehicles (SLS, SpaceX Starship, and Blue Origin New Glenn) are planned to have had at least one launch, solidifying the individual vehicles' expected performance as well as the likelihood that at least one of them will meet the requirements for LUVOIR's launch vehicle.

### 12.3.3 Technical margin philosophy
The LUVOIR team has adopted a margin philosophy for mass and power that is based on the following standards:

- ANSI/AIAA S-120A-2015, *"Mass Properties Control for Space Systems"*
- GSFC-STD-1000G, *"Goddard Space Flight Center Rules for Design, Development, Verification, and Operation of Flight Systems"* (a.k.a. Goddard GOLD Rules)
- ANSI/AIAA G-020-1992, *"Guide for Estimating and Budgeting Weight and Power Contingencies for Spacecraft Systems"*

### 12.3.3.1 Mass margin
While the overall philosophy of keeping extra mass "in reserve" to allow for future growth is universal, the specific details and terminology can differ from organization to organization, including how that extra mass is divided between mass growth allowance, margin, and reserve. Indeed, there can even be confusion as to what is called "margin" vs. "contingency" vs. "growth allowance" vs. "reserve."

Furthermore, no single standard sets bounds on all of the terms clearly. The American Institute of Aeronautics and Astronautics (AIAA) spells out requirements for growth allowance and a combination of growth allowance and margin. However, they do not set a requirement on reserve. Conversely, the Goddard GOLD rules spells out a clear requirement for margin and reserve, but not for growth allowance.

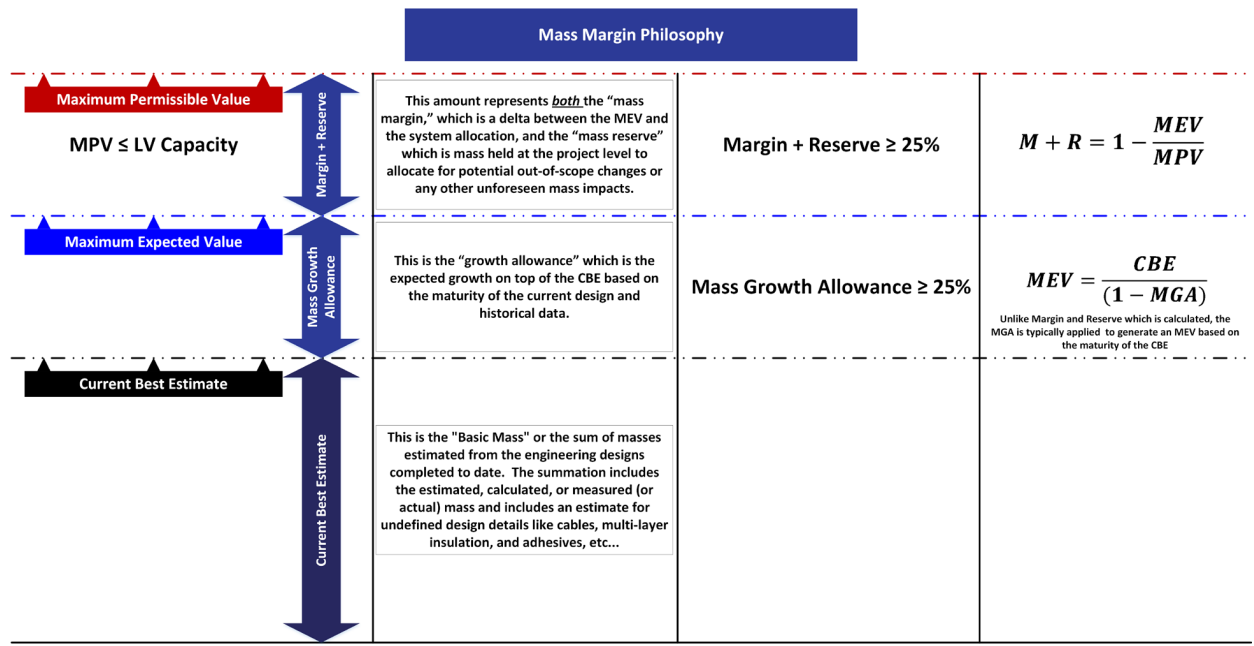

**Figure 12-8.** *LUVOIR mass margin philosophy. Power margins are computed similarly, except the power growth allowance percentage is 40%.*

Another difference is how the percentage for each quantity is computed, specifically with respect to which mass is used as the reference mass. The Goddard GOLD Rules (as of Rev. G) defines a relationship between the estimated resource and the available resource that uses the available resource as the reference. Using the definitions above, the current best estimate (CBE) and maximum expected value (MEV) masses are related by the equation: CBE = (1 − MGA)*MEV, where MGA is the fractional mass growth allowance. Similarly, the MEV and maximum permissible value (MPV) masses would be related by: MEV = (1 − Margin)*MPV.

The AIAA Standard uses the estimated resource as the reference. Under this framework, the CBE and MEV masses are related by the equation: (1 + MGA)*CBE = MEV. Furthermore, the MPV mass is related to both the MEV and CBE masses via the equation: MPV = Margin*CBE + MEV.

Under these different definitions, the LUVOIR team has adopted a standard, shown in **Figure 12-8**, that is consistent with the intent of both the AIAA Standard and the Goddard GOLD rules. In all cases of computing resource margins & growth allowance, LUVOIR adopts the Goddard GOLD Rules (Rev. G) method, using the "available resource" (i.e., higher value) as the reference (denominator) when computing percent difference. The mass growth allowance of 30% is drawn from the AIAA Standard. We apply this MGA uniformly to all components, assemblies, and sub-systems in each LUVOIR concept, with one exception; for the primary mirror segment assemblies and secondary mirror assembly we use MGA = 20%. This reduced MGA is justified by the fact that mirror assemblies similar to those envisioned for LUVOIR have been studied for over two decades, beginning with the Advanced Mirror System Demonstration (AMSD) that ultimately led to the JWST mirror segment design. The follow-up Multiple Mirror System Demonstration (MMSD) yielded several fully qualified ULE mirror segment assemblies that are similar in design to those imagined for LUVOIR.

**Table 12-3.** *Summary of mass estimates and mass margins for each LUVOIR concept. The mass growth allowance and margin and reserve are computed as percentages of the MEV and MPV masses, respectively. The Total Mass Contingency represents the total difference between the CBE mass and MPV mass.*

| | CBE Mass [kg] | MGA [kg] | MGA [%] | MEV Mass [kg] | Margin & Reserve [kg] | Margin & Reserve [%] | MPV Mass [kg] | Total Mass Contingency [kg] | Total Mass Contingency [%] |
|---|---|---|---|---|---|---|---|---|---|
| LUVOIR-A | 27,801 | 9,630 | 25.7% | 37,431 | 6,869 | 15.5% | 44,300 | 16,499 | 37.2% |
| LUVOIR-B | 15,132 | 5,470 | 26.6% | 20,602 | 16,398 | 44.3% | 37,000 | 21,868 | 59.1% |

The final mass margin and reserve percentage is computed based on the MPV mass, or launch vehicle capacity: 44,300 kg for SLS Block 2, and 37,000 kg for SLS Block 1B. **Table 12-3** summarizes the mass and mass margins for each LUVOIR concept.

### 12.3.3.2 Power margin

Similar principals apply for power that applied to mass. It should be noted that we designed the electrical power system to have an end-of-life capability that was equal to or exceeded the power MEV plus the desired margin and reserve.

For power, we again adopt the Goddard GOLD Rules recommended value of 25% for our margin + reserve, and we adopt the Aerospace Corporation recommended value of 40% for our power growth allowance, drawn from the ANSI/AIAA-G-020-1992 document. We believe that this approach to tracking mass and power growth, margins, and reserves is conservative and will protect against the risk of not having enough mass or power to allocate to the various systems, thus reducing overall mission risk.

### 12.3.4 Essential trade studies completed

The LUVOIR engineering team executed several trade studies in order to have the highest probability of successfully achieving LUVOIR's science. All of LUVOIR's design trades and decisions were carefully weighed and led to the concepts described in this report. Those trades that had the greatest impact on the LUVOIR concepts are discussed in **Appendix D**.

### 12.3.5 Future trade studies recommended

The LUVOIR engineering team performed many trades and every design choice was made with a carefully thought out decision process. However, with limited time and resources, we were unable to consider every possible trade study, especially as new information pertaining to launch vehicle capability and coronagraph performance became available. We have identified several trade studies that we believe to be high-value; specifically, alternate approaches to the concept that could lead to simpler, more cost-effective designs and help further reduce risks. We summarize three of those future recommended trade studies here.

### 12.3.5.1 Polarization aberration and sunshade size

The relationship between polarization aberration and the size of LUVOIR's sunshade is an example of both the need to study LUVOIR as an integrated system and of a trade that could be explored further during Pre-Phase A activities.

A major limitation of LUVOIR's coronagraph wavefront control system is that it can only control effectively the average of the wavefront error of the two polarization states. As unpolarized starlight passes through the optical system and reflects off the metallic mirror coatings, the two component polarization states accumulate aberrations differently. The more the two polarizations differ, the more starlight leaks through the coronagraph and degrades the contrast.

To mitigate this issue, we required the angle of incidence at any given optical surface to be less than 12° and that any 90° fold mirrors be used in compensated pairs. The angle-of-incidence requirement drives the optical system to be slower, increasing the separation between mirrors—specifically between the primary mirror (PM) and secondary mirror (SM). In turn, the distance between the PM and SM drives the size of the sunshade, as both mirrors must remain in shadow as the payload slews between targets.

Recently, we completed an analysis of the end-to-end polarization of the LUVOIR-A concept. We found that our 12° angle-of-incidence requirement was successful (Will & Fienup 2019); the coronagraph wavefront control is able to correct the polarization aberration and achieve its nominal design contrast. Analysis for LUVOIR-B is ongoing, but we expect it to yield similar results. Based on these results, it may be possible to relax the 12° requirement before the contrast degradation becomes unacceptable. This could allow for more flexibility in the system optical design, reducing PM to SM separation, and therefore reducing sunshade size.

Another alternative is to explore parallel coronagraph channels that each operates in a separate polarization state. By splitting polarization prior to entering the coronagraph optics, the wavefront control system in either channel can correct for its own polarization aberration. This effectively frees the optical design of the telescope from any angle-of-incidence requirement and might allow substantially reduced sunshade sizes. The downside to this approach is a more complicated coronagraph instrument. This trade study should explore whether a more complicated coronagraph instrument is less risky than the current large size of the sunshade.

## 12.3.5.2 Launch vehicle fairing & deployment complexity

Even though two of LUVOIR's top-five risks pertain to the potential unavailability of the baseline launch vehicle fairing, these risks are all trending in a positive direction. In the three years of the LUVOIR mission concept study, the number of likely launch vehicle candidates has increased. LUVOIR's scalable architecture allows it to potentially leverage the advent of commercially available heavy-lift vehicles such as SpaceX's Starship and Blue Origin's New Glenn. This opens options for optical system design and observatory packaging.

The overarching guidance on selecting the LUVOIR aperture sizes from the STDT was "the largest aperture that could fit in an SLS Block 2 8.4-m diameter fairing" for LUVOIR-A, and "the largest aperture that could be fit in an existing, commercial-off-the-shelf (COTS) 5-m fairing" for LUVOIR-B. Since that time, the SpaceX Starship fairing and the Blue Origin New Glenn fairing offer substantially more volume than the COTS 5-m faring. This opens up the option to launch LUVOIR-B partially deployed, substantially reducing its complexity, and therefore reducing its cost and risk.

During our study, the SLS program also redesigned its notional 8.4-m fairing, adding more height under the fairing ogive. It is possible that with very minor adjustments, the secondary mirror deployment on LUVOIR-A could be simplified, removing at least one joint in each leg of the secondary mirror support structure.

We therefore recommend future architecture studies do a more complete trade study of packaging LUVOIR concepts in a wider range of available launch vehicle fairings.

### 12.3.5.3  Near-UV starshade

Early in the LUVOIR concept study, the team decided to proceed with a coronagraph-only design to enable the exoplanet science objectives. This decision was based on the fact that the required starshade diameter (in excess of 100-m) for a telescope of LUVOIR's size and wavelength range would be well beyond what is currently being investigated by the Starshade Technology Development community.

During the course of the LUVOIR study, the community has continued to explore the relationship between starshade size, operational bandwidth, and the exoplanet observation process. This has opened the possibility of a more reasonably sized ($\lesssim 70$ m) starshade to execute LUVOIR's near-UV exoplanet science (**Appendix I.1**). Such a starshade could remove the necessity of the near-UV channel in the ECLIPS instrument. This would simplify the instrument, allowing for more flexibility in addressing the polarization issue addressed above. It would also allow for overall higher throughput in the coronagraph instrument, since reflective aluminum pre-optics within the coronagraph could be coated with higher-reflectivity silver instead.

The cost, of course, is the additional complexity of an independent starshade spacecraft. However, recent advances in the Starshade Technology Development community indicate that it may be worth exploring this trade more fully.

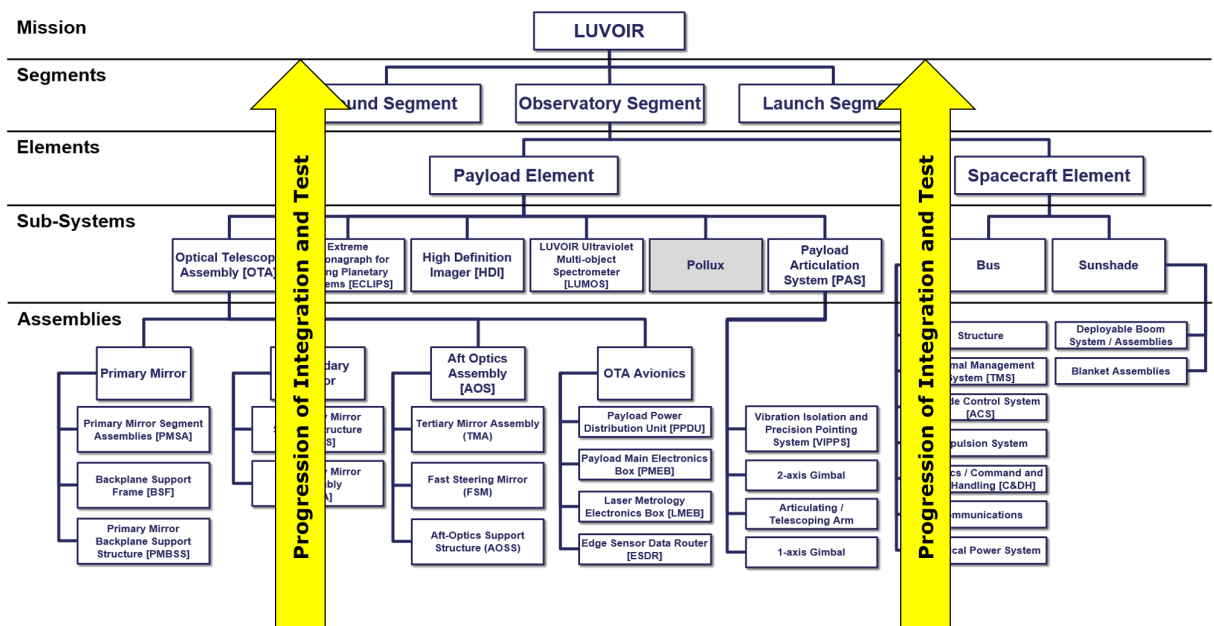

**Figure 12-9.** *The flow of integration and test relative to the architecture*

## 12.4   Integration and test

### 12.4.1   Integration and test flow

The LUVOIR team proposes that I&T follow the LUVOIR architecture as shown in **Figure 12-9**. Integration and qualification will be done at the lowest levels before being delivered to the next higher level of assembly as shown in **Figure 12-10** through **Figure 12-13**. While this minimizes the risk that a component, assembly, or sub-system problem will be identified after assembly through the highest level, mechanical testing may subject all hardware—especially the lowest level hardware—to significant amounts of testing, which is not benign. We will tailor this approach as designs evolve, much like the JWST team did for the NIRSpec microshutter exposure to the acoustic test environments.

### 12.4.2   Modularity and parallel I&T operations

In order for the LUVOIR team to efficiently manage the schedule, parallel I&T operations need to be factored into the design. One area where this is very relevant is the manufacturing of the PMSAs (Cole 2017), as well as the primary mirror backplane support structure that makes up much of the OTA. The current concept for LUVOIR-A requires 120 mirror segments, while LUVOIR-B requires 55 mirror segments. It is critical that manufacturing and assembly of these mirrors happen in parallel and not in serial, otherwise the overall mission development schedule could be lengthened by several years (**Figure 12-14**).

On one-of-a-kind projects like LUVOIR, it is not uncommon for a single expert to develop designs, plans, procedures, and then oversee the work themselves. On a flagship mission of this scope, the team can never be only one person deep for any level of assembly. In the case of PMSA assembly and integration, at least one expert cognizant engineer needs to train four other leads to oversee the parallel assembly of PMSAs. It is also critical that the design allow for integration of mirrors onto modular components of the PMBSS and that those modules be easily integrated over multiple cycles (**Figure 12-14** through **Figure 12-17**).

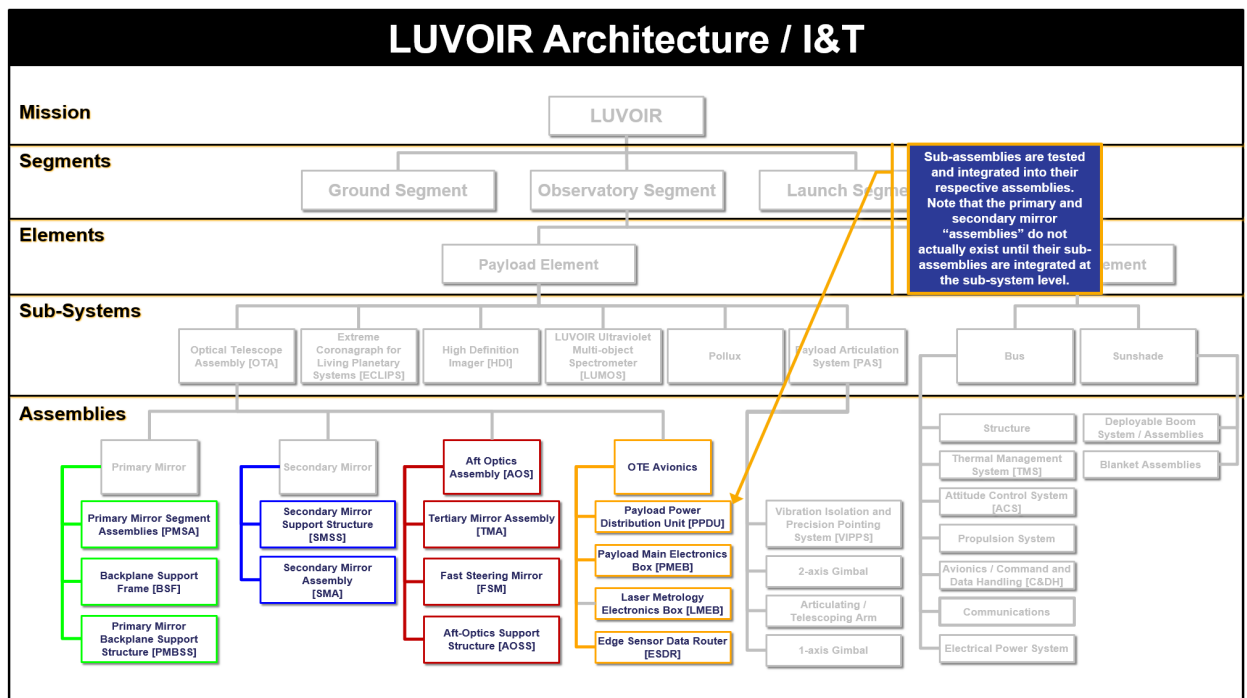

**Figure 12-10.** *Integration and test at the sub-assembly level*

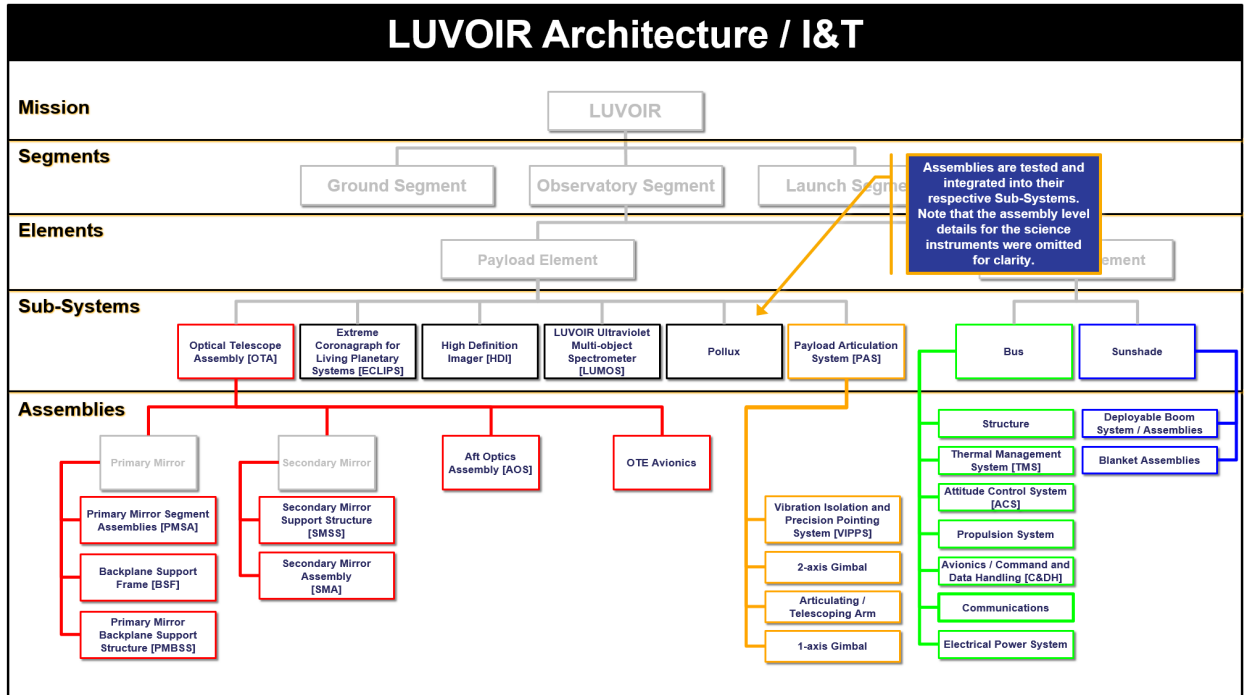

**Figure 12-11.** *Integration and test at the sub-system level*

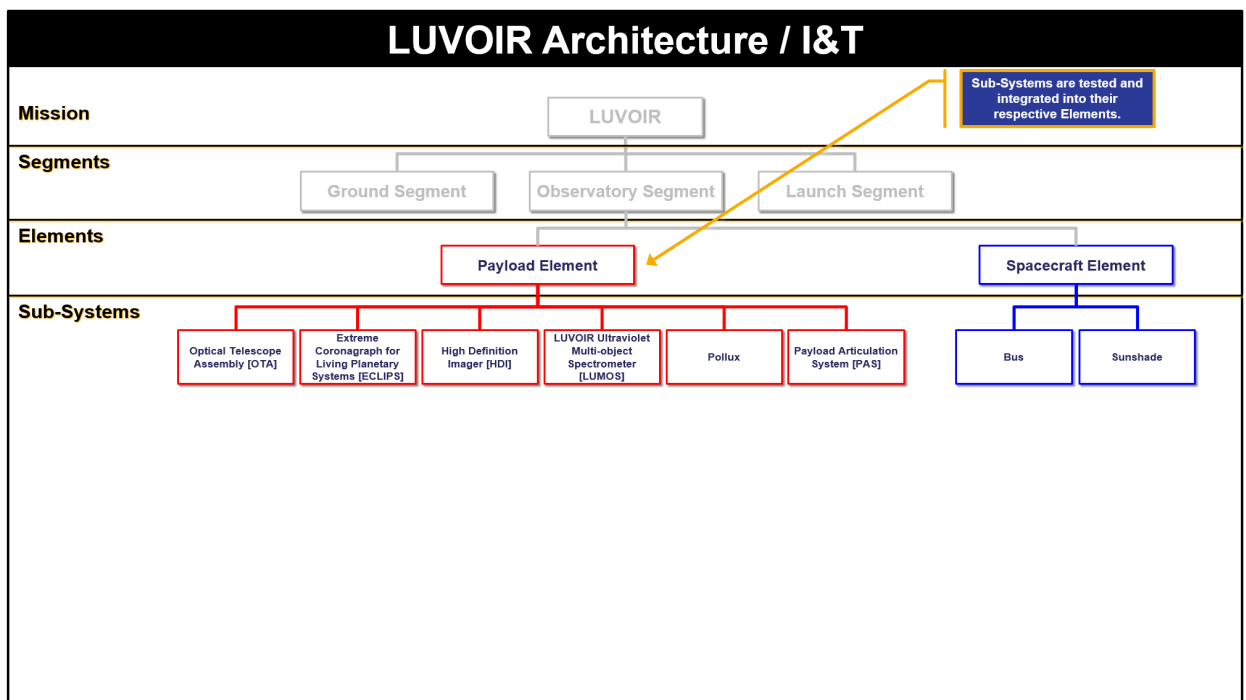

**Figure 12-12.** *Integration and test at the element level*

This is a prime example of why "storyboarding" the entire process and pre-visualizing the integration sequence is critical. Without that foresight, the process will be reduced to assembling in series, drastically increasing the risk of schedule overruns. Furthermore, modularity may be required to transport LUVOIR from an integration and test facility to a launch facility. Modularity allows LUVOIR to be separated into parts that are small enough to be readily transported with existing infrastructure, and then re-assembled at the destination facility.

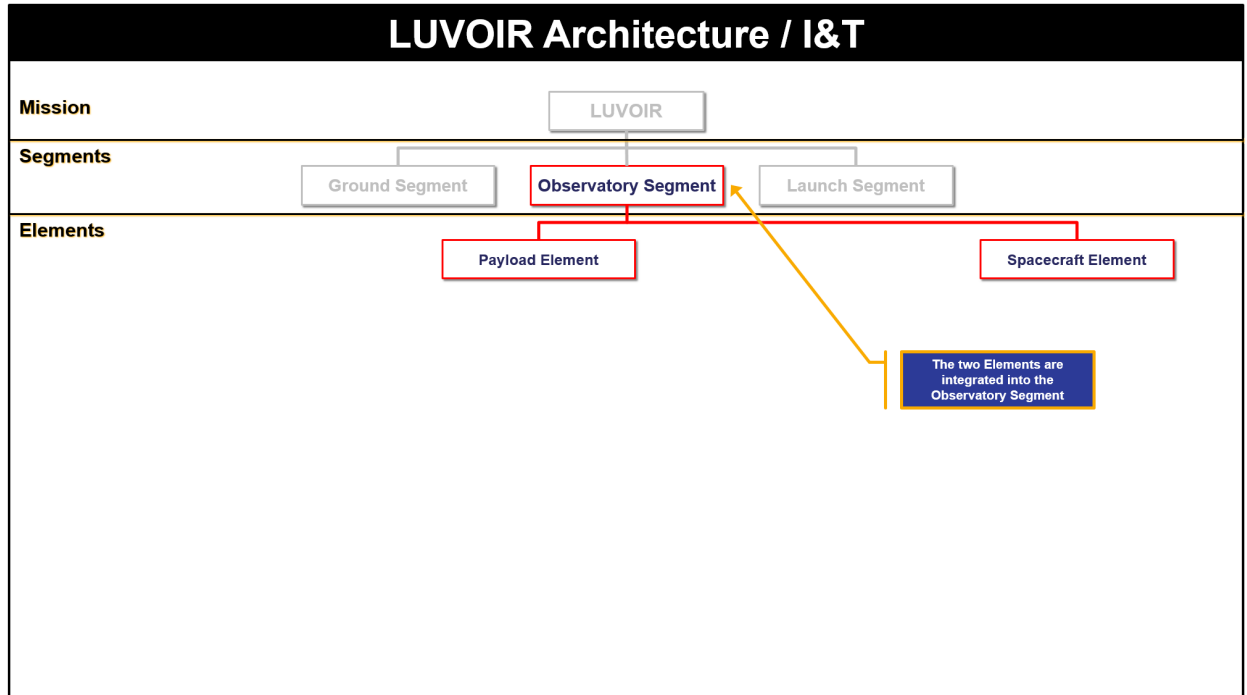

**Figure 12-13.** *Integration and test at the segment level*

### 12.4.3  Integration and test matrix

A rigorous test program will be instituted at every level, in part to minimize the risk of discovering problems at high levels of integration. At these levels of integration, some parts might require significant work to gain access to them, such as a filter wheel inside an instrument or an actuator on the back of a primary mirror segment. On a Class A flagship program like LUVOIR, this makes a rigorous testing process critical for minimizing unforeseen issues at higher levels of assembly. Indeed, JWST and other large missions have used a similar approach. A notional test matrix is shown in **Table 12-4**. This test matrix will need to be tailored as the architecture and conceptual designs evolve. However, it sets the stage for the forward vision that will be required to both create the ground support equipment (GSE) and find the facilities that will accommodate these tests.

### 12.4.4  Integration and test facilities

The size of LUVOIR will make the integration and test efforts at the payload and spacecraft element levels, as well as the observatory segment level, challenging. For reference, **Figure 12-18** shows the relative sizes of various space telescope mirrors including Hubble and JWST. **Figure 12-19** shows the relative sizes of various space telescopes. LUVOIR proposes that the payload and spacecraft elements as well as the observatory segment be integrated and tested in a single location (Crooke et al. 2006). Sub-systems would be integrated and tested at the location they are developed.

**Table 12-4.** *Conceptual figure showing parallel assembly os the PMSA's, and the engineers required to perform these parallel operations.*

LUVOIR Integration and Test Matrix

| Assy Level | Architecture Level | Entity | Modal Survey | Static Loads/Pull | Sine Burst | Sine Vibration | Random Vibration | Acoustics | Mechanical Shock | Torque Ratio | Life Tests | Deployment | Mass Properties | Leak | Thermal Vacuum/Cycles | Thermal Balance | Bake-out | Conductance | Alignment | Performance | Comprehensive Performance Test | EMC/EMI | Continuity | Propulsion | Modular Interface Performance |
|---|---|---|---|---|---|---|---|---|---|---|---|---|---|---|---|---|---|---|---|---|---|---|---|---|---|
| 1 | Segment | Observatory | | | | X | | X | X | | | X | X | | | X | | | | | X | X | | | |
| 2 | Element | Payload | | | | X | | X | X | | | X | X | X | X | X | X | | X | X | X | X | | | |
| 3 | Sub-system | Optical Telescope Assembly [OTA] | X | X | | | | | | | | X | X | X | X | X | X | | X | X | X | X | | | X |
| 4 | Assembly | Primary Mirror | The primary mirror "assembly" is not a stand-alone assembly and does not exist until the subassemblies are integrated to the OTA | | | | | | | | | | | | | | | | | | | | | | |
| 5 | Sub-Assembly | Primary Mirror Segment Assemblies | | | | X | | X | | | X | | | | X | | | | | | | | | | |
| 5 | Sub-Assembly | +P2 PMBSS #2 | X | X | | X | | | | | | | | | X | | | | X | | | | | | X |
| 5 | Sub-Assembly | +P2 PMBSS #1 | X | X | | X | | | | | | | | | X | | | | X | | | | | | X |
| 5 | Sub-Assembly | Center PMBSS + BSF | X | X | | X | | | | | | | | | X | | | | X | | | | | | X |
| 5 | Sub-Assembly | -P2 PMBSS #1 | X | X | | X | | | | | | | | | X | | | | X | | | | | | X |
| 5 | Sub-Assembly | -P2 PMBSS #2 | X | X | | X | | | | | | | | | X | | | | X | | | | | | X |
| 4 | Assembly | Secondary Mirror | The secondary mirror "assembly" is not a stand-alone assembly and does not exist until the two subassemblies are integrated to the OTA | | | | | | | | | | | | | | | | | | | | | | |
| 5 | Sub-Assembly | Secondary Mirror Support Structure [SMSS] | X | X | | | | | | | | | | | X | | | | | | | | | | |
| 5 | Sub-Assembly | Secondary Mirror Assembly [SMA] | | X | | | | | | | | | | | X | | | | | | | | | | |
| 4 | Assembly | Aft Optics Subsystem | | | | X | | X | | | | | | | | | | | X | X | | X | | | |
| 5 | Sub-Assembly | Fast Steering Mirror | | | | | | | | | X | | | | | | | | | X | | X | | | |
| 5 | Sub-Assembly | Aperture Plate | | | | | | | | | | | | | | | | | | X | | | | | |
| 5 | Sub-Assembly | Telescoping Tube | | | | | | | | X | X | | | | | | | | | | | X | | | |
| 5 | Sub-Assembly | Tertiary Mirror | | | | | | | | | | | | | | | | | | X | | | | | |
| 4 | Assembly | Thermal Management System | Components of the TMS will have conductance testing prior to integration | | | | | | | | | | | | | | | X | | | | | | | |
| 4 | Assembly | Electrical Subsystems | Components of the electrical subsystem will have | | | | | | | | | | | | | | | | | | | | | X | |
| 3 | Sub-system | Extreme Coronagraph for Living Planetary Systems [ECLIPS] | X | | | X | X | X | | | X | | X | X | X | X | X | | X | X | X | X | | | X |
| 3 | Sub-system | High Definition Imager [HDI] | X | | | X | X | X | | | X | | X | X | X | X | X | | X | X | X | X | | | X |
| 3 | Sub-system | LUVOIR Ultraviolet Multi-object Spectrometer [LUMOS] | X | | | X | X | X | | | X | | X | X | X | X | X | | X | X | X | X | | | X |
| 3 | Sub-system | Pollux* | X | | | X | X | X | | | X | | X | X | X | X | X | | X | X | X | X | | | X |
| 3 | Sub-system | Payload Articulation System [PAS] | X | X | | X | | | | | X | | X | | | | | | | X | X | X | | | |
| 4 | Assembly | Vibration Isolation and Precision Pointing System [VIPPS] | X | | | X | | X | | | | | | | X | | | | | | X | X | | | |
| 4 | Assembly | 2-axis Gimbal | | X | | | X | | | X | | | | | | | | | | | | | | | |
| 4 | Assembly | Articulating / Telescoping Arm | | | | | | | | X | | | | | | | | | | | | | | | |
| 4 | Assembly | 1-Axis Gimbal | | X | | | X | | | X | | | | | | | | | | | | | | | |
| 2 | Element | Spacecraft | X | | | X | | X | X | | | X | X | X | X | X | X | | | | X | X | | | X |
| 3 | Sub-system | Sunshade | | | | | | | | | | X | X | | | | | | | | | | | | |
| 4 | Assembly | Deployable Boom System / Assemblies | X | X | | | | | | | | X | X | | | | X | | | | | X | | | |
| 4 | Assembly | Blanket Assemblies | | | | | | | | | | | | | | | | X | | | | | | | |
| 3 | Sub-system | Bus | X | | | X | X | X | | | | | X | X | X | X | X | | | | X | X | | | |
| 4 | Assembly | Structure | | X | | X | | | | | | | X | | | | | | | | | | | | |
| 4 | Assembly | Thermal Management System [TMS] | | | | | | | | | | | | | | | | X | | | | | | | |
| 4 | Assembly | Attitude Control System [ACS] | | | | | | | | | | | | | | | | | | | X | X | | | |
| 4 | Assembly | Propulsion System | | X | | | | X | | | | | | | | | | | | | | | | X | |
| 4 | Assembly | Avionics / C & DH | | | | | | | | | | | | | | | | | | | X | X | | | |
| 4 | Assembly | Communications | | | | | | | | | | | | | | | | | | | X | X | | | |
| 4 | Assembly | Electrical Power System | | | | | | | | | | | | | | | | | | | X | X | X | | |

\* Pollux is only part of LUVOIR-A

Facility does not exist to do this testing.

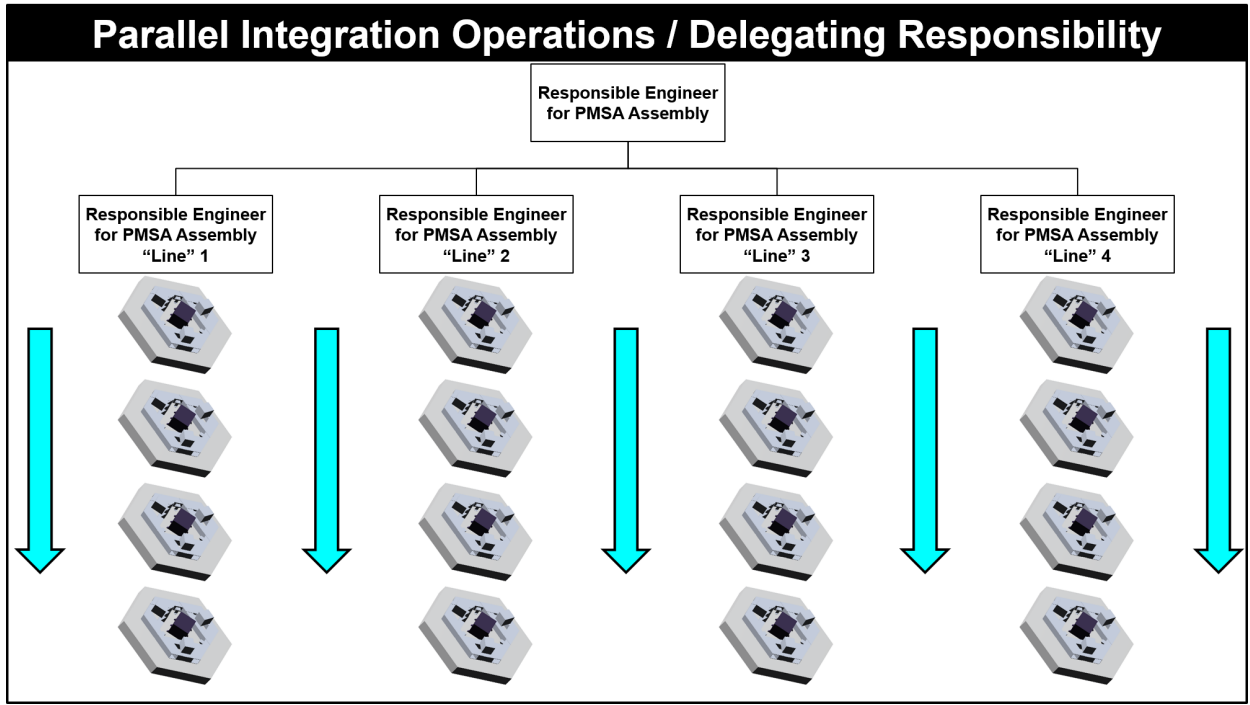

**Figure 12-14.** *Parallel assembly of the PMSAs*

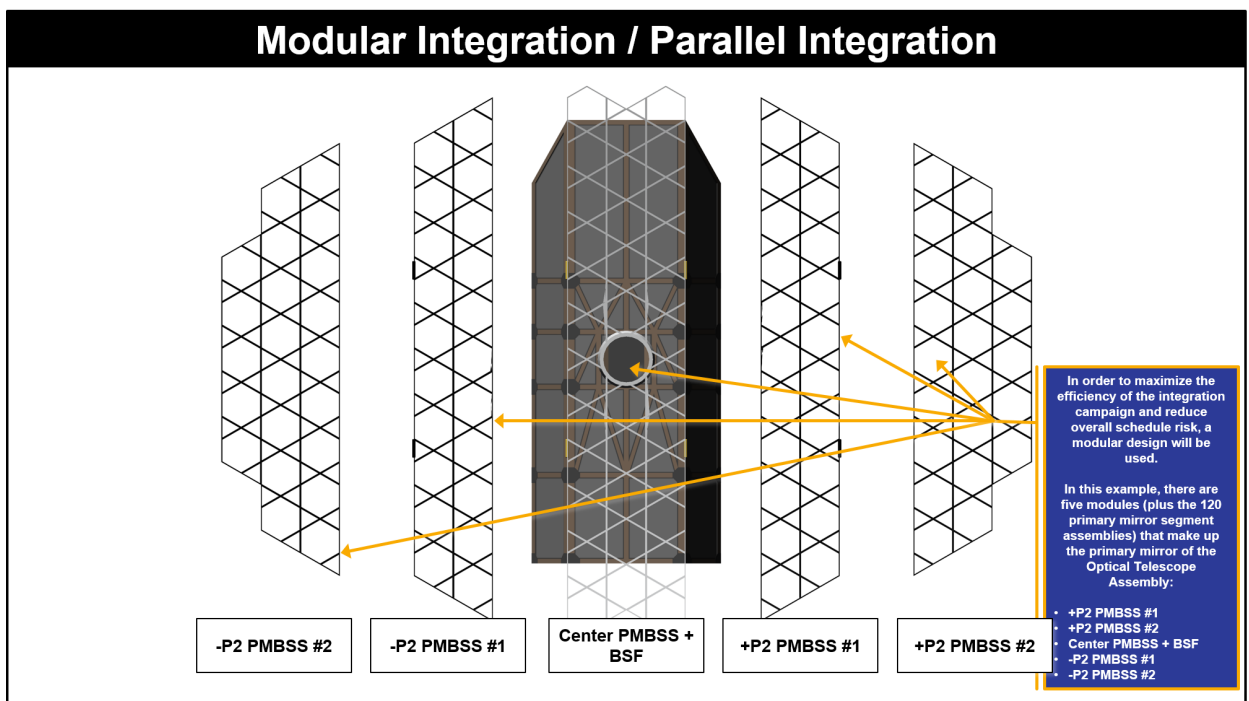

**Figure 12-15.** *The five modules that make up the OTA primary mirror. Segments are integrated to each module separately, in parallel, before the entire assembly is brought together as the primary mirror.*

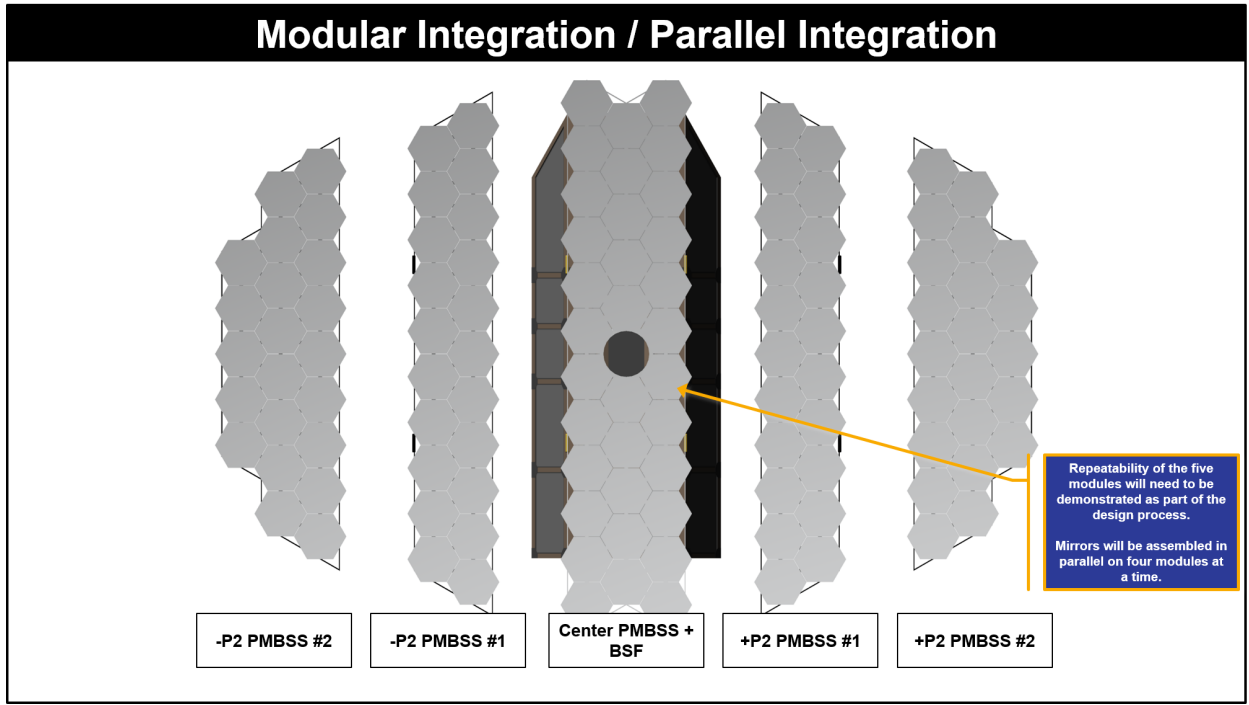

**Figure 12-16.** *Four primary mirror modules will be integrated in parallel to reduce the development schedule. Repeatability of de-integrating and re-integrating the primary mirror modules will be demonstrated as part of the design.*

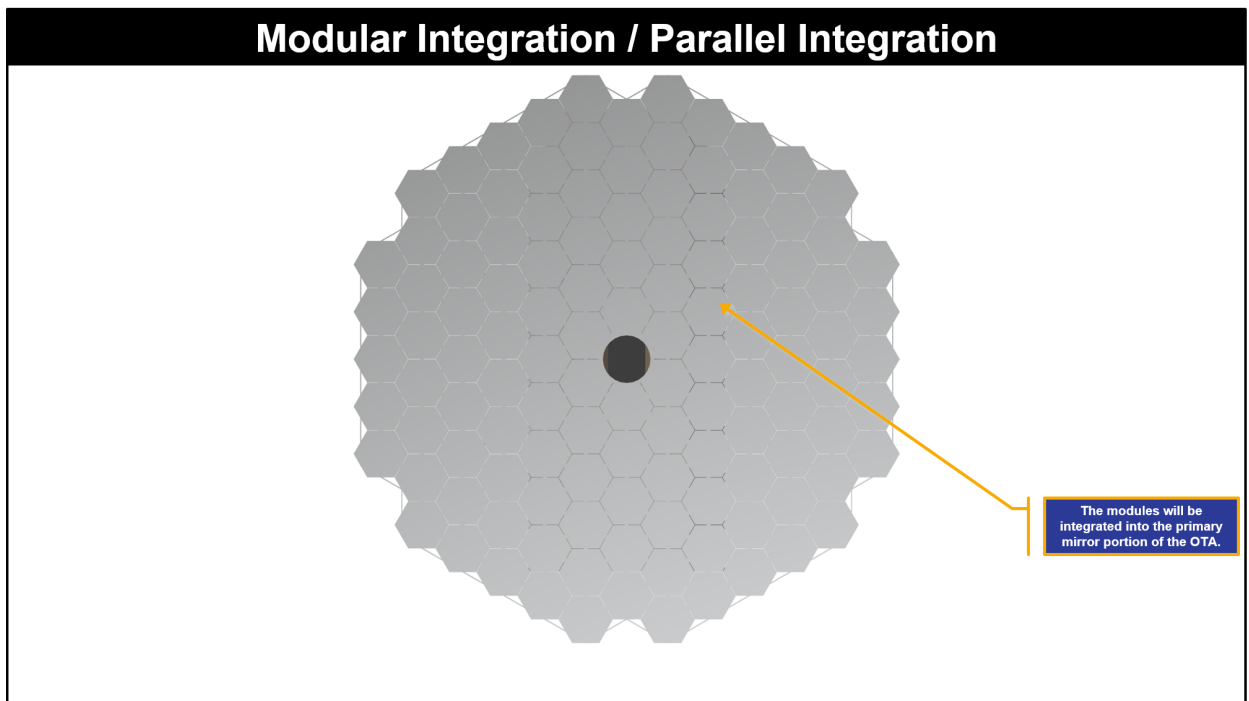

**Figure 12-17.** *The OTA primary mirror module interfaces will need to be very repeatable both for module integration and deployment.*

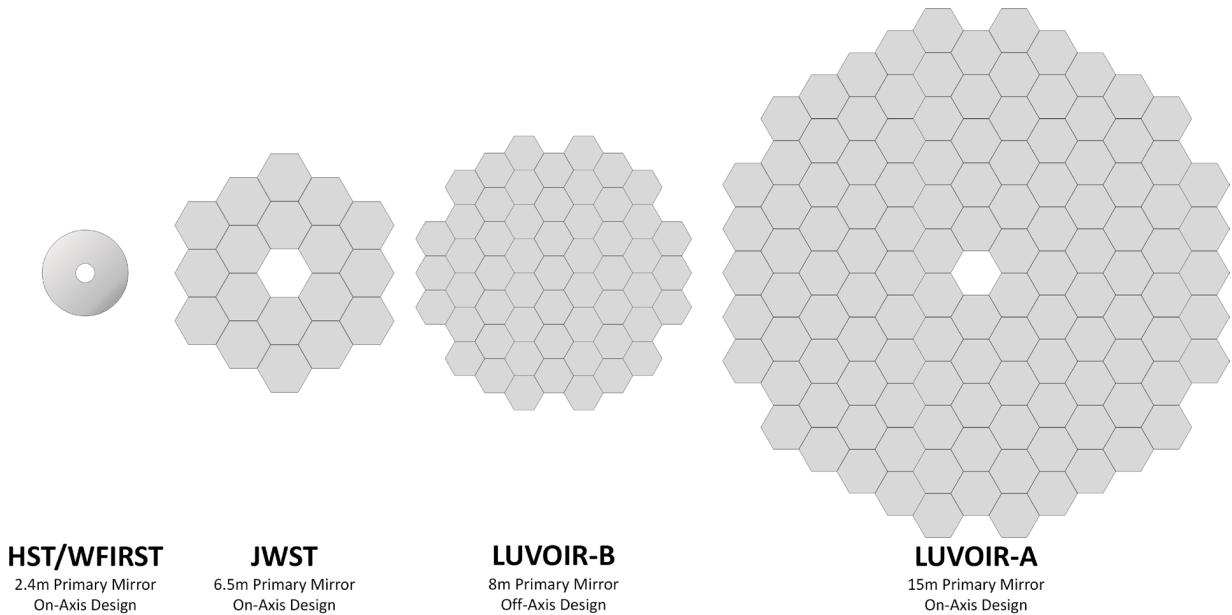

**Figure 12-18.** *Comparison of LUVOIR-A and -B primary mirrors with Hubble and JWST, shown to scale.*

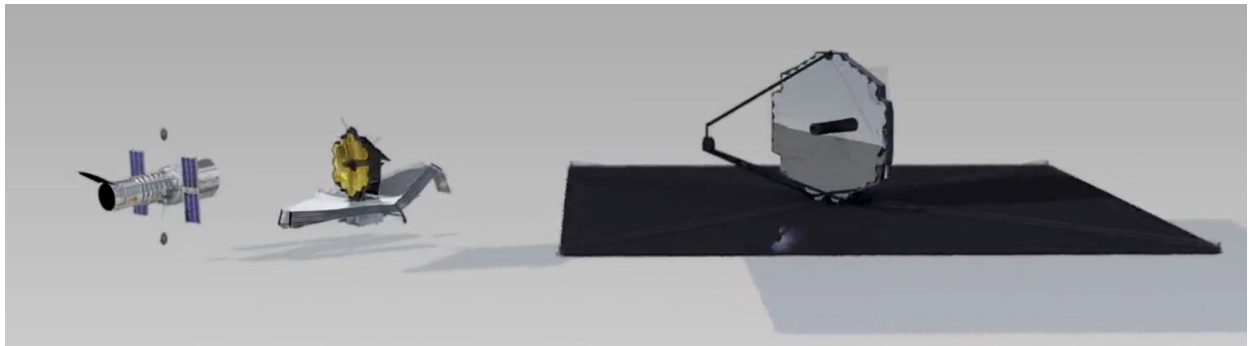

**Figure 12-19.** *Comparison of LUVOIR-A to Hubble and JWST, shown to scale.*

At this time, no single facility exists that is large enough to both integrate and test something as large as LUVOIR's payload, spacecraft element, or observatory segment. New facilities, potentially in combination with upgraded existing facilities, will be necessary to enable the integration and test of LUVOIR. This includes, but is not limited to, facilities capable of performing traditional acoustics, sine vibe, thermal testing, optical testing, or electromagnetic interference/compatibility (EMI/EMC) testing at the payload, spacecraft, or observatory levels of assembly. These facilities will also need to maintain the required cleanliness levels for an observatory that is sensitive in the far-UV (100 nm wavelength).

Existing facilities such as Glenn Research Center's Space Power Facility, Johnson Space Center's Chamber A, and Marshall Space Flight Center's Saturn V Dynamic Test Stand (which has been retired) demonstrate that large scale facilities are feasible even if ones with the required size and accessibility do not currently exist (Crooke et al. 2006).

Transportation to and from test facilities presents an additional set of challenges. Bridge height and load restrictions, along with other transportation aspects for air, land, or water transportation, need to be considered near each of the testing facilities that is used. A modular design, as envisioned by LUVOIR, helps mitigate some of these issues. Conventional transportation methods may be used if LUVOIR can be readily disassembled at a point of origin, and re-assembled at its destination.

This mission concept study is too preliminary to know exactly what the mechanical ground support equipment (MGSE), optical GSE (OGSE), and electrical GSE (EGSE) configurations will need to be, and these uncertainties could impact the needed size of the integration and test facilities. To put this into perspective, **Figure 12-20**, **Figure 12-21**, and **Figure 12-22** show both LUVOIR-A and LUVOIR-B in a volume cross-section of the Glenn Research Center (GRC) Space Power Facility Space Simulation Vacuum Chamber, the Johnson Space Center (JSC) Chamber A, the GRC Reverberant Acoustic Test Facility, and the European Space Agency's (ESA's) Large European Acoustic Facility. In the two vacuum chamber figures (**Figure 12-20** and **Figure 12-21**), the cyan blue area represents the approximate space required by MGSE and OGSE for cryogenic optical tests on JWST. On JWST, the flight hardware did not extend into this part of the chamber. While the MGSE and OGSE that LUVOIR will need have not yet been determined, this gives a perspective on the size required for these facilities.

As for acoustics testing, it is possible that a traditional chamber may not be required and that large banks of speakers could be moved into an integration facility to run a test in-situ. However, sine vibration testing will also provide challenges. LUVOIR-A is estimated to weigh ~37,400 kg, while LUVOIR-B is estimated to weigh just under 21,000 kg (MEV mass). At this time, there is no known mechanical shaker in the world that would be able to shake LUVOIR-A if it realized its maximum permissible mass of 44,300 kg. However, there is a clear need beyond LUVOIR for such a facility. NASA's SLS and SpaceX's Starship are both being designed with the capacity to launch payloads exceeding 45,000 kg, indicating that facilities required to verify a payload's ability to withstand the launch environment will need to be designed regardless of LUVOIR's development.

Given the size of LUVOIR-A, there may be a need to evaluate the requirement to perform a traditional vibration test at the payload, spacecraft, and observatory levels of assembly. Given the modularity of the system, it is clear that all modules could be vibration tested individually and that all interfaces could be strength tested. That would leave a quality control test to be completed following the assembly of the modules. These tests could potentially be done using some other dynamic test to check dynamic signatures of the assembly with all of the modules integrated. Testing of this kind would need to be carefully considered when planning future facilities and upgrades.

As mentioned earlier, Marshall Space Flight Center dynamically tested the Saturn V and the Space Shuttle in facilities that have since closed. New facilities for testing the components of the SLS are being developed, or exist today. While these facilities are outside and subject to the elements, it again demonstrates the feasibility of creating facilities that can test LUVOIR in a traditional manner. While the scope of the task may not be fully understood until decisions are made on the exact scale of LUVOIR, it is clear that new facilities will be required to test the largest of the options, and should be planned for early in the project lifecycle.

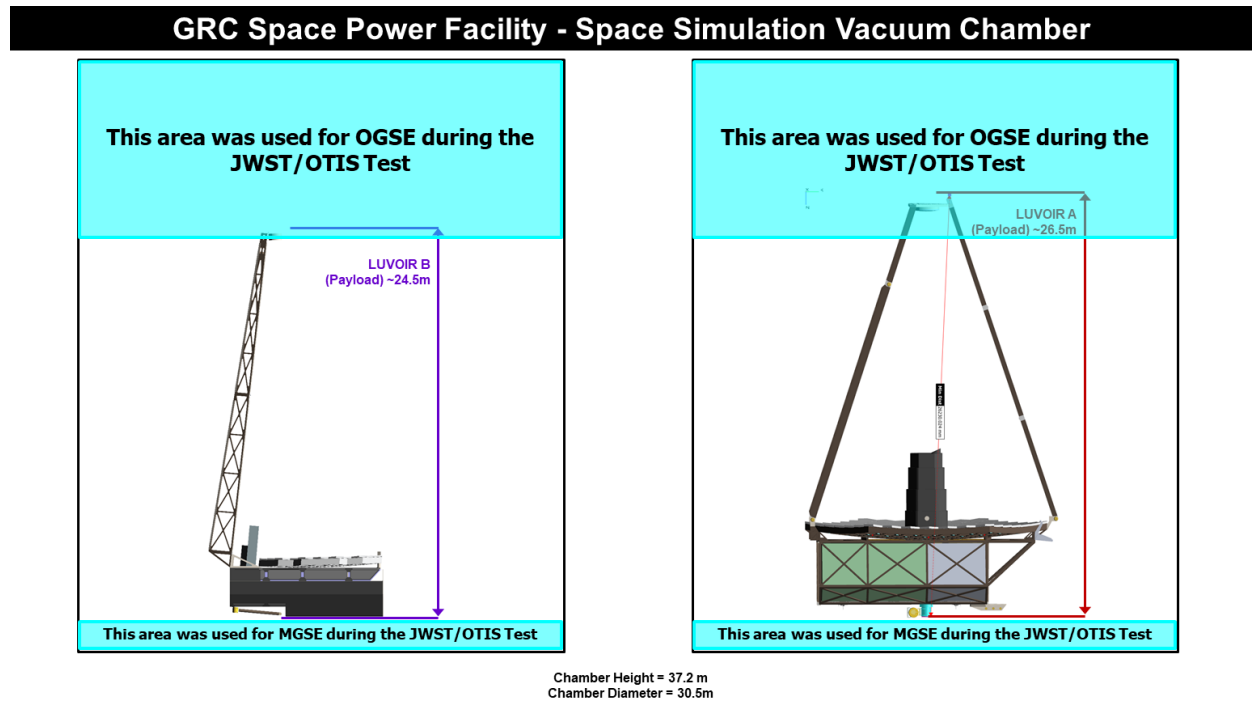

**Figure 12-20.** *LUVOIR-A (right) and LUVOIR-B (left) shown in the GRC Space Simulation Vacuum Chamber*

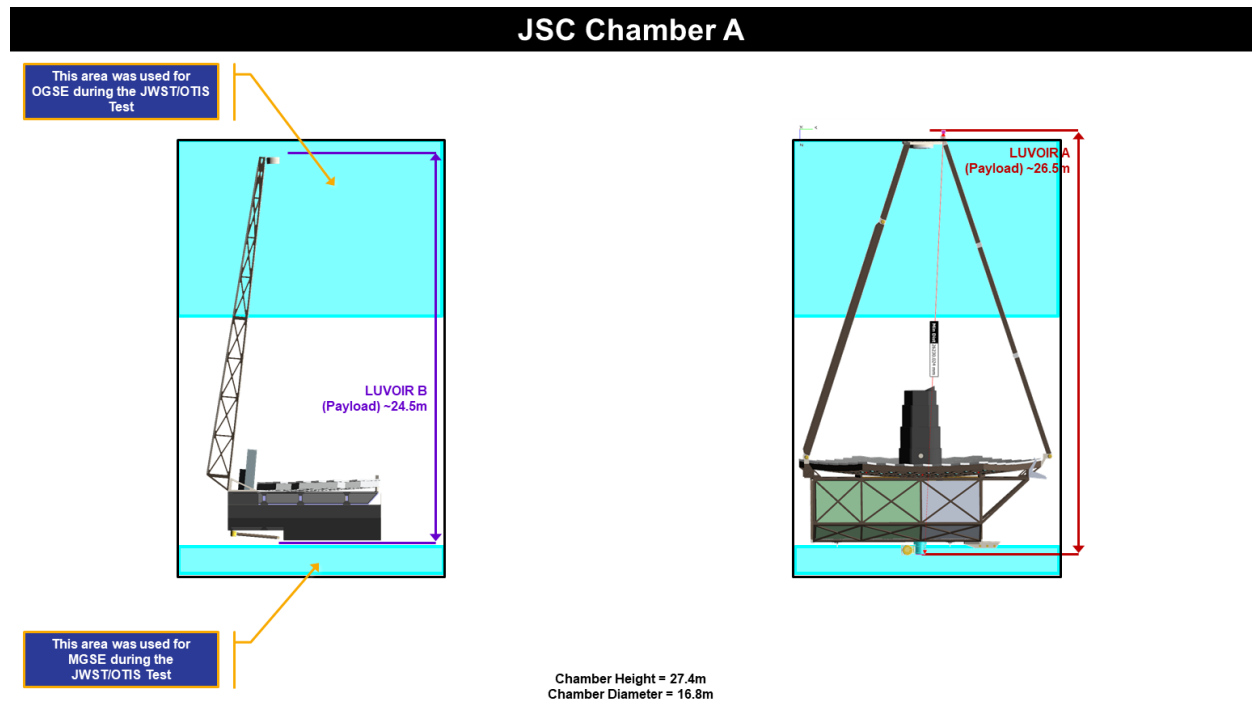

**Figure 12-21.** *LUVOIR-A (right) and LUVOIR-B (left) shown in the JSC Chamber A.*

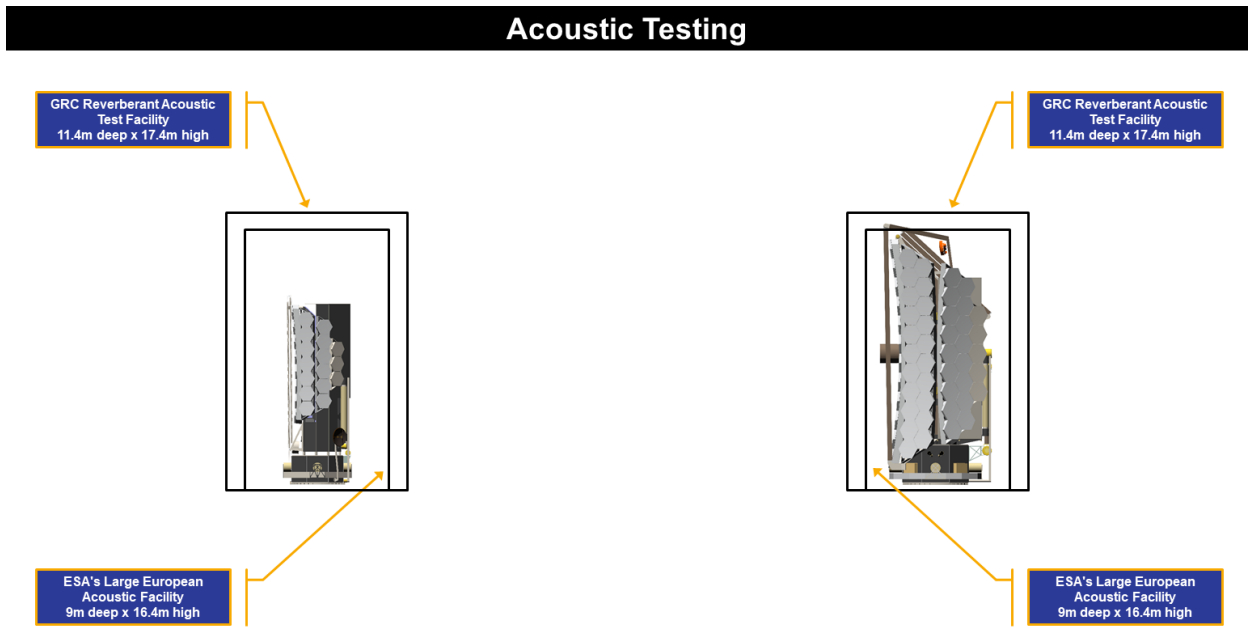

**Figure 12-22.** *LUVOIR-A (right) and LUVOIR-B (left) shown in NASA's and ESA's largest acoustics test chambers.*

### 12.4.5 Pathfinders

One of the major lessons learned during other large flagship missions such as Chandra (Arenberg et al. 2014), Cassini (e.g., Hagopian 1996; Crooke & Hagopian 1996; Jennings et al. 2017), and JWST (Lyon et al. 2004; Martin 2012; Windhorst et al. 2013; Feinberg et al. 2018; Bitten et al. 2019) is the need for engineering pathfinder tests. These are important for fleshing out integration and test processes. In a broader sense, engineering development units and engineering test units will also be used to vet design concepts. The intent is to use these pathfinders off the critical schedule path to help mature and optimize designs, plans, procedures, test setups, and integration sequences as much as possible for a one-of-a-kind design such as LUVOIR. Pathfinders can use a combination of ground support equipment, engineering test or development units, and smaller pathfinder hardware for lower level assemblies.

The following is a list of some of the pathfinders that we expect LUVOIR would use to reduce technical and schedule risk during the I&T at the payload, spacecraft, and observatory levels.

- PMSA construction strategies (design and procedure planning)
- PMBSS modular interface repeatability (design)
- PMSA to OTA modules (procedure planning)
- Sunshade (design)
- Sunshade (test)
- Payload element (testing)
- Spacecraft (testing)

This list is not meant to be exhaustive, only to highlight the types of pathfinder tests that will need to be planned. Pathfinders should be part of the mission storyboards at the earliest stages of planning so that they can be scoped and logically connected to flight integration and test events in the schedule. As with other early phase planning, engineering pathfinders should help to inform the architecture and the design concepts as they evolve.

## 12.5 Verification and validation

We intend the V&V approach for LUVOIR to be consistent with the processes outlined in the NASA Systems Engineering Handbook, NASA/SP-2016-6105 Rev 2. Through tests and modeling efforts, the LUVOIR team will show compliance with all requirements levied on the mission and all of the requirements flowed down to lower level systems. Additionally, we will perform tests and use models to validate that the observatory will actually perform as intended as a system. Much like other large, flagship missions—including JWST, Cassini, and Chandra—LUVOIR's size and operating environment will prevent its performance from being tested in a true flight configuration. Despite the mantra to "test as you fly," many missions are not tested exactly as they fly. LUVOIR will be no exception.

With that in mind, LUVOIR will rely on a three-pronged approach to validating its performance:

- Lower level testing and qualification as outlined in the integration and test section (**Section 12.4**)

- Verification and validation by analysis using validated analytical models

- Active control of the observatory

## 12.5.1 Lower level testing

As described in **Section 12.4**, testing of LUVOIR will not be left to the highest level of assembly. Each sub-system will arrive at the element level with requirements verified and hardware fully qualified and tested. We accept that some requirements will not be able to be verified at this level of assembly because they are not part of the larger system. These requirements will be agreed upon between the mission systems team and the product leads. For example, the science instruments may not include an end-to-end optical test at operating temperature because the OTA is not present. However, science instrument tests should include a detailed and validated analytical check of the optical performance with given boundary conditions that the integrated system must meet. These tests may include, for example, a high-fidelity OTA simulator that generates a source beam that matches the key performance parameters of the OTA (field-of-view, F/#, wavefront error, jitter stability, etc.)

## 12.5.2 Verification by analysis with validated models

Because the flight environment will be impossible to replicate on the ground, the LUVOIR engineering team will need to rely on verification by analysis with validated models to demonstrate performance in the operational environment. This process has been used extensively during the testing phases of many missions including Chandra and JWST. While JWST has not flown yet, validated analytical models have been used to accurately predict what will happen in tests with known boundary conditions—albeit ones that are different from flight. The ability to do this is what demonstrates that the analytical models can be used to predict what will happen in the operational environment (Crooke et al. 2006).

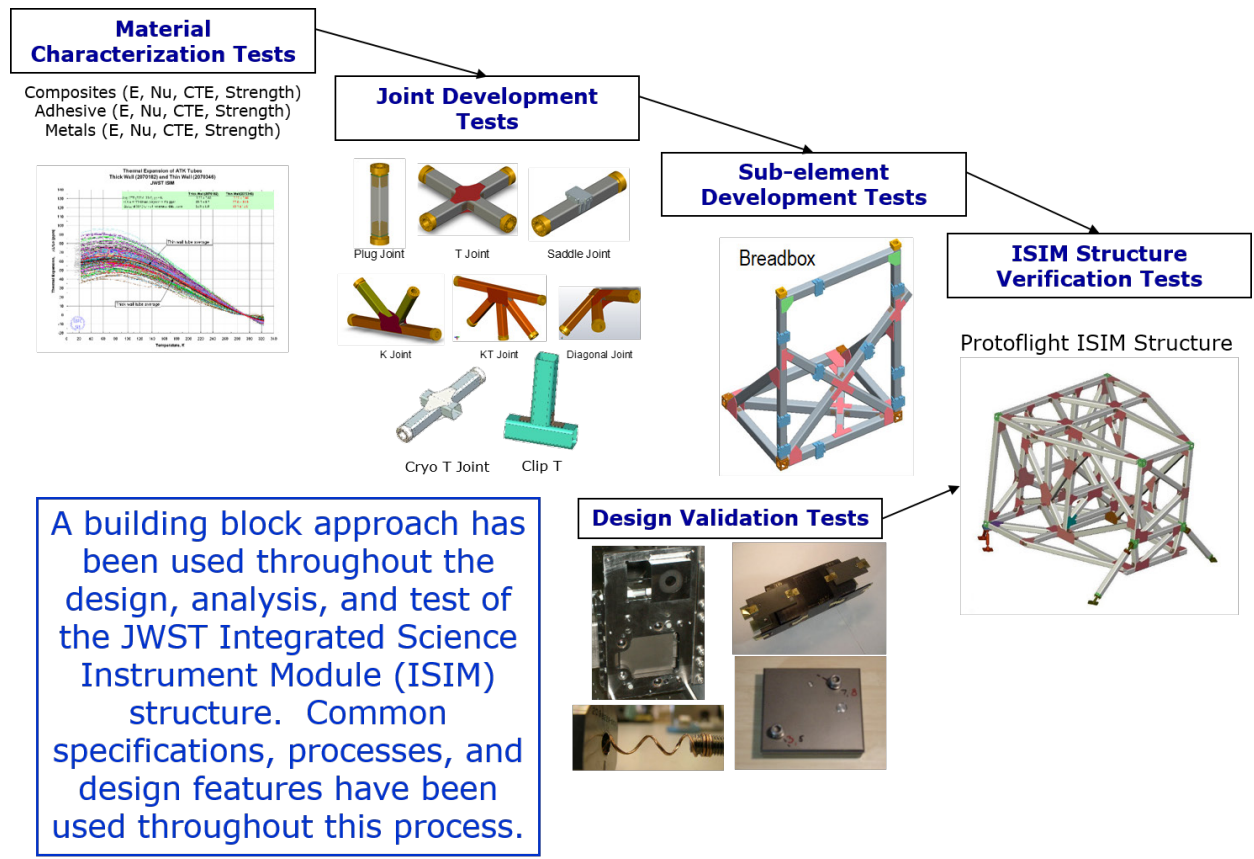

**Figure 12-23.** *The "building block" approach to model validation was used on different parts of JWST including the ISIM. This technique is similar to what will be needed to validate the models used to predict LUVOIR's performance in tests and on orbit.*

This process of developing validated analytical models requires a carefully developed integrated modeling plan. This plan starts with characterizing material properties at the coupon level, and then proceeds with a series of increasingly complex test articles to demonstrate that analytical models can predict what the test articles will do with various boundary conditions. Those validated analytical models are then used to predict performance of a system in a test. If successfully correlated to a system under test, the models should be able to accurately predict the flight performance. The JWST's Integrated Science Instrument Module (ISIM) team successfully implemented this process with the composite, cryogenic optical metering structure for JWST's science instruments. Tests started at a single coupon level, and proceeded to single lap shear test articles, then single flight-like joints, then more complex 3D joints, and ultimately a 3D composite structure with flight-like joint configurations. At every step, analytical models were validated against test data. **Figure 12-23** provides a top-level overview of this process.

Even though this process has been demonstrated on previous missions, those missions did not need to predict performance with picometer precision. Work is already being done to refine this type of a process so that smaller, more accurate measurements can be made in the model validation process (Saif et al. 2019). Improvements to software will need to continue so that picometer-level predictions can be accurately generated by the validated model.

### 12.5.3  Active control of the observatory

Even with advances in the ability to verify by analysis, something that distinguishes the design of LUVOIR from predecessor observatories is the active control that is built in by design. As an example, JWST uses a sunshield to passively cool it to a cryogenic temperature, and then relies on the stability of the environment to maintain the temperature within certain bounds. In contrast, while LUVOIR has a sunshade that will passively cool the observatory, LUVOIR is actively heated to a set temperature. Those heaters, along with the associated control system will be used to maintain the temperature of the observatory with precision on the order of milli-Kelvin.

LUVOIR will update its mirror alignment far more frequently than JWST. JWST will adjust the alignment of its primary mirror segments on a cadence of a few weeks. LUVOIR will update the alignment of its mirror segments several times per second. The system alignment will adjust to reject the influence of any outside disturbances, such as the effects of expanding or contracting materials.

These active controls will need to be demonstrated on their own in a test environment – as described in the technology development section (see **Chapter 11**). Once the technology is matured, these features will be integrated into the detailed design of the observatory and those designs will be tested in known—albeit not flight-like—conditions. Using the validated model technique described in the previous section, these active control systems will be tested to demonstrate that they will work in the flight environment.

In summary, while the requirements for thermal stability and mirror stability during certain observations are much tighter than previous flagship missions, the current concept design is also more adaptive to changing boundary conditions. This design should relieve some of the complexity in the verification and validation process.

## 12.6  Schedule overview

The schedules for both LUVOIR-A and LUVOIR-B have been developed following the guidelines outlined in NASA 7120.5 *NASA Space Flight Program and Project Management Requirements*, NASA/SP-2010-3403 *NASA Schedule Management Handbook*, and NASA/SP-2007-6105 *NASA Systems Engineering Handbook*.

### 12.6.1  Executive summary

The schedules outlined in **Figure 12-24** and **Figure 12-25**, and shown in detail in **Appendix G**, cover the pre-Decadal mission concept study described in this report, through the Pre-Phase A technology and concept development plan, to the end of a nominal 5-year operational mission lifetime (end of Phase E). As instructed by NASA HQ, the LUVOIR development schedule assumed Phase A start on January 1, 2025. For LUVOIR-A, the flight project development schedule for Phases A-D, including commissioning, is 15.6 years with a launch readiness date in November 2039. For LUVOIR-B, the flight project development schedule for Phases A-D, including commissioning, is 15.3 years with a launch readiness date in July 2039. As stated earlier in this chapter, we have taken the approach of optimizing the schedule with the intent of minimizing total mission cost. However, the schedule and development efforts are robust enough that they can be adjusted to accommodate many scenarios. Should individual year costs be constraining, the effort can be adjusted for that at the expense of total overall cost and schedule duration.

**Figure 12-24.** *LUVOIR-A Development Schedule Summary. See Appendix G for a more detailed version.*

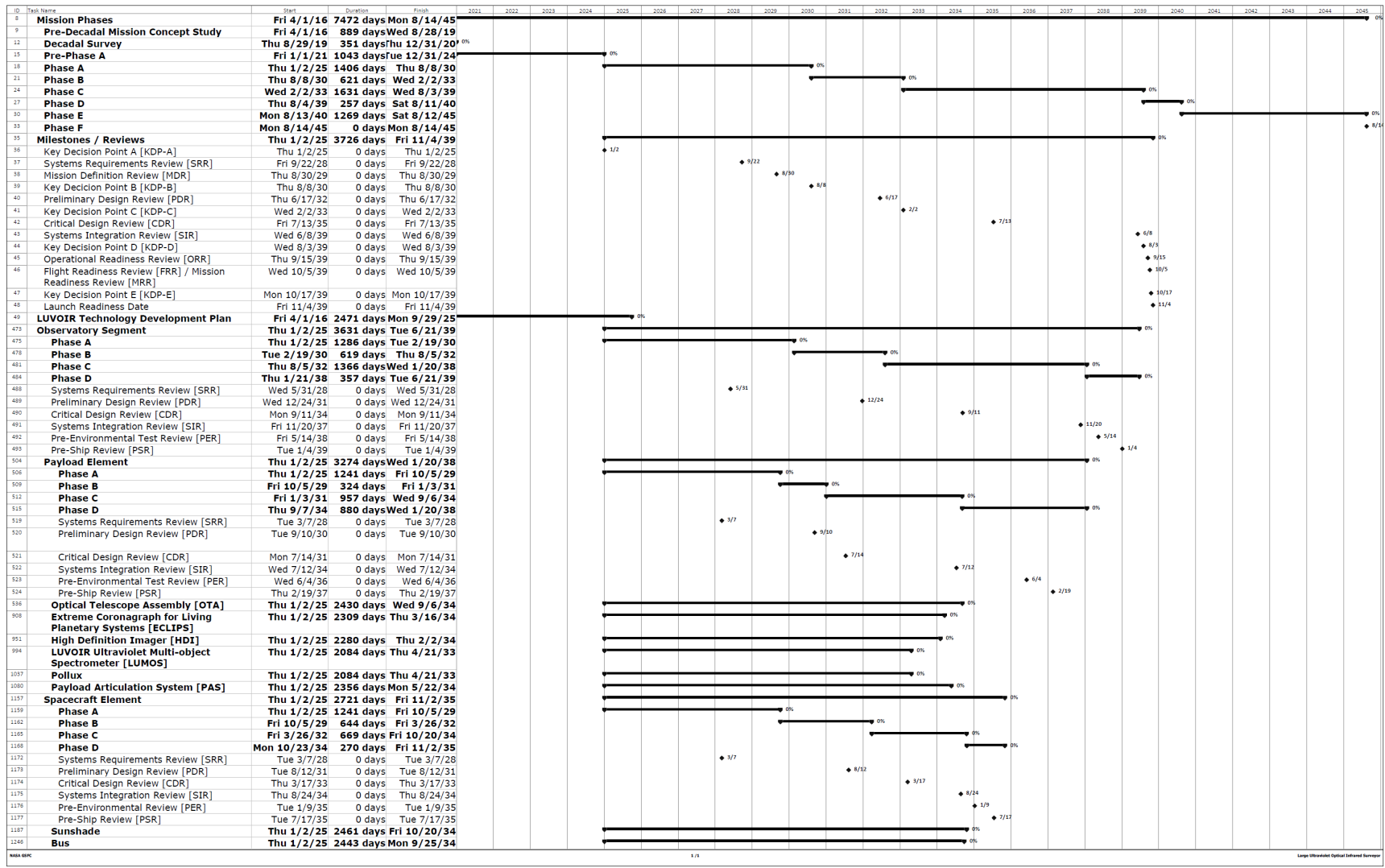

**Figure 12-25.** *LUVOIR-B Development Schedule Summary.* See **Appendix G** *for a more detailed version.*

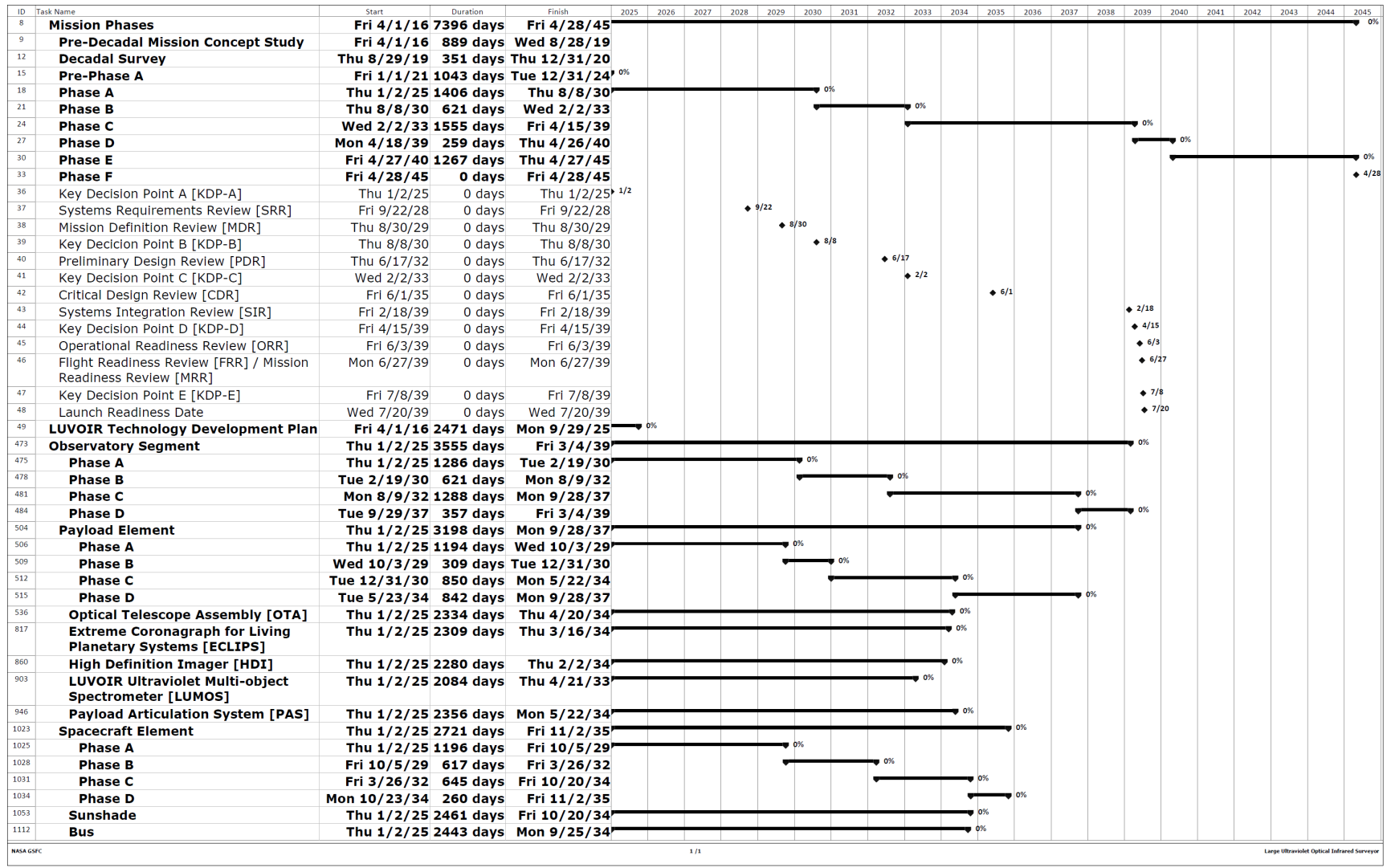

We assumed that the observatory segment is the critical path for the mission, not the ground support or the launch segments. As such, the ground and launch segments were ignored in this schedule development exercise. Furthermore, at this stage of concept development, we have based the schedules on the flight hardware only. We recognize that there are many other items that will need to be developed, including but not limited to: flight software, analysis software, pathfinder hardware, and mechanical, optical, and electrical ground support equipment. We also recognize that there will be many supporting tasks required to develop and integrate an observatory like LUVOIR, such as preparing facilities for integration and test (I&T). At this point in our study, we are assuming that the development of those other entities will be enveloped by the flight hardware development and that other preparatory tasks will be scheduled in a way to meet the currently scheduled events.

### 12.6.2 Developing the schedule

The schedules are built around the mission architecture as shown in **Figure 12-3**. The flight development portions of the schedule follow the same mission, segment, element, sub-system, and assembly organization in their layouts and logic. It is critical to LUVOIR's success that the management organization and the schedule both mirror the system architecture. The schedules were developed around the hardware development flow, from the ground up. We designed them assuming that the management strategies discussed throughout this chapter have been implemented. The development phases and major milestones were derived from the schedule.

The overarching phasing of any project is prescribed in the NASA references cited earlier and follows the NASA mission lifecycle shown in **Figure 12-26**. While these references provide overall guidance for the relationships between the phases and major reviews, they lack specifics on how they are linked to hardware development. The LUVOIR team studied both planned and as-executed schedules from several flight projects and generated our own metrics, shown in **Table 12-5**, for the reviews and milestones shown in **Figure 12-26**.

LUVOIR's schedules were developed using the as-run schedules of several projects, and distinguish between days necessary to do the work and days necessary to resolve unforeseen issues. Members of the LUVOIR team have extensive experience on JWST's I&T programs. That experience was used to better understand the differences between the nominal schedule based on the estimated time required to do the work, and the actual "as-run" schedule that required additional time because of other circumstances. Some of those differences have been translated into the management recommendations found earlier in this chapter. Some of those differences have been incorporated into better estimates of what the nominal schedule should have been. Funded schedule reserve is assumed to cover the remaining unforeseen problems (Bitten et al. 2014).

The schedule development process made us consider the way that Requests for Proposals (RFPs) are handled to better optimize the schedule. We recommend that RFPs only be issued at a single level—in our case, the "sub-system" level identified in the architecture—with the government acting as the system integrator for the mission elements and segments. This RFP approach distributes the engineering and fabrication work below the sub-system level to

**Table 12-5.** *Guidelines developed and used to determine review dates and phasing around the hardware development schedule.*

| Guidelines for Determining Review Dates and Phasing | | | | | | |
|---|---|---|---|---|---|---|
| **Architecture Level** | | **Mission** | **Segment** | **Element** | **Sub-System** | |
| | | | | | **with "assemblies" explicitly in the schedule** | **without "assemblies" explicitly in the schedule** |
| **Major Reviews** | **System Requirements Review (SRR) Must be in Phase A** | ~80 days AFTER the last **Segment** SRR | ~60 days AFTER the last **Element** SRR | ~40 days AFTER the last **Sub-System** SRR | Start ~2/3 through **Sub-System** "Requirements Refinement / Formal Interface Agreements" task. | Start ~2/3 through **Sub-System** "Requirements Refinement / Formal Interface Agreements" task. |
| | **Mission Design Review (MDR) Must be in Phase A** | (1/2) of the way between the Mission SRR and the end of Mission Phase A | n/a | n/a | n/a | n/a |
| | **Preliminary Design Review (PDR) Must be in Phase B** | (3/4) of the way through Element Phase B | (3/4) of the way through Element Phase B | (3/4) of the way through Element Phase B | Start ~40 days AFTER the last **assembly** PDR. Each **"assembly"** PDR is ~1/2 through "Design and Analysis" task in Phase B/C. | Start ~1/2 way through "Design and Analysis" task in Phase B. |
| | **Critical Design Review (CDR) Must be in Phase C** | (3/8) of the way through Element Phase C | (3/8) of the way through Element Phase C | (3/8) of the way through Element Phase C | Start ~40 days AFTER the last **assembly** CDR. Each **"assembly"** CDR is ~7/8 through "Design and Analysis" task in Phase B/C. | Start ~7/8 through Sub-System "Design and Analysis" task in Phase B. |
| | **System Integration Review (SIR) Must be in Phase C** | Start ~40 days prior to **Mission** "Integration" (at launch site) | Start ~40 days prior to **Element** "Integration" | Start ~40 days prior to **Element** "Integration" | Start ~40 days prior to **Sub-System** "Integration" | Start ~40 days prior to **Sub-System** "Integration" |
| | **Pre-Ship Review (PSR) Must be in Phase D** | Concurrent with the end of **Element** testing | Concurrent with the end of **Segment** testing | Concurrent with the end of **Element** testing | Concurrent with the end of **Sub-System** "testing" | Concurrent with the end of **Sub-System** "testing" |
| | **Operations Readiness Review (ORR) Must be in Phase D** | (1/2) of the way between the Mission Phase D start and the Launch Readiness Date | n/a | n/a | n/a | n/a |
| | **Launch Readiness Date (LRD) Must be in Phase D** | Concurrent with the end of Mission (launchsite) "testing" + FSR + MSR | n/a | n/a | n/a | n/a |

| | | | | | Sub-System | |
|---|---|---|---|---|---|---|
| Architecture Level | | Mission | Segment | Element | with "assemblies" explicitly in the schedule | without "assemblies" explicitly in the schedule |
| | | | | | | |

<div style="text-align:center"><strong>Guidelines for Determining Review Dates and Phasing</strong></div>

| | | Mission | Segment | Element | Sub-System with "assemblies" explicitly in the schedule | Sub-System without "assemblies" explicitly in the schedule |
|---|---|---|---|---|---|---|
| Phases | Start of A | Concurrent with the start of **Mission** "Architecture Development / Design & Analysis" | Concurrent with the start of **Segment** "Architecture Development / Design & Analysis" | Concurrent with the start of **Element** "Architecture Development / Design & Analysis" | Concurrent with the start of **Sub-System** "Architecture Development / Design & Analysis" | Concurrent with the start of **Sub-System** "Architecture Development / Design & Analysis" |
| | End of A | Duration of A is calucated as (3/2) * [Mission SRR – Mission Phase A start] | Duration of A is calucated as (3/2) * [Segment SRR – Segment Phase A start] | Duration of A is calucated as (3/2) * [Element SRR – Element Phase A start] | Set "Phase B Start" as the predecessor – (minus) one day | Set "Phase B Start" as the predecessor – (minus) one day |
| | Start of B | Set "Phase B Start" as Phase A end + 1 day | Set "Phase B Start" as Phase A end + 1 day | Set "Phase B Start" as Phase A end + 1 day | Concurrent with the end of **Sub-System** "Lower Level Requirements Definition" task. | Concurrent with the end of **Sub-System** "Lower Level Requirements Definition" task. |
| | End of B | Duration of B is [Latest Segment Phase B End - Earliest Segment Phase B Start] | Duration of B is [Latest Element Phase B End - Earliest Sub-System Phase B Start] | Duration of B is [Latest Sub-System Phase B End - Earliest Sub-System Phase B Start] | Duration of B is calculated as (4/3) * [SS PDR - SS Phase B Start] | Duration of B is calculated as (4/3) * [SS PDR - SS Phase B Start] |
| | Start of C | Set "Phase C Start" as Phase B end + 1 day | Set "Phase C Start" as Phase B end + 1 day | Set "Phase C Start" as Phase B end + 1 day | Set "Phase C Start" as Phase B end + 1 day | Set "Phase C Start" as Phase B end + 1 day |
| | End of C | Set "Phase D Start" as the predecessor - (minus) one day | Set "Phase D Start" as the predecessor - (minus) one day | Set "Phase D Start" as the predecessor - (minus) one day | Set "Phase D Start" as the predecessor - (minus) one day | Set "Phase D Start" as the predecessor - (minus) one day |
| | Start of D | Concurrent with the start of **Mission** "Integration." | Concurrent with the start of **Segment** "Integration." | Concurrent with the start of **Element** "Integration." | Concurrent with the start of **Sub-System** "Integration." | Concurrent with the start of **Sub-System** "Integration." |
| | End of D | Concurrent with the end of **Mission** "delivery." | Concurrent with the end of **Segment** "delivery." | Concurrent with the end of **Element** "delivery." | Concurrent with the end of **Sub-System** "delivery." | Concurrent with the end of **Sub-System** "delivery." |
| | Start of E | Set "Phase E Start" as Phase D end + 1 day | n/a | n/a | n/a | n/a |

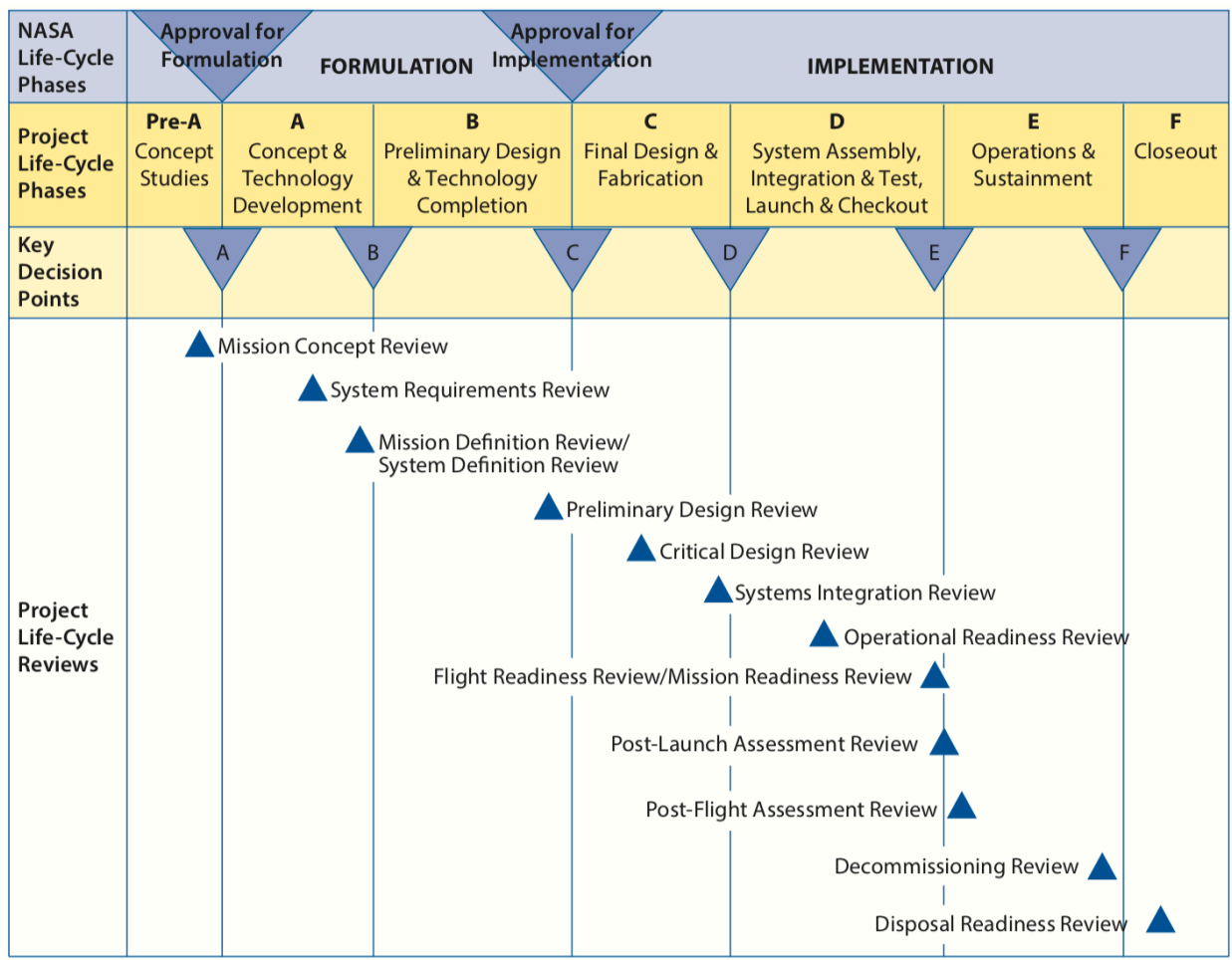

**Figure 12-26.** *The NASA mission lifecycle. From NASA/SP-2007-6105 NASA Systems Engineering Handbook*

multiple vendors. Each vendor would be responsible for any assemblies that are part of their sub-system. That is, the government would not issue separate RFPs for every assembly, only the top-level sub-systems. This helps to optimize the schedule by not having **serial** RFPs, which are time consuming processes.

Issuing RFPs in this way further helps optimize the schedule by allowing the mission to allocate requirements down to the sub-system level in a way that allows them to work as independently as possible from other sub-systems. This enables a single vendor to allocate requirements and resources from the sub-system down as they see fit, to make their sub-system compliant with mission-issued requirements, as opposed to requiring a prime contractor to allocate requirements down to the assembly or sub-assembly levels. The intent here is to enable as much parallel work as possible as early as possible.

The integration and test of the hardware is designed around the idea that the sub-systems (and some of the assemblies below that) will be designed to standardized interfaces. On all flight projects, things are designed to be modular in the sense that entities are designed, fabricated and integrated independently of each other and then integrated together. Here, we intend modular to refer to aspects of the design that enable "quick" integration of entities to their higher-level parent with minimal quality control testing. JWST's ISIM was not designed to be modular in this sense on purpose. JWST had four science instruments and three of them had to be installed in a particular sequence. That is, if instrument #3 had to be removed, instruments #2 and #1 had to be removed first. Before that could be done, a sequence of other hardware, including harness and heat-straps, had to be removed. This was as intended, because the ISIM was only meant to be integrated one time. However, circumstances required that the ISIM be integrated multiple times leading to schedule inefficiencies. Conversely, we are designing LUVOIR so that entities such as the science instruments can be installed and removed in any order. We are further designing the system in such a way that the payload infrastructure (harnesses, heat-straps, bracketry, etc…) can be installed independent of the science instruments. This is a critical assumption as it allows us maximum schedule flexibility. It also allows us to separate the time required to build the payload infrastructure from the time required to integrate entities that are modular.

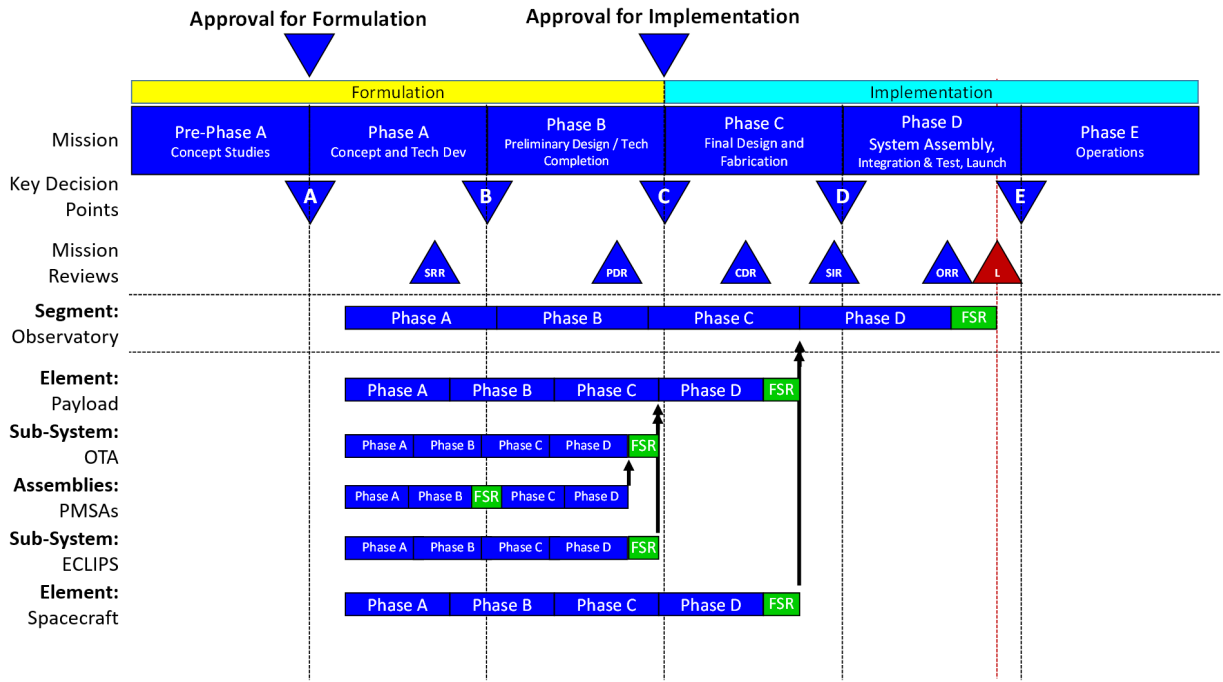

**Figure 12-27.** *Visual representation of how funded schedule reserve is applied to the LUVOIR schedule.*

### 12.6.3 Funded schedule reserve

Funded schedule reserve is included in the schedule and is accounted for as shown in **Figure 12-26** and **Figure 12-27**. The LUVOIR Study Team adopted the funded schedule reserve recommendation from the Aerospace Corporation (Bitten et al. 2014). The total funded schedule reserve on the critical path is 791 days for LUVOIR-A and 661 days for LUVOIR-B.

### 12.6.4 Schedule conclusion

We do acknowledge that there is a limit to the accuracy of any schedule at LUVOIR's current concept maturity level and that the schedule will need to evolve in parallel with the architecture and the concept should this mission proceed. Until such time as the schedule contains the details of each of the assemblies and their components, the full level of interdependencies cannot be truly understood. However, we have developed what we believe are ambitious yet achievable schedules, with the appropriate strategies at each of the project architecture levels described in this report.

**Table 12-6.** *Funded schedule reserve durations and where they are applied in the schedule.*

| Funded Schedule Reserve | | | |
|---|---|---|---|
| Architecture Level | FSR Rate | | Duration to Which FSR is Applied |
| Mission | 18 weeks per year | x | the duration between the start of Mission integration and the end of Mission testing |
| Segment | 18 weeks per year | x | the duration between the start of Element integration and the end of Element testing |
| Element | 18 weeks per year | x | the duration between the start of Element integration and the end of Element testing |
| Sub-System | 18 weeks per year | x | the duration between the start of Sub-System integration and the end of Sub-System testing |
| Assemblies | 15 weeks per year | x | the duration between the start of Assembly Design & Analysis and the end of Assembly Fabrication / Procurement |

## 12.7  Conclusion

NASA benefits from LUVOIR's plan to use management best practices and lessons learned from the Hubble servicing missions, Chandra, Cassini, Magnetospheric Multiscale Mission, MAVEN, and the OSIRIS-REx missions, as well as some aspects of the ongoing developments of JWST and WFIRST. The LUVOIR management approach also incorporates successful aspects of such large Department of Defense projects as aircraft carrier, submarines, and fighter jet procurements, and even NASA's Apollo program and Return to Flight after Challenger.

A critical aspect of LUVOIR's management approach is a detailed Pre-Phase A plan that:

- Matures all technologies to TRL 6 prior to initiating Phase A.

- Develops the science requirements, architecture, and concept in parallel with technology development, ensuring all converge to a single, coherent mission.

- Lays the foundation for long-term activities that will support the implementation of LUVOIR, including developing engineering test units, planning and developing pathfinders, and laying out plans for integration and test and verification and validation activities.

- Identifies facility gaps, and plans for the upgrade or construction of new facilities to support LUVOIR integration and test.

We have also identified several strategies that can be implemented at the project level, to better manage the scope and complexity of a mission such as LUVOIR. Furthermore, we propose a new method of funding large-scale NASA projects to address known issues with the phasing and stability of fund appropriations. Finally, we presented approaches to the integration and test and verification and validation of LUVOIR, and present a credible schedule for each LUVOIR concept. These recommendations, plans, and schedule incorporate fundamental lessons that are best summarized in a 2012 NASA OIG Report, "*NASA's Challenges to Meeting Cost, Schedule, and Performance Goals.*"

According to a 2012 NASA OIG Report, *NASA's Challenges to Meeting Cost, Schedule, and Performance Goals*, pg. 20,

*"Managing Technical Complexity and Cost Uncertainty. We acknowledge that space development projects are technically complex and their development costs are difficult to assess at the start of implementation when NASA managers are required to establish costs and schedule estimates. Nonetheless, in our judgment few projects should proceed to implementation unless requirements are well-defined and stable and the available resources—mature technologies, realistic schedule, and adequate funding—are set. In addition, the project's critical technologies should be matured to the extent that a prototype that closely approximates form, fit, and function is demonstrated in a relevant environment. Finally, adequate funding should be available to meet the project's requirements and account for its technical risks.*

*Over the years, the OIG, GAO, and others have reported extensively about the cost and schedule risks associated with projects that proceed to implementation with unproven technologies, inadequate funding, or unstable requirements. Collectively, those reports have identified measures that could help achieve more accurate cost estimates and minimize cost growth in NASA's projects, including: (1) maturing technologies prior to establishing baseline cost estimates; (2) appropriately funding management reserves to match technical risks; and (3) controlling changes to requirements."*

## CHAPTER 13.  POLLUX: EUROPEAN STUDY OF A UV SPECTROPOLARIMETER

## Executive summary

POLLUX is a high-resolution spectropolarimeter operating at UV wavelengths, designed for LUVOIR-A. POLLUX's study is supported by the French Space Agency (CNES), and developed by a consortium of European scientists (see **Appendix H.4**). POLLUX will operate over a broad spectral range (~100 to 400 nm), providing point-source spectroscopy at high spectral resolution (R $\geq$ 120,000). This will allow us to resolve narrow UV emission and absorption lines, enabling us to follow the baryon cycle over cosmic time—from galaxies forming stars out of interstellar gas and grains, and planets forming in circumstellar disks, to the various forms of feedback into the interstellar and intergalactic medium (ISM and IGM), and from active galactic nuclei (AGN).

The most innovative characteristic of POLLUX is its unique spectropolarimetric capability. This enables detection of the UV polarized light reflected from exoplanets or from their circumplanetary material, and moons, and characterization of the magnetospheres of stars and planets, and their interactions. Observations of planets in the solar system with POLLUX will be complementary to *in situ* observations, as they will inform us on the magnetospheric properties with the same levels of detail, but further provide a long-term survey capability.

The influence of magnetic fields on the Galactic scale and in the IGM will be measured. UV circular and linear polarization will provide a full picture of magnetic field properties and impact for a variety of media and objects, from AGN jets to all types of stars. Linear polarimetry is especially powerful to provide information on deviations from spherical symmetry, providing an extension of interferometry into a domain that is not restricted by the angular size of the objects but by their flux. This aspect of POLLUX will be a very powerful tool for studies of the physics and large-scale structure of accretion disks around young stars and white dwarfs, or supermassive black holes in AGNs, and to constrain the properties of stellar ejecta and explosions.

Since the parameter space opened by POLLUX is essentially uncharted territory, its potential for ground-breaking discoveries is tremendous. It will also neatly complement and enrich some of the cases advanced for LUMOS, the multi-object UV spectrograph for LUVOIR, by providing access to very high spectral resolution and spectropolarimetry. The very high spectral resolution of POLLUX is mandatory to clearly separate circumgalactic absorption lines from interstellar ones in circumgalactic halos, or to disentangle the volatile and refractory composition of planetesimals as functions of stellar mass and metallicity in edge-on debris disks, both strong science cases of LUMOS.

Similarly, while LUMOS is needed to map the launching region of jets and disk winds, and reveal their interaction, high-resolution spectropolarimetry with POLLUX is uniquely capable of constraining the strength and topology of stellar magnetic fields during pre-main sequence evolution, with important implications for stellar and disk formation and evolution. In planetary science studies, the large field of view of LUMOS can be used to map the surface and surroundings of a solar system body; for instance, spectral imaging of water plumes from icy/ocean moons (e.g., Europa). This would set the context for subsequent observations with POLLUX. The unique capability to perform UV polarimetry with POLLUX would then enable characterization of volcanism and/or plume activity in polarized solar continuum reflected light and the spectral UV albedo.

In this chapter, we outline a selection of key science cases driving the POLLUX design. We also introduce the current instrument concept and identify technological challenges that we will address in the coming years toward the advanced design and construction of POLLUX. Should LUVOIR-A be prioritized by the 2020 Astrophysics Decadal Survey, POLLUX will be proposed to ESA in the framework of a M-class call for proposals (cost at completion below 550 M€).

## 13.1 Science overview

### 13.1.1 Exoplanet atmospheres and interactions with the host stars

The characterization of exoplanets is key to our understanding of planet formation, including those in the Solar System. The unique, simultaneous, high-resolution and polarimetric capabilities of POLLUX in the UV are pivotal to unveil the origins of the huge range of chemical and physical properties found in exoplanetary atmospheres (e.g., Sing et al. 2016), and to understand how planets interact with their host stars (e.g., Cuntz et al. 2000).

#### 13.1.1.1 The characteristics of exoplanetary atmospheres

The line and continuum polarization state of starlight that is reflected by a planet depends on the star-planet observer phase angle and is sensitive to the optical properties of the planetary atmosphere and surface (**Figure 13-1**, left). Atmospheric gases are efficient scatterers at UV wavelengths making UV polarization a unique tool revealing the presence, coverage, particle size, and composition of aerosols (Rossi & Stam 2018; García Muñoz 2018). Furthermore, irregular temporal variations in polarization could indicate changing cloud patterns (i.e., weather), constraining heat transportation and distribution (e.g., García Muñoz 2015), while regular temporal variations could reveal planetary rings, trojans, and/or moons (Berzosa Molina et al. 2018). Because the polarized and unpolarized intensities originate at different altitudes in the exoplanet atmosphere, POLLUX observations will simultaneously probe two altitude ranges, thus being critical for testing models of exoplanet atmospheres. POLLUX can detect and spectrally resolve UV polarization signatures for close-in gas giants and brown dwarfs lying several tens of pc away from us, particularly if they orbit stars hotter than the Sun (**Figure 13-1**, right). Such a sample currently comprises about twenty targets, but several more will be found by Gaia, TESS, and PLATO. POLLUX will thus enable studying single planets in detail and gaining insights into the large diversity of planetary atmospheres using a sample large enough to be statistically valid.

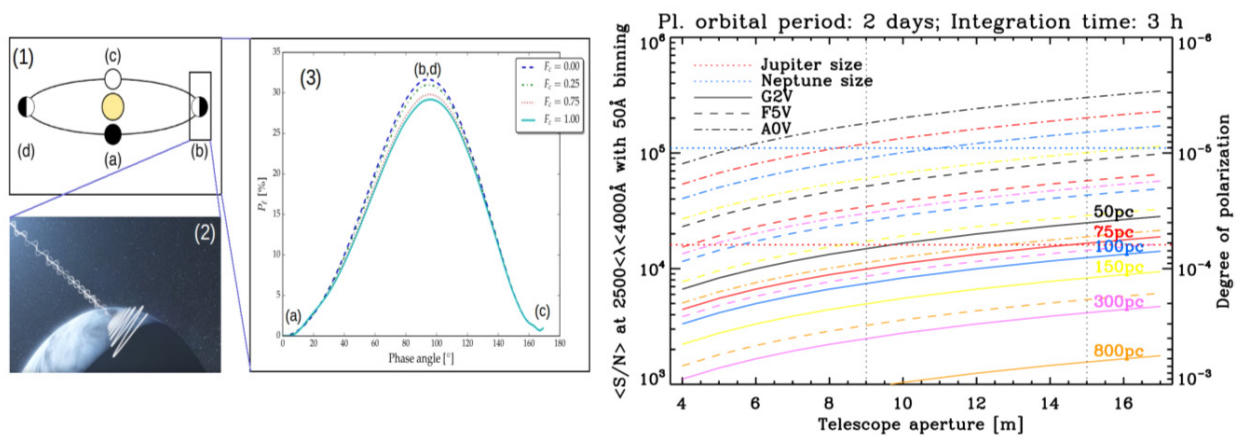

**Figure 13-1.** *Left (1) & (2): schematic of a planetary system in which the unpolarized stellar light becomes polarized through reflection by a planetary atmosphere. Middle (3): degree of polarization (in %) as a function of planetary orbital phase, that will be measured by POLLUX labeled as in panel (1), at 300 nm. The different lines indicate different atmospheric cloud coverages, where zero corresponds to the cloudless condition. Right: average SNR between 2500 and 4000 Å obtained with 3 hours of POLLUX shutter time as a function of the distance of the target star and size of the telescope primary mirror. Results are for a G2V (Sun-like; solid line), F5V (dashed line), and A0V (dash-dotted line) star. The red and blue dotted horizontal lines indicate the maximum UV polarization signal of respectively a Jupiter-radius and Neptune-radius planet in a two-day orbit around the host star. The black dotted vertical lines indicate the size of the primary mirror for LUVOIR's architectures A and B.*

The polarimetric capabilities of POLLUX will allow us to constrain the composition, optical thickness, and particle size of dust clouds resulting from disintegrating rocky planets, which could be the remnant cores of larger, gaseous, planets. The high spectral resolution of POLLUX, together with the large aperture of LUVOIR that allows to keep exposures short, further enables us to avoid blurring arising from the large orbital velocities of these objects. Kepler and K2 have already discovered such planets (e.g., Rappaport et al. 2012). The recent Dispersed Matter Planet Project already finds systems likely to host analogues and progenitors within 100 pc (Haswell et al. 2019), and TESS and PLATO will undoubtedly find more. These dust clouds provide the unique opportunity to analyze the composition of planetary cores, which is impossible even for Solar System bodies.

The high spectral resolution of POLLUX enables using cross-correlation techniques to detect and measure abundances of various key molecules in the atmospheres of nearby low-mass planets. This technique, applied to polarimetric high-resolution spectra, further enables the measurement of cloud coverages (García Muñoz 2018). These observations are not possible from the ground because the terrestrial atmosphere blocks the UV radiation. High-resolution spectropolarimetry is superior to broadband polarimetry to disentangle the planet and stellar signals through the corresponding planet-star Doppler shift. Key molecules presenting significant UV bands and possibly revealing Earth-like habitats are $O_2$, $O_3$, $SO_2$, $CH_2O$, and $NO_2$ (e.g., Schwieterman et al. 2018; Lammer et al. 2018). These molecules provide contextual information on a planet's habitability through their relations with the planet atmospheric composition, energy budget, volcanic activity, and the presence of hydrocarbons and lightning. Similar observations could be done for (young) planets in wide orbits around their stars, yielding for example planetary spin velocities, essential for understanding accretion and hence informing about planet formation, or UV auroral emission (e.g., for Proxima Cen, Barnard's Star b, and a handful of nearby brown dwarfs such as Luhman 16AB), revealing planetary magnetic fields and upper atmospheric composition (e.g., Ribas et al. 2018).

### 13.1.1.2 Tidal and magnetic star-planet interactions

Cuntz et al. (2000) theorized that star-planet interactions (SPI), either of gravitational or magnetic origin, could generate detectable signatures in exoplanetary systems. The repeated expansion and contraction of stellar tidal bulges produced by a close-in planet can lead to an increased level of stellar activity (Cuntz et al. 2000), and hence also planetary mass loss (Lanza 2013). Indeed, enhanced stellar activity has been seen for WASP-43 (Staab et al 2017), a system with parameters suggesting particularly vigorous tidal interactions. A fraction of the particles released in the magnetic star-planet reconnection funnels along the stellar magnetic field lines down to their foot-points on the star. The resulting condensations of material lost from the planet are expected to produce non-stationary narrow absorbing features across the profiles of stellar UV emission lines formed in the chromosphere and transition region, such as Mg II h & k, C IV, Si III and N V (e.g., Lanza 2009, 2014; Fossati et al. 2015a). Thanks to its high spectral resolution, POLLUX can detect the absorption signatures of the condensations of planetary-lost material for numerous nearby systems already known to host close-in giant planets. This will give the opportunity to observationally constrain the fate of the material that has escaped from the planet and to uniquely characterize the topology of the magnetic fields arising from SPI. Rocky planets orbiting late M dwarfs (Anglada-Escudé et al. 2016; Gillon et al. 2017; Ribas et al. 2018) are possibly subject to significant internal tidal and/or induction heating (e.g., Driscoll & Barnes 2015; Kislyakova et al. 2017, 2018; Barr et al. 2018). These heating mechanisms can be powerful enough to melt planetary mantles leading to strong volcanic activity or even magma oceans. The ejected volcanic material may then form a torus along the planetary orbit, similar to the plasma torus of the Jovian satellite Io. The high spectral resolution of POLLUX enables to reveal these structures through the detection of narrow absorption lines at the position of a variety of stellar emission features (e.g., Kislyakova et al. 2018, their Figure 6). The observations would uniquely yield knowledge on atmospheres and interior composition of rocky planets orbiting M dwarfs.

### 13.1.2 Testing fundamental physics and cosmology

Absorption-line systems produced by intervening gas in the spectra of background sources provide sensitive probes of fundamental physics and cosmology. These probes include (a) the measurement of the primordial abundance of deuterium, linked to cosmological parameters through the Big-Bang nucleosynthesis (BBN); (b) the redshift evolution of the cosmic microwave background (CMB) temperature (for which any departure from the linear relation could indicate a violation of the equivalence principle, or that the number of photons is not conserved, e.g., Uzan et al. 2004); and (c) the stability of fundamental constants over time and space. We note that while these are independent probes, they are intimately related by the underlying physics. For example, models of varying couplings (e.g., Avgoustidis et al. 2014) affect the temperature-redshift relation, $T_{CMB}(z)$, while BBN calculations of the D/H ratio also depend on the fundamental constants and can be altered if new physics is at play (e.g., Olive et al. 2012). Observations of electronic absorption lines with optical spectrographs (with R~50,000) on ground-based telescopes have put constraints on fundamental physics using the above three probes (e.g., Cooke et al. 2018, Noterdaeme et al. 2011, Ubachs et al. 2018). These were obtained for a handful of systems at high redshift, typically z~2–3, where the lines of interest are shifted redward of the atmospheric cutoff. A few additional constraints have been set at low and intermediate redshifts on both the variation of fundamental constants and $T_{CMB}(z)$, using radio observations of molecular ro-vibrational lines, but these remain very sparse.

The high spectral resolution UV coverage of POLLUX combined with the large collecting area of LUVOIR will make it a unique instrument for probing new physics through UV absorption lines in redshift ranges not accessible to future ground-based 30m-class telescopes. In particular, the z = [0.5–1] range corresponds to the end of matter domination and the start of acceleration, where, in some extensions of the standard model, we could expect the signatures of new physics, associated with the onset of domination by dark energy (e.g., Mortonson et al. 2009).

### 13.1.2.1  The D/H ratio

The relative abundances of primordial light elements depend directly on the product of $\Omega b$ and $h^2$ and on the physics considered in the BBN calculations. The measurement of D/H in primordial gas therefore places limits on possible departures from the standard physics, in particular through independent constraints on the cosmological parameters. Accurate measurements are possible by observing the Lyman series of DI and HI, which have rest-frame wavelengths in the range [911–1215] Å. POLLUX will enable precise measurements at 0.3 < z < 2.3 thanks to the simultaneous observations of the full Lyman series in its medium-UV and near-UV arms (see **Section 13.2**). At z < 0.3, the Lyman series extends over the far-UV and medium-UV arms (see **Section 13.2**), meaning that two observations would be required. The FUV alone can cover the full series for the local Universe (z~0). For typical gas temperatures of ~$10^4$ K (e.g., Noterdaeme et al. 2012), the DI lines are resolved at R~50,000, but a higher resolution remains mandatory to deblend components and properly assess the unabsorbed continuum regions in between Ly$\alpha$-forest lines. This is only achievable with POLLUX. The accuracy on D/H depends on factors such as the density of the Ly$\alpha$ forest (see Cooke et al. 2014) but, for a given system, it linearly correlates with the SNR. Cooke et al. reached a precision of about 0.01 dex on D/H at high-redshift using spectra with SNR~40 and R~50,000. Since the Ly$\alpha$ forest is less dense at lower redshift, such an accuracy should easily be reached with POLLUX in a few hours of observing time for 0.3 < z < 2.3 systems towards m~18 quasars. The main difficulty will be to identify low-metallicity systems for the measurement. Low metallicity is indeed important because it mitigates astration effects (destruction in interiors of stars) and provide potentially better constraints for BBN models (see also **Section 13.1.3**).

### 13.1.2.2  Evolution of the cosmic microwave background (CMB) temperature

Several interstellar species have transition energies in the sub-mm range and can be directly excited by photons from the CMB. A few, rare, CO absorption systems at z>1.6 have been used as a direct CMB thermometer in diffuse gas where collisional excitation is small (i.e., $T_{ex} \approx T_{CMB}$, Noterdaeme et al. 2011). The measurements are based on the relative column densities in different rotational levels, for which the corresponding lines in a given band are separated by only a few kilometers per second. The statistical uncertainty is currently of the order of 1K but this is expected to linearly correlate with the achieved SNR and the spectral resolution. Only POLLUX will allow full separation of the rotational lines and obtain high-precision measurement of the excitation temperature. We estimate that, for SNR~100 and R~120,000, the statistical uncertainty on $T_{ex}$ becomes less than 0.1 K. In addition to the main CO bands (at 100–150 nm rest-frame), the associated $H_2$ and the main CI fine-structure lines will be covered by POLLUX at z < 2.4. These are crucial to derive the physical conditions in the gas (e.g., Noterdaeme et al. 2017, Balashev et al. 2017) and correctly estimate the contribution by collisional excitation.

### 13.1.2.3 Variation of fundamental constants

Modern theories that try to unify the fundamental interactions predict the variation of fundamental constants over cosmological times and scales (e.g., Coc et al. 2006). Such constants determine the quantum mechanical energy-level structure of atoms and molecules. Comparing spectra of the same species in different places in the Universe and in the laboratory can therefore set strong constraints on any space-time variation of these constants (e.g., Ubachs et al. 2016). In particular, the proton-to-electron mass ratio ($\mu$) can be constrained in the distant universe by observations of the Lyman and Werner bands of molecular hydrogen (~900–1100 Å rest-frame). Similarly, combining the resonance UV metal absorption lines with 21-cm absorption lines, from the same absorber, leads to a stringent constraint on a combination of fundamental constants, say $x = gp\,\alpha^2/\mu$, where $\alpha$ is the fine-structure constant and gp is the proton gyromagnetic factor (e.g., Rahmani et al. 2012). The high resolution of POLLUX will provide better precision on single-component wavelength measurements, but also the de-blending of different kinematic components and the identification of hidden components, which otherwise produce systematic errors. As for D/H and the $T_{CMB}(z)$ relation, the precision depends almost linearly on the achieved SNR. For 100h exposure time, the precision on $\Delta\mu/\mu$ is of a few parts-per-million in each individual case, close to what has been obtained from radio observations of NH3 inversion transitions in rare systems (**Figure 13-2**). Finally, we note that POLLUX will probe lookback times not covered by any other observation.

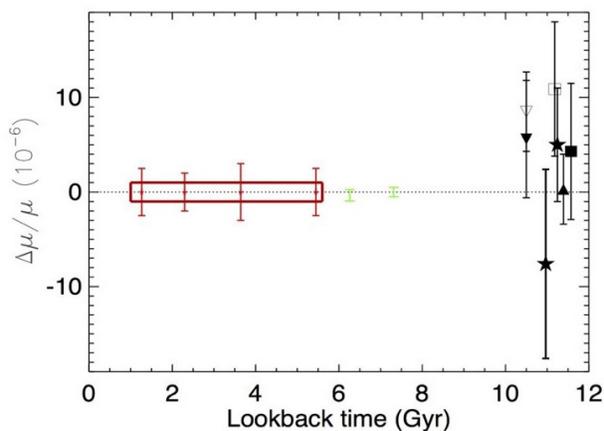

**Figure 13-2.** *Constraints on $\Delta\mu/\mu$ from existing observations at high-z using optical observations of $H_2$ (black points), intermediate redshifts from radio observations of $NH_3$ inversion and $CH_3OH$ rotational transitions (green points), see King et al. 2008, 2011, Malec et al. 2010, Rahmani et al. 2013, Albornoz Vasquez, D. et al. 2014. The red points show the precision expected with POLLUX in systems like those presented in Muzahid et al. (2015). The box corresponds to the combined POLLUX constraint, which will be about an order of magnitude better than that for the high-z counterparts.*

### 13.1.3 Interstellar and circumgalactic medium

In the UV, a wealth of absorption features arises from the interstellar medium (ISM) observed toward hot stars, or from the circumgalactic/intergalactic medium (CGM, IGM, respectively) observed toward quasars, yielding fundamental insights on chemical abundances, kinematics, and polarization of the gas across a wide range of environments.

### 13.1.3.1 The various phases of the ISM

Matter in the ISM is distributed in diverse, but well-defined phases that consist of the hot ($T \sim 10^{6-7}$ K) ionized medium, the warm (6000–$10^4$ K) neutral or ionized medium, and the cold (10–200 K) neutral medium and molecular clouds (e.g., McKee & Ostriker 1977; Cox 2005). Boundaries between phases can be quite abrupt and are believed to play a critical role in cooling the hot material, although it is not yet clear how they trade matter and entropy. Thermal conduction and turbulent mixing are, for instance, still poorly understood (e.g., Slavin et al. 1993 ; Scalo & Elmegreen 2004).

POLLUX will, for the first time, give access to all ISM phases at the same required high spectral resolution R > 120,000. It will enable simultaneous study of tracers of hot-gas (O VI, C IV **...**), warm gas (O I, N I and many singly-ionized species), as well as cold-H I gas like C I, and of the molecular phases through $H_2$ and CO lines. Together with the superior sensitivity reached by LUVOIR, this will open up a wide discovery space (some of it summarized in Jenkins & Wallerstein 1999), in particular (1) the dissipation of turbulence, the formation of $H_2$, and the role of $H_2$ as a coolant, through velocity-resolved $H_2$, CO, CH, CH+ **...** profiles in various rotational and vibrational levels, (2) the properties of cooling layers behind radiative shock fronts and the associated time-dependent ionization scheme through weak high-velocity components of S III, Si IV and C IV, (3) the interplay between energetic sources and the ISM through detailed ionization structure of the partly-ionized neutral gas and of the warm ionized gas (e.g., Gry & Jenkins 2017), (4) the identification and properties of tiny-scale ($10^{1-4}$ AU) atomic structures (Heiles et al. 1997; Stanimirović & Zweibel 2018) through tracking of spatial and temporal absorption variations, (5) the understanding of basic processes such as the $H^0$-to-$H_2$ and $C^0$-to-CO transitions, and the CO/$H_2$ ratio by reaching for the first time truly molecular clouds ($A_V \gtrsim 5$) where the chemistry is influenced by cosmic rays.

A revolutionary aspect of POLLUX is the possibility to study the magnetic field structure in relation with potentially all the ISM phases through the measurement of linear polarization of optically-thin UV absorption lines. Ground state alignment occurs when optical/UV pumping transfers angular momentum from the radiation field to the atoms/ions (Yan & Lazarian 2006), allowing us to trace 3D direction of magnetic field and to examine the specific role of magnetic fields in star/disk formation and, in general, in regulating the mass and energy circulation between stars and the ISM. Furthermore, unlike dust polarization measurements, POLLUX gives access to 3D magnetic field tomography by measuring its orientation in different locations along the lines of sight, on small spatial scales, and from multiple lines. Given the level of polarization expected, a signal-to-noise ratio (SNR) of a few hundreds is required (**Figure 13-3**).

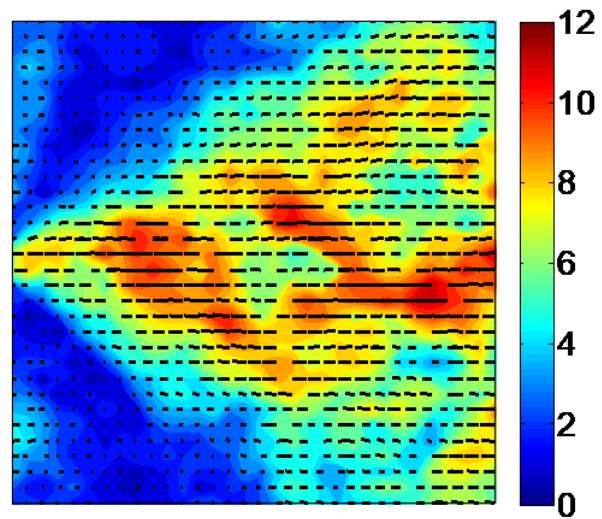

**Figure 13-3.** *Synthetic polarization (%) map (1 pc². An O/B star is located 0.1 pc at the left center. The color indicates the percentage of polarization induced in the S II 1250 Å absorption line and the orientation of the bars represents the direction of the polarization. The mean magnetic field is 3 μG, oriented along the X-axis. The expected polarization is mostly above 5%, within POLLUX capabilities (Zhang et al. 2015).*

#### 13.1.3.2 Extragalactic ISM

Beyond the Milky Way ISM, an important prospect for future UV observatories is to explore different environments. As such, LUVOIR is expected to extend the same level of detailed physical studies available in the Milky Way ISM to the ISM in external galaxies, in particular those with low metallicity or strong evidence of gas exchange with the IGM. In the UV, apart from a few lines of sight toward individual stars in the Magellanic Clouds, the extragalactic ISM has been mostly observed toward stellar clusters and at low spectral resolution, with inherent biases for column density determination (non-linear combination of column densities, "hidden" saturation, and mixed interstellar absorption and nebular emission along kpc-long lines of sight; e.g., Lebouteiller et al. 2013). LUVOIR and POLLUX will solve these technical issues, providing for the first time detailed and reliable knowledge of (1) the dust production/composition through depletion patterns, (2) the interplay between the ISM and compact sources, (3) the molecular gas content and distribution ($H_2$ clumps at ~pc size), (4) chemical abundance variations in different phases, and (5) the role of infalling and outflowing gas in the metallicity buildup.

#### 13.1.3.3 CGM: the link between galaxies and the large-scale cosmological component

No model of galaxy formation and evolution is complete without considering gas flows around galaxies, in particular in the CGM. The CGM is characterized by its extreme multiphase nature consisting of cold neutral/molecular gas clumps embedded in diffuse, highly-ionized gas filaments, with wide temperature ($50–5\times10^6$ K) and density ($10^{-5}–100$ cm$^{-3}$) ranges (e.g., Richter 2017; Tumlinson et al. 2017). High spectral resolution is necessary to disentangle the various components and to resolve the line thermal broadening while high sensitivity is required to increase the number of quasar lines of sight per halo that can probe the CGM of many galaxies at various redshifts. POLLUX will make it possible to probe gas flows, to map the distribution of metals, and, in general, to understand the role of CGM clouds in galaxy formation and evolution. Going back full circle, some of these questions can also be studied by observing the CGM of our own Milky Way toward a large sample of halo stars with known distances. Are CGM clouds disrupted and incorporated into the halo coronal gas or do they reach the disk where they can fuel star formation? It is worth noting that UV observations are well adapted to the measurement of much lower HI column densities ($\lesssim10^{13}$ cm$^{-2}$; Lehner et al. 2006) than those accessed with 21cm observations.

#### 13.1.4 Stellar magnetic fields across the HR diagram

We can obtain sophisticated tomographic mapping of structures on the surface of stars and their magnetic fields from current ground-based observations at visible wavelengths. However, to understand the formation and evolution of stars and their accompanying planets we also need to explore their circumstellar environments. POLLUX offers a powerful, high-resolution, full Stokes (IQUV) spectropolarimetric capability across the UV domain (100–400 nm) to uniquely trace these structures (see **Figure 13-4**). POLLUX will enable breakthroughs in several areas of contemporary stellar astrophysics.

#### 13.1.4.1 How do magnetic fields influence star formation and evolution?

Magnetic fields play a significant role in stellar evolution because they control the angular momentum budget (Townsend 2010) and mediate evolution of accretion processes and wind outflows (ud-Doula et al. 2008); they are also an important factor in both planet and star formation (e.g., McKee & Ostriker, 2007). For stars in the pre-main sequence (PMS) phase, UV spectropolarimetry is the best tool to characterize the strong magnetic fields arising at the sheared interface between the star and the disk (via UV lines like Mg II, Fe II and C II from the interface region), and to trace the flows via polarization measurements of the nearby continuum.

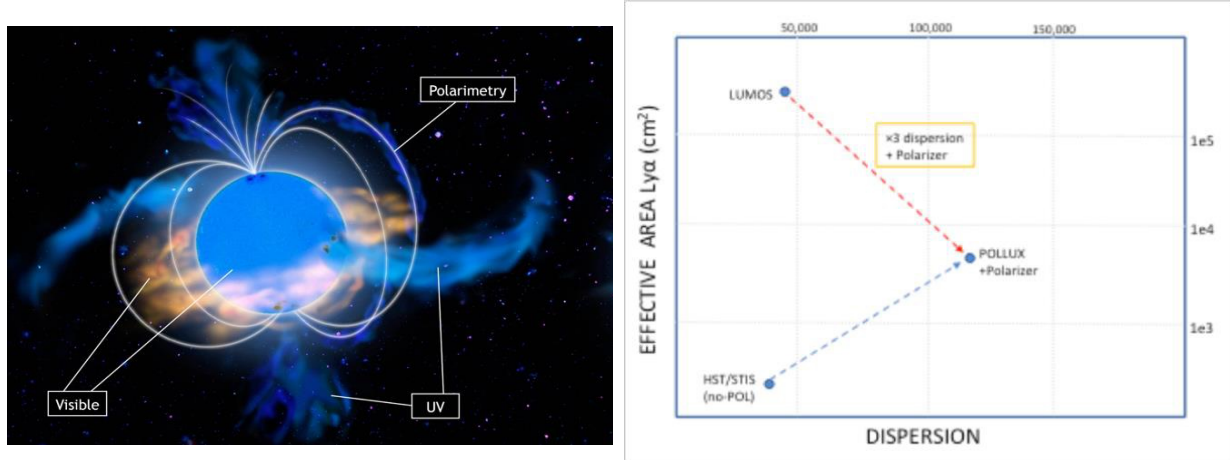

**Figure 13-4.** *Left: Schematic of the different features and regions traced at different wavelengths. Right: Parameter space covered by POLLUX and LUMOS compared to HST; POLLUX gives a huge gain in effective area and spectral resolving power.*

In the case of young low mass stars, interaction between the stars and their planetary disks are magnetically-mediated during dynamo formation and stabilization, when planets and planetary atmospheres form. As stellar rotation decouples from the young planetary disk, the magnetic field is predicted to get stronger and more complex (Emeriau-Viard & Brun, 2017). By investigating the propagation of magnetic energy through the stellar atmosphere into the uppermost coronal layers and stellar wind, POLLUX will reveal how the stellar dynamo builds-up and evolves.

### 13.1.4.2 Stellar magnetism in galactic massive stars, and beyond

A small but significant fraction of massive stars have strong magnetic fields (6-8%; e.g., Fossati et al. 2015b, Grunhut et al. 2017), but ultra-weak fields could be present in the other stars. POLLUX has the unique potential to detect these sub-Gauss magnetic fields in main-sequence massive stars. It will also be able to detect weak fields in hot evolved stars, providing a much needed understanding of the strong field-decay mechanisms and their impact on stellar evolution. Moreover, thanks to the unprecedented collecting power of LUVOIR, POLLUX will extend these studies of stellar magnetic fields beyond the Milky Way (MW) for the first time. This will open up observations in the metal-poor Magellanic Clouds (MCs), probing the evolutionary processes linked to magnetic fields at metallicities equivalent to those at the peak of star-formation in the Universe. This is crucial if we are to understand the progenitors of γ-ray bursts, pair-instability supernovae (SNe), and gravitational-wave sources (e.g., Petit et al. 2017). For that we need to test massive-star models (both for single stars and those in binaries) over a range of metallicity, from the local environs of the MW to metal-poor systems analogous to conditions in the early Universe. One big uncertainty in the latest models is the link between rotation and the mass lost via their intense stellar winds (e.g., Vink & Harries 2017), with a big impact on the ultimate explosions of the stars and the nature of their remnants. POLLUX observations in the Galaxy and MCs across all four Stokes parameters will map the density contrast with stellar latitude to assess the links between mass loss, rotation and environment, providing vital ingredients to calibrate evolutionary models.

### 13.1.4.3 Outflows & jets

The interaction between disks around PMS stars and their magnetospheres triggers powerful jets that are the earliest tracers of star formation, but the mechanism driving the jets is poorly constrained (von Rekowski et al. 2003, Gomez de Castro & von Rekowski 2011). UV spectropolarimetry with POLLUX of the base of such jets will allow determination of the dominant components of the magnetic field and their connection with disk rotation and ionization.

### 13.1.4.4 Extremely metal-poor stars

We can learn about the properties of the first generation of stars formed from primordial, metal-free gas via their chemical imprint on very metal-poor ([Fe/H] < –2.0) stars (Caffau et al. 2011, Keller et al. 2014). Space UV spectroscopy gives access to a vast range of diagnostic elements that cannot be measured from the ground (P, Ge, As, Se, Cd, Te, Lu, Os, Ir, Pt, Au), and enables more robust estimates of species that are otherwise very challenging (Fe, S, Cu, Zn, Pb). As these very metal-poor stars typically have low UV fluxes, the large collecting power of LUVOIR combined with the high resolution and efficiency of POLLUX is needed to probe the nucleosynthesis and early evolution of our Galaxy.

### 13.1.4.5 Novae & supernovae

High-resolution spectropolarimetry with POLLUX will constrain the dust lifecycle (formation and destruction) in the expanding ejecta in both classical novae and supernovae (SNe) remnants. We will dissect the ejecta structures from their line asymmetries and time variations during dust-forming events, allowing us to infer the effects of kicks imparted by core-collapse SNe, and the sites in the ejecta responsible for dust condensation. POLLUX spectropolarimetry will also tell us about the shapes/sizes of the grains involved, furthering our knowledge of this important channel of dust production.

### 13.1.4.6 White dwarfs

A few percent of WDs are known to be strongly (>1 MG) magnetic (Ferrario et al. 2015), but little is known of the incidence of weaker fields (Landstreet et al. 2012); UV spectropolarimetry with POLLUX will explore the presence (absence) of these. POLLUX will also provide estimates of the rotation periods for a large sample of WDs for the first time, needed to describe angular momentum transport in evolved stars and the processes responsible for magnetic-field generation, amplification of fossil fields and/or binary merger (Braithwaite & Spruit 2004; Tout et al. 2008).

### 13.1.5  Physics of active galactic nuclei

How much do we know about the UV polarization of active galactic nuclei (AGN)? Only 4 AGN were observed with WUPPE at low polarimetric resolution (see the Mikulski Archive for Space Telescopes). According to the Hubble Legacy Archive database, a total of 117 AGN were observed with different polarimetric instruments on board HST, out of which only 35 in the UV with HST/FOS. However, the spectral resolution offered by the instrument was $\lambda/\Delta\lambda \sim 1300$ at best. The polarized UV band of AGN is thus an almost uncharted territory. POLLUX will offer unique insight into the physics of AGN that is still little known, in particular by probing UV-emitting and absorbing material arising from accretion disks, synchrotron emission in jet-dominated AGN and large-scale outflows.

### 13.1.5.1  Probing the central AGN engine

Some key signatures of accretion disks can be revealed only in polarized light, and with higher contrast at UV than at longer wavelengths. Specifically, models of disk atmospheres usually assume Compton scattering in an electron-filled plasma, resulting in inclination-dependent polarization signatures (up to 10%, see e.g., Chandrasekhar 1960). Yet optical polarization is detected at less than a percent, and parallel to the radio jets if any (Stockman et al. 1979). Whether these low levels can be attributed to dominant absorption opacity (Laor & Netzer 1989) or complete Faraday depolarization (Agol & Blaes 1996) is unclear. This degeneracy can be broken by looking at the numerous UV lines that are formed in the innermost AGN regions (e.g., Ly$\alpha$ $\lambda$121.6, C II $\lambda$133.5, C IV $\lambda$154.9, Mg II $\lambda$280 **…**). These lines are the key to understanding UV polarization, and only observations with high SNR and the high spectral resolution of POLLUX can distinguish between the two effects.

At a parsec-scale distance from the accretion disk, an optically-thick dusty region probably forms along the equatorial region but its true geometry, size and composition remain unknown. A strong advantage for POLLUX is that the polarization induced by dust scattering rises rapidly toward the blue, peaking near 300 nm in the rest frame and remaining nearly constant at shorter wavelengths (see, e.g., Hines et al. 2001). Polarimetry at short wavelengths can thus discriminate between various grain models that provide different wavelength-dependent signatures and, if the grains are partially aligned, it can also help us to determine the magnetic fields topology and strength. Indeed, theory predicts that paramagnetic grains will be aligned with their longer axes perpendicular to the local magnetic field if exposed to magnetic fields (Lazarian & Hoang 2007). Therefore, POLLUX would not only selectively trace the smallest dust grains, allowing better characterization of AGN dust composition, but since the polarization degree is predicted to be proportional to the magnetic-field strength, polarimetry will also enable us to measure for the first time the intensity and direction of the magnetic field on parsec scales around the AGN core (Hoang et al. 2014).

Finally, trapped between the accretion disk and the circumnuclear dust is the broad emission line region. This dense, rapidly moving (> 1000 km/s) medium is responsible for the broad lines observed in the UV, optical and IR spectrum of AGN and their detection in the polarized light of dust-obscured AGN led to the bases of the unified model of AGN (Antonucci 1993). However, it was recently proven that a large fraction of non-detection of those broad lines in the scattering-induced polarized light was not due to a real absence of the region but rather to a non-sufficient spectral resolution. Thanks to the high-resolution capabilities of POLLUX, high SNR spectropolarimetric evidence (or absence) of those broad lines can be revealed in a large sample of AGN, allowing in-depth tests of the unified model of AGN (Antonucci, Hurt & Miller 1994; Ramos Almeida et al. 2016).

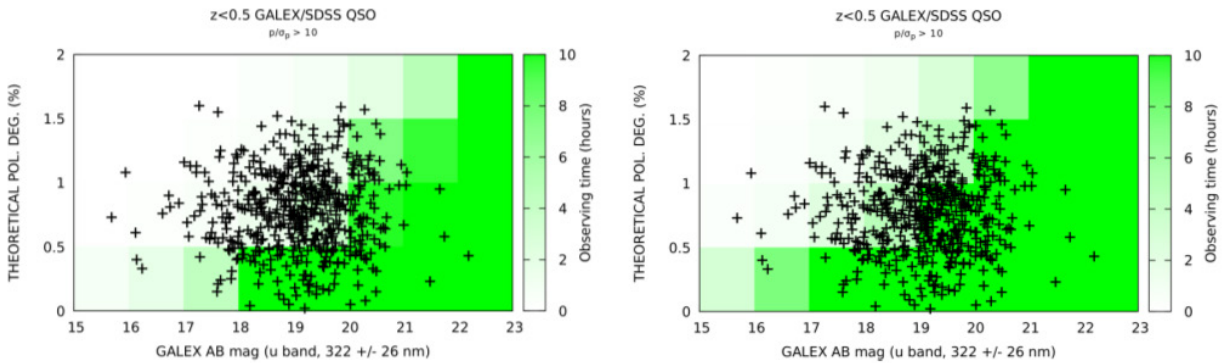

**Figure 13-5.** *Detectability of polarization in the GALEX/SDSS QSO catalog for a 15-m (left) and an 8-m LUVOIR aperture (right). The theoretical polarization degree in the u-band (322 ± 26 nm) assumes a Gaussian distribution centered around 0.8% with a standard deviation of 0.3%, similar to what was observed in nearby Seyfert-1 galaxies. The observing time needed to reach a SNR of $p/\sigma_p$ > 10 is color coded. About half the sample is beyond observation capabilities with an 8-m aperture.*

### 13.1.5.2  Outflows and jets

Magneto-rotational instabilities around the accretion disk are responsible for Poynting flux-dominated outflows (Blandford & Znajek 1977). The resulting jets tend to be collimated for a few parsecs and to dilute in giant lobes on kilo-parsec scales. Relativistic electrons traveling in ordered magnetic fields are responsible for the high polarization we detect (of the order of 40 - 60%, see e.g., Thomson et al. 1995). Interestingly, the continuum-polarization degree and angle are extremely sensitive to the strength and direction of the magnetic field, and to the charge distribution. By comparing the observed UV polarization of jetted AGN to leptonic, hadronic or alternative jet models, we will be able to better constrain the composition and lifetimes of particles in the plasma. Since jets are also responsible for ion and neutrino emission, they are valuable sources to understand how cosmic rays are produced.

In addition to jets, strong AGN outflows will be important targets for POLLUX. At redshift greater than 1.5–2, a sub-category of AGN, called Broad-Absorption-Line quasars (BAL QSO), exhibits very broad absorption features in UV resonant lines (Lyα, C IV, Si IV). BAL QSO are particularly interesting as they tend to have high polarization degrees (> 1%, e.g., Ogle et al. 1999), which can be used to constrain wind geometry (Young et al. 2007). These BAL QSO are believed to be the high-redshift analogues of more nearby, polar-scattered Seyfert galaxies, whose UV and optical emission can be explored by POLLUX. In particular, sensitive polarimetric observations would help investigate the dependence of broad absorption lines on bolometric luminosity and thus the role of radiative acceleration in the appearance of these lines (Arav & Li 1994; Arav et al. 1994).

### 13.1.5.3  Determining the impact of AGN on galactic evolution

At even larger scales (0.5 arcsec/kpc at z = 0.1), UV polarimetric studies of young star-forming regions will allow to constrain the relation between the triggering of star formation and the onset of nuclear activity, which is still unknown (Hough 2006). The radiative and kinetic power transferred from the AGN to the host can easily quench, suppress or re-activate star-formation, profoundly altering the whole galaxy (Wagner et al. 2012).

Polarization can help to quantitatively determine the AGN contribution to feedback by measuring the synchrotron (de)polarization signatures (Mao et al. 2014). In addition, high-resolution UV spectropolarimetry can bring additional formation on the star forming processes thanks to the Zeeman effect associated with the intense local magnetic fields. The star forming period can be probed by POLLUX thanks to spectral line polarization, such as advocated for NGC 1808 using the Hα line (Scarrott et al. 1993). Extended surveys of emission line polarization compared to other AGN activity indicators across the visible spectrum could measure how important the feedback is in the evolutionary path of galaxies.

### 13.1.6 UV spectropolarimetry in the solar system

#### 13.1.6.1 Surfaces of moons and small bodies

UV observations uniquely probe the surface of atmosphere-free telluric bodies of the solar system. They diagnose their volcanic and plume activity, their interaction with the solar wind and their composition in the frame of space weather and exobiology/habitability fields. When the Sun's unpolarized light is scattered by a rough surface or a dust covered surface of a solar system body, it becomes partially linearly polarized, and the polarization varies with the phase angle. This can be used to infer the properties of the surfaces of solar system bodies (Geake & Dolfus 1986).

The Wisconsin UV PhotoPolarimeter Experiment observations (WUPPE, Fox et al. 1997a) has revealed the Io surface as spatially covered by 25% $SO_2$ frost with polarization variations associated to different volcanic regions. POLLUX will primarily characterize volcanism and/or plume activity of icy moons from polarized solar continuum reflected light and the spectral UV albedo. Its high sensitivity is necessary to track any organic and ice composition of the crust of comets and Kuiper Belt objects from their UV spectrum. The UV spectral regime is fundamental to assess the presence of carbon on primitive bodies such as C, B, D asteroids and comets nuclei, since this element is nearly featureless both in the visible and in the IR. In the UV, carbon in various forms (amorphous, graphitized, hydrogenated, glassy) has a very important peak at 210–220 nm. To date, only *in situ* observations by spacecraft have been able to study these (e.g., the Rosetta mission at comet 67P, Feaga et al. 2015).

UV polarimetric observations of the Moon show a scattering from grains surface for wavelength < 220 nm, and volume scattering for wavelength > 220 nm (Fox et al. 1998). Such processes explain the change of polarization with phase angle at visible wavelengths (e.g., Steigmann et al. 1978). Systematic observations of the UV polarization of the surface of different objects in the solar system would open a new field of investigation to constrain the surface properties (refractive index, surface roughness, particle size, **...**). Such observations could be calibrated/validated with sample returns and then be used to interpret surface properties of exoplanets, or other interstellar objects that cannot be reached by spacecraft.

#### 13.1.6.2 Cometary comae

Comets show many emission features in their UV spectra, which allow the composition of their gas comae to be measured. At high resolution the composition can be determined at isotopic level, tracing the temperature structure of the solar system's proto-planetary disc (e.g., Yang et al 2018). Such observations are currently limited to the brightest comets; POLLUX would enable routine isotopic measurements to understand the variation across the full population. The sensitivity of a large UV space telescope would also allow POLLUX to investigate the drivers of cometary activity throughout the solar system. The UV has been identified as a key wavelength regime to investigate the enigmatic activity of the Main Belt comets, where sensitive observations could reveal evidence for the distribution of water within the asteroid belt (Snodgrass et al 2017).

### 13.1.6.3 Planet atmospheric properties

Several processes can polarize the light in planetary atmospheres: Rayleigh scattering, and scattering by aerosol/cloud particles, etc. The Rayleigh scattering cross section varies as ~ 1/$\lambda$4 while the scattering by aerosols/clouds is usually less wavelength dependent. Therefore, the Rayleigh scattering is generally dominant at short wavelengths while aerosols scattering is dominant at large wavelengths within the UV range. The transition between the two processes depends on the size of the aerosols and the atmospheric pressure.

On Mars, at a phase angle V = 21.7°, the linear polarized reflected light in the spectral range 200–400 nm is due to the tenuous atmosphere, while the reflected total flux in this spectral range is due to the surface and the atmosphere (Fox et al. 1997b). The WUPPE observations were used to retrieve simultaneously the surface albedo and the surface pressure of Mars. Such observations could be extended to derive the UV albedo from less studied bodies with optically thin atmospheres.

On Venus, the polarization is only due to the thick atmosphere. It has been measured at few wavelengths by Pioneer Venus Orbiter for different phase angles showing a contrast in the polarization of the dark and the bright region of the atmosphere of Venus (Esposito & Travis 1982). The UV spectrum at high spectral resolution coordinated with spectral imaging of Venus could be used to derive new information on the unknown absorber responsible for the dark UV regions in which 10% of the incident sunlight is absorbed.

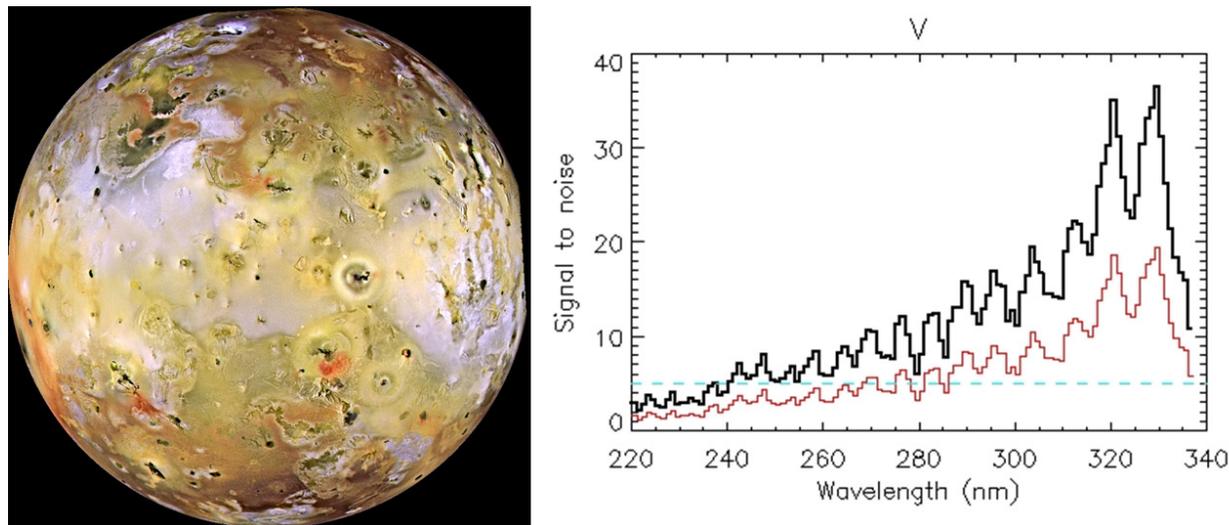

**Figure 13-6.** *Left: Volcanic activity of the Io surface. Right: Simulated signal-to-noise ratio for the Stokes parameter V for an observation of Io surface during an integration time of one hour and a resolution of 1 nm by POLLUX. Mirror aperture is 15 meters (black line) and 8 meters (red line). The light blue dashed line marks SNR = 5. Absorption by $SO_2$ ices causes the large decrease of the SNR below 320 nm.*

### 13.1.6.4 Magnetospheres and aurorae of the giant planets

The giant planets' UV aurorae are mainly due to atmospheric H atoms and $H_2$ molecules that are collisionally-excited by accelerated charged particles precipitating along the auroral magnetic field lines. Aurorae thus directly probe complex interactions between the ionosphere, the magnetosphere, the moons and the solar wind. Precipitation of aurora particles is a major source of atmospheric heating, knowledge of which is needed to assess the energy budget, the dynamics and the chemical balance of the atmosphere. POLLUX will measure the bright complex aurorae of Jupiter and Saturn, the fainter ones of Uranus, and catch those of Neptune, only seen by Voyager 2 (Lamy et al., 2017). The high spectral resolution will be used to map the energy of precipitating electrons from partial spectral absorption of $H_2$ by hydrocarbons (Gustin et al., 2017) and the thermospheric wind shear from the H Ly-$\alpha$ line (Chaufray et al. 2010). In addition, POLLUX observations of the persistently powerful planetary-like aurorae detected on several nearby cool brown dwarfs would provide critical insights into their magnetospheric/atmospheric interfaces, and into the origin of the precipitating particle streams above these substellar objects that do not involve an external stellar wind (Hallinan et al. 2015).

## 13.2 Overview of the instrument

The POLLUX study started in January 2017, and is supported by CNES (France). To define the present baseline configuration, we adopted the telescope parameters provided by the LUVOIR-A study, as of April 2018. No plans to build POLLUX for LUVOIR-B 8-m architecture were considered (see also **Section 13.2.12**). Indeed, the SNR below 130 nm of a putative POLLUX for LUVOIR-B would be so low that several science cases would become unfeasible. Furthermore, the off-axis telescope architecture of LUVOIR-B makes the demodulation process much less precise, eliminating science cases that require very high polarization accuracy.

POLLUX is a spectropolarimeter working in three channels. For practical reasons we refer to these as NUV (200–400 nm), MUV (118.5–200 nm), and FUV (90–124.5 nm). While the FUV channel extends down to 90 nm, the current sensitivity below 100 nm is low due to the LUVOIR-A mirror coatings (**Figure 13-22**). Each channel is equipped with its own dedicated polarimeter followed by a high-resolution spectrograph. This design allows to achieve high spectral resolving power with feasible and affordable values of the detector length, the camera optics field of view, and the overall size of the instrument. It also allows us to use dedicated optical elements, coatings, detector, and polarimeter for each band, hence gaining in efficiency. The MUV + NUV channels are recorded simultaneously, while the FUV is recorded separately.

POLLUX can be operated in pure spectroscopy mode or in spectropolarimetric mode. The full polarimeters are thus retractable in the MUV and NUV to allow the pure spectroscopic mode. In the FUV only the modulator is retractable. The analyzer is kept in the optical path to direct the beam towards the collimator. Full details about POLLUX can be found at https://mission.lam.fr/pollux/. The science goals described in **Section 13.1** lead to the technical requirements for POLLUX presented in **Table 13-1**, with comments in the last column.

**Table 13-1.** *POLLUX high-level requirements and goals*

| Parameter | Requirement | Goal | Reasons for Requirement/Goal |
|---|---|---|---|
| Wavelength range | 100–400 nm (design 90–400 nm) | 90–650 nm | 90 nm to go down the Ly series 400 nm to reach Ca II lines |
| Spectral resolving power | 120,000 | 200,000 | Resolve ISM line profiles, solar system and cosmology science cases |
| Length of the spectral order | 4 nm | ≥5 nm | To avoid having broad spectral lines spread over multiple orders |
| Polarization mode | Circular+linear (= IQUV) | | Magnetic field measurements Geometric shapes |
| Polarization precision | $10^{-4}$ | $10^{-6}$ | Detect weak stellar magnetic fields ($10^{-4}$); measure the fields of hot Jupiters ($10^{-6}$) |
| Aperture size | 0.03″ | 0.01″ | Avoid contamination by background stars in Local Group galaxies |
| Observing modes | spectropolarimetry and spectroscopy | | Optimize SNR as a function of science cases |
| Radial velocity stability | Absolute = 1 km/s and relative = 1/10 of pixel | | Absolute for line variations relative (within a spectropolarimetric sequence) to avoid spurious polarization signature |
| Flux stability | 0.1% | | Probe flux and polarization correlation in WDs |
| Calibration | Dark, bias, flat-field, polarization and wavelength calibration | Flux calibration | Flux calibration for spectroscopy is standard, but effective procedure is yet to be defined for spectropolarimetry |

## 13.2.1  Design implementation

The baseline configuration of POLLUX presented here allows fulfilment of all the requirements for the instrument performance needed to reach the science goals. Most of the technologies required for a complete implementation present technology readiness levels (TRLs) compatible with a Phase 0 study. We did not find fundamental restrictions or physical limitations preventing its implementation. We will discuss what research and development (R&D) is needed to allow us to realize the baseline configuration by the time of LUVOIR implementation. The cost of POLLUX has been assessed, details are available in **Appendix H.2**.

## 13.2.2  Optical design

The major assumptions that we adopted to design the baseline optical architecture of POLLUX are listed below. They are illustrated in **Figure 13-7**. Quantitative and qualitative data for all components of POLLUX are listed in **Table 13-2**.

- Although the current baseline wavelength range for LUVOIR-A starts at 100 nm, R&D work is ongoing to improve coating performance below 100 nm. Therefore, the spectral range used for POLLUX's design is 90 to 400 nm. Splitting into three channels allows to achieve high spectral resolving power with feasible values of the detector length, the camera optics field of view, and the overall size of the instrument. It also allows us to use dedicated optical elements, coatings, detector and polarimeter for each band, hence to obtain a gain in efficiency.
- The FUV and MUV boundaries are set relative to the Lyman-α line. The lower limit for the MUV band is set at Lyman-α minus roughly 3 nm, that is 118.5 nm, while the upper one for the FUV is Lyman-α + roughly 3 nm, i.e., 124.5 nm. Hence, Lyman-α is always observed (at redshift zero), whatever the POLLUX mode.

**Table 13-2.** *POLLUX components data*

| Channel | NUV | MUV | FUV |
|---|---|---|---|
| **Wavelength range, nm** | **200–400** | **118.5–200** | **90–124.5** |
| Polarimeter | | | |
| **Modulator** | 2 pairs of MgF$_2$ plates | | 3 mirrors in SiC |
| **Thicknesses, mm** | ~0.6 | | 20 mm x 2 |
| Analyzer | MgF$_2$ Wollaston prism | | ta-C mirror |
| **Prism angle, deg** | 8.302 | 9.815 | 33.365 |
| Wedge thickness center/edge, mm | 1/0.71 | 1/0.69 | N/A |
| Ray separation at detector, μm | 192 to 204 | 34 to 228 | N/A |
| Collimator | | | |
| **Focal length, mm** | 1600 | 1739.08 | 3255.73 |
| **Decenter, mm** | 111.61 | 121.3 | 227.11 |
| **Deviation angle, deg** | 4.0 | | |
| Echelle | | | |
| Frequency, 1/mm | 134.02 | 288.6 | 528.7 |
| Angle, deg | 60.587 | 58.362 | 39.927 |
| Orders | 33–65 | 30–50 | 20–27 |
| Order length, nm | 6.06–11.94 | 3.94–6.56 | 4.5–6.1 |
| Typical order separation, mm | 0.849 | 0.956 | 2.62 |
| Cross-disperser | | | |
| **Focal length, mm** | 1200 | | 2100 |
| Substrate | Sphere | | |
| Auxiliary mirror asphericity RMS/PTV, μm | 6.6/20.3 | 4.4/14.2 | 0.6/1.3 |
| Frequency, 1/mm | 133.3 | 204.5 | 267.98 |
| Recording wavelength, nm | 488 | | |
| Recording beams | Quasi-collimated in back-illumination + Aberrated in front illumination | | |
| Detector | | | |
| Type | δ−doped EMCCD | | |
| Active area, mm | 129.9 × 31.5 | 129.9 × 21.2 | 174.9 × 19.2 |
| Detector Size, mm | 129.9 × 63.0 | 129.9 × 42.4 | 174.9 × 38.4 |
| Pixel size, μm | 13 | | |
| Sampling, pixels | 2.5 | 2.3 | 2.15 |

- The shortest effective wavelength for the FUV channel strongly depends on the telescope throughput and may be reconsidered in the future, depending on coatings developments. The current design was developed for the lower limit equal to 90 nm. It would be easy to adapt the design to a cut-off at longer wavelength in the future if necessary.

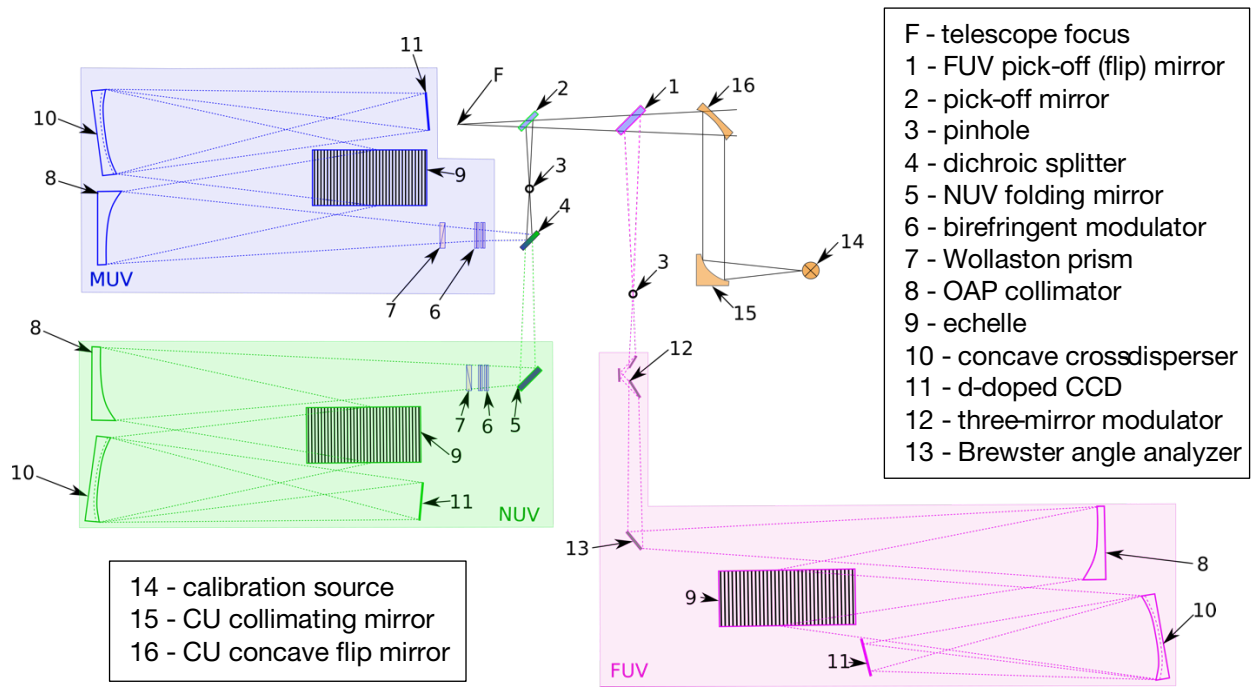

F - telescope focus
1 - FUV pick-off (flip) mirror
2 - pick-off mirror
3 - pinhole
4 - dichroic splitter
5 - NUV folding mirror
6 - birefringent modulator
7 - Wollaston prism
8 - OAP collimator
9 - echelle
10 - concave crossdisperser
11 - d-doped CCD
12 - three-mirror modulator
13 - Brewster angle analyzer

14 - calibration source
15 - CU collimating mirror
16 - CU concave flip mirror

**Figure 13-7.** *POLLUX baseline architecture schematic diagram. The elements in colored boxes are folded perpendicularly to the plane of the drawing, but are shown here to the left or right of the drawing for readability. NOTE: some folds are shown in 2D but actually they are in 3D.*

- In order to reach as high performance as possible, POLLUX uses two pick-off mirrors with optimized coating for the FUV and MUV+NUV channels, respectively.
- The FUV pick-off mirror is coated with SiC. In the telescope beam, it is located right in front of the MUV+NUV pick-off mirror, but is mounted such that it can be rotated off the beam, when the FUV channel does not need to be fed (see **Figure 13-8** and **Figure 13-9**).
- The MUV+NUV pick-off mirror is coated with Al+MgF$_2$.
- The instrument entrances are pinholes, rather than slits, for simpler aberration correction, and better polarimetric stability. Also, it allows rotation of each channel with respect to the chief ray.

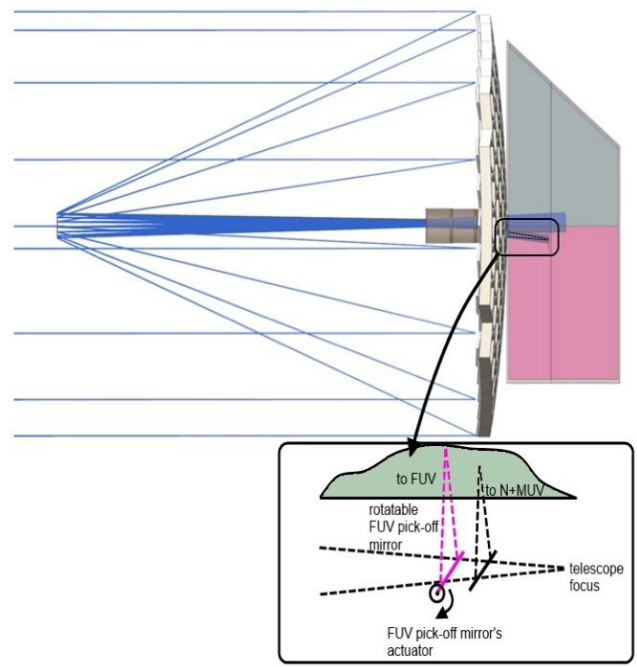

**Figure 13-8.** *Sketch of the mounting of POLLUX in the telescope payload, showing the instrument' pick-off mirrors.*

- The MUV and NUV channels are separated by means of a dichroic splitter. Such splitters can have a high efficiency (see **Section 14.5**). The dichroic splitter allows the instrument to work in two bands simultaneously and use the full aperture thus achieving the high resolving power with relatively small collimator focal length. In the present design, we set the MUV/NUV boundary at 200 nm, to have a maximum of one full octave in the NUV channel.

- The polarimeters are placed as close to the focal point as possible to reduce their size and influence on the image quality. The NUV and MUV polarimetric units are placed 30 mm away from the pinhole. For the FUV the distance is 20 mm.

- In each channel the beam is collimated by an ordinary off-axis parabolic (OAP) mirror. The off-axis shift and the corresponding ray deviation angle are chosen in such a way that the distance between the entrance pinhole and the echelle grating is large enough to place the polarimeter and corresponding mechanical parts. The MUV and NUV mirrors have slightly different geometry to provide the necessary sampling per pixel. They also have different orders separation and, subsequently, different cross-disperser diffraction and blazing angles.

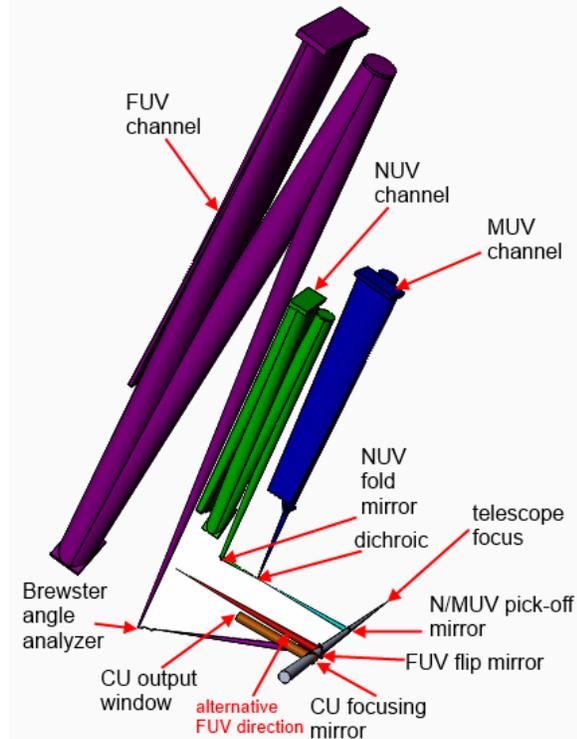

**Figure 13-9.** *3D rendering of POLLUX optical architecture*

- Echelle grating works in a quasi-Littrow mounting. The exact values of the groove frequency and the blaze angle are computed to obtain the target dispersion and subsequently the required spectral resolving power.

- The cross-disperser in each channel operates also as a camera mirror, so it is a concave reflection grating. This approach allows minimization of the number of optical components and increases the throughput. In order to correct the aberrations, the cross-disperser is a complex pattern of grooves formed by holographic recording. Each of the gratings is recorded by interference of two beams. One is collimated and another is aberrated by a customized freeform mirror. It allows to correct the aberrations over the extended linear field. The beams are oppositely directed, so it is possible to obtain triangular grooves with the necessary blazing angle without additional processing like ion-beam etching. The MUV and NUV camera parts are similar, which facilitates the manufacturing, testing and assembly.

- Adopted coatings on the optical elements of POLLUX are those used for the telescope, except for the polarizers. In the future, they will be optimized for each element of each channel (**Section 13.3**).

- Polarimeters are located immediately after the splitters in each channel to avoid instrumental polarization by the spectrograph elements. The polarimeters are retractable in the MUV and NUV to allow the pure spectroscopic mode. In the FUV only the modulator is retractable. The analyzer is kept in the optical path to direct the beam towards the collimator.

- Change of the optical path caused by removing the polarimeter from the beam is compensated by replacement of the collimating mirror. In each channel two OAP mirrors are mounted next to each other on a translation stage.

- The polarimeter design was optimized for each channel accounting for the technological feasibility (see **Section 13.2.3**). The polarimeters have the minimal size in order to decrease their influence on the image quality. Firstly, transparent plates introduce some aberrations. Secondly, due to polarization ray splitting the collimator may operate in an unusual mode and have considerable aberrations. Thirdly, the shorter the optical path inside the polarimeter, the smaller the difference between the spectropolarimetric and the pure spectral observation modes.

- Spot maximum sizes close to the detector center are 26.7 x 46.1 μm for the NUV, 38.5 x 62.6 μm for the MUV, and 37.9 x 70.5 μm for the FUV. The second dimension corresponds to the order separation direction. It may be increased to simplify the aberration correction, since it does not affect the spectral resolution.

It is necessary to switch beams in POLLUX in order to feed the detectors with light coming either from the telescope, or from sources in the calibration unit. Furthermore, in order to compensate the optical path difference and maintain the same beam position and the angle of incidence at the echelle when switching from the spectropolarimetric mode to the spectroscopic mode (done by removing the polarimeters from the optical train), it is necessary to change the collimator mirror (see #8 in **Figure 13-7**). Due to the focal length change, the collimated beam and therefore the theoretical resolution limit are also changed. On the other hand, the pinhole projection size is also changed, so the resolution values found with account for the aberrations is re-scaled.

### 13.2.3  Polarimeters

Each channel (FUV, MUV and NUV) has its own specific polarimeter in order to increase polarimetric performances. All polarimeters use temporal modulation. Each polarimeter is then divided in two components: a modulator and an analyzer. The modulator is rotating around the optical axis and rotates the polarization of the incoming light. The analyzer is a linear polarizer that filters the light in a fixed position. By rotating the polarization and filtering it, we create a temporal modulation which allows us to measure the input polarization. We use polarimetric efficiencies to characterize the performances of the polarimeters. We remind that to extract all Stokes parameters at the same time, efficiency optimum is at 57.7% for each Stokes parameter. Full details about the development studies of the three polarimeters are available in **Appendix H.3**.

### 13.2.3.1  NUV polarimeter

The NUV polarimeter works with birefringent material: $MgF_2$. Since plates of a few μm thickness cannot be built, the modulator is made of two double plates with each single plate thickness around 0.3 mm and a few μm thickness difference in each pair. In each double plate, the fast axis angle of plates are perpendicular and only the thickness difference plays a role in the modulation. For the first double plate, the thickness difference is 12.8 μm and the angle of the global fast axis is 32.6°. For the second double plate, the thickness difference is 3.7 μm and the angle of the global fast axis is 147.3°. The analyzer is made of an MgF$_2$ Wollaston prism. This design has been tested and validated in the visible, and is being tested in the UV.

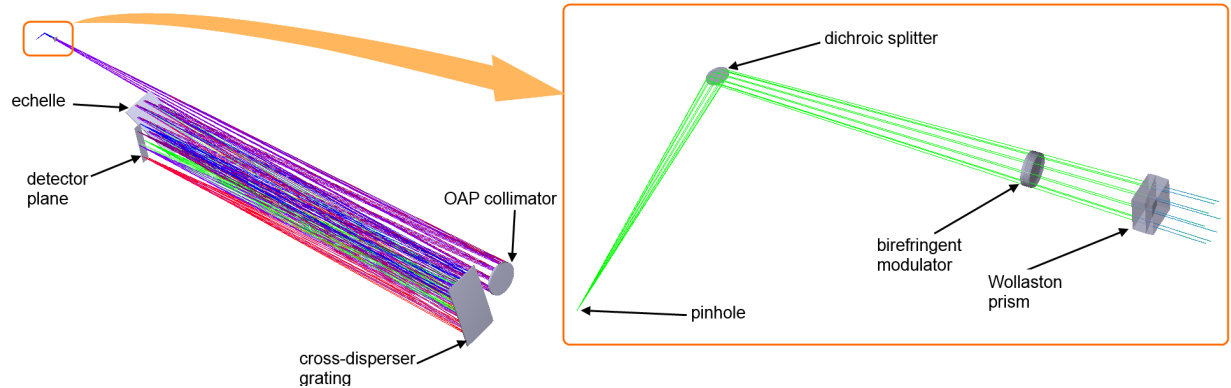

**Figure 13-10.** *Example optical scheme, here for the NUV channel, including a zoom on the NUV polarimeter unit.*

### 13.2.3.2  MUV polarimeter

From a polarimetry standpoint, the wavelength domain of MUV is remarkable by the absence of materials presenting birefringence all over the domain. MgF$_2$ loses its birefringence around 120 nm. A possible solution would be to use a reflective polarimeter as for the FUV (see below). While materials are being tested, a MgF$_2$ design is proposed as the baseline, excluding polarization measurements at wavelengths around 120 nm, which will be measured by the FUV channel. This design is similar to the NUV one. For the first double plate, the thickness difference is 9.6 μm and the angle of the global fast axis is 6.4°. For the second double plate, the thickness difference is 3.3 μm and the angle of the global fast axis is 70.0°.

### 13.2.3.3  FUV polarimeter

No known material transmits in the FUV domain. Therefore, POLLUX proposes the use of the retardance intrinsic to any reflection as a modulation mechanism. A 3-mirror device that rotates around the optical axis has been studied satisfactorily. Retardance is introduced by the reflections on the 1st and 3rd mirrors and the equivalent of a fast axis for birefringent crystals is given by the incidence plane onto the 1st mirror. Altogether, the device mimics a rotating waveplate. A plate at an analog of the Brewster angle will analyze the polarization (see **Figure 13-11**). An experiment is on-going to choose the best materials between SiC, ta-C and B4C. The proposed design focuses on SiC as basic material for the 3 mirrors in the K-device, and on a ta-C for the analyzer.

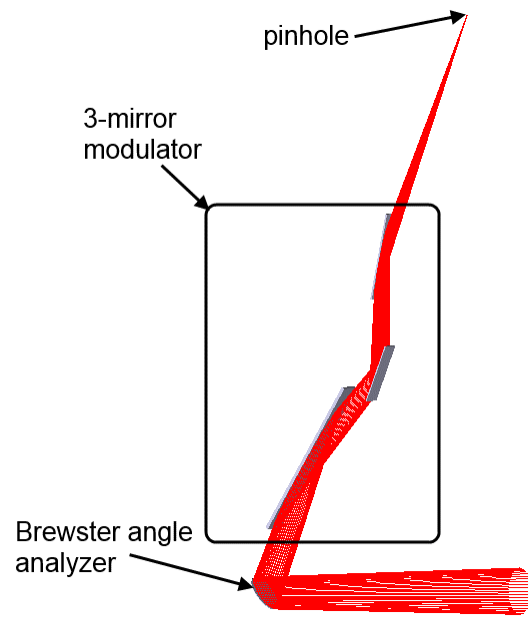

**Figure 13-11.** *Optical scheme for the polarimeter units of the FUV channel.*

### 13.2.4 Thermal and mechanical design

#### 13.2.4.1 Mechanical layout

The preliminary mechanical layout illustrates a feasible concept of the structural design with the necessary mechanisms that fit within the available space and mass envelope (**Figure 13-12**). This includes:

- Central support structure consisting of a central chassis that provides mechanical interfaces for the MUV, FUV, NUV arms, calibration unit, polarimetry and beam-switching mechanisms. This also provides the mechanical interface to the telescope.
- The MUV, FUV, NUV arms are generically similar optical benches constructed of rectangular or triangular box sections. They provide flat mounting interfaces for the collimator, disperser, camera and detector subassemblies. The benches are connected to the central support via box section brackets with angled bolted flanges.
- The collimator mechanism is implemented by linear slides using recirculating ball-bearing carriages. Two slides are used to better control tilt error and for robustness. The slides are positioned via a leadscrew driven by a stepper motor.

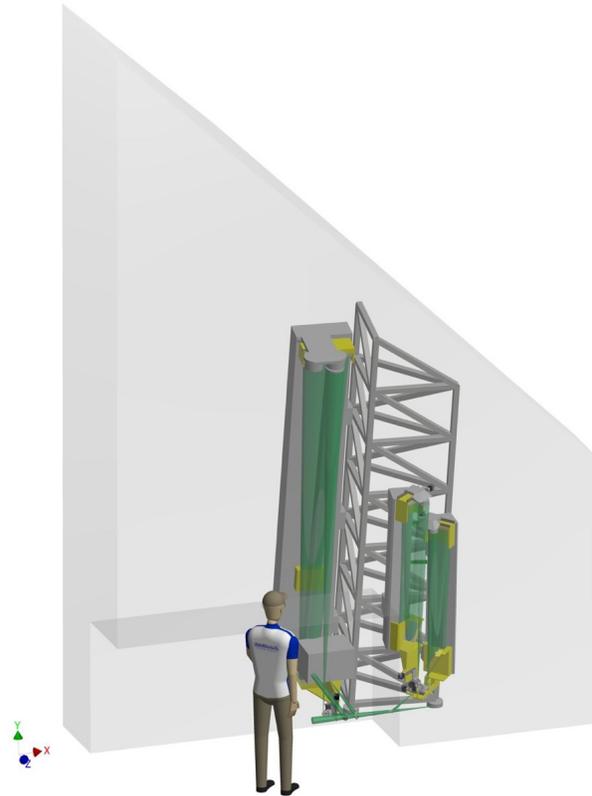

**Figure 13-12.** *Rendering of POLLUX within the allocated volume.*

- The FUV and calibration unit flip-mirrors are mounted on radius arms attached directly to the motor shaft to rotate them into the beam. To achieve the repeatability required, the mirror positions will be compliantly mounted and their positions defined kinematically against a stop.
- The MUV and NUV polarimeters and FUV modulator insert mechanisms are implemented by linear slides using recirculating ball-bearing carriages. Two slides are used to better control tilt error and for robustness. The slides are positioned via a leadscrew driven by a stepper motor.
- The polarimeter rotators are implemented by worm and gear mechanisms driven by a stepper motor. The axis is defined by a pre-loaded angular contact bearing pair.

For the FUV, the polarimeter modulator is rotated to four positions, while for the NUV and MUV, the modulators are rotated to six positions (**Figure 13-13**).

- The preliminary mass estimate of the current design, prior to optimization/ light-weighting, is ~375 kg, which includes a 25% overhead on the current CAD model to account for additional parts. This overhead will include items such as the necessary wiring/cabling and on-board computer (with the latter estimated at 8 kg; see **Appendix H.4.7** for more details). This is comfortably within the notional mass allocation of 480 kg.

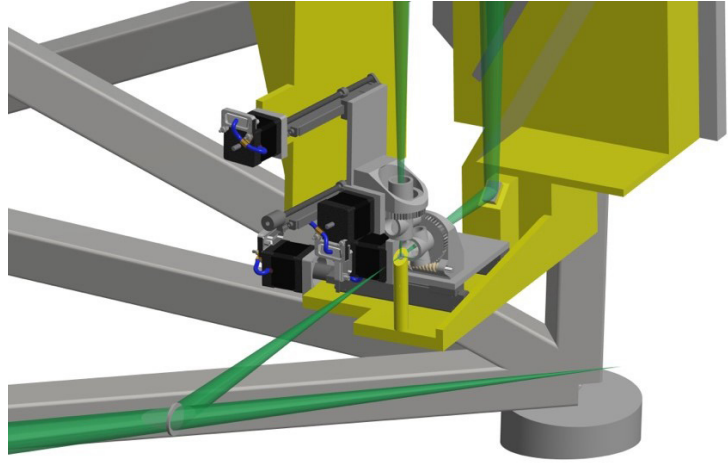

**Figure 13-13.** *Polarimeters and associated mechanisms for NUV/MUV channels.*

### 13.2.4.2 Thermal management

- POLLUX can operate with a 270 K housing, in line with the requirements of LUVOIR. The heat loads to 270 K will be conducted to the BSF support structure through the instrument chassis via the mechanical interface. The heat loads to 170 K (detectors) are assumed to be conducted to one of the radiators emitting to an 87 K background via copper wicks for each channel (**Figure 13-14**). The heat loads from the electronics and mechanisms in the current design are summarized in **Table H-4**, in **Appendix H.4**.

- As a conservative case to investigate the cooling wicks, we considered an optimum detector temperature of ~170 K. If the heat load is assumed not to raise the radiator temperature (infinite area), the wick must conduct the heat from this temperature to 87K.

- Integrated conductivity for tough pitch copper from 170 K to 87 K = 35.73 W/m. Contact resistance at joint is ignored as to first order bulk conduction will dominate.

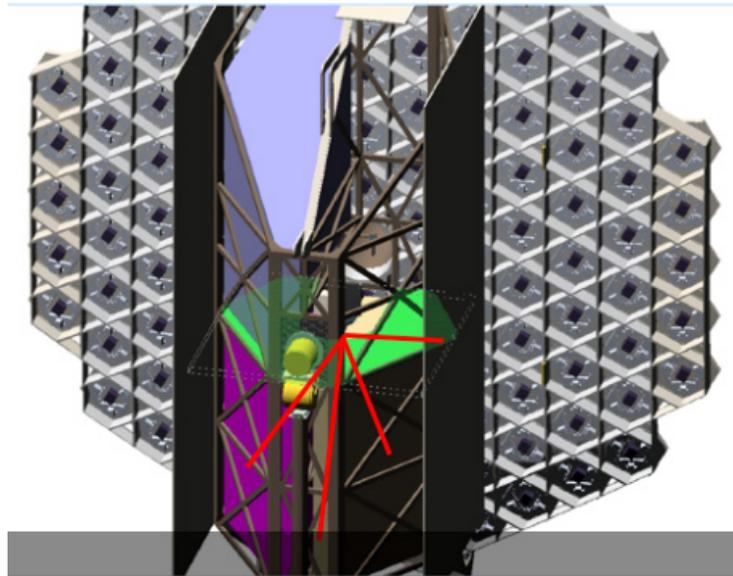

**Figure 13-14.** *Cooling wick route.*

- The estimated cross-section for the wicks is shown in **Table H-5** (**Appendix H.4**).
- Given the detector heat loads, there is only a modest impact on the radiator temperature, as shown for different areas in **Table H-6** of **Appendix H.4** (assuming radiator emissivity of 1.0). A rise in temperature of the radiator could be compensated for using wicks of large cross-section.
- Cold particle traps are also planned to reduce contamination of the optics (see **Section 13.2.11**).

### 13.2.5  Detectors and electrical design

The detectors of POLLUX are based on the technology of surface processing of thinned, back-side illuminated EMCCDs (e.g., "δ-doping"). EMCCDs (see **Figure 13-15**) have now become competitive with MCPs in the FUV to NUV range. They combine the linearity of CCDs with the photon-counting ability, which is a key capability enabling detection of faint UV signals.

Furthermore, these detectors now deliver high quantum efficiency (QE > 60%, see http://www.mdpi.com/1424-8220/16/6/927), thus offering the possibility to reach very high SNRs. Recent develop-ment show that visible-blindness can be achieved with proper treatment (e.g., anti-reflection coatings). Detectors with 13 μm pixels will be used for POLLUX (**Table 13-3**). They will be passively cooled to ~ 170 K (to reduce dark current level etc.). The three detectors can fit within one single wafer of 8″ each. Such wafer will be available by the mission Preliminary Design Review (PDR) which is anticipated to happen in 2025 (communication from Dr. S. Nikzad, JPL).

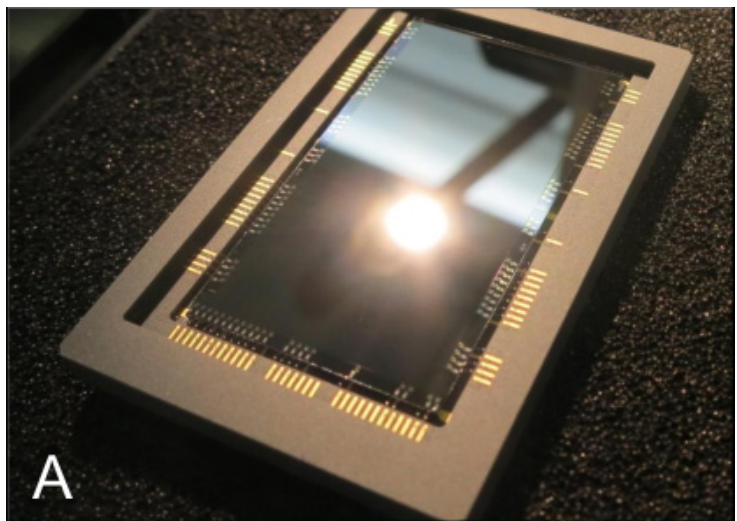

**Figure 13-15.** *Packaged EMCCD with custom AR coating for FIREBALL-2*

In order to operate the POLLUX detectors, a front-end electronics (FEE) has to be implemented to process their analog output signals and to generate clocks and biases. To achieve optimum performances of the sub-assembly (detector plus FEE), the location of the electronics unit allows short electrical connections between both elements; typically 10 to 15 centimeters depending on detector performances and readout rates. Internally the FEE is divided into two main blocks: the analog chain(s) for the processing of the output signal(s) of the CCD, from input coupling to the signal digitization. Height analog chains, corresponding to detector with height outputs were considered when conducting this preliminary design study. The second block implements the clocks and biases generation necessary to operate the detector and achieved readout of the accumulated photo-charges from the pixels.

**Table 13-3.** *Summary of the main characteristics of POLLUX EMCCDs*

| Detector Type | δ-doped EMCCD |
|---|---|
| Readout Mode | Split Frame Transfer |
| Pixel pitch | 13 μm |
| Number of outputs | 12 |
| CVF | 1.4 μV/e- |
| Amplifier noise | 50 e- |
| Max Data Rate | 10 Mpix/s |
| Op. data rate | 1 Mpix/s |

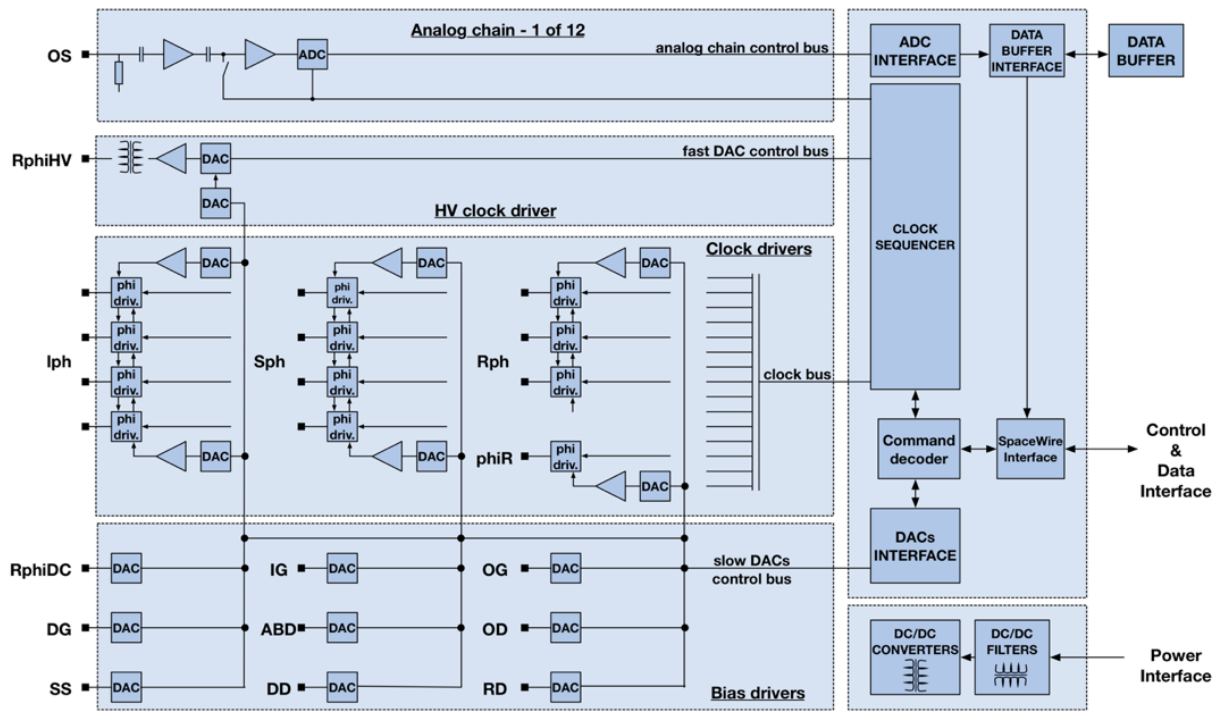

**Figure 13-16.** *FEE block*

In turn this block is divided into two parts: a digital clock sequencer that is generating different clock patterns for detector readout and analog driver circuits. Transmission of the digitized detector raw data is implemented as a bidirectional SpaceWire link. **Figure 13-16** depicts the internal block diagram of this unit based on the interface definition of the CCD 201-20 from e2v. Overall electrical architecture is given in **Figure 13-17** for the POLLUX instrument channels.

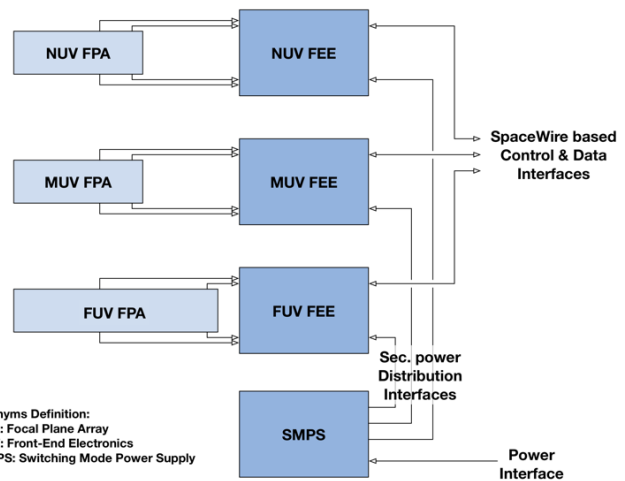

**Figure 13-17.** *POLLUX electrical block diagram*

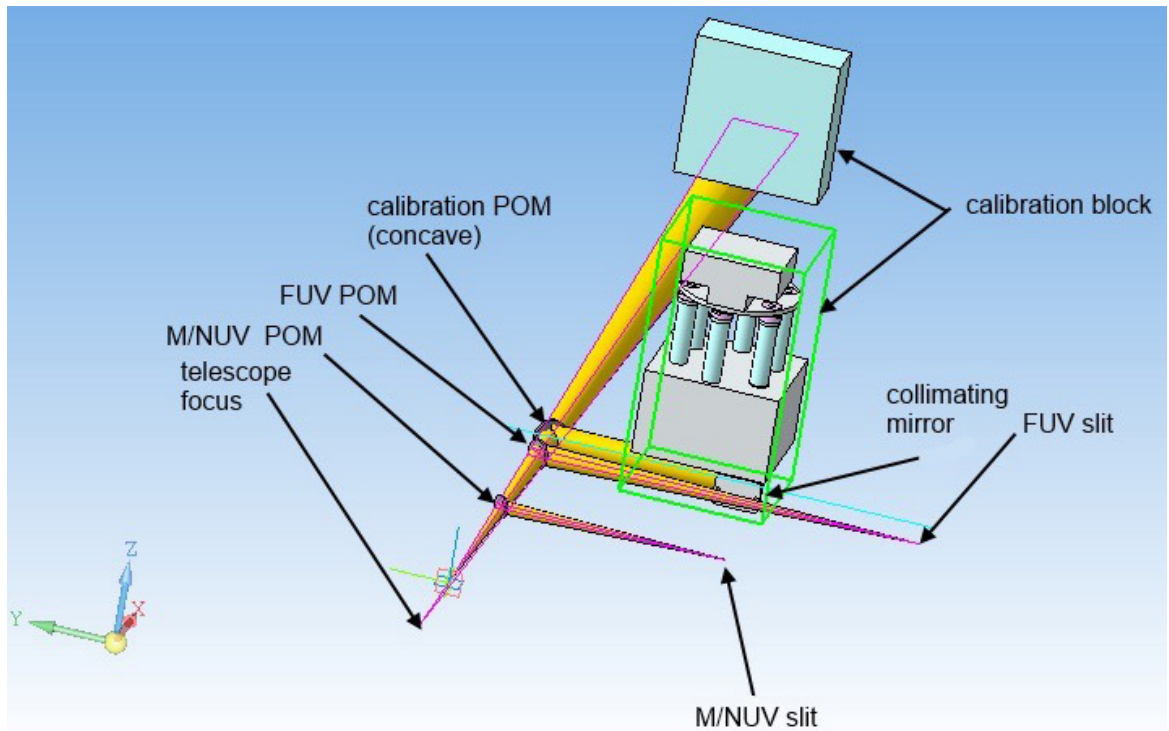

**Figure 13-18.** *Schematic rendering of the calibration unit of POLLUX*

### 13.2.6  Operating modes

POLLUX can be operated in pure spectroscopy mode or in spectropolarimetric mode.

- In pure spectroscopy mode, one measurement produces either one spectrum in the FUV, or two spectra, one in the MUV and one in the NUV. It is expected that in a majority of science cases, two measurements (one in the FUV and one in the MUV+NUV) will be obtained for a target.
- In spectropolarimetric mode, one full-Stokes polarimetric measurement requires four spectra in the FUV, or twelve spectra in the MUV+NUV channels (six spectra in the NUV and six spectra in the MUV). Here again, we anticipate that in most cases two measurements (one in the FUV and one in the MUV+NUV) will be obtained for a target.

In addition to the science modes, POLLUX can work in calibration mode. This consists of inserting one of the calibration lamps in the light path, instead of the stellar light, to acquire calibration images. The calibration unit (see **Appendix H.5**) groups the necessary light sources, including a cold redundancy. In order to minimize the impact of the instrumental polarization on the final calibration, the light from the calibration sources is injected before the polarimeters of each channel as early as possible in the optical chain, i.e., at the level of the mirror mechanism directing the light towards the MUV+NUV or FUV channels (see **Figure 13-18**).

Standard calibration images will be collected once per day for each detector through a fixed sequence: 10 flat-field images, 5 bias images, 1 dark image. In addition, wavelength calibration should be obtained. In pure spectroscopy mode: 2 wavelength calibration images will be downloaded at each new pointing of the telescope, one once the telescope is pointed and before the science acquisition start, and the other after the science acquisition is finished and the telescope moves away. In spectropolarimetric mode: 1 wavelength calibration image will be obtained not only before and after the acquisition but also between each spectropolarimetric measurement. In addition, one flux calibration image and one polarimetric calibration image should be obtained once per month by observing spectrophotometric standard stars (typically white dwarfs) and spectropolarimetric standard stars.

### 13.2.7 Main electronics

#### 13.2.7.1 Instrument control and power unit

The Instrument Control and Power Unit (ICPU) provides the interface between the dedicated FEE and the spacecraft interfaces. The main tasks for the ICPU are: providing the power for the FEEs and the ICPU, configuring the FEEs and the mechanisms for the different operation modes, collecting the scientific data, compressing the science data in case this is necessary, generating the telemetry packets and collecting the housekeeping data from the FEE and the power supply.

The Power Supply Unit (PSU) shall accept the 28 V spacecraft bus voltage. It provides the galvanic isolation between the spacecraft bus and the secondary voltages. The ICPU will be based on 3.3V logic. In addition, 2.5V might be needed to supply the core of Field Programmable Gate Array (FPGA) or processor. Additional core voltages, 1.8 V and 1.5 V, shall be generated by point of load converters inside the Digital Processing Unit (DPU). For the analogue interface stage presently a +5V bipolar supply is foreseen. A dedicated converter in the PSU shall supply the FEEs. The PSU shall provide all individual raw voltages and post regulation will be performed inside the FEE.

The POLLUX instrument contains several mechanical functions, collimator exchange mechanism, calibration fold mirror, polarimeter insert and rotation mechanisms and the flip mirrors. All these mechanisms need power and control for their activation as well as position sensors to enable a closed loop controlled operation. Stepper motors and similar actuators might be operated by H-bridges located inside the PSU, controlled via the control link between DPU and PSU. More complex operations might need additional electronics to be placed in the Analogue and Digital Interface board (ADI). A more detailed concept will be developed in the upcoming future. The signals from the feedback sensors will be handled by the ADI. In case of more complex operations an additional control circuitry (small FPGA with IP core) might become part of the ADI, to decouple data processing and instrument configuration (**Figure 13-19**).

Except for the SSDP, all other components are state of the art and have been used before in instrument control units for other missions (CHEOPS, Solar Orbiter). A similar design, using almost identical components, is presently under development for the Wide Field Imager as part of the ATHENA mission. Therefore, it is planned to purchase in the near future an evaluation board for the SSDP to run performance tests. At the time when the design for the POLLUX ICPU shall be started, a TRL level 5/6 will already be fulfilled.

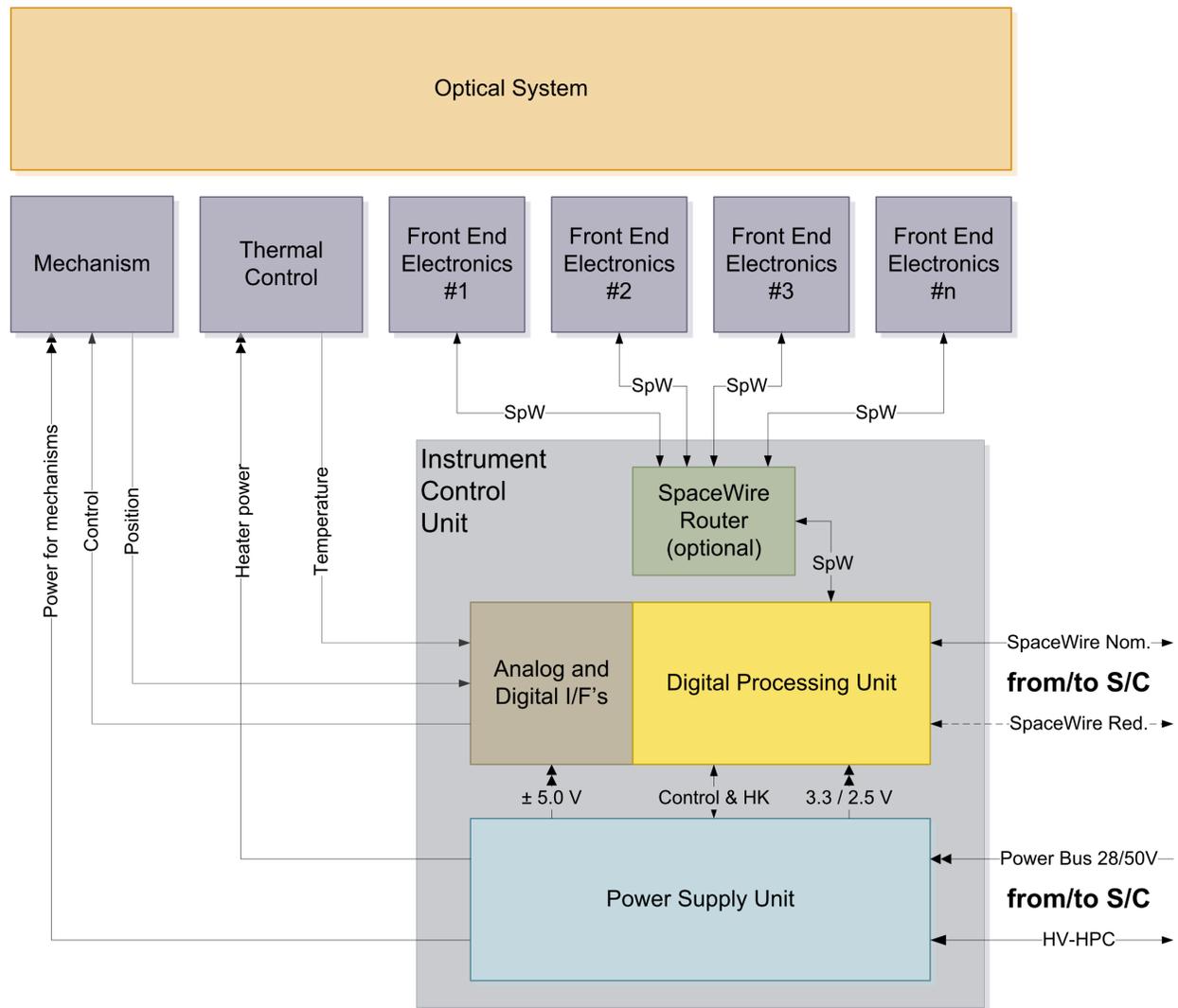

**Figure 13-19.** *Block Diagram for the Instrument Controller Unit*

### 13.2.7.2  Software

The POLLUX flight software, embedded in the DPU board, is in charge of all the monitoring and control aspects of the instrument as well as the data acquisition and on-board processing. The software can be broken down in the following sets of functions (see also **Figure 13-20**):

- Communication with the spacecraft: telecommand packet reception, telemetry packet transmission.
- Standard services: request verification, housekeeping reporting, event reporting, memory management, time management, on-board monitoring, are-you-alive connection test, parameter management, on-board operations procedure.
- Mode management: safe mode, standby mode, service mode, calibration modes, science modes (pure spectroscopy mode, spectropolarimetric mode).
- Instrument configuration management: exposure times, acquisition sequences, read-out frequencies, etc.
- Control of the instrument mechanisms: polarimeter retractability, UV channel selection, etc.

- Management of the FEE driving the detectors: configuration, control, monitoring, failure detection, recovery procedures.
- Housekeeping acquisition from the mechanisms and the FEE
- Data acquisition (full images) from the three detectors: FUV, MUV, NUV
- If needed, data processing: averaging performed on series of identical calibrations, windowing focused on the relevant zone of the detectors, lossless compression, meta-data reporting, telemetry packetization.

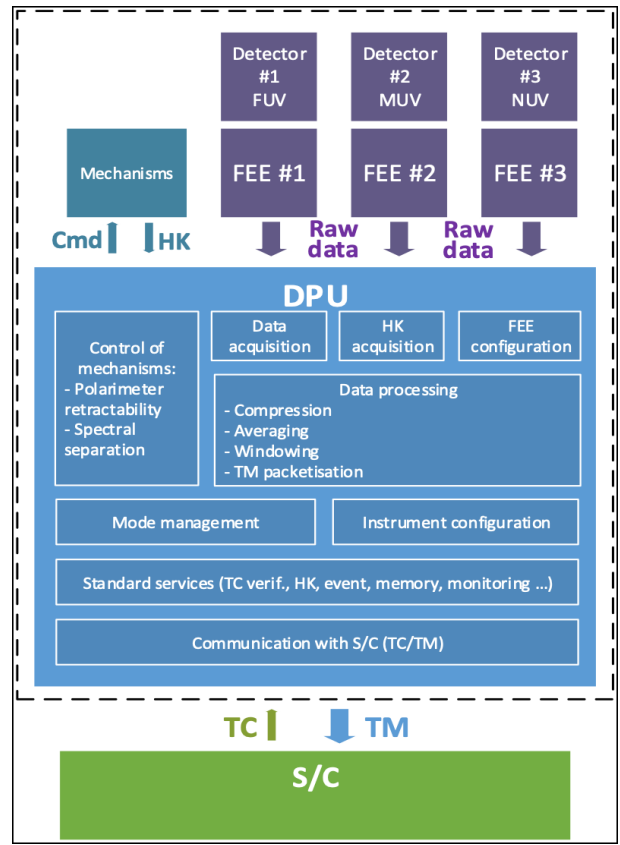

**Figure 13-20.** *POLLUX on-board software diagram*

### 13.2.8 Performance

The optical design is optimized for the conditions described in **Section 13.2.2**. In order to account for possible misalignments due to switching from the pure spectroscopic mode to the spectropolarimetric one, the target spectral resolving power was set to 123,000 rather than the required 120,000. The design of the spectrographs proposed here allows coverage of the wavelength range from 90 to 400 nm. In addition, polarization can be measured everywhere. The Lyman-α line is not covered in polarimetry in the MUV channel (because the birefringence of $MgF_2$ is around 0 at this wavelength), but is in the FUV. The Point Spread Function (PSF) is contained within 4 x 4 pixels (80% of encircled energy), thereby confirming that the image quality is good.

The polarimetric efficiency for the NUV and MUV channels (**Figure 13-21**) meets the expected requirements from the science cases for POLLUX. For the FUV channel, the efficiency decreases towards the shortest wavelengths but may be improved if better materials are available. The overall efficiency of POLLUX was computed under the following set of assumptions:

- Although the telescope is currently coated with Al+LiF+MgF$_2$, reflectivity data are for Al+eLiF (recommendation from LUVOIR's Lead Engineer).
- There are 4 bounces inside the telescope, before the light beam enters POLLUX.
- Coatings for the pick-off mirrors are optimized: Al+eLiF for MUV and NUV channels, and SiC for FUV channel.
- The dichroic has reflectivity and transmissivity properties from the study by Safran REOSC (see **Section 13.2.2** and **Section 13.3**).
- NUV and MUV polarimeters are made of MgF$_2$.
- The 3-mirror modulator of the polarimeter in the FUV is SiC. The analyzer is ta-C and it is assumed that the Brewster reflection is perfect (it does not introduce flux losses).
- The MUV and NUV polarimeters and FUV modulator can be retracted entirely (although in the design the FUV analyzer remains in the beam).
- OAP collimators are coated with the same material as the pick-off mirrors, respectively.
- For the echelle gratings, theoretical values of the reflectivity have been used. The coating is Al+eLiF for the MUV + NUV channels, and SiC for the FUV channel.
- Cross-dispersers represent reflective concave holographic gratings. Each holographic element is recorded by interference of 2 beams, one of them is reflected from a flat freeform mirror. It is necessary to correct the aberrations, so in the spectral resolution modelling the exact wavefronts corresponding to the recording beams are considered. However, in the efficiency modelling they are replaced by the nearest spherical wavefronts. The difference in efficiency caused by this simplification is negligible.
- EMCCDs have anti-reflection (AR) coatings. Quantum efficiencies from Nikzad et al. (2017) were used (see **Figure 13-15**). Since the wavelength coverage of each POLLUX channel is broader than in this paper, two AR coatings for NUV and MUV detectors have been used, each coating covering one half of the sensitive area. On the other hand, we could not find reliable data for AR coating reaching 90 nm, hence the curve for bare EMCCD was used.
- The total efficiency (**Figure 13-22**) is the product of efficiencies of all the elements. The result was also converted into the effective area (adopting a geometrical coefficient A$_{geom}$ = 155 × 10$^4$ cm$^2$ for LUVOIR).

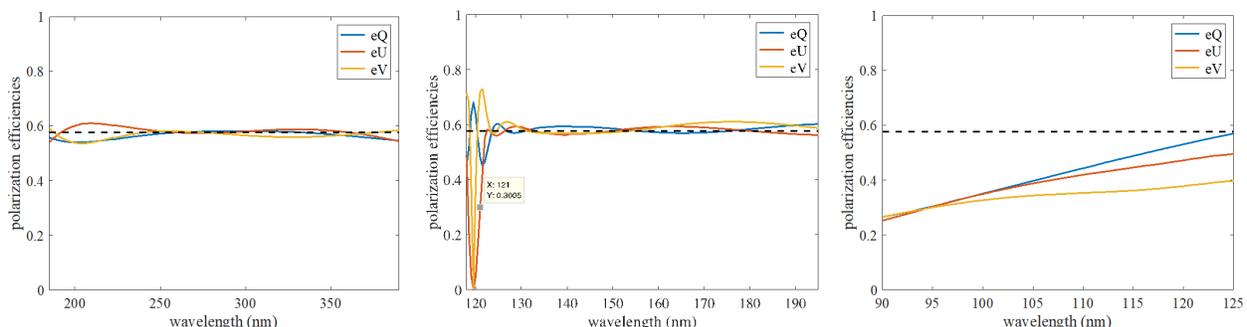

**Figure 13-21.** *Polarization efficiencies for the NUV (left), MUV (middle), and FUV (right) polarimeters. Results are excellent (close to the optimum 57.7% value) for the NUV and MUV except at wavelengths around 120 nm which will be covered in the FUV channel; in the FUV channel, the efficiency drops towards shorter wavelengths with the current design but we expect improvement following current developments on materials (**Appendix H.3**).*

**Appendix H.6** presents graphics illustrating the assumptions above, as well as more details about the performances of POLLUX in terms of spectral resolution or comparison to HST/COS. **Appendix H.7** presents the simulator of the instrument that is available upon request.

### 13.2.9  Required resources
The budget of the resources has been evaluated at the level of each component of POLLUX. The full detail is presented in **Appendix H.4**. **Table 13-4** summarizes the volume, mass, power, telemetry budgets.

**Table 13-4.** *Summary of main quantities budget for POLLUX*

|  | Allocation | Current Best Estimate |
|---|---|---|
| Mass (kg) | 449 | 375 |
| Volume (m$^3$) | 30 | 5.2 |
| Power (W) | 450 | 130 |
| Telemetry (Gb/day) | 350 | < 100 |

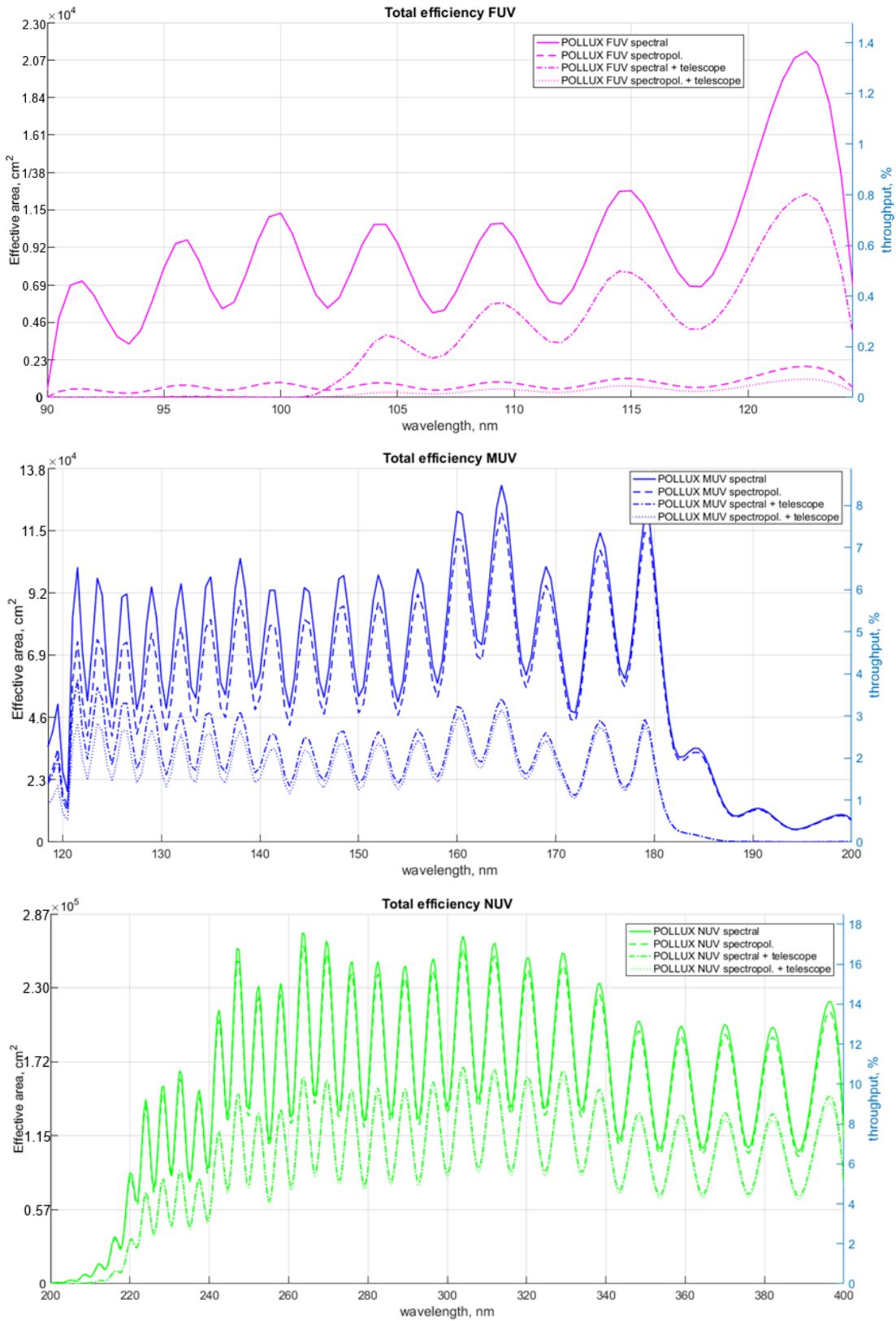

**Figure 13-22.** *Throughput and effective area of POLLUX channels. See text for underlying assumptions.*

### 13.2.10 AIT/AIV

The focal plane instrumentation AIT/AIV program is the responsibility of the POLLUX consortium and the integration of this with the telescope will be carried out in the USA. The AIV process for the instrument will center on a model build philosophy that specifies four levels of development (plus a spare):

- Structural thermal model (STM)—replicates opto-mechanical mountings, subsystem structures, mechanical interfaces and masses, thermal hardware and controls, representative heat dissipations
- Engineering model (EM)—replicates electrical systems, sensors and interfaces
- Qualification model (QM)—full optical bench, electrical systems and mechanical interfaces
- Flight model (FM)—full flight ready system— only exposed to acceptance testing
- Flight spare—hold stock of critical and long-lead qualified flight spare subsystems, replace on any failure.

Key interfaces for the AIV program will be with all subsystems, which will specify functional and calibration test parameters for the AIV program. The model build philosophy allows parallel development programs, to address structural and electrical functionality separately, before combining these in a qualification build. The qualification model will be subject to full examination of the boundaries of qualification regarding, in particular, vibration and shock levels, temperature extremes, and electro-magnetic exposure. The FM will be exposed to lower acceptance / workmanship level to avoid over stressing critical sensitive elements of the payload. We will operate a subsystem flight spare policy to allow rapid change out of any items that fail during test. Overall this approach is easier to schedule and minimizes the impact of technical/build problems on the focal plane instrument delivery schedule. This also optimizes the program cost by avoiding the need to build a complete flight spare focal plane.

### 13.2.11 Contamination and cleanliness

The spectral range of the POLLUX mission covers the far-UV through to the near-UV. At these wavelengths POLLUX will be particularly sensitive to molecular contamination of the main telescope optics and other optical components within the polarimeters and spectrographs. Therefore, stringent contamination control procedures required for the UV will drive the overall mission requirements. The main risk from contamination is a loss of sensitivity, through reductions in transmission or transparency. Procedures will need to be applied equally to all subsystems, not just optical ones, to avoid contamination since such material can easily be transferred to sensitive elements in the vacuum that will be established in testing and flight within the instruments.

The cleanliness program will apply at subsystem and instrument levels and must be established from the design stage through to AIT/AIV. A number of design rules must be applied. For example, rigid control of materials selection will be a requirement (according to ECSS-Q-ST-70, ranging from wire harnesses and printed wiring boards to epoxies), to minimize outgassing. These will need to be prepared with a careful cleaning and bake-out (where possible) before assembly. Electrical components should be separated from optical housings. Similar considerations will need to be applied to the design of test equipment such as vacuum and other calibration facilities. Use of oil/grease free pumping and air systems is absolutely essential.

Furthermore, the instrument must be integrated in an environment with a very low level of humidity and particle and molecular contamination. Even with the most stringent controls, it is almost impossible to completely avoid contamination. Therefore, in-flight mitigation mechanisms will be required, implementing contaminant traps or heaters to provide temperature control of sensitive parts to prevent accumulation or redistribution of contaminants or drive off accumulated material if it occurs.

### 13.2.12 POLLUX and LUVOIR-B

POLLUX as presented here is designed for LUVOIR-A architecture. No plans to build POLLUX for LUVOIR-B have been considered. The first and most obvious reason for this is that LUVOIR-B is designed to have a bay hosting 3 instruments only, which are already occupied by ECLIPS-B, LUMOS-B, and HDI-B. Furthermore, the collecting area of LUVOIR-B is much smaller than LUVOIR-A (43.8 m² versus 155 m²), and the SNR in the FUV channel of a putative POLLUX-B would be so low that the FUV channel (and likely part of the MUV channel) would have to be dropped, removing several science cases altogether.

Several other instrumental reasons imply that significant parts of the science cases would become unfeasible with LUVOIR-B. The strongest issue is the off-axis telescope architecture of LUVOIR-B, which makes the demodulation process much less precise, hence further eliminating science cases that require very high polarization accuracy. POLLUX-B would thus not significantly add science to LUMOS-B beyond polarimetric capability. For all these reasons, and in consultation with the LUVOIR team, we therefore suggest a **polarimetric module** (delivered by Europeans) for LUMOS-B rather than POLLUX-B. This would enhance the science delivered by LUMOS-B, and allow LUVOIR to maintain some polarimetric capacity.

### 13.3 Technology development

The high-level science requirements of POLLUX drove the design of the instrument. The technological solutions adopted in the present design have different levels of maturity (TRL). In this section, we discuss the technology development we have devised to reach a TRL = 6 by the end of the PDR, which we assume would be in 2025. The solution path to go from today's TRL to TRL =6 are summarized in **Table 13-5**.

**Table 13-5.** *POLLUX components technology status and path ahead. Green shaded options are in the POLLUX baseline. Yellow options indicate alternative or enhancing solutions given enough development.*

| Element/ component | Driving requirements | Technical challenges | Solutions paths | Development status and path ahead | | | | Notes |
|---|---|---|---|---|---|---|---|---|
| | | | | Current TRL | Path to TRL 4 | Path to TRL 5 | Path to TRL 6 | |
| Dichroic | **Reflectance**: > 60% in MUV **Transmittance**: > 85% in NUV | Thickness control Material deposition Adherence Interfaces | Multilayer AlF3, LaF3, LiF | 2 | Validation of fabrication with chosen materials | Make sample in flight-like configuration | Full qualification of flight prototype device | R&D (REOSC, Zeiss) GALEX heritage |

| Element/component | Driving requirements | Technical challenges | Solutions paths | Development status and path ahead | | | | Notes |
|---|---|---|---|---|---|---|---|---|
| | | | | Current TRL | Path to TRL 4 | Path to TRL 5 | Path to TRL 6 | |
| FUV coating | Reflectance as high as possible, 50% and above | Identification of a coating meeting the > 50% requirement, and check space qualification | Si | >6 | ✓ | ✓ | ✓ | SiC is at TRL=9 (FUSE heritage) with efficiency > 45% |
| MUV & NUV coatings | Reflectance as high as 90% over the MUV+NUV spectral range | high-uniformity coatings (< 1%), low polarization, and stability | Al/eLiF, Al/AlF3, Al/LiF/MgF2, Al/LiF/AlF3 | 3 | Coating deposition in laboratory | Coating validation in vacuum, with relevant radiation conditions | Full qualification of flight prototype device | Dual-layer coatings are already qualified. |
| FUV polarimeter | Mirrors with high-reflectance and retardance, and high incidence Brewster angle | Identification of a proper coating + characterization of its polarimetric properties | SiC, ta-C and B4C | 3 | Ongoing experience at LATMOS | Ongoing tests at LATMOS | Build full prototype in selected coating | See **Appendix H-3** for more details |
| MUV polarimeter | Efficient separation of polarized beams + efficiency of polarization extraction at 57.7% | Thin plates of MgF2 and resistance to thermal changes | Thermal cycling tests | 4 | ✓ | Ongoing tests at Paris Observatory | Cubesat experiment CASSTOR led by CNES | HINODE and CLASP heritage, and ARAGO study. See **Appendix H-3** for more details |
| | Mirrors with high-reflectance and retardance, and high incidence Brewster angle | Identification of a proper coating + characterization of its polarimetric properties | SiC, ta-C and B4C | 3 | Ongoing experience at LATMOS | Ongoing tests at LATMOS | Build full prototype in selected coating | See **Appendix H-3** for more details |
| NUV polarimeter | Efficient separation of polarized beams + efficiency of polarization extraction at 57.7% | Thin plates of MgF2 and resistance to thermal changes | Thermal cycling tests | 4 | ✓ | Ongoing experiences at Paris Observatory | Cubesat experiment CASSTOR led by CNES | HINODE and CLASP heritage, and ARAGO study. See **Appendix H-3** for more details |
| FUV echelle grating | **Frequency (1/mm)**: 590 **Blazing angle**: 43d **Max diagnonal dimension:** 154.7 × 210.6 mm | | | | | Make full format grating on a flight-like substrate + coating + optimization of groove profiles | Full qualification of flight prototype device | See R. McEnaffer's (PSU) work. Test gratings flew on sounding rockets |
| MUV echelle grating | **Frequency (1/mm)**: 290 **Blazing angle**: 59d **Clear aperture:** 87.4 × 167.4 mm | Wafer/write size Blaze angle accuracy Groove roughness | Electron beam lithography + ion etching on Si wafer | 4 | ✓ | | | |
| NUV echelle grating | **Frequency (1/mm)**: 134 **Blazing angle**: 61d **Clear aperture:** 81.5 × 165.2 mm | | | | | | | |

| Element/ component | Driving requirements | Technical challenges | Solutions paths | Development status and path ahead | | | | Notes |
|---|---|---|---|---|---|---|---|---|
| | | | | Current TRL | Path to TRL 4 | Path to TRL 5 | Path to TRL 6 | |
| FUV Cross-disperser | **Frequency (1/mm):** 272.6 **Clear aperture:** 416.1 × 165.4 mm **Asphericity (μm):** 2.4/3.7 | Concave freeform holographic Wafer/write size | | | ✓ | Manufacture and test the gratings with proper coatings, relevant size, and groove sharpness | Full qualification of flight prototype device | Planned R&D in collaboration with Jobin-Yvon HORIBA (France) |
| MUV Cross-disperser | **Frequency (1/mm):** 212.3 **Clear aperture:** 215.4 × 98.3 mm **Asphericity (mm):** 2.27/3.36 | | | | | | | |
| NUV Cross-disperser | **Frequency (1/mm):** 105.4 **Clear aperture:** 216 × 99.5 mm **Asphericity (μm):** 2.15/3.21 | | | | | | | |
| FUV detector | **Max. diagonal dimension:** 180 mm **Quantum Efficiency:** > 50% | Sensitivity Dynamic range Array size (8" wafer) Radiation hardness | delta-doped EMCCD | 4 | ✓ | Production of detector prototype | Full qualification of detector prototype + FEE subsystem | Sounding rocket heritage devices |
| | | | UV-enhanced CMOS | 4 | ✓ | | | |
| MUV detector | **Max. diagonal dimension:** 180 mm **Quantum Efficiency:** > 50% | Sensitivity Dynamic range Array size (8" wafer) Radiation hardness | delta-doped EMCCD | 4 | ✓ | Production of detector prototype | Full qualification of detector prototype + FEE subsystem | Sounding rocket heritage devices |
| | | | UV-enhanced CMOS | 4 | ✓ | | | |
| NUV detector | **Max. diagonal dimension:** 180 mm **Quantum Efficiency:** > 50% | Sensitivity Dynamic range Array size (8" wafer) Radiation hardness | delta-doped EMCCD | 4 | ✓ | Production of detector prototype | Full qualification of detector prototype + FEE subsystem | Sounding rocket heritage devices. |
| | | | UV-enhanced CMOS | 4 | ✓ | | | |

## Dichroic

The MUV and NUV channels are separated by means of a dichroic splitter. According to a study performed by Safran REOSC, the dichroic splitter can achieve RMS transmittance in the NUV wavelength range of 89.82% with a maximum of 94.87% (see **Figure 13-23**). The reflectance in the MUV wavelength range is 58.02% RMS, and 69.45% at the maximum. The development for this component foresees to further increase the efficiency in both channels of POLLUX will be done in partnership with the REOSC company.

## Coatings

A CNES research & development (R&D) study to obtain and select coatings with excellent UV and FUV efficiency, particularly in the 90 to 400 nm range, started in October 2017. The goal is to reach above 90% for as much as possible of this spectral range, high-uniformity (< 1%), low polarization (< 1%), and pre-launch stability of ultra-thin coatings.

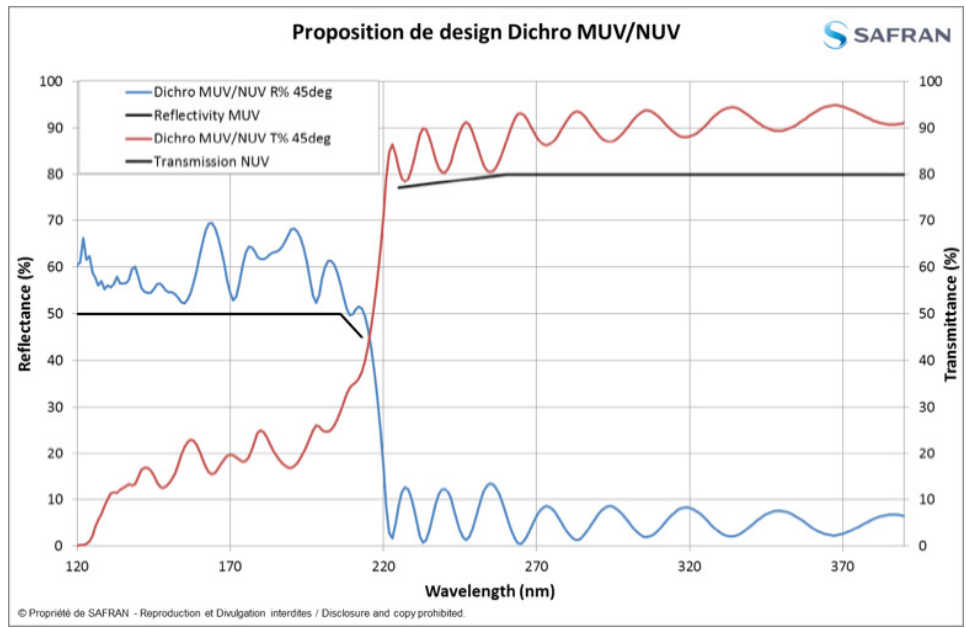

**Figure 13-23.** *Design of a dichroic for MUV/NUV spectral range.*

Materials with best perspectives are $MgF_2$, LiF, and $AlF_3$. Single and double layers of these materials have been investigated, e.g., Al/eLiF, Al/$AlF_3$, Al/LiF/$MgF_2$, Al/LiF/$AlF_3$, and combinations with different process conditions. At the short wavelength edge, presently a top coating with $AlF_3$ gives the best reflectance results, as published most recently by Del Hoya and Quijada (NASA GSFC). A reflectance of more than 30% above 100 nm wavelength and more than 90% from 120 to 130 nm has been reported.

**Polarimeters**
Two types of polarimeter designs were considered for POLLUX. For the longest UV wavelengths, the challenges reside in demonstrating the resistance of stacks of thin plates for space applications. For the shortest UV wavelengths, the choice of material for the mirrors will dominate the performances. An R&D funded by CNES is on-going to study these points. See **Appendix H.3** for more details.

**Echelle gratings**
In the current design, the MUV and NUV echelle gratings have almost identical sizes and blaze angles. The groove frequencies differ approximately by a factor of two. Both the groove frequencies and the angles are non-standard. For the FUV echelle grating, the groove frequency and blazing angle are even more unusual. We have no evidence that such a grating has ever been produced for the UV domain. For the three channels, however, feasibility is in the range of today's technological limits (communication from Dr. Randall McEntaffer, Penn State University (PSU); see **Appendix H.8**). Echelle grating prototypes for the MUV channel manufactured by PSU have been tested (mainly reflected efficiency) in May 2018 at the facilities of the Center for Astrophysics and Space Astronomy in Colorado. Intrinsic reflectivity (i.e., coating independent) is above 70% throughout the MUV wavelength range, with a maximum of ~94% near 160 nm (Grisé et al. in preparation). Other open issues include the accuracy of the groove profiles and the coating optimization for each of the channels.

Another high technological risk is the clear aperture of the gratings. Today's process tools are limited to 200 mm diameter wafers (~140x140 mm clear aperture), so at least one of the dimensions on each of these gratings is a challenge (e.g., Milles et al. 2018, 2018, *ApJ*, 869, 95). These points will be addressed in the framework of an R&D plan between PSU and Colorado University, and the POLLUX consortium.

**Cross-dispersers**

Each of the cross-dispersers of the present design is a concave freeform holographic grating on a spherical substrate. This is a novel type of optical elements having high aberration correction capabilities. For the cross-dispersers, computations show that a sufficient aberration correction is possible only if the spectral components at the cross-disperser's surface are separated, although it leads to an increase in aperture. An R&D project has been set up in collaboration with Jobin-Yvon HORIBA (France) to study the feasibility of the required holographic recording geometry parameters, develop and test prototypes. According to current technology, no show stopper has been identified.

**Detectors**

The technology of $\delta$-doped EMCCDs is not fully mature. More R&D is required to further demonstrate the dynamic range and how low spurious noise is, in a realistic, end-to-end environment for spectroscopy. We need to assess how the detectors behave around and below Lyman-$\alpha$ for instance. Implementation of antireflection coatings, or of metal-dielectric stacks as visible rejection filters must be improved. A last point will be to demonstrate feasibility of detector wafers large enough to accommodate our needs (typical detector size is 15k × 2k for POLLUX). This size appears achievable in the coming years (communication from Dr. Shouleh Nikzad, JPL). As a promising alternative, we will also consider CMOS, $\delta$-doped for enhanced UV performance (see **Appendix H.9**). These devices are rapidly developing and offer larger formats than CCD devices. Further developments are needed to reach sub-electron read noise throughout their pixels, in which case they could represent a credible alternative for POLLUX. These are specific studies that will be led by Leicester Univ. (UK).